\documentclass[12pt,english,letterpaper]{kuthesis}
\usepackage{amsmath}
\usepackage{mathptmx}

\usepackage[T1]{fontenc}
\usepackage[utf8]{inputenc}
\usepackage{geometry}
\usepackage{color}
\usepackage{babel}
\usepackage{url}
\usepackage{graphicx}
\usepackage{setspace}
\usepackage{esint}
\usepackage{tikz}
\usepackage{tikz-feynman}
\usepackage{feynmp-auto}
\usepackage{rotating}
\usepackage[unicode=true,
 bookmarks=true,bookmarksnumbered=false,bookmarksopen=false,
 breaklinks=true,pdfborder={0 0 0},pdfborderstyle={},backref=false,colorlinks=true]
 {hyperref}
\hypersetup{pdftitle={University of Kansas Thesis Template},
 pdfauthor={Anonymous},
 pdfsubject={A Thesis},
 urlcolor={blue},citecolor={blue},allcolors={blue}}

\DeclareMathAlphabet{\mathcal}{OMS}{cmsy}{m}{n} 
\DeclareSymbolFont{newfont}{OML}{cmm}{m}{it}
\DeclareMathSymbol{\Epsilon}{3}{newfont}{15}
\renewcommand{\epsilon}{\Epsilon}

\makeatletter

\providecommand{\LyX}{\texorpdfstring%
  {L\kern-.1667em\lower.25em\hbox{Y}\kern-.125emX\@}
  {LyX}}

\newcommand {\beq} {\begin{equation}}
\newcommand {\eeq} {\end{equation}} 
\usepackage{float}
\usepackage{textcomp}
\usepackage{multirow}
\usepackage{multicol}
\usepackage[toc]{glossaries}
\usepackage[toc,page]{appendix}

\usepackage{dcolumn}
\usepackage{booktabs}
\usepackage{lineno}

\def \SUc {\text{SU}\!(3)_\text{C}}
\def \SUl {\text{SU}\!(2)_\text{L}}
\def \Uy {\text{U}\!(1)_\text{Y}}
\def \sqe {\sqrt{s_\text{E}}}
\def \alr {A_{\mathrm{LR}}}
\def \ee {\mathrm{e}^{+}\mathrm{e}^{-}}
\def \electron {\mathrm{e}^{-}}
\def \positron {\mathrm{e}^{+}}

\def \suf {SU(3)_\text{F}}

\def \invfb {\mathrm{fb}^{-1}}
\def \invab {\mathrm{ab}^{-1}}

\def \fccee {\mathrm{FCC}\text{-}\mathrm{ee}}
\def \sqp {\sqrt{s}_\mathrm{p}}

\def \eemmg { \mathrm{e}^{+}\mathrm{e}^{-} \to \mu^{+} \mu^{-} (\gamma)}

\def \tinn {\theta_\text{inner}}

\def \alr {A_\text{LR}}
\def \afb {A_\text{FB}}
\def \gv {g_\text{V}}
\def \ga {g_\text{A}}
\def \ee {\mathrm{e}^{+}\mathrm{e}^{-}}
\def \electron {\mathrm{e}^{-}}
\def \positron {\mathrm{e}^{+}}

\def \eemmg { \mathrm{e}^{+}\mathrm{e}^{-} \rightarrow \mu^+ \mu^- (\gamma)}

\def \invfb {\mathrm{fb}^{-1}}
\def \invab {\mathrm{ab}^{-1}}

\def \g1z {g_{1}^{\mathrm{Z}} }

\def \p12mag {|\vec{p}_{12}|}

\title{Integrated Luminosity with 100 ppm Precision, Methods for $\sqrt{s}$ Precision of 1 ppm, and Beyond Standard Model Sensitivity using Photonic Events, at $\mathrm{e^{+}e^{-}}$ Higgs Factories}
\author{Brendon Cory Madison}

\priorcreds{B.Sc. Pre-Professional Physics, University of Kansas, 2017}
\priorcreds{M.Sc. Physics, University of Kansas, 2020}

\dept{Department of Physics and Astronomy}
\degreetitle{Doctor of Physics}
\papertype{Dissertation} 

\committee{Graham W. Wilson}{Jin Feng}{K.C. Kong}{Jenny List}{Chris Rogan}{}{}
\role{Chairperson}{External Review}{}{DESY}{}{}{}

\@printd@testrue
\datedefended{May 7, 2025}
\dateapproved{June 4, 2025}

\@ifundefined{showcaptionsetup}{}{%
 \PassOptionsToPackage{caption=false}{subfig}}
\usepackage{subfig}
\makeatother

\usepackage{listings}

\makeglossaries

\newglossaryentry{DiLepton}
{
    name=dilepton,
    description={Two same flavor leptons that are associated with each other. As in, two of $e^\pm$, $\mu^\pm$ or $\tau^\pm$}
}

\newglossaryentry{dielectron}
{
    name=dielectron,
    description={Two electrons, $e^\pm$, usually used when they are associated with each other. Unlike Bhabhas it includes other methods of production}
}

\newglossaryentry{DiFermion}
{
    name=difermion,
    description={Two fermions, $f\bar{f}$, usually used when they are associated with each other}
}

\newglossaryentry{recoil mass}
{
    name=recoil mass,
    description={Estimate of the invariant mass of the recoiling particle(s) from the particles that have been measured. Usually under the assumption that the original system was in the center-of-mass frame and not boosted. Used in measuring Higgs in the $ZH$ process}
}

\newglossaryentry{FTD}
{
    name=FTD,
    description={Forward Tracker Detector (FTD) is a sub-detector of a collider detector that is typically placed at low polar angles, the forward region, and close to the beam pipe. It is common for the FTD to start close to the interaction point}
}

\newglossaryentry{LHC}
{
    name=LHC,
    description={Large Hadron Collider (LHC) is proton-proton collider situated at CERN outside of Geneva, Switzerland.}
}

\newglossaryentry{SUSY}
{
    name=SUSY,
    description={Supersymmetry (SUSY) is an extension of the Standard Model that introduces a new quantum number, R-parity, that characterizes the commutation relationships of particles. This causes super-partners and super-particles, with $R_\text{p}=-1$ of the Standard Model particles, with $R_\text{p}=+1$, and leads to lots of new physics}
}

\newglossaryentry{SM}
{
    name=SM,
    description={The Standard Model of particle physics (SM) is a collection of theories that have been, in many ways, exhaustively proven to high precision by experimental evidence. Underlying theories can be summarized as Quantum Chromodynamics (QCD), Quantum Electrodynamics (QED), Glashow-Salam-Weinberg Electroweak Theory (GSW EW) and the Higgs mechanism}
}

\newglossaryentry{THDM}
{
    name=THDM,
    description={Two-Higgs-Doublet Model (THDM) is a fairly simple extension of the Standard Model to include an additional Higgs boson that mixes with the existing Higgs boson. Thus adding an additional Higgs field and parameters for the mixing of the two fields}
}

\newglossaryentry{GEANT4}
{
    name=GEANT4,
    description={The GEometry ANd Tracking (GEANT) software has long been used for the effects particles have on materials as they traverse them~\cite{GEANT4}. It is in its 4th iteration. It is primarily used for the simulation of particle detectors but also applications of particle physics, such as nuclear medicine}
}

\newglossaryentry{MIP}
{
    name=MIP,
    description={Minimum Ionizing Particle, or MIP, is typically referred to as an energy of which is the minimum amount of energy deposited in a material by a charged particle passing through it}
}

\newglossaryentry{MSSM}
{
    name=MSSM,
    description={The Minimally Super-symmetric Standard Model (MSSM) is an extension of the Standard Model to include super-fields and super-partners (sparticles) of existing Standard Model fields and particles}
}

\newglossaryentry{GP}
{
    name=GuineaPig++,
    description={GuineaPig, which was then upgraded to use C++ and thus GuineaPig++, is a general purpose lepton beam collision simulation software~\cite{Guinea-PIG}. It is mostly used for calculating instantaneous beam luminosities but can also be used for creating differential beam luminosity, simulating beam pairs, simulating beam deflection}
}

\newglossaryentry{BHLUMI}
{
    name=BHLUMI,
    description={BHLUMI is software for generating small-angle Bhabha scattering events with specific focus on the precision theory developments relevant to small-angles~\cite{BHLUMI}. It was developed during LEP to aid in the integrated luminosity measurement and to provide a small-angle Bhabha scattering event generator}
}

\newglossaryentry{MC}
{
    name=MC,
    description={A Monte Carlo (MC) scheme is one where statistical distributions and randomness are used to solve a physics problem or simulate a physics scenario}
}

\newglossaryentry{WHIZARD}
{
    name=WHIZARD,
    description={An event generator for \textbf{W}, \textbf{Hi}ggs, \textbf{Z}, \textbf{a}nd \textbf{r}espective \textbf{d}ecays, or WHIZARD, is a general purpose MCEG used by numerous experiments in high energy physics. WHIZARD can be found \href{https://whizard.hepforge.org/}{here}}
}

\newglossaryentry{dimuon}
{
    name=dimuon,
    description={Two muons, $\mu^\pm$, usually used when they are associated with each other}
}

\newglossaryentry{FSR}
{
    name=FSR,
    description={Final State Radiation (FSR) is when a radiative correction, typically a photon, is emitted by a charged final state particle. Since radiative corrections scale with the mass of the emitting particle, a final state of particles more massive than $\ee$ will emit FSR that is typically lower energy than the ISR of the $\ee$}
}

\newglossaryentry{DiTau}
{
    name=ditau,
    description={Two taus, $\tau^\pm$, usually used when they are associated with each other}
}

\newglossaryentry{diphoton}
{
    name=diphoton,
    description={Two photons, $\gamma\gamma$, usually used when they are associated with each other and their energy is comparable to the beam energy or center-of-mass energy}
}

\newglossaryentry{ISR}
{
    name=ISR,
    description={Initial State Radiation, refers to a type of radiative correction where photons can be emitted by the initial state particles before undergoing their higher energy `hard' interaction}
}

\newglossaryentry{Bhabha}
{
    name=Bhabha,
    description={Final state dielectron coming from t-channel interaction with a initial state dielectron}
}

\newglossaryentry{tchan}
{
    name=$\text{t-channel}$,
    description={Indicating a process came from a Feynman diagram that would use Mandelstam $t$. Generally refers to scattering processes, which generally increase in cross-section with energy}
}

\newglossaryentry{schan}
{
    name=$\text{s-channel}$,
    description={Indicating a process came from a Feynman diagram that would use Mandelstam $s$. Generally refers to annihilation/creation processes, which generally decrease in cross-section with energy}
}

\newglossaryentry{neutral}
{
    name=$X^0\gamma$,
    description={Also known as neutral events with photons. Explained in more depth in subsect~\ref{subsec-Simulating}}
}

\newglossaryentry{GLIP}
{
    name=GLIP,
    description={Granular Long Instrument for Precision (GLIP) is a proposed forward calorimeter for future collider experiments}
}

\newglossaryentry{LumiCal}
{
    name=LumiCal,
    description={Refers to a luminosity calorimeter though it may refer to specifically the ILD's Luminosity Calorimeter (LumiCal) too}
}

\newglossaryentry{ILD}
{
    name=ILD,
    description={International Large Detector, a proposed detector for ILC, though there exist proposals to deploy equivalent detectors at FCC-ee and CLIC. It is considered one of the most, if not the most, mature of the Higgs factory detector proposals}
}

\newglossaryentry{ECAL}
{
    name=ECAL,
    description={Electromagnetic Calorimeter (ECAL) is a type of particle calorimeter that focuses on measuring the electromagnetic component of a particle shower as well as possible}
}

\newglossaryentry{HCAL}
{
    name=HCAL,
    description={Hadronic Calorimeter (ECAL) is a type of particle calorimeter that focuses on measuring the hadronic component of a particle shower as well as possible. Typically bulkier than ECAL due to how hadronic showers differ from electromagnetic showers}
}

\newglossaryentry{TPC}
{
    name=TPC,
    description={Time Projection Chamber (TPC) is a type of particle tracker that is well regarded for its momentum resolution. Used in ILD as the main tracker}
}

\newglossaryentry{ILC}
{
    name=ILC,
    description={International Linear Collider, a proposed future $e^+e^-$ linear collider and Higgs factory}
}

\newglossaryentry{CLIC}
{
    name=CLIC,
    description={Compact Linear Collider (CLIC), a proposed future $e^+e^-$ linear collider and Higgs factory that uses plasma wakefield accelerator technology to allow for a more compact design. This also allows for potential Multi-TeV upgrade}
}

\newglossaryentry{TESLA}
{
    name=TESLA,
    description={TeV–Energy Superconducting Linear Accelerator (TESLA) was a proposed linear $e^+e^-$ collider that was to be built in Germany. Much of the collaboration became the ILC collaboration after efforts for constructing TESLA failed}
}

\newglossaryentry{DESY}
{
    name=DESY,
    description={Deutsches Elektronen-Synchrotron (DESY) is a research facility in Germany that is responsible for funding and maintaining numerous physics researchers}
}

\newglossaryentry{XFEL}
{
    name=XFEL,
    description={X-ray Free Electron Laser (XFEL) is a type of x-ray laser source that uses electrons and magnetic fields, typically undulators, to generate high luminosity and short time length bunches of x-rays. This leads to incredibly high energy density x-ray pulses that allow for investigation of the strong-field regime of QED}
}

\newglossaryentry{ERL}
{
    name=ERL,
    description={Energy Recovery Linac (ERL) are linear accererators, and/or colliders, that are able to recycle either the colliding particles or their energy so that the overall power efficiency of the experiment is improved}
}

\newglossaryentry{ReLiC}
{
    name=ReLiC,
    description={Recycling $e^+e^-$ Linear Collider (ReLiC) is a proposed ERL that plans to recycle both the electrons and positrons and their energies so that lower power is used by the facility and so that the experiment can achieve higher instantaneous luminosity}
}

\newglossaryentry{SMEFT}
{
    name=SMEFT,
    description={Standard Model Effective Field Theory (SMEFT), which is an extension of the Standard Model of particle physics that allows for examining potential new physics by treating new vertices (interactions) with their own new parameters. Instead of the alternative, creating a comprehensive field theory and deriving or constraining the parameters in some way}
}

\newglossaryentry{BSM}
{
    name=BSM,
    description={Beyond Standard Model (BSM) is a term used to describe new physics, be it measured or theoretical, that either cannot be described by the existing Standard Model of particle physics or is not described well by existing Standard Model calculations
    }
}

\newglossaryentry{IP}
{
    name=IP,
    description={Interaction point (IP) is the point where beams in a collider experiment cross and, as a result, collisions between the beam particles occur
    }
}

\newglossaryentry{EW}
{
    name=EW,
    description={Electroweak, or EW for short, is used to describe the high energy regime in which the photon, Z boson, $\text{W}^{\pm}$ bosons and Higgs boson all contribute in significant ways to the possible physics interactions
    }
}

\newglossaryentry{Snowmass}
{
    name=Snowmass,
    description={Snowmass process, a high energy physics conference named after the first place it was convened in, Snowmass, Colorado. The papers and presentations of Snowmass are generally considered important to the planning of the future of the field}
}

\newglossaryentry{FCC}
{
    name=FCC,
    description={Future Circular Collider, a proposed future $e^+e^-$ circular collider with an expected upgrade to a hadron circular collider. To be built near Geneva, Switzerland}
}

\newglossaryentry{CEPC}
{
    name=CEPC,
    description={Circular Electron Positron Collider (CEPC) is a proposed $e^+e^-$ circular collider and Higgs factory to be built in China.}
}

\newglossaryentry{HLLHC}
{
    name=HL-LHC,
    description={High Luminosity Large Hadron Collider (HL-LHC) is the high luminosity (high data rate) upgrade to the LHC that is expected to start as early as 2028 and take as much as 3 $\text{ab}^{\text{-1}}$ of data}
}

\newglossaryentry{LEP}
{
    name=LEP,
    description={Large Electron-Positron Collider (LEP) was a circular electron-positron collider that used the same beam tunnel as the LHC. LEP primarily ran below 200~GeV center-of-mass energy and precisely measured the Z and W bosons}
}

\newglossaryentry{SLC}
{
    name=SLC,
    description={Stanford Linear Collider (SLC) was a polarized linear electron-positron collider that was built at SLAC. It had overlap with LEP in terms of energies and measurements but operated with much lower luminosity}
}

\newglossaryentry{SLD}
{
    name=SLD,
    description={The detector of the SLC experiment}
}

\newglossaryentry{WIMP}
{
    name=WIMP,
    description={Weakly Interacting Massive Particles (WIMPs) are the dark matter candidate that is preferred by the existing cosmological measurements and a core component of the prevailing $\Lambda \mathrm{CDM}$ model of cosmology}
}

\newglossaryentry{insLumi}
{
    name=$\text{instantaneous}$ $\text{luminosity}$,
    description={Refers to the luminosity of a collider experiment at one instant. Usually for one collision of one beam bunch with another beam bunch and has units of area/time, such as $\frac{\text{cm}^2}{\text{s}}$}
}

\newglossaryentry{ingLumi}
{
    name=$\text{integrated}$ $\text{luminosity}$,
    description={Refers to the luminosity of a collider experiment at over an entire run of taking data. Usually has units of inverse area (cross-section)}
}

\newglossaryentry{MCEG}
{
    name=MCEG,
    description={A Monte Carlo Event Generator (MCEG) is a simulation tool that is used to generate final start particles from their underlying scattering processes and initial state particles}
}

\newglossaryentry{SLHA}
{
    name=SLHA,
    description={The SUSY Les Houches Accord format is a formatting standard used for formatting beyond standard model parameter values, particularly from types of SUSY models, in a way that can be read and used by various software and people with minimal issues}
}

\newglossaryentry{GUT}
{
    name=GUT,
    description={A Grand Unified Theory, or GUT, is a theory that aims to describe all aspects of all fields of physics relevant to this universe and, in some cases, all universes. In the Standard Model of particle physics it is postulated that at a sufficiently high energy all of the fundamental forces become unified. This energy is typically referred to as the GUT energy}
}

\newglossaryentry{CMB}
{
    name=CMB,
    description={Cosmic Microwave Background (CMB) is the universal photon background originating from an event known as ``last scattering'' wherein photons decoupled from the rest of the thermal universe, fated to travel in the expanding universe and red-shift to increasingly longer wavelengths until arriving at the microwave regime that we measure them at today}
}

\newglossaryentry{SABS}
{
    name=SABS,
    description={Small Angle Bhabha Scattering (SABS) is the regime of Bhabha scattering ($e^+e^-\to e^+e^-$) where the scattering angle is small. This makes the $t$-channel dominant and the cross-section becomes large}
}

\newglossaryentry{LABS}
{
    name=LABS,
    description={Large Angle Bhabha Scattering (LABS) is the regime of Bhabha scattering ($e^+e^-\to e^+e^-$) where the scattering angle is large, comparable to 10 degrees. This makes all of the different contributions comparable and relevant}
}

\newglossaryentry{invFb}
{
    name=$\text{fb}^{\text{-1}}$,
    description={Known as inverse femtobarns. Unit of integrated luminosity. Equivalent to $10^{15}$ inverse barns or $10^{39}$ inverse centimeters}
}

\newglossaryentry{BERB}
{
    name=BERB,
    description={Beam-pipe Escape Radiative Bhabhas, or BERB for short, are Bhabha scattering events where the outgoing charged particles, electron and positron, escape reconstruction by going down the beam pipe. The event is then measured from any photon(s) that said charged particles emitted before escaping}
}

\newglossaryentry{HEP}
{
    name=HEP,
    description={High Energy Physics, or HEP for short, is a common short-hand for a field that encompasses much of particle physics, astro-particle physics and nuclear physics}
}

\newglossaryentry{BDT}
{
    name=BDT,
    description={Boosted Decision Tree (BDT) is a type of machine learning classification algorithm that is, in general, generated using a rule set of mathematical operations and conditions which are then used on data that is given to said algorithm. Due to this, it is closer to a mathematical robot than a brain.}
}

\begin{document}
\begin{romanpages}

\maketitle

\begin{abstractlong}

Future electron-positron ($\ee$) colliders, operating as Higgs factories or Z factories, promise unprecedented precision electroweak measurements that are vital to testing the Standard Model (SM) and exploring physics beyond it. Here we present work on the precision of integrated luminosity and center-of-mass energy, measurements that are needed to make future precision measurements possible. We also conduct these studies for the International Linear Collider (ILC) at center-of-mass energies ($\sqrt{s}$) from the Z pole ($m_{Z}$) to 1~TeV to provide a comprehensive study of these issues for future $\ee$ colliders. Multiple paths to 100 parts-per-million (ppm) precision on integrated luminosity are presented, with focus on small-angle Bhabha scattering (SABS) and two-photon production (diphotons, $\gamma\gamma$). Previous studies found that beam deflection of SABS events introduce biases on integrated luminosity of $10^{-2}$. To address this, we present a novel method that uses M\o{}ller scattering with SABS to measure beam deflection and minimize the effects of it on integrated luminosity. We present a proposal for a Highly Granular Luminosity Calorimeter, the GLIP LumiCal, and its design considerations. We demonstrate that the GLIP LumiCal can achieve $\approx35$ ppm precision on integrated luminosity precision with $\gamma\gamma$ for all values of $\sqrt{s}$ and that the current LumiCal design is insufficient for reaching 1000 ppm. Multiple methods for precision $\sqrt{s}$ estimation are presented. We introduce the use of Kernel Density Estimation (KDE) to ensure that $\sqrt{s}$ is measured accurately and precisely. The GLIP LumiCal allows for precise measurement of diphoton energies, adding an additional precision $\sqrt{s}$ measurement method and tool for precision energy scale calibration of the LumiCal. We find that $\gamma\gamma$ can achieve $\sqrt{s}$ precision of $\sim1$ ppm at various center-of-mass energies, assuming one calibrates the LumiCal to the $\sqrt{s}$ estimate of tracker dimuons. The utility of photon measurements is extended to photon with invisible ($X^0\gamma$) events where we demonstrate methods to measure left-handed and right-handed neutrino couplings and put constraints on Beyond Standard Model (BSM) physics.

\end{abstractlong}

\tableofcontents{}




\end{romanpages}

\chapter{Introduction}

\pdfoutput=1

High-energy physics (\Gls{HEP}) is proceeding to a phase transition, driven by the need to extend the frontiers of precision measurements and to explore physics beyond the Standard Model (\Gls{BSM}). The discovery of the Higgs boson at the Large Hadron Collider (\Gls{LHC}) marked another such transition, solidifying existing Large Electron-Positron Collider (\Gls{LEP}) and Stanford Linear Collider (\Gls{SLC}) measurements into a definitive new piece of \Gls{SM} physics. While the LHC and its high-luminosity upgrade (\Gls{HLLHC}) continue to probe the Higgs, their capability to perform precision measurements is constrained by the inherent complexity, noise, and sources of uncertainty inherent to hadronic colliders. To achieve the level of precision required for the next HEP phase, proposals for electron-positron Higgs factories have emerged as the successors to the LHC phase.

These Higgs factories, such as the proposed International Linear Collider (\Gls{ILC}) and Future Circular Collider (\Gls{FCC}), promise an unprecedented level of precision in luminosity and center-of-mass energy determination. These facilities aim to study Higgs boson couplings, electroweak interactions, and rare Standard Model processes with great accuracy; accurate measurements are made possible by leveraging the cleaner environment of $\ee$ colliders. These precision goals are contingent on innovative methodologies and technologies for measuring luminosity, calibrating beam energies, and the underlying systematic uncertainties. These efforts are not merely refinements of existing techniques done at LEP, SLC and LHC, but represent a fundamental change in how we approach these problems.

This work contributes to the broader effort of preparing for the next-generation Higgs factory by addressing key challenges in precision luminosity and energy measurements. By developing novel techniques for beam energy calibration and luminosity determination, this study lays the foundation for ensuring that future Higgs factories reach their ambitious precision targets. Such advancements will not only enable the most stringent tests of the Standard Model but also open new paths for discovering physics beyond our current theoretical framework. We present this chapter, and the next chapter, as providing further context and introduction.

\section{High Energy Physics Setting}

This section will discuss the current status of High Energy Physics (HEP) research and the Standard Model (SM) of particle physics. This is not meant to be a primer for thoroughly understanding the work that follows. To make this report on the current status of HEP more tangible we will focus on the experiments, and their measurements, that have set the current standards and understanding for HEP. This will not go over the theory of quantum field theory or the Standard Model beyond what is necessary to discuss this work. For a primer on quantum field theory, or for the most accurate and precise representation of high energy physics, we recommend other sources.

As of this work the majority of measurement standards for HEP have been set by the numerous international collaborations at either the Large Electron-Positron (LEP) collider or the Large Hadron Collider (LHC); both of which used the same beam tunnel located at Conseil Européen pour la Recherche Nucléaire (CERN) near Geneva, Switzerland~\cite{Myers:1991ym}~\cite{CERN:1991gja}. These experiments worked hand-in-hand for discovering the Higgs boson in 2012, with the precision measurements of LEP providing evidence that a particle like the Higgs existed at energies in the range of roughly 50 to 250 GeV~\cite{LEPHiggs}.

Previous to the age of LEP and LHC, the Tevatron at Fermilab, in the suburbs of Chicago, Illinois, discovered the top quark. Prior to the Tevatron, the Super Proton Synchrotron (SPS), again at CERN, had discovered the W and Z bosons. Going further back in time, the most massive charged lepton, the $\tau$ lepton, was discovered at Stanford Linear Accelerator Center (SLAC) outside of San Francisco, California. SLAC, particularly at the Stanford Linear Collider (SLC) and SLC Large Detector (SLD), also made contributions to Z boson measurements alongside LEP. All of these experiments shared a common goal of increasing energy and precision as a method of discovering new fundamental physics.

These measurements and experiments, prior to the LHC, were foundational to the current setting of HEP and where most of the current methods and personnel of the field started. Even though these experiments have taken place over the past five decades, the HEP field is still relatively immature. Only a handful of high energy collider experiments have been constructed and performed. Compared to other fields of science, where experiments can be iterated hundreds or thousands of times.

There are numerous reasons for this lack of iteration in HEP. Colliders are costly and consume large amounts of electricity, on the order of 1 billion USD and 100 MWhr, or 1TJ, of electrical energy~\cite{ZurbanoFernandez:2020cco}. Most of this does not go into ``interesting events'' and instead things like the facilities, personnel and generating the collider's instantaneous luminosity ($L$) and integrated luminosity ($\mathcal{L}$.) The instantaneous luminosity, which is a measure of collision areal density per unit time, is effectively the data rate of the experiment too. At LEP the luminosity was $\sim10^{31}$ $\frac{\mathrm{cm^2}}{\mathrm{s}}$ while at LHC the luminosity is $\sim10^{34}$ $\frac{\mathrm{cm^2}}{\mathrm{s}}$. Here we will not discuss the fine details of how the differential luminosity per center-of-mass energy, often called the luminosity spectrum, is different between lepton colliders and hadron colliders. We will note though that, in general, the luminosity spectrum at lepton colliders goes as $\sim\sqrt{s}$ while it goes as $\sim1/\sqrt{s}$ at hadron colliders, meaning that the majority of events at hadron colliders are at lower energies. Due to this and small cross-sections, similar to the probability of an event, the total number of ``interesting events'' at the LHC is in the thousands to millions~\cite{ATLAS:2022hro}. For example, the cross-section and luminosity expected thousands of Higgs boson events but the Higgs discovery papers only saw hundreds of events above the background~\cite{CMS:2012qbp}~\cite{ATLAS:2012yve}. The resources and labor for these events, and the colliders to make them, require collaboration across numerous sub-fields and, generally, approval and funding from numerous national governments. All of this is difficult and time consuming, especially compared to some fields of science where all that is needed is the proverbial ``leave no stone unturned''.



\subsection{The Standard Model and Gravity}

Here we discuss the prevailing model of high energy physics and particle physics, the Standard Model (SM). We include measurements and theories that are accessible in the near future, or have been considered as alternatives in the present or recent past. Since this work includes measurements done in the near future this inclusion is aspirational for what the status of the Standard Model may be at the time of execution of these measurements.

\begin{figure}[h]
\centering
\includegraphics[width=15cm,clip]{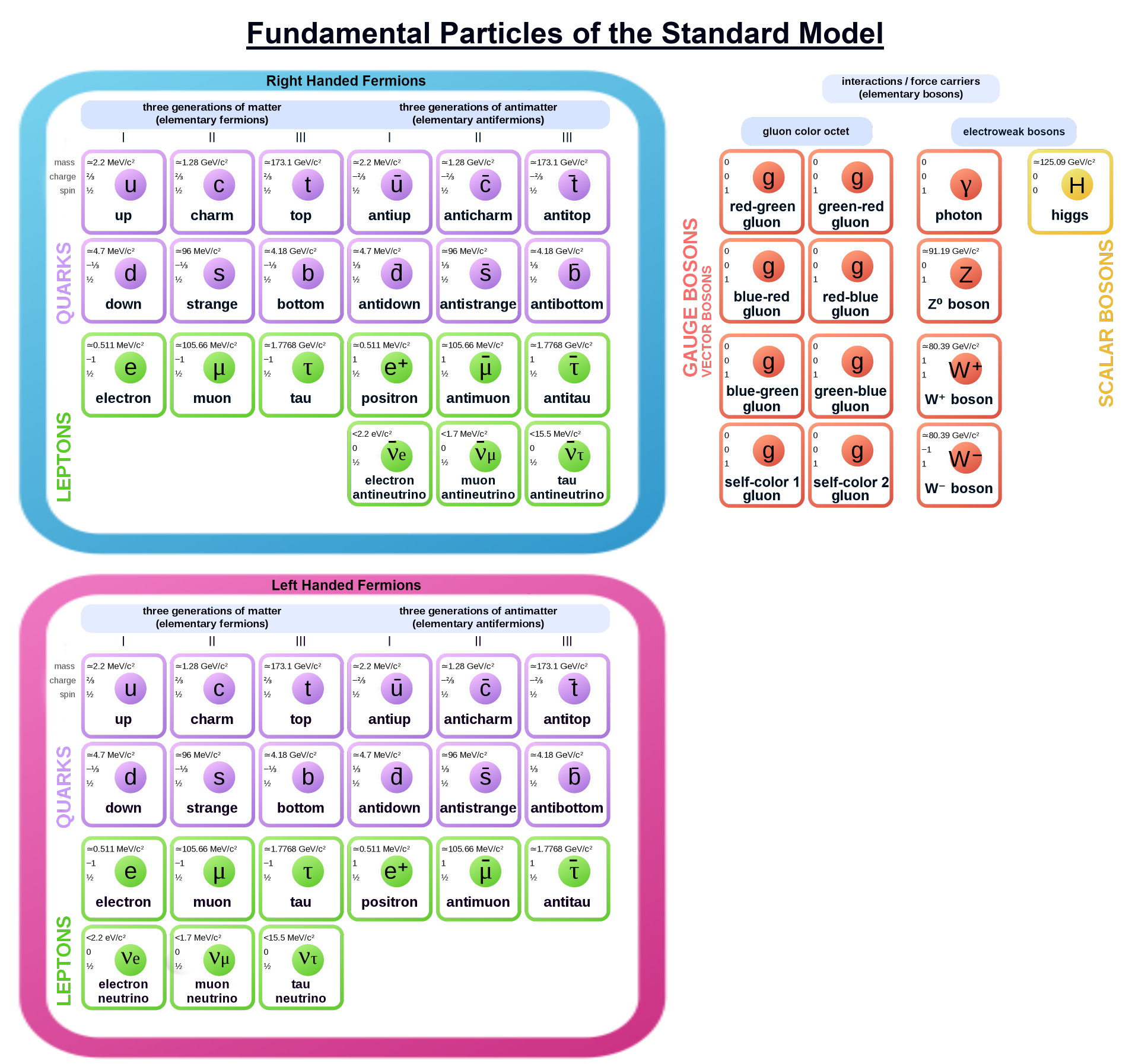}
\caption{Diagram showing the particles of the Standard Model as have been predicted and proven by experiment thus far. Note that particles that are plausible with minimal changes to the Standard Model, such as right-handed neutrinos or the colorless gluon state, are not shown.}
\label{fig-StdMod}       
\end{figure}

As seen in figure~\ref{fig-StdMod}, there are numerous particles in the Standard Model. Variations of this diagram get presented but here we have gone through the process of including all fermion and boson variants of the fundamental particles. Many of these particles share qualities and quantities. For example, the left-handed and right-handed variants of charged fermions, are thought to share mass, flavor, charge and spin states.

The vector bosons of the particle zoo of figure~\ref{fig-StdMod} are described in group theory with the $\SUc \times \SUl \times \Uy$ representation embedded within a general relativistic field theory. This representation is motivated by experimental results. Given that this is embedded within general relativity we expect that there should be a massless spin 2 boson. This is commonly known as the graviton but it has not been experimentally measured, in a way that would exhaustively establish its existence, so it is not in figure~\ref{fig-StdMod}. 

\subsubsection{Spacetime}\label{sec-space}

We note, as many presentations of the Standard Model in other work and theses repeat a common error, that the Standard Model is not separate from, or incompatible with, general relativity. This is repeated as many consider the quantum field theories to require flat spacetime. However, this is simply not true. The spacetime trajectory of charged particles in an electromagnetic field is curved in a way that is, by construction, indistinguishable from curved spacetime. Furthermore, this curvature from electromagnetism is additive to gravitational curvature. This additivity of curvature of both electromagnetism and gravity is why charged particles can be trapped in radiation belts above celestial bodies, like Earth. The spacetime trajectory of these particles becomes closed such that there is no longer a path that can escape without changing the external gravitational and/or electromagnetic fields.

\begin{figure}[h]
\centering
\includegraphics[width=10cm,clip]{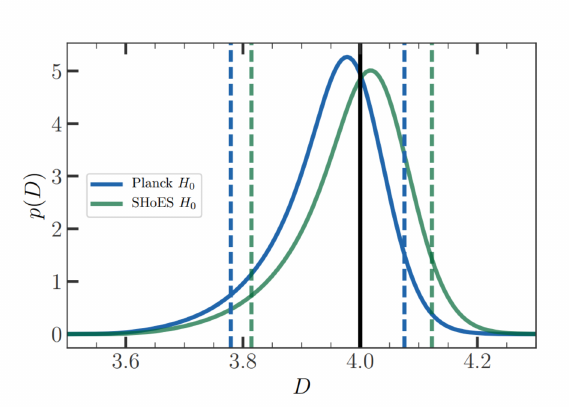}
\caption{Plot showing the posterior probability distribution (p) for the number of spacetime dimensions (D) from the neutron star merger GW170817. Photon observations from Hubble space telescope and the Planck observatory were used alongside gravitational wave data from LIGO and Virgo to compute two possible distributions for the number of spacetime dimensions. Figure credit from~\cite{Pardo_2018}. Some alterations were applied to make the figure clearer and more presentable for this context.}
\label{fig-GravWav}       
\end{figure}

This is in contrast to Kaluza-Klein theory wherein there is a proposed 5D spacetime, with the electromagnetic and gravitational parts not being additive but instead being distinct components of the 5D spacetime~\cite{KLEIN_1926}. Kaluza-Klein theory has been heavily disfavored by experimental results that indicate that gravity and electromagnetism exist in the same 4D spacetime. Experimentally this has been measured through multi-messenger observation of gravitational waves and electromagnetic waves from neutron star and black hole mergers~\cite{Pardo_2018}. As seen in figure~\ref{fig-GravWav}, two different approaches find that the 4D spacetime of general relativity is preferred and that electromagnetism and gravity share this spacetime.

To understand why many believe general relativity and the Standard Model are incompatible we must look at gauge theory. Quantum electrodynamics (QED), which is a part of the Standard Model, is a gauge theory. Due to this, QED has what is known as gauge freedom. Gauge freedom means that one can choose a form of the gauge field to aid in creating solutions. However, a given gauge field may have assumptions that aren't always true. QED also requires computing or choosing a metric, a spacetime curvature. Usually the Minkowski metric is assumed to be true, which corresponds to flat spacetime. As such, there is no need to compute Christoffel symbols, the corresponding metric or, the associated gravitational forces as they are constant. However, this is not a requirement of the underlying theory. It is a requirement chosen by those who are creating solutions to the problems to make solutions easier to attain.

\begin{figure}[h]
\centering
\includegraphics[width=16cm,clip]{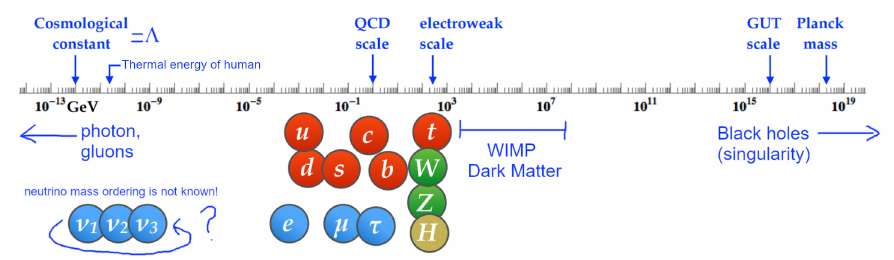}
\caption{Diagram showing the energy scales of the Standard Model and other measurements of cosmology for comparison. The thermal energy of a human is provided to give a comparison to something that is common and understandable.}
\label{fig-EneScl}       
\end{figure}

The Minkowski metric is almost always justified as the gravitational field strength required to significantly bend spacetime on the quantum scale would only be expected with blackholes. Examining figure~\ref{fig-EneScl}, it is observed that the majority of the universe, which is dark energy dominated, is at energy scales far below where most of the Standard Model is used; the low energy scale of the universe mean that the locally flat spacetime approximation is valid. This also means that, conversely, blackholes serve as important tests for the quantum connection of gravity and quantum field theory (QFT).

In the past the quantum link between gravity and QFT has been attempted with measuring the Unruh effect of blackholes, which manifests as Hawking radiation. These are expected for certain blackholes and are experimental signatures coming from electromagnetic radiation that could take place due to the extreme gravitational field strength of a blackhole. However, similar to QFT's flat space approximation, blackhole solutions are typically done using the Schwarzschild approximation to make solutions more attainable~\cite{schwarzschild}. The Schwarzschild approximation assumes there is no charge, electromagnetic field, angular momentum or dark energy. The requirements of Hawking radiation are for the Unruh effect to be possible, and for the Schwarzschild approximation to be valid~\cite{Hawking:1974rv}. The first issue being that the Unruh effect, of any form, has never been experimentally confirmed to high precision. The second issue is the validity of the Schwarzschild approximation becomes questionable once radiation takes place. So Hawking radiation as an experimental signature is itself questionable. Existing measurements of blackholes, particularly those done by the Event Horizon Telescope (EHT), have also proven that the Schwarzschild approximation is, in all measured cases, false~\cite{EHT}. More work needs to be done to determine how quantum effects may look in an experiment on blackholes. One proposal is to use the time dependence of the accretion disks and shadows of blackholes; these signatures are expected to be measurable by EHT with the blackhole of the M87 galaxy~\cite{Giddings_2018}.

\begin{figure}[h]
\centering
\includegraphics[width=12cm,clip]{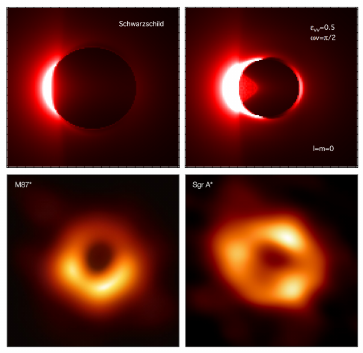}
\caption{Compilation of blackhole accretion disk and shadow images from simulation (top) and EHT (bottom). The top left image is in the Schwarzschild approximation. The top right image includes quantum effects. This figure is based on experimental results of EHT and from modification of a sourced figure~\cite{Giddings_2018}. The orientation of the blackholes here are not consistent so rotate as needed.}
\label{fig-BlkHol}       
\end{figure}

A simulation of blackhole images, along with experimental comparisons from EHT, at a snapshot in time, can be seen in figure~\ref{fig-BlkHol}. The experimental results favor structure that is more complex than the Schwarzschild approximation. Since no experiment has yet been able to exhaustively measure these effects, we have no reason to include the graviton in the Standard Model or to consider general relativity and the Standard Model as equivalent at the quantum level. Still, despite the commonly repeated error that they do not agree, we have no evidence that they do not agree and that they describe two different universes. It is also possible that, from EHT measurements, the graviton could soon be proven and included in the Standard Model.

Dark matter and dark energy are not present within the Standard Model, but there are constraints for them in experimental results. From cosmology measurements the prevailing model is $\mathrm{\Lambda CDM}$, wherein dark matter is in the form of Weakly Interacting Massive Particles (WIMPs)~\cite{Planck:2018vyg}. The most popular form of WIMPs are neutralinos from supersymmetry (\Gls{SUSY})~\cite{Jungman:1995df}~\cite{Bertone:2004pz}. Dark energy is characterized well by the cosmological constant, which is an additive term in the Einstein field equations of general relativity. At the quantum level, the Einstein field equations are propagated by the graviton, a spin-2 particle~\cite{Clifton:2011jh}. The results of the LIGO experiment indicate that gravitons are likely massless, with $\mathrm{\leq~10^{-23}}$~eV~\cite{PhysRevD.103.122002}. Furthermore, there is currently no evidence for new cosmological physics beyond general relativity and $\mathrm{\Lambda CDM}$~\cite{PhysRevD.103.122002}~\cite{Planck:2018vyg}.

When attempting to measure astrophysical dark matter, measurements are usually done alongside neutrino measurements. This is to say, the energy and cross-section of an incoming particle interacting with a nucleon, proton or neutron, of an atom. Since measuring neutrinos or neutrino-like particles at colliders is difficult, most dark matter searches at colliders are within the scope of \Gls{BSM} physics and \Gls{SUSY}~\cite{CMSDarkMatter}. Detectors at the LHC have looked for decays of neutralinos, composite \Gls{SUSY} particles that are similar to neutrinos~\cite{CMSDarkMatter}. No evidence for a neutralino has been found~\cite{CMSDarkMatter}. 

\begin{figure}[h]
\centering
\includegraphics[width=12cm,clip]{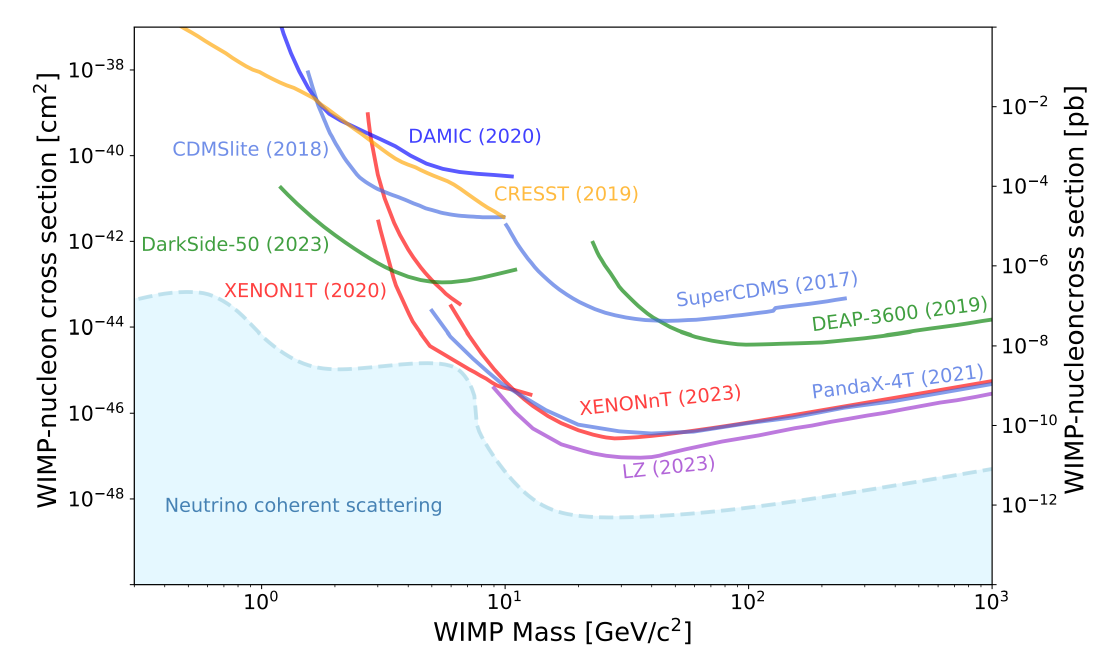}
\caption{Compilation of neutrino-like, WIMP, dark matter searches in terms of cross-section and particle rest mass. Results from collider experiments are not included since collider experiments look for different forms of dark matter. Figure, with some edits to improve labels, from~\cite{RefPDG}.}
\label{fig-DrkMat}       
\end{figure}

Measurements for dark matter that falls within the scope of the Standard Model have been done at large volume experiments like XENON~\cite{XENON:2018voc}. At these experiments emphasis is usually given to WIMP dark matter. XENON uses a 260~$\mathrm{m^3}$ tank of liquid xenon in a time projection chamber (TPC) setup~\cite{XENON:2018voc}. As seen in figure~\ref{fig-DrkMat}, the constraints for dark matter are approaching values consistent with neutrinos~\cite{XENON:2018voc}~\cite{RefPDG}.

\subsubsection{Group Representation and Bosons}\label{sec-rep}

Further examining the group representation of $\SUc \times \SUl \times \Uy$, from $\Uy$ we know that there must be $1^2$, one, massless, chargeless, colorless boson that follows space-time curvature and preserves the probability of quantum interactions and the counting of weak hypercharge. Weak hypercharge, written as $\mathrm{Y}$, is a quantity that is a sum of electromagnetic charge and weak isospin. The massless particle known for the aforementioned properties is the photon ($\gamma$). The photon is also a part of the electroweak sector, which comes from $\SUl \times \Uy$. The photon acts as a force mediator for electromagnetic charge. This includes the electric charge, magnetic dipoles and other forms of electromagnetic moments. It is also expected, and measured, that the photon would follow paths of least time since it is a massless boson embedded in general relativistic spacetime.

The equivalent of electromagnetic charge for $\SUc$ is known as color and is covered by Quantum Chromodynamics (QCD). Given the representation of $\SUc$ it requires that there must be a massless color transmitting boson that has $3^2-1=8$ variants. We know that it must be $\SUc$ and not $U(3)_\text{C}$ as $U(3)_\text{C}$ would require $3^2=9$ and thus a ninth, colorless, boson. This octet of particles is the gluon and since the colorless gluon has never been observed the Standard Model must be $\SUc$. The gluon transmits color current in a way that is similar to how the photon transmits electromagnetic current, but there are more distinct forms of color current than electromagnetic current. Since the strong force has a stronger coupling strength than the electromagnetic force the distances involved are smaller, on the scale of femtometers ($\mathrm{10^{-15}}~\text{m}$) to nanometers ($\mathrm{10^{-9}}~\text{m}$).

\begin{figure}[h]
\centering
\includegraphics[width=12cm,clip]{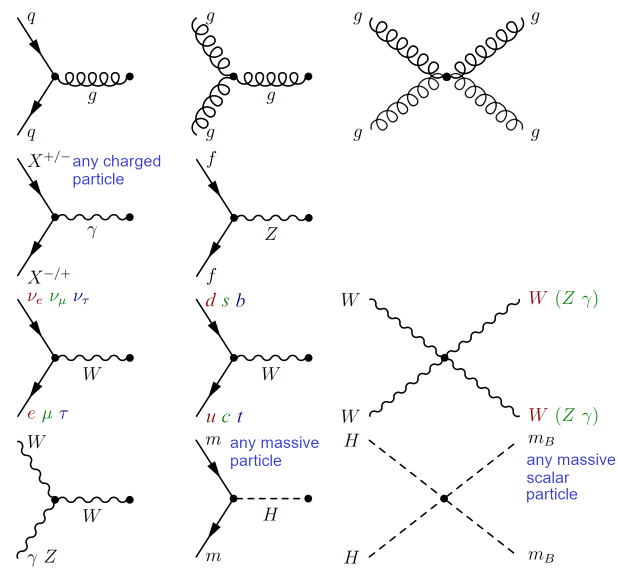}
\caption{Standard Model vertices used in Feynman diagrams. No loop level vertices are shown. Diagrams made by and organized by author of this work.}
\label{fig-FeynVert}       
\end{figure}

We also know that the gluon and photon are distinct massless force mediators as the gluon, as seen in the top right of figure~\ref{fig-FeynVert}, has additional self-coupling vertices that the photon does not. These self-couplings have been directly measured at LEP using Z decays to four jets~\cite{L3:1990jlf}. Thereby proving that the measured gluon is not ``QED-like'' and, instead, the $\SUc$ gluon of QCD is favored. The LHC and the Electron Ion Collider are expected to perform measurements which will further improve upon these measurements~\cite{Dainton:2006wd}~\cite{Surrow:2007zz}.

For $SU(2)$ things are more complex as the electroweak symmetry of $SU(2) \times U_{Y}(1)$ is partially broken. Due to this symmetry breaking, known as electroweak symmetry breaking, some of the bosons of $SU(2)$ must be massive. This is also why we only have $SU_{L}(2)$, the special unitary group for left-handed particles, and no $SU_{R}(2)$, the special unitary group for right-handed particles. Since the photon is covered by $U_{Y}(1)$ the electroweak symmetry breaking does not generate a massive photon. We still have a requirement of $2^2-1=3$ bosons, but things are more complicated.

Electroweak symmetry breaking, and its cause(s), are still not exhaustively measured. The prevailing experimental results indicate that the cause is from the Higgs mechanism and the Higgs boson. Still, the Higgs boson has not been measured precisely enough so the possibility of BSM Higgs cannot be ruled out~\cite{CMS:2019ekd}. For simplicity we will assume here that the simplest case of a single Higgs is sufficient in describing reality. From this we know that there must be three particles from $SU_{L}(2)$ and one for the Higgs mechanism. Of these four particles one must be scalar, spin zero, and three must be vector, spin one. Of the vector particles two must be charged and one is neutral. These have been measured as the Z boson, the W bosons and the Higgs boson. The vector bosons, the Z and W bosons, couple the handedness and electromagnetic charge states. Due to electroweak symmetry breaking this coupling is different for right-handed particles. Such that only left-handed particles can participate in interactions with the W boson and the coupling for the Z boson depends on the particle's handedness. This then covers the bosons of the Standard Model.

\subsubsection{Standard Model Fermions}

For reasons not proven beyond measurement, there are three generations of fundamental fermions. These generations follow a mass hierarchy that also has no proven motivation beyond measurement. In the mathematical formulation these flavors can be considered as $\suf$, but it is known that this symmetry must be broken as the flavor states and mass states of fermions are different. The massive nature of fermions requires this broken flavor symmetry. Still, there is no clear reason why this is $\suf$ and not, for example, $SU(2)_\text{F}$ or $SU(4)_\text{F}$. 

\begin{figure}[h]
\centering
\includegraphics[width=12cm,clip]{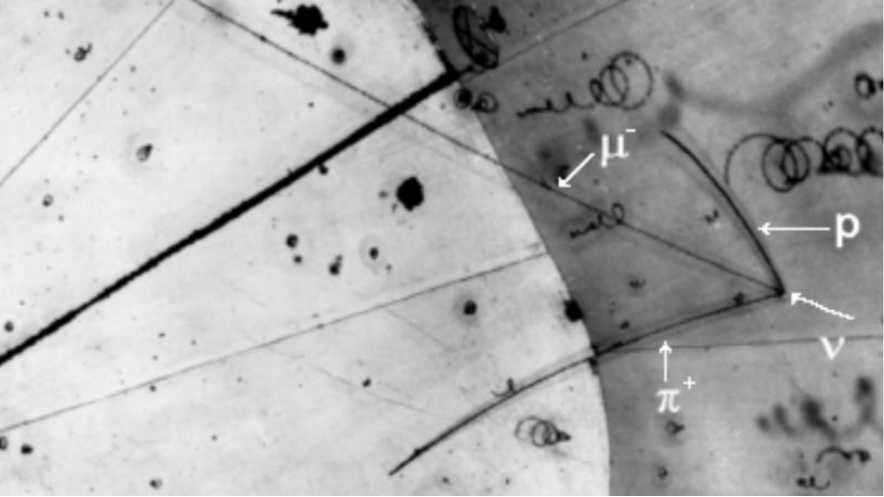}
\caption{Image of a muon neutrino inelastically scattering with a proton from a camera in Argonne's 12-foot bubble chamber, filled with hydrogen and deuterium, being bombarded by the Zero Gradient Synchrotron's neutrino beam~\cite{Barish:1977qk}. The result is a particle shower that kicks the proton but also creates a muon and a pion, a bound state of quarks. The spiraling tracks are typically created by secondary electrons which spiral as they lose energy.}
\label{fig-BubNeu}       
\end{figure}

Since the Standard Model has both weak hypercharge and color charge there are two types of fermions. Fermions that can transmit with both charges are known as quarks. Fermions that transmit with only weak hypercharge are known as leptons. Neutral leptons are known as neutrinos. Due to their neutrality neutrinos are difficult to detect and do not show signatures unless they collide with something in the detector. As seen in figure~\ref{fig-BubNeu}, the neutrino has an invisible track and then collides with a proton that is inside the detector, here a bubble chamber, to produce charged particles. These charged particles are then visualized by the bubble chamber as the dark tracks. The different shapes of tracks correspond to different particles. These sorts of bubble chamber measurements were used in some of the first imaging of neutrinos and charged leptons as well as discovering that neutrinos could interact with nuclei.

Though we do not necessarily know why the fundamental fermions exist relative to the underlying theory group structure, it can be argued that the fundamental fermions are possible as extensions of the fundamental bosons. The clearest example of this is that each matter and anti-matter pair can be expressed as one or two bosons. The Feynman vertices of the Standard Model, which give a theoretical representation of how particles relate to each other, can be seen in figure~\ref{fig-FeynVert}. The motivation for connecting bosons with fermion pairs was first grounded in experimental results from cosmic rays, circa the 1920s and 1930s. This connection was predicted by theories written earlier, most notably the Dirac equation.

Extending to chromodynamics, a quark and anti-quark pair can be expressed as one or two gluons. A left-handed fermion with a right-handed anti-fermion can be expressed as one $\mathrm{W^{\pm}}$ boson or $\mathrm{Z}$ boson, depending on the net charge of the two fermions. Finally, the connection of fermions to the Higgs boson comes from electroweak symmetry breaking, which generates the Yukawa coupling, or mass coupling. The vertex representing this mass coupling can be seen in the bottom left of figure~\ref{fig-FeynVert}. This coupling gives all of the fermions mass. 

This coupling is dependent on the mass of the fermion, and, as such, lighter fermions are harder to measure as their cross-section is smaller. Currently, the lightest particle coupling to Higgs measurements is of the muon and charm quark~\cite{Sirunyan_2019}~\cite{atlas_muon2019}~\cite{Aad_2022}. Coupling to the lighter up, down, and strange quarks has not been measured conclusively, nor has coupling to the electron or any of the neutrinos. These are expectedly harder measurements as they involve smaller cross-sections and larger backgrounds. In the case of neutrinos it is especially difficult as the cross-section, if it is non-zero, is expected to be extremely small. The small cross-section compounded with the difficulty of measuring a neutrino at a collider experiment results in a measurement that is incredibly challenging.

Since neutrinos are chargeless they can only interact with themselves, through the Z boson, the Higgs boson and, with charged leptons through the W boson. Technically neutrinos may also interact through the photon if they have a non-zero magnetic moment, but current measurements favor increasingly smaller values of neutrino magnetic moment~\cite{Borexino:2017fbd}. Since the W boson only couples to left-handed particles then right-handed neutrinos can only be produced from decays of the Z boson or Higgs boson. This differentiation between the Z and W boson is important to measurements of the neutrino left-handed and right-handed couplings as W boson with neutrino measurements is sensitive only to the left-handed coupling while the Z boson with neutrino measurements are sensitive to the left-handed and right-handed couplings.

Neutrinos are already difficult to measure, due to their small cross-section, so often they are measured from astronomical sources, like the Sun or active galactic nuclei~\cite{Borexino:2017fbd}. This isn't to say that colliders and their detectors are not useful for neutrino measurements. The number of light neutrinos was measured at LEP using the Z boson's line shape~\cite{L3:1998uub}~\cite{ALEPH:2005ab}. A similar measurement, though less precise, has been done using measurements from the CMB and cosmological Large Scale Structures~\cite{Rossi:2014nea}. Searches for effects from the Higgs boson in astronomical settings have been formulated but rely on measurements of inflation and of energies of $10^{6}$ GeV or higher at colliders~\cite{Graham_2015}~\cite{Espinosa_2015}. Neither of which have Higgs interactions with neutrinos. This means that measuring right-handed Standard Model neutrinos may be uniquely resolvable at colliders; methods for measuring right-handed neutrinos at future $\ee$ colliders have already been proposed~\cite{Carena:2003aj}.

\begin{figure}[H]
\centering
\includegraphics[width=14cm,clip]{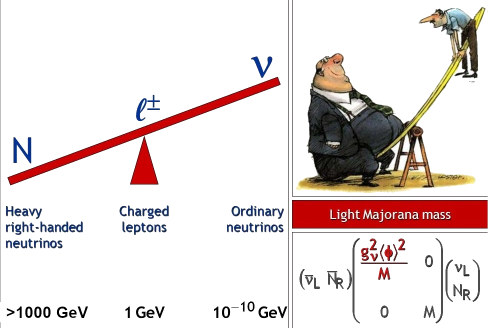}
\caption{Diagram to show, in an illustrative way, how a seesaw mechanism can generate massive right-handed neutrinos while keeping left-handed neutrinos with relatively small mass. Figure, with some alterations for presentation, from Georg Raffelt.}
\label{fig-SeeSaw}       
\end{figure}

Higgs boson events at current experiments are already rare and measuring neutrinos at collider experiments is also rare. As should be expected, there are no conclusive measurements of right-handed neutrinos. Further exacerbated by the fact that current constraints on neutrino Yukawa couplings put them at very small values ($\approx\mathrm{10^{-9}}$), meaning that any Higgs-neutrino events would have a small rate~\cite{Garcia_Cely_2013}. Current measurements of neutrinos also make it unclear if they couple to the Higgs boson at all. There may be mechanisms like the Seesaw mechanism, which includes Majorana mass, for generating the total neutrino mass. As seen in figure~\ref{fig-SeeSaw}, this Majorana mass can be large, and dominant, for the right-handed neutrino. The Z boson and CMB measurements can be respected if the bulk of Majorana mass goes into making the right-handed neutrinos super massive, ~$\mathrm{\geq1~TeV}$ .

This result is similar to the claim that there are no right-handed neutrinos in the Standard Model. However, we have distinguished things here by providing the experimental context. That there is no direct evidence for right-handed neutrinos. Similarly, there is no direct evidence that right-handed neutrinos do not exist.

\subsection{Status of Electroweak Measurements}

We now focus on the current status of electroweak measurements. Meaning measurements relevant to the photon, $\mathrm{W^{\pm}}$, Z and Higgs bosons. For the case of photon, $\mathrm{W^{\pm}}$ and Z measurements we will focus on methods and results from LEP since they are more relevant to later sections. For Higgs measurements we will use LHC methods and results alongside planned methods to be used at the International Linear Collider (ILC). The following sections are not meant to be exhaustive.

\subsubsection{Photon Measurements}

For photon measurements of mass and charge, which are often assumed zero, the most exclusive results are done with astronomical measurements. Therefore, we will not discuss these. For testing QED predictions for photons at electron-positron ($\mathrm{e^+e^-}$) colliders, the $\mathrm{e^+e^-\rightarrow\gamma\gamma}$, or digamma, production may be used. The digamma channel is unique in that it is high cross-section and relatively clean in terms of higher order and radiative corrections. For energies near and above 200 GeV the expected corrections are $\approx$1\%~\cite{OPALGamGam}. It is also sensitive to new physics that can couple to photons or electrons and, due to the aforementioned reasons, measurements and limits of this new physics can be done. 

\begin{figure}[h]
\centering
\includegraphics[width=10cm,clip]{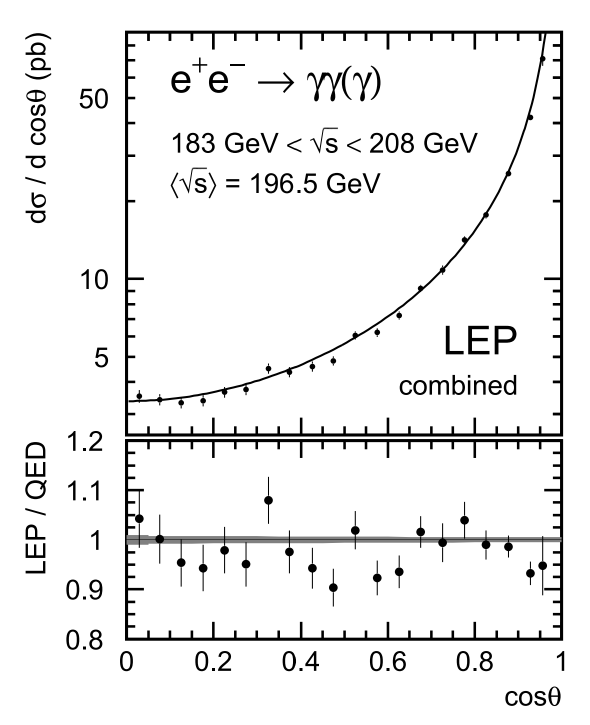}
\caption{Plot of differential cross-section from QED theory calculation of digamma production compared to that measured at LEP for a given center-of-mass ($\mathrm{\sqrt{s}}$) energy range of 183 GeV to 208 GeV. No significant deviation was measured. Figure, with minor edits for presentation, from~\cite{ALEPH:2013dgf}.}
\label{fig-GamGam}       
\end{figure}

As seen in figure~\ref{fig-GamGam}, digamma production favors forward angles. To measure photons well calorimeters must be used as trackers, which rely on charge curvature or charged interactions, do not work well for neutral particles. Considering both of these, measurements of the digamma channel would benefit from a forward calorimeter (FCAL) which can measure the energy of the photons well. This property of calorimeters is referred to as energy resolution. For the LEP experiments most of the digamma events were measured in their respective electromagnetic calorimeters (ECALs) and not a FCAL~\cite{ALEPH:2013dgf}.

From diphoton production one can put limits on extra dimensions, such as Kaluza-Klein discussed in Sect.~\ref{sec-space}, on excited states for the electron and, on neutral resonances. This sensitivity is possible as extra dimensions could result in massive gravitons or graviton-like particles. Depending on the underlying theory, either the virtual exchange or propagator of the graviton-like particle produces perturbations to the diphoton production, particularly at high center-of-mass energies and high invariant mass diphotons~\cite{Giudice:1998ck}. No measurements of extra dimensions or excited states was found, but significant limits were set~\cite{ALEPH:2013dgf}. At LEP the neutral resonance was a proposed channel for measuring the Higgs boson~\cite{Eboli:1998vg}. There was not a significant excess of events to prove the existence of a neutral resonance nor sufficient statistics to measure the Higgs boson~\cite{OPALGamGam}.

\subsubsection{$W^{\pm}$ Couplings}

Given the vertices of figure~\ref{fig-FeynVert} we know that there are different coupling constants, like a statistical weight, for different processes. For the W boson we can assign three electroweak couplings, denoted $\kappa_{\gamma},\lambda_\gamma,g_{4,5}$, and one scalar boson coupling for coupling to the Higgs boson. We will not address the quartic W boson coupling here. By better measuring these couplings the measurements become more sensitive to new physics since new physics would manifest small corrections to these couplings. Particularly loop effects from new physics may lead to perturbations in these couplings.

The W boson couplings, like other couplings, can be found in the Lagrangian for the SM. From an experimental point of view these couplings manifest as changes in differential cross-section. These changes are possible as couplings can have both constant, i.e. multiplicative, components and tensor components such as Dirac matrices or Levi-Cevita symbols. The latter of which can exchange, rotate or boost quantitites in the differential cross-section. We note the electroweak couplings as $\kappa_{\gamma}$, $g_{4,5}$ and $\lambda_\gamma$~\cite{QuantumILC}. We also note that $\kappa_{\gamma}$ is sometimes written as $g_1$~\cite{QuantumILC}. The SM values for these are $\kappa_{\gamma}~=~1$, $g_{4,5}~=~1$ and $\lambda_\gamma~=~0$~\cite{QuantumILC}.

\begin{figure}[h]
\centering
\includegraphics[width=10cm,clip]{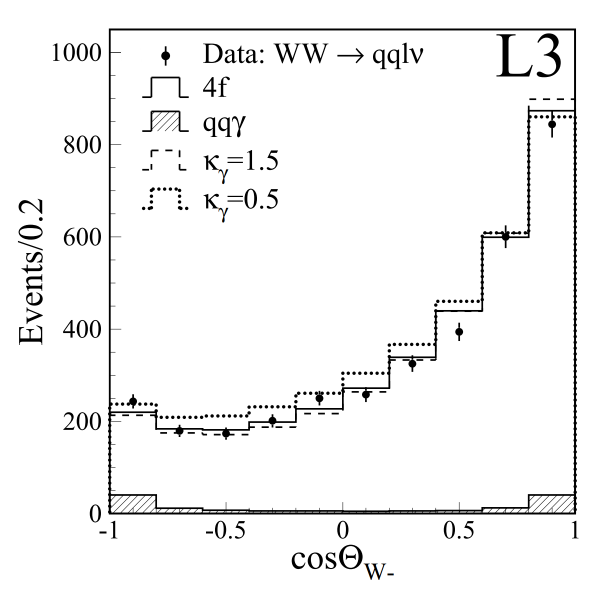}
\caption{Plot of reconstructed $\mathrm{W^{-}}$ production angle from the L3 experiment at LEP~\cite{L3:2004ulv}. The angle is in terms of $\cos(\theta)$ and for the semi-leptonic channel. Three histograms are added for results assuming $\kappa_{\gamma}~=~1$, $\kappa_{\gamma}~=~0.5$ or $\kappa_{\gamma}~=~1.5$. The main background, from $\mathrm{qq\gamma}$ is also shown. Figure, with minor edits for presentation, from~\cite{L3:2004ulv}.}
\label{fig-WCoup}       
\end{figure}

For measuring these couplings the reconstructed production angles of W boson pairs is used alongside the angles of the quarks and leptons from their decays~\cite{L3:2004ulv}. For similar reasons to W boson mass measurements the semi-leptonic channel is preferred. As can be seen in figure~\ref{fig-WCoup}, the semi-leptonic channel has a preference for forward angles in W boson production~\cite{L3:2004ulv}. Multiple values of coupling are shown for demonstration purposes.
These angular measurements are then used in a multi-dimensional fit with the couplings used as parameters~\cite{L3:2004ulv}.

Current results indicate that the SM values are preferred but the $\kappa_{\gamma}$ fit performs the worse. Since W bosons are created and measured differently between $e^+e^-$ and hadron colliders there is a need for future $e^+e^-$ colliders to better measure $\kappa_{\gamma}$. The polarization dependence of the electroweak bosons makes it especially advantageous to use polarized beams as, in general, statistics are larger and uncertainties are smaller~\cite{QuantumILC}. A polarized $e^+e^-$ collider is uniquely sensitive to the channels with coupling between a neutrino and the W boson since the W boson contributions can be effectively enhanced and suppressed depending on the chosen beam polarization~\cite{L3:2004ulv}~\cite{QuantumILC}.

\begin{figure}[h]
\centering
\includegraphics[width=10cm,clip]{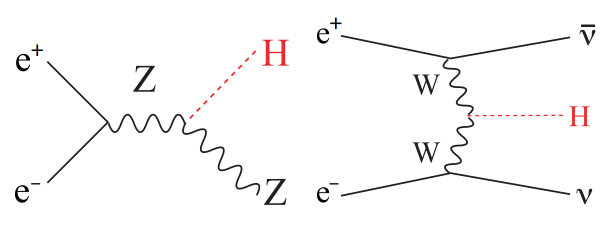}
\caption{The main processes for production of Higgs bosons at a $\ee$ collider, Higgstrahlung (HZ) and WW fusion ($\mathrm{\nu\nu H}$). Figure credit~\cite{asner2018ilchiggswhitepaper}.}
\label{fig-HZ_WWH}       
\end{figure}

Measurement of the W boson and Higgs boson coupling can be done using the W boson fusion process~\cite{asner2018ilchiggswhitepaper}. As seen in figure~\ref{fig-HZ_WWH}, this process is well suited for $\ee$ colliders. Due to energy conservation, the Higgs can only decay into W boson pairs off-shell, or with one W boson being off-shell, which reduces the cross-section by either two or one additional vertex factor respectively. These processes, available to the LHC, decrease with center-of-mass energy while their respective processes at a $\ee$ collider increase with center-of-mass energy~\cite{asner2018ilchiggswhitepaper}~\cite{Binoth:2005ua}. As a part of the Higgs sector this channel has merit for constraining the SM and BSM physics. We will cover this in greater detail in section~\ref{sec-Higgs}.

\subsubsection{Z Couplings}\label{sec-ZCoup}

\begin{figure}[h]
\centering
\includegraphics[width=10cm,clip]{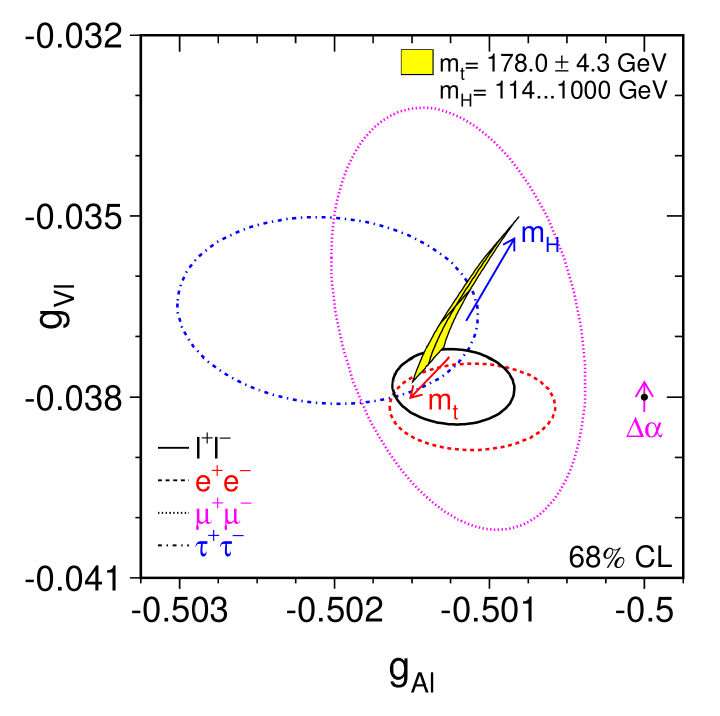}
\caption{Constraints on the vector and axial couplings, seen in equation~\ref{eqn-zcouplings}, as done using the three different leptonic decays. Additional constraints from what is allowed by the Standard Model for then plausible values of top quark and Higgs boson mass are included. Figure credit~\cite{ALEPH:2005ab}.}
\label{fig-ZCoup}       
\end{figure}

The couplings of the Z boson to fermions are among the most precisely measured parameters in the Standard Model. At electron-positron colliders, particularly LEP and SLC, measurements of difermion production around the Z resonance allowed high precision determination of these couplings~\cite{ALEPH:2005ab}. The interaction between the Z boson and fermions can be expressed through the neutral current interaction term in the electroweak Lagrangian
\begin{equation}\label{eqn-zcouplings}
\mathcal{L}_{Zff} = \frac{ig}{4\cos\theta_W}\sum_f \bar{f}\gamma^\mu(g_V^f - g_A^f\gamma_5)f Z_\mu,
\end{equation}
where $\theta_W$ is the Weinberg angle, and $g_V^f$ and $g_A^f$ represent vector and axial-vector couplings for the fermion spinors respectively~\cite{PDG2024}. A plot of the constraints on these values, as constrained by LEP, can be seen in figure~\ref{fig-ZCoup}.

Experimentally, the effective weak mixing angle $\sin^2\theta^{}_{\text{eff}}$ was measured to high accuracy at LEP/SLC as $0.23153 \pm 0.00016$~\cite{ALEPH:2005ab}. More recently, the LHC experiments ATLAS and CMS confirmed these couplings using forward-backward asymmetry measurements in dilepton events~\cite{ATLAS2018}~\cite{Shalaev2018}. Though they did not achieve the same level of precision.

High-precision measurements of $Z$ couplings also serve as a stringent test of lepton universality. Deviations would signal new physics beyond the SM. Currently, measurements across LEP, SLC, and LHC experiments have shown remarkable consistency with the SM predictions, reinforcing the robustness of the electroweak framework.


\subsubsection{Z Mass}\label{sec-zmass}

\begin{figure}[h]
\centering
\includegraphics[width=14cm,clip]{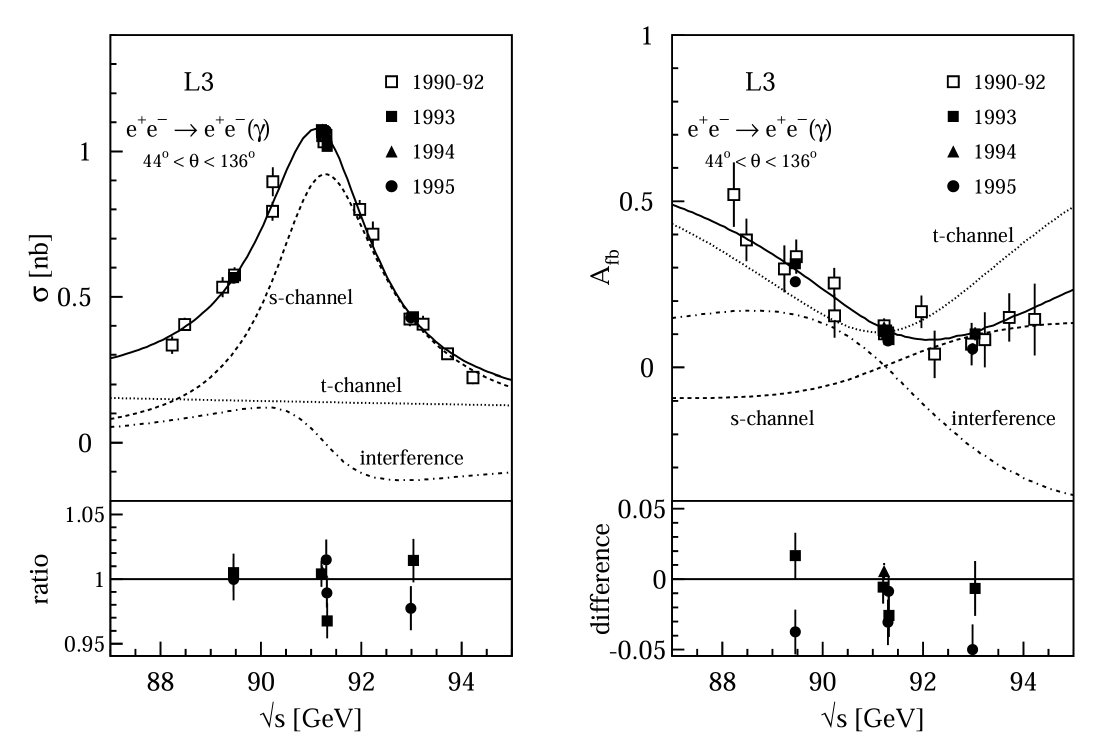}
\caption{(Left) Scan of the Z-pole resonance from the dominant, Bhabha, channel. The components of the resonance are broken down, which shows how the resonance is dominantly from the \Gls{schan} and interference contributions. (Right) A similar breakdown but for the Forward-Backward asymmetry, $A_{\text{FB}}$. Figure credit~\cite{ALEPH:2005ab}.}
\label{fig-ZScan}       
\end{figure}

The $Z$ boson mass, $m_\text{Z}$, is fundamental to electroweak theory and has been measured with exceptional precision at LEP and SLD. Combined they measured the $Z$ boson mass from resonance scans at center-of-mass energies near $m_\text{Z}$~\cite{ALEPH:2005ab}. An example of this scan can be seen in figure~\ref{fig-ZScan}, with the various \Gls{schan}, \Gls{tchan} and interference components of the resonance scan broken down. These experiments combined to produce an averaged mass measurement
\begin{equation}\label{eqn-zmass}
m_\text{Z} = 91.1875 \pm 0.0021,\text{GeV}.
\end{equation}
which was also consistent across individual experiments~\cite{OPAL:2000ufp}. Recent reviews by the Particle Data Group (PDG) reaffirm this precision and highlight the importance of the LEP beam energy calibration in reducing systematic uncertainties~\cite{PDG2024}.

The precise measurement of $m_\text{Z}$ plays a critical role in electroweak fits, constraining new physics. For example, before its discovery, the Higgs boson mass was constrained to roughly 125 $\pm$ 50~GeV by electroweak measurements~\cite{ALEPH:2005ab}. This level of precision on the Z mass and some electroweak observables is unlikely at LHC or HL-LHC, which have primarily been designed for high-energy frontier physics and discovering the Higgs boson. Still, electroweak measurements involving hadrons and gluons may be better measured at the LHC because the LHC is a hadron collider.

\subsection{Neutrinos}\label{sec-neutrinos}

\begin{figure}[h]
\centering
\includegraphics[width=12cm,clip]{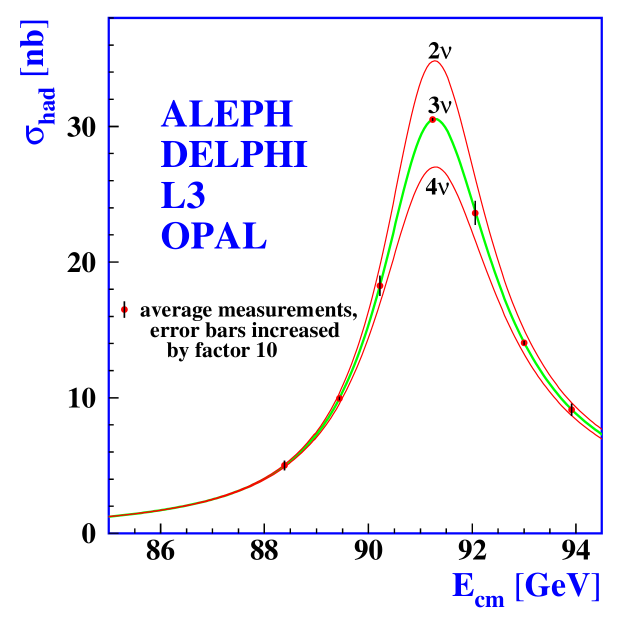}
\caption{Plot of the Z-pole scan with Standard Model theory expectations for different numbers of light neutrino flavors~\cite{ALEPH:2005ab}.}
\label{fig-NeuWid}       
\end{figure}

Neutrinos interact via the weak force, and their properties are associated with $Z$ boson measurements. At LEP, the invisible width of the $Z$ boson ($\Gamma_{\text{inv}}$), attributed to neutrino decays, provided the strongest evidence for three active neutrino flavors. The invisible decay width is expressed as:
\begin{equation}\label{eqn-invisiblewidth}
\Gamma_{\text{inv}} = \Gamma_Z - \Gamma_{\text{had}} - 3\Gamma_{\ell\ell},
\end{equation}
where $\Gamma_Z$ is the total width, $\Gamma_{\text{had}}$ is the hadronic decay width, and $\Gamma_{\ell\ell}$ is the width to a single species of charged lepton or the combined width of each lepton flavor~\cite{ALEPH:2005ab}. The results of LEP found that $\Gamma_{\text{inv}}$ was 499.0 $\pm$ 1.5~MeV. This corresponded to the number of light neutrino species to be
\begin{equation}\label{eqn-nnu}
N_{\nu} = 2.984 \pm 0.008
\end{equation}
and confirmed the existence of three light neutrino species. Later this measurement was updated to include a refined theory calculation and included additional experimental considerations to
\begin{equation}\label{eqn-nnu2}
N_{\nu} = 2.9963 \pm 0.0074
\end{equation}
be even closer to the Standard Model expectation~\cite{Janot_2020}.
As had been previously measured with astrophysical observations on Solar neutrinos and then later re-confirmed by neutrino oscillation experiments. A plot of how this is associated to the Z boson width can be seen in figure~\ref{fig-NeuWid}.

At the LHC, ATLAS and CMS recently provided a precise measurement of the $Z$ invisible width through mono-jet events. Their measurement, $\Gamma_{\text{inv}} = 506 \pm 1.4$~MeV is close to the LEP result but outside of $\text{3}\sigma$. We caution that this is a newer measurement that may, as was the case at LEP, have new systematics in the invisible width measurement that are yet to be discovered. Still, the LHC measurements on neutrinos reinforce the current understanding of neutrino flavors~\cite{ATLAS:2024vqf}.

Neutrino oscillation experiments further complement collider-based results, providing independent confirmation of neutrino properties, as well as providing insight into neutrino masses and mixing angles that are inaccessible at colliders~\cite{Esteban_2024}. Similarly, neutrino couplings and widths are not accessible in neutrino oscillation experiments. So there is a significant complementarity in a two-sided approach for the purpose of best characterizing neutrinos. By exploring from both angles we can solidify our current neutrino theory within the Standard Model and, hopefully, find new physics in the neutrino sector.

\newpage








\chapter{Proposed Higgs Factories}
The discovery of the Higgs boson at the \Gls{LHC} in 2012 has motivated proposals for new “Higgs factory” colliders dedicated to precision studies of the Higgs particle. A Higgs factory generally refers to a high-luminosity $\ee$ collider operating at energies around the Higgs production threshold, typically for the ZH channel. This enables the production of large samples of Higgs bosons. With the presence of the \Gls{HLLHC}, this is not a new environment. Instead, a Higgs factory focuses on producing large samples of Higgs bosons while also being a clean and precise environment, one that cannot be found in a hadron collider due to their significant backgrounds and pileup.

\begin{figure}[h]
\centering
\includegraphics[width=14cm,trim=0.5cm 0.5cm 0.5cm 0.5cm]{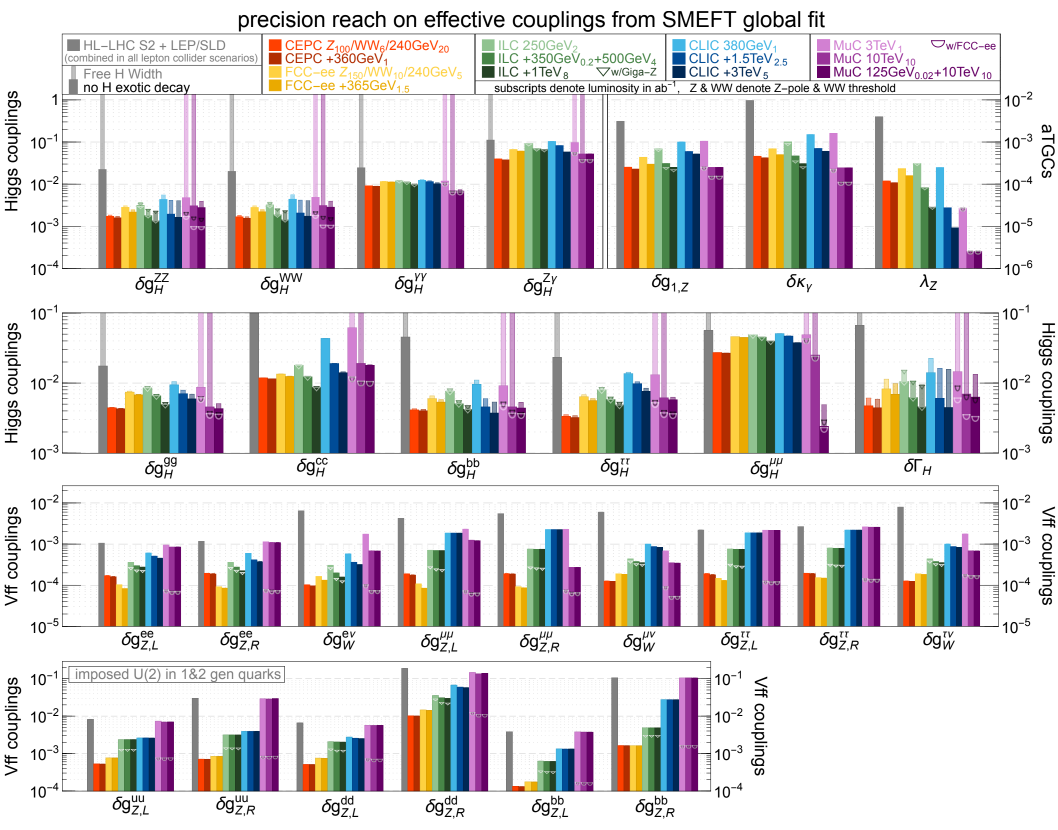}
\caption{A compilation, by the \Gls{SMEFT} community and presented at \Gls{Snowmass} 2021, of the aspirational precisions of various couplings at various Higgs factories~\cite{deblas2024globalsmeftfitsfuture}. Polarized couplings, accessible to experiments like \Gls{ILC} and \Gls{CLIC}, are not shown so that comparisons can be made with unpolarized experiments like \Gls{FCC}-ee and \Gls{CEPC}.}
\label{fig-SMEFTCoup}       
\end{figure}

For improving precision over \Gls{HLLHC} the community working on Higgs factory proposals has advocated for numerous types of lepton colliders. Several proposals have been put forward, which can be broadly categorized into linear $e^+e^-$ colliders and circular $e^+e^-$ colliders. These machines aim to surpass the precision of the \Gls{HLLHC} in measuring Higgs boson parameters, and to provide improved sensitivity to electroweak observables and potential new physics. A compilation of the degree these parameters, and more, may be measured can be seen in figure~\ref{fig-SMEFTCoup}. In the text below we summarize the major proposed Higgs factory projects by their designs, advantages, and key physics goals and contextualize their performance to that expected from the LHC and HL-LHC. 

\section{Linear Colliders} 

Linear colliders accelerate electrons and positrons in straight linacs that collide at a single interaction point, as opposed to bending them in a ring. The primary advantage of a linear $e^+e^-$ collider is the avoidance of synchrotron radiation energy losses, which become prohibitive for electrons in circular machines at high energies. This allows linear colliders to reach center-of-mass energies in the several-hundred GeV to TeV range, extending well beyond the limits of circular $e^+e^-$ colliders. The power loss of the facility per particle generated is also, in general, lower due to not having to compensate for synchrotron losses. This is not a simple comparison as circular colliders can recycle particles, while only \Gls{ERL}s, like \Gls{ReLiC}, are capable of doing this in a linear collider environment~\cite{Litvinenko:2022qbd}.

\begin{figure}[h]
\centering
\includegraphics[width=14cm]{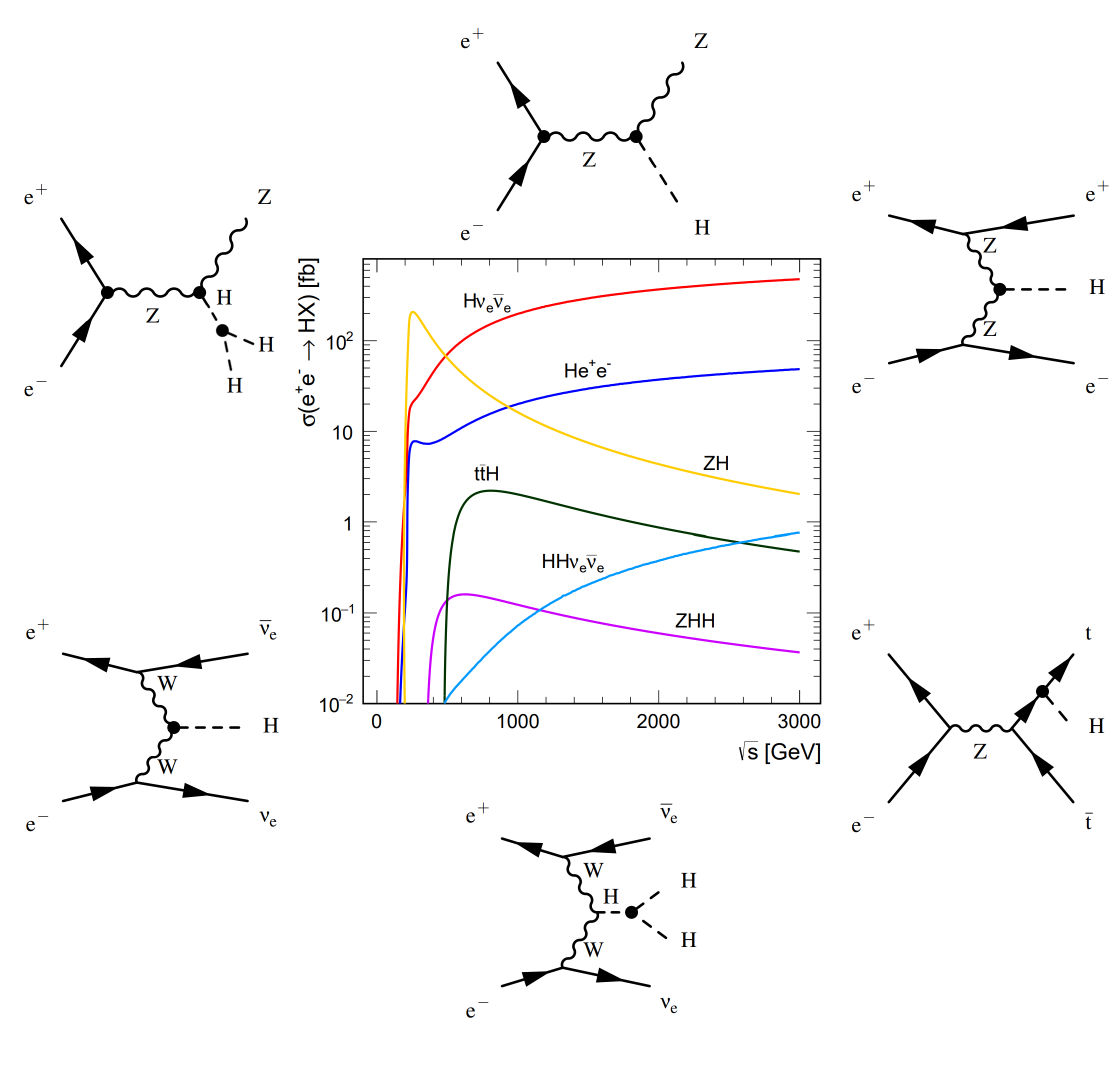}
\caption{A plot, compiled by the \Gls{CLIC} collaboration, of various contributions to the Higgs production cross-section accessible to $\ee$ colliders. Figure, with edits for presentation and consolidation of figures, from ~\cite{Abramowicz_2017}.}
\label{fig-HiggsXSecDiag}       
\end{figure}

Linear $e^+e^-$ colliders provide a clean environment, with minimal underlying events or pileup as there is at hadron colliders. Polarized beams, which enhance precision and allow for additional observables, are also possible since there is minimal synchrotron depolarization~\cite{Assmann:1994wt}. The two leading linear collider Higgs factory proposals are the International Linear Collider (\Gls{ILC}) and the Compact Linear Collider (\Gls{CLIC}), described below. Both are conceived as multi-stage projects, starting at a Higgs-factory energy, at 250~GeV to 380~GeV, and subsequently upgrading to higher energies for expanded physics reach. The motivation for the starting energy range can be found from the Higgs production modes, as seen in figure~\ref{fig-HiggsXSecDiag}, which shows how the total Higgs cross-section threshold is about 250~GeV and increases, but also changes composition, above that energy.
For this reason, it may be advantageous to have different runs in this starting energy range, so that different Higgs couplings can be focused on.

Each linear collider would operate with one or two large multipurpose detectors, analogous to ATLAS and CMS at the LHC, but optimized for lepton collisions. Later, in section~\ref{sec-ILD}, we will summarize one possible detector, \Gls{ILD}, for the sake of demonstration.

\subsection{International Linear Collider}

The International Linear Collider (\Gls{ILC}) is a proposed electron-positron linear collider with an initial center-of-mass energy of 500~GeV or 250~GeV, depending on the starting scenario. The 250~GeV start is optimized for Higgs boson production via the $\ee \to ZH$ process while 500~GeV is better suited for overall Higgs statistics~\cite{Behnke2013}. The project envisions a $\sim31$km dual-linear-accelerator complex based on superconducting radio-frequency (SCRF) technology, similar to that pioneered at \Gls{DESY} for \Gls{TESLA}/European \Gls{XFEL}. The ILC design can be extended in length to reach 1~TeV in staged upgrades~\cite{Brau2015}. This allows studies of top-quark pairs, via $t\bar{t}H$ production, as well as double Higgs production from $e^+e^- \to ZHH$ and $e^+e^- \to HH\nu_e\bar{\nu_e}$~\cite{Bechtle2024}. A 1~TeV ILC would also provide a unique precision study of TeV scale particle physics, complementing the existing work of the LHC.

\begin{figure}[h]
\centering
\includegraphics[width=14cm]{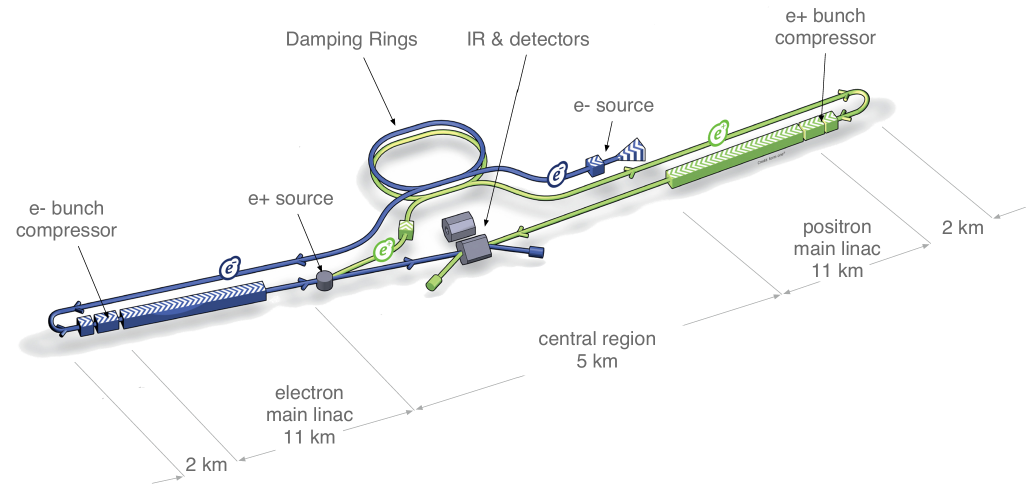}
\caption{Diagram of ILC at 500~GeV, also known as ILC500, showing the main systems needed for the starting energy range. Systems are not to scale. Credit with slight editing of background color from ~\cite{Behnke2013}.}
\label{fig-ILCSchem}       
\end{figure}

The proposed host for the ILC is Japan, with a preferred site in the Kitakami mountains of the Tohoku region, where the linear tunnel would be constructed underground. There have also been proposals for ILC at FermiLab and ILC at CERN~\cite{Behnke2013}. A schematic of ILC and its systems can be seen in figure~\ref{fig-ILCSchem}. The schematic shows the design as well as how ILC will use an electron undulator, denoted ``$e^+$ source'' in figure~\ref{fig-ILCSchem}, to generate the polarized positron beam.

The ILC’s primary physics goal is to serve as a Higgs factory, producing of order $10^5$–$10^6$ Higgs bosons in a clean environment to enable precise measurements of the Higgs boson’s properties. At the 250~GeV scenario the ILC would produce Higgs bosons predominantly through the Higgsstrahlung process, $ZH$. This process offers an indirect measurement of the Higgs by reconstructing the recoiling $Z$ boson. This allows a direct determination of the $HZ$ coupling and the total Higgs production rate without assuming specific Higgs decays. It also allows for an indirect measurement of the Higgs mass since $e^+e^-$ colliders have well-known initial particles. With an expected integrated luminosity on the order of 2ab$^{\text{-1}}$ at 250~GeV, the ILC can measure the $e^+e^-\to ZH$ cross section to about 2–3\% precision, translating to percent level measurements of the $HZZ$ coupling~\cite{Fujii2019}.

By combining measurements of multiple decay channels, the couplings of the Higgs to gauge bosons and fermions can be extracted with uncertainties at the few-percent level or better. For example, projections show that the ILC, in combination with HL-LHC data, can achieve $\sim0.5\%$ uncertainty on the $HZZ$ and $HWW$ couplings, and on the order of 1\% for the $Hbb$ coupling~\cite{Fujii2019}. This precision is a significant improvement over the $5-10\%$ level expected from HL-LHC measurements alone. The clean environment and well-constrained kinematics at the ILC also facilitate the measurement of difficult channels, e.g. $H\to c\bar{c}$ or $H\to b\bar{b}$, that are challenging at the LHC.

\begin{figure}[h]
\centering
\includegraphics[width=10cm]{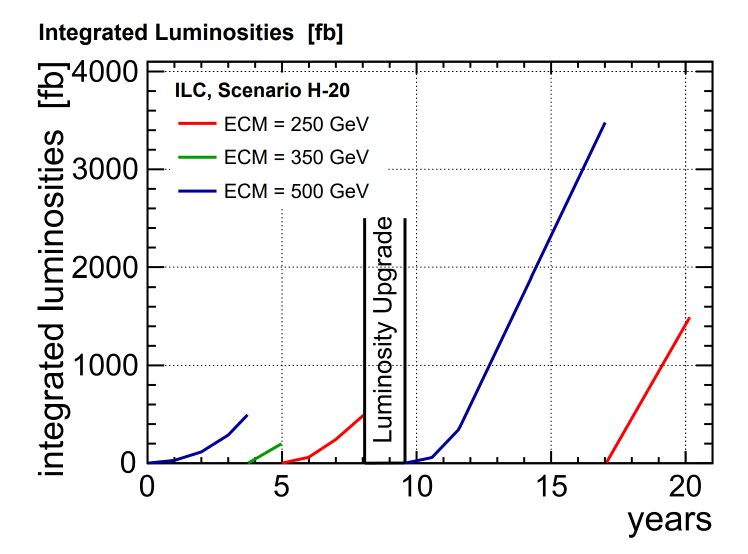}
\caption{A possible timeline for ILC, referred to as scenario H-20~\cite{Brau2015}. This does not include possible runs at lower energies, like the Z-pole or $WW$ threshold. It also does not include runs past any possible second upgrade. Which would allow for even higher luminosity and energies, such as 1~TeV.}
\label{fig-ILCTimeline}       
\end{figure}

Polarization of the beams at the ILC is a notable advantage, as it can be used to enhance signal production rates and to separately probe the chiral structure of electroweak couplings in both Higgs and other processes. Beyond Higgs physics, the ILC offers a broad program of precision electroweak and top-quark measurements. At 250~GeV, ILC would collect large samples of $W^+W^-$ pairs and fermion pairs, e.g. $e^+e^-\to f\bar{f}$, events, allowing for improved determinations of $W^\pm$ and $Z$ couplings. An energy upgrade to $\sim500$~GeV would enable pair-production of top quarks, providing a clean environment to measure the top quark mass, electroweak couplings, and the top Yukawa coupling via $t\bar{t}H$ events~\cite{Fujii2019}. As seen in figure~\ref{fig-ILCTimeline}, possible timelines of ILC allow for luminosity of about 4ab$^{-1}$ at 500~GeV. Such a scenario would measure the top Yukawa coupling to around 5–10\% precision, with improvements should there be a later 1~TeV run.

In the 500~GeV–1~TeV runs, the ILC would also be sensitive to double Higgs production via vector boson fusion, $e^+e^- \to \nu\bar{\nu}HH$ and $ZH H$, which permits direct extraction of the Higgs self-coupling $\lambda_{HHH}$. Although the cross sections are small, studies indicate that a 1~TeV ILC with several ab$^{-1}$ could measure the Higgs self-coupling with about 10–20\% precision~\cite{Fujii2019}, whereas the HL-LHC is only expected to constrain $\lambda_{HHH}$ to roughly 50\% uncertainty.

The ILC is envisioned as an international collaboration with significant contributions from laboratories worldwide. As a green-field project in Japan, it has the advantage of starting fresh with tailored design and infrastructure, but it also faces challenges. Substantial construction cost, estimated on the order of 6.8 billion ILCU, the ILC currency equivalent to \$1 as of January 2012, for the 250~GeV stage and, the need for global funding, and securing political approval~\cite{LCVision}. As of this writing, the ILC has undergone a full technical design report and extensive R\&D~\cite{Behnke2013}. The Japanese high-energy physics community has expressed strong interest in hosting it, and international peer reviews, from the European Strategy Update and American P5 Report, have recognized ILC in Japan as a mature option for a Higgs factory. If realized, the ILC would be the first new energy-frontier $e^+e^-$ collider since LEP, and would provide a rich physics program well into the mid-21st century.


\subsection{Compact Linear Collider}

\begin{figure}[h]
\centering
\includegraphics[width=14cm]{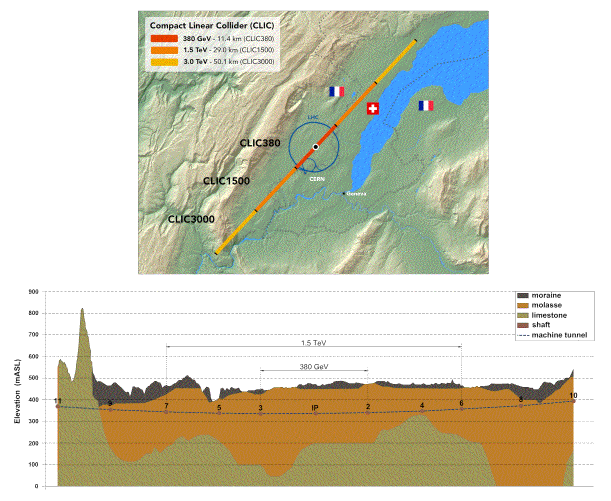}
\caption{(Top) View of the proposed CLIC site from the sky with three proposed deployments at 380~GeV, 1.5~TeV and 3~TeV respectively. (Bottom) Side-view of the proposed CLIC site with respect to metrology and geology of the site~\cite{CLIC2018}.}
\label{fig-CLICView}       
\end{figure}

The Compact Linear Collider (\Gls{CLIC}) is another leading proposal for a linear $e^+e^-$ Higgs factory and energy-frontier machine. The CLIC accelerator would be situated at or near the CERN site, potentially starting at the existing SPS complex and extending outward, making use of CERN’s infrastructure and expertise in accelerator operation. A top-down view of CLIC, as well as a profile view with geological information, can be seen in figure~\ref{fig-CLICView}.

\begin{figure}[h]
\centering
\includegraphics[width=14cm]{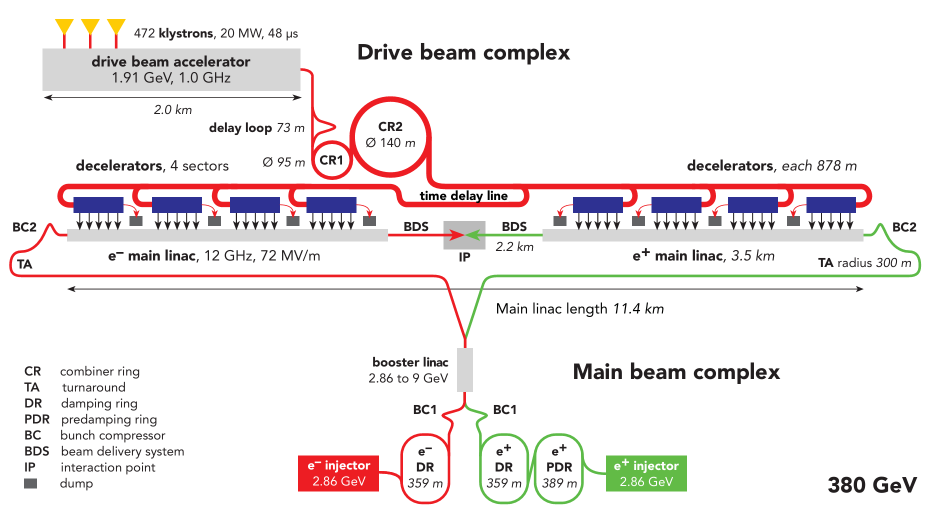}
\caption{Diagram of CLIC at 380~GeV with emphasis on the accelerator stages and technology~\cite{CLIC2018}.}
\label{fig-CLICSchem}       
\end{figure}

CLIC aims to reach multi-TeV center-of-mass energies in a relatively compact linear geometry by leveraging novel drive-beam acceleration technology. The distinguishing feature of CLIC is its two-beam acceleration scheme, which can be seen in figure~\ref{fig-CLICSchem}. One high-intensity, low-energy drive beam is used to generate accelerating fields for the main beam, allowing the use of normal-conducting accelerating structures at gradients of 100 MV/m~\cite{Aicheler2012}. This approach enables CLIC to attain very high energy in a linear collider of manageable length, on the order of 10~km for the initial stage and up to about 50~km for the final 3~TeV stage~\cite{Aicheler2012}.

The CLIC design is modular and staged: the first stage would operate at $\sqrt{s}\approx 380$~GeV, sufficient to study the Higgs boson and top quark. The second stage at around 1.5~TeV, and the ultimate stage at 3TeV center-of-mass~\cite{CLIC2018}. Each stage would likely run for several years to accumulate the target luminosity of a few ab$^{-1}$ per stage. An upgrade path between stages is built into the design. At its initial 380~GeV stage, CLIC acts as a Higgs factory much like the 250~GeV ILC, albeit at slightly higher energy. The Higgsstrahlung process and $WW$-fusion, for $e^+e^- \to H\nu\bar{\nu}$, are both accessible at 380~GeV. This allows for a similar statistics sampling of the $HZZ$ and $HWW$ coupling, which is complementary to ILC's run at 250~GeV and measurement of $HZZ$~\cite{CLIC2018}. 

With an expected luminosity around $1.0-1.5~\text{ab}^{-1}$ at 380~GeV, the precision achievable for Higgs couplings at CLIC is comparable to ILC’s. Which is on the order of a few percent for major channels~\cite{CLIC2018}. For instance, studies project the $HZZ$ and $HWW$ couplings can be measured to about 0.5–1\% , similar to ILC500 results. After all CLIC stages are completed, the $Hbb$ coupling, as well as $Htt$ and $Hcc$, are planned to be measured to, at least, the 1\% level~\cite{CLIC2018}. At the initial stage CLIC plans to perform a top-quark threshold scan around 350~GeV to measure the top mass with an expected uncertainty of order of 50~MeV or 300 ppm~\cite{CLIC2018}. The ability to run slightly above the $ZH$ threshold means CLIC can simultaneously tackle Higgs and top physics in its first phase.

The real distinction of CLIC comes with its higher-energy phases. The 1.5~TeV second stage extends the physics reach to phenomena beyond the Standard Model. Multi-TeV center-of-mass energies at a lepton collider substantially increases sensitivity to new particles, such as heavy superpartners, neutralinos, or exotic resonances up to masses near the TeV scale. This will be complementary to existing LHC and HL-LHC searches for new heavy particles. At these energies multi-Higgs production is guaranteed. Allowing the first direct observation of double Higgs production in $e^+e^-$ collisions via $ZHH$~\cite{CLIC2018}. Observation of $e^+e^- \to ZHH$ at 1.5~TeV would provide a handle on the Higgs self-coupling, even if the event rate is low. Across all of the CLIC runs one can achieve an initial measurement of $\lambda_{HHH}$ to potentially $\mathcal{O}(30\%)$ uncertainty~\cite{CLIC2018}. 

At the 3~TeV stage, CLIC would produce Higgs bosons predominantly through the vector-boson fusion and, due to the rise in cross-section with center-of-mass energy, a 3~TeV CLIC would produce a large number of Higgs bosons. The Higgs self-coupling could then be measured more precisely and be expected to reach the 10--20\% level. This can be further enhanced by exploiting the dependence of single Higgs production, particularly at higher energies, on $\lambda_{HHH}$ via loop effects~\cite{CLIC2018}. The 3~TeV collision energy also opens discovery potential for new physics well beyond the direct reach of the LHC. New electroweak states up to $\sim1.5$~TeV or more, or indirect effects of new physics at multi-TeV scales, become accessible. In this sense, later-stage CLIC acts not only as a Higgs factory but as a true “Terascale” explorer. 

From an experimental standpoint, CLIC faces challenges common to linear colliders. Single-pass collisions necessitating very high instantaneous luminosity and nanometer-scale beam focus, plus additional complications from beam-induced backgrounds due to its high beam energy density. Particular focus is usually given to beamstrahlung, a type of short-range synchrotron radiation that occurs as two charged beams collide and their particles interact with their fields. The CLIC beamstrahlung is significant at TeV energies, leading to a spread in the collision energy and production of beam backgrounds. The CLIC detector concepts, CLICdet, adapted from ILC’s ILD and SiD detector proposals, plans to deal with these by employing fast timing and high granularity to distinguish physics events from background~\cite{CLICdetector}. Nonetheless, extensive R\&D has shown that these issues are manageable, and the projected detector performance meets the requirements for Higgs and top measurements~\cite{CLICdetector}.

\begin{figure}[h]
\centering
\includegraphics[width=14cm]{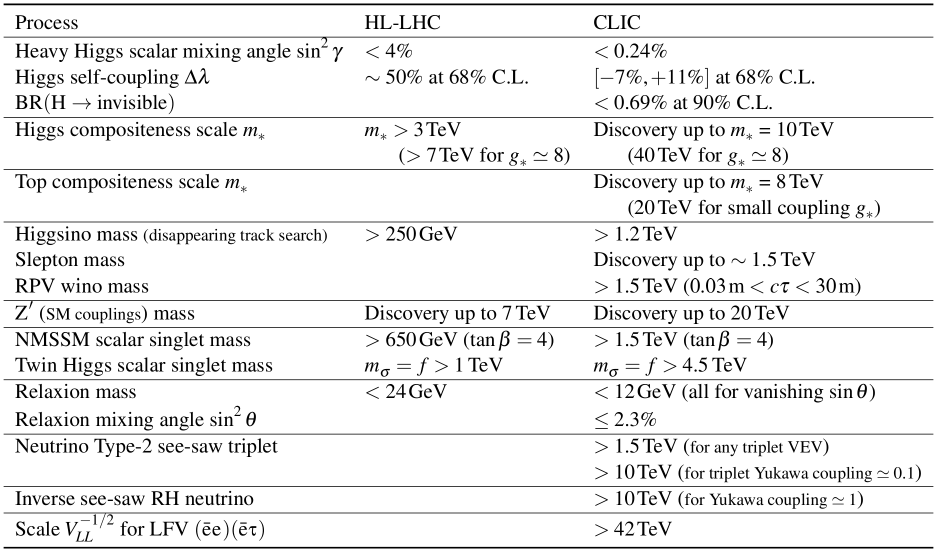}
\caption{Table of BSM observables with expected performance from HL-LHC and CLIC. The CLIC performance assumes combining three runs, at 380~GeV, 1.5~TeV and 3~TeV with 80:20 ratio of data for +80\% and -80\% electron beam polarization respectively~\cite{CLIC2018}.}
\label{fig-CLICDisc}       
\end{figure}

The ability to maintain high luminosity at 3~TeV with acceptable background is a major achievement of the CLIC design studies. In summary, CLIC offers an attractive but technologically ambitious path: by reaching energies up to 3~TeV, it promises not only high-precision Higgs measurements, at the percent level for couplings, but also a window to new physics complementary to the LHC. Its staged approach allows an initial focus on the Higgs, at $\sim$380~GeV, and then an evolution into a multi-TeV discovery machine. The utility of CLIC as a discovery platform can be seen in figure~\ref{fig-CLICDisc}, where the CLIC collaboration has provided some \Gls{BSM} measurements and how HL-LHC and CLIC constrain them. In general, all measurements are improved relative to HL-LHC in addition to being able to measure new quantities unavailable to the HL-LHC.

The primary drawbacks are the reliance on yet proven two-beam acceleration technology and the need for a large new linear tunnel infrastructure. Cost estimates for the full CLIC program, for all stages, are moderate, $\sim$\$6 billion, in the same scale as ILC. Global collaboration and funding would be required. From the current collaboration, this is already seen across much of the world. CLIC remains under active development at CERN, and is often compared and contrasted with the ILC in strategic plans. Though they need not be competitive or exclusive. It has been suggested building CLIC as ILC's successor in the same beam tunnel, after ILC finishes running and CLIC's technology is well proven.

The choice between a linear collider like ILC/CLIC and a circular approach, discussed in the next section, involves trade-offs between energy reach and luminosity/precision, as well as different timelines and technical risks.

\section{Circular Colliders}

An alternative approach to a linear $e^+e^-$ collider is a circular $e^+e^-$ collider, in which electrons and positrons circulate in opposite directions in a storage ring and collide at one or more interaction points (\Gls{IP}). Circular $e^+e^-$ colliders have a long legacy in particle physics, e.g. LEP at CERN and KEKB in Japan. Circular colliders benefit from the ability to achieve very high instantaneous luminosities and to host multiple detectors. By storing beams for many turns, they can reuse the particles for multiple collisions, and with continuous top-up injection, can sustain high collision rates. 

The primary limitation of a circular collider is synchrotron radiation. As the beam energy or ring size increases, energy loss per orbit scales as $E^4/(m^4R)$, which limits the maximum feasible energy. For a Higgs factory, circular designs are feasible in the range up to about 250~GeV to 360~GeV center-of-mass. Beyond that, the required power to compensate synchrotron losses becomes impractical even for very large rings. The polarization of beams is also more complicated in circular colliders as resonant depolarization, from synchrotron radiation, can (de)polarize beams and, resultingly, affect the data precision. Still, a circular collider could produce Higgs bosons in amounts similar to linear colliders.

In summary, a circular Higgs factory is often seen as the first stage of a larger project: after finishing the $e^+e^-$ physics program, the same tunnel could be used to host a future high-energy proton-proton collider to reach energy frontiers beyond the LHC. The leading proposal in this category is the Future Circular Collider (\Gls{FCC}) concept at CERN, specifically the FCC-$ee$ which is a dedicated $e^+e^-$ machine~\cite{Abada2019}. A similar proposal is the Circular Electron Positron Collider (\Gls{CEPC}) in China, which has broadly similar goals and design parameters as FCC-$ee$~\cite{CEPC2018}. Here we focus on the FCC as a representative circular Higgs factory concept.

\subsection{Future Circular Collider}

\begin{figure}[h]
\centering
\includegraphics[width=14cm]{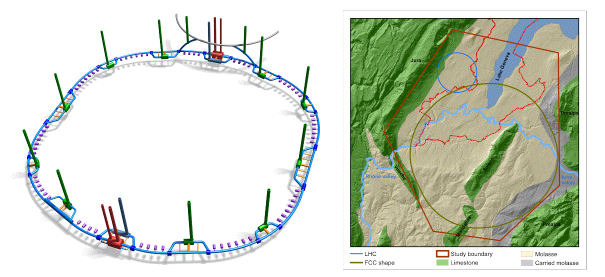}
\caption{(Left) Schematic of the underground structures of FCC, with IP facilities in red. (Right) A view of the proposed FCC tunnel site, as seen from the sky, with topological and geological information included~\cite{Abada2019}.}
\label{fig-FCCDiag}       
\end{figure}

The Future Circular Collider is a proposed new accelerator complex at CERN, centered on a $\sim90$~km circumference tunnel that would be built as a successor to the LHC tunnel and have an expected start time of 2048. By comparison, the future circular collider built in China, CEPC, would be $\sim100$~km and has an expected start time of 2038. The \Gls{FCC} proposal is actually a two-stage plan: first, build an $e^+e^-$ collider, FCC-$ee$, in the new tunnel to serve as a Higgs and \Gls{EW} factory, followed later by a proton-proton collider, FCC-$hh$, in the same tunnel. FCC-$hh$ would then reach center-of-mass energies around 80~TeV for discovery physics~\cite{Abada2019}. A diagram of the FCC beam tunnel, as seen from the sky, can be found in figure~\ref{fig-FCCDiag}.

The first stage, FCC-$ee$, is the one relevant as a Higgs factory. Its design envisions a circular $e^+e^-$ collider with two to four interaction points and detectors. It plans on running at various center-of-mass energies corresponding to important thresholds and precision measurements: $\sqrt{s}=m_Z$, for a $Z$ boson factory, $\sqrt{s}=2m_W$, for $W^+W^-$ threshold studies, $\sqrt{s}\approx 240$~GeV for Higgs production via Higgsstrahlung ($ZH$), and $\sqrt{s}\approx 350$ for $t\bar{t}$ threshold and enhanced $WW$-fusion Higgs production~\cite{Abada2019}. By covering this range the FCC-$ee$ would not only be a Higgs factory but a comprehensive electroweak factory, improving precision on many aspects of the Standard Model. The amount of statistics collected at these points is to-be-determined but there are proposals for Giga-Z, $\text{10}^{\text{9}}$ Z bosons, or even Tera-Z, $\text{10}^{\text{12}}$ Z bosons, runs.

\begin{figure}[h]
\centering
\includegraphics[width=14cm]{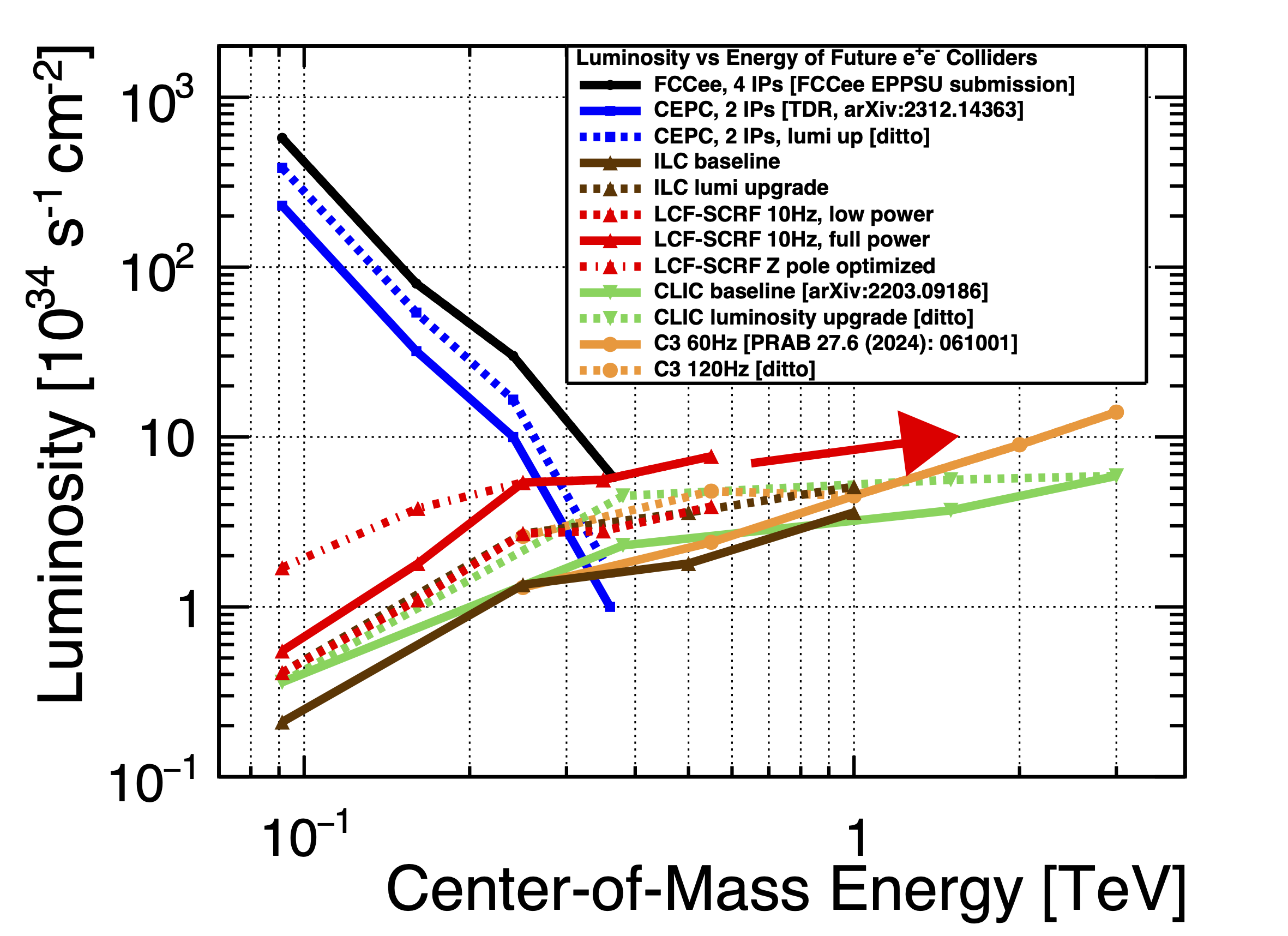}
\caption{Comparison of the instantaneous luminosity as a function of center-of-mass energy of proposed Higgs factories. The expected $1/E^4$ trend in the proposed circular colliders is observable. From this trend, and the data presented here, the circular colliders all under-perform the linear colliders around 380~GeV center-of-mass energy~\cite{LCVision}.}
\label{fig-FCCLumi}       
\end{figure}

The ability to host multiple detectors is a major strength that echoes the LEP collider, which had four experiments, but with higher luminosities per experiment and a greater range of energies to explore. The FCC-$ee$ design instantaneous luminosity is extremely high, on the order of $L \sim 10^{36}\space\text{cm}^{-2}\text{s}^{-1}$ at the $Z$ pole and $10^{34}$–$10^{35}\space\text{cm}^{-2}\text{s}^{-1}$ at the $ZH$ threshold, 240~GeV, per interaction point~\cite{Abada2019}. This would produce an integrated luminosity of roughly 5ab$^{-1}$ \emph{per experiment} at 240~GeV over a typical run length of roughly 4 years. On the order of one million $ZH$ events in total
would be measured~\cite{Abada2019}. As seen in figure~\ref{fig-FCCLumi}, FCC has a much higher instantaneous luminosity than any other proposed experiment when at the Z-pole.
The high luminosity dedicated runs for the $Z$, $W^\pm$ and $t$ enables an unparalleled program of precision measurements. For example, with $10^{12}$ $Z$ decays, the statistical error on many $Z$ properties would be negligible, pushing systematic and theoretical uncertainties to the forefront. The FCC design is being optimized to control such systematics, e.g. beam energy calibration with precision better than $10^{-6}$ to measure the $Z$ and $W$ masses extremely accurately. Current estimates suggest the $Z$ mass and width could be measured to a few keV, and the $W$ mass to 0.5 to 1~MeV precision, representing an improvement by factors of 20 to 50 over LEP and SLD era results~\cite{Abada2019}. These measurements would provide incisive tests of the Standard Model at loop level, in some cases sensitive to TeV-scale new physics through radiative corrections.

Building a 90~km tunnel and the associated infrastructure is a massive project. The FCC study indicates that the tunnel could be positioned in the Geneva region, spanning parts of Switzerland and France as seen in figure~\ref{fig-FCCDiag}, likely requiring deep drilling under the Jura mountains for certain sections. The cost of FCC-$ee$ has been estimated in the range of 15 billion CHF (Swiss-Franc) for the collider and four detectors, plus additional costs of 3 billion CHF for the later upgrade to FCC-$hh$~\cite{Abada2019}. The construction timeline would be on the order of 10 years once approved, with the earliest operation in the late 2040s.

The FCC proposal has gained traction through the CERN led design studies and was highlighted in the 2020 update of the European Strategy for Particle Physics, which recommended a feasibility study for FCC-$ee$ as the next step. Concurrently, China’s CEPC proposal, a similar 100~km $e^+e^-$ ring aimed at a 250~GeV Higgs factory, followed by a possible 100~TeV $pp$ collider, indicates a global interest in the circular collider route. If either FCC-$ee$ or CEPC proceeds, it would mark the construction of the largest collider ever built and usher in a new era of precision measurements. So, be it for record breaking physics or record breaking construction, there is a race for a new circular collider.

\section{International Large Detector}\label{sec-ILD}

\begin{figure}[h]
\centering
\includegraphics[width=14cm]{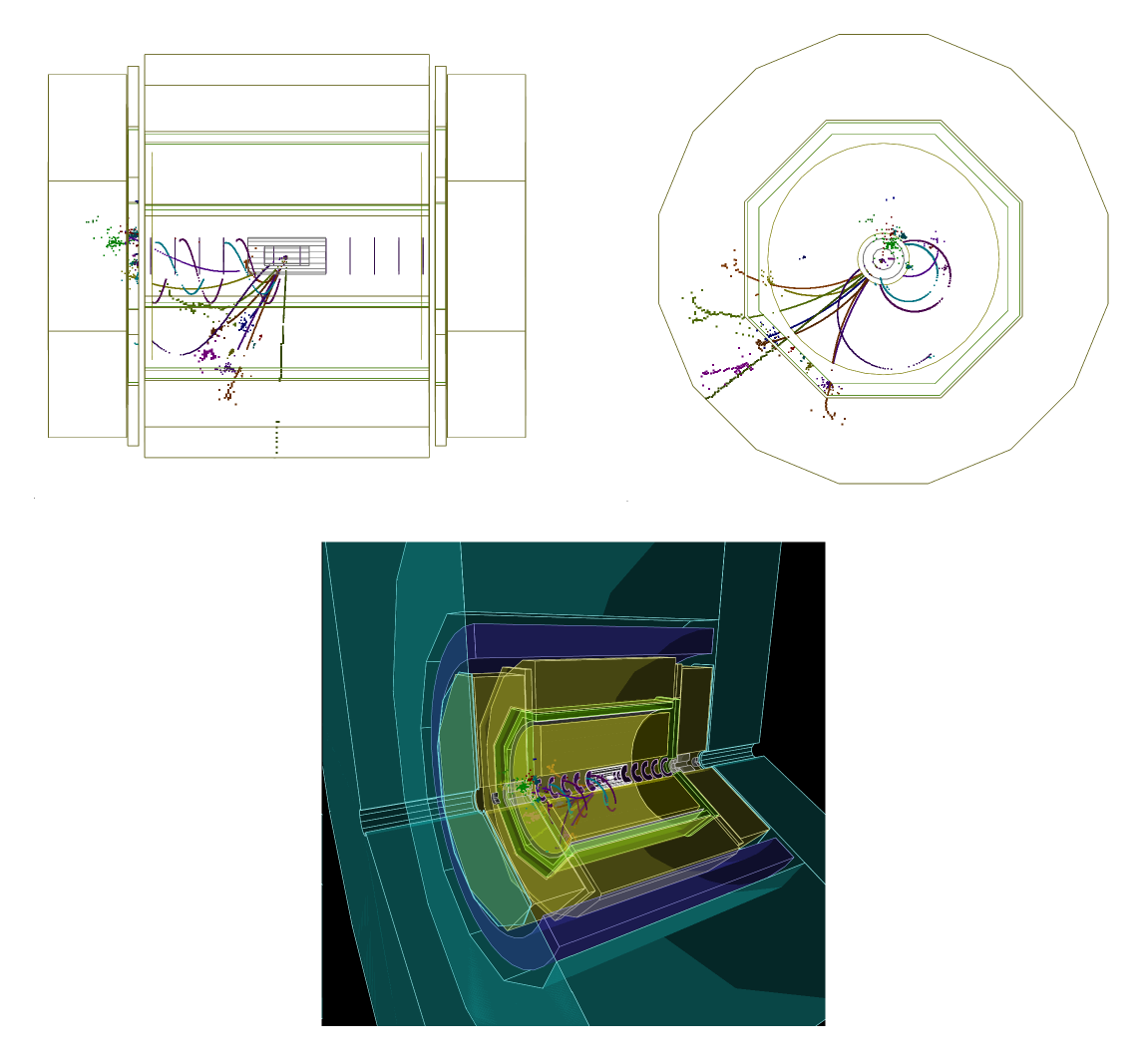}
\caption{ILD event display of a 250~GeV $e^+e^-\to ZH\to\nu\bar{\nu}H$ event. The asymmetry of the event is attributed to the escaping, and invisible, $\nu\bar{\nu}$. The top plots are, from left to right, the side view and forward view. The bottom plot is an isometric 3D view. Event displays generated, and made public, by Akiya Miyamoto (akiya.miyamoto@kek.jp).}
\label{fig-ILDReco}       
\end{figure}

The International Large Detector (\Gls{ILD}) is a proposed multipurpose detector concept designed for use at a future $e^+e^-$ collider, in particular the ILC but there are also proposals for FCC~\cite{ILD2020}. Though there are proposals for ILD-like deployments to other Higgs factories. ILD is one of two complementary detector designs that were developed for the ILC experiments, the other being the Silicon Detector (SiD) concept. The ILD design philosophy emphasizes excellent tracking resolution and full solid-angle coverage combined with high-granularity calorimetry. This allows for superior particle flow reconstruction, an example of which can be seen in figure~\ref{fig-ILDReco}. The detector is optimized for the clean environment of $e^+e^-$ collisions, where events typically have lower particle multiplicities and where achieving extreme precision in momentum and energy measurements is crucial for the physics goals.

In terms of layout, ILD features a large superconducting solenoid with a nominal magnetic field of 3 to 4T surrounding a central tracking system and calorimeters. The tracking system includes a high-resolution vertex detector near the interaction point to achieve excellent impact parameter resolution, followed by a combination of silicon trackers and a large-volume Time Projection Chamber (\Gls{TPC}) as the main tracker~\cite{Behnke2013}. The TPC provides continuous tracking with thousands of measured points per track, yielding very good momentum resolution, on the order of $\delta(1/p_T) \sim 2\times10^{-5}$~GeV$^{-1}$, and efficient track finding even in complex event topologies. The use of a TPC, as opposed to an all-silicon tracker as in SiD, reflects ILD’s emphasis on robust pattern recognition and particle flow algorithms (PFA)~\cite{ILD2020}. 

Surrounding the TPC, ILD has an electromagnetic calorimeter (\Gls{ECAL}) and a hadronic calorimeter (\Gls{HCAL}), both of which are highly segmented to optimize for PFA and energy resolution. In a particle flow approach, the goal is to individually reconstruct particles in a jet by combining tracker measurements, for charged particles, and calorimeter information, for neutral particles, thereby achieving superior energy resolution. ILD’s calorimeters are designed so that the particle tag error, mis-assigning calorimeter hits to the wrong particles in reconstruction, is minimized, with a target jet energy resolution of $\sigma_E/E \sim 3$ to 4\% for 100~GeV jets. This is about a factor of two better than what was achieved at LHC detectors, enabling clean separation of heavy particle decays. 

\begin{figure}[h]
\centering
\includegraphics[width=14cm]{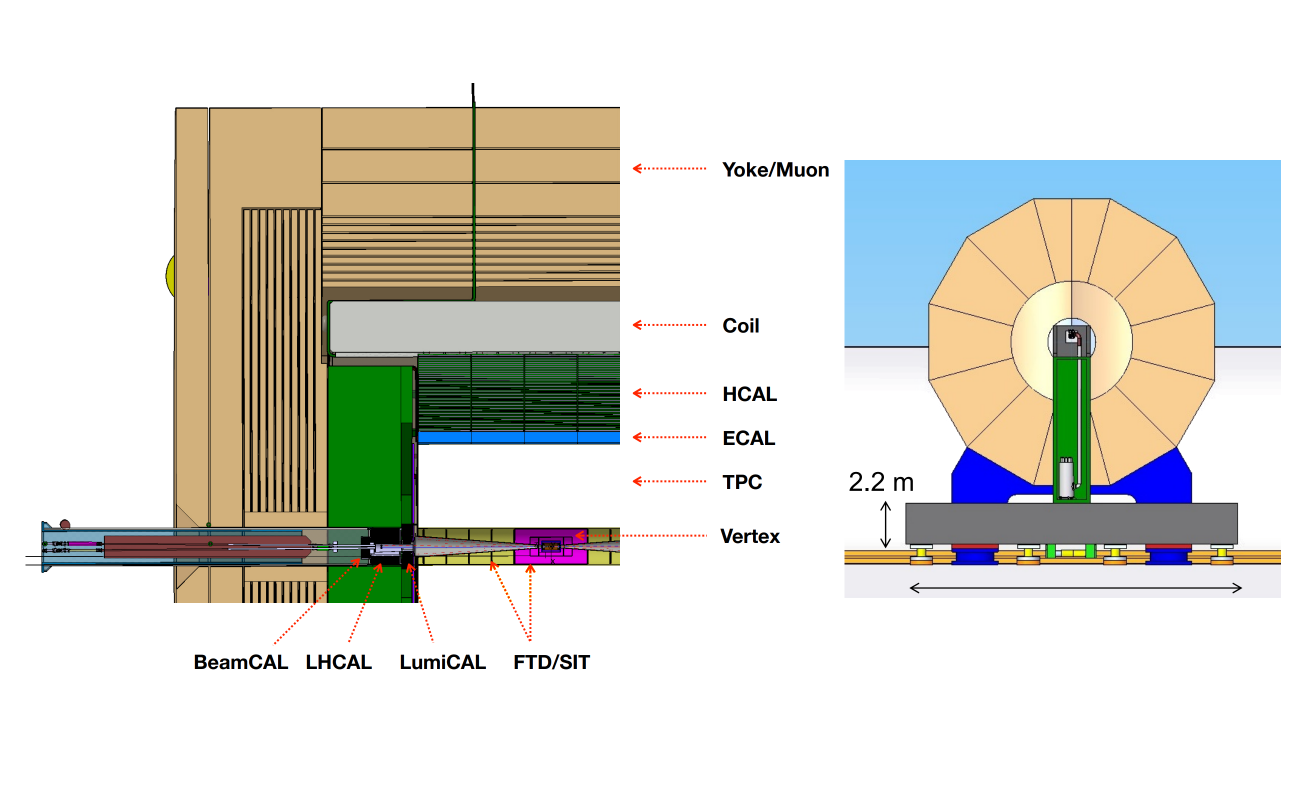}
\caption{(Left) ILD as viewed from a side cross-section with the sub-detector regions labelled. (Right) ILD as viewed from the front, or back, with the concrete base and rail, for the push-pull mechanism, displayed below.}
\label{fig-ILDDiag}       
\end{figure}

The outermost part of ILD includes a muon tracking system to identify muons and provide additional measurement of penetrating particles. The span of this outermost part is large, on the order of 14~m with two views of ILD seen in figure~\ref{fig-ILDDiag}. Making it comparable in scale to a detector like The Compact Muon Solonoid (CMS) but smaller than A Toroidal LHC Apparatus (ATLAS), both of which are detectors at the LHC. The choice of a large radius is driven by the needs of the TPC and the calorimeter granularity. A larger detector yields better separation of particles and a longer lever arm for momentum measurement, at the cost of more material and monetary budget.

ILD is designed to cope with the beam structure of the ILC, which has bunch trains separated by 200~ms intervals, meaning the detector can use power-pulsing to reduce heat and can read out between trains. The beam-induced backgrounds at the ILC, such as pair production in the beam crossing, are relatively moderate compared to LHC. ILD includes a forward calorimeter system, comprised of BeamCal and LumiCal, close to the beampipe to measure small-angle particles, assist with luminosity measurements. These forward detectors ensure hermetic coverage down to very low angles, which is important for measuring low mass t-channel processes, such as Bhabha scattering, and for measuring initial state radiation.

A notable aspect of the ILC experimental plan is the “push-pull” mechanism: since only one interaction region is likely, the ILD and SiD detectors would not run simultaneously but rather alternate taking data by moving in and out of the beam position on a platform~\cite{Behnke2013}. This will be achieved using a rail system, as seen in figure~\ref{fig-ILDDiag}. This approach allows two independent detectors to share the beam time (for example, switching every few weeks or months) without duplicating the collision halls. ILD’s design therefore also considered the engineering of being movable, including a self-shielding structure and alignment systems to ensure it can be repositioned reproducibly. The push-pull scheme is motivated by maximizing the scientific output since two detectors provide cross-checks and complementary designs, reducing systematic uncertainties on key measurements.

In summary, ILD represents a state-of-the-art detector concept tailored for an $e^+e^-$ Higgs factory. Its strengths lie in precision tracking and granular calorimetry, which together enable excellent reconstruction of the complex final states expected in Higgs decays. The performance studies for ILD indicate that it can achieve or exceed the requirements for the ILC physics program. ILD, together with SiD, has been validated in full simulation as part of the ILC Technical Design Report~\cite{Behnke2013} and in subsequent updates~\cite{ILD2020}. While the exact detector to be built would be decided closer to project approval, and could incorporate further technology advances, ILD provides a realistic and robust template. 

\section{Z-pole Measurements} 

One of the major advantages of proposed Higgs factories, especially circular $e^+e^-$ colliders and ILC, should it  dedicate a run, is the ability to produce huge samples of $Z$ bosons for precision measurements at the Z-pole, $\sqrt{s}\approx m_{\text{Z}}$. These measurements of the Z boson’s properties and couplings are of critical importance for refining our understanding of electroweak interactions. The \Gls{LEP} and \Gls{SLC} experiments in the 1990s performed a comprehensive scan of the $Z$ resonance, determining quantities such as the $Z$ boson mass, $m_\text{Z}$, width, $\Gamma_Z$, total and partial decay widths, and asymmetry parameters with high precision. 

The data from LEP and SLD was on the order of $10^7$ $Z$ decays at LEP and $10^5$ polarized $Z$ decays at SLD. The combined analysis of this data forms the cornerstone of precision electroweak tests of the Standard Model still as of writing of this document. However, the future colliders discussed here can far exceed those sample sizes, thereby improving the precision by orders of magnitude, and potentially revealing subtle discrepancies that could hint at new physics. ILC would see to improve the SLD count of polarized $Z$ decays to $10^{9}$, or four orders of magnitude, with the ILC Giga-Z run. The FCC-$ee$ in particular plans a dedicated $Z$ pole run, often dubbed Tera-Z, where it would collect on the order of $10^{12}$ $Z$ decays over a few years of running~\cite{Abada2019}. This staggering statistics, about five orders of magnitude beyond LEP, would reduce statistical uncertainties on many observables towards $10^{-5}$ relative uncertainty.

\begin{figure}[h]
\centering
\includegraphics[width=14cm]{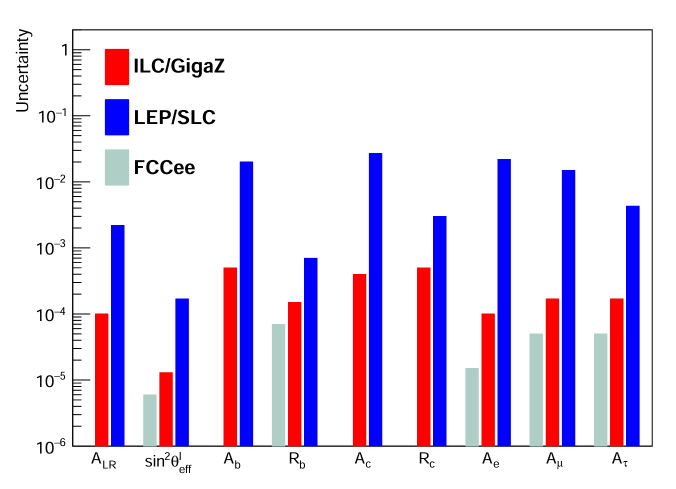}
\caption{Plot of Z pole observables from LEP/SLC, Giga-Z observables for ILC, and Tera-Z observables for FCC-$ee$~\cite{Fujii2019}.}
\label{fig-ZObs}       
\end{figure}

For instance, the LEP measurement of the left-right polarization asymmetry $A_{LR}$, from SLD, or the forward-backward asymmetry $A_{\rm FB}^{\ell}$ for leptons had uncertainties corresponding to $\sim10^{-3}$ in $\sin^2\theta_W$. Both ILC and FCC-$ee$ could, in principle, reach $10^{-5}$ level on these observables after averaging millions of events~\cite{ILC2013}~\cite{Abada2019}. We provide figure~\ref{fig-ZObs} to provide further context on possible precision on Z-pole observables. Note that, due to ILC's beam polarization, there are additional observables. 

The measurement of the $Z$ mass, which is currently $m_Z = 91.1876 \pm 0.0021$~GeV from a combined LEP and SLC analysis, could be improved to the scale of hundreds of keV by either ILC or FCC-$ee$. With limitations coming from the energy calibration precision needing to be known to, at least, 4 ppm, and beam modeling of the experiments to the percent or sub-percent level. Both of which have already been demonstrated as possible for ILC with ILD and will be discussed further in chapter~\ref{ch-EPre}~\cite{Wilson2023}~\cite{Madison2022}~\cite{Madison2023}. The $Z$ width, presently known to $\sim2.3$~MeV accuracy, could be determined to a few tens of keV. As a part of this, the invisible width of the $Z$, which is directly related to the number of light neutrino species under the assumption of no anomalous $Z$ couplings, could be pinned down to 0.002\% or better. Precisely measuring this opens the possibility of additional light neutrino flavors or exotic, incredibly weak, $Z$ couplings.

Crucially, many of these $Z$ measurements test the radiative corrections of the Standard Model. For example, the effective weak mixing angle $\sin^2\theta_W^{\rm eff}$, extracted from asymmetries in $Z$ decays, is sensitive to loop corrections involving the top quark and Higgs boson. The precision of LEP and SLC, around $2\times10^{-4}$ on $\sin^2\theta_W^{\rm eff}$, was enough to indirectly infer both the top quark and Higgs boson masses before their discovery. Pushing this precision by an order of magnitude or more could potentially reveal loop effects of new particles up to scales of tens of TeV, even if those particles are too heavy to be produced directly. Any significant deviation from the Standard Model predictions in the $Z$ pole observables, once both experimental and theoretical uncertainties are shrunk, would be a clear sign of new physics. For example, a deviation in the $Zb\bar{b}$ coupling, parameterized by $R_b$ or the forward-backward asymmetry $A_{\rm FB}^{0,b}$, could indicate the presence of new dynamics affecting third-generation quarks. LEP observed a slight anomaly in $A_{\rm FB}^{0,b}$, though not statistically definitive. A future Giga-Z or Tera-Z measurement could definitively confirm or refute such anomalies and, optimistically, provide a road map for future experiments~\cite{ALEPH:2005ab}.

ILC also has the option of running at the $Z$ pole, often referred to as the Giga-Z, by reducing its beam energy. In the Giga-Z scenario, ILC would have higher luminosity, since the energy is lower, and collect of order $10^9$ $Z$ bosons~\cite{Brau2015}. With polarization of the electron beam, ILC Giga-Z could measure left-right asymmetries extremely precisely. This would already significantly sharpen tests of the electroweak theory. In the Tera-Z scenario the circular colliders, FCC-$ee$ or CEPC, have a far greater potential in terms of statistics. In general, as illustrated by figure~\ref{fig-ZObs}, there is no clear superior experiment. Both experiments have value for improving precision measurements.

\begin{figure}[h]
\centering
\includegraphics[width=14cm]{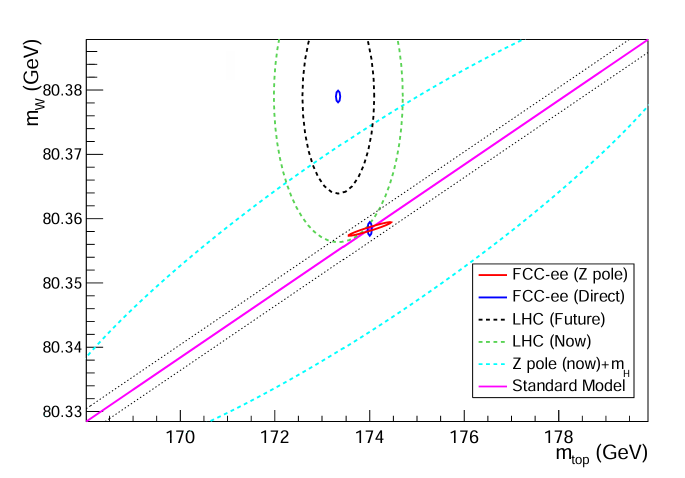}
\caption{Plot of the allowed experimental contours, considering underlying uncertainty, of the W boson and top quark masses. The current LHC measurements do not agree well with the SM but are not precise enough to exclude the SM. Regardless of if the LHC measurements regress after HL-LHC, the plot showcases the importance of increasing precision of mass measurements~\cite{Abada2019}.}
\label{fig-WTop}       
\end{figure}

In addition to the $Z$ pole, future colliders should improve measurements at the $W$ boson pair production threshold. A dedicated run for ILC around 161~GeV would improve precision of $m_{\text{W}}$ to 2~MeV by using a polarized threshold scan~\cite{Wilson2016}. FCC-$ee$ also plans a threshold scan to measure $m_\text{W}$. FCC-$ee$ expects precision on $m_\text{W}$ comparable to $\sim0.5$~MeV, slightly improving over ILC~\cite{Abada2019}. As seen in figure~\ref{fig-WTop}, improving $m_W$ and $m_Z$ precision is critical because together with the top quark mass and Higgs mass, these parameters test the consistency of the Standard Model’s electroweak sector. For instance, one can relate  $m_\text{W}$, $m_\text{Z}$, $\sin^2\theta_W$ and loop corrections with precision measurements to infer deviations caused by new physics, like supersymmetry, etc.. 

Another interesting aspect of a large $Z$ sample is the potential to search for rare or exotic decays of the $Z$. For example, within the SM, lepton flavor-violating decays $Z\to e\mu$ or $Z\to \mu\tau$, which are forbidden in the SM at tree level but allowable at loop level. Decays of the $Z$ into new light particles, such as dark sector candidates or axion-like particles or new neutrino states, could also be explored by looking for small deviations in the $Z$ invisible decay width. Thus, the $Z$ factory capability of future colliders provides not only precision tests but also discovery potential in its own right. In summary, the $Z$ measurements at a Higgs factory will extend the legacy of LEP and SLC to a new level of precision. They are a key part of the physics case for FCC-$ee$ and an optional but valuable part of the ILC program, especially considering that ILC is uniquely posed to make polarized measurements. Indeed, deploying Z-pole runs at both ILC and FCC-$ee$ would repeat the history and successes of LEP and SLC. If successfully executed, these measurements could either further cement the Standard Model or point to new physics via tiny deviations. In either case, they complement the Higgs measurements by providing a broader context of precision electroweak measurements to constrain or reveal new physics.

\section{Higgs Factory Measurements}\label{sec-Higgs}

The central objective of a Higgs factory is to scrutinize the Higgs boson with unprecedented precision. This means measuring its production rates, decay branching fractions, couplings to other particles, and internal properties like mass, width, as precisely as possible. The LHC has established the existence of the Higgs and roughly measured many of its couplings, at the 10 to 20\% so far, expected to improve to the 5–10\% level with the HL-LHC. A dedicated $e^+e^-$ collider can significantly improve on these results. The cleaner environment, lower backgrounds, and well-defined initial state of $e^+e^-$ collisions allow for systematically superior measurements than those at a hadron collider. 

\begin{figure}[h]
\centering
\includegraphics[width=14cm]{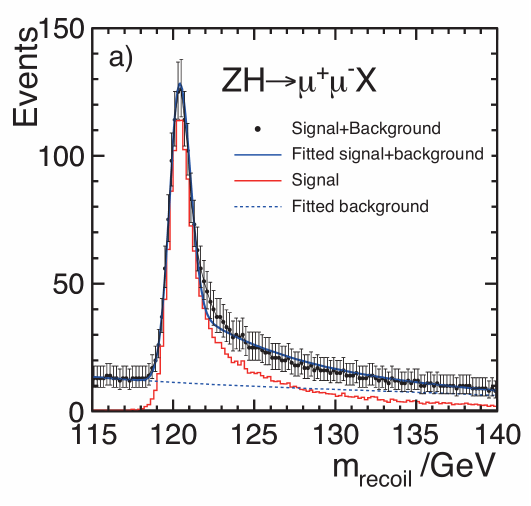}
\caption{Histogram plot of the recoil mass distribution as measured by ILD full simulation at ILC250 for $ZH\to\mu^+\mu^-H$~\cite{asner2018ilchiggswhitepaper}.}
\label{fig-RecMass}       
\end{figure}

Consider Higgs production and couplings in $e^+e^-$ collisions around 250~GeV. Reviewing figure~\ref{fig-HiggsXSecDiag}, we see the most important production process for Higgs is the Higgsstrahlung process, $e^+e^- \to ZH$. By tagging the $Z$, usually with leptonic decays or hadronic decays constrained with modeling and event kinematics, one can identify Higgs production without needing to directly observe the Higgs. The \gls{recoil mass} for $Z$ decay to fermion-antifermion pair (\gls{DiFermion})
\begin{equation}\label{eqn-recmass}
    m_{\text{rec.}} = \sqrt{(s - 2\sqrt{s}E_{\text{ff}} + m_{\text{ff}}^2 )}
\end{equation}
can be computed from the center-of-mass energy and the invariant mass and energy of the difermion. From equation~\ref{eqn-recmass}, and under the assumption that the initial system is characterized well by $\sqrt{s}$ and has no significant boost, one can construct a spectrum of recoil mass values. Under these assumptions the spectrum will show a clear signal of the Higgs mass, as seen in figure~\ref{fig-RecMass}. 

This technique also allows for a direct measurement, in a model-independent way, of \Gls{SM} and \Gls{SMEFT} $ZH$ couplings~\cite{Barklow2017}. At ILC250 and FCC-$ee$, as seen in figure~\ref{fig-SMEFTCoup}, these couplings are expected to be measured to about 0.3\% accuracy~\cite{deblas2024globalsmeftfitsfuture}. By contrast, the LHC cannot perform an absolute Higgs cross section without assuming underlying modeling. Thus $e^+e^-$ data provides a reference point for all relative Higgs measurements done elsewhere. 

Another production mechanism, which dominates at higher energies, is $WW$-fusion, $e^+e^- \to H\nu_e\bar{\nu}e$, where a virtual $W$ from each beam fuses into a Higgs. This process, being dominantly \Gls{tchan}, has a rising cross section with energy, while $ZH$, being dominantly \Gls{schan}, falls off at higher energies. Measuring the $WW$-fusion rate gives direct information on the $HWW$ coupling. Similar to $ZH$ coupling, the results of ILC250 and FCC-$ee$ expect $HWW$ coupling to be measured to about 0.3\%. The difference between the experiments being that ILC can upgrade to higher energies. The combined results of which will approach 0.1\% precision on the $HWW$ and $HZZ$ couplings. 

Confirming these couplings at the per-mille level and that they obey the Standard Model relations is a crucial test of the Higgs mechanism. These couplings, in the SM Lagrangian

\begin{equation}\label{eqn-VVHCouplings}
    \mathcal{L}_{VVH} = gm_{\text{W}}W_{\mu}^+W^{\mu-}H + \frac{gm_\text{Z}}{\cos{\theta_{W}}}Z_\mu Z^\mu H
\end{equation}

depend on their respective vector boson fields and the couplings. In the BSM of 2-Higgs-doublet (\Gls{THDM}) the Higgs field, $H$, of equation~\ref{eqn-VVHCouplings} is promoted

\begin{equation}\label{eqn-HiggsDoubVVH}
    \mathcal{L}_{VVH} = (gm_{\text{W}}W_{\mu}^+W^{\mu-} + \frac{gm_\text{Z}}{\cos{\theta_{W}}}Z_\mu Z^\mu)q_{k1}h_k
\end{equation}

to $q_{k1}$, combinations of the Higgs mixing angles, and $h_k$, the two Higgs fields~\cite{asner2018ilchiggswhitepaper}. In the case where one of these Higgs bosons is dominant over the other then small deviations from equation~\ref{eqn-VVHCouplings} would indicate that a model like equation~\ref{eqn-HiggsDoubVVH} is preferential. We note that this structure is not unique to THDM and can be found in the \Gls{MSSM} too.

\begin{figure}[h]
\centering
\includegraphics[width=14cm]{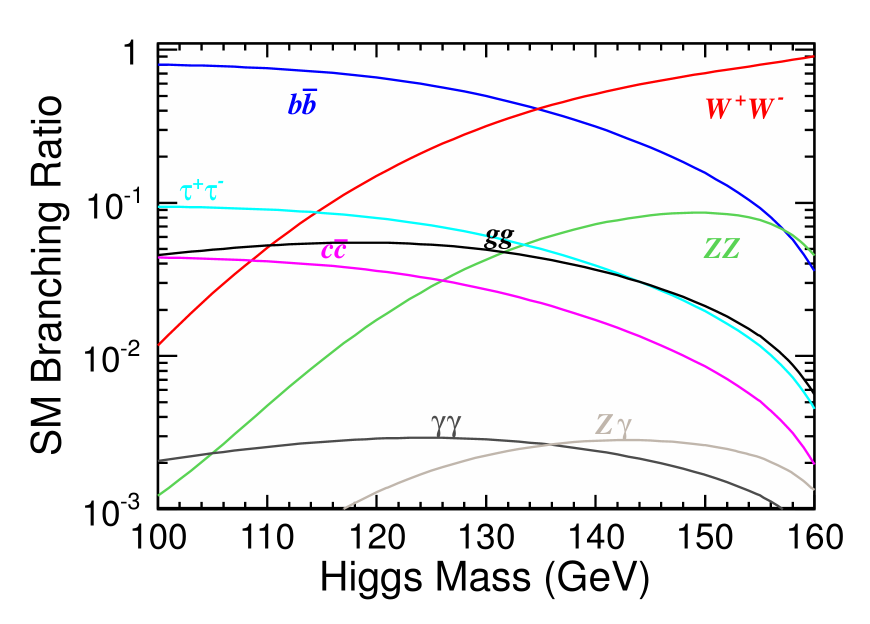}
\caption{Plot of the expected Higgs branching ratios as computed from SM theory for potential Higgs masses. For the currently measured Higgs mass, $m_{\text{H}}\approx$~125~GeV, the $H\to b\bar b$ decay is dominant and the $H \to Z\gamma$ decay is the smallest shown decay. Though other decays, such as to DiMuons, are even smaller~\cite{asner2018ilchiggswhitepaper}.}
\label{fig-HDec}       
\end{figure}

Once a Higgs event is identified, the Higgs decay products can be analyzed to measure branching fractions, often labelled as $\mathcal{B}$. Lepton colliders offer the advantage of relatively low backgrounds for many decay modes. As seen in figure~\ref{fig-HDec}, the dominant decay is $H \to b\bar{b}$, which is about 58\% in the SM. At the LHC, $H\to b\bar{b}$ is challenging to isolate except in associated production with a $W/Z$ or $t\bar{t}H$. At ILC and FCC-$ee$, $H\to b\bar{b}$ events can be tagged with the recoil $Z$ or in $\nu\bar\nu H$ with missing mass techniques. The precision on $\mathcal{B}(H\to b\bar{b})$ is expected to be $\sim1\%$ at ILC250 and improved to $\sim0.5\%$ with higher energy runs~\cite{Fujii2019}. Similarly, $H\to c\bar{c}$ and $H\to gg$ can be measured whereas these are very hard at LHC due to enormous backgrounds. The charm coupling extraction is particularly improved by using charm-tagging algorithms in the clean electron-positron collider environment and by the use of PFA, such as those in ILC's ILD. ILC projections for $Hcc$ coupling uncertainty are on the order of 2\% with full program~\cite{Fujii2019}. The gluon coupling can be determined from the total width subtraction once other dominant modes are known. The precision on $\mathcal{B}(H\to gg)$ is $\sim2\%$ percent at ILC250 and possibly approaching 1\% at FCC-ee, due to larger sample sizes.

The rare decay $H \to \gamma\gamma$, which has a branching ratio $\approx 0.23\%$ as measured by the LHC, is expected to reach 10\% precision at HL-LHC. A Higgs factory can also measure $H\to \gamma\gamma$, although the rate is low compared to the massive vector bosons. The precision might be on the order of 20\% with 250 GeV data alone, but combining higher energy runs and HL-LHC data could improve the global fit to a few percent~\cite{Fujii2019}. FCC-$ee$, assuming it took enough statistics, could measure $H \to \gamma\gamma$ to about 1\% or so~\cite{Abada2019}. This decay is loop-induced, particularly with the top quark and W boson, so it serves as a check on consistency of direct coupling measurements. 

Another loop-induced channel is $H \to Z\gamma$, which is even rarer, $\mathcal{B}(H \to Z\gamma) \sim 0.15\%$. This channel is hard to measure anywhere due to its rarity and background processes, like $ZZ\gamma$, which look very similar if the underlying event is $ZH$. HL-LHC will have limited sensitivity, likely enough to conclusively measure its existence. Future colliders may not dramatically improve precision, expected $\sim10\%$ or so, unless very large data samples, 10s or 100s of $\text{ab}^{\text{-1}}$, are taken.

Decays to leptons like $H\to \tau^+\tau^-$ and $H\to \mu^+\mu^-$ are also of interest. With $\mathcal{B}(H\to \tau\tau)\approx6.3\%$, lepton colliders can cleanly measure decays to \gls{DiTau} using leptonic and 1-prong hadronic tau decays. ILC projections are 2–3\% precision on the $H\tau\tau$ coupling~\cite{Fujii2019}, compared to ~5\% at HL-LHC. The $H\to \mu\mu$, 0.022\% BR, is very challenging even with millions of Higgs decays. HL-LHC aims to conclusively observe it and measure it to 20 to 30\%. An $e^+e^-$ collider, with a run with 10s or 100s of $\text{ab}^{\text{-1}}$, might confirm this observation and improve the global fit to, perhaps, 15\%~\cite{Abada2019}. A second option is to construct a muon collider Higgs factory. A muon collider would have acces to direct \Gls{schan} production, since any Higgs propagator in the \Gls{schan} would include the $H\to\mu\mu$ vertex. We leave further discussion of muon colliders to other studies.

Measuring the visible Higgs decay channels is crucial: an unexpected deviation in the total width could indicate additional decay channels, e.g. Higgs decays to invisible. Higgs factories also plan to search for these invisible decays, e.g. $H \to$ dark matter candidates, like neutralinos. By looking at the recoil $Z$ in $ZH$ events and noting if the Higgs decay products are missing, one can directly measure $\mathcal{B}(H \to \text{invisible})$. The sensitivity is very high because backgrounds for $Z + \text{nothing}$ final states are low, with  $ZZ\to Z\nu\bar\nu$ being a significant, but small, background. At ILC, the 95\% CL upper limit on $\mathcal{B}(H \to \text{invisible})$ could be as low as 0.3\% with 250 GeV data~\cite{Fujii2019}. FCC-$ee$ would be able to confirm this measurement at a similar precision~\cite{Abada2019}. The current prospects for the HL-LHC envision a few percent at best. A Higgs factory would be able to discover an invisible mode, if one exists at a fraction of a percent level, or strongly limit theories where the Higgs acts as a portal to hidden sectors.

The Higgs self-coupling $\lambda_{HHH}$, probed via double Higgs production, is a crucial parameter for understanding the scalar potential shape. As seen in figure~\ref{fig-HiggsXSecDiag}, at 250~GeV double Higgs production is essentially negligible. Only at higher energies does it become somewhat accessible. For example, even at 500~GeV, $ZHH$ and $HH\nu_e\bar{\nu_e}$ have a very small cross-section. A Higgs factory at 500~GeV thus cannot significantly measure $\lambda_{HHH}$ directly, but an indirect constraint could be obtained by including the lack of deviation in single Higgs processes~\cite{Fujii2019}.

\begin{figure}[h]
\centering
\includegraphics[width=14cm]{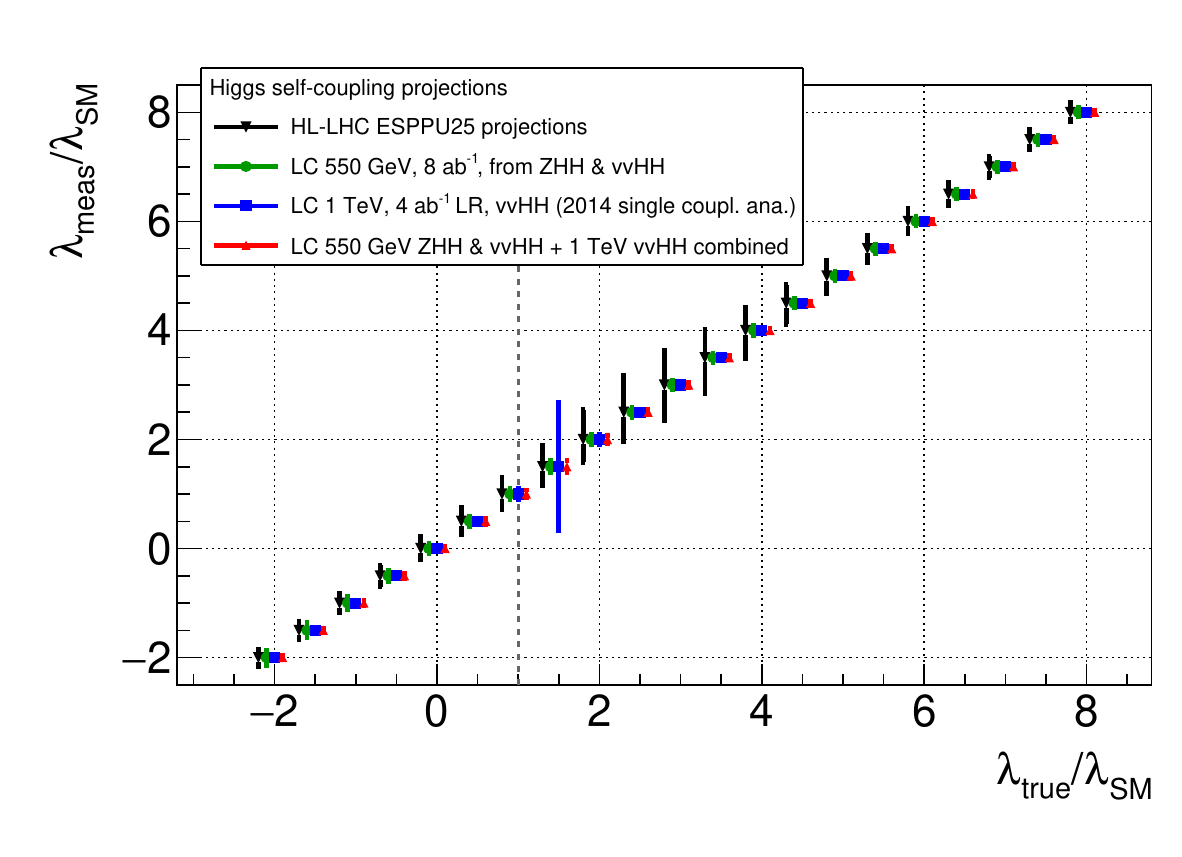}
\caption{Plot of various plausible values of the Higgs self-coupling precision, $\lambda_{HHH}$ or $\lambda$ here, as compared to its Standard Model expectation ~\cite{torndal2023,LCVision}.}
\label{fig-TripHiggs}       
\end{figure}

With a TeV scale ILC or CLIC upgrade, double Higgs production, $ZHH$ and $HH\nu_e\bar{\nu_e}$, yields hundreds to thousands of events, enough to get 10 to 20\% measurement of $\lambda_{HHH}$~\cite{Fujii2019}~\cite{torndal2023}. This is also more complicated as the underlying precision on $\lambda_{HHH}$ depends on its value. Various scenarios of precision, taken with-respect-to the SM expectation, can be seen in figure~\ref{fig-TripHiggs}.

CLIC at 3 TeV would significantly improve this to around 10\% precision with 5 ab$^{-1}$ of data~\cite{CLIC2018}. FCC-ee cannot directly produce two Higgs due to its max energy of 365~GeV being below the $\sim450$ GeV threshold for $ZHH$. So in the near term, among $e^+e^-$ Higgs factories, only a TeV scale linear collider, like later ILC or CLIC stages, would significantly pin down the Higgs self-coupling. These measurements from ILC and/or CLIC would be valuable information of the Higgs potential. It could reveal anomalies if the true self-coupling deviates, as some BSM physics predicts.

By the time of constructing a Higgs factory the Higgs boson mass will be very well measured already by the LHC, to around 0.1 GeV. Lepton colliders can measure it in the $ZH$ recoil spectrum with precision of a few MeV at ILC and/or FCC-$ee$~\cite{Abada2019}. This is a clean determination that is independent of decay channels and underlying modeling. Such precision is important for Standard Model consistency checks. The spin and parity of the Higgs, currently measured as being dominantly $0^+$ by LHC experiments, can be further tested via angular distributions in $ZH$ events or in decay products with better precision. Any small admixture of CP-odd in the Higgs coupling, which could happen if Higgs is not a perfect CP eigenstate as predicted by some BSM physics, could be probed by observables like the angular correlations of decay products. With polarized $e^+e^-$ beams, one can construct CP-odd asymmetries in $e^+e^- \to ZH$ to constrain any CP-violating $HZ\gamma$ or $HZZ$ form-factors. This puts ILC and CLIC in an advantageous position to reach sensitivities well below the percent level for any CP-violating couplings, improving what is achievable at the LHC~\cite{Fujii2019}.

All of the proposed Higgs factories would greatly advance our knowledge beyond what the LHC can do. Their results would be complementary due to different underlying methods and systematics. Constructing a Higgs factory is seen by the community as the essential next step to fully explore the nature of the Higgs boson and expand our knowledge of the universe of the fundamental. Though, considering this section, it is clear that constructing multiple Higgs factories is even better.












---------------------
\chapter{Theory and Techniques of Relevant Processes}\label{ch-Theory}

For the sake of simulating events with massive fermions and beam polarization, we must consider the different polarization states of the initial state beam particles. For massive fermions this includes transverse and longitudinal polarization states. However, from here forward, we will not discuss transverse polarization and instead focus on longitudinal polarization. Assume that any polarization is longitudinal unless specified otherwise. 

From the polarization states of the initial particles we have four different possible cross-sections given by the helicity combination of the polarization of each beam. This can be done by relating the electron longitudinal beam polarization, $P_{-}$, and the positron longitudinal beam polarization, $P_{+}$, to each cross-section. We use the standard notation such that $+1$ is for a purely right-handed beam and $-1$ is for a purely left-handed beam. The resulting weights for each helicity of cross-section

\begin{equation}\label{eqn-polcombo}
\begin{gathered}
    \sigma_{\text{RR}} \propto \frac{1+P_-}{2}\frac{1+P_-}{2}\\
    \sigma_{\text{LL}} \propto \frac{1-P_-}{2}\frac{1-P_-}{2}\\
    \sigma_{\text{RL}} \propto \frac{1+P_-}{2}\frac{1-P_-}{2}\\
    \sigma_{\text{LR}} \propto \frac{1-P_-}{2}\frac{1+P_-}{2}\\
\end{gathered}
\end{equation}

is entirely dependent on the chosen beam polarizations~\cite{Moortgat_Pick_2008}. Given a desired combination of $P_-$ and $P_+$, we may use an event generator to get events. However, this would not be time efficient if we plan on testing lots of combinations. Instead, generating each of the various helicity permutations separately and then using equation~\ref{eqn-polcombo} to calculate their weights, is more efficient. Such that there is a repository of events for each of $\sigma_{LL}$, $\sigma_{RR}$, $\sigma_{RL}$ and $\sigma_{LR}$ that can be used to generate any permutation of beam polarization. To compare equation~\ref{eqn-polcombo} to the unpolarized case, we consider a vector boson mediated process. In such case, the relationship of cross-sections 

\begin{equation}\label{eqn-sigpol}
    \sigma_{\text{pol.}} = \sigma_{\text{unpol.}}(1-P_-P_+)[1-P_{\text{eff.}}A_{\text{LR}}]
\end{equation}

depends on the effective polarization and left-right asymmetry factors. Such that 

\begin{equation}\label{eqn-peff}
    P_{\text{eff.}} = \frac{P_- - P_+}{1-P_-P_+}
\end{equation}

the effective polarization is dependent only on the individual polarizations. The left-right asymmetry

\begin{equation}\label{eqn-alr}
    A_{\text{LR}} = \frac{\sigma_\text{LR}-\sigma_\text{RL}}{\sigma_\text{LR}+\sigma_\text{LR}}
\end{equation}

depends on the ratio of cross-sections from the $\text{J}=1$, spin 1, helicity combinations. The value of equation~\ref{eqn-alr} for a pure QED process is zero. In said limit, equation~\ref{eqn-sigpol} does not depend on equations~\ref{eqn-peff} or ~\ref{eqn-alr} 

\begin{equation}\label{eqn-sigpolQED}
    \sigma_{\text{QED} \space, \space \text{pol.}} = \sigma_{\text{unpol.}}(1-P_-P_+)
\end{equation}

and therefore is only dependent on the beam polarizations. To leading order, equation~\ref{eqn-sigpolQED} is what we expect for diphoton production and similar QED processes. This makes a simple example for showing the benefit of polarized beams and, for simulation purposes, the benefit of separating the simulation of each helicity combination. Once an integrated luminosity and beam polarization combination is chosen, then equation~\ref{eqn-polcombo} can be used to choose the correct amount of each combination and also calculate a weighted cross-section. Thus, the user only needs to generate events four times, once for each helicity combination, and then they can generate whatever combination from those.

It is well known that the cross-section for diphotons changes with the polarization of the beams such as in equation~\ref{eqn-sigpolQED}. This is not entirely true once ISR is allowed. With the presence of ISR, there can be helicity conserving and helicity changing ISR. This allows for a helicity flip of the normally forbidden helicity combinations into an allowed combination.

For demonstration we will derive the ratio of helicity changing ISR to helicity conserving ISR within the collinear limit. This is a common limit for ISR calculations as it leads to one leading order term. From this the differential cross-section

\begin{equation}\label{eqn-isrdiff}
    \frac{d\sigma}{dE_{\gamma}\space d\text{cos}\theta} \propto 1 + \frac{m_e^2}{E_\gamma^2(1-\cos\theta)^2}
\end{equation}

depends on the helicity conserving term, which is unity at leading order, and the helicity changing term~\cite{Kuraev:1985hb}. The helicity changing terms depend on the mass of the radiating particle, here assumed electron mass, and the energy of the radiated photon, here $E_{\gamma}$, and the polar angle of the photon with-respect-to the electron. Using equation~\ref{eqn-isrdiff}, the ratio of the helicity conserving ISR and helicity changing ISR

\begin{equation}\label{eqn-fliprat}
    \frac{\sigma_{\text{flip}}}{\sigma_{\text{cons.}}} \bigg|_{e}\approx \frac{m_e^2}{E_\gamma^2(1-\cos\theta)^2}
\end{equation}

follows from the second term of equation~\ref{eqn-isrdiff}. For typical values expected in diphoton production at center-of-mass energies of 250 GeV this ratio is expected to be $\sim10$~ppm or smaller. Checking this result with the general purpose \Gls{MCEG} \Gls{WHIZARD} finds that the helicity flip ISR contribution is of $\sim100$~ppb. Therefore, in this higher energy regime, it is safe to assume that processes like diphoton production can be effectively turned on and off by changing beam polarization. The use of pure $\text{LL}$ and $\text{RR}$ beam helicity combinations will turn off diphoton and similar vector boson vertex processes. This is advantageous for both precision measurements and searches for new physics~\cite{Moortgat_Pick_2008}. The background processes for diphoton, such as Bhabhas, still contribute in the $\text{LL}$ and $\text{RR}$ combinations, so these combinations can be used to characterize backgrounds. This modelling is then used to subtract out the backgrounds when data is taken at $\text{LR}$ and $\text{RL}$ helicity combinations. Therefore, allowing for more precise, and less correlated, signal tagging.

We note that this ability to effectively control helicity contributions is only true for polarized electron-positron colliders. For a polarized muon collider equation~\ref{eqn-fliprat} is

\begin{equation}\label{eqn-flipmu}
    \frac{\sigma_{\text{flip}}}{\sigma_{\text{cons.}}} \bigg|_{\mu} \approx \frac{m_\mu^2}{E_\gamma^2(1-\cos\theta)^2}
\end{equation}

dependent on the muon mass, which is much larger than the electron mass. The result of equation~\ref{eqn-flipmu}, is that the helicity flip ISR is comparable to the helicity conserving ISR even at center-of-mass energies of 250 GeV or greater. This means that a collider using polarized muon beams cannot effectively turn on or off different helicity contributions as the ISR flips will contaminate the sampling. This contamination is also angular dependent, with a preference for the forward region, so it would need to be carefully accounted for. Therefore, for simplicity and precision purposes, polarized electron and positron beams are preferred. Especially when looking at processes like $X^0\gamma$, where the measurements are being done on polarization dependent radiative processes.

\section{Bhabha Scattering}\label{sec-BhaThe}
Bhabha scattering refers to the electron-positron scattering process $e^+(p_1) e^-(p_2) \to e^+(p_3) e^-(p_4)$. At lowest order two Feynman diagrams contribute: an \Gls{schan} diagram and a \Gls{tchan} diagram. Due to the complex geometry in QFT, these diagrams interfere. Yielding matrix elements, $\mathcal{M}$, that are squared and then used to compute a differential cross-section. This differential cross-section then has contributions from the two diagrams and their interference. The $t$-channel matrix elements can be written

\begin{equation}
    \mathcal{M} = \frac{e^2}{t}(\bar{v}_{\text{p1}}\gamma^\mu v_{\text{p3}})(\bar{u}_{\text{p4}}\gamma_\mu u_{\text{p2}})
\end{equation}\label{eqn-bhabhaTMat}

where the matter spinors are written using $v$ and the anti-matter spinors are written using $u$ with subscripts denoting their four-momenta. We use the Mandelstam variables of $s=(p_{1}+p_{2})^2$, $t=(p_{1}-p_{3})^2$, and $u=(p_{1}-p_{4})^2$ to simplify. We can promote the dirac matrix, $\gamma_\mu$, to exclusively handle left-handed or right-handed contributions by use of

\begin{gather}
    \gamma_\mu \mathcal{P}_\text{L} = \gamma_\mu \frac{1-\gamma_5}{2}\\
    \gamma_\mu \mathcal{P}_\text{R} = \gamma_\mu \frac{1+\gamma_5}{2}
\end{gather}\label{eqn-projOp}

where the projection operators of $\mathcal{P}_\text{R}$ and $\mathcal{P}_\text{L}$ are for projecting out the right-handed and left-handed polarization components respectively. Then we can rewrite equation~\ref{eqn-bhabhaTMat} using equation~\ref{eqn-projOp} to decompose into 

\begin{equation}
    \mathcal{M}_{\text{i,j}} = \frac{e^2}{t}(\bar{v}_{\text{p1}}\gamma^\mu \mathcal{P}_\text{i} v_{\text{p3}})(\bar{u}_{\text{p4}}\gamma_\mu\mathcal{P}_\text{j} u_{\text{p2}})
\end{equation}\label{eqn-bhabhaTMat2}

the different helicity combinations and their contributions. From the structure of equation~\ref{eqn-bhabhaTMat2} we can see that the final state and initial state particles of the same charge must have the same helicity because the opposite helicity combinations vanish. We prove this by squaring and summing over all possible helicity combinations

\begin{align}
    \sum_{\text{i,j,k,l}}|\mathcal{M}|^2 = \frac{e^4}{t^2}\text{Tr}\left[ 
\sum_\text{i} (\bar{v}_\text{p1} v_\text{p1}\gamma^\mu \mathcal{P}_\text{i}) \sum_\text{j} (\bar{v}_\text{p3} v_\text{p3}\gamma^\nu \mathcal{P}_\text{j}) \right] \\
\text{Tr}\left[ 
\sum_\text{k} (\bar{u}_\text{p2} u_\text{p2}\gamma_\mu \mathcal{P}_\text{k}) \sum_\text{l} (\bar{u}_\text{p4} u_\text{p4}\gamma_\nu \mathcal{P}_\text{l}) \right]
\end{align}\label{eqn-bhabhaTMat3}

and then use trace technology to simplify. By the polarization projections, the only non-vanishing terms of equation~\ref{eqn-bhabhaTMat3} will be where $\text{i}=\text{j}$ and $\text{k}=\text{l}$. Terms where the same charge particles undergo a helicity flip will be forbidden, at least at leading order. Using completeness relations and trace identities we can simplify this to

\begin{equation}
    \sum_{\text{hel.}}|\mathcal{M}|^2 = \frac{32e^4}{t^2}\left[ (p_3\cdot p_4)(p_1\cdot p_2) + (p_3 \cdot p_2)(p_1 \cdot p_4) - m_{\text{e}}^2(p_4 \cdot p_2) - m_{\text{e}}^2(p_3 \cdot p_1) + 2m_e^4  \right]
\end{equation}\label{eq:bhabha-tchan}

be in terms of the four-momentum, the electron mass, after being summed over allowed helicity combinations. Using the high-energy limit we can simplify equation~\ref{eq:bhabha-tchan} 

\begin{equation}
    \sum_{\text{hel.}}|\mathcal{M}|^2 = 32e^4\frac{s^2 + u^2}{t^2}
\end{equation}\label{eq:bhabha-tchan2}

to only depend on the Mandelstam variables.

We can repeat this process for the $s$-channel contribution by writing

\begin{equation}
    \mathcal{M} = \frac{e^2}{s}(\bar{v}_{\text{p1}}\gamma^\mu u_{\text{p2}})(\bar{u}_{\text{p4}}\gamma_\mu v_{\text{p3}})
\end{equation}\label{eqn-bhabhaSMat}

the matrix element out. We see that the structure is slightly different, coupling the initial and final states together instead of the shared charge states together, and that the denominator is now $s$ instead of $t$. We can proceed with equation~\ref{eqn-bhabhaSMat} similarly until we get to

\begin{align}
    \sum_{\text{i,j,k,l}}|\mathcal{M}|^2 = \frac{e^4}{s^2}\text{Tr}\left[ 
\sum_\text{i} (\bar{v}_\text{p1} v_\text{p1}\gamma^\mu \mathcal{P}_\text{i}) \sum_\text{j} (\bar{u}_\text{p2} u_\text{p2}\gamma^\nu \mathcal{P}_\text{j}) \right] \\
\text{Tr}\left[ 
\sum_\text{k} (\bar{u}_\text{p4} u_\text{p4}\gamma_\mu \mathcal{P}_\text{k}) \sum_\text{l} (\bar{v}_\text{p3} v_\text{p3}\gamma_\nu \mathcal{P}_\text{l}) \right]
\end{align}\label{eqn-bhabhaSMat3}

the helicity sum. From equation~\ref{eqn-bhabhaSMat3} we can see that the helicity of the initial particles and final state particles must be the same, but that they do not have to be otherwise related. Since the helicity of anti-matter is flipped with respect to that of matter, this means that we only have opposite sign contributions in the initial and final states. Using the same process as before, and the same high-energy limit

\begin{equation}
    \sum_{\text{hel.}}|\mathcal{M}|^2 = 32e^4\frac{t^2 + u^2}{s^2}
\end{equation}\label{eq:bhabha-schan}

we can, like before, write equation~\ref{eqn-bhabhaSMat3} in terms of Mandelstam variables. Combining equation~\ref{eq:bhabha-schan} with equation~\ref{eq:bhabha-tchan2} to write out

\begin{equation}
    \sum_{\text{hel.}}|\mathcal{M}|^2 = 32e^4\left[\frac{t^2 + u^2}{s^2} + \frac{s^2 + u^2}{t^2} + \frac{2u^2}{st}\right]
\end{equation}\label{eq:bhabha-all}

the combined solution which includes a term from interference. The resulting Born-level differential cross-section is then

\begin{equation} 
\frac{d\sigma}{d\Omega}\Big|_{\text{Born}} = \frac{\alpha^2}{64e^4s}\sum_{\text{hel.}}|\mathcal{M}|^2 = \frac{\alpha^2}{2s}\left(\frac{u^2 + s^2}{t^2} + \frac{u^2 + t^2}{s^2} + \frac{2u^2}{st}\right) 
\label{eq:bhabha-born} 
\end{equation} 

written to only depend on the Mandelstam variables and the fine-structure constant. Expanding equation~\ref{eq:bhabha-born} in terms of the scattering angle $\theta$ yields 

\begin{equation} 
\frac{d\sigma}{d\Omega}\Big|_{\text{Born}} = \frac{\alpha^2}{4s}\left(3+\cos^2\theta + \frac{8 + (1+\cos\theta)^2}{\cos\theta-1} + \frac{4}{\sin^4(\theta/2)}\right) 
\label{eq:bhabha-born2} 
\end{equation} 

a characteristic $(1+\cos^2\theta)$ term from the $s$-channel and two terms that divergence at small angles, from the $t$-channel. This limit assumed that no helicity flips were possible, which is not true since the electron has mass. In particular, we choose to use Dirac spinors here

\begin{equation}\label{eq:DiracSpinor}
u_\text{L} = \sqrt{E + m}
\begin{pmatrix}
\chi_\text{L} \\
\frac{\vec{\sigma} \cdot \vec{p}}{E + m} \chi_\text{L}
\end{pmatrix}
\end{equation}

and see that a left-handed spinor would include an off-handed component that is dependent on the Pauli matrix vector, $\vec{\sigma}$, the momenta and mass. From equation~\ref{eq:DiracSpinor}, we can derive the ratio of helicity flip to helicity conserving, in the high-energy limit, is

\begin{equation}
    \frac{\mathcal{P}_\text{R}u_\text{L}}{\mathcal{P}_\text{L}u_\text{L}} \approx \frac{m^2}{s}
\end{equation}

dependent on $s$. Including this term clearly leads to increasingly small corrections at increasingly higher center-of-mass energies. Still, we write out the correction for a massive electron with a single helicity flip in equation~\ref{eq:bhabha-born2}

\begin{equation} 
\frac{d\sigma}{d\Omega}\Big|_{\text{Born}} = \frac{\alpha^2}{4s}\left(1+\frac{m_\text{e}^2}{s} \right)\left(3+\cos^2\theta + \frac{8 + (1+\cos\theta)^2}{\cos\theta-1} + \frac{4}{\sin^4(\theta/2)}\right) 
\label{eq:bhabha-born4} 
\end{equation} 

as having an additional correction to the overall cross-section. This correction does not adjust the angular dependence of the differential cross-section.

\begin{figure}[h]
\centering
\includegraphics[width=10cm]{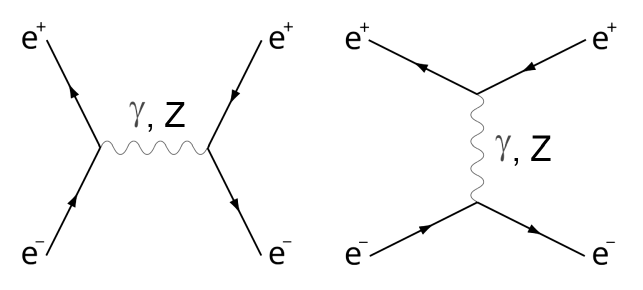}
\caption{Bhabha scattering Feynman diagrams at leading order in the Standard Model.}
\label{fig-bhadiag}       
\end{figure}

At wide angles, the contributions from annihilation and scattering are of comparable magnitude. These events, large angle Bhabha scattering or \Gls{LABS}, are a valuable tool for precision QED and electroweak measurements due to this overlap and their high cross-section. Due to the presence of the Z boson, as seen in figure~\ref{fig-bhadiag}, LABS contains electroweak effects at leading order.

At $\sqrt{s} \ll m_\text{Z}$, Bhabha scattering is dominantly photon exchange so equation~\ref{eq:bhabha-born} provides a good approximation of the differential cross-section. At higher energies, especially near the Z-pole, $\sqrt{s}\sim M_Z$, the Z boson exchange becomes considerable. We will cover this later in section~\ref{sec-LABS}.

At $\sqrt{s} \gg m_\text{e}$, radiative corrections from the emission of photons, and from internal photon exchange, become considerable. These are typically broken down into two classifications, hard photon corrections and soft photon corrections. For LABS soft photon corrections are not as important as the change in angle is not large enough to make the resulting Bhabha scattering event large angle. Instead, LABS has hard photon corrections. With the most common being radiative return to $m_\text{Z}$ or, less common, radiative return to $m_\text{e}$. For SABS soft photon corrections and hard photon corrections are important. Soft photon corrections are important as the small angle deviation means that the Bhabha scattering even can still be small angle. For hard photon corrections there can be multiple hard photon corrections that balance out, resulting in the overall scattering angle still being small. We cover these effects, and more, in the next section.

Over the decades, many theoretical and computational tools have been developed to predict Bhabha scattering with high precision. Notable examples include BHLUMI, a \Gls{MCEG} tailored for SABS, with multi-photon radiative corrections exponentiated to achieve $\sim0.1\%$ accuracy~\cite{BHLUMI}. A counterpart for LABS is BHWIDE, which emphasizes electroweak corrections while also striving for high accuracy~\cite{BHWIDE}. Going forward, we will use these two simulation tools for the generation of SABS and LABS as they are well regarded, and tested, for this purpose. In the following subsections, we will go further in depth on the two different regimes of Bhabha scattering, SABS and LABS, as well as derive a new, and interesting, regime of Bhabhas.

\subsection{Small Angle Bhabha Scattering (SABS)}\label{sec-SABS}
Small angle Bhabha scattering (SABS) refers to the regime where both the $e^+$ and $e^-$ emerge at shallow angles relative to the beam axis. They are typically measured within the forward detectors, such as a Luminosity Calorimeter (LumiCal) or Beam Calorimeter (BeamCal). For detectors with Forward Tracker Detectors (FTD) they may also be measured there. Though we note, and will demonstrate later, that tracker measurements degrade in the forward region due to the small radius of curvature and short lever arm. Typical scattering angles for SABS are tens of milliradians, or less than 5 degrees. At leading order, these events are dominated by the $t$-channel diagram. For electron-positron colliders, SABS is typically the dominant physics process. The SABS cross-section is typically a couple orders of magnitude larger than any other physics process. For example, for ILC250, using a forward calorimeter with an inner acceptance of 1 degree, this is of order 10 nb. The next largest processes are of order 10 pb in cross-section.  As another comparison, LEP's OPAL experiment had a SABS sample at the Z-pole that was roughly 79 nb for an inner acceptance of 1.1 degrees~\cite{OPALLumi}. For this reason, SABS is considered an ideal collider luminosity process because the large sample size can be used to minimize statsitical uncertainty.

We can derive the differential cross-section in this regime by using the small angle approximation for equation~\ref{eq:bhabha-born}

\begin{equation}
\begin{gathered}
\frac{d\sigma}{d\Omega}\Big|_{\text{Born}} \approx \frac{\alpha^2}{4s}\left(\frac{8 + (1+\cos\theta)^2}{\cos\theta-1} + \frac{4}{\sin^4(\theta/2)}\right)
\end{gathered} \label{eq:bhabha-born3}
\end{equation} 

there is a sharp rise, of $\sim\theta^{-4}$, of the differential cross section at small angles. If we integrate equation~\ref{eq:bhabha-born3}

\begin{equation}
\begin{gathered} 
\sigma_{\text{Born}} = \frac{\pi\alpha^2}{2s}\bigg( \frac{\cos^2\theta}{2} + \frac{2\cos\theta}{\sin^2\theta} + 3\cos\theta 
+ \log((1-\cos\theta)^{-1}) \\ + 12\log(\cos\theta-1) + \log(\cos\theta+1) - \frac{7}{2} \bigg)\\
\approx \frac{\pi\alpha^2}{2s}\left[\frac{2}{\theta^2}+\log\left(\frac{2}{\theta^2}\right)\right]
\end{gathered}\label{eqn-bhabha-sig}
\end{equation}

we see that there are a few terms that dominate at small angles and that the cross-section goes as $\sim\theta^{-2}$. This is expected as the integration contributes a factor of $\sin\theta$ to the $\sin^{-4}(\theta/2)$ term, and then integrates the term a-la power law. 

\begin{figure}[h]
\centering
\includegraphics[width=12cm]{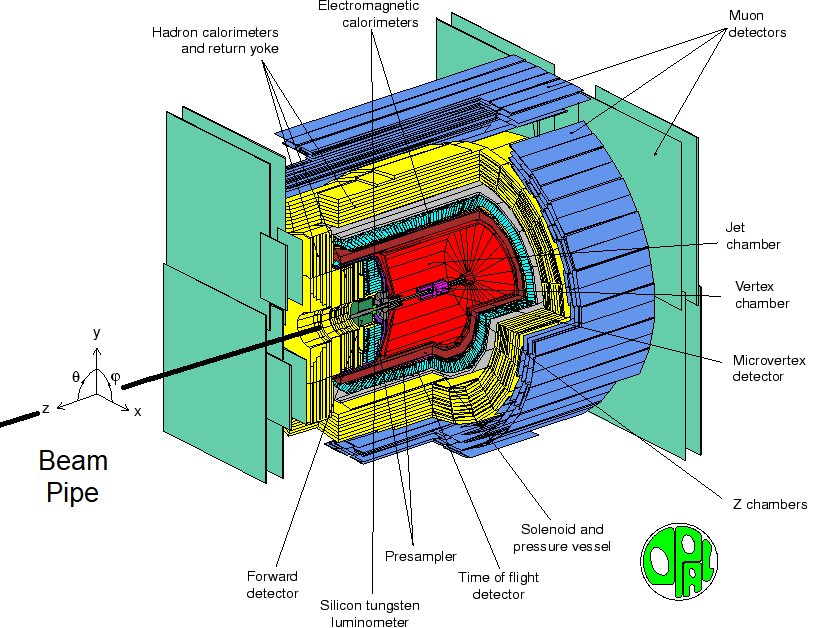}
\caption{Diagram of the OPAL detector that was used as one of the interaction points of the LEP experiment. The beam pipe geometry of OPAL resulted in $\tinn$ of roughly 1 degree.}
\label{fig-OPALDet}       
\end{figure}

The extrapolation of the cross-section dependence on angle for small angles is important because, in reality, SABS must be measured by a detector. Despite the aim of collider detectors to be fully hermetic, therefore covering all angles and measuring all particles, this is simply not true. For instance, long lived neutral particles, like neutrinos and neutrons, are poorly measured, if at all. However, this is not a concern here. There is an inner acceptance angle, $\tinn$, that defines the smallest angle that the detector can fully measure things at. As seen in figure~\ref{fig-OPALDet} at LEP's OPAL detector, this cannot be extended to $0^{\circ}$ because collider detectors require a beam pipe to go through the cylindrical center of the detector.
Due to the imperfect hermiticity of collider detectors there is a contribution to the uncertainty on the cross-section from $\tinn$ which will follow the trend established in equation~\ref{eqn-bhabha-sig}. We use equation~\ref{eqn-bhabha-sig}

\begin{equation}\label{eqn-SABS_Inner_Unc}
    \delta\sigma(\theta_{\text{inner}}) \propto \left[ \frac{4}{\tinn^3} + \frac{2}{\tinn} \right] \delta\theta
\end{equation}

to show that the dependence of the uncertainty in the SABS cross-section on the uncertainty of $\theta$, $\delta\theta_{\text{inner}}$, goes as $\sim\tinn^{-3}$ at leading order, with a small correction from $\sim\tinn^{-1}$. If we consider that an experiment would likely want to maximize the SABS cross-section in order to minimize the statistical uncertainty on the SABS cross-section then equation~\ref{eqn-bhabha-sig} shows that we must reduce $\tinn$. However, this is not an easy design consideration because doing so, by equation~\ref{eqn-SABS_Inner_Unc}, will put a stricter design requirement on the metrology of the detector from $\delta\tinn$. For example, at LEP, the various experiments chose values of $\tinn$ that were $\sim$1$^\circ$ to $\sim$3$^\circ$. With these values of inner acceptance the LEP experiments achieved a systematic uncertainty of about $0.05$--$0.1\%$~\cite{Brock:1996ty}~\cite{OPALLumi}.

\begin{figure}[h]
\centering
\includegraphics[width=12cm]{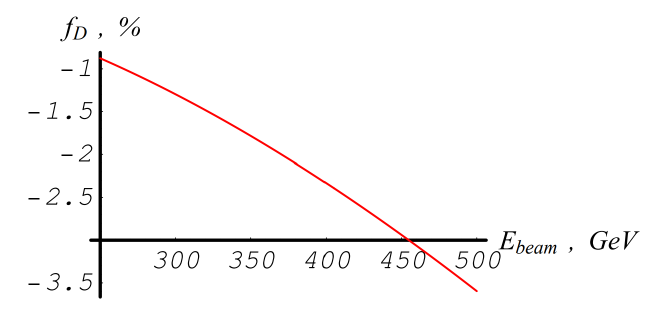}
\caption{Plot of the double-spin asymmetry, $f_\text{D}$, as a function of the electron (positron) beam energy. Figure source~\cite{grahFCAL}.}
\label{fig-DoubleAsym}       
\end{figure}

\begin{figure}[h]
\centering
\includegraphics[width=12cm]{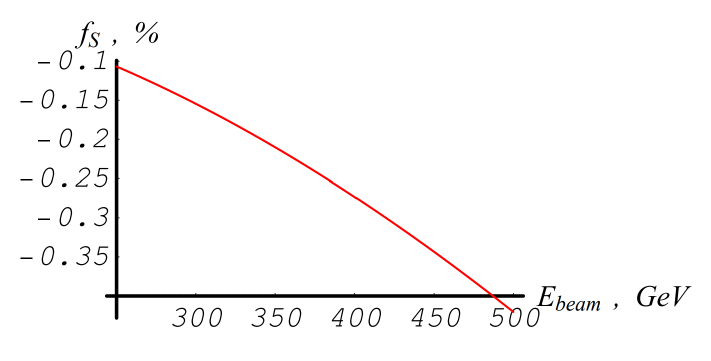}
\caption{Plot of the single-spin asymmetry, $f_\text{S}$, as a function of the electron (positron) beam energy. Figure source~\cite{grahFCAL}.}
\label{fig-SingleAsym}       
\end{figure}

At leading order the beam polarization dependence of SABS is zero. When including higher order corrections this is no longer true. We may approximate the beam polarization dependence
\begin{equation}\label{eqn-SABSPol1}
    \sigma = \sigma_0(1+P_-P_+f_{\text{D}})
\end{equation}
in terms of the double-spin asymmetry, $f_{\text{D}}$, and the leading order cross-section, $\sigma_0$~\cite{grahFCAL}. The value of $f_{\text{D}}$ changes as a function of center-of-mass energy since the contribution or higher order corrections changes. At a beam energy of 250~GeV, as seen in figure~\ref{fig-DoubleAsym}, the double-spin asymmetry is roughly -1\%.
We may also include the single-spin asymmetry, $f_{\text{S}}$, in equation~\ref{eqn-SABSPol1}
\begin{equation}\label{eqn-SABSPol3}
    \sigma = \sigma_0(1+(P_- - P_+)f_\text{S} + P_-P_+f_{\text{D}})
\end{equation}
such that there is an additional correction from the difference in the beam polarizations. As seen in figure~\ref{fig-SingleAsym}, the single-spin asymmetry also depends on the beam energy and is, in general, smaller by a factor of roughly ten.
Using equation~\ref{eqn-SABSPol1} and~\ref{eqn-SABSPol3} and error propagation 
\begin{equation}\label{eqn-SABSPol2}
    \delta\sigma = \sigma_0\sqrt{(\delta PP_-f_\text{D})^2 + (\delta PP_+f_\text{D})^2 + 2(\delta Pf_\text{S})^2} = \sigma_0 \delta P f_\text{D}\sqrt{P_-^2 + P_+^2 + 2\frac{f_\text{S}^2}{f_\text{D}^2}}
\end{equation}
we derive the dependence of the SABS cross-section on the beam polarization uncertainty, $\delta P$, double-spin asymmetry and single-spin asymmetry. We assume the beam polarizations are uncorrelated and share an uncertainty and that the spin asymmetry values have no uncertainties.

For the purpose of extrapolating out a collider luminosity measurement, there must be control on the measurement region and quality through acceptance, quality cuts, energy/momentum and metrology calibrations. The underlying theory must also be known to an equivalent precision since 

\begin{equation}\label{eqn-SABSLumi}
    N_\text{SABS} = \mathcal{L}\sigma_\text{SABS}
\end{equation}

the luminosity, $\mathcal{L}$, will be derived from the count of measured SABS, $N_\text{SABS}$, and the theory value of the cross-section. The experimental strategy is to count the number of $e^+e^-$ events within a well-defined angular range in the forward calorimeters and then extract the absolute luminosity by normalizing to the theoretical cross section predicted for that angular range. This demands extremely precise theoretical calculations. As an example, the LEP experiment's $0.05$--$0.1\%$ luminosity had theory uncertainty as a large source of uncertainty. Meaning that the theory prediction for the SABS cross-section was known to similar precision, but was still not precise enough to be negligible. 

Theory precision requires inclusion of radiative corrections through $\mathcal{O}(\alpha^2)$, or greater order, and resummation of leading logarithms of the form $\alpha^n \ln^n(s/m_e^2)$, which accounts for the enhanced probability of collinear ISR photons. We now outline the key radiative effects for SABS. 

For interactions with charged initial particles \Gls{ISR} allows the incoming chared particles to radiate one or multiple photons before the hard scattering. \Gls{ISR} reduces the $\sqrt{s}$ of the hard scattering of a $e^+e^-$ collision, shifting some SABS events to lower $\sqrt{s}$. The greatest deviation for this is observed at wide-angles where Bhabhas can radiative return to the Z-pole. However, this is rare in the SABS regime. For small-angle luminosity events, ISR and \Gls{FSR} yields additional particles predominantly along the beam directions; many radiative photons escape undetected down the beam pipe, affecting the kinematics of the measured $\ee$ system. To account for multi-photon ISR, exponentiation techniques and structure-function methods, where QED particles are treated like partons, are employed~\cite{Kuraev:1985hb}~\cite{YFS1961}. Interference between ISR and FSR also contributes, though for small-angles, whichever emission is wide-angle is usually suppressed. 

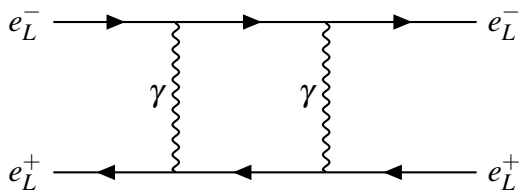
\begin{figure}[h]
\centering
\begin{tikzpicture}
  \begin{feynman}[large]
    \vertex (i1) {\( e^-_L \)};
    \vertex [below=2cm of i1] (i2) {\( e^+_L \)};
    \vertex [right=2cm of i1] (a);
    \vertex [right=2cm of a] (c);
    \vertex [right=2cm of i2] (b);
    \vertex [right=2cm of b] (d);
    \vertex [right=2cm of c] (f1) {\( e^-_L \)};
    \vertex [right=2cm of d] (f2) {\( e^+_L \)};
    

    \diagram* {
      (i1) -- [fermion] (a) -- [fermion] (c) -- [fermion] (f1),
      (i2) -- [anti fermion] (b) -- [anti fermion] (d) -- [anti fermion] (f2),
      (a) -- [photon, edge label'=\(\gamma\)] (b),
      (c) -- [photon, edge label'=\(\gamma\)] (d),
    };
  \end{feynman}
\end{tikzpicture}
\caption{Feynman diagram for box diagram of SABS where an initial state right-handed electron and positron have four internal lines, propagators, that result in the same final state as the leading order SABS diagrams.}
\label{fig-EleBox}       
\end{figure}

Virtual loop corrections from vertex corrections, self-energy corrections, modify the cross section, but not the differential, at the few-percent level. Instead, it is radiative corrections and box diagrams that modify the differential cross-section. Notably, the electron box diagram, seen in figure~\ref{fig-EleBox}, where two photons are exchanged between the $e^-$ and $e^+$ in the scattering process. 

The box diagram of figure~\ref{fig-EleBox} gives an $\mathcal{O}(\alpha^2)$ contribution that must be calculated explicitly as it does not factorize and skews the differential cross-section with a term of $\approx|\log(t/u)|^2\approx|\log(\tan^2(\theta/2))|^2$. This results in a cross-section that is preferentially more forward and more backward at the order of $1\%$~\cite{Hollik:1988ii}.

\begin{figure}[h]
\centering
\includegraphics[width=12cm]{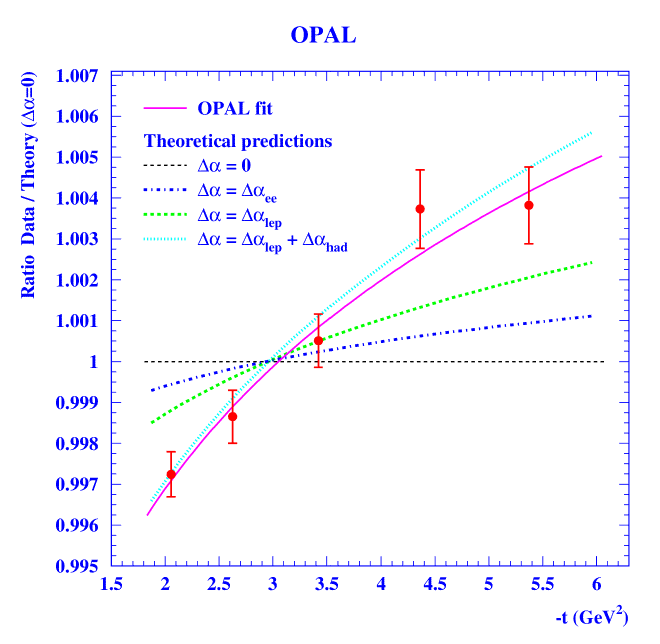}
\caption{Plot of the running of $\alpha$ as measured by OPAL using SABS with BHLUMI being used to model the Standard Model calculation for various contributions given by dashed and dotted lines. The OPAL results agree with the result that includes hadronic and leptonic contributions. Figure from~\cite{OPALALpha}.}
\label{fig-OPALAlpha}       
\end{figure}

Vacuum polarization of the photon propagator, where $\alpha$ is promoted to $\alpha(q^2)$ and the photon coupling changes with the value of momentum transfer, $q^2=t$, also adjusts the cross-section. At small values of $t$ $\alpha$ is the standard value of $\approx1/137$ but at large values of $t$ the coupling increases. Results from OPAL, where SABS was used to measure the deviation in $\alpha$, can be seen in figure~\ref{fig-OPALAlpha}.

They find that both leptonic and hadronic vacuum polarization contributions must be included to achieve a valid fit and, likewise for theory calculations, precise theoretical evaluations~\cite{OPALALpha}. One-loop electroweak corrections from diagrams involving $W$ or $Z$ loops are generally very small for SABS since the momentum transfer is very low and the $\gamma$ exchange dominates. However, for future experiments, these contributions may not be negligible.

For the simulation of SABS the primary generator of choice at LEP, and still to this day, is BHLUMI for its inclusion of numerous radiative and higher order corrections~\cite{BHLUMI}. The BHLUMI 4.04 program, developed for LEP, included exact $\mathcal{O}(\alpha)$ corrections and leading logarithmic $\mathcal{O}(\alpha^2)$ terms with exponentiation of higher orders, claiming a theoretical uncertainty of $\delta\sigma/\sigma \approx 0.06\%$ for LEP~\cite{OPALLumi}. Alternative \Gls{MCEG} include BABAYAGA which uses a parton shower method and has reached the $0.1\%$ precision level for SABS with LEP settings~\cite{babayaga}. These independent calculations and their comparisons provide crucial cross-checks and will be needed for future experiments. For future $e^+e^-$ colliders, further improvements will be required, but there is a general belief that the current tools and theories are capable of meeting future theory precision requirements~\cite{Jadach:2018jjo}. In summary, SABS is a simple, yet important, process for past and future experiments. It is crucial for precision measurements of QED and $\ee$ luminosity.

\subsection{Large Angle Bhabha Scattering (LABS)}\label{sec-LABS}
\begin{figure}[h]
\centering
\includegraphics[width=12cm]{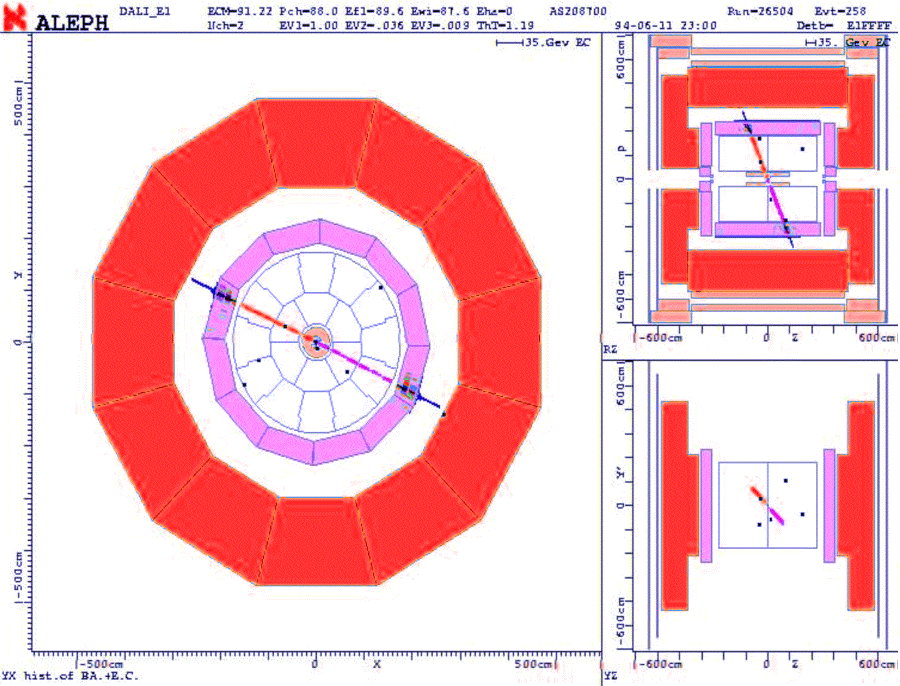}
\caption{Event display from ALEPH, one of LEP's detector experiments, of a LABS event. Taken at the Z-pole, with $\sqrt{s}\approx91.22$~GeV , but the center-of-mass energy of the LABS is estimated to be 88~GeV. Indicating that this event likely had some soft photon effects that went unaccounted for or unmeasured.}
\label{fig-LABSDisp}       
\end{figure}

Large angle, or wide-angle, Bhabha scattering refers to $e^+e^- \to e^+e^-$ events where the outgoing $e^-$ and $e^+$ are detected at wide angles. These are typically detected in the main tracker, like a \Gls{TPC}, or the main \Gls{ECAL}. This is to say, in a region of $10^\circ \to170^\circ$ in polar angle. We provide an event display, as measured at LEP's ALEPH detector, in figure~\ref{fig-LABSDisp}. 

For LABS both the $t$-channel and $s$-channel contribute significantly. At higher energies these contributions are also sensitive to electroweak corrections. In terms of the differential cross-section, the $t$-channel still dominates in the forward part of the large angle acceptance while the $s$-channel dominates at the widest angles. This is a result of the $1-\cos^2\theta$ form of $t$-channel and the $1+\cos^2\theta$ form of the $s$-channel. Since the $t$-channel divergence at small channels is not at play here, one can integrate the differential cross-section over the inner acceptance. This gives a finite and relatively smaller fraction compared to the SABS cross section. For example, at ILC250, the cross-section for LABS is $\sim1$ nb while the cross-section for SABS is $\sim10$ nb.

\begin{figure}[h]
\centering
\includegraphics[width=14cm]{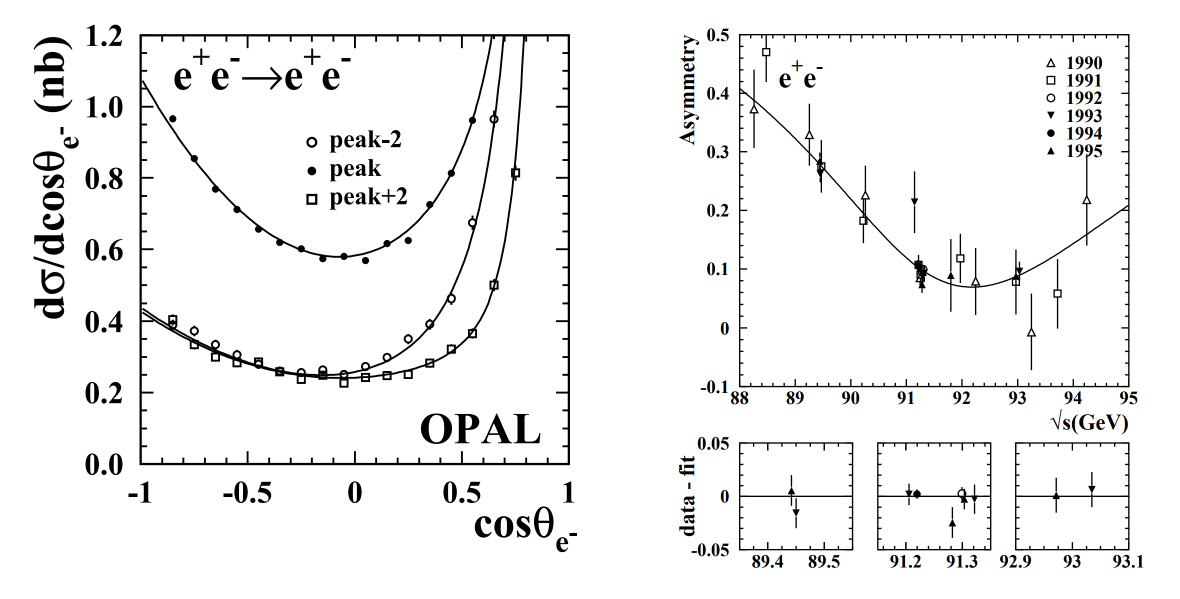}
\caption{(Left) Differential cross-section for LABS events at OPAL from measuring at $\sqrt{s}=m_\text{Z}$ and 2~GeV above and below $m_\text{Z}$.(Right) Plot of LABS $\afb$ as the center-of-mass energy changes. Plots combined from~\cite{OPAL:2000ufp}.}
\label{fig-LABS-FB}       
\end{figure}

For the purpose of precision measurements, LABS is a crucial process. LABS, due to their large cross-section, can serve as an additional luminosity channel, a momentum calibration source for a tracker and, an energy calibration for the \Gls{ECAL} and the $\sqrt{s}$ of the experiment. These are all driven by the simplicity and large statistics of LABS events. LABS events are also probes of the electroweak sector due to the significant contribution of the Z boson and its interference with the photon. This leads to the ``Zedometry'' observables, such as forward-backward asymmetry, 

\begin{equation}\label{eqn-forback}
    A_\text{FB} = \frac{\sigma_\text{F}-\sigma_\text{B}}{\sigma_\text{F}+\sigma_\text{B}}
\end{equation}

and left-right asymmetry, as defined in equation~\ref{eqn-alr} earlier, which are sensitive to electroweak physics~\cite{OPAL:2000ufp}. For transparency, the forward and backward regions are defined by the bisection of $\theta$ along the beam axis, $\theta=0^\circ$, while the left and right regions are defined by the bisection of $\phi$. For the purposes of experimentation the cross-sections of equations~\ref{eqn-forback} and~\ref{eqn-alr} may be replaced by the product of counts and efficiencies, since measuring the raw cross-section is not typically the practice. When the $s$-channel is largest, such at the Z-pole, $\afb$ for LABS dips lower and approaches zero, as seen in figure~\ref{fig-LABS-FB}. 

The manifestation of asymmetry is correlated to the polarization structure of the underlying physics and final state particles. In particular, from the Standard Model Lagrangian, we can identify two terms

\begin{equation}\label{eqn-PhoZCoup}
    \mathcal{L} = -ig\sin\theta_\text{W}A_\mu(e^{-\lambda}\gamma^\mu e^{\lambda}) + \frac{ig}{2\cos\theta_\text{W}}Z_\mu(e^{-\lambda}\gamma^\mu(2\sin^2\theta_\text{W} - \mathcal{P}_\text{L})e^\lambda)
\end{equation}

which come from the photon and Z-boson coupling to leptons. We observe that the lepton fields, $e^{\pm\lambda}$, couple identically to the photon field, $A^\mu$, given handedness but that the coupling to the Z boson field, $Z^\mu$, changes due to the $\mathcal{P}_\text{L}$ term. Using equation~\ref{eqn-PhoZCoup} we can observe that the left-handed coupling

\begin{equation}\label{eqn-leftcoup}
    c_\text{L} = I_3 - Q\sin^2\theta_\text{W}
\end{equation}

will come from a quantum number, noted as weak-isospin or $I_3$, the lepton charge, $Q$, and the Weinberg angle $\theta_\text{W}$. Equation~\ref{eqn-PhoZCoup} can give the right-handed coupling

\begin{equation}\label{eqn-rightcoup}
    c_\text{R} = - Q\sin^2\theta_\text{W}
\end{equation}

which does not depend on weak-isospin. This results in different couplings for left-handed and right-handed particles that participate in $Z$ mediated physics. Common notation is to decompose this into axial and vector couplings such that

\begin{equation}\label{eqn-VecCoup}
    g_\text{V} = c_\text{L} + c_\text{R} = I_3 - 2Q\sin^2\theta_\text{W}
\end{equation}

is the vector coupling as related to the left and right-handed couplings and the weak-isospin, $I_3$, and the lepton charge, $Q$. The axial coupling

\begin{equation}\label{eqn-VecCoup2}
    g_\text{A} = c_\text{L} - c_\text{R} = I_3 
\end{equation}

is the difference of equations~\ref{eqn-leftcoup} and~\ref{eqn-rightcoup} and depends only on the weak isospin. We can relate $\alr$ to theory couplings in $\ee\to f \bar{f}$, \Gls{DiFermion}, processes

\begin{equation}\label{eqn-AlrTheory}
    \alr = \frac{s G_\text{F}}{\sqrt{2}\pi\alpha Q}g_\text{A}g_\text{V}
\end{equation}

such that the electromagnetic coupling, $\alpha$, the Fermi constant, $G_\text{F}$, the center-of-mass energy and the axial and vector couplings all contribute to $\alr$~\cite{Moortgat_Pick_2008}. In an experiment with polarized beams this is modified by a factor of the effective polarization

\begin{equation}\label{eqn-Peff}
    P_\text{eff.} = \frac{\mathcal{L}_\text{RL} - \mathcal{L}_\text{LR}}{\mathcal{L}_\text{RL} + \mathcal{L}_\text{LR}}
\end{equation}

which depends on the luminosity of the beams for $\text{LR}$ or $\text{RL}$ helicity combinations of the electron and positron respectively. Assuming that $P_\text{eff.}$, $G_\text{F}$, $\sqrt{s}$, $\alpha$ are all well constrained by existing measurements then $\alr$ can be used to precisely measure $g_\text{V}$ and $g_\text{A}$. Using LABS and methodology related to this \Gls{SLD} measured the $\gv$ and $\ga$ for electrons as well as $\sin^2\theta_\text{W}$ to roughly 10 parts per mille~\cite{SLD:1994kvj}. This was later improved upon by LEP to roughly 700 ppm~\cite{ALEPH:2005ab}.

Radiative corrections in LABS include all those present in SABS but now the virtual electroweak corrections are important as well. One-loop diagrams involving exchange of virtual Z bosons and $W^\pm$ bosons contribute at the level of a few $\times 10^{-3}$ to $10^{-2}$ at energies relevant to LEP. At energies relevant to Higgs factories these contributions increase by roughly an order of magnitude. Two-loop QED corrections provide sub-leading contributions but also grow at higher energies. At $\sqrt{s}\sim 1$~TeV, some two-loop electroweak Sudakov logarithms, particularly the W-boson $\log(s/M_W^2)$ terms, are on the order of 1\% of the cross section~\cite{Hollik:1988ii}. These have been studied analytically but need additional work to be ready for the precision and energy of a TeV-scale $e^+e^-$ collider, like ILC or CLIC. For the purpose of precision electroweak measurements, this is a fortunate challenge. Even though theory calculations will become more diverse and numerous, the underlying measurements are more sensitive to electroweak physics. This, in turn, makes TeV-scale LABS more attractive for signals or discoveries of BSM as derived from precision measurements.

For simulation of LABS the \Gls{MCEG} BHWIDE, from the creators of BHLUMI, is considered state-of-the-art~\cite{BHWIDE}. BHWIDE was developed during the LEP-era to simulate LABS and, as such, it incorporates exact $\mathcal{O}(\alpha)$ QED corrections and YFS exponentiation so that it can meet the demands of LEP's electroweak and QED precision measurements. Cross-validation has been done between BHWIDE and other MCEG for LABS, TOPAZ0 and ALIBABA~\cite{Montagna:1998kp}~\cite{Beenakker:1990mb}. This cross-check is important and gives confidence in the theoretical machinery needed for precision LABS measurements. Ongoing improvements in theory are aimed at matching the ever-increasing demands of proposed Higgs factories and future $\ee$ colliders. Assuming that the demand is met, LABS will continue to be an important precision electroweak and QED channel.

\subsection{Beam-pipe Escape Radiative Bhabhas (BERB)}\label{sec-BERB}
\begin{figure}[h]
\centering
\begin{tikzpicture}
  \begin{feynman}[large]
    \vertex (i1) {\( e^-_L \)};
    \vertex [below=2cm of i1] (i2) {\( e^+_L \)};
    \vertex [right=2cm of i1] (a);
    \vertex [right=2cm of a] (c);
    \vertex [right=4cm of i2] (b);
    \vertex [right=2cm of c] (f1) {\( e^-_L \)};
    \vertex [right=2cm of b] (f2) {\( e^+_L \)};
    
    \vertex [above right=1.0cm of a] (gamma) {\(\gamma_{\text{ISR}}\)};

    \diagram* {
      (i1) -- [fermion] (a) -- [fermion] (c) -- [fermion] (f1),
      (i2) -- [anti fermion] (b) -- [anti fermion] (f2),
      (c) -- [photon, edge label'=\(\gamma\)] (b),
      (a) -- [photon] (gamma),
    };
  \end{feynman}
\end{tikzpicture}
\caption{Feynman diagram for BERB where an initial state left-handed electron emits an ISR photon, which conserves helicity here.}
\label{fig-BERBDiag}       
\end{figure}
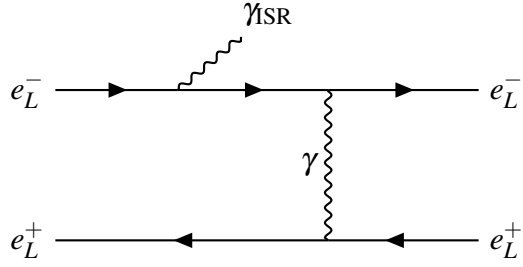

Processes like Bhabha scattering, which are dominantly forward scattering, are also capable of having their charged particles escape down the beam-pipe. Due to this, events that have charged particles can look like events that only have neutral particles if a charged particle emits photon(s) that are reconstructed in the detector and the charged particles escape down the beam-pipe. This can be especially troublesome for diphoton measurements in the case where the electron-positron pair escapes down the beam-pipe after each emits a photon as the final state is two photons with missing energy. In the context of Bhabha scattering we refer to these events as Beam-pipe Escape Radiative Bhabhas (BERB). Notable are the small angle t-channel contributions of Bhabha scattering, where we use equation~\ref{eqn-bhabha-sig}

\begin{equation}\label{eqn-bhaxsec}
\begin{gathered}
   \sigma_{\text{LL}\to\text{LL}}=\sigma_{\text{RR}\to\text{RR}} =    \sigma_{\text{LR}\to\text{LR}} =    \sigma_{\text{LL}\to\text{RL}} \approx \frac{\pi\alpha^2}{2s}\left[\frac{2}{\theta^2} + \log\left(\frac{2}{\theta^2}\right) \right]
\end{gathered}
\end{equation}

to write out the leading order non-zero helicity dependent contributions. We note that, as derived in section~\ref{sec-BhaThe}, there are no leading order contributions from helicity combinations where the final state is different from the initial state. Examining equation~\ref{eqn-bhaxsec} we see that the integration bounds lead to infinite results. To remedy this, an integration regularization is applied. We choose the Weizsäcker-Williams method (WWM), also known as Equivalent Photon approximation (EPA), where

\begin{equation}\label{eqn-WWApprox}
    \theta_\text{EPA}(m_e) = \frac{m_e^2}{s}
\end{equation}

a minimum angle is applied that is derived from the mass of the scattered particle and the center-of-mass energy~\cite{Peskin:1995ev}. From equation~\ref{eqn-WWApprox} we update equation~\ref{eqn-bhaxsec}, using an inner acceptance, $\tinn$

\begin{equation}\label{eqn-bhasmallEPA}
\begin{gathered}
    \sigma = \frac{\pi\alpha^2}{2s}\bigg[\frac{2}{\theta^2} + \log\left(\frac{2}{\theta^2}\right)  \Bigg|_{\tinn}^{\theta_\text{EPA}}
\end{gathered}
\end{equation}

to reflect the new bounds that are inside their previous bounds. Finishing the integration of equation~\ref{eqn-bhasmallEPA}

\begin{equation}\label{eqn-bhasmallVal}
\begin{gathered}
    \sigma = \frac{\pi\alpha^2}{2s} \left[\frac{2}{\theta_\text{EPA}^2} + \log\left(\frac{2}{\theta_\text{EPA}^2}\right) - \frac{2}{\tinn^2} - \log\left(\frac{2}{\tinn^2}\right)  \right]
\end{gathered}
\end{equation}

we find that the cross-section does not diverge to infinity. We now proceed to investigating radiative Bhabha scattering. We proceed with a calculation with one radiative photon, such as the diagram seen in figure~\ref{fig-BERBDiag}. 

\begin{figure}[h]
\centering
\includegraphics[width=16cm]{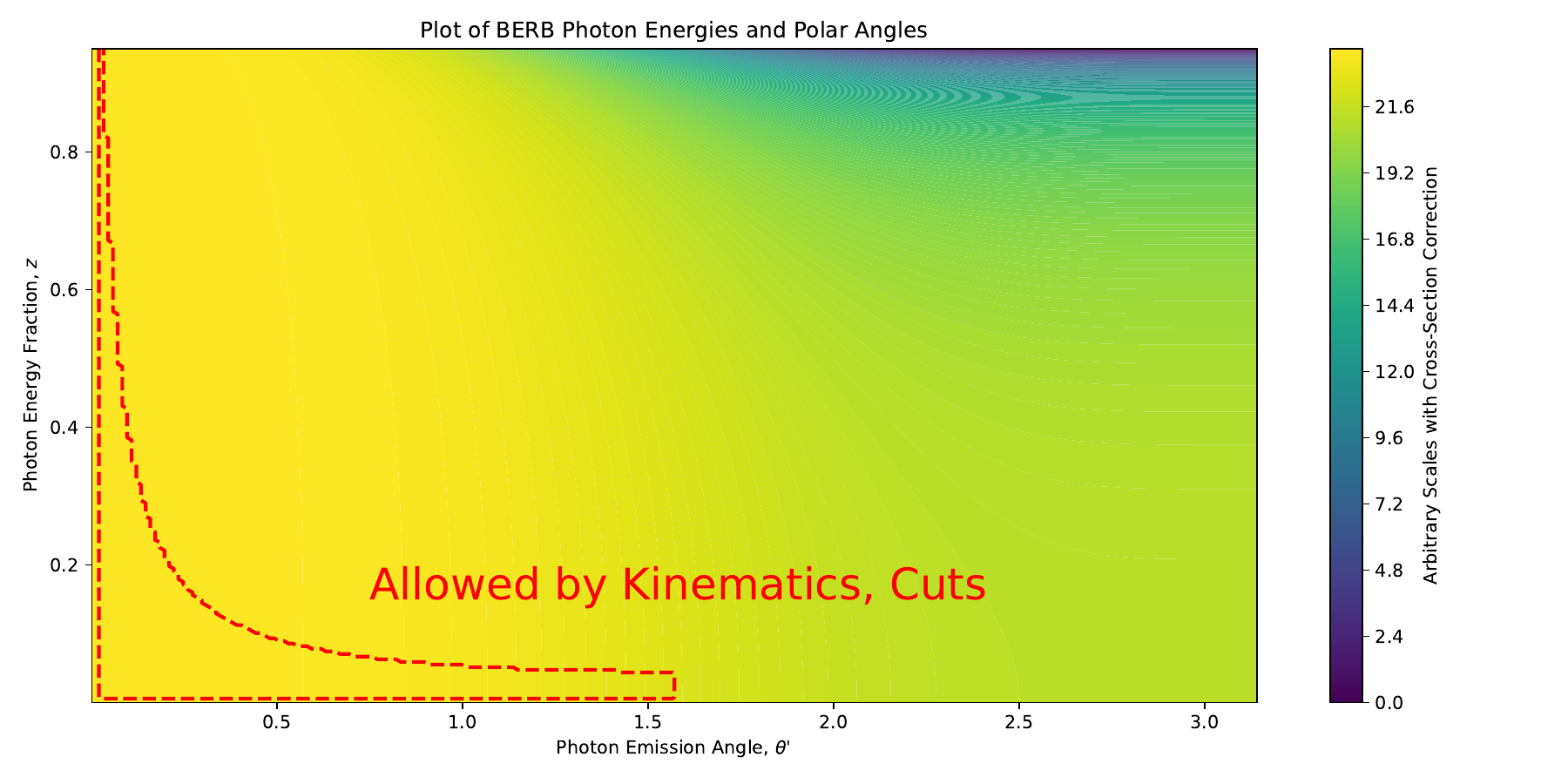}
\caption{Plot of the $z$ and $\theta'$ dependence of the BERB radiative correction for single photon emission at 250 GeV, for an inner acceptance of $1^\circ$, a crossing angle of 14 mrad and a minimum photon energy of 1 GeV. The cross-section favors forward angles. A red line which outlines the constraints from kinematics and cuts is included. The region bounded by these lines includes the kinematically allowed, by momentum and energy conservation, phase space, and after the cuts, for BERB photon emission.}
\label{fig-BERBPlot}       
\end{figure}

For one radiative photon in Bhabha scattering there are two permutations for Initial State Radiation (ISR) and two permutations for Final State Radiation (FSR). For each possible helicity contribution,

\begin{equation}\label{eqn-bharad}
    \frac{d\sigma_{\text{rad.}}}{d\cos\theta d\cos\theta'} = \sum_{\text{ii}} \frac{d\sigma_{\text{ii}}}{d\cos\theta}\alpha\sin^2\theta'\left[\frac{1+(1-z)^2}{z^2(1-z)}\right]
\end{equation}

the cross-section picks up an additional splitting term depending on the photon scattering angle, $\theta'$, and the energy fraction, $z$, of the outgoing electron~\cite{Peskin:1995ev}. We note the radiative cross-section as $\sigma_{\text{rad.}}$ and the pure Bhabha cross-section as $\sigma_{\text{ii}}$. The typical dependence on $s^{-1}$ seen in cross-sections cancels out due to the radiative fragmentation function depending on $s$ itself. It is also easy to see that, in the limit of no radiation, $z=0$, the cross-section experiences an infrared divergence to infinity. This divergence is usually handled with exponential resummation techniques such as those demonstrated in the Sudakov method or the YFS method~\cite{Sudakov1954}~\cite{YFS1961}. We integrate equation~\ref{eqn-bharad} and plot the cross-section radiative correction in terms of $z$ and $\theta'$ in figure~\ref{fig-BERBPlot} so as to see the geometry. The dependence of the radiative correction to $z$ is fairly flat at forward angles. In the regime of hard photon emission, $z\to1$, there is a preference of roughly two orders of magnitude for forward emission.

These results indicate that, in the case of requiring that the recoiling electron or positron stays in the beam pipe, the photon to escape the beam pipe, and that the photon is hard, $>1$~GeV, that the single photon emission is preferentially forward. Therefore, from a measurement perspective, BERB single photon events should dominantly be in the forward calorimeter. In figure~\ref{fig-BERBPlot} we have included contours for the constraints due to cuts and kinematics to show the allowed phase space. We have required a 1 GeV photon energy cut and $1^\circ$ inner acceptance. The hard photon emission has a thin corridor along the forward region and then some phase space at wider angles. For cross-section reasons, the narrow forward region of the allowed phase space is preferred. Outside this region the recoiling particle will escape the beam pipe, the photon will be too low energy, or the photon will also escape down the beam pipe. For this test case the ratio of the cross-section of the allowed phase space to the total phase space was 7.5\%. This is larger than $\alpha^{-1}$ by about a factor of roughly ten. This indicates that double radiative correction, which could allow most of the phase space by having two photons that balance each other, may be smaller than single radiative correction.

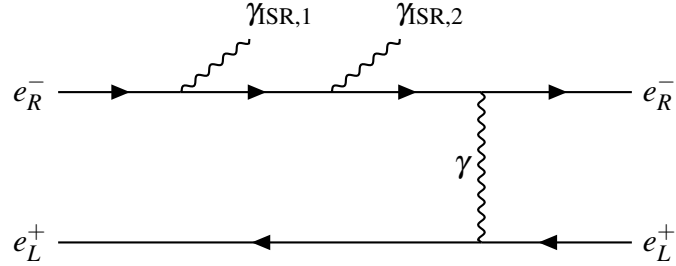
\begin{figure}[h]
\centering
\begin{tikzpicture}
  \begin{feynman}[large]
    \vertex (i1) {\( e^-_R \)};
    \vertex [below=2cm of i1] (i2) {\( e^+_L \)};
    \vertex [right=2cm of i1] (a);
    \vertex [right=2cm of a] (c);
    \vertex [right=2cm of c] (d);
    \vertex [right=6cm of i2] (b);
    \vertex [right=2cm of d] (f1) {\( e^-_R \)};
    \vertex [right=2cm of b] (f2) {\( e^+_L \)};
    
    \vertex [above right=1.0cm of a] (gamma1) {\(\gamma_{\text{ISR,1}}\)};
    \vertex [above right=1.0cm of c] (gamma2) {\(\gamma_{\text{ISR,2}}\)};
    \diagram* {
      (i1) -- [fermion] (a) -- [fermion] (c) -- [fermion] (d) -- [fermion] (f1),
      (i2) -- [anti fermion] (b) -- [anti fermion] (f2),
      (d) -- [photon, edge label'=\(\gamma\)] (b),
      (a) -- [photon] (gamma1),
      (c) -- [photon] (gamma2),
    };
  \end{feynman}
\end{tikzpicture}
\caption{Feynman diagram for BERB where an initial state right-handed electron emits two ISR photons. The net effect of which conserves helicity.}
\label{fig-BERBDiag2}       
\end{figure}

We test the case where a single particle undergoes double radiative correction, like in figure~\ref{fig-BERBDiag2}, to compare phase space and estimate cross-section ratio.

\begin{figure}[h]
\centering
\includegraphics[width=16cm]{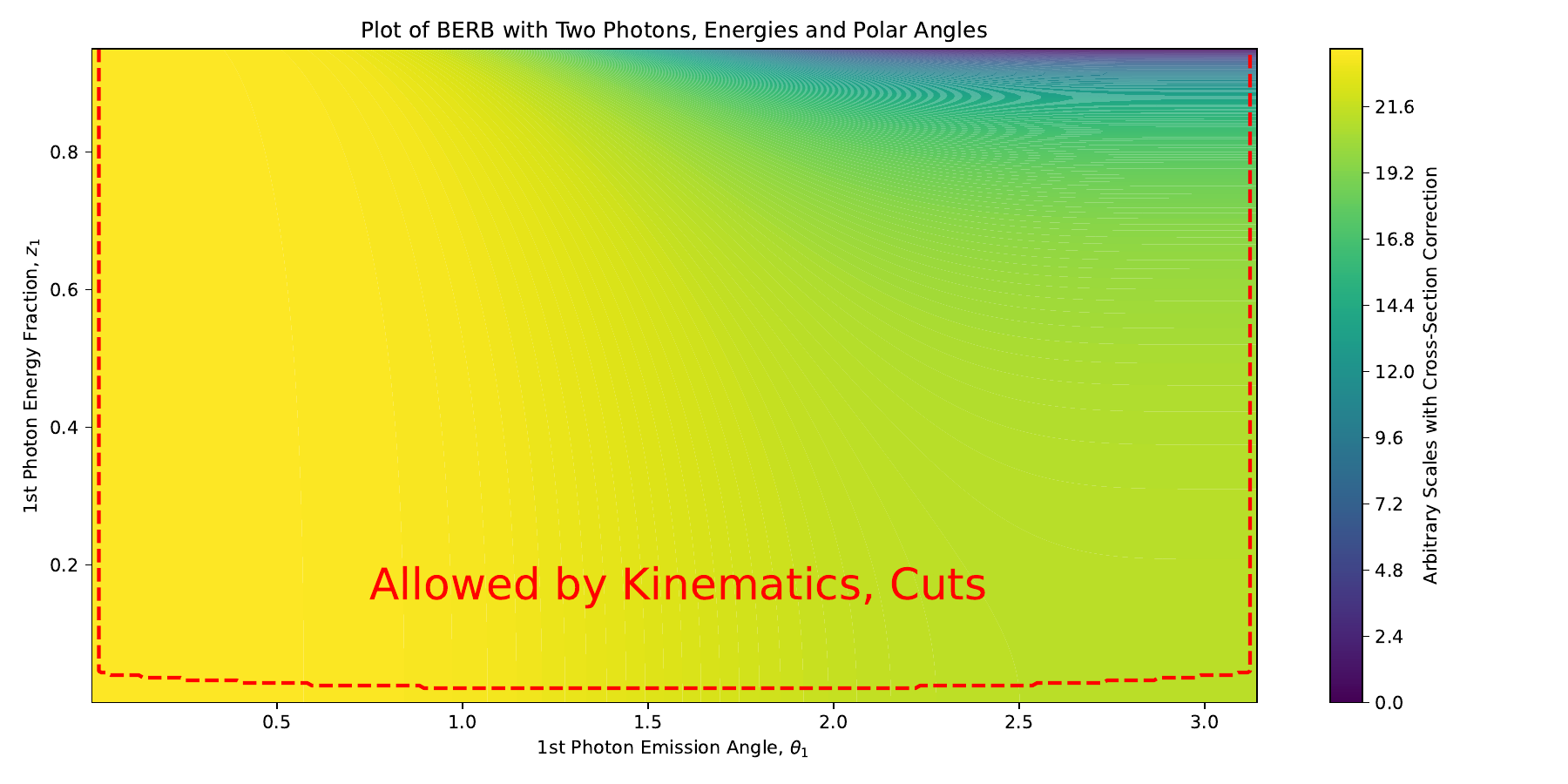}
\caption{Plot of the $z_1$ and $\theta_1$ dependence of the double BERB radiative correction for single photon emission at 250 GeV, for an inner acceptance of $1^\circ$, a crossing angle of 14 mrad and a minimum photon energy of 1 GeV. The photon cuts were applied to both photons. A red line which outlines the constraints from kinematics and cuts is included. The allowed region is above the red-line and comprises 95.1\% of the phase space.}
\label{fig-BERBPlot2}       
\end{figure}

By including a second radiative correction we adjust the notation to promote $z$ to $z_i$ and $\theta'$ to $\theta_i$ for each respective additional radiative correction. From this promotion we can use the same constraints with a promoted version of equation~\ref{eqn-bharad} to generate the phase space for the first emitted photon, $\gamma_\text{ISR,1}$. We do this, as seen in figure~\ref{fig-BERBPlot2}, to provide an easy to visualize comparison to figure~\ref{fig-BERBPlot}.

We observe that the inclusion of a second radiative correction increases the available phase space considerably. This also means that doubly radiative BERB has significant coverage into the wider angles of the detector, not just the forward calorimeter of the single radiative case. The resulting ratio of available phase space to unavailable is now 95.1\%, or a factor $\approx\times13$ the available phase space in single emission. The number of permutations is also greater, resulting in the relative weight, after factoring in $\alpha^{-1}$ from the additional photon, being close to 23\% of the single photon cross-section. This does not consider radiative interference effects and only includes the leading order small angle Bhabhas, so we suspect that this ratio is not a good representation of a full simulation. It merely demonstrates that the double radiative correction should be comparable to the single radiative correction.

\section{Diphotons (Two Photons)}\label{sec-DiGam}

\begin{figure}[h]
\centering
\begin{tikzpicture}
  \begin{feynman}[large]
    \vertex (i1) {\( e^-_L \)};
    \vertex [below=2cm of i1] (i2) {\( e^+_R \)};
    \vertex [right=2cm of i1] (a);
    \vertex [right=2cm of i2] (b);
    \vertex [right=2cm of a] (f1) {\( \gamma \)};
    \vertex [right=2cm of b] (f2) {\( \gamma \)};

    \diagram* {
      (i1) -- [fermion] (a) -- [fermion] (b),
      (i2) -- [anti fermion] (b),
      (a) -- [photon] (f1),
      (b) -- [photon] (f2),
    };
  \end{feynman}
\end{tikzpicture}
\caption{Feynman diagram for $e_\text{R}^+ e_\text{L}^- \to \gamma \gamma$, also known as diphoton production. Here a left-handed electron and right-handed positron annihilate into the final state photons.}
\label{fig-DiGamDiag}       
\end{figure}
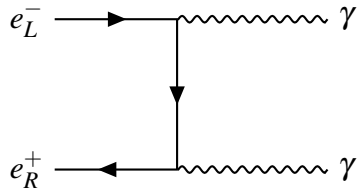

The process $e^+ e^- \to \gamma \gamma$, referred to as diphoton or \Gls{diphoton} production, is the annihilation of an electron and positron into two photons. This is a pure QED process at tree level and, as seen in figure~\ref{fig-DiGamDiag}, is a \Gls{tchan} process with a fermionic propagator.

We can write the matrix element for diphoton diagrams, using slash notation

\begin{equation}\label{eqn-diphotonTMat}
    \mathcal{M} = -e^2(\bar{v}_{\text{p1}}\gamma_\mu \epsilon^\mu_{\text{p3}})\frac{\not{p_1} - \not{p_3}+m_\text{e}}{t-m_\text{e}^2+i\epsilon}(\epsilon^\nu_\text{p4}\gamma_\nu u_{\text{p2}})
\end{equation}

we write the matrix element for diphoton production using $\epsilon_\mu$ to denote the polarization vectors of the outgoing photons. We can now prove why, in the massless, helicity conserving, limit that only the $\text{LR}$ and $\text{RL}$ helicity combinations contribute. In the massless limit we can rewrite equation~\ref{eqn-diphotonTMat} to

\begin{equation}\label{eqn-diphotonTMat2}
    \mathcal{M} = \frac{-e^2}{t}(\bar{v}_{\text{p1}}\gamma_\mu \epsilon^\mu_{\text{p3}})(\not{p_1} - \not{p_3})(\epsilon^\nu_\text{p4}\gamma_\nu u_{\text{p2}}) = \frac{-e^2}{t}S_{\mu\nu} \epsilon^\mu_{\text{p3}}\epsilon^\nu_\text{p4}
\end{equation}

where $S_{\mu\nu}$ is the scattering matrix for the $\ee$ components of the matrix element. From the scattering matrix of equation~\ref{eqn-diphotonTMat2} we can create a test case of assuming that both particles are left-handed and promote the dirac matrices

\begin{equation}\label{eqn-diphotonSMatLeft}
    S_{\mu\nu} = \bar{v}_{\text{p1}}\gamma_\mu (\not{p_1} - \not{p_3})\gamma_\nu u_{\text{p2}} \rightarrow  \bar{v}_{\text{p1}}\mathcal{P}_\text{L}\gamma_\mu (\not{p_1} - \not{p_3})\mathcal{P}_\text{L} \gamma_\nu  u_{\text{p2}}
\end{equation}

to their left-handed projections. From this we can move the left-handed projections through the center slashed momenta and use the identity

\begin{equation}\label{eqn-DiGamId}
    \mathcal{P}_\text{i}\gamma_\mu\mathcal{P}_\text{i}=0
\end{equation}

to rewrite equation~\ref{eqn-diphotonSMatLeft}

\begin{equation}\label{eqn-diphotonSMatLeft2}
    S_{\mu\nu} = \bar{v}_{\text{p1}}\mathcal{P}_\text{L}\gamma_\mu\mathcal{P}_\text{L} (\not{p_1} - \not{p_3}) \gamma_\nu  u_{\text{p2}} = 0
\end{equation}

and show that the scattering matrix for same-handedness initial particles vanishes. Starting from equation~\ref{eqn-diphotonTMat2} and using methods similar to section~\ref{sec-BhaThe}, we can proceed to solving for the differential cross-section 

\begin{equation}\label{eqn-DigamDiffXsec}
    \frac{d\sigma}{d\cos\theta} = \frac{2\pi \alpha^2}{s}\left(\frac{u}{t}+\frac{t}{u} \right) \approx \frac{\alpha^2}{s} \frac{1+\cos^2\theta}{1-\cos^2\theta}
\end{equation}

to find that it is preferentially forward and has an infrared divergence, similar to SABS. Integrating equation~\ref{eqn-DigamDiffXsec} over an inner acceptance

\begin{equation}\label{eqn-DigamXsec}
    \sigma_{\gamma\gamma} = \frac{2\pi\alpha^2}{s} \bigg[ \log(1+\cos\theta) - \log(1-\cos\theta) - \cos\theta \bigg|_{\tinn}^{\pi-\tinn}
\end{equation}

shows that the infrared divergence is no longer a concern. For a center-of-mass energy of 250~GeV and inner acceptance of $1^\circ$ equation~\ref{eqn-DigamXsec} evaluates to 35.4~pb. This makes diphotons one of the largest cross-section processes at an $\ee$ collider, especially in the forward region. To extrapolate out the uncertainty of $\sigma_{\gamma\gamma}$ on the inner acceptance 

\begin{equation}\label{eqn-DiGamTinn}
    \delta\sigma_{\gamma\gamma} = \sin(\tinn)\frac{\cos^2\tinn + 1}{\cos^2\tinn - 1}\delta\tinn \approx \left[\frac{2}{\tinn} + \tinn\right]\delta\theta
\end{equation}

we can use the small angle approximation to determine the leading order dependence on the inner acceptance is $\tinn^{-1}$. This dependence is less stringent than SABS, where equation~\ref{eqn-SABS_Inner_Unc} was of order $\tinn^{-3}$. This result also makes relative sense; SABS has a differential dependence of $\sin^{-4}\theta$ while diphotons only has a $\sin^{-2}\theta$ dependence. We can also add in the beam polarization dependence into equation~\ref{eqn-DigamXsec} by using equation~\ref{eqn-sigpol}
\begin{equation}\label{eqn-DigamXsecPol}
    \sigma_{\gamma\gamma} = \frac{2\pi\alpha^2}{s}(1-P_-P_+) \bigg[ \log(1+\cos\theta) - \log(1-\cos\theta) - \cos\theta \bigg|_{\tinn}^{\pi-\tinn} = \frac{2\pi\alpha^2}{s}(1-P_-P_+) f(\theta,\tinn)
\end{equation}
and see that we expect an enhancement of the diphoton cross-section when polarized beams are used. We have introduced the polar angle structure function, $f(\theta,\tinn)$, in order to make notation more compact. We can also use this to estimate the dependence of the precision of the diphoton cross-section on beam polarization uncertainty such that
\begin{equation}\label{eqn-DiGamPolUnc}
    \delta\sigma_{\gamma\gamma} = \sqrt{(P_+\delta P)^2 + (P_-\delta P)^2} = \delta P\sqrt{P_+^2 + P_-^2}
\end{equation}
we observe that the beam polarization dependence couples with the beam polarization fraction of the opposing beam. As before, in section~\ref{sec-SABS}, we assume that the beam polarization uncertainty is shared between the beams and are not correlated.

For the purposes of precision measurements in the forward region equation~\ref{eqn-DiGamTinn} makes diphotons attractive as they do not have as strict metrology requirements. The theory calculations of diphotons are also, in general, simpler and cleaner than Bhabha scattering as electroweak corrections do not appear in the tree level diagrams. The diphoton process is also clean and simple from a detector perspective; diphoton events show two energetic, back-to-back electromagnetic showers in the calorimeters, with little to no tracks. This also aids in electro-photon discrimination, so that diphoton samples are high purity. For these reasons diphotons are considered an additional precision QED and precision luminosity channel at $\ee$ colliders. This has already been done at BELLEII and is planned for future $\ee$ colliders due to the increasing complexities of using SABS~\cite{Adachi_2025}. Currently, there is no dedicated precision MCEG software for diphotons, but BABAYAGA is considered to be the most in depth~\cite{babayaga}. Still, there is no reason to believe that future theory calculations and simulations could not achieve a theory precision of $10^{-3}$ to even $10^{-5}$. 

\section{Dimuons (Two Muons)}\label{sec-DiMu}

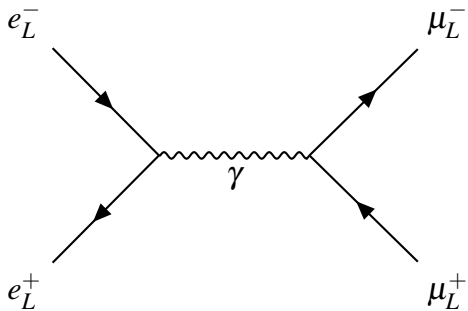
\begin{figure}[h]
\centering
\begin{tikzpicture}
  \begin{feynman}[large]
    \vertex (a);
    \vertex [above left=2cm of a] (i1) {\( e^-_L \)};
    \vertex [below left=2cm of a] (i2) {\( e^+_L \)};
    \vertex [right=2cm of a] (c);
    \vertex [above right=2cm of c] (f1) {\( \mu^-_L \)};
    \vertex [below right=2cm of c] (f2) {\( \mu^+_L \)};

    \diagram* {
      (i1) -- [fermion] (a), 
      (i2) -- [anti fermion] (a),
      (f1) -- [anti fermion] (c), 
      (f2) -- [fermion] (c),
      (a) -- [photon, edge label'=\(\gamma\)] (c),
    };
  \end{feynman}
\end{tikzpicture}
\caption{Feynman diagram for the leading order $s$-channel diagram for dimuon production from a $\ee$ collision.}
\label{fig-dimu}
\end{figure}

The process of a matter-antimatter pair of muons in being created in $\ee$ collisions, $e^+ e^- \to \mu^+ \mu^-$, is known as \gls{dimuon} production. Dimuons are a benchmark electroweak process and historically one of the primary channels for studying the interface of QED and the weak interaction. Unlike Bhabhas and diphotons, dimuons are a simpler \Gls{schan} only process, as can be seen in figure~\ref{fig-dimu}. 

This simplicity is extended to the detector as well; Dimuons offer one of the cleanest and simplest measurements since muons are higher in mass and lower interaction strength than electrons. Dimuons are measured and tagged exceptionally well in particle trackers and muon end-caps. Due to their exceptional measurement, dimuons are candidates for tracker calibration, especially when produced by well known particles like the $J/\psi$, as well as center-of-mass energy calibration from methods that either use beam-energy dimuons or use well established models to attain the center-of-mass energy from radiative return dimuons~\cite{Madison2022}~\cite{Wilson2023}. Dimuons are also candidates for many of the same precision measurements of QED and electroweak physics that can be conducted with Bhabhas. However, in general, dimuons can be measured more precisely. The addition of dimuons with Bhabhas also allows for some of the strictest tests of Lepton Universality, a particle physics principle that establishes that leptons of different flavor and mass should have identical QED and electroweak couplings.

\begin{figure}[h]
\centering
\begin{tikzpicture}
  \begin{feynman}[large]
    \vertex (a);
    \vertex [above left=2cm of a] (i1) {\( e^-_L \)};
    \vertex [above left=1cm of a] (b);
    \vertex [below left=2cm of a] (i2) {\( e^+_L \)};
    \vertex [above right=1cm of b] (gamma) {\( \gamma_{ISR} \)};
    \vertex [right=2cm of a] (c);
    \vertex [above right=2cm of c] (f1) {\( \mu^-_L \)};
    \vertex [below right=2cm of c] (f2) {\( \mu^+_L \)};

    \diagram* {
      (i1) -- [fermion] (b) -- [fermion] (a), 
      (i2) -- [anti fermion] (a),
      (f1) -- [anti fermion] (c), 
      (f2) -- [fermion] (c),
      (a) -- [photon, edge label'=\(\gamma\)] (c),
      (b) -- [photon] (gamma),
    };
  \end{feynman}
\end{tikzpicture}
\caption{Feynman diagram for the leading order $s$-channel diagram for dimuon production from a $\ee$ collision.}
\label{fig-dimuRadRet}
\end{figure}
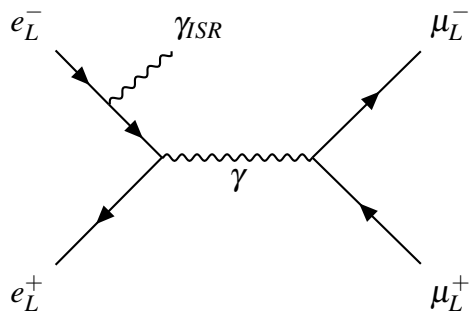

The dominant correction to the leading order tree diagram is for single \Gls{ISR}, due to the radiative return to $\sqrt{s}\approx m_\text{Z}$. The diagram for this can be seen in figure~\ref{fig-dimuRadRet}.

\begin{figure}[h]
\centering
\includegraphics[width=12cm]{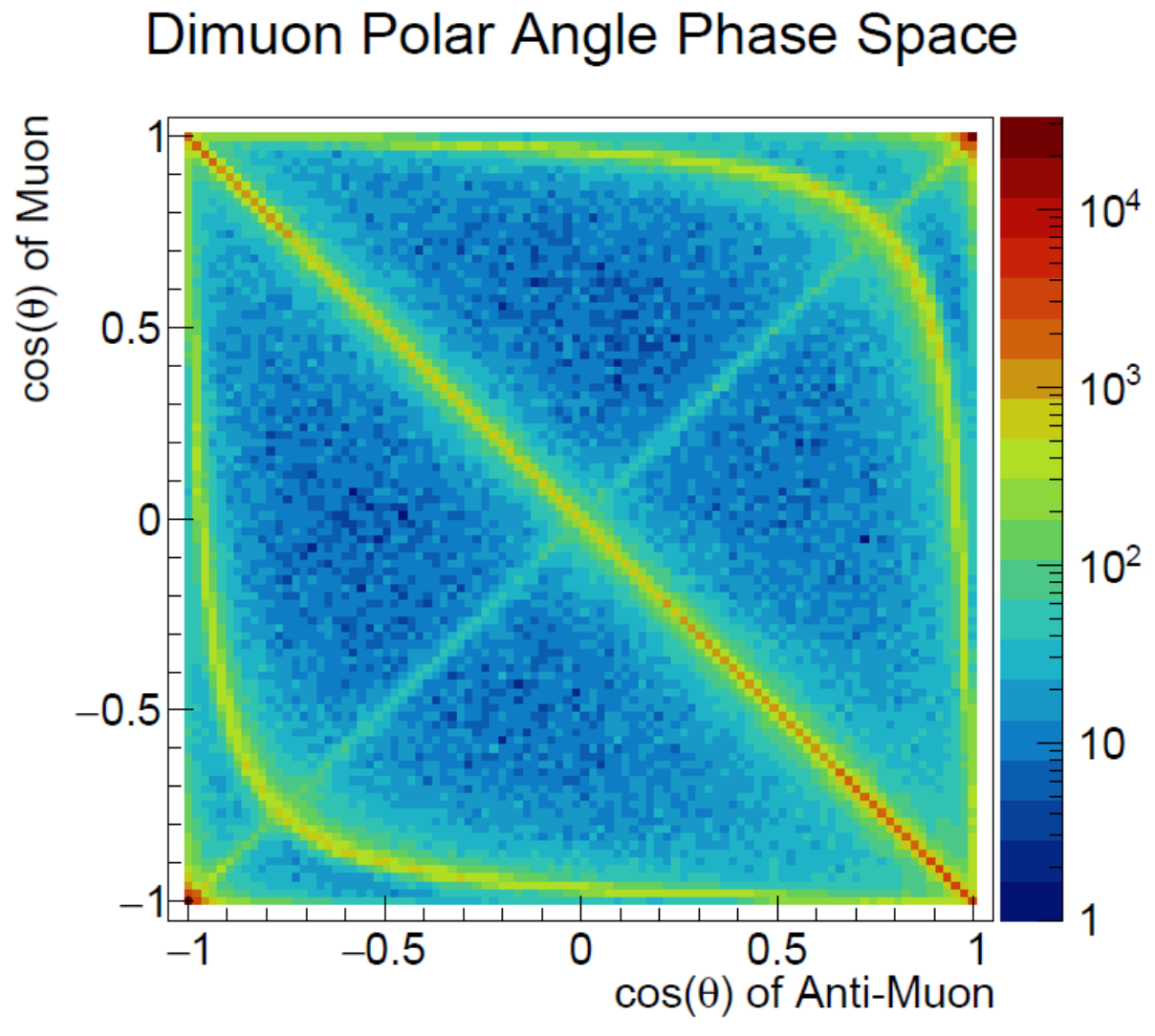}
\caption{Generator level distribution of the cosines of the 
muon polar angles for $\sqrt{s}=250$~GeV for $\eemmg$ 
with $P(\electron)=-0.8$ and $P(\positron)=+0.3$. The color axis represents counts of events in the bin. The radiative return events form an envelope shape. The helicity conserving contribution makes the dominant diagonal that runs top left to bottom right. This diagonal has the most observable foreward-backward asymmetry. The other diagonal comes from contributions from single or double ISR to very low energy.}
\label{fig-DiMuAng}
\end{figure}

The differential cross-section for when the initial and final state particles retain their helicity and the Z boson is the propagator

\begin{equation}\label{eqn-DiMuZRatRet}
\begin{gathered}
    \frac{d\sigma}{d\cos\theta}(s) \approx \frac{\alpha^2}{4s}\bigg|\frac{s}{(s-m_\text{Z}^2 + im_\text{Z}\Gamma_\text{Z})}\bigg|^2(|g_\text{V,e}|^2 + |g_{\text{A,e}}|^2)(|g_{\text{V,}\mu}|^2 + |g_{\text{A,}\mu}|^2)(1+\cos^2\theta \\
    +8g_\text{V,e}g_{\text{A,e}}g_{\text{V,}\mu}g_{\text{A,}\mu}\cos\theta)
\end{gathered}
\end{equation}

depends on the axial and vector couplings of the fermions involved as well as the Z boson mass and width. The last term in equation~\ref{eqn-DiMuZRatRet}, scaling with $\cos\theta$, is responsible for the $\afb$ measurements as the structure leads to a forward-backward asymmetry. A plot of dimuon polar angles against each other, as seen in figure~\ref{fig-DiMuAng}, shows how the forward-backward asymmetry and the radiative return processes look in measurements. 

Radiative return dimuons, due to their different invariant mass, are boosted to different angles. The forward-backward asymmetry also shows up as an imbalance in events in the corners of figure~\ref{fig-DiMuAng}. The non-dominant diagonal comes from radiative events where the emitted photons carry off almost all of the energy, leaving the muon(s) with energies comparable to their rest mass.

Using the same radiative correction techniques developed in section~\ref{sec-BERB} we can rewrite equation~\ref{eqn-DiMuZRatRet}

\begin{equation}\label{eqn-DiMuZRatRet2}
    \frac{d\sigma}{d\cos\theta d\cos\theta'} = \frac{d\sigma}{d\cos\theta}(s') \alpha\sin^2\theta'\left[\frac{1+(1-z)^2}{z^2(1-z)}\right]
\end{equation}

where the center-of-mass energy after the ISR radiation is given as $\sqrt{s'}$. The result is that an ISR radiative correction as described by equation~\ref{eqn-DiMuZRatRet2} can emit an ISR photon to go to a center-of-mass energy at $\sqrt{s}\approx m_\text{Z}$ and then participate in the large cross-section made possible by the Z boson mass resonance term seen in equation~\ref{eqn-DiMuZRatRet}. Due to the radiative return to the Z boson mass resonance

\begin{equation}\label{eqn-PhotonZRatRet}
    z_{Z} = 1-\frac{m_\text{Z}^2}{s}
\end{equation}

we label this special case of $z$ as $z_\text{Z}$, the value of photon energy fraction when a single ISR photon participates in radiative return to the Z boson mass resonance. Consider two example cases; at the Z-pole this contribution essentially vanishes while at 250~GeV the photon energy is roughly 108.4~GeV. From equation~\ref{eqn-PhotonZRatRet} we can rewrite equation~\ref{eqn-DiMuZRatRet2}

\begin{equation}\label{eqn-DiMuZRatRet3}
\begin{gathered}
    \frac{d\sigma}{d\cos\theta d\cos\theta'} = \frac{d\sigma}{d\cos\theta}(m_\text{Z}) \alpha\sin^2\theta'\left[\frac{1+(\frac{m_\text{Z}^2}{s})^2}{(1-\frac{m_\text{Z}^2}{s})^2(\frac{m_\text{Z}^2}{s})}\right]\\
    \approx \frac{d\sigma}{d\cos\theta}(m_\text{Z}) \alpha \sin^2\theta'\frac{s}{m_\text{Z}^2}
\end{gathered}
\end{equation}

and observe that the dependence on $s$, at leading order and the high energy limit, $\sqrt{s} \gg m_\text{Z}$, is linear. Given equation~\ref{eqn-DiMuZRatRet3}, the radiative correction for return to the Z boson mass resonance increases in fraction, scaling as $s$, of the total dimuon cross-section while the leading order, tree level, process decreases as $s^{-1}$. In reality this dependence is more complicated as the ISR structure function used here is insufficient. Earlier work has derived that, at leading order, the dimuon ISR structure in equation~\ref{eqn-DiMuZRatRet3} can be updated

\begin{equation}\label{eqn-DiMuZRatRet4}
\begin{gathered}
    \frac{d\sigma}{d\cos\theta d\cos\theta'} = \frac{d\sigma}{d\cos\theta}(m_\text{Z}) \alpha\sin^2\theta'\frac{s}{m_\text{Z}^2}\left[1+(1-\frac{m_\text{Z}^2}{s})^2\right]\left[\log\frac{s}{m_\text{e}^2}-1 \right]\\
    \approx \frac{d\sigma}{d\cos\theta}(m_\text{Z}) \alpha \sin^2\theta'\frac{2s}{m_\text{Z}^2}\left[\log\frac{s}{m_\text{e}^2}-1 \right]
\end{gathered}
\end{equation}

to use a slightly different dependence on $z$ and $s$ which includes the log-leading term~\cite{Berends:1987ab}. The electron mass in the log-leading term is due to the ISR emitting particles being electron/positron; the mass dependence of the log term in equation~\ref{eqn-DiMuZRatRet4} is determined by the emitting particle. From equation~\ref{eqn-DiMuZRatRet4} we can see that the electroweak contribution to dimuon production, by the radiative return, becomes increasingly dominant at increasing center-of-mass energies. Solidifying dimuons as a important channel for precision electroweak measurements.

Dimuons, with various radiative and electroweak corrections, have been simulated using the MCEG of KKMC, or, more recently, WHIZARD~\cite{Jadach_2023}~\cite{Kilian_2011}. KKMC was developed during the LEP-era and has in depth handling of higher order corrections that served as theory counterparts for the precision LEP measurements of DiFermions. While WHIZARD is a general purpose event-generator, it still can act as a cross-validation to KKMC for dimuons and has the ability to have higher order corrections implemented by external libraries. Due to this, we believe that there is a path forward to continue use of dimuons for precision measurements. The combination of KKMC and WHIZARD should be suitable for future cross-validations of dimuons at the future benchmarks of precision.

\chapter{Luminosity Proposal}\label{ch-LumiProp}

In this chapter we start by providing some historical context and methodology of luminosity measurements at $\ee$ colliders by using \Gls{LEP} as a case study. Later the currently proposed methodology of luminosity measurements at future $\ee$ colliders, with emphasis on work from ILC and ILD's FCAL collaboration, will be presented. Lastly, new methods and insights on existing methods are presented in a proposal-esque format. Particularly, we will provide more in-depth analysis on the beam deflection effect and demonstrate how \Gls{diphoton} and dimuon events can be collectively used with SABS to achieve more precise luminosity measurements. Later, in chapter~\ref{ch-Lumi}, we will go further in depth on the execution of these new methods and insights and the results that they yield.

\section{Luminosity at LEP}

\begin{figure}[h]
\centering
\includegraphics[width=12cm]{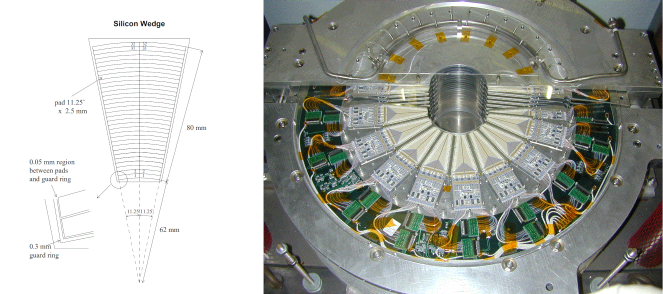}
\caption{(Left) Diagram of the internal structure of one of the detector pads of the OPAL luminometer.(Right) Picture of the OPAL luminometer from a top-down perspective.}
\label{fig-OPALFCal}       
\end{figure}

At \Gls{LEP} the precision on integrated luminosity was determined to the $10^{-4}$ level, with OPAL claiming $3.4\times10^{-4}$ precision~\cite{OPALLumi}. This level of integrated luminosity precision enabled precision electroweak measurements at the Z boson resonance and at higher center-of-mass energies too. At LEP the ALEPH, DELPHI, L3, and OPAL experiments all primarily relied on \Gls{SABS} as the physics process to measure integrated luminosity. These were measured in forward calorimeters of the various experiments. As an example, we provide a diagram of the OPAL luminometer, which was used for measuring SABS events, in figure~\ref{fig-OPALFCal}.
For thorough explanation of the theory component of integrated luminosity see chapter~\ref{ch-Theory}. To summarize, the use of SABS, from a theory perspective, is motivated by its theory simplicity and it being the dominant physics process and therefore statistically plentiful. With the theory and detectors, each experiment underwent similar analyses that carefully characterized their respective experiments.

As a part of this careful characterization, control regions, where the detectors and `upstream' material, material that would degrade the response function of detected particles, were precisely known and modeled. These control regions were then used for making closer to absolute measurements and then enabling other regions to make their measurements relative to the controls. The control regions also allowed for cross-validation and precision position control for the experiments.

\begin{figure}[h]
\centering
\includegraphics[width=12cm]{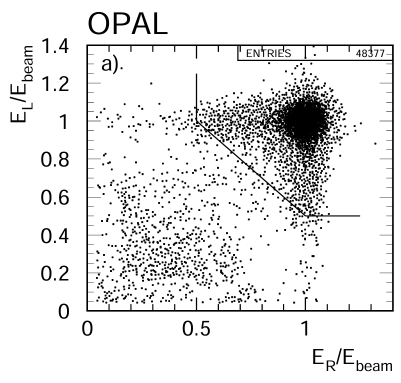}
\caption{Plot of the energy in the left and right calorimeter from SABS events before and after energy cuts, with the cut being indicated with the bold line. The events outside of the line are likely events with significant radiative corrections. The axes represent the energy of the Bhabhas with respect to the underlying beam energy. Figure from~\cite{OPALLumi}.}
\label{fig-OPALCut}       
\end{figure}

As a part of measuring SABS events the experiments employed cuts on on acollinearity of the electron-positron pairs, similar to how back-to-back they are, as well as cuts on energy. Such that the analyses focused on SABS that were close to the beam energy, $\ge 0.5 E_\text{beam}$ and the pairs were $E_1+E_2\ge 0.75 \sqrt{s}$. A plot of energy deposited in two parts of the forward calorimeter from SABS events before cuts, with a line denoting the energy cuts, can be seen in figure~\ref{fig-OPALCut}.
By requiring these cuts the analysis is less dependent on radiative corrections, therefore improving the quality of measurements. However, there is no free lunch, as it introduces dependence in the integrated luminosity on how well the angle and energy are measured. This requires precision metrology and angular measurements as well as precision energy calibration. Angular precision was characterized using detector anchors, well known positions, detector boundaries, test beams to essentially scan the detector and cosmic rays to do similar analyses. Some precisely measured particles, like muons, were also used. The overall position resolution contributed $1.5\times10^{-4}$ to integrated luminosity precision. Energy calibration was handled using test beams, internal calibration sources and measurements of well known particle decays. The overall energy calibration lead to an underlying energy precision of $1.8\times10^{-4}$ on the integrated luminosity. In summary, the combination of precision energy and position measurements, alongside the underlying precision in theory and statistics, are what enabled LEP to achieve integrated luminosity precision of the $10^{-4}$ level.

\section{Current Status of Luminosity at Future $\ee$ Colliders}\label{sec-Existing}

In this section we summarize the existing plans for precision integrated luminosity measurements at future $\ee$ colliders. We focus on \Gls{ILC} and \Gls{ILD} as their plans and proposed detectors are the most mature and tested. 

Plans to precisely measure integrated luminosity at future $e^+e^-$ colliders include upgraded versions of the LEP strategy: a compact forward calorimeter, typically of 1$X_0$ per layer, to measure \Gls{SABS} and employing careful calibration of energy and position as well as high-precision metrology~\cite{Abramowicz_2010}. The goal is to achieve similar or greater precision on integrated luminosity that LEP achieved at the Z-pole, $\sim10^{-4}$, as that is what is estimated to be required for precision measurements at the Z-pole at future $\ee$ colliders~\cite{Abramowicz_2010}. This precision requirement is less demanding at other center-of-mass energies, such as at the $ZH$ threshold where only $\sim10^{-3}$ is needed for the planned precision goals for the Higgs couplings available with $ZH$~\cite{Abramowicz_2010}.

\begin{figure}[h]
\centering
\includegraphics[width=10cm]{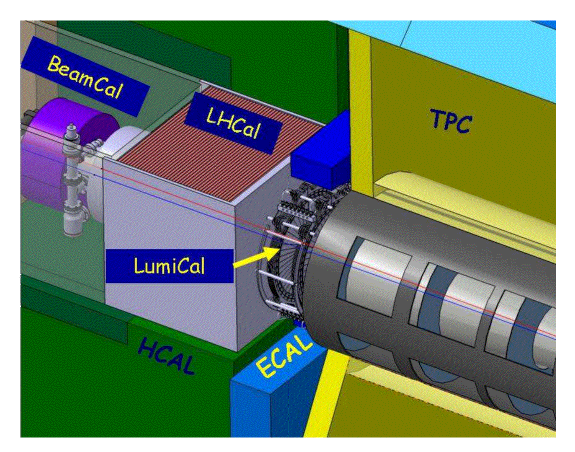}
\caption{Diagram of the sub-detectors of the forward region of ILD. Figure from~\cite{Abramowicz_2010}.}
\label{fig-ForReg}       
\end{figure}

\begin{figure}[h]
\centering
\includegraphics[width=8cm]{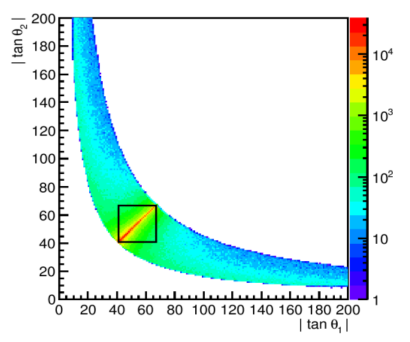}
\caption{Plot of the tangent of polar angle of electrons (positrons) of SABS events that are integrated luminosity candidates. A box demonstrating the $5^\circ$ acoplanarity cut is provided. Figure from~\cite{bozovicjelisavcic2014}.}
\label{fig-LumiCut}       
\end{figure}

In terms of instrumentation, both ILC and \Gls{CLIC} plan to use a compact forward calorimeter for integrated luminosity measurement. These are often referred to as luminometers or luminosity calorimeter or \Gls{LumiCal}. The existing design for the LumiCal involves the use of a $1X_0$ per layer Silicon-Tungsten, SiW, design with 320~$\mu$m thick silicon~\cite{Abramowicz_2010}. This design is, like the OPAL luminometer before it, divided into pads that are separated by divisions in $\text{r}\phi$. The angular acceptance of the LumiCal is $31<\theta<77$~mrad or about 1.75$^\circ$ to 4.40$^\circ$. In addition to the LumiCal, the forward region of ILD includes a LHCAL, which measure forward hadronic events, and the BeamCal. BeamCal is a forward calorimeter that measures pairs produced by the beam, which are numerous, low energy, and very forward angles. BeamCal may also measure SABS. BeamCal has an angular acceptance of $5<\theta<40$~mrad or about 0.30$^\circ$ to 2.30$^\circ$. A diagram of the forward region of ILD can be found in fig~\ref{fig-ForReg}.
To ensure simplicity and purity in event samples, due to contamination from \Gls{diphoton}s and four-fermion background, the current methodology proposes that SABS events must undergo preliminary cuts on energy, $80\%$ of nominal center-of-mass energy, and acoplanarity of $5^\circ$ or smaller~\cite{bozovicjelisavcic2014}. As seen in figure~\ref{fig-LumiCut}, this makes the dominant source of SABS events along the back-to-back topology, which is the diagonal line in the plot. 
This cut also introduces biases from beamstrahlung and ISR, due to the boosting of the initial state causing acoplanarity biases, and energy calibration, due to the energy cut. The result is that these cuts and beamstrahlung influence the precision of the integrated luminosity by roughly $0.4\times10^{-3}$ at ILC at all energies~\cite{bozovicjelisavcic2014}. The effect on precision from energy calibration depends on the quality of energy calibration, but is expected to be $1\times10^{-4}$ for 4.5~MeV precision on energy calibration~\cite{smiljanić2024metrology}.

Investigations into the polarization dependence of SABS integrated luminosity have been performed. When the helicity asymmetric contributions are significant there is a bias in the SABS cross-section of $(1+P_+P_-A)$, where $A$ is the asymmetry. However, as derived and discussed in section~\ref{sec-BhaThe}, this is only non-zero for the \Gls{schan} and electroweak contributions. Which current theory calculations predict are negligible above $10^{-5}$ for SABS. The asymmetry approaching zero has also been reproduced in other work, such as~\cite{grahFCAL}. This, combined with the excellent beam polarization monitoring of ILC, leads to negligible dependence of integrated luminosity precision on beam polarization~\cite{SBoogert_2009}.

Metrology of the forward region, and the beam interaction point, also play key roles in the integrated luminosity precision. In-depth work has been performed to determine the precision of various aspects of metrology needed to achieve $10^{-3}$ and $10^{-4}$ precision in integrated luminosity~\cite{smiljanić2024metrology}. For a breakdown of specific contributions of metrology we encourage examining table~1 of~\cite{smiljanić2024metrology}. We note that there is a need for precision on the radial position ($\theta$) of roughly 25~$\mu$m (10 $\mu$rad). Precision angular reconstruction in the forward region can help with this by allowing for the use of well reconstructed events to determine detector boundaries and other well-known positions. Aiding in the minimization of metrology effects on the integrated luminosity precision.

In summary, the current ILC methodology of using SABS in the LumiCal is capable of factors of $10^{-3}$ precision at all center-of-mass energies. The largest contributors to the integrated luminosity precision come from the event cuts used to tag SABS events, particularly their dependence on beam effects, as well as the metrology of the detectors and the beams.

\section{New Luminosity Methods Proposal}\label{sec-NewLumi}

While \Gls{SABS} remains the primary method for luminosity measurement, there is increasing interest in alternative and complementary processes to both cross-validate and improve the precision. Recently at Belle~II, a 10.58~GeV $e^+e^-$ B-factory, they have demonstrated the feasibility of using multiple channels simultaneously to measure luminosity~\cite{Adachi_2025}. Due to using multiple channels one has to modify the experimental measurement of integrated luminosity
\begin{equation}\label{eqn-IntLumi}
    \mathcal{L} = \frac{N_{\text{Obs}}}{\sigma_\text{sig}\epsilon_\text{sig} f_\text{tag}}
\end{equation}
to express in terms of the observed data, $N_{\text{Obs}}$, the signal cross-section within the acceptance, $\sigma_\text{sig}$, the signal selection efficiency, $\epsilon_\text{sig}$ and the tagging of each luminosity channel, $f_\text{tag}$. The BELLE~II reasoning is that, even if you introduce new systematics from $f_\text{tag}$, the use of multiple processes, SABS, \gls{dimuon}s, and \gls{diphoton}s, have different systematics and theoretical considerations that can collectively lead to superior precision. There is also more reason to be confident in the integrated luminosity precision as, if each measure converges on the same integrated luminosity, then there likely are no unknown and confounding effects like there was at LEP. Moreover, some processes can be used to reduce dependence on particular detector components. For example, using muon pair events can measure integrated luminosity based mainly on tracking detectors instead of calorimeters. In this section we discuss two such approaches, dimuons, diphotons, and weigh how each provides unique strengths and complementarity to SABS and each other as luminosity channels. We also touch on the experimental challenges and demands that each of these methods might see. The results of these proposals can be found in chapter~\ref{ch-Lumi}.

\subsection{New SABS Luminosity Proposal}

We propose the continuance of the use of SABS as a luminosity channel but to change the methodology to decrease some of the underlying systematics. We propose to not use the current cuts, which introduced dependence on beamstrahlung, as this systematic was $0.4\times10^{-3}$, a considerable contribution on the integrated luminosity precision. Instead, we propose the use of a combination of detector and reconstruction level decision trees. Which must be trained to achieve precision that is acceptable for the integrated luminosity precision goals. We also propose that the beam deflection effects be investigated further, so that their effects on integrated luminosity precision are better understood and more up-to-date. In the following section, section~\ref{sec-BDE}, we will conduct this investigation. 

For the purpose of minimizing integrated luminosity precision, we propose that the LumiCal of future $\ee$ colliders must be highly granular, i.e. less than $X_0/2$ per layer, SiW calorimeters. By maintaining the use of SiW the energy resolution is expected to be comparable, or better, than the existing LumiCal design~\cite{Madison:2024jak}. Increasing granularity significantly increases position resolution, by a factor of $\sim10$~\cite{Madison:2024jak}. This improved position resolution is able to reach $10\mu$m, meaning that it can reach, or even surpass, the metrology demands for integrated luminosity precision of $10^{-4}$~\cite{Madison:2024jak}~\cite{smiljanić2024metrology}. The increase in granularity also allows for the acceptance for SABS to increase to include narrower and wider angles. With an acceptance of $\approx1^\circ<\theta<5^\circ$ compared to the $\approx1.75^\circ<\theta<4.40^\circ$ of the current LumiCal design. This increase in statistics means that quality cuts on SABS can be stricter, and therefore more precise. It also means that, for the purpose of training a decision tree or other machine learning technique, there is more data to learn from.

\begin{figure}[h]
\centering
\includegraphics[width=12cm]{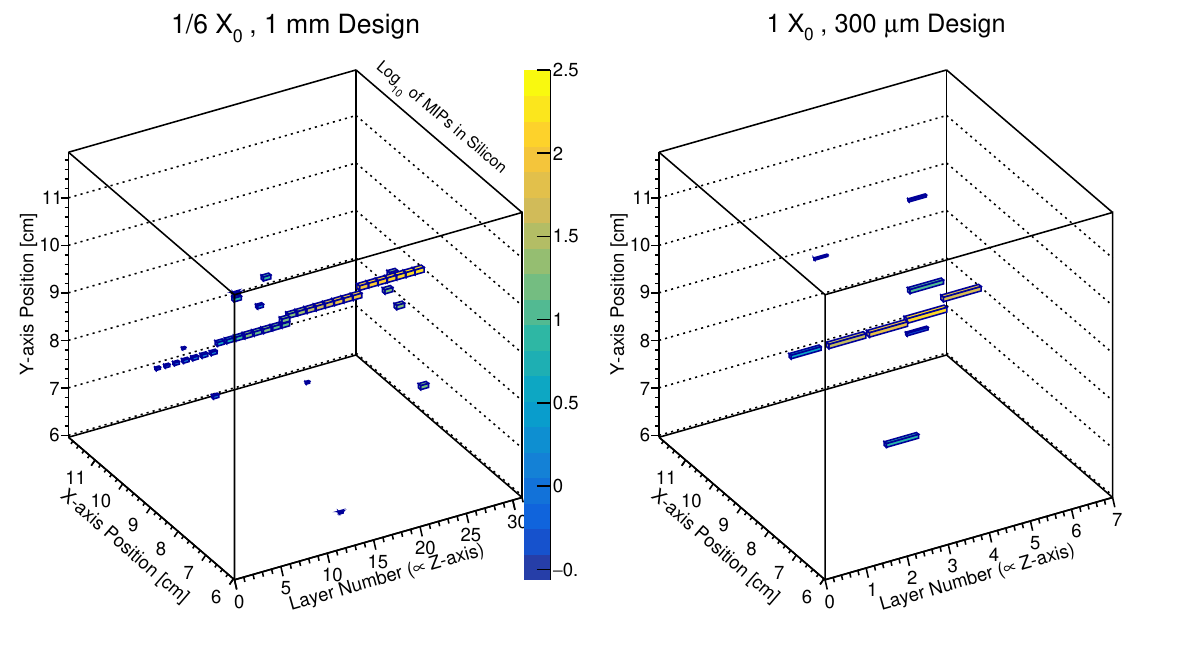}
\caption{Event display of two different forward calorimeter designs measuring a single 125~GeV photon.}
\label{fig-HighGran}       
\end{figure}

Higher granularity calorimeters also benefit from simply having more information from the increased number of layers. A demonstration of this difference can be seen in figure~\ref{fig-HighGran}. 
This benefit is two-fold as the particle shower structure, especially in the initial shower, are able to be isolated from the bulk, more stochastic, parts of the shower. This allows detector level cuts and tagging to be more precise to the underlying event types and makes `Big Data' techniques, like machine learning, feasible at either the \emph{in situ} or offline level. These techniques already exist at the LHC for the trigger as well as for tagging events~\cite{ATLAS:2022rkn}~\cite{Duarte:2018ite}. Due to the goal of achieving $10^{-3}$ ($10^{-4}$) precision on integrated luminosity, any tagging method would need to achieve similar or better precision on tagging SABS events. Highly granular calorimeters are also better at measuring and tagging photons. This is demonstrated in greater detail in chapter~\ref{ch-Lumi}. This gives better control on separation of the various physics processes that occur in the forward region. It also allows for radiative SABS events, where the photons are also reconstructed in the LumiCal, to be dressed to their respective electron (positron) to get more precise energy and position resolution.

In summary, we propose updating the existing SABS methods to remove some of the larger systematics introduced by the cuts. We also propose a highly granular LumiCal so that the position and energy resolutions are improved and so that the event cuts and tagging can be improved. With these changes we are certain that $10^{-3}$ precision on integrated luminosity can be met and hopeful that it could even be extended to 100~ppm precision.

\subsubsection{Beam Deflection Effect}\label{sec-BDE}

Since we are presenting an alternative integrated luminosity method for SABS it is worthwhile to determine how much the beam deflection effect influences the integrated luminosity precision. Previous studies into precision measurements and modeling of Bhabha scattering have found that the interaction of outgoing electrons or positrons with the beam can produce non-negligible deflection. Initial studies of this for ILC indicated effects at the $10^{-3}$ to $10^{-4}$ level~\cite{CRimbault_2007}. The beam deflection effect was initially discovered years after LEP was converted to LHC and, therefore, numerous papers about adding this correction to update LEP measurements have been published~\cite{Voutsinas_2020}. In particular to LEP the deflection was found to have a bias in the mean of roughly 10~$\mu$rad~\cite{Voutsinas_2020}. This bias leads to a shift in the number of events measured. If, instead, there is no bias in the mean and the distribution of deflection is symmetric about zero, then there is no net influence on integrated luminosity. For example, if we use the angular dependence of SABS on $\tinn$ as derived in equation~\ref{eqn-SABS_Inner_Unc},
\begin{equation}\label{eqn-SABSShift}
    \frac{\delta N}{N} = \frac{-2}{\theta_\text{min}^{-2}-\theta_\text{max}^{-2}}\left[ \frac{\delta\theta_\text{min}}{\theta_\text{min}^3} - \frac{\delta\theta_\text{max}}{\theta_\text{max}^3} + \frac{\delta\theta_\text{min}}{2\theta_\text{min}} - \frac{\delta\theta_\text{max}}{2\theta_\text{max}} \right]
\end{equation}
the values of minimum and maximum polar angle acceptance for SABS, and the angular deflection at the minimum and maximum angle, lead to a roughly $-0.1$\% effect on integrated luminosity precision. Previous studies of this have been done for ILC, but the published studies are over a decade old and designs and software have changed significantly since then~\cite{CRimbault_2007}. In this section we will investigate and model beam deflection as well as redo the tests of SABS deflection at ILC. We will also extend this work to include other center-of-mass energies.

\begin{figure}[h]
\centering
\includegraphics[width=12cm]{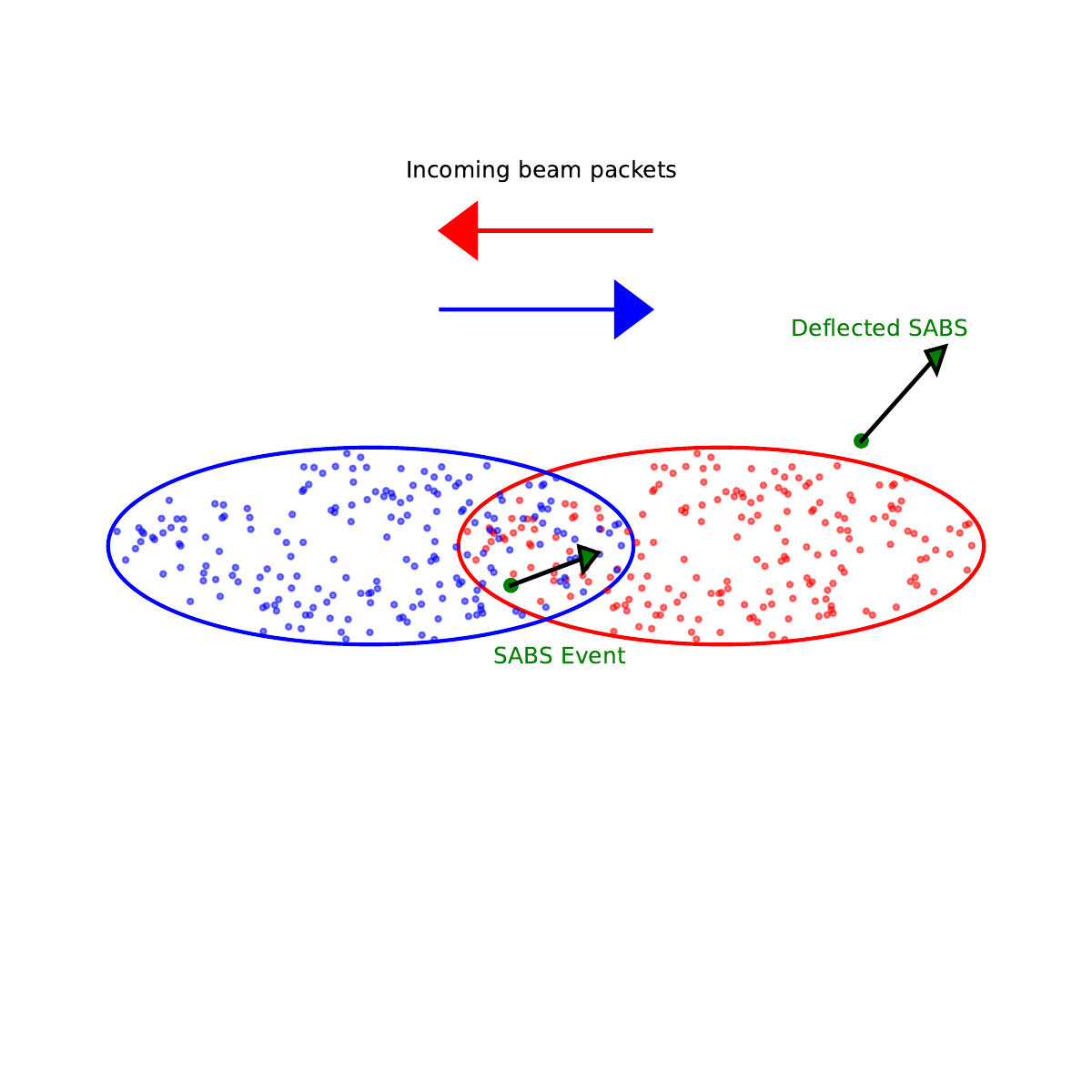}
\caption{Diagram of how a SABS event can be deflected by the field of the beam particles. In this hypothetical a SABS event that would normally fall below the inner acceptance is deflected up and into the acceptance.}
\label{fig-DeflDia}       
\end{figure}

To demonstrate one possible form of deflection, a Bhabha scattering event, seen in figure~\ref{fig-DeflDia} and usually restricted to SABS, is produced at the front of a beam packet and then must travel through the remaining beam packet.
The strong electric field of the beam packet will then create a force upon the out-going particle, which then is deflected.

By using GuineaPig++, to handle particle deflection by the beam, and \Gls{BHLUMI}, to generate SABS events, we can simulate the effect of beam deflection on SABS events given accelerator parameters for a given experiment or run. To do so, one must convert the output of BHLUMI, which is a text file of particle values per event, to a format that is acceptable to GuineaPig++'s \textbf{bhabha.ini} file. The format of this can be found in a GuineaPig++ manual. The input parameter card for GuineaPig++, usually named some variation of \textbf{acc.dat}, must indicate that the setting for Bhabhas and pairs are turned on, but that pair generation is turned off. The same functions within GuineaPig++ that handle transport of beam pairs through the beam fields handles the transport of Bhabhas, but they are done exclusively so only one can be turned on at a time. After running GuineaPig++ the output file of interest is \textbf{pairs.dat}, which contains the Bhabhas after all simulation operations have been handled and the Bhabhas have been transported outside the beam field. The \textbf{pairs.dat} contains matching output particles for the input Bhabha events supplied in \textbf{bhabha.ini}, but it is not sorted. After sorting, the user can do calculations on beam deflection from the kinematic values from the pre-deflection and post-deflection values.

\begin{figure}[h]
\centering
\includegraphics[width=16cm]{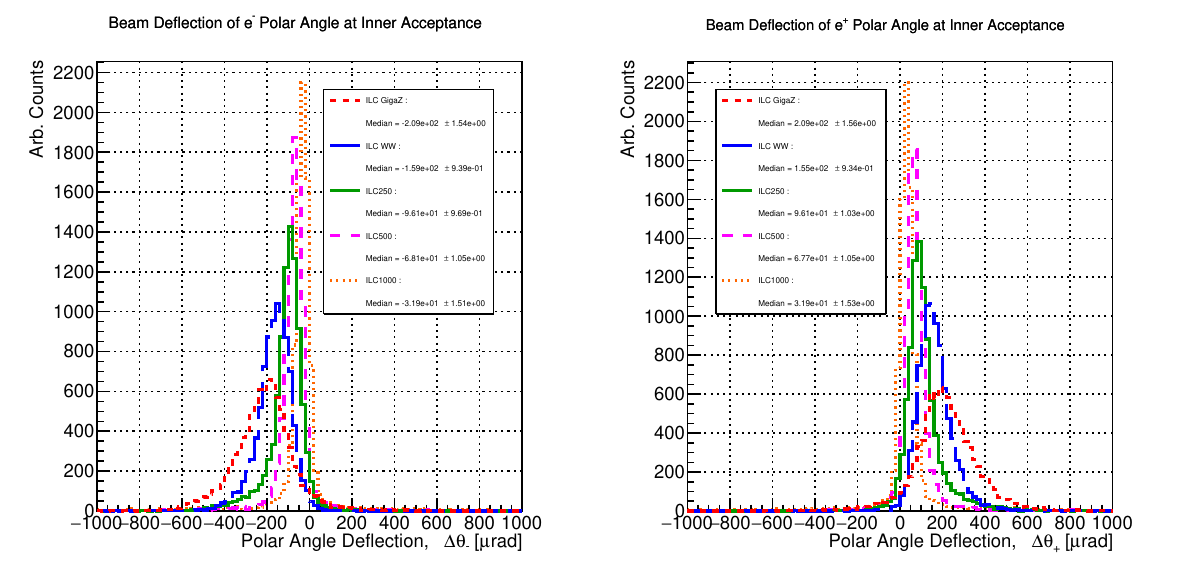}
\caption{(Left) Histograms of the polar angle deflection for the out-going electron at the inner acceptance. (Right) Histograms of the polar angle deflection for the out-going positron at the inner acceptance. Median of the distributions are provided for various beam energies tested.}
\label{fig-DefPolPlot}       
\end{figure}

\begin{figure}[h]
\centering
\includegraphics[width=16cm]{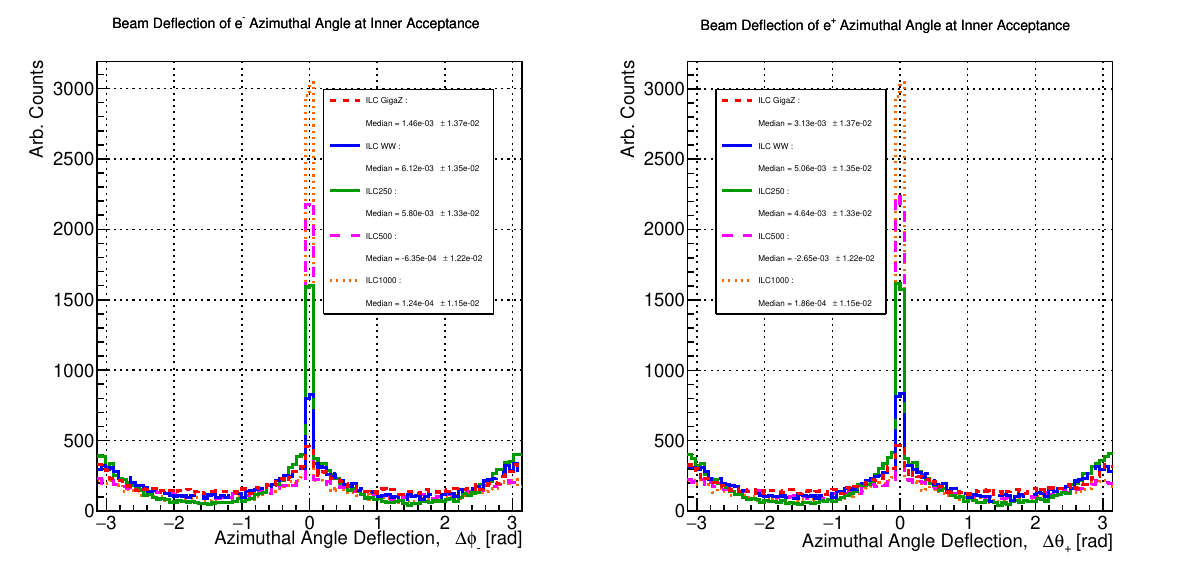}
\caption{(Left) Histograms of the azimuthal angle deflection for the out-going electron at the inner acceptance. (Right) Histograms of the azimuthal angle deflection for the out-going positron at the inner acceptance. Median of the distributions are provided for various beam energies tested.}
\label{fig-DefAziPlot}       
\end{figure}

\begin{figure}[h]
\centering
\includegraphics[width=14cm]{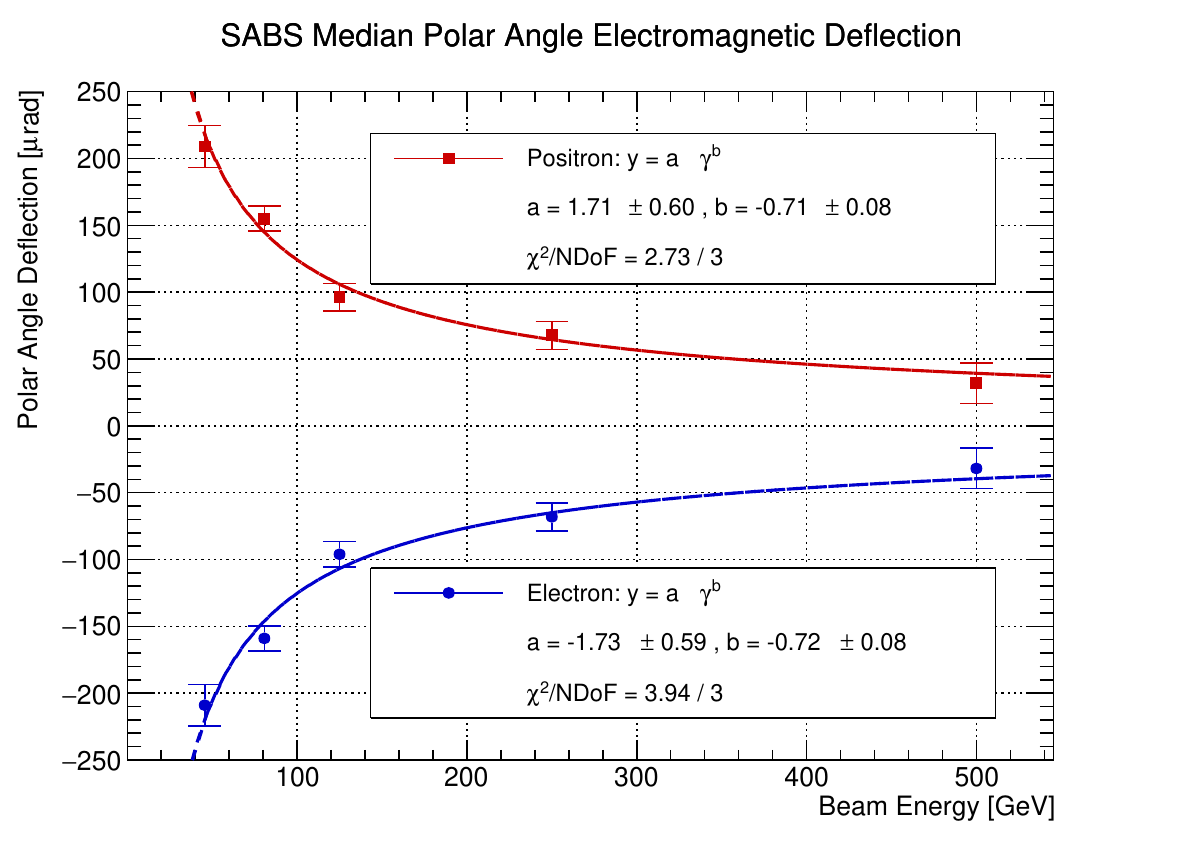}
\caption{Plot of the SABS median polar angle deflection at ILC at an inner acceptance angle of $1^\circ$ for various beam energies. A fit for the trend in the Lorentz factor is done and finds that the trend fits well for both electron and positron.}
\label{fig-DefTrendPlot}       
\end{figure}

\begin{figure}[h]
\centering
\includegraphics[width=14cm]{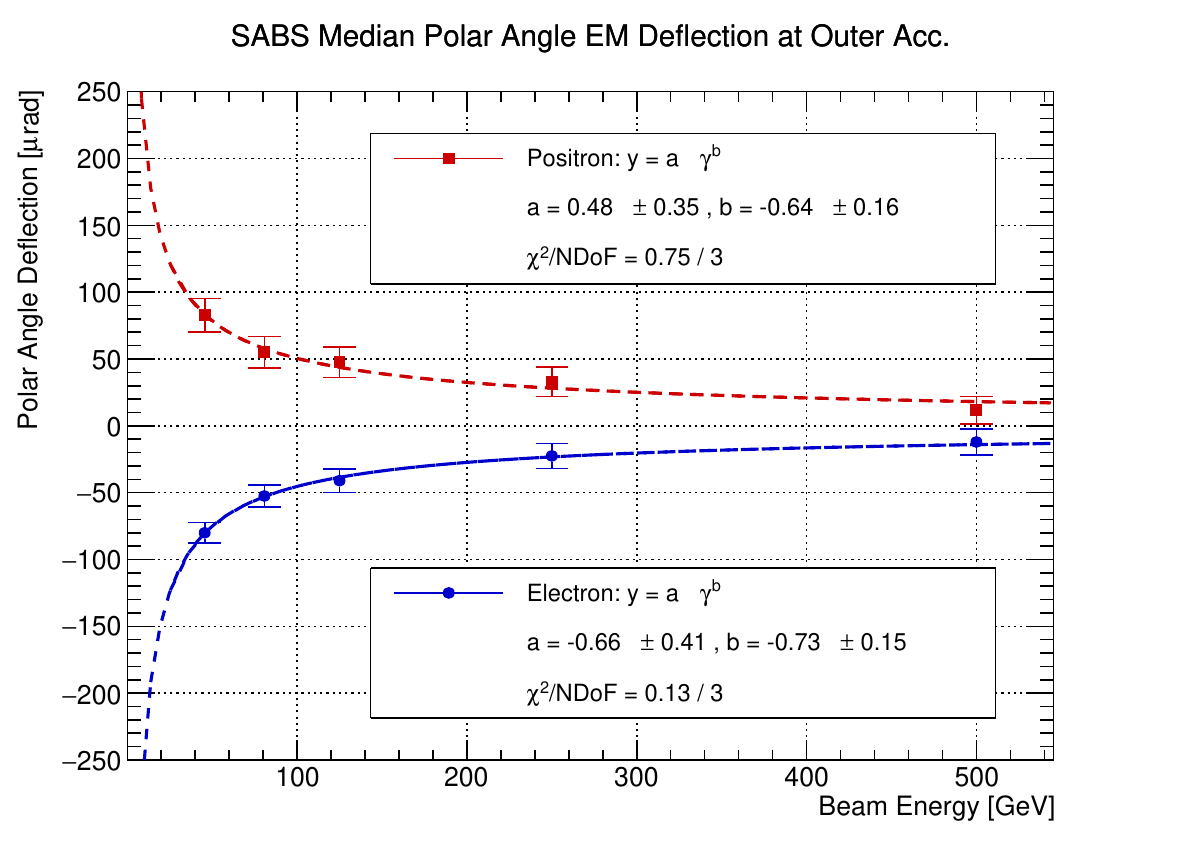}
\caption{Plot of the SABS median polar angle deflection at ILC at an outer acceptance angle of $6^\circ$ for various beam energies. A fit for the trend in the Lorentz factor is done and finds that the trend fits well for both electron and positron.}
\label{fig-DefOutTrend}       
\end{figure}

We simulate 2M SABS events at the various energies for ILC, from GigaZ to ILC1000, within an acceptance of $1^\circ$ to $6^\circ$. We investigate the deflection values in polar angle and azimuthal angle at the inner and outer acceptance. We compute the deflection angle in terms of the initial angle subtracted from the final angle. A plot of the polar angle deflection at the inner acceptance angle, of $\tinn=1^\circ$, and plot it according to figure~\ref{fig-DefPolPlot}.
While it is not the focus of this study, we also present the azimuthal angle deflection in figure~\ref{fig-DefAziPlot} at the inner acceptance angle of $1^\circ$.
Returning to the polar angle deflection, we find that the polar angle deflection roughly follows the inverse of the Lorentz factor, $\sim\gamma^{-1}$. This agrees with math and modeling done in previous work which expects a $E^{-1}$ dependence~\cite{Voutsinas_2019}. To better test this we fit for a model of the form of $a\gamma^{b}$, where $a$ and $b$ are fit parameters. To put the fit in terms of the Lorentz factor we divide the beam energy values by the electron mass. During this test of fitting we found that the fit fails unless one assumes that the statistical error on the median, as reported in the values of figure~\ref{fig-DefPolPlot}, is not the dominant uncertainty source and that instead there is a systematic uncertainty that was $\sim\times10$ the statistical uncertainty reported. We find that testing this at all beam energies finds a similar level of systematic uncertainty. In terms of precision, the systematic uncertainty is $\sim3$\%. We investigated this factor by running repeated runs of BHLUMI and GuineaPig++ to determine what the change from run to run was. Going forward, we will use this factor of systematic uncertainty in quadrature with statistical uncertainty unless we have noted that the result is from an ensemble of runs. The fit of Lorentz factor after including systematic uncertainty, as seen in figure~\ref{fig-DefTrendPlot}, is a plausible fit with a trend of $\approx\gamma^{-0.7}$.
We use a fit for the power of the Lorentz factor such that $a\gamma^{b}$ is the fit function form. This fit is motivated by the underlying math which, as mentioned previously, expects a trend of $\gamma^{-1}$. This result indicates that the trend in the Lorentz factor is slightly more complicated. We can continue this trend fitting for the polar angle deflection at an outer acceptance of $6^\circ$ too. The results, seen in figure~\ref{fig-DefOutTrend}, indicate that the outer acceptance deflection is about 36\% of the inner acceptance deflection but still follows the general trend of $\approx\gamma^{-0.7}$.
The fits and data at the inner and outer acceptance have different sample sizes as we chose here to use the raw BHLUMI data instead of generating artificial data that looks like SABS but has fixed angles for both particles. We mention this as some previous studies use artificial data while others use the output of a \Gls{MCEG}, as done here~\cite{CRimbault_2007}~\cite{Voutsinas_2019}. This also means that these results include radiative effects that were already present in the BHLUMI data. We expect that the fit quality of figure~\ref{fig-DefOutTrend} and figure~\ref{fig-DefTrendPlot} would change significantly if an ensemble of runs were used at each data point in energy and if more energy points were tested. We suspect that the collider being investigated will change the fit parameters; therefore, for the purpose of comparing colliders with respect to their beam deflection, we recommend that this fit form be used. So comparisons can be made in a quantitative manner over the fitted parameter values. Given the values measured here, and using equation~\ref{eqn-SABSShift}, we expect that the electromagnetic beam deflection effect on integrated luminosity precision is 1\%-6\%, depending on the beam energy.

\subsubsection{M\o{}ller Scattering To Measure Beam Deflection Efffect}\label{sec-Moller}

Given the results of section~\ref{sec-BDE}, which indicate that the electromagnetic beam deflection effect would be a dominant source of integrated luminosity uncertainty, we search for ways to minimize this uncertainty. A simple perspective would be to measure the beam deflection in some way. Previous studies have suggested using beam test bunches or monitoring the beam deflection as the beam `ramps up'~\cite{Voutsinas_2019}. Instead, we propose a new method of measuring the beam deflection by using M\o{}ller scattering. The M\o{}ller scattering process involves $e^{-}e^{-}\to e^{-}e^{-}$, so it is essentially inaccessible to an $\ee$ collider. Therefore, in order to use it, one would need to do a dedicated run where the collider is converted to an $\text{e}^-\text{e}^-$ collider. This is already among the proposals for future linear colliders, as there are interesting physics reasons to run as an $\text{e}^-\text{e}^-$ collider~\cite{LCVision}. M\o{}ller scattering is an additional, and clean, measure of the weak mixing angle, it is sensitive to $W^-W^-$ production and therefore any Majorana contributions that may be present in $\nu_e$, and it is sensitive to selectron production via t-channel neutralino propagators~\cite{LCVision}. This is not feasible for a circular collider as it would require multiple rings or other technology to allow for same-sign particles in the same ring in opposite directions.

\begin{figure}[h]
\centering
\includegraphics[width=14cm]{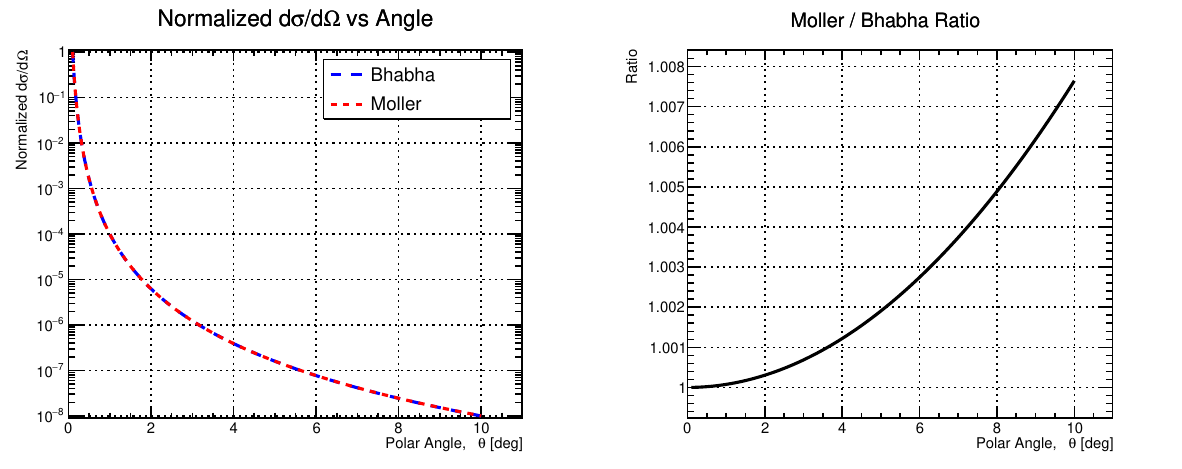}
\caption{(Left) Plots of the differential cross-section, with respect to the polar angle in degrees and normalized so that the maximum is unity, of SABS and M\o{}ller scattering at leading order.(Right) The ratio of differential cross-sections of M\o{}ller scattering and Bhabha scattering.}
\label{fig-SABSMolRat}       
\end{figure}

At leading order and at small angles M\o{}ller scattering is nearly kinematically identical to Bhabha scattering. This is not surprising, as the t-channel formalism for both is identical, with the main difference being that the identical particles in $\text{e}^-\text{e}^-$ allows for interference of the t-channel and u-channel and therefore an additional term for M\o{}ller scattering. To demonstrate this, we plot the leading order SABS and M\o{}ller differential cross-section as well as the ratio of the two in figure~\ref{fig-SABSMolRat}.
The difference between SABS and M\o{}ller scattering in a LumiCal is roughly $10^{-3}$. We expect that a larger effect will be seen from the beam effects since the beam packet of positrons is now replaced with a beam packet of electrons. From studies on beam effects and the beam collision software \Gls{GP}, we expect that the beamstrahlung and instantaneous luminosity will be less~\cite{Guinea-PIG}. We also do not expect the pinch effect, which enhances opposite charge sign collisions, to enhance instantaneous luminosity~\cite{Guinea-PIG}. The desire for this approach is that these differences in beam effects will not result in significantly different polar angle deflections. Instead, by changing the positron bunch to an electron bunch, the field sign and charge sign should both change. This should then flip the direction of the polar angle deflection. Such that, across the Bhabha scattering and M\o{}ller scattering events, there would be measurements of deflection, $\delta\theta$, at $\tinn+\delta\theta$ and $\tinn-\delta\theta$. This would then allow for solving for the amount of polar angle beam deflection. This can be done by using the small-angle approximation of M\o{}ller scattering and Bhabha scattering to establish a system of equations. Such that we have
\begin{equation}\label{eqn-ShiftSys}
\begin{gathered} 
    N_\text{Mol,F} \propto \sigma_{\text{Mol,F}} \propto \frac{1}{(\theta+\delta\theta)^{3}} \\
    N_\text{Bha,F} \propto \sigma_{\text{Bha,F}} \propto \frac{1}{(\theta-\delta\theta)^{3}}
\end{gathered}
\end{equation}
two equations for the number of M\o{}ller scattering electrons in the forward direction, $N_\text{Mol,F}$, and the number of Bhabha scattering electrons in the forward direction, $N_\text{Bha,F}$. Using the count of events, we can solve for $\delta\theta$ by using a cross-section weighted midpoint calculation to estimate the midpoint between $\theta+\delta\theta$ and $\theta-\delta\theta$. By extending this to a fitting method one can also incorporate uncertainty values, allowing for an estimation of both the amount of beam deflection and the uncertainty in the amount of beam deflection.

\begin{figure}[h]
\centering
\includegraphics[width=16cm]{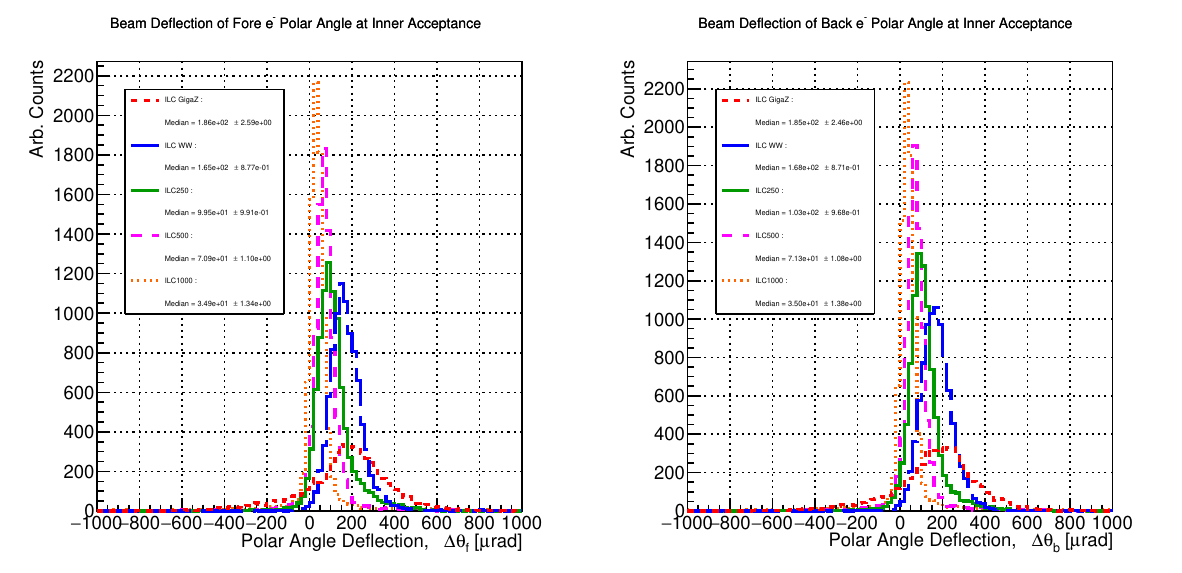}
\caption{(Left) Histograms of the polar angle deflection for the out-going electron at the inner acceptance. (Right) Histograms of the polar angle deflection for the out-going positron at the inner acceptance. These tests were conducted using small-angle M\o{}ller scattering. The median of the distributions are provided for various center-of-mass energies tested using ILC beam specifications.}
\label{fig-MolDef}       
\end{figure}

For simulation of M\o{}ller scattering and the beam effects we will use \Gls{BHLUMI} and \Gls{GP}. We recognize that BHLUMI generates Bhabha scattering events; the results of figure~\ref{fig-SABSMolRat} shows that the difference between SABS and small-angle M\o{}lller scattering is small. Still, future work should incorporate dedicated M\o{}ller scattering generators or use a general purpose \Gls{MCEG}, like WHIZARD, with higher order corrections~\cite{Kilian_2011}. We also want this analysis to focus on changes that are purely from beam-effect differences between $\text{e}^-\text{e}^-$ and $\ee$ collisions, so we will continue with BHLUMI as the generator of choice. For running GuineaPig++ one must edit the \textbf{acc.dat} such that the value of \emph{charge\_sign} is switched from $+1$ to $-1$. This changes the simulation to $\text{e}^-\text{e}^-$ collisions from $\ee$ collisions. After this change the simulation of electromagnetic beam deflection proceeds as it did in section~\ref{sec-BDE}, with no other changes to beam specifications. The result, seen in figure~\ref{fig-MolDef}, indicates that the sign of electron beam deflection flips and the magnitude of beam deflection for M\o{}ller scattering is close, within at least $\sim$10\%, of the values observed for Bhabha scattering in figure~\ref{fig-DefPolPlot}.
As discussed earlier, we observed that the GuineaPig++ simulation of the electromagnetic beam effect is roughly 3\% consistent between runs. We are unsure how much the change between SABS and M\o{}ller scattering is from this effect or genuine differences in beam dynamics.

\begin{figure}[h]
\centering
\includegraphics[width=14cm]{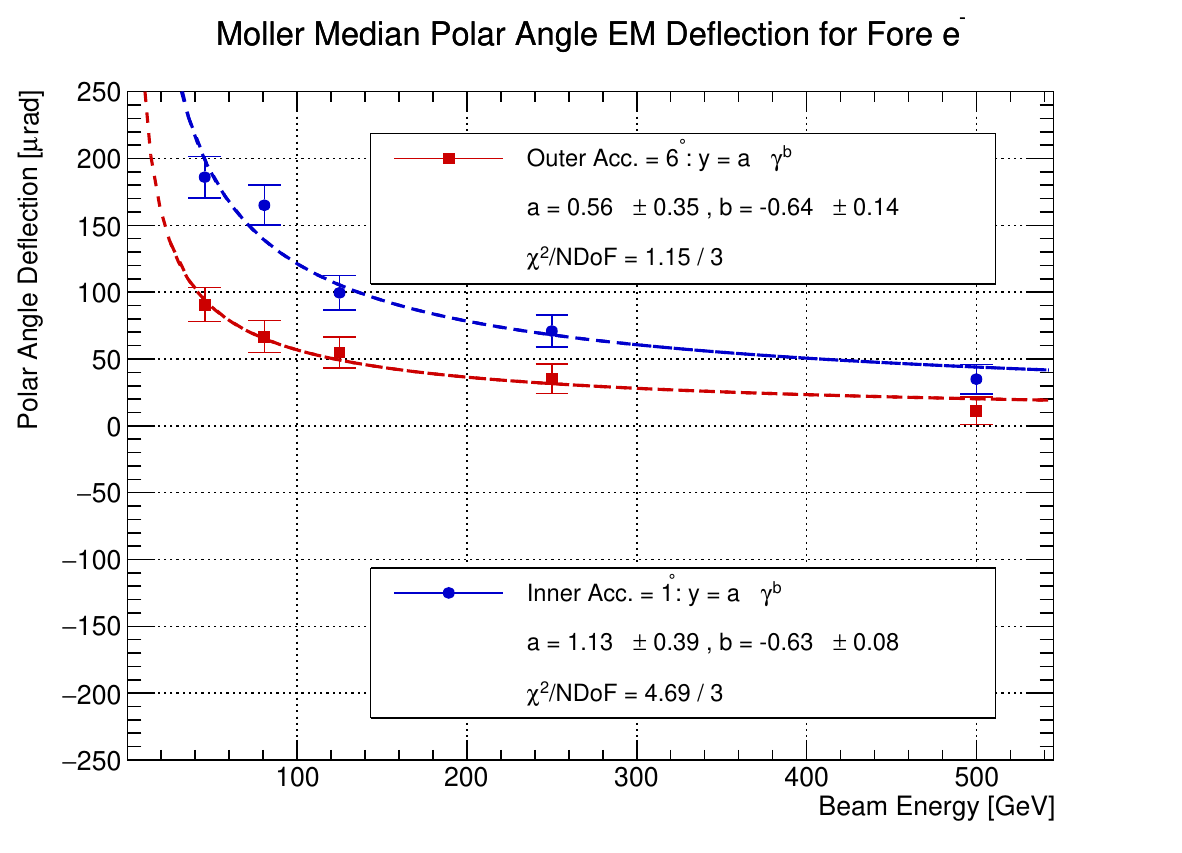}
\caption{Plot of the small-angle M\o{}ller scattering median polar angle deflection for the foreward electron at ILC at the inner acceptance angle and outer acceptance angles for various beam energies. A fit for the trend in the Lorentz factor is done and finds that the trend fits well for both the inner and outer acceptance.}
\label{fig-MolTrend}       
\end{figure}

\begin{figure}[h]
\centering
\includegraphics[width=14cm]{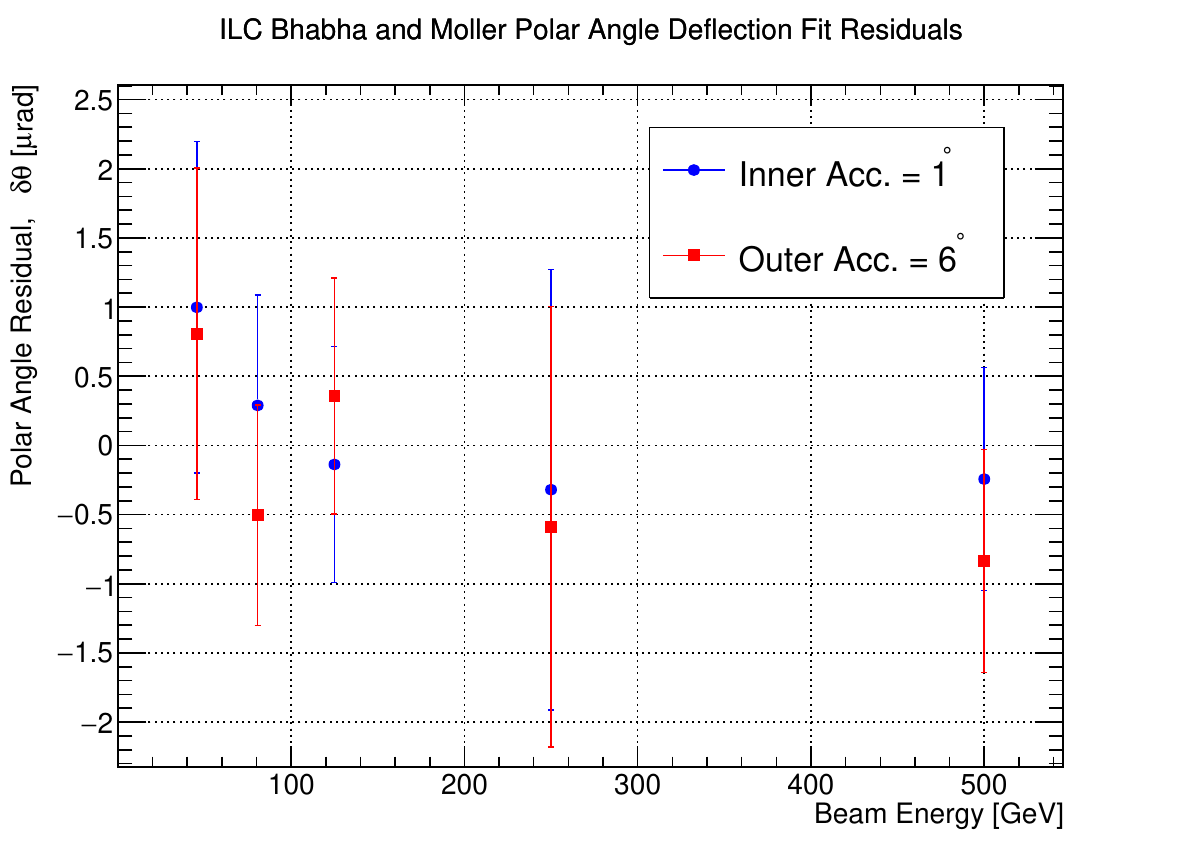}
\caption{Plot of the residual of the fitted deflection angle with respect to the true deflection angle from electromagnetic deflection of SABS electrons. The result for the inner and outer acceptance are included. Results do not indicate that performance changes much with changes in the beam energy.}
\label{fig-FitTrend}       
\end{figure}

Since we only need the values for the polar angle deflection of the forward electron at the inner acceptance and outer acceptance to solve for equations~\ref{eqn-ShiftSys} and ~\ref{eqn-SABSShift}, we will restrict the fitting the trend to these cases. Due to the change in beam dynamics we hypothesize that the trend will change with respect to the results seen in figures~\ref{fig-DefTrendPlot} and~\ref{fig-DefOutTrend}. The result, seen in figure~\ref{fig-MolTrend}, indicates that there is a change in both the coefficient and power terms, but that it is not significant enough to be discernible with the current sample size and accuracy.
This confirms the hypothesis that the beam dynamics are different between the two scenarios, but they are not different enough to make this approach fail. To prove this we proceed to solving for the deflection angle. First we start with the fits from figures~\ref{fig-DefTrendPlot},~\ref{fig-DefOutTrend} and~\ref{fig-MolTrend} to generate the expected values of deflection at each beam energy value. We then compute the number of events for a given luminosity and the theory cross-section at leading-order and small-angles for a given angle and deflection amount. Particularly at the inner and outer polar angle acceptance angles. In the experimental setting, this step would be replaced by simply counting the number of events in a given acceptance around the inner and outer acceptance angles. We then inject noise into these counts to reflect the underlying reality that there was significant systematic uncertainty in the original dataset. Then we proceed to fitting using Minuit and a system of equations like equation~\ref{eqn-ShiftSys}, and propagating the uncertainty from the results of the Minuit fit~\cite{James:1994vla}. This then gives us an estimate and uncertainty for $\delta\theta$, the amount of deflection in the polar angle, at both the inner and the outer acceptance angles. For this fit we use the injected noise and statistical uncertainties, in quadrature, as the uncertainty on the counts and 10~$\mu$rad as the uncertainty on the polar angle measurement. Our initial test used 200M events for Bhabha scattering in the LumiCal. From the fit result we can compare to the true value from the SABS simulation results, as seen in figure~\ref{fig-FitTrend}, and find that the fitted values are able to be within a few $\mu$rad of the true value.
Given these results we are confident that one can minimize the effect of electromagnetic beam deflection on the precision of integrated luminosity.

As last remarks, we recognize that there may be pitfalls in this technique. The underlying theory cross-section for both Bhabha scattering and M\o{}ller scattering must be known fairly well so that the ratio of Bhabha and M\o{}ller events are correct. This also extends to the control of integrated luminosity; the integrated luminosity must be the same, within some amount, between the two data samples, otherwise this method will not reconstruct the correct deflection angle. The data sample must also be large enough. We tested data samples of 2M and 200M total events, with the results above reflecting the 200M case. The results of both were similar, with the 200M result slightly improving, by $\sim10\%$, in both accuracy and uncertainty. Both are fairly small data samples compared to the billions of SABS events from a full run, with the 200M data sample being comparable to 4~$\invfb$ of data at ILC250. As a caveat, please note that the instantaneous luminosity when running as an $\text{e}^-\text{e}^-$ collider for M\o{}ller scattering is $\approx1/10$ the instantaneous luminosity when running as an $\ee$ collider, so it would take longer to obtain similar statistics. Finally, we note that recent developments in theory and simulation of $\ee$ and $\text{e}^-\text{e}^-$ colliders have shown that strong-field QED effects can create an anomalous pinch effect in $\text{e}^-\text{e}^-$ collisions that is not seen in $\ee$ collisions~\cite{Zhang:2024wob}. It was also found that this is mostly relevant to CLIC and other colliders with similar density or even denser beams. Using M\o{}ller scattering to diagnose beam deflection at ILC or a similar facility may be a simpler starting point for such an experiment since it has not been done before.

\subsection{Diphoton Luminosity}\label{sec-dipho}

As a leading background process for SABS, \gls{diphoton}s can contribute to the systematics of using SABS for precision integrated luminosity measurements. Since diphotons have a large cross-section in the forward region, comparable to 10~pb at 250~GeV or $0.1\%$ of SABS, there is a need to develop methods to separate SABS and diphotons. Here we propose to go one step further; we propose to develop separate tags and cuts for diphotons and SABS so that both event topologies can be identified and separated from other background processes and from each other. Using diphotons also has three additional benefits in terms of systematics. diphotons, as derived in section~\ref{sec-DiGam}, have a shallower angular dependence compared to SABS so the dependence on metrology and position resolution of the inner and outer acceptance is smaller by a factor of $\sim\theta^{-3}/\theta^{-1}$. diphotons are neutral events, they involve no charged particles at leading order, so there is no beam deflection or other electromagnetic deflection. Regardless of how small or large the beam deflection effect is for a given experiment, diphotons are always an alternative. There is also no multiple scattering of the particles until the photons undergo photon conversion and the particle shower starts. So diphotons are better suited for precision position reconstruction of the particles as well as reconstructing the position of the \Gls{IP} and effects that may arise from the changes in the IP.

\begin{figure}[h]
\centering
\includegraphics[width=12cm]{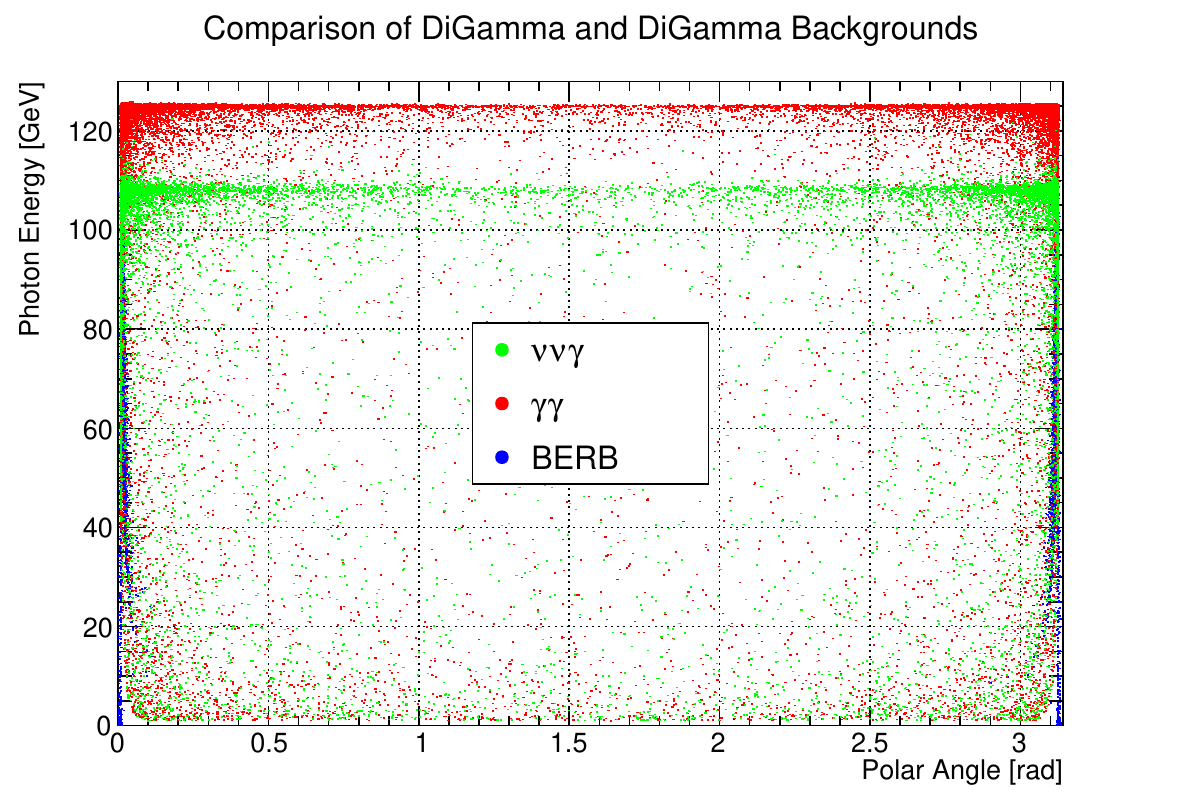}
\caption{Plot of the energy and polar angle of all photons in any of the given physics processes. An energy of 125~GeV and minimum acceptance of $1^\circ$ was used here and the number of events was controlled to 10000 events per process. The BERB photon emission contains two components, from single and double emission, and so there are two trends in the distribution of energy and polar angle.}
\label{fig-EneThtGam}       
\end{figure}

In addition to the background from SABS, diphotons have additional backgrounds from \Gls{BERB} and radiative return events. For the radiative return events there can be a veto of events with reconstructed charge particles but there are still neutrino radiative return events that will pass this veto. Similarly, for BERB, the photon(s) will almost always be in the LumiCal with no additional particles in the event. This can be seen in figure~\ref{fig-EneThtGam}, where the BERB events are almost exclusively at the left and right edges of the plot, in the forward region. As can be seen in figure~\ref{fig-EneThtGam}, the polar angle and energy distributions of the photons from these three processes occupy similar areas in the forward region but they occupy different energy scales.
Radiative return events dominantly emit single photons with an energy fraction close to that calculated by equation~\ref{eqn-PhotonZRatRet}. Considering this, we propose that radiative return photon events be vetoed by requiring two photons in the forward calorimeter with a calorimeter energy sum cut similar to SABS, being $80\%$ of the nominal center-of-mass energy. BERB events dominantly emit one or two photons with an energy that is comparable to $\sqrt{s}/4$, or half the beam energy, so requiring this energy cut should also reject BERB events.

Using diphotons as a luminosity channel has two weaknesses; the cross-section of diphotons is not large enough to ensure that it will not be statistics limited and it introduces a systematic from the beam polarization. Due to the $(1+P_-P_+)$ dependence on the diphoton cross-section and given the $0.1\%$ precision on polarization monitoring at ILC, the resulting effect on diphoton integrated luminosity is $6\times10^{-4}$~\cite{Karl:424633}. This systematic can be avoided if one combines two different runs of opposite beam polarizations and can control their systematics in a relative way to be equal and opposite. This has yet to be tested or demonstrated, but we do not consider it to be a difficult task, partly because it is aligned with the already existing plan to run two opposite polarization runs. In terms of statistics, for the high-granularity LumiCal proposed here, the statistical precision for the full ILC250 $2\text{ab}^{-1}$ run would be comparable to $2\times10^{-4}$. Using the existing LumiCal would give a statistical precision comparable to $4\times10^{-4}$. For this reason, it is crucial that the forward calorimeter be able to provide as much polar angle coverage as possible and/or that some fraction of wide-angle diphotons can be included in the integrated luminosity measurement. By doing so it is possible, though likely difficult, to reach $1\times10^{-4}$ for a $\sim2\text{ab}^{-1}$ sample. We present two alternatives, one could take more data and reach the integrated luminosity precision benchmark that way. The other would be to perform a thorough study on the possible utility of BERB events, which has a larger cross-section, $\approx100$~pb at 250~GeV, as an integrated luminosity channel.

\subsection{LABS and Dimuon Luminosity}

\Gls{dimuon}s and \Gls{LABS} are two fairly simply physics processes that occur dominantly at wider-angles and, because of this and their charged nature, are typically resolve in the tracker. With dimuons having arguably the best tracker performance, and corresponding position and momentum resolution, of any physics process. This makes dimuons and LABS attractive for any precision measurement, as discussed in chapter~\ref{ch-Theory}. At BELLE~II the motivation for using dimuons was to reduce the dependence of the integrated luminosity on the clustering of the \Gls{ECAL}, which was a dominant systematic in their integrated luminosity precision~\cite{Adachi_2025}. This addition is not free as it does require the separation of dimuons and LABS from each other. However, as noted by BELLE~II, this can be done using tracker and ECAL information since the energy deposited per hit by a muon is typically much different from an electron. Using this methodology BELLE~II was able to get separation of $0.23\%$ between dimuons and the dominantly Bhabha background, which is acceptable for their overall $\sim1\%$ precision on integrated luminosity.

At future $\ee$ colliders, the demands on integrated luminosity precision are far greater, needing $10^{-3}$ or $10^{-4}$ precision instead of $10^{-2}$. As such, separation between dimuons and LABS will need to be improved beyond what BELLE~II achieved, with separation of $10^{-4}$ being a target goal. We believe that this is achievable because the higher energies of future $\ee$ colliders make the path-length of muons and electrons much different, with muons penetrating much deeper into the detector. The beam pairs that are generated by the beams at BELLE~II, which contributed $0.13\%$ ($0.26\%$) to the Bhabha (dimuon) integrated luminosity precision, are also less of a concern at future $\ee$ colliders because the beam pairs are much lower energy, $\sim1$~GeV, than the integrated luminosity processes, $\sim100$~GeV.

\begin{figure}[h]
\centering
\includegraphics[width=16cm]{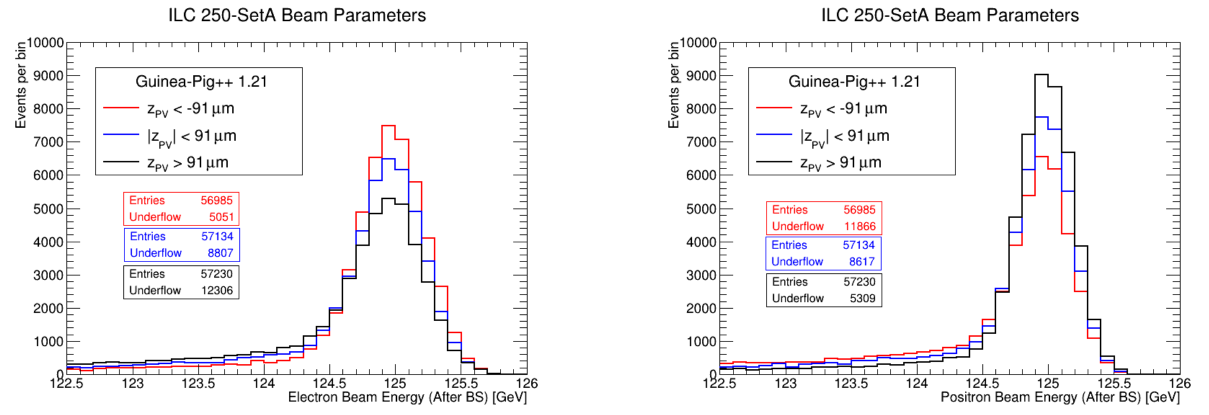}
\caption{(Left) Plot of the electron beam energy after beamstrahlung for three cuts on the primary vertex along the z-axis, $z_\text{PV}$. (Right) Plot of the positron beam energy after beamstrahlung for the same three cuts. The different in energy spread is a design choice of ILC250. The beam energy distribution for electrons (positrons) is broader (narrower) when $z_\text{PV}$ is late and greater than 91 microns.}
\label{fig-ZVert}       
\end{figure}

As a part of each of the proposed integrated luminosity channels there is an underlying luminosity spectrum that can influence the precision of the integrated luminosity. This luminosity spectrum is usually treated in terms of its differential in center-of-mass energy. This is not entirely accurate, as recent studies, as seen in figure~\ref{fig-ZVert}, have found that the luminosity spectrum varies as a function of the IP vertex along the z-axis~\cite{GWILCCon21}. 
This effect, along with the beam deflection effects that introduce a luminosity spectrum bias along the polar angle, result in a luminosity spectrum that is not a constant differential in $\sqrt{s}$. Instead of $d\mathcal{L}/d\sqrt{s}$, the full differential treatment should be as $d^3\mathcal{L}/d\sqrt{s}dz_\text{PV}d\theta$ or $d^4\mathcal{L}/dx_-dx_+dz_\text{PV}d\theta$ if one wishes to decompose the center-of-mass into beam energy fractions of the two beams. dimuons and LABS can help model this differential due to their different polar angles, and wider polar angle coverage, as compared to SABS. dimuons, due to having the finest resolution in energy, could provide the strictest limits on the energy dependence on the underlying luminosity differential. We propose that the precision position reconstruction of dimuons and LABS can also help with reconstructing the $z_\text{PV}$ dependence.

As in the case for diphotons, the limit for dimuons will likely be statistical due to the cross-section, $\approx9$~pb as calculated by \Gls{WHIZARD}, and data sample sizes. While LABS has larger cross-section, $\approx2.3$~nb as calculated by BHWIDE, but is expected to have worse systematics and performance than dimuons. Still, both channels are promising for aiding in the measurement of the integrated and differential luminosity. This, in turn, may reduce systematics in other integrated luminosity channels through improved knowledge of the differential luminosity, as well as adding an additional measure of the integrated luminosity.

%

\chapter{LumiCal Proposal}\label{ch-FCAL}

In this chapter we will focus on the LumiCal of future $\ee$ colliders, specifically focusing on \Gls{ILD}'s LumiCal. After summarizing ILD's LumiCal and forward tracker we shall present an alternative LumiCal that is a SiW design that is highly granular in terms of radiation lengths per layer as well as the transverse cell size. The existing forward tracker, or one similar to it, is included with the newly proposed LumiCal. To see additional documentation on this newly proposed LumiCal we recommend previous work~\cite{Madison:2024jak}~\cite{BMLCWS24}. The work presented here on energy resolution and position resolution is mostly sourced from previous work, while the particle identification (PID) work is new.

\section{Existing LumiCal and Forward Tracker Detector}
\begin{figure}[h]
\centering
\includegraphics[width=14cm,clip]{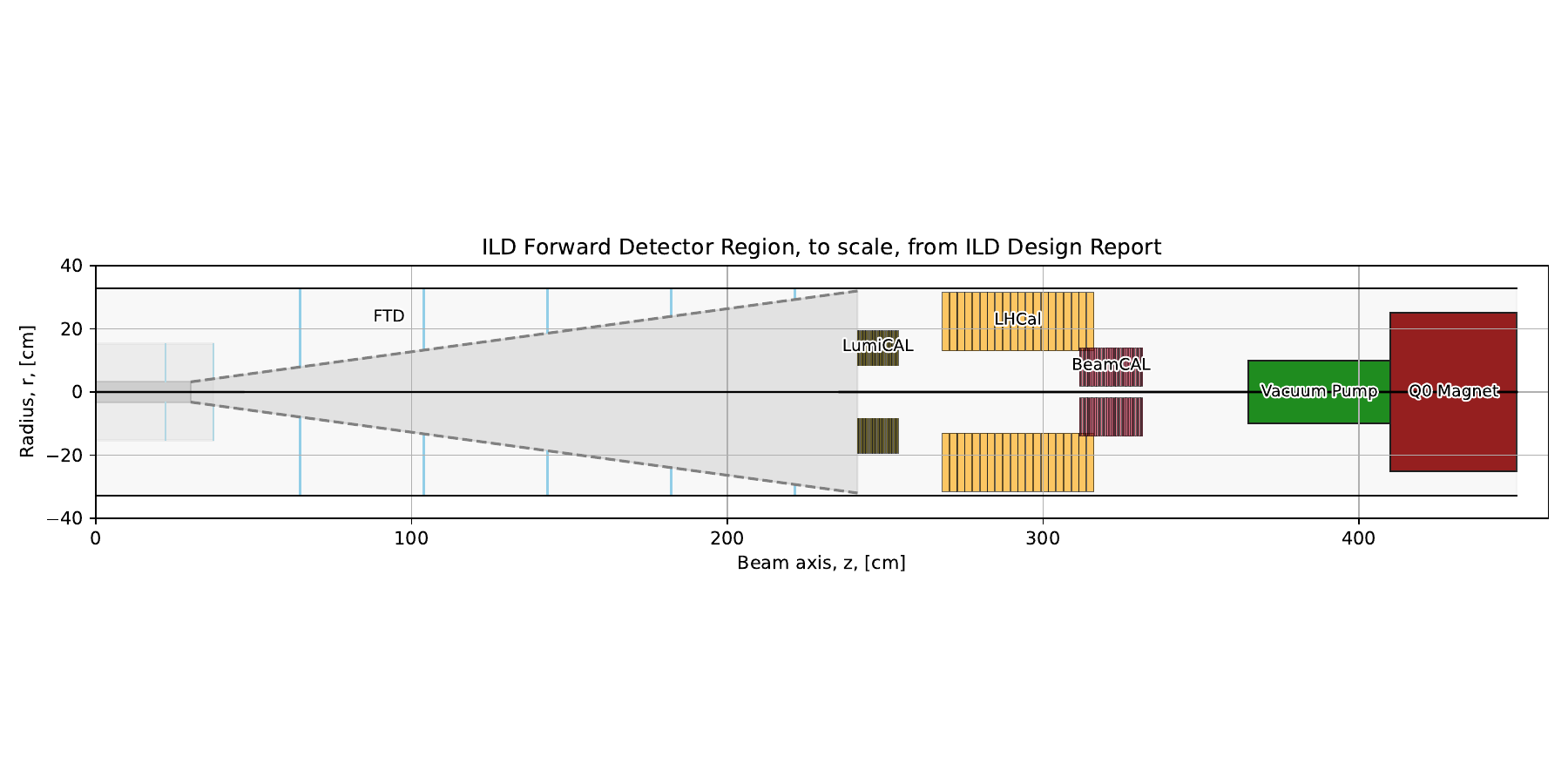}
\caption{A diagram of the ILD forward region generated using the geometry values given in the ILD interim design report. In particular, see the tables of chapter five of the design report~\cite{ILD2020}.}
\label{fig-ILDForDiag}       
\end{figure}

\begin{figure}[h]
\centering
\includegraphics[width=14cm,clip]{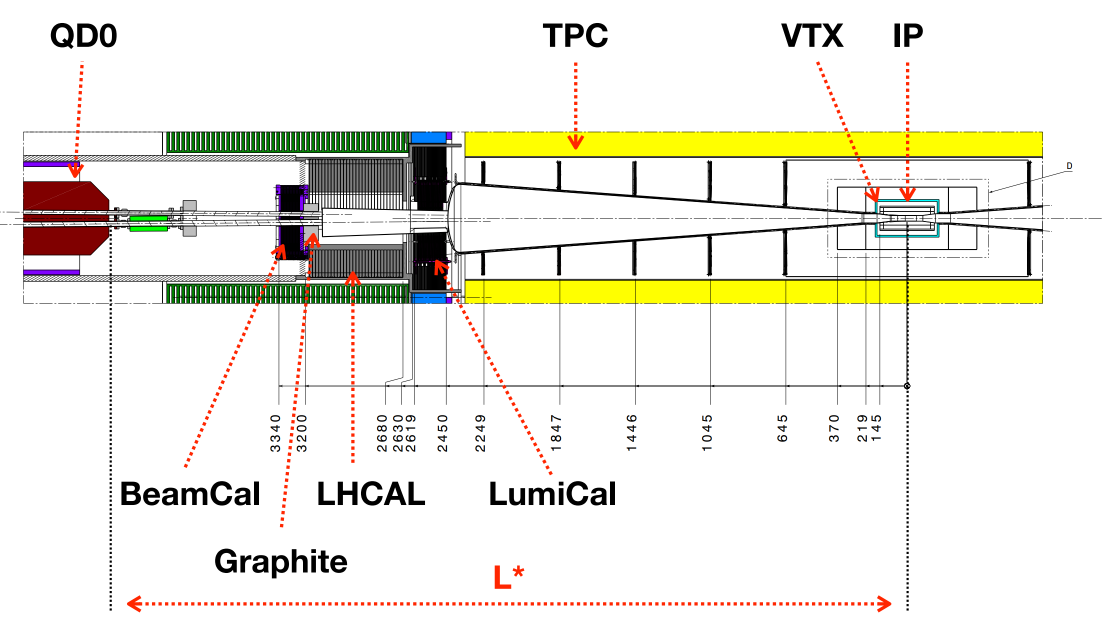}
\caption{An image of a diagram of the ILD forward region as supplied by the authors of the interim design report~\cite{ILD2020}.}
\label{fig-ILDForImage}       
\end{figure}

In this section we will provide context by summarizing the existing LumiCal and Forward Tracker Detector (FTD). A diagram of the forward region of ILD can be seen in figure~\ref{fig-ILDForDiag}, where geometry from the ILD design report, verified as accurate using the key4HEP k4geo repository for the ILD geometry, was used to create a scale 2-D model~\cite{ILD2020}. We also provide a comparison image, as seen in figure~\ref{fig-ILDForImage} and from the design report, because the ILD forward region has various versions, it is a work-in-progress and the image has additional details.
The ILD LumiCal is a compact silicon-tungsten sampling calorimeter in a cylindrical geometry, comprising 30 absorber disks of 3.5~mm-thick tungsten, each about one radiation length $X_0$, interleaved with $\sim300$~$\mu$m silicon sensor planes and a 1~mm gap~\cite{Abramowicz_2010}. This structure yields a total depth of roughly 30~$X_0$, which is claimed to be sufficient to contain most high-energy electromagnetic showers that originate from charged particles. However, it is shallow enough that longitudinal leakage will occur with some electromagnetic showers, especially those originating from photons. The overall compact design, along with the tungsten absorbers’ small Molière radius, ensures compact lateral shower size which is critical for isolating SABS events against background. The calorimeter covers a polar angle range from approximately $1.75^\circ<\theta<4.40^\circ$ around the outgoing beam line~\cite{Abramowicz_2010}~\cite{FCAL:2017uhw}. Each silicon sensor plane is segmented into pads segmented both radially and azimuthally, $r\phi$ segmentation, to provide fine granularity~\cite{Abramowicz_2010}. The baseline design divides the azimuth into 12 identical sectors, each sensor module spanning $30^\circ$, while each sector is segmented radially into 64 rings of pads, yielding a polar angular pad pitch of $\Delta\theta \approx 0.8$~mrad~\cite{Abramowicz_2010}. This segmentation was chosen to minimize bias and migration of events across the fiducial volume; simulations have shown that this angular resolution is adequate for keeping the resulting luminosity bias under $10^{-3}$. It does introduce bias in position measurement, but this can be calibrated and made to be as small as 3$\mu$rad, ensuring high precision in position.

Using detailed Geant4-based simulations, the energy deposition and reconstruction performance of ILD's LumiCal has been evaluated. Each shower’s energy is summed over all hit pads and calibrated to the incident energy; a combination of clustering and logarithmic weighting of pad energies is used to determine the shower’s centroid position in polar angle. When evaluating energy resolution, it is common to decompose into two components such that 

\begin{equation}\label{eqn-eneres}
    \sigma_\text{E} = \frac{a_1}{\sqrt{E}} + a_2
\end{equation}

the energy resolution, $\sigma_\text{E}$ or sometimes $\sigma/E$, depends on a stochastic term, $a_1$, and a constant term, $a_2$. For ILD LumiCal $a_{1}=21\%\pm2\%~\big[\sqrt{GeV}\big]$ and there was no constant term observed~\cite{Abramowicz_2010}. In conventional terms, this corresponds to a relative energy resolution $\sigma_E/E \approx 21\%/\sqrt{E(\text{GeV})}$, with no significant constant term observed. Previous studies found that there was no significant longitudinal leakage for the majority of electron showers up to 300~GeV. Above this energy a non-zero $a_2$ would likely become evident. The transverse shower shape is also well contained; due to the small Molière radius, is is claimed that 99\% of a shower’s energy falls within a radius of the LumiCal. This ensures that electromagnetic showers remain compact over the entire polar angle range, minimizing overlaps between neighboring showers and allowing for precise fiducial acceptance.

LumiCal’s granularity and reconstruction algorithm also provide excellent angular resolution. Using clustering and a pad energy log-weighting method, the polar angle of an incident electron can be measured with a resolution of order $\sigma_\theta \sim20$~$\mu$rad. This is made possible by the $r\phi$ segmentation, as the resolution using the same methods with square $xy$ segmentation is closer to 1~mrad~\cite{Madison:2024jak}. Meaning that the gain in position resolution from switching from $xy$ segmentation to $r\phi$ segmentation is $\sim\times50$. Each sensor plane has on the order of $N_{\text{pad}} = 12 \times 64 = 768$ pads, totaling $\sim9.2\times10^4$ readout channels after combining all hemispheres of the full calorimeter~\cite{Abramowicz_2010}. To provide a contemporary comparison, the ATLAS inner pixel detector has $\sim8.0\times10^7$ channels.

The ILD Forward Tracking Detector (FTD) refers to the system of silicon tracking disks that cover the forward angular region beyond the acceptance of the central tracker. Together with a Silicon Internal Tracker (SIT) and Silicon External Tracker (SET) in the barrel-endcap transition, the FTD ensures efficient tracking and vertexing for polar angles roughly in the range $5^\circ < \theta < 30^\circ$~\cite{Aplin:2013oca}. The FTD is therefore critical to bridge the gap and provide track measurement points in the forward cones. ILD’s FTD consists of seven disk-shaped tracking planes per endcap, spaced along the $z$-axis between the vertex detector and the endcap calorimeters. The support structure of each disk is engineered for minimal material. The mechanical design employs lightweight petals; typical FTD petals consist of thin silicon sensors, thickness $\sim200$~$\mu$m, glued to a low-mass support such as carbon fiber or printed circuit board. The cumulative material of each disk is low; the innermost FTD layers contribute only about 0.12\%$X_0$ each and the outer strip disks are 0.6\%$X_0$ each~\cite{ILD2020}. This ultra-thin design preserves the performance of both tracking, by reducing multiple scattering, and calorimetry, by having less dead material.

\begin{figure}[h]
\centering
\includegraphics[width=10cm,clip]{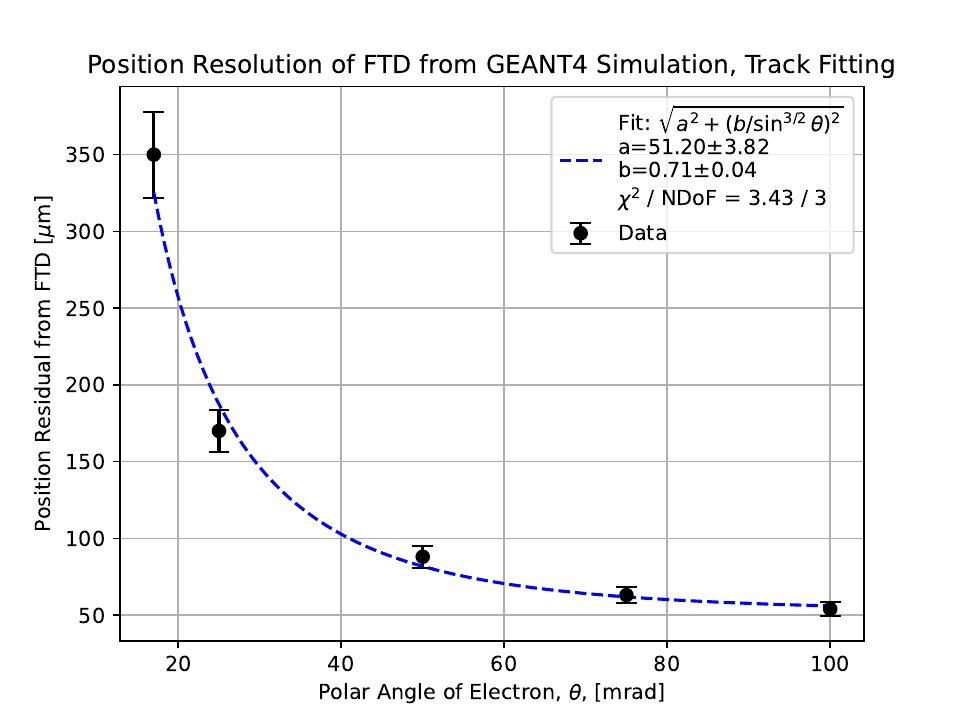}
\caption{A plot of FTD position resolution using a simple linear $z\phi$ fit for track reconstruction. An improved algorithm would likely outperform this, so we propose that this performance be considered modest.}
\label{fig-FTDVert}       
\end{figure}

With the combined system of forward trackers, ILD is designed to achieve nearly uniform tracking performance over the full solid angle, approaching the goal of tracking coverage down to $\approx 5^\circ$. The reduced lever arm and fewer measurement layers in the forward region inevitably lead to some degradation in track parameter resolution for shallow-angle tracks. Momentum resolution is the key metric; the superb momentum resolution of ILD in the central region, $\sigma_{1/p_t} \sim 2\times10^{-5}$~GeV$^{-1}$, degrades to $\sigma_{1/p_t} \sim 10^{-3}$~GeV$^{-1}$ in the forward region~\cite{ILD2020}~\cite{Abramowicz_2010}. This is due to having a shorter lever arm and smaller radius of curvature from the magnetic field deflection. For a demonstration of the position resolution of ILD's forward trackers we provide figure~\ref{fig-FTDVert}, where a fit of the position residuals is performed. We use the standard tracker position resolution equation
\begin{equation}\label{eqn-trackres}
    \sigma_\text{pos} = a \oplus \frac{b}{\sin^{3/2}(\theta)} 
\end{equation}
which depends on two fit parameters, $a$ and $b$, used to characterize the constant position resolution and the polar angle dependent resolution.
The position resolution was derived using a GEANT4 simulation of 125~GeV electrons passing through the FTD trackers, which were extended to cover $1^\circ<\theta<6^\circ$, and using a linear fit of $z\phi$ and then extrapolating to a fit in $r\phi$ to get the position with respect to the true position. We observe that, for the region covered by the forward trackers, there is a position resolution of 50$\mu$m to 350$\mu$m. The trend of position resolution closely follows equation~\ref{eqn-trackres} well. This is comparable to the position resolution achievable by the ILD LumiCal. One key differentiation between the LumiCal and FTD position measurements is that tracker position measurements are sensitive to $d\phi/dz$, which is indicative of the charge sign of the particle being measured. Therefore, by combining the forward tracker with any LumiCal, one can have precision charge identification, position resolution and energy resolution. Combining these should also allow for particle identification, which will be discussed later in this chapter, in section~\ref{sec-PID}.

\section{Proposed LumiCal, Granular Long Instrument for Precision (GLIP)}\label{sec-GLIP}
We present a new design for a future $\ee$ collider LumiCal that focuses on taking advantage of highly granular, in layering, designs as well as fine transverse cell sizes. Such a design is motivated by the CMS High-Granularity Calorimeter (HGCAL), and Fermi Large Area Telescope (Fermi-LAT), both of which take advantage of fine layering to achieve superior energy and position resolution~\cite{Magnan:2017exp}~\cite{Fermi-LAT:2009ihh}. We also note that, like demonstrated with Fermi-LAT, the use of high-granularity calorimeters allows for better sensitivity to photons. Work on HGCAL has also shown promise in making use of the high information content that comes from high granularity by developing more complicated and precise analyses. 

We propose two possible implementations of the Granular Long Instrument for Precision (GLIP). The standard (short) implementation will be as a 40$X_0$ (30$X_0$) radiation length luminosity calorimeter that uses 240 (180) layers of $1/6X_0$ thickness with a 1~mm thick Silicon active layer and Tungsten passive layer. We propose this design as previous studies found it to have superior energy and position resolution and that it would be acceptable given the constraints of the forward region~\cite{Madison:2024jak}. For simulation purposes we have used $xy$ segmentation but, due to the gain in position resolution from using $r\phi$ segmentation, we propose future designs use $r\phi$ segmentation. In order to increase angular acceptance, and therefore statistics of integrated luminosity processes, we also propose that GLIP cover as much radial space as possible. For the scenarios proposed here we have decided to use an inner radius of 4~cm. As a part of this process we have also adjusted the LHCAL inner radius so that it no longer overlaps with the BeamCal.

\begin{figure}[h]
\centering
\includegraphics[width=14cm,clip]{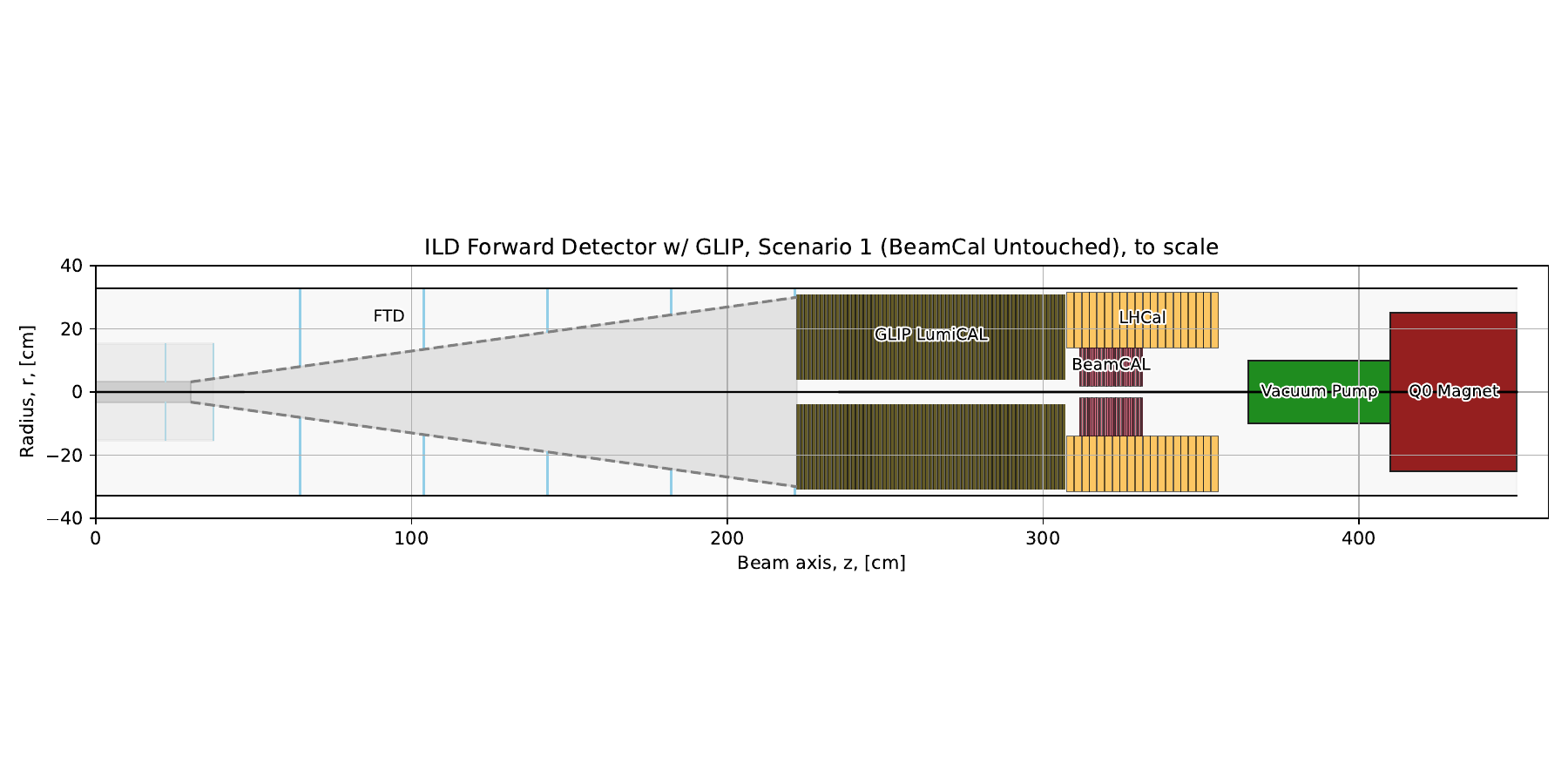}
\caption{A diagram of the ILD forward region assuming deployment of a GLIP LumiCal in Scenario~1, where the BeamCal is left untouched.}
\label{fig-GLIPSc1}       
\end{figure}

\begin{figure}[h]
\centering
\includegraphics[width=14cm,clip]{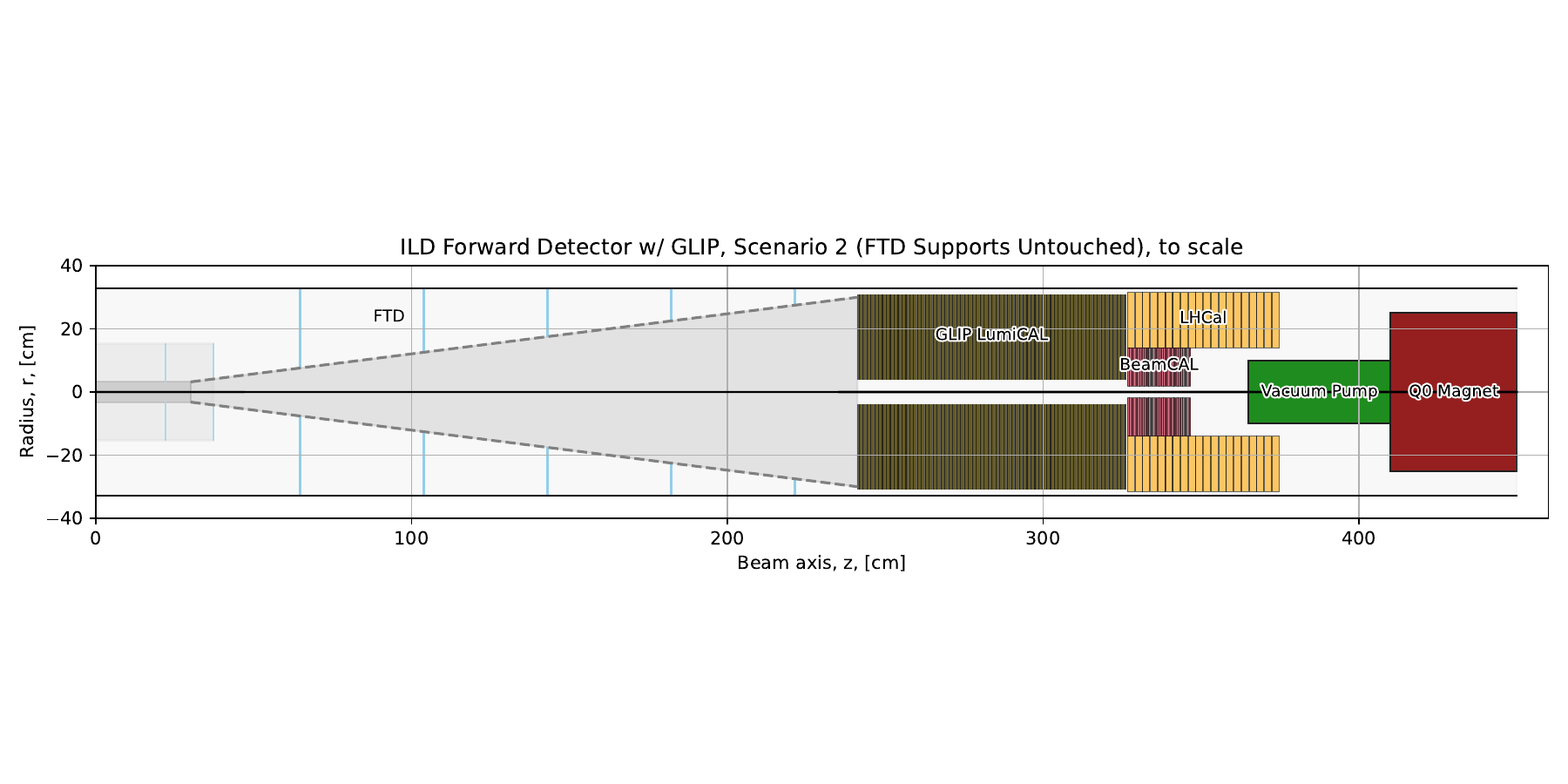}
\caption{A diagram of the ILD forward region assuming deployment of a GLIP LumiCal in Scenario~2, where the FTD internation support structure is left untouched.}
\label{fig-GLIPSc2}       
\end{figure}

\begin{figure}[h]
\centering
\includegraphics[width=14cm,clip]{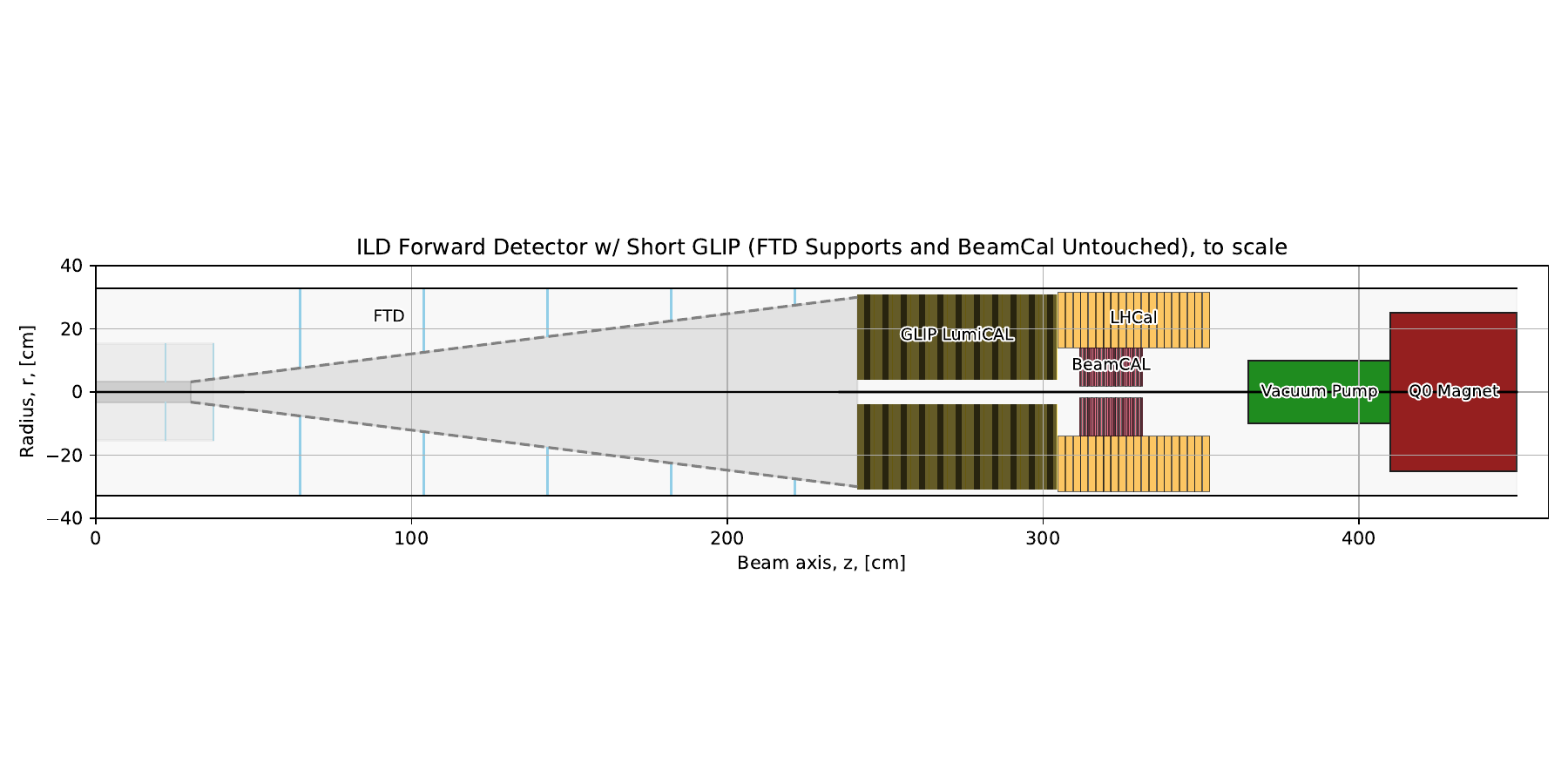}
\caption{A diagram of the ILD forward region assuming deployment of a GLIP LumiCal in Scenario~3, a.k.a. the short scenario.}
\label{fig-GLIPSc3}       
\end{figure}

For GLIP deployment, in the context of ILD, we have provided three scenarios. Scenario~1, as seen in figure~\ref{fig-GLIPSc1}, where the \Gls{FTD} internal support structure is cut short such that the beginning of GLIP would be more or less as close to the last tracker in the FTD as possible. 
Given Scenario~1 BeamCal is left untouched and LHCAL is moved backwards by roughly 40~cm. The angular acceptance for this deployment is $1^\circ<\theta<6^\circ$. Scenario~2, as seen in figure~\ref{fig-GLIPSc2}, where the FTD internal support structure is left untouched but the BeamCal is moved back to be at the end of GLIP.
For Scenario~2 the angular acceptance is $0.93^\circ<\theta<5.6^\circ$, having the shallowest polar angle coverage of all scenarios presented here. Scenario~3, also known as the `short scenario' as discussed previously, can be found in figure~\ref{fig-GLIPSc3}. Scenario~3 represents only 30$X_0$ and leaves the FTD internal support structure and the BeamCal untouched.
In Scenario~3 the LHCAL is moved back by about 35~cm. Given the maturity of the ILD design more study on implementing a highly granular LumiCal, like GLIP, is needed. We stress that these scenarios are likely not optimal for the other sub-detectors and therefore further collaboration and optimization is needed to see a viable deployment strategy. Hence the motivation for presenting multiple deployment scenarios. From this point forward we will proceed with the assumption that, unless said otherwise, the performance of GLIP is being done in Scenario~1.

\subsection{Energy Resolution}\label{sec-EneRes}

A more in depth treatment of energy resolution for GLIP LumiCal can be found in other work~\cite{Madison:2024jak}. Energy resolution, $\sigma_\text{E}$, typically scales with the energy sampling fraction, $f_\text{samp.}$, which is the fraction of energy in the active layer compared to the other calorimeter components. This relation can be written in terms of

\begin{equation}\label{eqn-EneResFrac}
    \sigma_\text{E} = \frac{d}{\sqrt{f_\text{samp.}}}
\end{equation}

the sampling fraction and $d$, an empirical parameter. From testing the design of the current ILD LumiCal the energy fraction is $\sim0.95\%$ and the energy resolution is $\sim19\%/\sqrt{E}$, which is within $1\sigma$ of the energy resolution that the ILD LumiCal quotes~\cite{Madison:2024jak}. By comparison, the GLIP LumiCal would have an energy fraction of $\sim21\%$ and an energy resolution of $~\sim3.6\%/\sqrt{E}$~\cite{Madison:2024jak}. This is a significant improvement over the existing LumiCal design. For evaluating the energy resolution at each energy we have used two datasets, one with one calorimeter and no magnetic field with 1000 events and one a forward and backward calorimeter with a respective beam for each and a magnetic field and 100 events as generated by \Gls{GEANT4}. We use a full suite of electromagnetic, nuclear and radioactive physics effects as used by the \textbf{TestEM3} and radioactive decay examples of GEANT4. The number of events is limited by both computation time and memory. Especially at higher energies, the amount of RAM needed quickly becomes 10s of GB and the computation time becomes hours per event. We found that the primary time and memory consumption comes from bremsstrahlung and multiple scattering. This can be reduced by using GEANT4's reweighting and Russian roulette methods, but we found that doing so alters the physics quality, something that is undesirable for seeking a precision simulation. We generated a collection of energies from 8~GeV to 250~GeV both with and without a 3.5~T magnetic field as outlined by the ILD design report~\cite{ILD2020}. During this we found that using a magnetic field in the simulation degrades the energy resolution by a small amount, comparable to 5\% decrease on the stochastic and constant energy resolution terms both. We have also chosen to use transverse dimensions for calorimeter tests that are similar to their potential deployed values, comparable to 30~cm. We chose to use a constant polar angle value of 35~mrad and to let the azimuthal angle vary over the entire range of azimuthal angle.

\begin{figure}[h]
\centering
\includegraphics[width=14cm,clip]{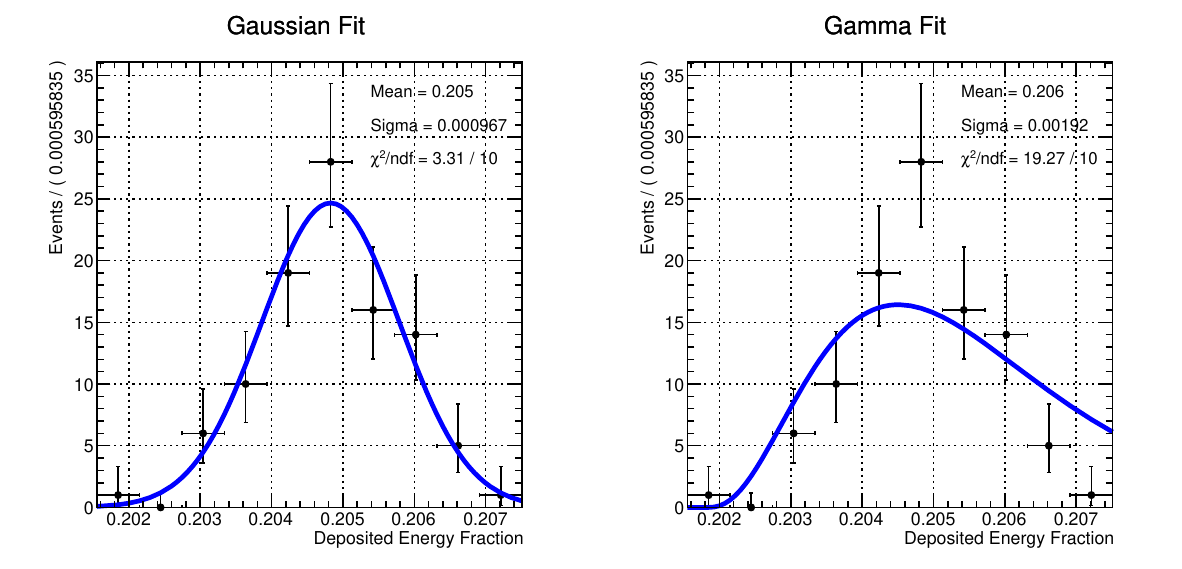}
\caption{(Left) Gaussian distribution fit of the electromagnetic energy deposited fraction of in the Silicon (active) layer of GLIP LumiCal as simulated by GEANT4 and plot created with RooFit.(Right) A similar fit to the same data but for a Gamma distribution. We find that the Gaussian distribution is of higher fit quality here. The binning is not the same despite the same data as the original fits are unbinned and the binning is done after and depends on the fit parameters so as to include enough range to adequately show both the data and the fit.}
\label{fig-GammaEneResFit}       
\end{figure}

For each energy, we evaluated the mean and spread using both an unbinned Gaussian distribution fit and an unbinned Gamma distribution fit, using RooFit, as motivated by previous studies we performed on energy resolution fitting found that the Gamma distribution can be a better fit for some energies and detector designs~\cite{Madison:2024jak}~\cite{Verkerke:2003ir}. From these two fits we keep the result of the fit with the better $\chi^2/\text{NDoF}$ value. An example result, for 125~GeV photons on GLIP, can be found in figure~\ref{fig-GammaEneResFit}.
We have converted the x-axis of the fits to the electromagnetic energy deposited fraction with respect to the total energy. We provide both the Gamma distribution and Gaussian distribution fits as a comparison. We also found during this process that using the Guassian distribution fit to constrain the parameters for the Gamma distribution is preferential as otherwise the Gamma distribution can find odd parameter values that are poor fits yet somehow reach convergence. We find that the energy fraction in figure~\ref{fig-GammaEneResFit} is slightly lower than that expected from the sampling fraction and equation~\ref{eqn-EneResFrac}.

\begin{figure}[h]
\centering
\includegraphics[width=16cm,clip]{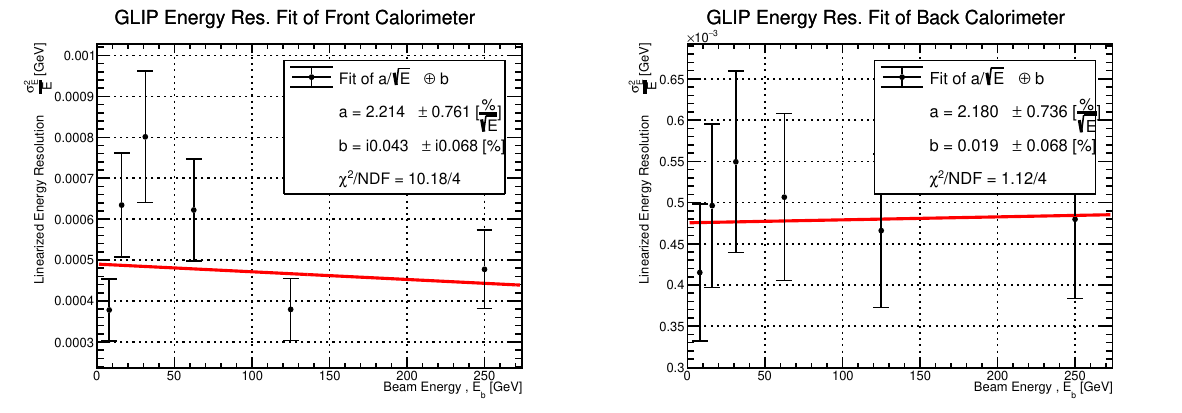}
\caption{(Left) Fit of the linearized energy resolution of the proposed GLIP LumiCal given electrons with 1~mm cells and in the front calorimeter, tested at various energies from 4~GeV to 250~GeV.(Right) Same fit and simulation but using positrons in the back calorimeter.}
\label{fig-GLIPeneres}       
\end{figure}

We provide a fit of the energy resolution for energies of 8~GeV to 250~GeV, for electrons in the front half of GLIP LumiCal and positrons in the back half of GLIP LumiCal, in figure~\ref{fig-GLIPeneres}.
We do this so as to provide a result that would be characteristic of \Gls{SABS}. In figure~\ref{fig-GLIPeneres} we use a linearized form of the energy resolution as the y-axis. This is done so that a constant term can be fit such that
\begin{equation}\label{eqn-EneResLine}
\begin{gathered}
    \frac{\sigma_\text{E}}{f_\text{samp.}E} = \frac{a}{\sqrt{E}} \oplus b \rightarrow \\
    \left(\frac{\sigma_\text{E}}{f_\text{samp.}E}\right)^2 = \frac{a^2}{E} + b^2 \rightarrow \\
    \frac{\sigma_E^2}{f_\text{samp.}^2E} = a^2 + b^2E
\end{gathered}
\end{equation}
we end up with an equation for a line if we use the particle energy, $E$, to linearize the equation. The other parameters are the stochastic term, $a$, the constant term $b$ and the energy fraction deposited in the active layer, $f_\text{samp.}$. Due to the quadratic nature of equation~\ref{eqn-EneResLine} we may have imaginary values of the parameters if the fit prefers them to be negative. This is a benefit of using the form of equation~\ref{eqn-EneResLine}, that the sign of the contributions can be fitted for and, if they are imaginary valued, this can be determined. Here we see that the value of $b$ is statistically equivalent to zero but that the fit does get a best-of-fit value that is imaginary for electrons in the front calorimeter. By using equation~\ref{eqn-EneResLine}, the fit quality will, generally, improve and the uncertainties on the fit parameters will also improve. We observed that trying to fit the unlinearized energy resolution with a constant term almost always resulted in mis-estimating the stochastic term and having a constant term that is zero or close to zero. This, as well as having no magnetic field, led to previous estimations of the GLIP LumiCal energy resolution measuring a stochastic term of roughly $4\%/\sqrt{E}$.

\begin{figure}[h]
\centering
\includegraphics[width=16cm,clip]{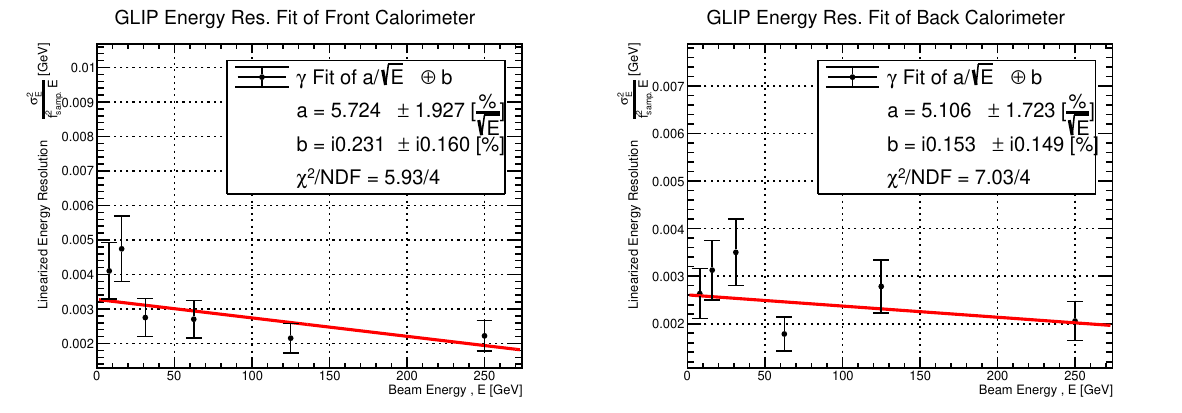}
\caption{(Left) Fit of the linearized energy resolution of the proposed GLIP LumiCal given a photon in the front half of GLIP LumiCal with 1~mm transverse cell size at various energies from 8~GeV to 250~GeV.(Right) Same fit but done with an equal and opposite photon in the back half of the GLIP LumiCal.}
\label{fig-GLIPPhoRes}       
\end{figure}

We repeat this process for photons, as seen in figure~\ref{fig-GLIPPhoRes}, and find that photons have a similar energy resolution performance.
In the case of all particles tested, we were able to refine earlier values of $\approx4\%/\sqrt{E}$ for the stochastic term to closer to $\approx5\%/\sqrt{E}$. From these results we suspect that the approximation of equation~\ref{eqn-EneResFrac}, which predicts a $3.6\%/\sqrt{E}$, is not entirely correct here. Previous work on calorimetry has found that this approximation stops being valid when the fraction of energy deposited exceeds 10\%. Since we see energy fractions for GLIP LumiCal of 20.5\%, which is close to the value expected from the calorimeter material data-sheets, we do not expect the energy fraction values are incorrect and we do expect that equation~\ref{eqn-EneResFrac} is no longer valid~\cite{RefWig}. Regardless of the origin of these results, we have exhaustively tested them with improved methodology and various tests of physics effects. The results show that the GLIP LumiCal will precisely measure the energies of both electrons and photons at future experiments.

\begin{figure}[h]
\centering
\includegraphics[width=14cm,clip]{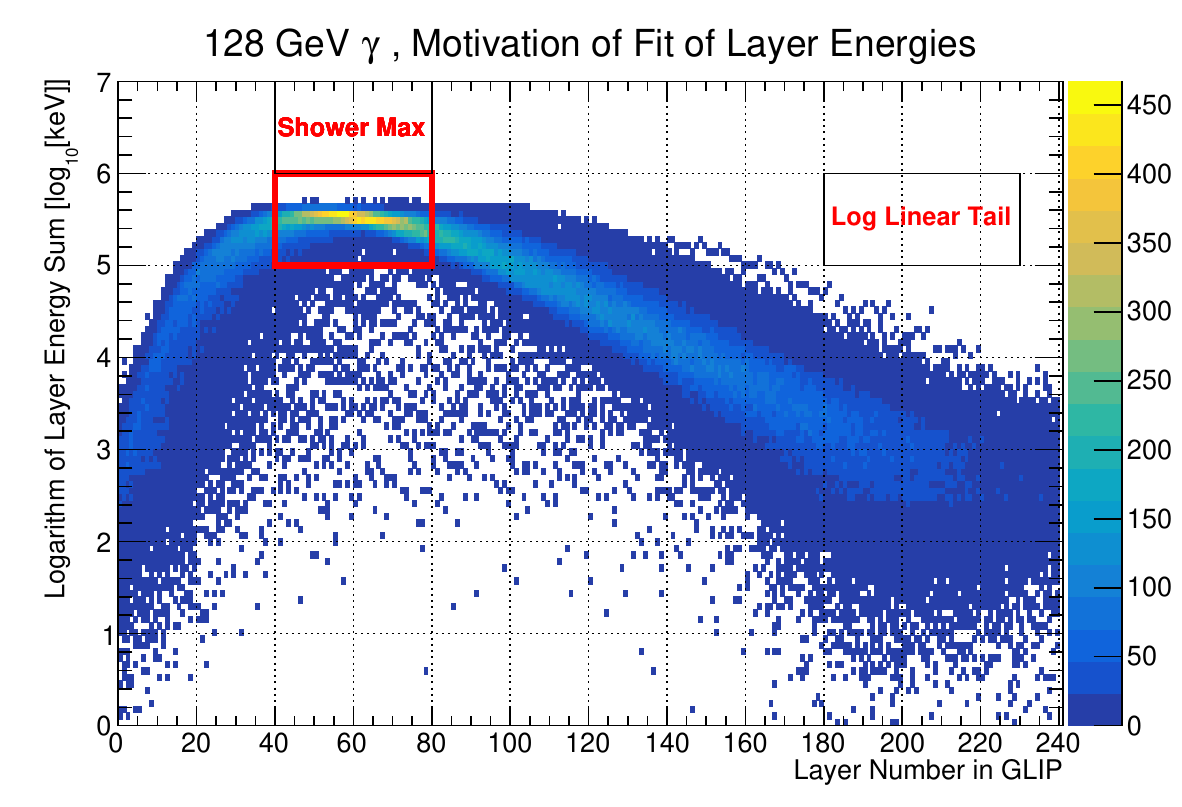}
\caption{2D histogram plot of the calorimeter layer number, proportional to the distance along the beam axis, and the logarithm of the sum of energy in a given layer. This is a compilation of roughly 1000 events.}
\label{fig-FitMoti}       
\end{figure}

\begin{figure}[h]
\centering
\includegraphics[width=16cm,clip]{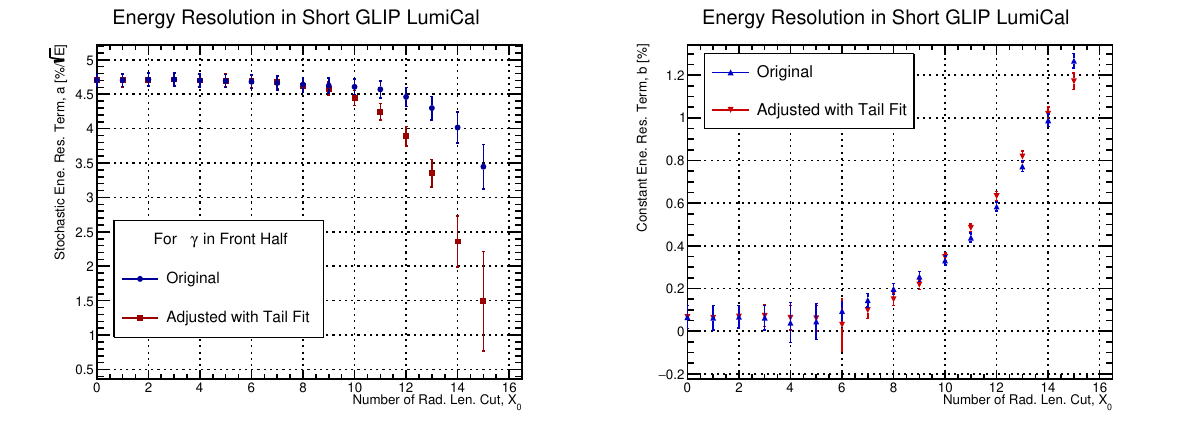}
\caption{(Left) Change in the stochastic energy resolution term as the length of the GLIP LumiCal is reduced using the log-linear tail fit adjusted and unadjusted values.(Right) Change in the constant energy resolution term as the length of GLIP LumiCal is reduced. This test was conducted using photons in the front calorimeter.}
\label{fig-GRIPeneres}       
\end{figure}

Due to the increased length of GLIP, we previously proposed a short scenario. We propose that the energy resolution of the short scenario, which would likely suffer leakage, could be ameliorated by use of fits of the shower energy deposited per layer. As seen in figure~\ref{fig-FitMoti}, the shower energy deposited per layer follows a fairly consistent but stochastic pattern. 
The distribution after the shower max, which is the layer with the most energy deposited, follows a log-linear trend in terms of depth and energy deposited. This trend can then be fitted to determine how much energy the shower would have deposited if the calorimeter continued for additional radiation lengths. We conducted a test of this using a log-linear fit and a GEANT4 simulation using photons hitting the GLIP LumiCal at 35~mrad. The log-linear fit was applied in each event and started 10 layers, or roughly 1.5~$X_{0}$, after the shower maximum layer, i.e. the layer with the most energy deposited in it and then went to the end of the calorimeter. The result, where the stochastic energy resolution term is evaluated from the fitted result and result with no correction up to removing the last $15X_0$, can be seen in figure~\ref{fig-GRIPeneres}.
We find that the loss in energy resolution is minor over the loss of the first 6~$X_0$ of layers cut off. We find that the tail fit does improve the energy resolution as compared to not using the tail fit particularly for the stochastic term. The constant term, which usually scales with energy leakage, increases steadily after 6~$X_0$ of layers cut off regardless of whether or not a fit is used. This result indicates that a better fit algorithm, or some other approach, is needed to make a short GLIP LumiCal that does not degrade in energy resolution performance or energy leakage.

\subsection{Position Resolution}\label{sec-PosRes}

\begin{figure}[h]
\centering
\includegraphics[width=16cm,clip]{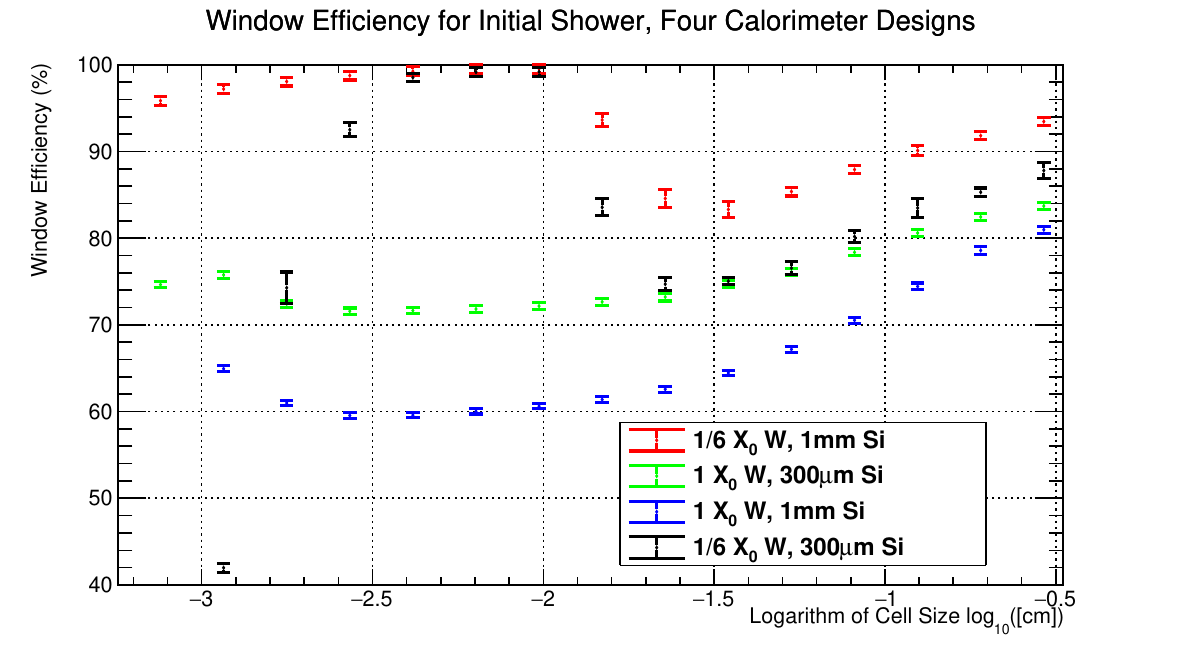}
\caption{Window efficiency for the polar angle window used for the initial shower
position reconstruction methods.}
\label{fig-WEff}       
\end{figure}

\begin{figure}[h]
\centering
\includegraphics[width=16cm,clip]{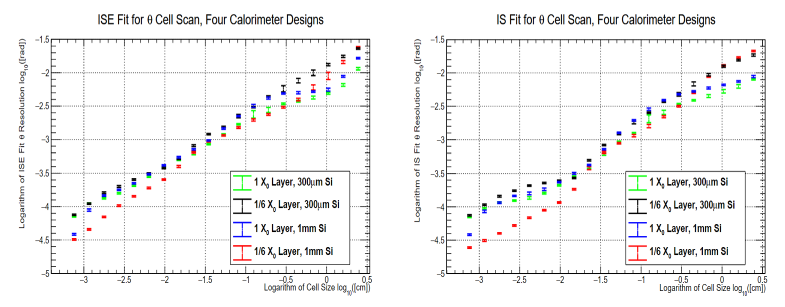}
\caption{(Left) Results of cell scan of $\theta$ resolution for the ISE fit.(Right) Results of the unweighted initial shower fit method, IS fit. At cell sizes smaller than 100~microns the $1/6X_0$ and 1~mm silicon design is best. Similar performance is observed for both methods.}
\label{fig-ISFits}       
\end{figure}

As done in previous work, we do not propose to use the same clustering with log-weighting for position resolution that ILD LumiCal uses~\cite{Madison:2024jak}. This is motivated by poor results of said algorithm. Instead, GLIP LumiCal plans to use two new methods; initial shower fits and the single hit method both were able to achieve the best performance in angular resolution~\cite{Madison:2024jak}. The initial shower fit methods work as follows. First a window is used in terms of the z-axis, or layer number, and the value of $\theta$ corresponding to hits in the active layer with respect to the beginning of the shower. This windowing cut was scaled according to the transverse cell size such that it would be a three cells in diameter. The remaining hits are then discarded if they exist outside the window. A second cut is applied to exclude hits that occur deeper than the shower maximum. Of the remaining hits an algorithm, which starts at the earliest hit along the z-axis,
checks the subsequent layer for hits. If there are no hits then the algorithm assumes that the initial shower has scattered out of the window and truncates the sample to those which are before this point. If there are hits in the next layer then it includes these hits with the starting hit and then repeats the process, looking at the next layer for hits. This repeats until either no new hits are found in the window or the layer of the shower maximum is reached. At this point the algorithm stops and the collected hits are used as representing the initial shower. The efficiency of this windowing process can be seen in figure~\ref{fig-WEff}, where we see that the GLIP LumiCal design, 1~mm and $1/6X_0$, has the best efficiency, especially with cell sizes comparable to 100~micron.
Once the initial shower sample is collected we perform one of two possible fitting algorithms. One for energy weighted (ISE) and one for unweighted (IS) of the initial shower position. The results, seen in figure~\ref{fig-ISFits}, indicate that there is a logarithmic dependence on the silicon cell size, which is expected given previous results and models of position resolution~\cite{Madison:2024jak}~\cite{Serpukhov-Brussels-AnnecyLAPP:1981yhn}. 
We find that, for this method and silicon cell sizes tested, that there is no cell size where diminishing returns begin.

\begin{figure}[h]
\centering
\includegraphics[width=14cm,clip]{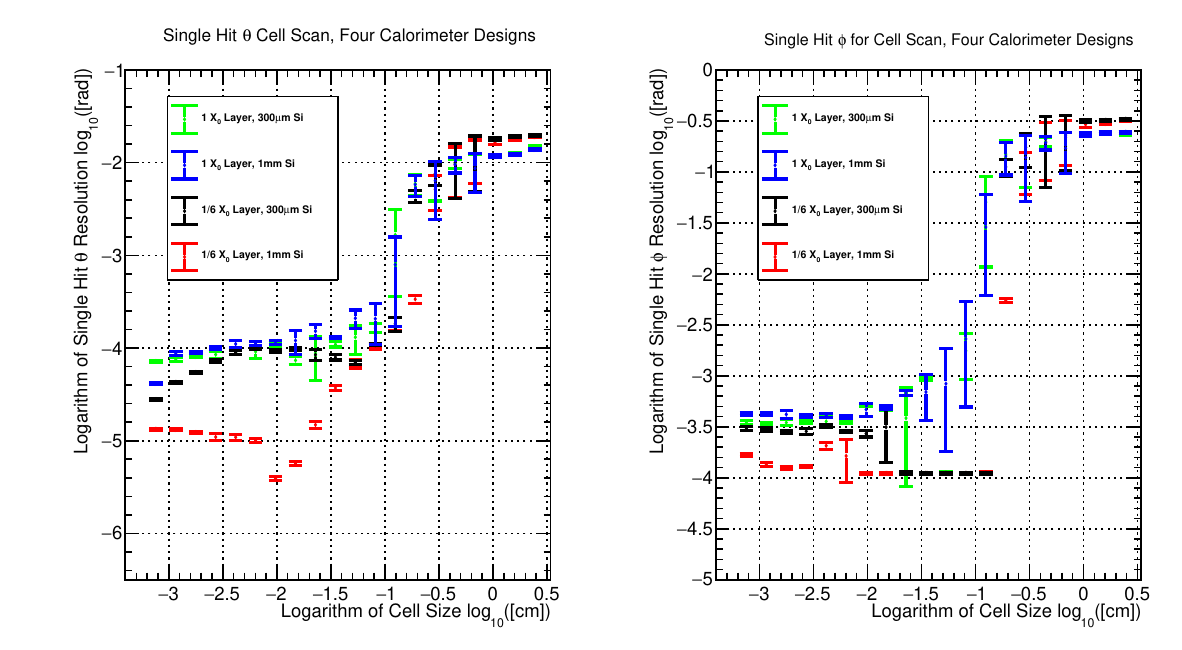}
\caption{Results of cell scan of $\theta$ and $\phi$ resolutions for the single hit method. At cell size of roughly 100~microns the $1/6X_{0}$ and 1~mm silicon design is best.}
\label{fig-singlehit}       
\end{figure}

As an alternative approach to position reconstruction we test a `single hit' method, where the initial shower candidate from the IS fit methods is used and all of the hits are tested for their residual with the true position. The hit with the minimal residual is kept and used as the position estimator. Since this is not possible to due in an experimental setting it is assumed that a future algorithm, or machine learning tool, is capable of finding which hit is the `single hit'. From the `single hit' method, as seen in figure~\ref{fig-singlehit}, we find that the position resolution approximately follows an S-curve but is otherwise unpredictable in its trend.
The optimal design for optimizing the polar angle resolution was found to be 100~micron cell size with the $1/6X_0$ layers and 1~mm thick silicon, which achieved roughly 10~$\mu$rad polar angle resolution. We propose that if $r\phi$ segmentation was used, the position resolutions presented in this section would improve by a similar $\times10\to\times50$ factor of improvement that the ILD LumiCal saw.  It is also the case that by using $r\phi$ segmentation the number of cells can likely be reduced. This is key as the $xy$ segmented GLIP design would feature roughly $10^{8}$ cells, a value that is large even compared to contemporary LHC detectors. Given the improvement of using $r\phi$ segmentation on position resolution, the results of the IS fits for the optimal design would be $\sim10$~$\mu$rad polar angle resolution and the `single-hit' resolution would be $\sim1$~$\mu$rad. Both methods are limited by the roughly 4~$\mu$rad opening angle of a 125~GeV photon shower~\cite{Madison:2024jak}. Therefore, it is likely not possible to improve much beyond this point, unless higher energies are used.

\subsection{Particle Identification (PID)}\label{sec-PID}

For the purpose of integrated luminosity measurement, especially when both SABS and diphotons are being used, it is key that these two processes be separable. This is often done by using Particle Identification (PID) techniques that can identify which particle has been detected. In particular to the integrated luminosity precision, as discussed in section~\ref{sec-NewLumi}, the PID precision must be comparable or less than the desired integrated luminosity precision. The PID precision is also important for the purpose of rejecting background events caused by beam pairs. At high energies, $>10$~GeV, PID can be difficult for electromagnetic particles that are light, that is to say less than 1~GeV of mass, such as electrons, photons, muons. This is reflective of the energy loss per distance, $dE/dx$, being in the radiative regime for these particles. Due to this, the particles look similar in calorimeters because the energy loss per layer of the calorimeter look similar. This is made even more difficult at forward regions because the charged particles have small lever arms over which to measure their magnetic deflection, $d\phi/dz$, so their charge or mass is difficult to reconstruct.

\begin{figure}[h]
\centering
\includegraphics[width=12cm,clip]{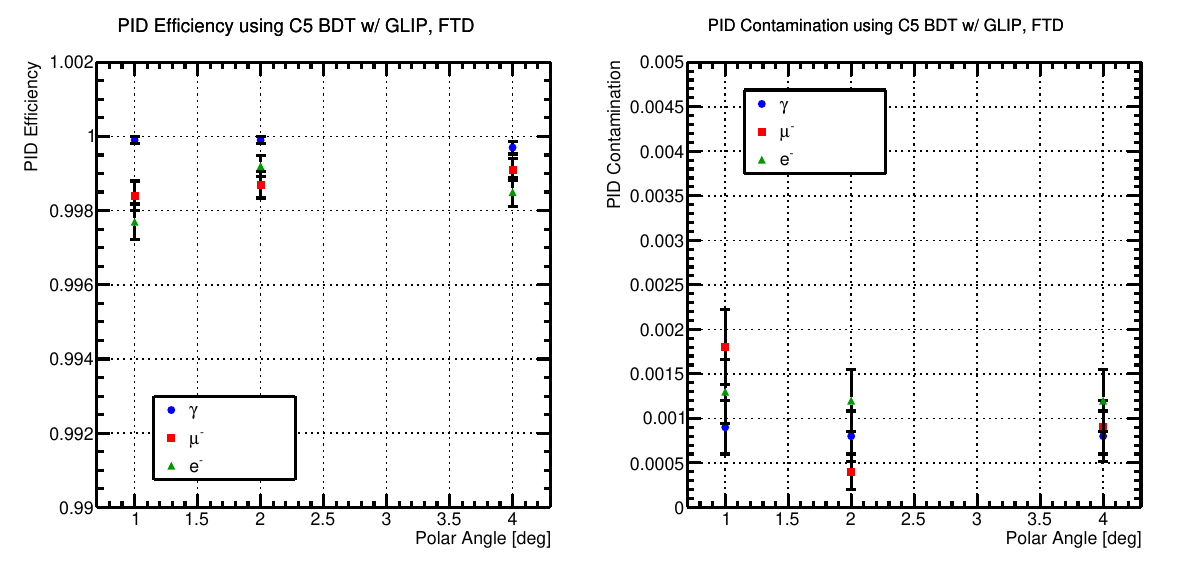}
\caption{(Left) Plot of the PID tagging efficiency for 125 GeV particles and the standard GLIP LumiCal and FTD design.(Right) Plot of the PID contamination for the same particles and design.}
\label{fig-PIDAng}       
\end{figure}

\begin{figure}[h]
\centering
\includegraphics[width=12cm,clip]{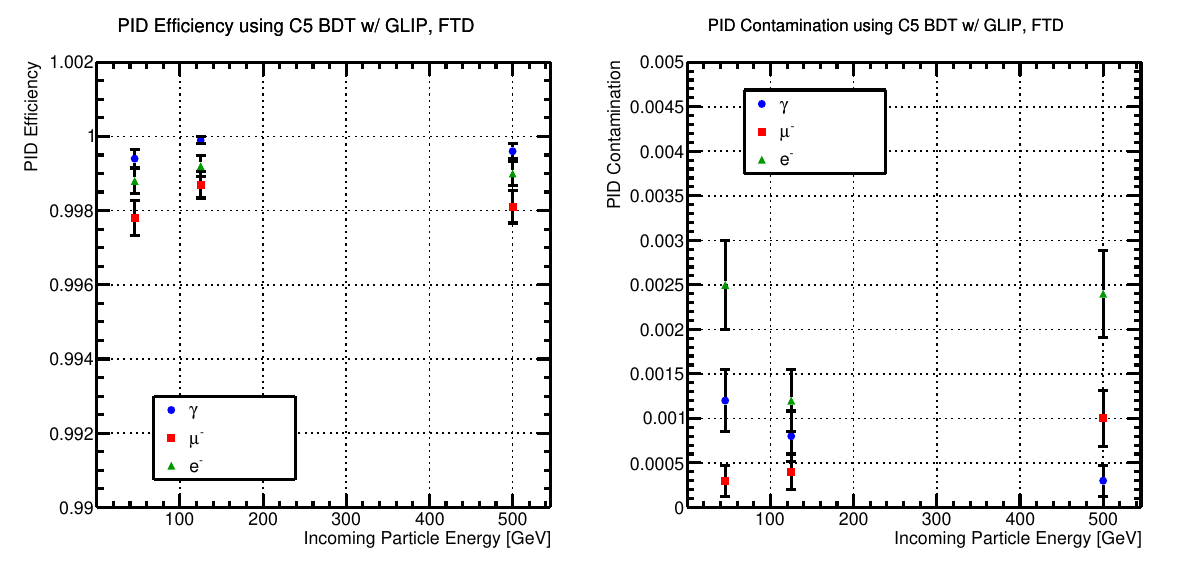}
\caption{(Left) Plot of the PID tagging efficiency for particles with varying energies and incident angle of $2^\circ$ with the standard GLIP LumiCal and FTD design.(Right) Plot of the PID contamination, with varying energy, for the same particles and design.}
\label{fig-PIDEne}       
\end{figure}

\begin{figure}[h]
\centering
\includegraphics[width=12cm,clip]{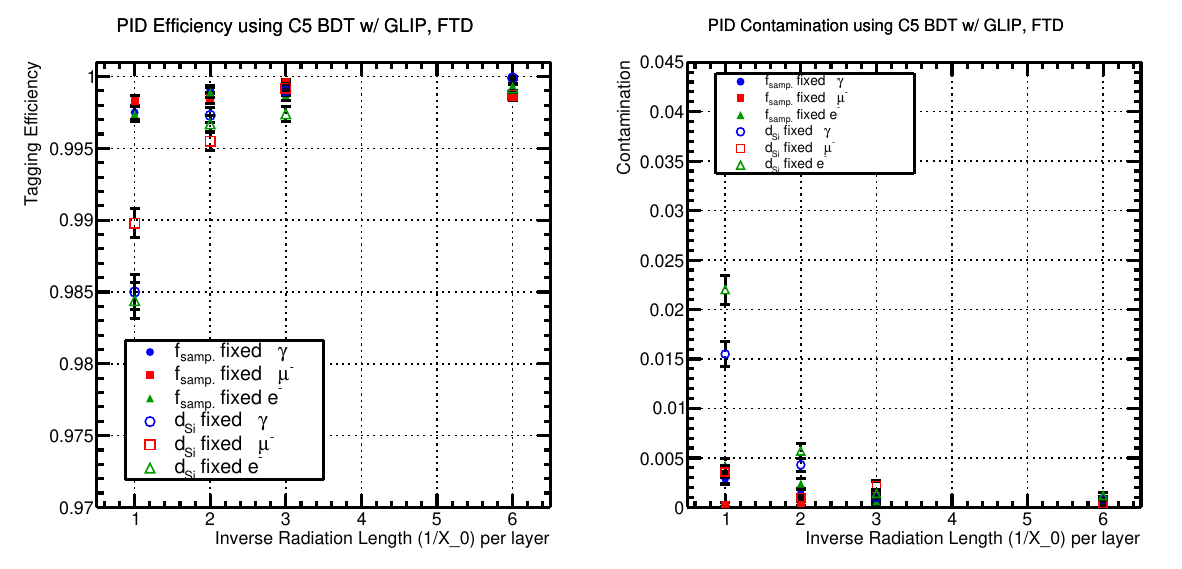}
\caption{(Left) Plot of the PID tagging efficiency for 125~GeV particles at $2^\circ$ degrees of polar angle with varying GLIP LumiCal designs by sampling fraction or silicon thickness.(Right) Plot of the PID contamination for the same data samples.}
\label{fig-PIDInvX}       
\end{figure}

To remedy this demand for precision PID in the forward region, we propose using Boosted Decision Trees (BDT), using the C5.0 automated classifier software, with the FTD and GLIP LumiCal design to identify electrons, photons and muons~\cite{C5}. An initial test of this was done using only the FTD trackers and the first $1.5X_0$ of the GLIP LumiCal, to reduce the long simulation time needed to generate large sample sizes. We gave the BDT averages and standard deviations for the energy deposited, polar and azimuthal angles for each layer of the first $1.5X_0$ of the GLIP LumiCal and all of the FTD trackers. The results of the BDT were also expected to pass cross-validation tests with a data sample of similar size to their training data. Each data set was 10000 samples and the uncertainties and precisions quoted here reflect 10000 samples. In general, we find that the BDT preferentially uses calorimeter information to identify particles at a roughly 1:10 ratio to the tracker information. Results for varying the incidence angle of 125~GeV particles can be seen in figure~\ref{fig-PIDAng}, where the PID precision for photons was observed to be $\approx1\times10^{-4}$ for the forward region.
We find that the PID precision for electrons is $\approx4\times10^{-4}$ and $\approx3\times10^{-4}$ for muons. We provide a plot for the contamination percentage of each particle PID as well. Note that the BDT is able to return confidence percentage for each tag, and contamination events often have low confidence. So this contamination can be reduced further by requiring cuts on BDT confidence. As seen in figure~\ref{fig-PIDEne}, a run which controlled the incident angle to $2^\circ$ and varied the energy, found that the current GLIP LumiCal and FTD are optimal at a center-of-mass energy of 250~GeV.
At ILC GigaZ the PID precision for photons decreased to $\approx2.5\times10^{-4}$ and at ILC1000 it decreased to $\approx2\times10^{-4}$. A similar relative degredation of PID performance was observed for the other particle species. A test was also conducted with a control of angle, again $2^\circ$, and energy, here 125~GeV, but varying the silicon thickness, $d_\text{Si}$, or the sampling fraction, $f_\text{samp.}$, as seen in figure~\ref{fig-PIDInvX}.
Both of these tests were performed so that a comparison between the current GLIP LumiCal design and other possible designs that have different sampling fractions or silicon thicknesses could be made. We note that the run that controlled sampling fraction to being the same as the standard GLIP LumiCal, resulted in the $1X_0$ design having 6~mm thick silicon, which was still used as a test point even if it is not experimentally viable. The results, seen in figure~\ref{fig-PIDInvX}, indicate that decreasing the number of layers, regardless of method, results in a degradation of PID precision. The degradation of PID precision is more severe when the silicon thickness is fixed but the sampling fraction is allowed to get smaller. This makes sense as the smaller the sampling fraction the less sensitive a calorimeter tends to be to the underlying particle shower. 

The results we present here on the PID precision of GLIP LumiCal indicate that it is feasible to use multiple physics processes in the forward region for integrated luminosity precision and still achieve, at least, $10^{-3}$ precision if not $10^{-4}$ precision with some minor improvements. We find that this level of precision in tagging is lost when using less granular designs; using a design similar to the current LumiCal results in a PID precision $\approx1.5\times10^{-3}$, which is incapable of meeting integrated luminosity precision goals. The PID precision of the GLIP LumiCal will also be key in rejecting beam pairs from the integrated luminosity measurements. Future studies which include the entire calorimeter, and therefore more information and particle back-splash from the deeper layers, will be key to ensuring that this PID precision is robust. We also plan to extend this study to include charge differentiation of particles, potentially by including layer $d\phi/dz$ values for the BDT, such that electrons and positrons can be separated. Charge separation would improve beam pair rejection and provide an additional tool for integrated luminosity measurement.

%

\chapter{Center-of-Mass Energy Precision}\label{ch-EPre}

This chapter recaps existing work measuring center-of-mass energy using dimuons and wide-angle Bhabhas. New results on refining these measurements is presented. A new numerical method, the Kernel Density Estimate, for estimating $\sqrt{s}$ and its spread, is presented. This is done with a focus on ensuring $\sqrt{s}$ accuracy in addition to verifying precision. Discussion on the effects beam deflection, as discussed in section~\ref{sec-BDE}, has on center-of-mass energy distributions and precision is also given. A new method of estimating the center-of-mass energy, using the LumiCal to measure SABS and diphotons, is presented as an alternative to check the other methods in a way that is sensitive to different sub-detectors and uncertainties.

\section{Existing $\sqrt{s}$ Precision Work}\label{sec-CoMPrev}

The precise determination of the center-of-mass energy, $\sqrt{s}$, at future $\ee$ colliders is needed for the precision measurement of any energy or mass measurements performed. Of interest to these experiments is the precision measurement of the masses and widths of the Z boson, $\text{W}^{\pm}$ bosons, Higgs boson and the top quark. It is also relevant to any energy cut used in an analysis as the precision of the value used in the cut will be determined by some underlying center-of-mass energy calibration. For this reason the center-of-mass energy precision is also directly relevant to the integrated luminosity measurement and precision as, in general, the integrated luminosity measurement uses cuts on energy. This is discussed more extensively in chapter~\ref{ch-LumiProp}.

Three recent papers cover efforts on determining the center-of-mass energy precision using dilepton events at future $\ee$ colliders~\cite{Madison2022}~\cite{Madison2023}~\cite{Wilson2023}. In this section we will summarize this work and the precision expected from the methods used. We will provide a general summary and then a more detailed summary on the use of dimuons measured in the general tracker and then the use of \Gls{LABS} measured in the general ECAL. These are done under the assumption of being at ILC250 using the ILD large design. Both methods are capable of approaching or even surpassing 1~ppm precision on the center-of-mass energy. There is no reason to assume that these methods are specific to these designs; similar studies have also been conducted for similar methods at FCC-$\text{ee}$~\cite{FCCeeCoMPrecision}. While the FCC-$\text{ee}$ approach also uses tracker dimuons it instead focuses on the use of angles. Given that the FCC-$\text{ee}$ method is yet to undergo a thorough simulation or extrapolation of their results to an estimate of center-of-mass energy precision, we will not present their method further.

\subsection{Tracker Based dimuon $\sqrt{s}$ Estimate}\label{sec-sqp}

\begin{figure}[h]
\centering
\includegraphics[width=16cm]{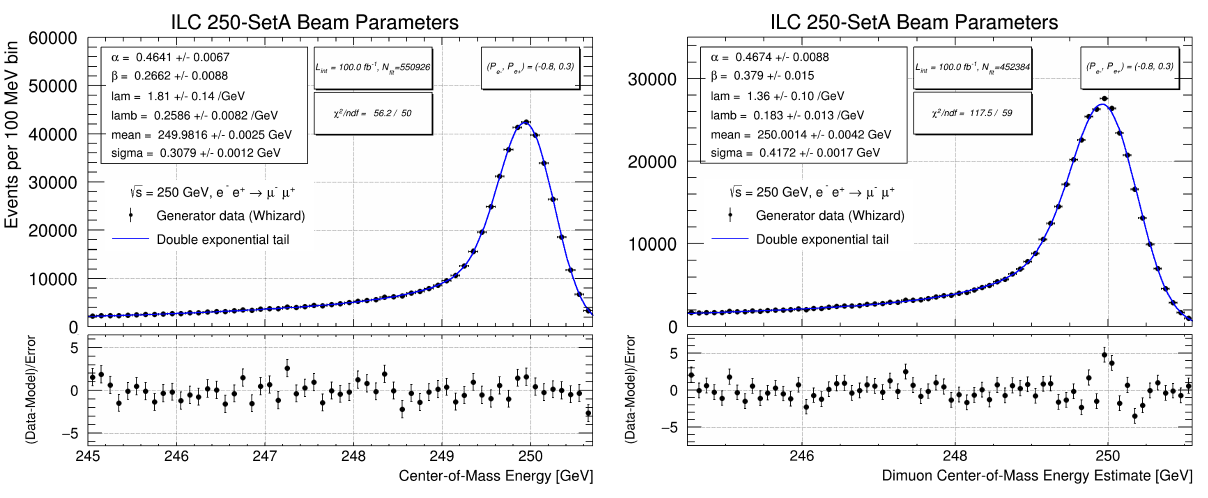}
\caption{(Left) True generator level center-of-mass energy of dimuons produced at ILC250.(Right) Generator level center-of-mass energy estimator $\sqp$ for dimuons at ILC250. Both were done with 100~$\invfb$ of data.}
\label{fig-SqspComp}       
\end{figure}

For the dimuon tracker method it is not efficient to simply use the invariant mass of the muon pairs because a large fraction of dimuon events will have an ISR photon from the radiative return to the Z-pole, as discussed in section~\ref{sec-DiMu}. This can be remedied in two ways, inferring the missing momentum from the kinematics of the dimuon system and then adding it back in using some formalism or by assuming you know the ISR photon energy exactly from the mass of the Z boson. The implementation of the latter method makes the center-of-mass measurement relative to the Z boson mass and, considering the Z boson mass is an observable for future $\ee$ colliders, it would result in a measurement that is \emph{ipso facto} circular. The first method is at the core of the momentum based $\sqp$ method. In a two-body to three-body interaction where the initial two-body system is balanced in momentum and has a center-of-mass energy of $\sqrt{s}$ then the center-of-mass energy
\begin{equation}\label{eqn-sqp1}
    \sqrt{s} = \sqp = E_1 + E_2 + |\bf{p_1}+\bf{p_2}|
\end{equation}
is exactly equal to $\sqp$, which is solved from the energy, $E_\text{i}$, and momenta, $\textbf{p}_\text{i}$, of two of the final-state particles. Here we denote the vectors as bold to make it clear that they are three-vectors and not four-vectors. In this case we choose to use the muon and anti-muon of the final-state dimuon because they are measured better than the ISR photon is. Given this constraint, and considering that since beam effects and crossing angles result in the initial two-body system being boosted, we rewrite equation~\ref{eqn-sqp1}
\begin{equation}\label{eqn-sqp2}
    \sqrt{s} \approx \sqp = \sqrt{p_1^2 + m_\mu^2} + \sqrt{p_2^2 + m_\mu^2} + |\bf{p_1}+\bf{p_2}|
\end{equation}
to depend only on the momenta and change the case to an approximate one. For a detailed derivation of how to recover the exact solution of $\sqrt{s}$ from equation~\ref{eqn-sqp2} by introducing additional observables from the the boost of the system and the mass of the photonic system, which can be non-zero if there are multiple photons, see previous work~\cite{Madison2022}. Of particular interest was the event-by-event beam energy difference, $\Delta E_\text{b}$ or Ediff, which introduces a boost along the z-axis and the invariant mass of all the photons emitted, $M_3$. The main issue with including these terms to achieve the exact center-of-mass energy is that $\Delta E_\text{b}$ is not well known and that the photonic system is measured poorly compared to the dimuons and can have components that are either unmeasured or partially measured due to occurring at very forward angles. This leads to the $\sqp$ estimator of $\sqrt{s}$ being inherently biased, as can be seen in figure~\ref{fig-SqspComp}, where the true value of center-of-mass of dimuons is provided next to the $\sqp$ value of center-of-mass of the dimuons. 

\begin{figure}[h]
\centering
\includegraphics[width=12cm]{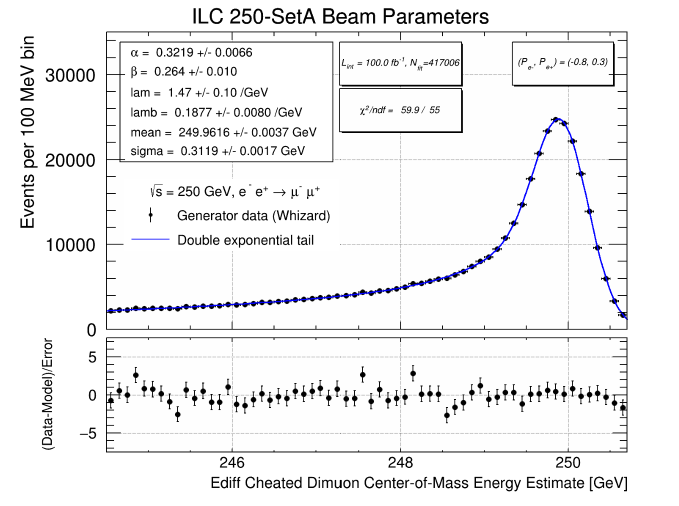}
\caption{Generator level center-of-mass energy estimator $\sqp$ with a correction for the cheated value of $\Delta E_\text{b}$, the beam energy difference. Data was generated with 100~$\invfb$ of data with settings for ILC250.}
\label{fig-SqspBeamCorrected}       
\end{figure}

Previous studies found that, in general, the momentum correction for the unmeasured third body in the $\sqp$ method tends to slightly over-correct and therefore biases the $\sqp$ method high. It was found that knowing the beam energy difference for a particular event is sufficient for improving the fit quality and making the bias in $\sqp$ comparable to the underlying uncertainty~\cite{Madison2022}. As can be seen in figure~\ref{fig-SqspBeamCorrected}, the inclusion of the energy difference term results in a plausible fit quality and similar fitted mean to the true center-of-mass energy seen in figure~\ref{fig-SqspComp}.
The fitting of the center-of-mass energy and its estimators in this section is done using a six parameter double exponential convolved with a Gaussian distribution. The peak region is the normally distributed component, with the double exponential being used to allow the fit to be viable in the tails of the distribution where radiative effects have pulled the center-of-mass energy to lower, and rarely higher, values. A derivation of this fit can be found in previous work~\cite{Madison2022}. 

This study was performed a second time but with changes. We used an acceptance of $5^\circ<\theta<175^\circ$, which is slighly outside the tracking coverage of the FTD, for the \Gls{MCEG} generation of events. A software package, Guinea-Pig-2-X (GP2X), was written. The utility of GP2X is that it can combine the full results of beam and luminosity effects from GuineaPig++ with the results from a MCEG. This is done by first running GuineaPig++ to generate the differential luminosity with respect to center-of-mass energy, also known as the 1D luminosity spectrum. Then an MCEG is run in bins of center-of-mass energy, which are typically 100 MeV bins, over all the center-of-mass energy values covered by the 1D luminosity spectrum for the physics process of interest. This is done with sufficient statistics so that no single bin will contribute more statistics to the final data sample than it has. GP2X then takes data from the GuineaPig++ output and pairs the position, four-momenta values with values from the MCEG according to their center-of-mass energies. In order for this pairing to be time efficient GP2X uses an Orthagonal Sampling method, which reduces pairing time from $\mathcal{O}(N^2)$ using a brute force method to $\mathcal{O}(N)$~\cite{Madison2023}. Since the number of events here is large, typically more than one million, this reduction is significant. After pairing, boosting and rescaling using Newton-Raphson method to the second order are used to ensure energy and momenta are correct and conserved~\cite{Madison2023}. After boosting and rescaling, GP2X uses parameterized models of the ILD \Gls{TPC} and \Gls{ECAL} to apply detector effects. This gives a final dataset that has all of the beam effects, even the position of the primary vertex, from GuineaPig++ and all of the physics processes and particles from the MCEG. Typically, this is approximated with CIRCE2, which parameterizes the 2D luminosity spectrum, where the differential is for the energies of the two beams~\cite{Ohl:1996fi}. To summarize, the use of GP2X gives a more accurate dataset, at the expense of generating and pairing two datasets, one from GuineaPig++ and one from a MCEG.

\begin{figure}[h]
\centering
\includegraphics[width=12cm]{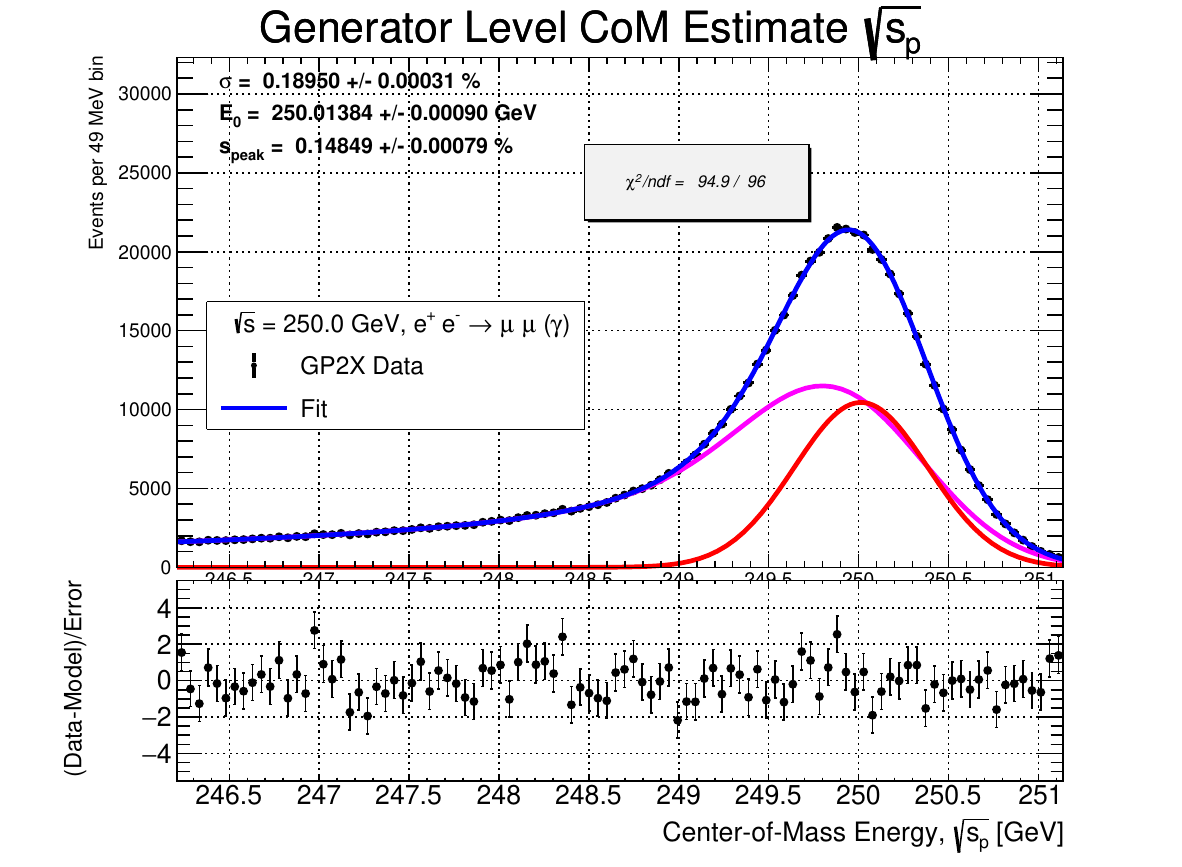}
\caption{KKMC and GP2X were used to generate the 200~$\invfb$ generator level dataset. Fitting was done using the beta convolution with Gaussian added method and RooFit. The $\chi^{2}$ indicates a plausible fit and the $s_\text{peak}$ value is within 1.9\% of the ILC250 design beam energy spread}
\label{fig-BetaConv}       
\end{figure}

\begin{figure}[h]
\centering
\includegraphics[width=12cm]{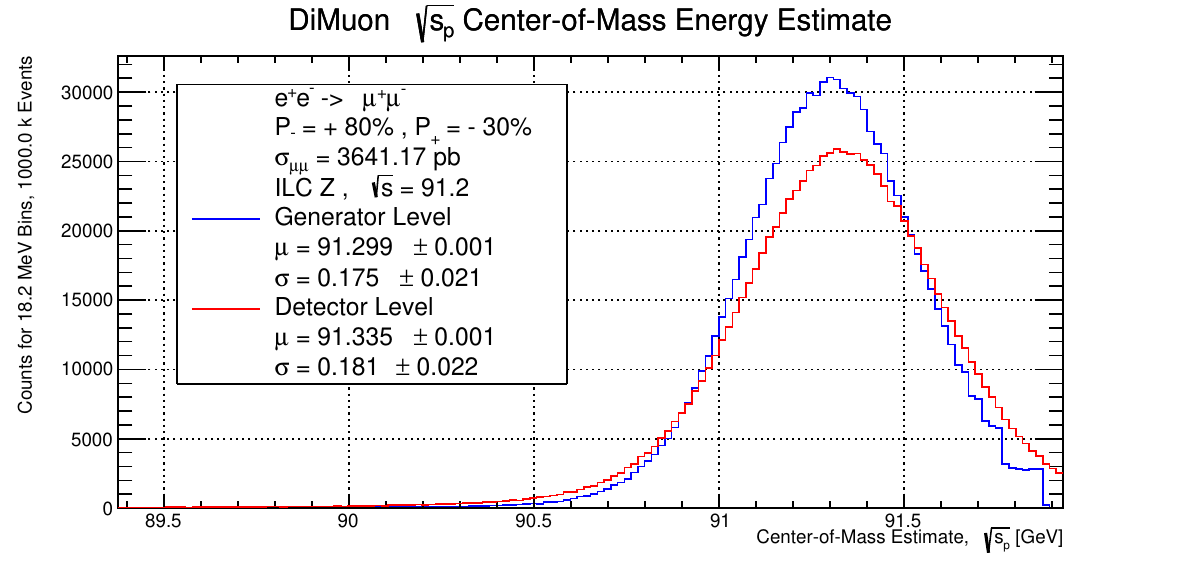}
\caption{WHIZARD was used to generate a dimuon data sample of 1M events and then detector smearing was done using a parameterized form of the ILD TPC momentum resolution. From this sample we computed the KDE mode per methodology described in section~\ref{sec-KDE}.}
\label{fig-DiMusqspZ}       
\end{figure}

\begin{figure}[h]
\centering
\includegraphics[width=15cm]{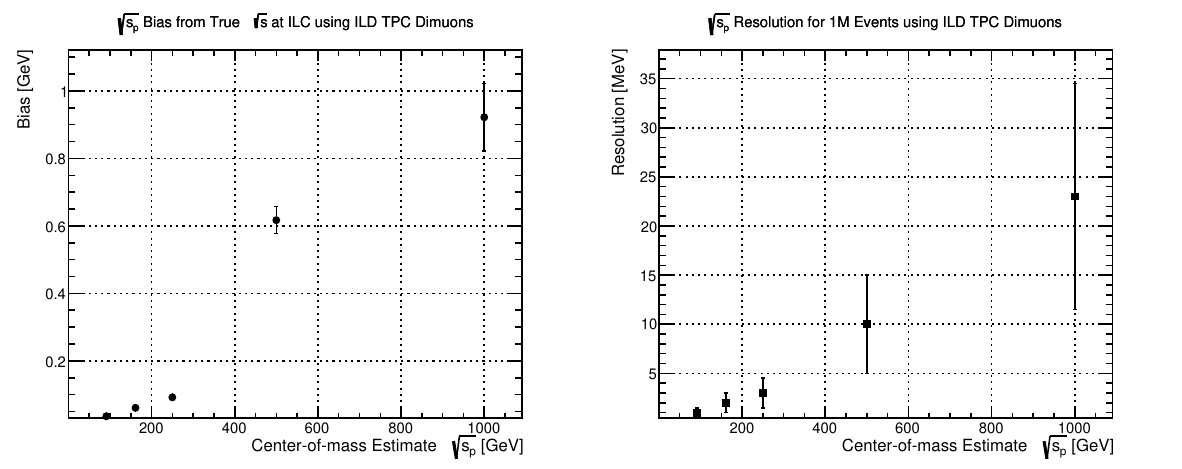}
\caption{(Left) Bias of the center-of-mass estimator $\sqp$ from dimuons measured in the ILD TPC.(Right) Resoltuion of the center-of-mass estimator $\sqp$ from 1M dimuons measured in the ILD TPC. Both values were assessed using the KDE method as outlined in section~\ref{sec-KDE}.}
\label{fig-DiMuBiasRes}       
\end{figure}

For the generation of the dimuon dataset we choose to use KKMC as the MCEG to pair with GuineaPig++ in GP2X due to KKMC's renown for providing as accurate a simulation of dimuons as possible~\cite{Madison2023}~\cite{Jadach_2023}. From the result of GP2X we compute the value of $\sqp$ of the dimuon. A new type of fit, which uses a beta distribution convolved with a Gaussian, is then used~\cite{Madison2023}. As seen in figure~\ref{fig-BetaConv}, the fit improved in quality compared to the right plot of figure~\ref{fig-SqspComp}.
By using the beta convolved fit the fit quality and precision on center-of-mass energy are both improved. The fit was also able to recover the percent energy spread, $s_\text{peak}$, that is expected from the ILC250 design parameters. The bias in the center-of-mass energy estimated by $\sqp$ is seen as was seen in the previous method. We proceed to testing this with the ILC Z parameters. Here we have used WHIZARD with up to two radiative corrections and CIRCE2 for handling beam effects~\cite{Kilian_2011}~\cite{Ohl:1996fi}. As seen in figure~\ref{fig-DiMusqspZ}, we also observe a bias in the $\sqp$ mode here.
We have used the KDE method to evaluate the mode here, which will be further explained in section~\ref{sec-KDE}. From these results the bias in $\sqp$ has been verified with two different methods. The resulting precision on center-of-mass for ILC Z dimuons using the $\sqp$ method and sample size of 2~$\invab$ is 130 parts-per-billion (ppb). We extend this to other center-of-mass energy runs for ILC to determine both the bias and resolution per 1M events. To do this GuineaPig++ had to be run for the various beam parameters and then the \textbf{lumi.ee} files were used with CIRCE2 to generate \textbf{.circe} files to use with WHIZARD to generate event samples with luminosity spectrum effects~\cite{Guinea-PIG}~\cite{Ohl:1996fi}. We chose to generate events with ($+80\%$ , $-30\%$) beam polarization. The results, seen in figure~\ref{fig-DiMuBiasRes}, indicate that the bias and resolution both trend roughly linearly with the increasing center-of-mass energy.
The level of precision of the momentum based center-of-mass energy measurements is then dependent on the resolution and the sample size.

Using these fits with the momentum based center-of-mass energy estimator with tracker dimuons allows for precision of the center-of-mass of roughly 5~ppm for 100~$\invfb$ or 1.6 ppm for 2~$\invab$ at ILC250. When the correlation of fit parameters is removed, by fixing shape parameters, the center-of-mass precision further improves, and the performance becomes comparable to the performance expected for a statistics dominated result. Future studies which can measure or constrain the beam energy difference and the photonic mass can likely improve the systematics. These studies should also include larger data sets, as there may be confounding systematics that degrade quality that only become relevant at larger statistics.

\section{Kernel Density Estimate (KDE) for $\sqrt{s}$ Accuracy}\label{sec-KDE}

When conducting experiments, one often encounters measurements with an underlying distribution that is not known \emph{a priori}. A common practice is to report the distribution in terms of its mean or most probable value, also known as the mode. There are two traditional approaches: to numerically compute the mean under the approximation that the distribution is a normal distribution, or to fit the distribution to some parameterized model distribution and then extract the mean, or mode, from the parameters. In the limit where the data does not approximate or approach a Gaussian then the first approach is simply incorrect. In the limit where the data is not fit well by the model, the second approach is also not correct. When both methods fail it is possible to use Kernel Density Estimation (KDE) as a numerical, model-independent, method to estimate the mode of the data.

\begin{figure}[h]
\centering
\includegraphics[width=14cm]{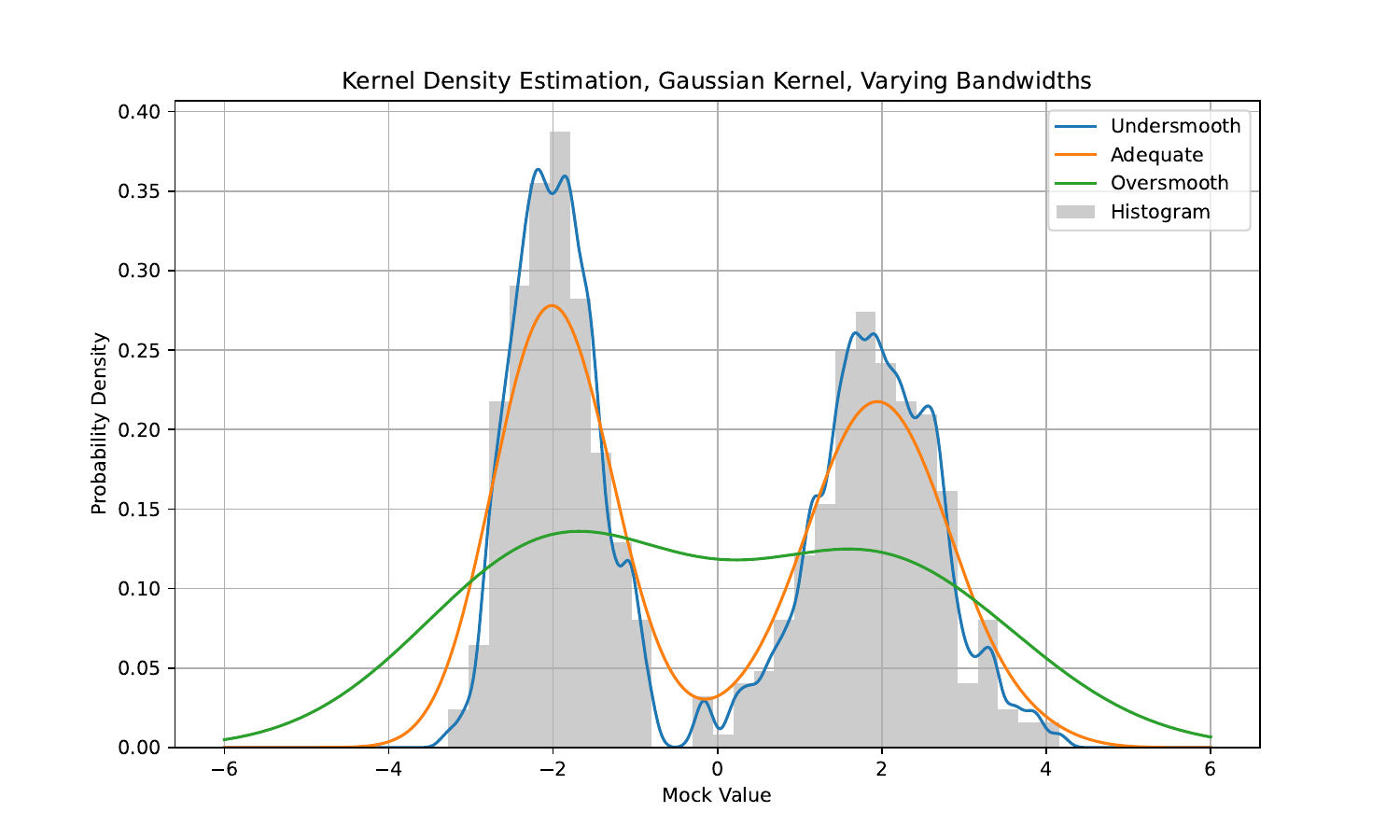}
\caption{Demonstration of the Kernel Density Estimator (KDE) method for a toy dataset and various levels of bandwidth for a Gaussian kernel.}
\label{fig-KDEDemo}       
\end{figure}

The mathematics of KDE are as follows: assume that there is a non-parametric technique to estimate the probability density function of a random variable based on a finite sample of data. Given a data sample $x_1, x_2, \dots, x_n$ drawn from some unknown density $f(x)$, the KDE constructs a smooth estimate $\hat{f}(x)$ of the density~\cite{ParzenKDE}~\cite{RomanoKDE}. The estimator is built by using a smooth, as in differentiable, kernel function, $K(x)$, at each data point and summing their contributions. Then the estimate  
\begin{equation}\label{eqn-KDE}
\hat{f}(x) \;=\; \frac{1}{n\,h}\sum_{i=1}^{n} K\!\Big(\frac{x - x_i}{\,h\,}\Big)~
\end{equation}
depends on the data set, the kernel function, and a bandwidth parameter, $h$. The chosen kernel function is typically a smooth and symmetric function like a Gaussian. This leads to each data point becoming its own Gaussian within some bandwidth, or amount of smoothing, determined by $h$. As seen in figure~\ref{fig-KDEDemo}, a small value of $h$ produces a fluctuating set of narrow peaks while a large value of $h$ leads to a less structured, smoothed, distribution.

Choosing an appropriate bandwidth is critical to the performance of the KDE method. Common practice is to perform KDE across a grid of bandwidth values and then find the bandwidth value that minimizes the uncertainty on the KDE mode estimate~\cite{RomanoKDE}. Here we have devised a renormalizing iterative grid method, that is similar to arithmetic encoding, for finding the minimal uncertainty bandwidth. The minimal uncertainty point of the first grid, $h_{\text{min},0}$, is used as the center of the next grid and then the grid size is renormalized to span the distance from the next-smallest, $h_{\text{-1},0}$, and next-largest, $h_{\text{1},0}$, grid point with respect to the minimal point. We repeat this for the next grid with $h_{\text{min},1}$ and so on. We finish after four iterations such that $h_{\text{min},3}$ is the bandwidth value that is kept. We then proceed to construct $\hat{f}(x)$ is numerically constructed, finding the mode of the underlying data is done by finding the value of $x$ that corresponds to the maximum of $\hat{f}(x)$. Since $\hat{f}(x)$ is, by construction, smooth, the maximum can be verified through differentiation of the peak. Research into the validity of the KDE estimate of the mode has found that it is consistent with the true mode as the number of members of a dataset grows towards infinity~\cite{RomanoKDE}. Using the mode estimate comes with a trade-off in terms of accuracy and precision. In terms of accuracy, the mode can always be identified by the topology of the distribution, the global maximum inflection point. This, in turn, means that the mode is easier to reproduce and verify with multiple methods. Making the mode an overall more accurate estimator than the mean. In terms of precision, the precision of the mode generally scales as $~\sim n^{-1/3}$ whereas the precision of the mean generally scales as $n^{-1/2}$~\cite{VenterMode}. For large data sets, and under the assumption that the mode and mean estimates have similarly negligible uncertainties, the mean outperforms the mode in precision.

After using KDE to estimate the mode it is standard practice to estimate the uncertainty on the estimated mode through the use of bootstrapping~\cite{RomanoKDE}. Bootstrapping is the process of re-running a numerical process with randomized inputs, within a given window of acceptable values, and then finding the uncertainty of the results~\cite{EfronBootstrap}. During bootstrapping we also estimate the standard deviation of the peak so that we can report a standard deviation in addition to the KDE mode. We are confident that the bootstrapping method is reliable here because previous studies have found that bootstrapping, when datasets are large enough, gives a robust estimate of the true underlying statistics~\cite{EfronBootstrap}.

\begin{figure}[h]
\centering
\includegraphics[width=14cm]{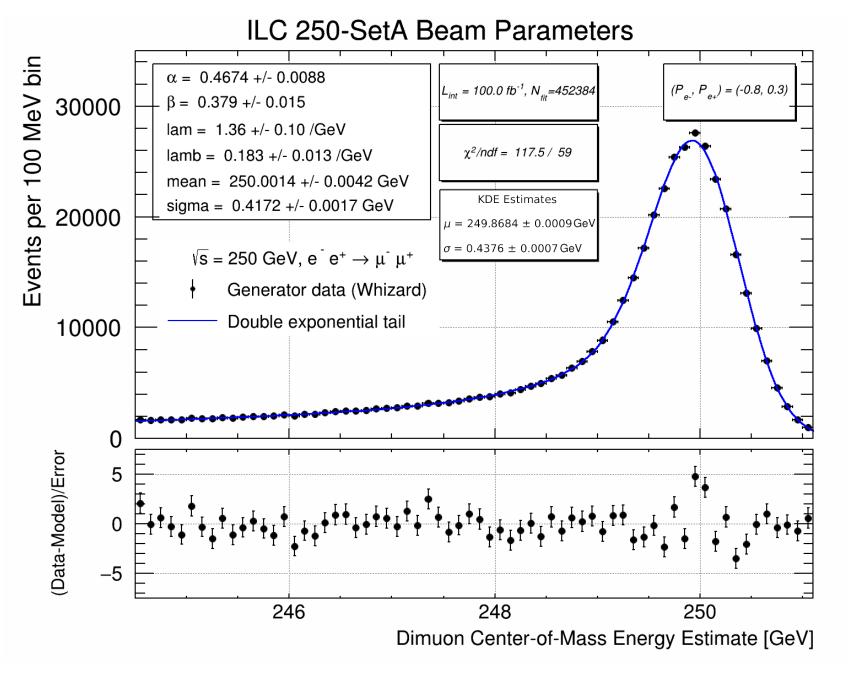}
\caption{Generator level $\sqrt{s_\text{p}}$ distribution evaluated with dimuons at ILC250. A 6-parameter fit is conducted as done previously. For comparison to the fit, we include the estimation of the KDE mode and spread. The results of the KDE method are more accurate, being closer to the true mode and true center-of-mass energy, and more precise than the fit.}
\label{fig-KDEBenchMark}       
\end{figure}

We proceed to implementing the KDE to test against a previously published and verified result~\cite{Madison2022}. In previous studies we found that the we were able to achieve acceptable fits for 100$\invfb$ datasets of the beam energy of each beam, after beamstrahlung and ISR, using either a 6-parameter fit or a convolution fit~\cite{Madison2022}. We were able to achieve comparable to 1~MeV, 8 ppm, energy precision for these distributions and fits. This fit quality was not seen when trying to fit $\sqrt{s_\text{p}}$, the tracker momentum estimate of center-of-mass energy, from tracker \Gls{dimuon}s. Instead, as seen in figure~\ref{fig-KDEBenchMark}, we find a degraded energy precision of 4.2~MeV, or 17 ppm.

An additional issue of the previous model fitting is that the fitted mean is observationally different from the mode as the maximum bin seen in figure~\ref{fig-KDEBenchMark} is clearly below 250~GeV, not above it as the fit would suggest if the mode and mean are similar. This difference is important as any tracker or energy calibration needs to be accurate. Consider, for example, if the tracker is calibrated to the mass of a resonance, such as the $J/\psi$, and a fit is performed that finds a value that is 10~MeV higher than the $J/\psi$ mass but is otherwise exceptionally precise. If the tracker is then calibrated for this shift, when the shift doesn't actually exist and is an artifact of the fit, then the experiment has erroneously introduced a $\sim10$~MeV shift in all of its tracker measurements. Using the KDE method, as shown in figure~\ref{fig-KDEBenchMark}, we find an energy precision of 0.9~MeV, or 4 ppm. In this instance the KDE mode outperforms the fitted mean in terms of precision, likely because the poor fit results in the mean estimate being systematics limited, not statistics limited. We also find that the mode is plausibly in agreement with the observed mode from the histogram bins. We note that the KDE method reconstructed a broader standard deviation, which implies that the KDE method, compared to a high-quality fit, would be less precise. Given this result, and previous discussions in this section regarding the mode precision scaling as $n^{-1/3}$, the KDE mode likely cannot out-perform a high-quality fit in terms of precision. Instead, we advocate the use of both approaches. The KDE mode should be used to give the most accurate estimate, while a high-quality fit should be used for precision.

\section{LumiCal Based Diphoton Energy $\sqrt{s}$ Estimate}

In this section, we will perform an estimation of the center-of-mass energy using diphotons as measured in the current ILD LumiCal and the proposed GLIP LumiCal. We will use a fast simulation to create the effects of the detector based on the specifications of position and energy resolutions outlined in chapter~\ref{ch-LumiProp}. We choose to do this instead of a full simulation as a full simulation implementation would involve a significantly larger time investment and troubleshooting. We use the same acceptance of $1^\circ$ to $6^\circ$ for each LumiCal design so that this analysis is not sensitive to differences in acceptance. We generate diphoton events with up to two radiative corrections using WHIZARD with ILC beam effects done using CIRCE~\cite{Kilian_2011}~\cite{Ohl:1996fi}. For ILC runs that do not have standard CIRCE files already publicly available, we generated CIRCE files using the \textbf{lumi.ee} output of GuineaPig++ that was run with beam specifications provided by ILC references and correspondence with other ILC collaborators~\cite{ILCSnowmass}. We will evaluate the precision using the KDE method, even if it is likely to give a worse precision than an adequate fitting method.

\begin{figure}[h]
\centering
\includegraphics[width=14cm]{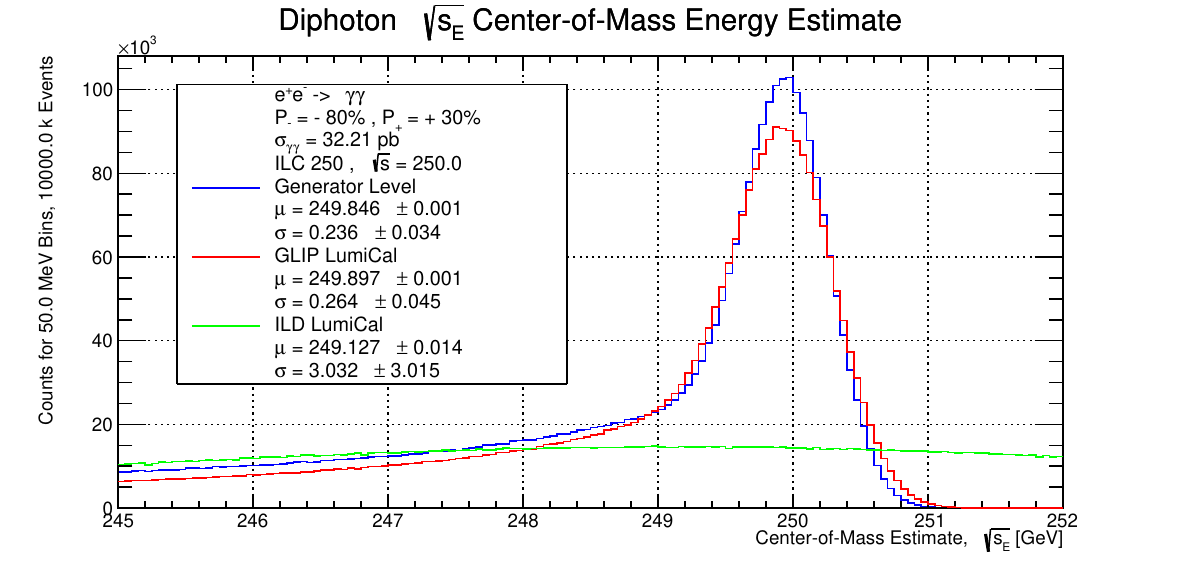}
\caption{Plot of the energy sum center-of-mass energy estimate, $\sqe$, for diphotons measured in the GLIP LumiCal and ILC LumiCal. This sample used 10M diphotons, of which roughly 5M diphotons were measured in the LumiCal. This corresponds to $\approx158$~$\invfb$ of integrated luminosity.}
\label{fig-Diphosqe}       
\end{figure}

\begin{figure}[h]
\centering
\includegraphics[width=14cm]{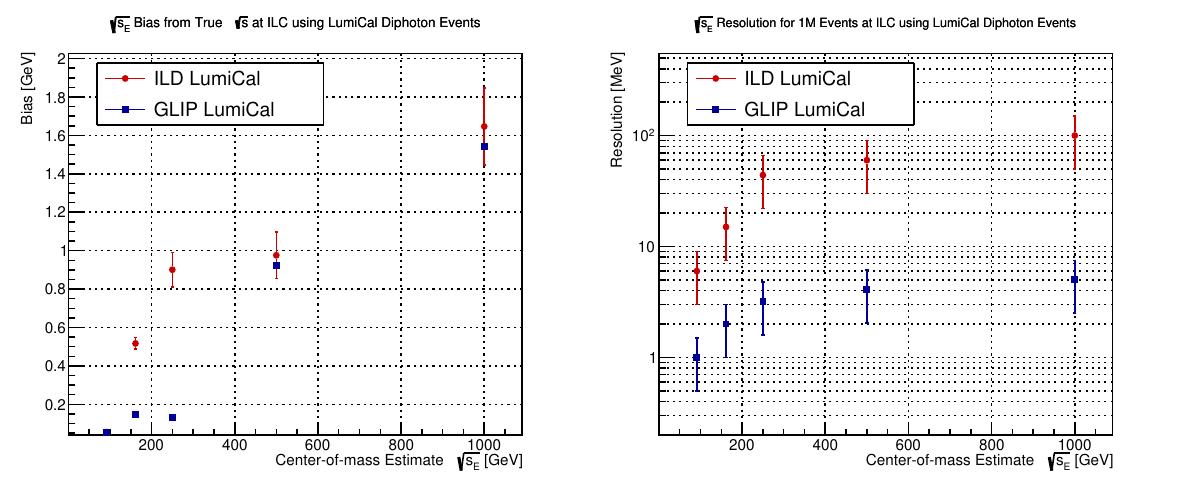}
\caption{(Left) Plot of the center-of-mass energy bias of using diphotons in the LumiCal at ILC for various center-of-mass energies. (Right) Plot of the center-of-mass energy resolution for 1M events of diphotons in the LumiCal at ILC for various center-of-mass energies. We evaluate these using the KDE methods outlined in section~\ref{sec-KDE}.}
\label{fig-DiphoBiasRes}       
\end{figure}

As a starting point and comparison to the results using $\sqp$ with dimuons we use ILC250. Since the diphotons measured in the LumiCal are not momentum based measurements and there is no radiative return in diphotons, thereby removing the motivation of the momentum balancing of $\sqp$, it is not correct to use $\sqp$ for diphotons. Instead, we use a simpler center-of-mass estimate
\begin{equation}\label{eqn-sqe}
    \sqe = E_1 + E_2
\end{equation}
that we dub the `energy sum center-of-mass estimate' which depends on the energy sum of $E_1$ and $E_2$, the calorimeter energies of the two photons. Due to the kinematics, this will typically manifest as a photon shower in the front section of the LumiCal and a photon shower in the back section. Using equation~\ref{eqn-sqe} we plot the results of simulations for ILC250 and, as seen in figure~\ref{fig-Diphosqe}, we find that the center-of-mass energy estimate with GLIP LumiCal achieves a similar precision to the dimuon $\sqp$ method.
The results of fig~\ref{fig-Diphosqe} also indicate that the current ILD LumiCal has poor performance for reconstructing the center-of-mass energy using the $\sqe$ method. In addition to evaluating the resolution we can also evaluate the bias, that is the energy difference from the true center-of-mass energy. We find that the energy bias is better in the GLIP LumiCal as compared to the current ILD LumiCal. We then extrapolate this methodology to the other center-of-mass energy runs of ILC, as can be seen in figure~\ref{fig-DiphoBiasRes}.
These results are promising, but without a method for calibrating the LumiCal with comparable precision they are without utility. Existing calorimeter calibration methods using test beams, well known processes and radioactive sources can achieve calibration on the 0.1\% to 1\% level of precision~\cite{RefWig}. Here we instead use the center-of-mass energy estimate of the tracker dimuons. Since the tracker can be calibrated to comparable to 1~ppm using well known resonances like the $J/\Psi$, the center-of-mass energy from tracker dimuons becomes a high precision calibration tool for calibrating the calorimeters of the detector. Then the limitation in precision is in how well the calorimeter can measure the center-of-mass distribution. From the results of section~\ref{sec-sqp} and figure~\ref{fig-DiphoBiasRes} we can see that there is promise for precision energy measurements in the forward region should the GLIP LumiCal be deployed. This has great promise for precision measurements and would allow unique sensitivity to both very narrow and very small energy processes or new particles. Take, for example, a Higgs decay to two photons where the photons are both measured in the GLIP LumiCal. The narrow width of the Higgs, of $\sim4$~MeV, with the energy resolution of the GLIP LumiCal, could give an incredibly precise measurement of the Higgs mass. This precision could also be used to measure ISR photons precisely and then dress them to dimuon events, or other radiative return events, for increased precision in said events. Previous studies of this with the current ILD LumiCal found that the poor energy resolution resulted in a net loss of performance when including the ISR photon measurements~\cite{MadisonMasters}. This benefit is unique to the GLIP LumiCal, or a LumiCal design with comparable performance.






\chapter{Measuring Integrated Luminosity}\label{ch-Lumi}

This chapter continues the work started in chapter~\ref{ch-LumiProp} and uses results from chapters~\ref{ch-FCAL} and~\ref{ch-EPre} to simulate realistic results and estimate the overall integrated luminosity precision. We also identify sources of uncertainty and separate them into non-detector and detector sources of uncertainty so that the sources of uncertainty relevant to this work, the detector uncertainties, can be evaluated separately. We focus this work on the \Gls{SABS} and \gls{diphoton} processes as measured in either the current ILD LumiCal or GLIP LumiCal designs. The designs will be compared in terms of their integrated luminosity precision. This study will be conducted at center-of-mass energies of 91.2~GeV, 161.4~GeV, 250~GeV, 500~GeV and 1000~GeV so as to provide a comprehensive study of integrated luminosity precision for a future linear $\ee$ collider.

\section{Non-Detector Uncertainties}\label{sec-NonDet}

In this section we discuss the sources of uncertainty on integrated luminosity precision that arise from effects not relevant to the detector. We disregard transient effects here though note that, considering the results of LEP, there are likely to be transient effects. We also note that there are metrology requirements such that the measurements from the LumiCal can be made with respect to the interaction point with a high degree of accuracy. In general, we require that the metrology for polar angle measurement be better than 5~$\mu$rad to be negligible, which corresponds to measuring the inner and outer radius to $\approx20$~microns and length along the beam axis to a similar $\approx20$~microns. This agrees with previous work on metrology for ILC for the precision of integrated luminosity of $10^{-4}$~\cite{Smiljanic:2024twn}. This demand is within the realm of feasibility even with current methods and technology as multiple LHC detectors already achieve these values using laser alignment, as well as cosmogenic or lab-generated muons~\cite{CMS:2014pgm}~\cite{ATLAS:2010nca}. We also note that upstream material, particularly the beam pipe cone of the ILD detector design, influences the integrated luminosity measurements of the LumiCal by roughly $10^{-4}$~\cite{Sadeh2008}. However, as done at LEP, effects from upstream material can be corrected for and made negligible~\cite{OPALLumi}.

\subsection{Theory and Statistics}\label{sec-lumitheo}

As a part of determining the integrated luminosity one must solve for the expected cross-section with respect to the underlying theory as evident by the $\mathcal{L} = N/\sigma$ relationship. From this it is clear that the precision on the cross-section calculation, as well as the statistical precision, both play direct parts in the precision of integrated luminosity. Considering this, in this section we will use references from previous work done by theorists and values from \Gls{MCEG} to determine the theory precision and statistical precision at each operating point of center-of-mass energy. This will also be done for the two different LumiCal designs, which have different polar angle acceptance and therefore difference theory uncertainties and statistics. This will also be done for the two different integrated luminosity processes, \Gls{SABS} and \gls{diphoton}s.

\begin{figure}[h]
\centering
\includegraphics[width=14cm]{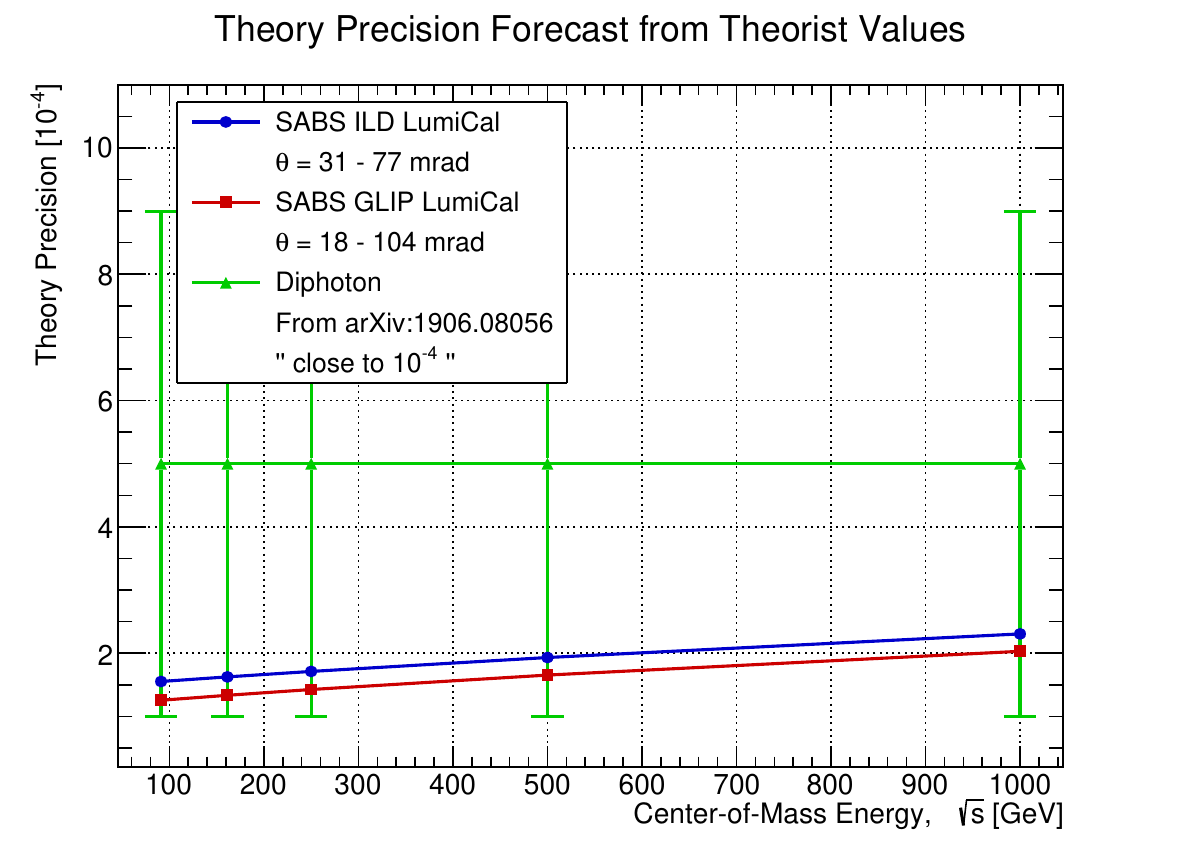}
\caption{Forecast of the theory precision for the various ILC runs and the two different proposed LumiCal designs, particularly their different polar angle acceptance.}
\label{fig-TheoryForecast}       
\end{figure}

At LEP, the theory precision of SABS was evaluated to being a few factors of $10^{-4}$ and later updates have been able to reduce this to smaller factors of $10^{-4}$~\cite{OPALLumi}~\cite{Jadach:2021ayv}. As a part of evaluating the theory precision various factors must be considered, as seen in table~\ref{tab-TheoryTab}, where the expected and forecasted theory precision of SABS for various aspects of theory are evaluated for ILC500.
\begin{table}[h]
\begin{centering}
  \includegraphics[width=14cm]{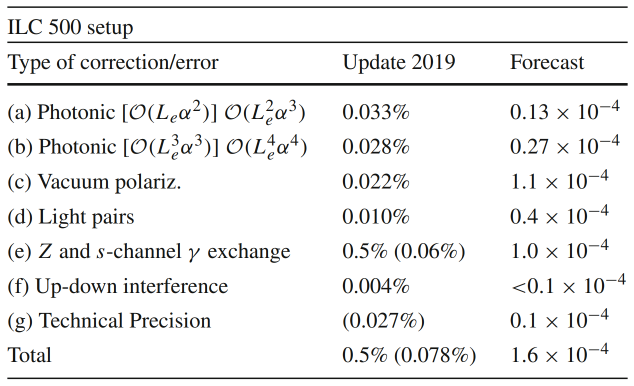}
  \caption{Table of the various contributions to the SABS theory precision at ILC500. The current values and forecasted, aspired values for once ILC500 is taking data, are shown. The polar angle acceptance here is for the ILC LumiCal of 31 to 77 mrad. Table source~\cite{Jadach:2021ayv}. }
  \label{tab-TheoryTab}  
\end{centering}
\end{table}
The forecasted values are those that the theorists claim to be possible by the time ILC500 would be taking data. For extrapolating the theory precision of SABS to different runs we have used references from theorists that describe the running of the various contributions, written in their notation, in terms of the inner acceptance, $\theta_\text{min}$ outer acceptance, $\theta_\text{max}$, scattering energy at inner acceptance, $t_\text{min}$, scattering energy at outer acceptance, $t_\text{max}$, center-of-mass energy, $\sqrt{s}$ and average scattering energy, $\bar{t}=\sqrt{t_\text{min}t_\text{max}}$~\cite{Jadach:2021ayv}. To do this, we have written python code to approximate the various contributions and terms of table~\ref{tab-TheoryTab}, and then computed the relative changes of them with respect to their reported LEPZ values. We found that this approximation was, in general, able to reproduce their values to within 5\%. We then adjusted these according to their forecasted values for ILC500 and ILC1000, per each of the items in table~\ref{tab-TheoryTab}, so that we may report forecasted values here. This was done for each center-of-mass energy and for the two different polar angle acceptance values for the two different LumiCal designs. For the diphoton theory we use a different theorist work where they claim that the forecasted diphoton theory precision should be factors of $10^{-4}$ and therefore will use a value of $(5\pm4)\times10^{-4}$~\cite{Carloni_Calame_2019}. For this reason we were unable to determine the diphoton theory precision for different center-of-mass energies or different polar angle acceptances. The results, seen in figure~\ref{fig-TheoryForecast}, indicate that theory precision will be factors of $10^{-4}$ and that the theory precision will be slightly superior at GLIP LumiCal compared to the current ILD LumiCal.
We estimate that the dominant factor for the difference between the two LumiCal designs is derivative of the different inner acceptance angles. When the inner acceptance is wider, there is considerably more up-down QED interference and Z-$\gamma$ electroweak interference, leading to the theory precision degrading~\cite{Jadach:2021ayv}~\cite{Jadach:1990zf}. This also poses problems for the proposed LumiCal design of $\fccee$, which starts at an even wider angle of 64~mrad~\cite{Jadach:2021ayv}~\cite{Dam:2021sdj}.

\begin{figure}[h]
\centering
\includegraphics[width=14cm]{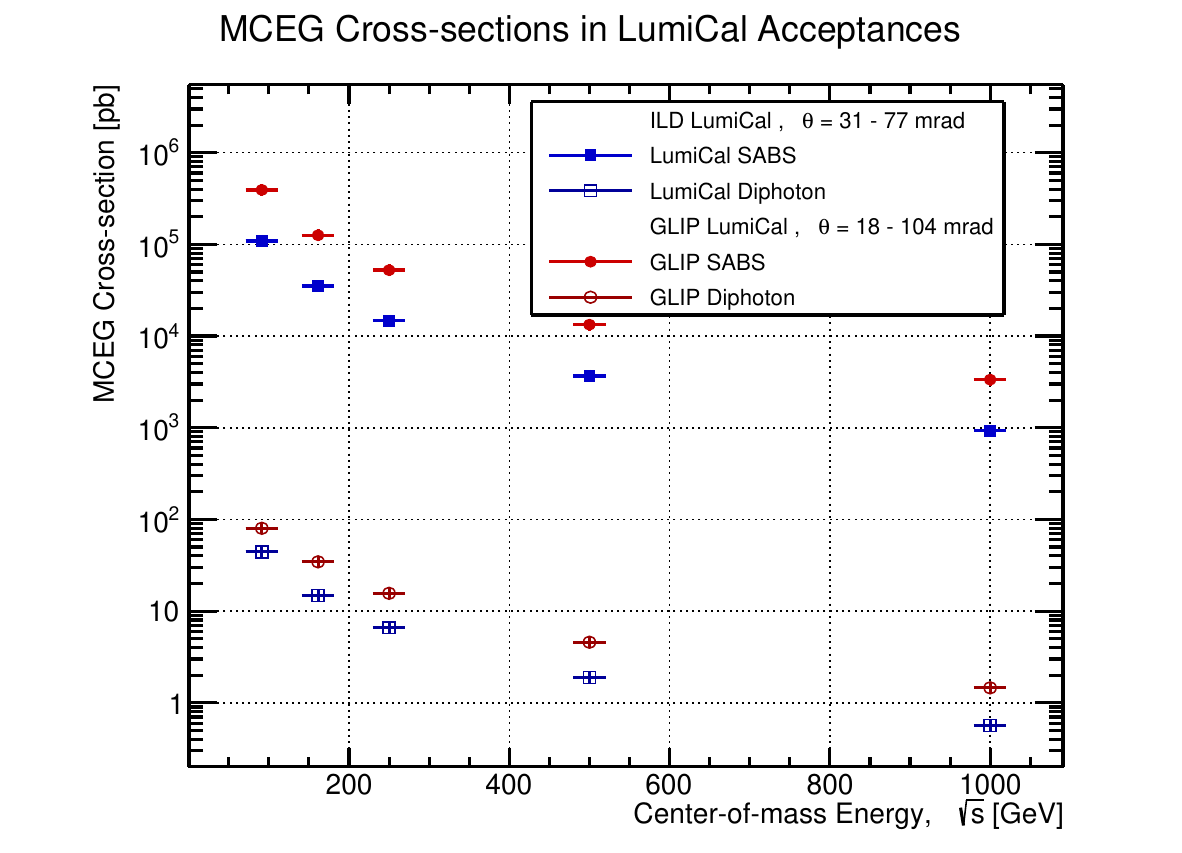}
\caption{Results for the calculation of the cross-section for the two integrated luminosity processes of SABS and diphotons for the polar angle acceptances of the two different LumiCal designs. The error bars are reflective of the reported accuracy of the MCEG and are smaller than the markers in all cases.}
\label{fig-MCXSec}       
\end{figure}

\begin{figure}[h]
\centering
\includegraphics[width=14cm]{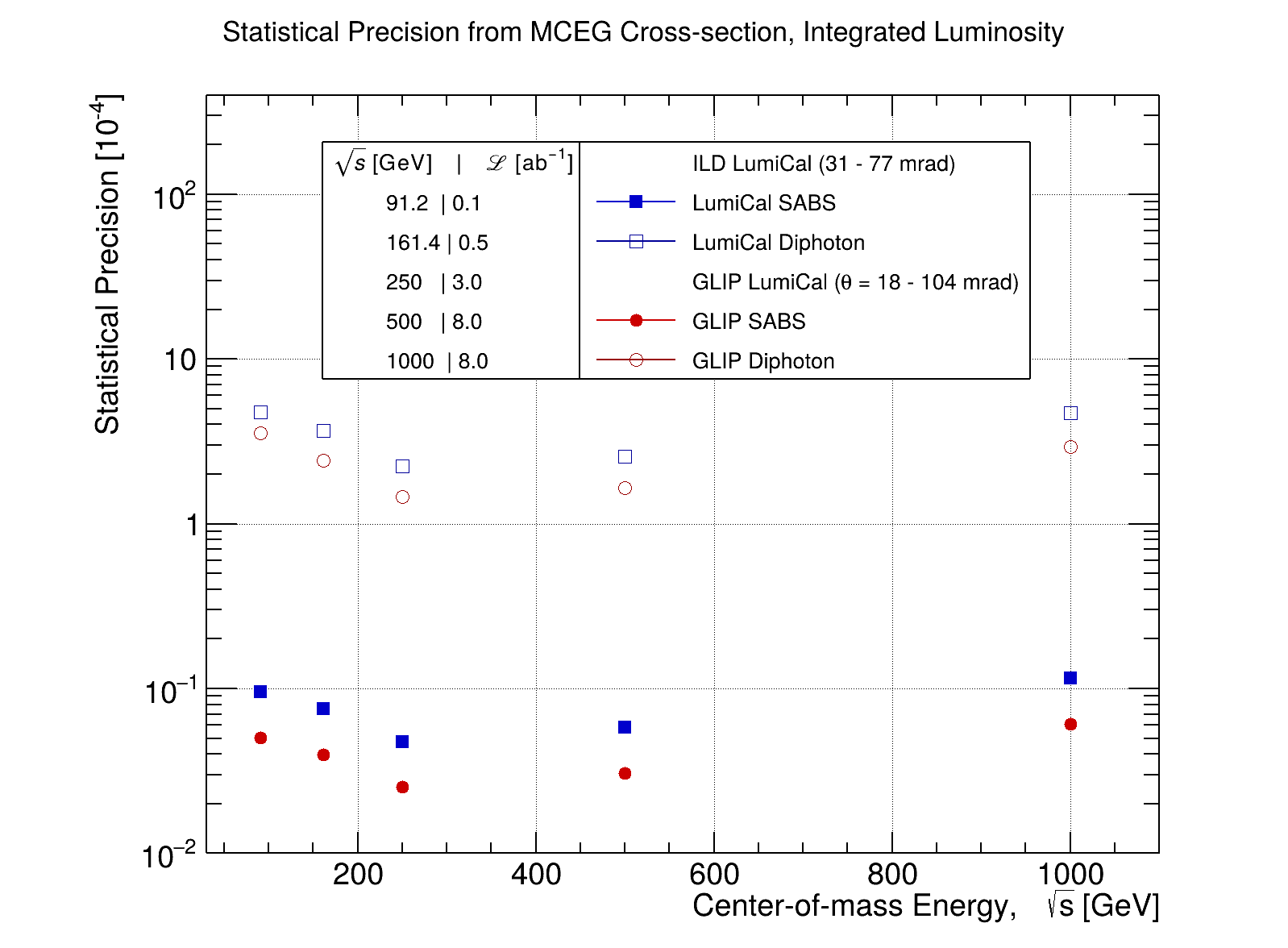}
\caption{Estimation of the statistical precision of the two integrated luminosity processes of SABS and diphotons for the polar angle acceptances of the two different LumiCal designs. We do not include error-bars. Values for the integrated luminosities can be found here~\cite{LCVision}}
\label{fig-LumiStat}       
\end{figure}

We now switch to computing the cross-sections for the integrated luminosity processes of SABS and diphotons for the polar acceptances of the two LumiCal designs. For SABS we use BHLUMI, using the setting used for LEP era simulations, and for diphotons we use WHIZARD with CIRCE to handle beam effects, a crossing angle of 14~mrad, and up to three additional photon radiative corrections~\cite{BHLUMI}~\cite{Kilian_2011}. The results, seen in figure~\ref{fig-MCXSec}, show how the SABS cross-section is much larger than the diphoton cross-section and how the GLIP LumiCal, due to having a smaller inner acceptance and wider outer acceptance, has a consistently larger sample size.
The BHLUMI values were computed with a run of 2M events while the WHIZARD values were computed using integration over a weighted grid of 2M data points for 5 iterations. We chose these values for consistency and because these settings achieved consistent stability, at the $10^{-3}$ level, in the results across the various run settings. From these values, and plausible integrated luminosity values for these runs from recent work, we can estimate the statistical precision of each variation, as seen in figure~\ref{fig-LumiStat}~\cite{LCVision}.
We find that, as expected from the polar angle acceptance values, that the GLIP LumiCal will have improved statistical precision compared to the current ILD LumiCal. The statistical precision for SABS at all energies and designs is not a dominant source of uncertainty for the integrated luminosity. By contrast, the diphoton statistical uncertainty, for the currently proposed sample sizes, will be a dominant source of uncertainty on the integrated luminosity, especially if the precision goal is $10^{-4}$. We note that the statistical precision will be even worse for $\fccee$, as the $\fccee$ LumiCal polar acceptance of $64-86$~mrad gives cross-sections at the Z pole that are 14 nb for SABS and 13 pb for diphotons~\cite{Dam:2021sdj}. These are roughly one order of magnitude smaller than the cross-sections of either ILC LumiCal design, meaning that $\fccee$ will need to take $\approx\times10$ the data of ILC to reach a similar statistical precision on integrated luminosity.

\subsection{Luminosity Spectrum and Beam Effects}\label{sec-LumiSpec}

Previous studies have investigated the effects that beamstrahlung and other beam effects as well as electromagnetic deflection have on the precision of integrated luminosity and found them to be relevant at the $10^{-4}\to10^{-3}$ level~\cite{CRimbault_2007}. Beamstrahlung is particularly relevant for linear colliders such as ILC or CLIC, where the intense fields at the collision point enhance the instantaneous luminosity but also enhance the beam disruption and energy loss. Resulting in a loss of events from the peak region of the luminosity spectrum to the tail region, which is at lower energies. Since beamstrahlung is an initial-state effect, in the classical sense and not in the quantum sense, it applies universally to all final states by modifying the momenta of the incoming electron and positron. The effect applies to both \Gls{SABS} and \gls{diphoton}s in the same manner. We define $N_\text{init.}$ as the expected event count in the absence of luminosity spectrum effects, and $N_\text{fin.}$ as the observed event count after such effects. The luminosity spectrum effects result in a loss of events, and therefore a bias in the count of events of
\begin{equation}\label{eqn-BSCnt}
    N_\text{LS} = N_\text{init.} - N_\text{fin.}
\end{equation}
where we define the events loss due to luminosity spectrum effects from the initial count, before said effects, and the final count. This effect can be minimized if one is able to model the luminosity spectrum effects and calculate the value of equation~\ref{eqn-BSCnt} to a passing degree of accuracy. Then the value of $N_\text{LS}$ can be added to the experimental value, $N_\text{fin.}$, to attain an estimate of $N_\text{init.}$. This correction still has effects on the precision of integrated luminosity based on how well the luminosity spectrum can be measured and/or modeled. Such that
\begin{equation}\label{eqn-BSLumi}
    \delta\mathcal{L} = \frac{N_\text{init.} - N_\text{fin.}}{N_\text{init.}}\delta_\text{LS}
\end{equation}
the effect on the precision of integrated luminosity, $\delta\mathcal{L}$, from the luminosity spectrum modeling precision, $\delta_\text{LS}$, is scaled by the ratio of events that are lost. We can do a consistency check of equation~\ref{eqn-BSLumi} by considering the case where no events are lost. In such a case, no knowledge of luminosity spectrum is needed to provide a correction. Any value of $\delta_\text{LS}$ performs the same; the precision on the integrated luminosity is invariant of $\delta_\text{LS}$.

\begin{figure}[h]
\centering
\includegraphics[width=14cm]{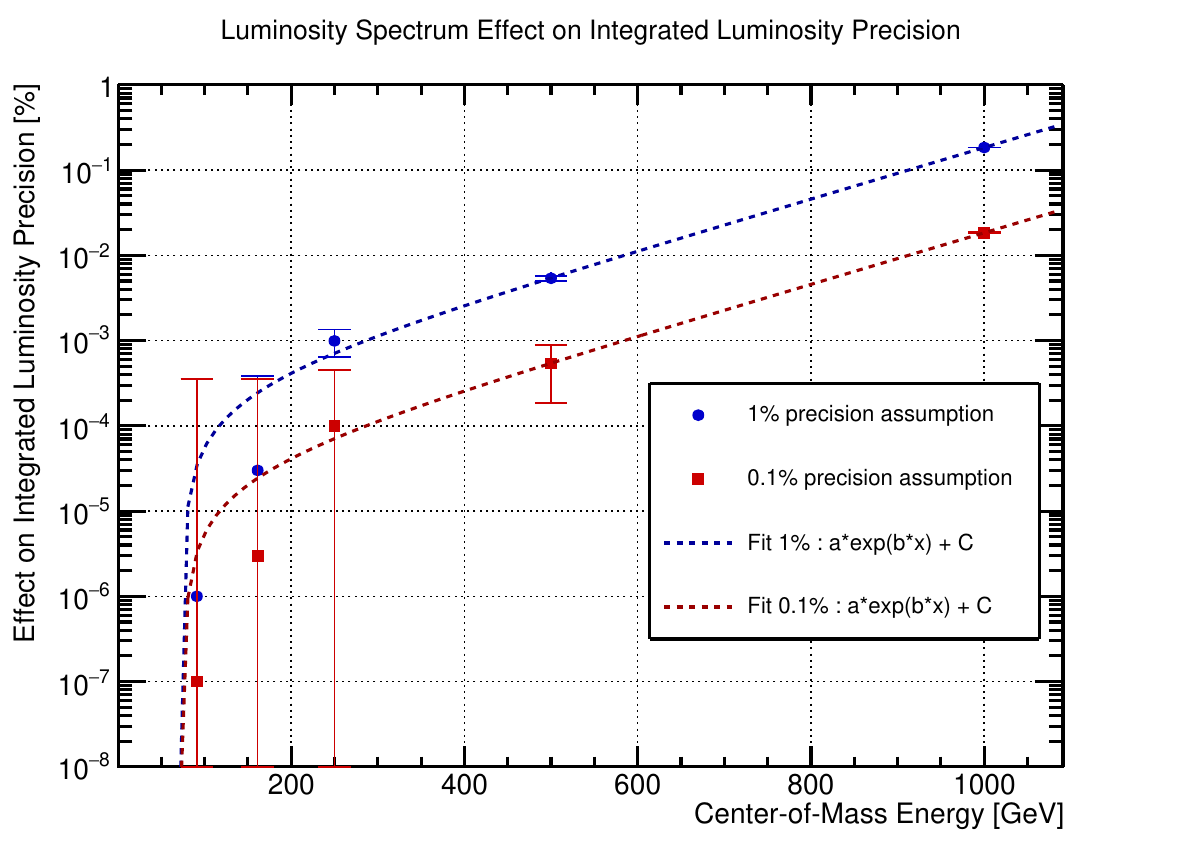}
\caption{Plot of the percent effect on the integrated luminosity precision resulting from loss of events into the tail region of the luminosity spectrum. Here we use energy cuts similar to previous studies~\cite{Abramowicz_2010}. We plot for two different scenarios of quality of luminosity spectrum measuring and modeling, 1\% and 0.1\%. In both cases the effect on the precision of integrated luminosity does not become significant until 500~GeV.}
\label{fig-BeamssLumi}       
\end{figure}

We now look at extending equation~\ref{eqn-BSLumi} to a plausible future $\ee$ collider scenario of measuring integrated luminosity. Using the energy cut of a previous study, of 80\% on the calorimeter energies of each particle and the sum of their energies, we can proceed~\cite{Abramowicz_2010}. We entertain two plausible scenarios for the quality of luminosity spectrum modeling of 1\% and 0.1\%. The precision of the luminosity spectrum will depend on how well it can be modeled and measured, but we are confident, given previous studies, that the 1\% to 0.1\% range are both plausible~\cite{Wilson2023}~\cite{Sailer:2009zz}~\cite{Poss:2013oea}. Using equation~\ref{eqn-BSLumi} and the values outlined in previous work, we can compute the effect of the luminosity spectrum on precision of integrated luminosity arising from energy cuts. The result, seen in figure~\ref{fig-BeamssLumi}, indicates that the luminosity spectrum effects are negligible at the lower center-of-mass energy runs, but becomes significant at 1000~GeV.
The data in figure~\ref{fig-BeamssLumi} is reflective of ILC design values and 200k samples of the luminosity spectrum for each run as simulated in \Gls{GP}. We provide fits to an exponential function, both of which indicated a $\chi^2$/NDoF value comparable to 1, although we note that the predictive power of a $\chi^2$ value becomes less when the number of data points is small. Therefore, while we find that the exponential does visibly and statistically match the trend of the effect on the precision of integrated luminosity, the fit should be taken lightly.

\begin{figure}[h]
\centering
\includegraphics[width=14cm]{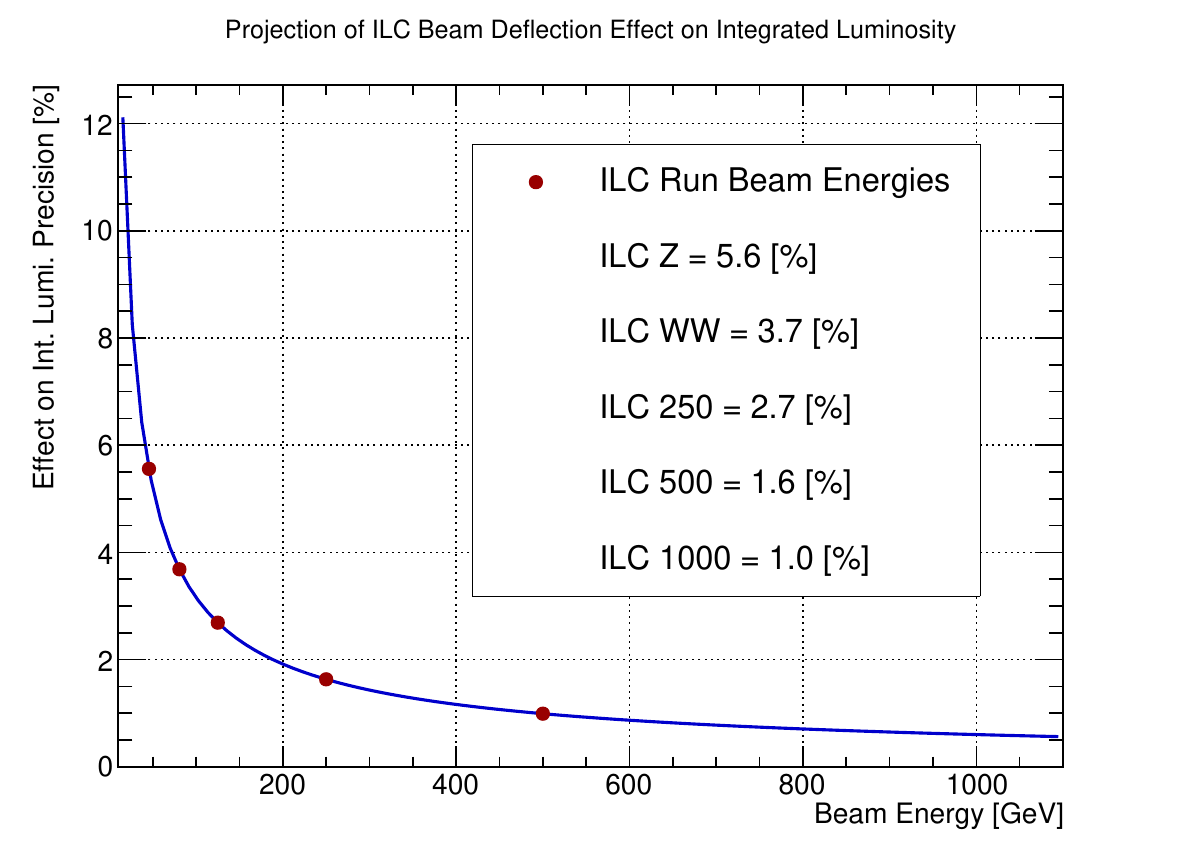}
\caption{Plot of the projection of the effect of electromagnetic beam deflection on the integrated luminosity precision as calculated from using equation~\ref{eqn-SABSShift} with the Lorentz factor fits for electromagnetic beam deflection as done in section~\ref{sec-BDE}.}
\label{fig-DefLumiPlot}       
\end{figure}

\begin{figure}[h]
\centering
\includegraphics[width=14cm]{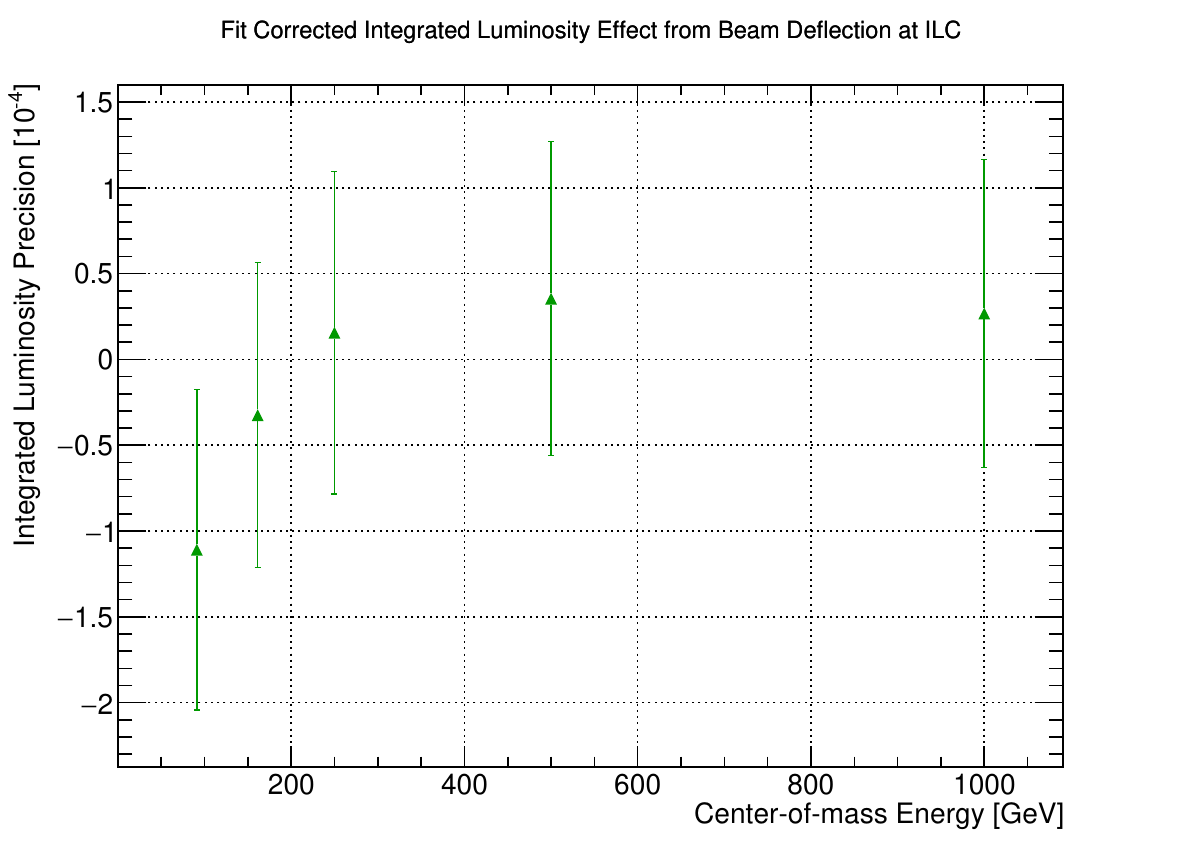}
\caption{Plot of the result of using the fit corrected values of acceptance angles and electromagnetic beam deflection, from using the methods of section~\ref{sec-Moller}. Negative values are possible for this effect as the amount of luminosity can increase or decrease depending on the sign of the amount of polar angle deflection. Error bars reflect a 200M sample size and the propagated uncertainties from the fitting method and an assumed 10 $\mu$~rad polar angle resolution.}
\label{fig-DefLumiFit}       
\end{figure}

Electromagnetic deflection from the beam packets on the emission of charged particles also occurs. This effects SABS but not diphotons as diphotons are neutral. More in-depth work on this topic can be found in section~\ref{sec-BDE}. We continue the work from that section here by using equation~\ref{eqn-SABSShift} with the Lorentz factor fits of section~\ref{sec-BDE} to construct a functional result for the effect on integrated luminosity precision. This is possible as the fit allows for the $\delta\theta$ terms in equation~\ref{eqn-SABSShift} to be solved for and then the remainder of the equation is merely substitution of values. We plot the result of this, along with the results for the relevant ILC runs, in figure~\ref{fig-DefLumiPlot}.
From figure~\ref{fig-DefLumiPlot} we can see that the electromagnetic beam deflection effect is significant and, potentially, the leading-order effect on integrated luminosity precision. Considering this result, it is integral to determine a method for solving for the beam deflection or negating this effect in some manner. Previous studies have proposed using pilot or test bunches as a diagnostic for beam deflection~\cite{Voutsinas_2019}. We will instead expand on the proposal of section~\ref{sec-Moller}, where we run the collider as a $\text{e}^-\text{e}^-$ collider and take measurements of M\o{}ller scattering. By using the results of figure~\ref{fig-FitTrend} we can use equation~\ref{eqn-SABSShift} but now under the assumption that we use the fit results to adjust the inner and outer acceptance values. As an example, if we originally used 18~mrad as an inner acceptance but measured that the inner acceptance was being deflected by -100~$\mu$rad then we will instead use a fit corrected value of inner acceptance of 17.9~mrad. These fit corrected values are then used with equation~\ref{eqn-SABSShift} to evaluate an estimate of the effect of any residual electromagnetic beam deflection on the integrated luminosity precision. The result, seen in figure~\ref{fig-DefLumiFit}, indicates that the fit corrected values are able to keep the effect on integrated luminosity precision at the $10^{-4}$ level.
We propagate the uncertainties of the underlying fit and measurements to allow for a range in the effect of integrated luminosity precision so that we can see how well this effect is constrained. In general, the uncertainty is comparable to the size of the effect itself. Given these results we are confident that the electromagnetic beam deflection effect can be well measured and constrained using the methodology of section~\ref{sec-Moller} such that it does not become a dominant source of uncertainty in the integrated luminosity precision for all energies tested here for ILC.

\subsection{Beam Polarization}\label{sec-BeamPol}

As discussed in chapter~\ref{ch-Theory}, the cross-section of processes may depend on the polarization of the electron and positron beams. We restrict this work to longitudinal polarization, though we note that transverse polarization effects are relevant for circular colliders. For SABS this dependence is small and only becomes relevant at higher orders and higher center-of-mass energies. As derived in equation~\ref{eqn-SABSPol1}, the higher order corrections have a roughly 1\% polarization dependent effect at a center-of-mass energy of 250~GeV. For diphotons this dependence is at leading order and is of the form of $(1-P_+P_-)$. This leads to a dependence of the precision of the cross-section and, similarly, the precision of the integrated luminosity, on beam polarization and the uncertainty of beam polarization. The form for the dependence of diphotons on the beam polarization uncertainty was derived in equation~\ref{eqn-DiGamPolUnc} while the form for the dependence of SABS on the beam polarization uncertainty was derived in equation~\ref{eqn-SABSPol2}. To do these computations, we need a value for the beam polarization uncertainty. We attain a value of beam polarization uncertainty from previous work and studies on the beam polarimeters, which have an aspirational value of 0.2\%~\cite{SBoogert_2009}. We use this value with the aforementioned equations~\ref{eqn-DiGamPolUnc} and~\ref{eqn-SABSPol2} and compute that the precision on integrated luminosity for diphotons and beam polarization uncertainty is $\approx2\times10^{-3}$ while the  precision on integrated luminosity for SABS and beam polarization uncertainty is $\approx4\times10^{-6}$. While this is essentially negligible for the SABS integrated luminosity precision for current experiments, it is significant for the diphotons. 

\begin{figure}[h]
\centering
\includegraphics[width=14cm]{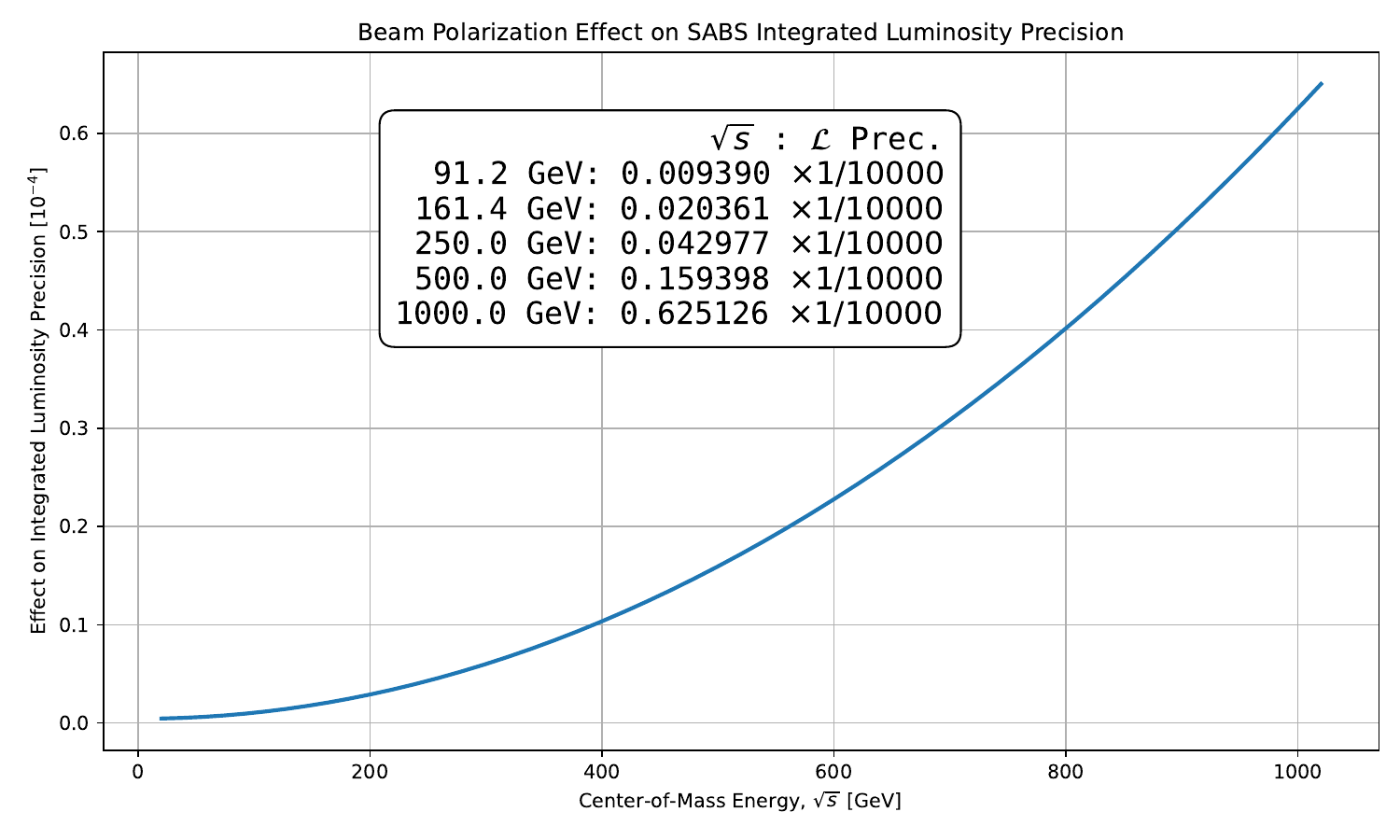}
\caption{Plot of the forecast of the precision of integrated luminosity from the beam polarization dependence of SABS. We refer to this as a forecast because it is derived from fitting existing trends in the double-spin and single-spin asymmetries and forecasting them to other energies. Precision is quoted in units of $10^{-4}$.}
\label{fig-SABSLumiPol}       
\end{figure}

By fitting for the trend in the double-spin and single-spin asymmetries in figure~\ref{fig-DoubleAsym} and~\ref{fig-SingleAsym} and then using equation~\ref{eqn-SABSPol3}, we can extrapolate out the SABS integrated luminosity precision to other center-of-mass energies, as seen in figure~\ref{fig-SABSLumiPol}.
From figure~\ref{fig-SABSLumiPol} we see that, even at higher center-of-mass energies, the effect of beam polarization on the SABS integrated luminosity precision is not significant for the current precision goals. From this result we will assume that any additional polarization dependence that diphotons may accrue from higher order corrections is likely much smaller than the leading order dependence. Therefore, we will not include any corrections to the diphoton polarization dependence or change its effect on integrated luminosity precision for different center-of-mass energies.

Given the current state of polarization uncertainty, we recommend that future $\ee$ colliders determine a way to improve the beam polarization uncertainty. Especially considering that, if this is improved, it ensures the feasibility of a $10^{-4}$ goal on the precision of diphoton integrated luminosity. Even in the case of unpolarized beams, the lack of polarization needs to be controlled and known; the polarization dependence cannot simply be disregarded for an experiment designed to be unpolarized. For this reason, we recommend that unpolarized experiments use beam polarimeters or similar technology to ensure that no outstanding source of uncertainty from polarization occurs. In the hypothetical case of a 1\% drift on the control of polarization in an unpolarized collider, assuming that it is controlled at a similar 1\% level, would have an effect on the precision of the diphoton integrated luminosity of $1.4\times10^{-4}$. Thus, the need to improve beam polarization uncertainty is not unique to polarized beam colliders.

We now propose three methods to further develop beam polarization measurement. The first is to use polarization-dependent particle physics processes, like WW production, to measure beam polarization. A review of the literature will show that this has already been proposed several times in different investigations of beam polarization~\cite{BaBar:2023upu}~\cite{karl2017polarimetryilc}~\cite{GWWTalk}. The gist of this method is to use physics processes with well-known beam polarization dependence to reconstruct the beam polarization. This has been proposed for tau pair production, as tau polarization can be reconstructed well, W pair production due to its well-known polarization dependence, and single boson production due to their large statistics and well-known polarization dependence. Previous work has shown that a fit of the WW threshold with a polarized $\ee$ collider can reconstruct the polarization of the beams with an accuracy and precision of factors of $10^{-4}$ for a full 500~$\invfb$ WW threshold dataset~\cite{GrahamWW}~\cite{LCVision}. Similarly, fits at other energies for the single boson production can achieve, in a best-case scenario, factors of $10^{-4}$ and worst-case scenario similar to $10^{-3}$. This puts these proposed particle physics methods at levels of polarization uncertainty that are comparable to or even an order of magnitude better than what the beam polarimeters currently achieve~\cite{SBoogert_2009}. The downside to using particle physics processes for beam polarization measurement is that it turns precision measurement processes into calibration processes. Previous work has gotten around this by doing simultaneous fits to beam polarization and the physically interesting measurements~\cite{GrahamWW}.

Our second proposal is mainly applicable to using \gls{diphoton}s as an integrated luminosity process, though, as we will show, it also helps the uncertainty in the \Gls{SABS} integrated luminosity measurement. The leading order polarization dependence of diphotons, from equation~\ref{eqn-DigamXsecPol}, is additively inverse when one flips only one beam polarization. Such that averaging an equal statistics set of $(P_-,P_+)=(+80\%,-30\%)$ with $(P_-,P_+)=(-80\%,-30\%)$ would yield no leading order dependence on polarization since $1+P_-P_+ + 1 - P_-P_+ = 2$. Therefore, to minimize the diphoton integrated luminosity uncertainty from beam polarization, one could average two data sets that have one polarization flip difference between them. We dub this method as polarization flip luminosity (PFL) method. Then, assuming we can control the correlation of beam polarization across runs to $\delta_\text{P}$
\begin{equation}\label{eqn-polsum}
    \frac{\delta\sigma}{\sigma_0} = \sqrt{2}\delta_\text{P}|P_-P_+|
\end{equation}
the polarization dependence of the precision of the diphoton cross-section, at leading order, is dependent only on the control of the covariance of the beams. If they vary independently then $\delta_\text{P}\to0$ and diphoton dependence on beam polarization will come from higher-order corrections. As part of precision Z boson measurements, SLC performed a similar investigation and found $\delta_\text{P}\approx0.2\%$, which is non-zero but small~\cite{SLD:2000leq}. For ILC previous studies require that this value be $<0.1\%$, and with proper methodology it can achieve this level of precision, so we will use a value of $5\times10^{-4}$ for the PFL method for diphotons~\cite{karl2017polarimetryilc}.

\begin{figure}[h]
\centering
\includegraphics[width=14cm]{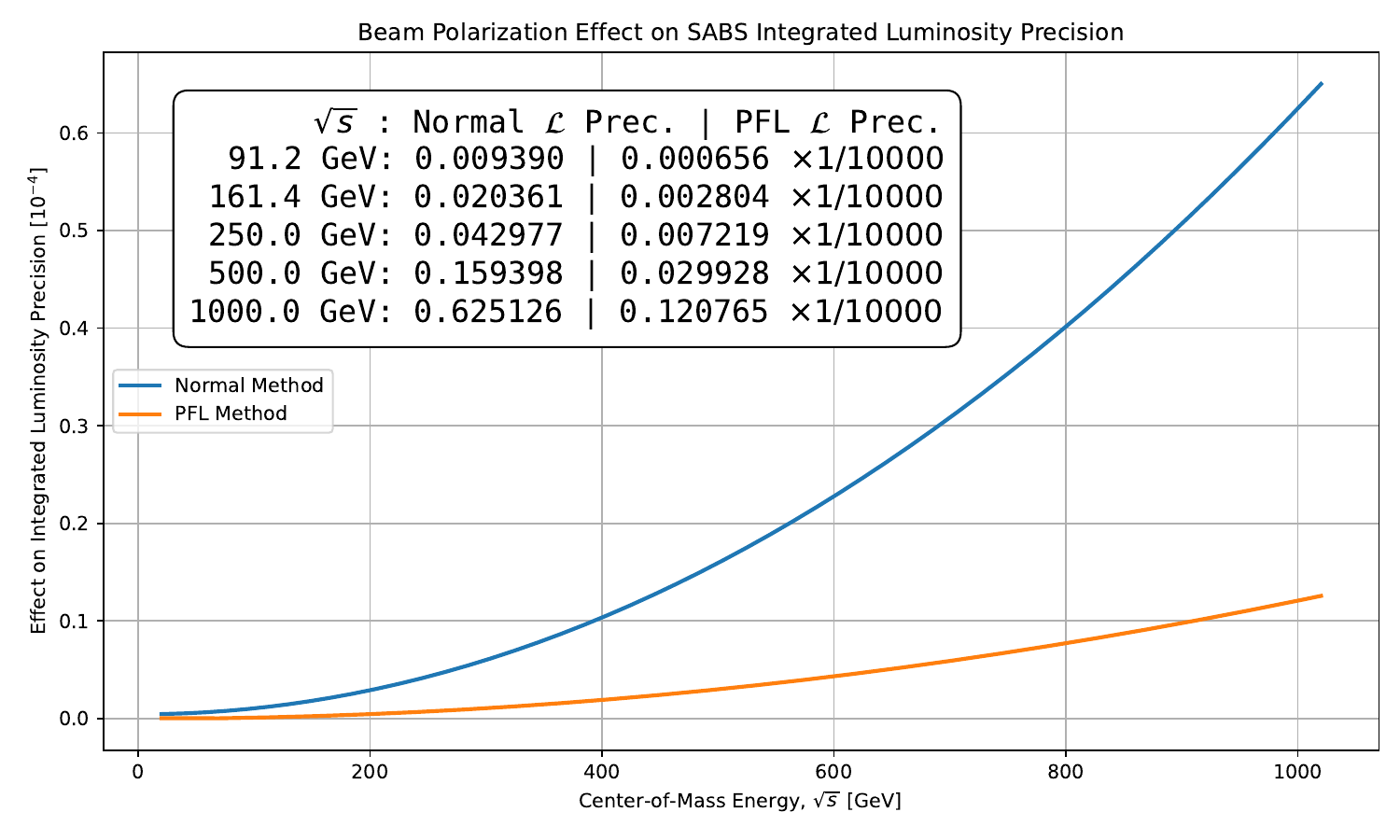}
\caption{Plot of the forecast of the precision of integrated luminosity from the beam polarization dependence of SABS. We refer to this as a forecast because it is derived from fitting existing trends in the double-spin and single-spin asymmetries and forecasting them to other energies. We also include the uncertainty if one uses the PFL method, wherein one sums two runs, one which has a flip in the beam polarization with respect to the other. Precision is quoted in units of $10^{-4}$.}
\label{fig-SABSLumiPol2}       
\end{figure}

Using the sum of single beam polarization flip runs benefits \Gls{SABS} as an integrated luminosity process too, but from the higher-order double-spin and single-spin asymmetry terms. Starting from equation~\ref{eqn-SABSPol3} we can re-asses the polarization dependence when summing single beam polarization flip runs. We observe that, for flipping the electron beam polarization,
\begin{equation}\label{eqn-SABSPolFlip}
    \sigma = \sigma_0(1-2P_+f_\text{S})
\end{equation}
the cross-section then depends only on the positron beam polarization and the single-spin asymmetry. The extrapolated cross-section precision from equation~\ref{eqn-SABSPolFlip}
\begin{equation}\label{eqn-SABSPolFlip2}
    \frac{\delta\sigma}{\sigma_0} = \sqrt{2}(\delta P)f_\text{S}
\end{equation}
shows that we have managed to remove all but one of the terms of equation~\ref{eqn-SABSPol2}, thereby reducing the uncertainty. We use this to update figure~\ref{fig-SABSLumiPol2} to what can be seen in figure~\ref{fig-SABSLumiPol2}.

The PFL method requires taking more data at the same-sign helicity combinations than is currently allotted in planned runs~\cite{LCVision}. Considering that this would maintain the superior sensitivities of polarized beam experiments, we argue that taking more data on the same-sign helicity combinations is worthwhile for performing a precise measurement of integrated luminosity with diphotons.

The third method we propose is more of a proposal to investigate possible technology upgrades. In particular, we propose the investigation of newer optical diagnostics of plasmas. The existing methodology, to use Compton polarimeters, is decades old and, over this same time, there have been numerous advances in optical and plasma physics. These advances have allowed for revolutionary measurements in chemistry and condensed matter physics. We are confident that investigating these methods within the context of electron and positron beams could discover new, and more precise, methods of reconstructing polarization.

\section{Detector Uncertainties}

In this section we will cover the sources of uncertainty on the integrated luminosity that are derivative of the detector, specifically the LumiCal design being tested. We will also restrict this section to the case where the larger experiment is running in a triggerless configuration. Furthermore, we will assume that there is no uncertainty from the trigger, or lack thereof, or uncertainty from read-out electronics. Even during the LEP era, which did use a trigger, the uncertainty from the trigger was $\approx1\times10^{-5}$, so it was not a leading source of uncertainty in the integrated luminosity precision~\cite{OPALLumi}. We will also not consider transient effects, like structural degradation or radiation damage, nor will we consider thermal effects.

\subsection{Position Resolution and Bias}

\begin{figure}[h]
\centering
\includegraphics[width=12cm]{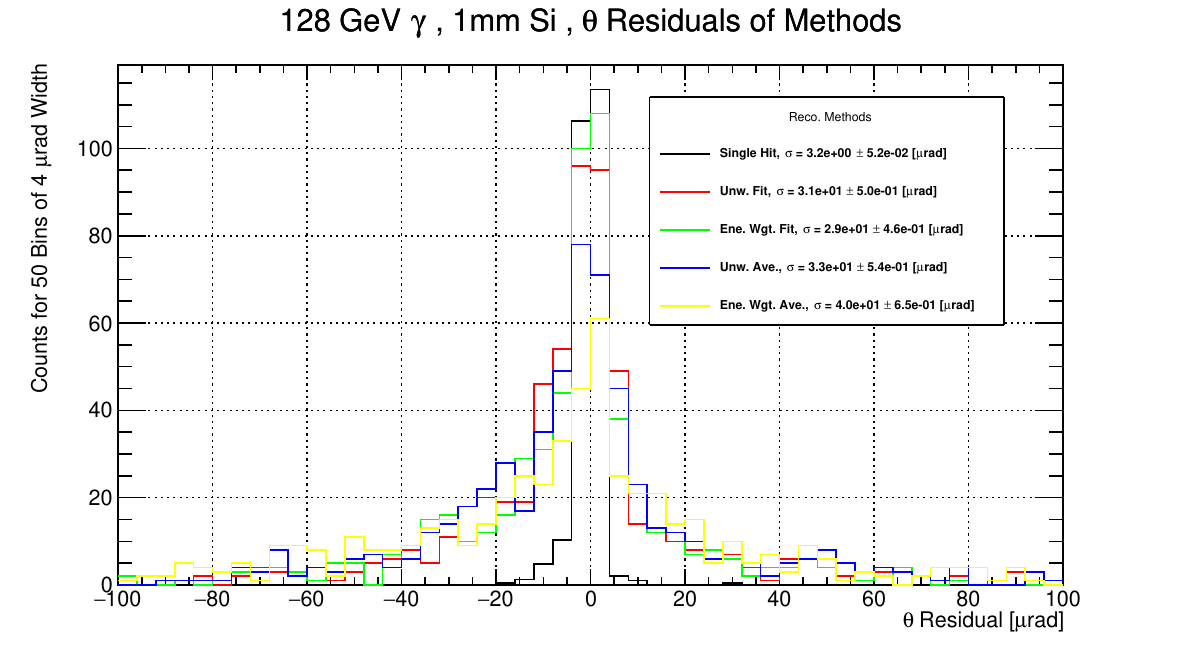}
\caption{Plot of the polar angle residual for different reconstruction (reco.) methods, as outlined in section~\ref{sec-PosRes}, for the standard GLIP LumiCal design. We find no statistically significant bias. This was done using 128~GeV photons at an incident polar angle of $2^\circ$. The reconstruction methods tested, omitting the single-hit method, are similar to the ISE and IS methods outlined in section~\ref{sec-PosRes}. There are two versions of each used, one where a fit of the hit positions with a 2-dimensional gaussian was used and one where the average of the hit positions were used.}
\label{fig-BiasMeasure}       
\end{figure}

\begin{figure}[h]
\centering
\includegraphics[width=12cm]{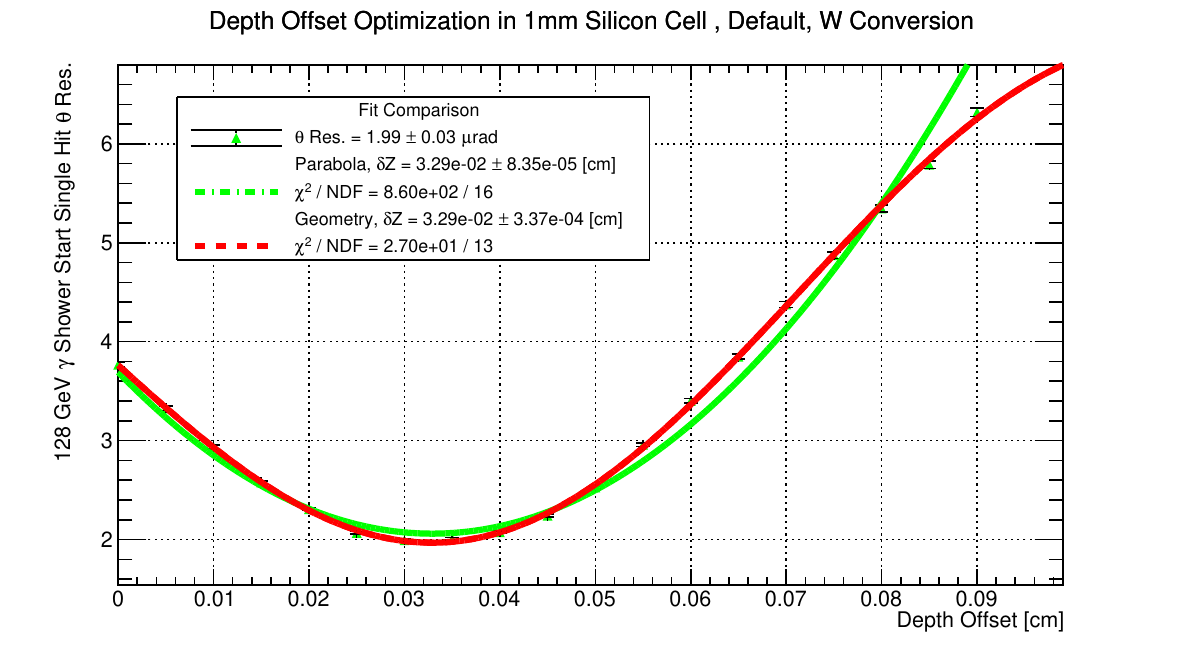}
\caption{Plot of the polar angle resolution, calculated from the polar angle residual distribution, along the y-axis, as the offset along the z-axis position value used for polar angle calculation is changed. This was done using the single-hit reconstruction method and for the standard GLIP LumiCal design where the photon converts in the passive tungsten layer, with 128~GeV photons at an incident polar angle of $2^\circ$.}
\label{fig-WConv}       
\end{figure}

\begin{figure}[h]
\centering
\includegraphics[width=12cm]{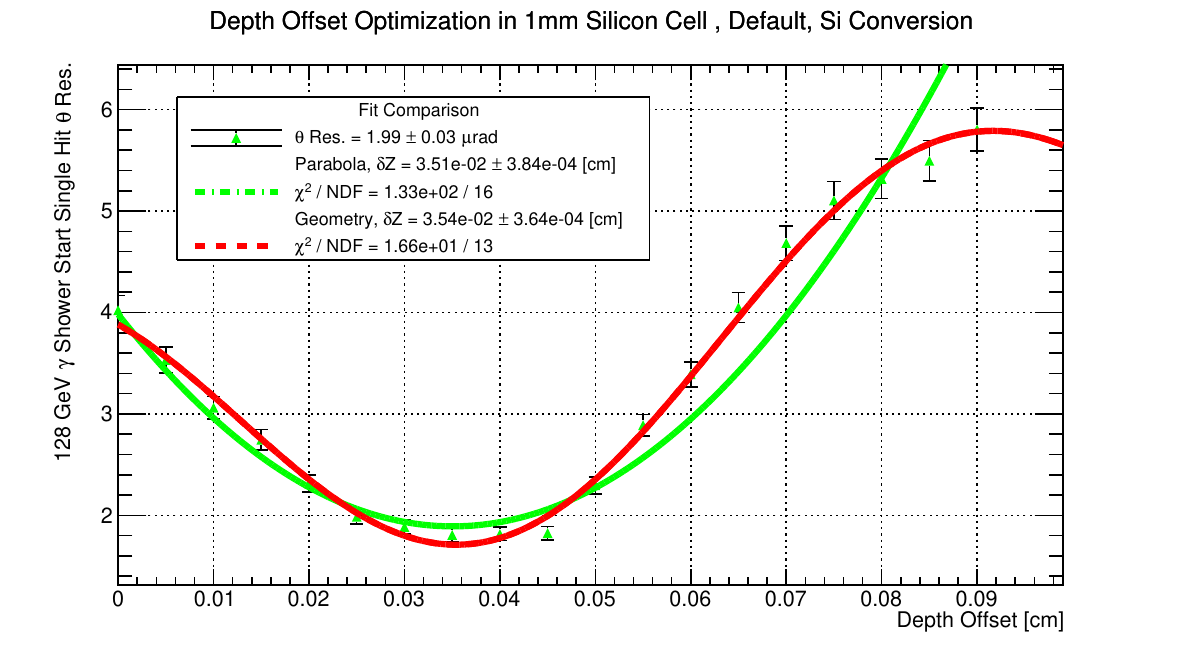}
\caption{Plot of the polar angle resolution, calculated from the polar angle residual distribution, along the y-axis, as the offset along the z-axis position value used for polar angle calculation is changed. This was done using the single-hit reconstruction method and for the standard GLIP LumiCal design where the photon converts in the active silicon layer, with 128~GeV photons at an incident polar angle of $2^\circ$.}
\label{fig-SiConv}       
\end{figure}

Previous studies of position bias have found that integrated luminosity precision depends on
\begin{equation}\label{eqn-lumpos}
    \frac{\Delta \mathcal{L}}{\mathcal{L}} \approx \frac{2\Delta\theta}{\tinn}
\end{equation}
the inner acceptance angle and the bias of the estimate of the polar angle~\cite{Abramowicz_2010}. This is independent of the physics process being used in integrated luminosity. For the current ILD LumiCal design the polar angle bias is found to be roughly 3.2~$\mu$rad~\cite{Abramowicz_2010}. For the GLIP LumiCal design we observed that the polar angle estimate did not have any observable bias, as seen in figure~\ref{fig-BiasMeasure}, where multiple methods of reconstruction were used.
From these we particularly focus on the single-hit reconstruction method and measure a bias of $\approx-0.05\pm0.05$~$\mu$rad, which is statistically insignificant. During follow-up testing we discovered a type of depth effect. The depth effect is when a particular value of depth, z-axis position, that is chosen for calculating the polar angle differs from the true value. In particular we examined this using the `single hit' reconstruction method discussed in section~\ref{sec-PosRes} and found that, as seen in figure~\ref{fig-SiConv} and figure~\ref{fig-WConv}, the depth bias is dependent on the particular material with which the particle shower starts in.
For this depth effect we fit for the depth offset, $\delta Z$, for the minimum polar angle residual using two different models. A parabolic model, which, as the name implies, used a parabola to fit. A `geometry model', which uses a standing wave, $\cos()$, function to fit. We find that both models are not satisfactory but that the geometry model is preferred. 

This depth effect is particularly problematic for diphoton measurements as the shower does not begin at the surface of a material, like it can with a charged particle. Instead a photon must undergo photo-conversion and produce charged particles. It can do this in any of the materials of the calorimeter, be it the active silicon layer, the passive tungsten layer, or the passive structural elements. We find that the difference in bias between photo-conversion in the Tungsten and photo-conversion in the Silicon is $23\pm5$~microns. This means that, even with a calibration run to characterize this depth effect, the correction factor would still be of order of 10~microns off without the knowledge of which layer the photon converted in. Further studies of this, extended to all angles from 0.1~mrad to 0.02~mrad, found that the depth effect varies by roughly 10~microns from the incident polar angle of the photon. It may be possible to use the information of a highly granular calorimeter to infer which material the photon converted in, alongside the polar angle measurement, to improve on this bias, but for now we will assume that the depth effect results in 25~micron bias. For the depth of the GLIP LumiCal this results in a polar angle bias of $\approx-0.3$~$\mu$rad and an effect on the precision of the integrated luminosity of $3.4\times10^{-5}$. For comparison, the current ILD LumiCal reports a value of 3.2~$\mu$rad in polar angle bias and a resulting effect on the precision of the integrated luminosity of $2.6\times10^{-4}$.

For the polar angle resolution similar effects arise but instead from how well one can accept or cut out a particular event. The form of this therefore depends on the physics process being measured. We previously derived, in chapter~\ref{ch-Theory}, the equations for this dependence of the cross-section of SABS on the inner angle and polar angle resolution, as in equation~\ref{eqn-SABS_Inner_Unc}, as well as for the cross-section of diphotons, as in equation~\ref{eqn-DiGamTinn}. From these, and using values for the current ILD LumiCal design from section~\ref{sec-Existing} and values for the GLIP LumiCal from section~\ref{sec-PosRes} we can calculate the effect on the precision of the integrated luminosity. For SABS we calculate that the effect on integrated luminosity is $1\times10^{-4}$ for the current ILD LumiCal and $3\times10^{-4}$ for the GLIP LumiCal. With the significant difference being from the shallower acceptance that the GLIP LumiCal has. This can be adjusted by only accepting SABS for the GLIP LumiCal at the same acceptance, at which point the effect on the precision of the integrated luminosity would be $5\times10^{-5}$. For diphotons both designs see a similar effect on the precision of integrated luminosity of $1\times10^{-5}$.

\subsection{Energy Resolution Bias and Energy Scale Effects}\label{sec-EneLumi}

\begin{figure}[h]
\centering
\includegraphics[width=12cm]{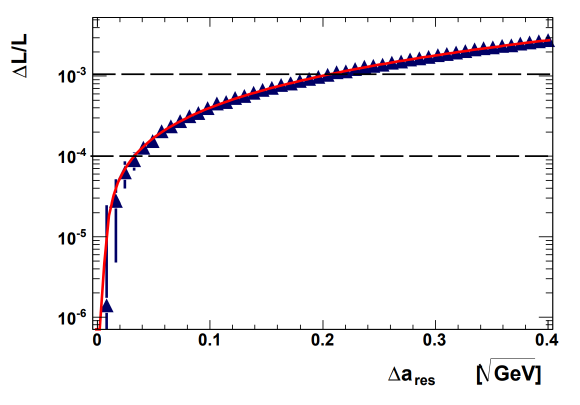}
\caption{Plot of the effect on integrated luminosity precision from energy resolution bias of a given calorimeter design. Figure credit~\cite{Abramowicz_2010}.}
\label{fig-EneBias}       
\end{figure}

Previous studies of energy resolution bias have found that integrated luminosity precision depends on the bias in energy resolution in a way seen in figure~\ref{fig-EneBias}~\cite{Abramowicz_2010}.
For the current ILD LumiCal the plan is to control the bias in energy resolution to 15\% of the energy resolution of $21\%/\sqrt{E}$. This results in an effect on integrated luminosity of $\approx8\times10^{-5}$. For the GLIP LumiCal, with an energy resolution of $4\%/\sqrt{E}$ the energy resolution bias at the same level of control of 15\% would be $\approx1\times10^{-6}$. For the purposes of the runs planned here, this is negligible, but we will still report it in the final sum. This effect is not expected to change at higher or lower energies as the energy resolution already incorporates energy scaling.

The energy scale calibration of the luminosity calorimeter also effects the integrated luminosity precision as the precision and bias of the energy cuts used to accept and reject integrated luminosity events is affected by the energy scale calibration. This bias in the energy scale calibration is not the primary issue as it can be measured and corrected for. The issue arises from how well the bias is known. For example, if the bias is only known to 10\% then the resulting energy cut can only be known to 10\% and some fraction of that 10\%, depending on the shape of the luminosity spectrum where the energy cut is made, will contribute as a systematic uncertainty on integrated luminosity. Using the tracker dimuons as the calibration source and the generator level data, that is to say no detector effects, we can evaluate the bias and resolution on the bias of the energy scale calibration. 

\begin{figure}[h]
\centering
\includegraphics[width=12cm]{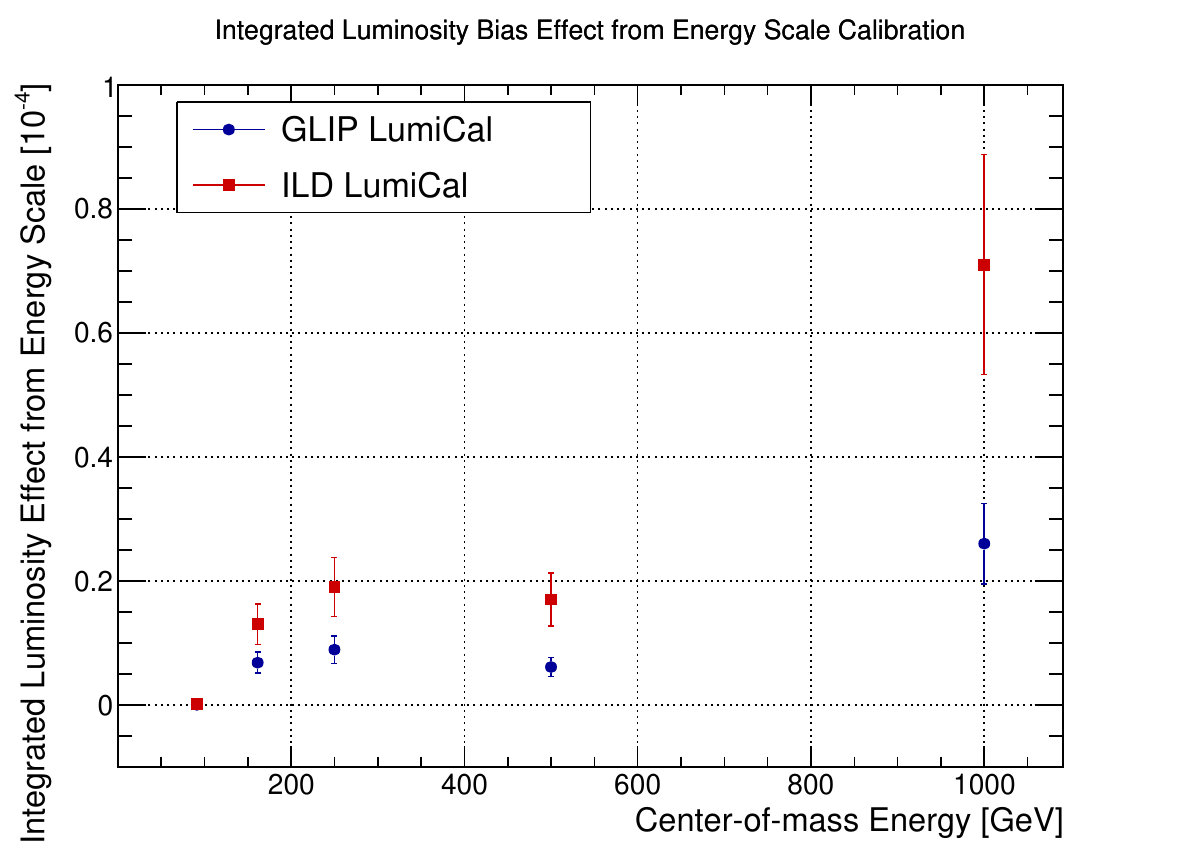}
\caption{Plot of the energy scale bias effect on the integrated luminosity at ILC for the current ILD LumiCal and the proposed GLIP LumiCal. This does not include corrections for the luminosity spectrum effects.}
\label{fig-EneBiasLumi}       
\end{figure}

These resolution and bias values are taken from the measurements and fits done in chapter~\ref{ch-EPre}. In general, we find that the bias and resolution is comparable to 5~MeV for all center-of-mass energies and both the GLIP LumiCal and ILD LumiCal designs. We also find that, in general, the bias is negative and resultingly the sign of the bias in integrated luminosity for this effect is positive. Using these values we then conduct a Monte Carlo simulation which uses a varying threshold cut of the 80\% on calorimeter hemisphere energy and total energy cuts. Each detector design and center-of-mass energy is tested in the Monte Carlo using the \Gls{MCEG} data that has been smeared with its respective detector effects. The bias in integrated luminosity that this creates is evaluated by evaluating the difference for a Monte Carlo run that is done with an idealized threshold cut that has no bias or resolution effects. The results find that the integrated luminosity bias is always positive and less than $10^{-4}$ here, as can be seen in figure~\ref{fig-EneBiasLumi}.

For comparison, the current ILD LumiCal operating at ILC250 the energy scale calibration is envisioned to see a 400~MeV calibration as a worst-case scenario~\cite{Abramowicz_2010}. We find that using the dimuon center-of-mass energy calibration method discussed in this work will, in general, do better than 400~MeV. Using this method with the GLIP LumiCal does especially well and indicates that a $10^{-4}$ goal on integrated luminosity precision is reasonable. We also find that, at the Z-pole, the uncertainty for both LumiCal designs is incredibly small, $<10^{-7}$. We suspect that changes in the luminosity spectrum as center-of-mass energy increases considerably influence the precision seen here. Correcting for luminosity spectrum effects and improving this precision further may be possible but would require developing knowledge and measurements of the luminosity spectrum and would be outside the scope of this work.

\subsection{Separating Integrated Luminosity Processes and their Backgrounds}\label{sec-LumiPID}

For the purpose of measuring an integrated luminosity process, one must be able to differentiate it from other integrated luminosity processes and from background processes. Any amount of contamination, the accidental inclusion of incorrect events, as well as the loss of events from inefficient tagging, both influence the integrated luminosity precision. As outlined in chapter~\ref{ch-LumiProp}, we can use equation~\ref{eqn-IntLumi} to describe how tagging of a process influences the integrated luminosity precision. In particular, there is a linear dependence, so any tagging and background rejection precision applies directly to the integrated luminosity precision.

Of particular interest is the tagging of our two integrated luminosity processes, SABS and diphotons, as well as the rejection of the background processes. The background processes that are relevant are four-fermion production. Even though the BERB process has a large cross-section, of tens to hundreds of pb, the kinematic constraints lead to a negligible background when one requires that two photons, both of which are above 80\% of the beam energy, are reconstructed. For more information on the kinematics of BERB, see section~\ref{sec-BERB}. 

For an $\ee$ collider the dominant four-fermion channels are leptons that are in association with $\ee$, as the t-channel Bhabha-like process still dominates at small angles. At LEP, the Monte Carlo programs FermisV and KoralW were used to simulate four-fermion production and estimate the effects of four-fermion production on precision measurements and luminosity~\cite{fermisV}~\cite{koralw}. It was found then, and we have reproduced this result here, that at small angles the Bhabha-like process dominates. One observes an $\ee$ pair that is mostly back-to-back and carrying most of the energy of the event, while the other fermion-antifermion pair is soft, less than 10\% of the center-of-mass energy. In this case, the four-fermion production is essentially a soft radiative correction to Bhabha scattering. For this reason some collaborators for LEP luminosity measurements treated this case of four-fermion production as a radiative correction to Bhabha scattering. We propose that future $\ee$ colliders do the same as, we will soon show, four-fermion production limits the precision of SABS as an integrated luminosity channel. Combining the two significantly improves the uncertainty from event separation for measuring integrated luminosity using SABS.

Since four-fermion production can look similar to SABS, cuts in energy and acollinearity are used to refine the SABS selection to be less sensitive to four-fermion production. Here we continue to use the standard 80\% of beam energy cut on luminosity calorimeter hemisphere energy as well as a cut on the total luminosity calorimeter energy of 80\% of the center-of-mass energy. We then used WHIZARD for diphoton and four-fermion simulation and BHLUMI for SABS simulation to generate 100k events of each type. For WHIZARD generation we included ISR corrections and one radiative photon correction but no beam or polarization effects. We applied the previously mentioned cuts and then used C5 to generate a \Gls{BDT} to differentiate each process~\cite{C5}. The measurements given to C5 were the invariant mass, acollinearity, acoplanarity and the energy and angles of both particles. From this the BDT was given ten boosting trials. It was also cross-validated on an equal statistics data sample to ensure validity. The resulting confusion matrices, which determine how each event type is identified for each classification, can be found in appendix~\ref{app-conf-rawff}. We find that the BDT, at all center-of-mass energies, is able to identify each luminosity process at around 80\% purity, meaning that 20\% of the final sample are other processes that have been incorrectly identified and included.

Combining this with the ability for the luminosity calorimeter to differentiate particle species gives two levels of classification for signal differentiation, the BDT of calorimeter data used in section~\ref{sec-PID}, and now a BDT of reconstruction level data to differentiate physics processes. Using both of these in consort, and weighting for the cross-section of each process, we have created the weighted confusion matrices in appendix~\ref{app-conf-glip} and appendix~\ref{app-conf-lumical} which are for the GLIP LumiCal and ILD LumiCal respectively. We also observe that, since the $\ee$ pair of the four-fermion events is what is typically measured, the calorimeter-level PID does not contribute much to the differentiation of SABS and four-fermion events as both measure an $\ee$ pair. From these values we observe that the largest uncertainty for SABS, for both luminosity calorimeter designs, is from separating SABS and four-fermion events. For the GLIP LumiCal the contamination of four-fermions in the SABS sample was $\approx1.2\times10^{-4}$ at each center-of-mass energy. For the ILD LumiCal the contamination of four-fermions in the SABS sample was slightly larger at $\approx1.7\times10^{-4}$. The largest difference in performance between the two designs is the level of contamination of SABS events in the diphoton sample. For the ILD LumiCal the contamination of SABS in the diphoton sample is $\approx40\times10^{-4}$ while, for the GLIP LumiCal, the contamination is $\approx0.1\times10^{-4}$. This result indicates that existing, not high-granularity, luminosity calorimeter designs are not able to sufficiently separate SABS and diphotons to make diphotons a feasible process for precision integrated luminosity measurement.

\begin{figure}[h]
\centering
\includegraphics[width=14cm]{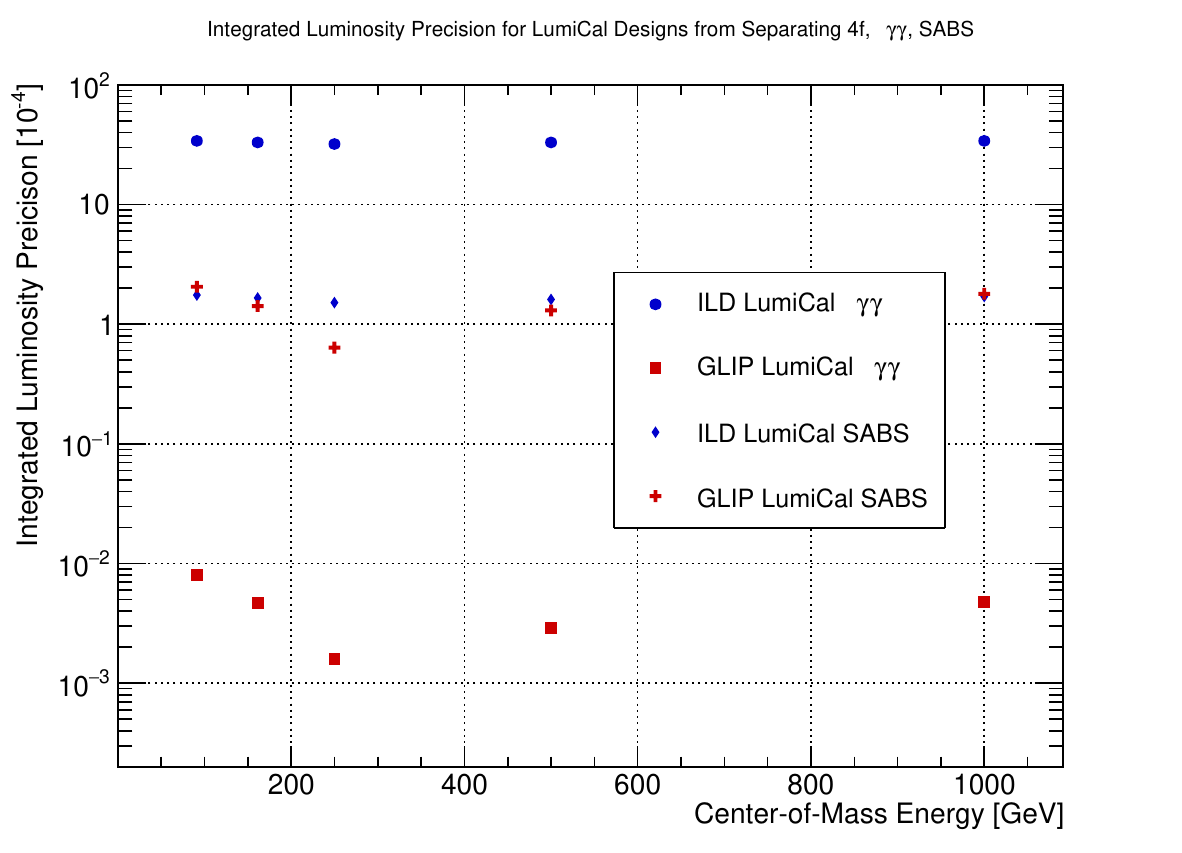}
\caption{Plot of the systematic uncertainty from physics process tagging and separation of the SABS, diphoton and four-fermion processes for the two different LumiCal designs using the calorimeter level and reconstruction level BDTs to do tagging.}
\label{fig-PIDPrec}       
\end{figure}

By comparison, the existing estimations for the current ILD LumiCal, puts SABS separation from four-fermion at factors of $10^{-3}$~\cite{Abramowicz_2010}. This work has fleshed-out the estimation and has shown that it can be improved to $\approx1.7\times10^{-4}$. However, it has also shown that the existing ILD LumiCal is not suitable for using diphotons to make precision measurements of integrated luminosity. Upgrading to a high-granularity design is a requirement for using diphotons as a precision integrated luminosity measurement. This means that any future $\ee$ collider will need to update their luminosity calorimeter designs and, potentially, their accelerator and forward region designs, if they wish to use diphotons as a precision integrated luminosity channel. We have compiled the results of these values for the center-of-mass energies tested in this work and for the GLIP LumiCal and current LumiCal designs, as well as for diphotons and SABS, in figure~\ref{fig-PIDPrec}.

\begin{figure}[h]
\centering
\includegraphics[width=14cm]{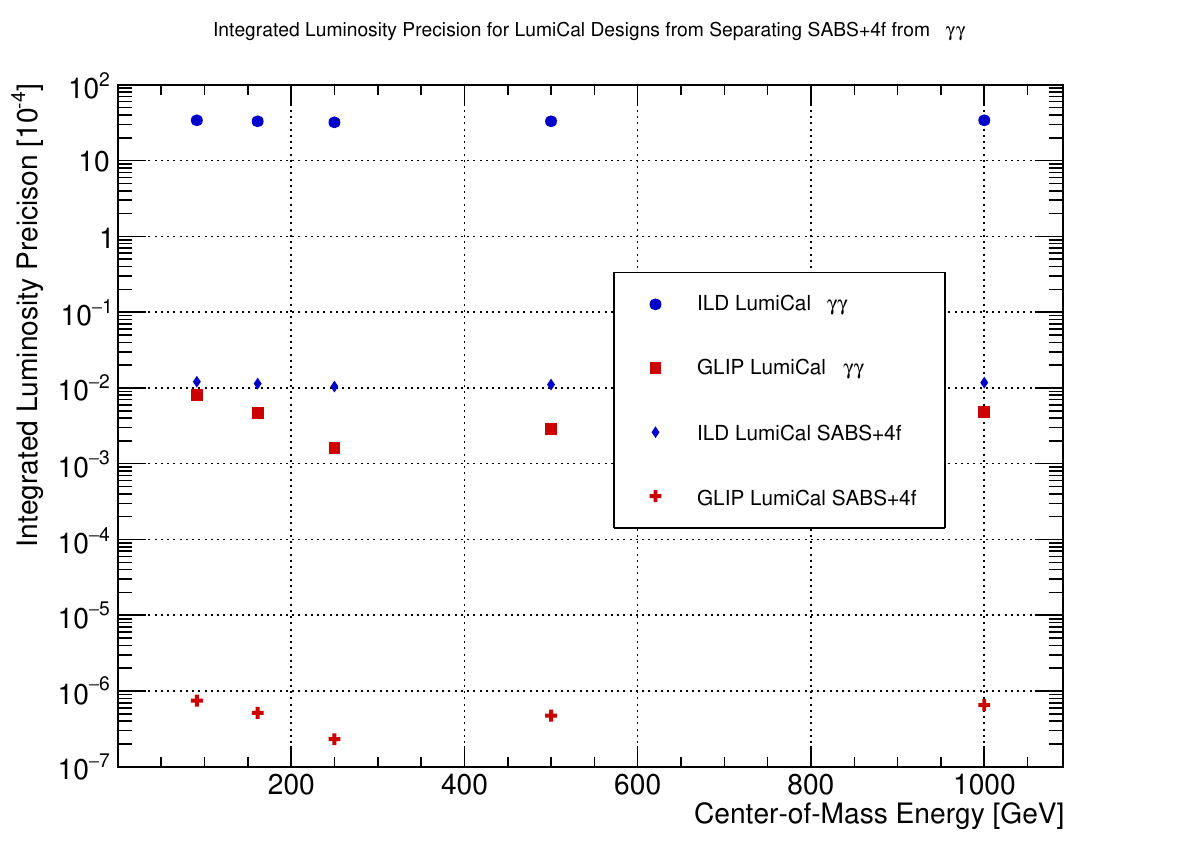}
\caption{Plot of the systematic uncertainty from physics process tagging and separation of the SABS+4f and diphoton processes for the two different LumiCal designs using the calorimeter level and reconstruction level BDTs to do tagging.}
\label{fig-PIDPrecNo4f}       
\end{figure}

We also include a scenario where the Bhabha-like four-fermion production is included as a part of SABS, dubbed SABS+4f. As seen in figure~\ref{fig-PIDPrecNo4f}, the performance for SABS increases significantly such that the effect on integrated luminosity precision from contamination in the SABS data sample is completely negligible. 
We suspect that there are additional physics processes, such as two-fermion processes, that could introduce contamination that would need to be considered to get a more realistic estimate. Still, this highlights the strength of the proposal to combine Bhabha-like four-fermions with SABS.

\section{Results of Integrated Luminosity Precision}\label{sec-LumiTabs}

\begin{figure}[h]
\centering
\includegraphics[width=16cm]{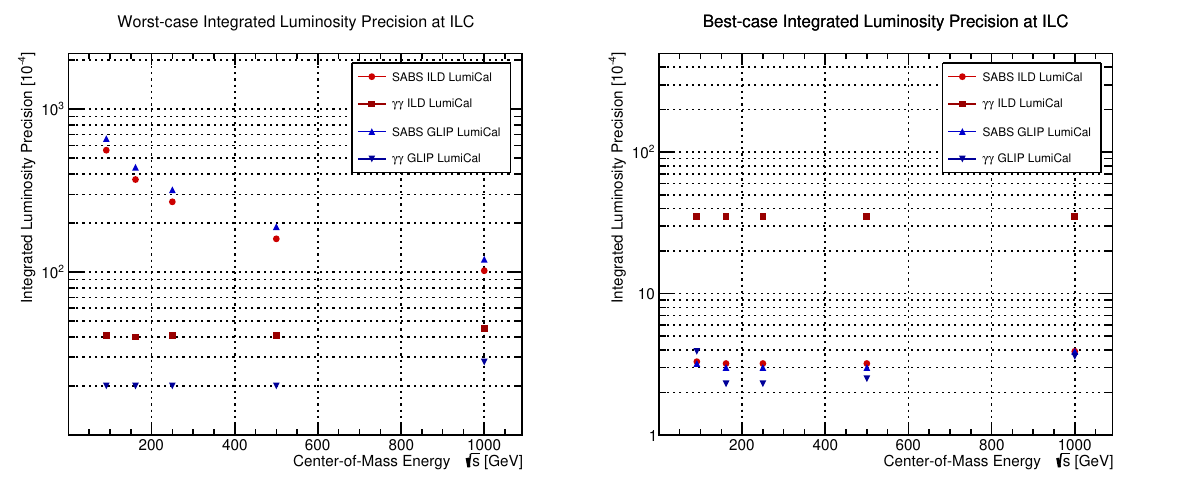}
\caption{(Left) Plot of the worst-case scenario for integrated luminosity precision at ILC for various center-of-mass energies as outlined by the tables of section~\ref{sec-LumiTabs}. These are dominated by the electromagnetic beam deflection effect. (Right) Plot of the best-case scenario for integrated luminosity precision at ILC for various center-of-mass energies as outlined by the tables of section~\ref{sec-LumiTabs}. We provide results for the current ILD LumiCal, using the improved methods outlined in this work, alongside the GLIP LumiCal. We provide results for SABS and diphotons.}
\label{fig-BestWorse}       
\end{figure}

\begin{figure}[h]
\centering
\includegraphics[width=16cm]{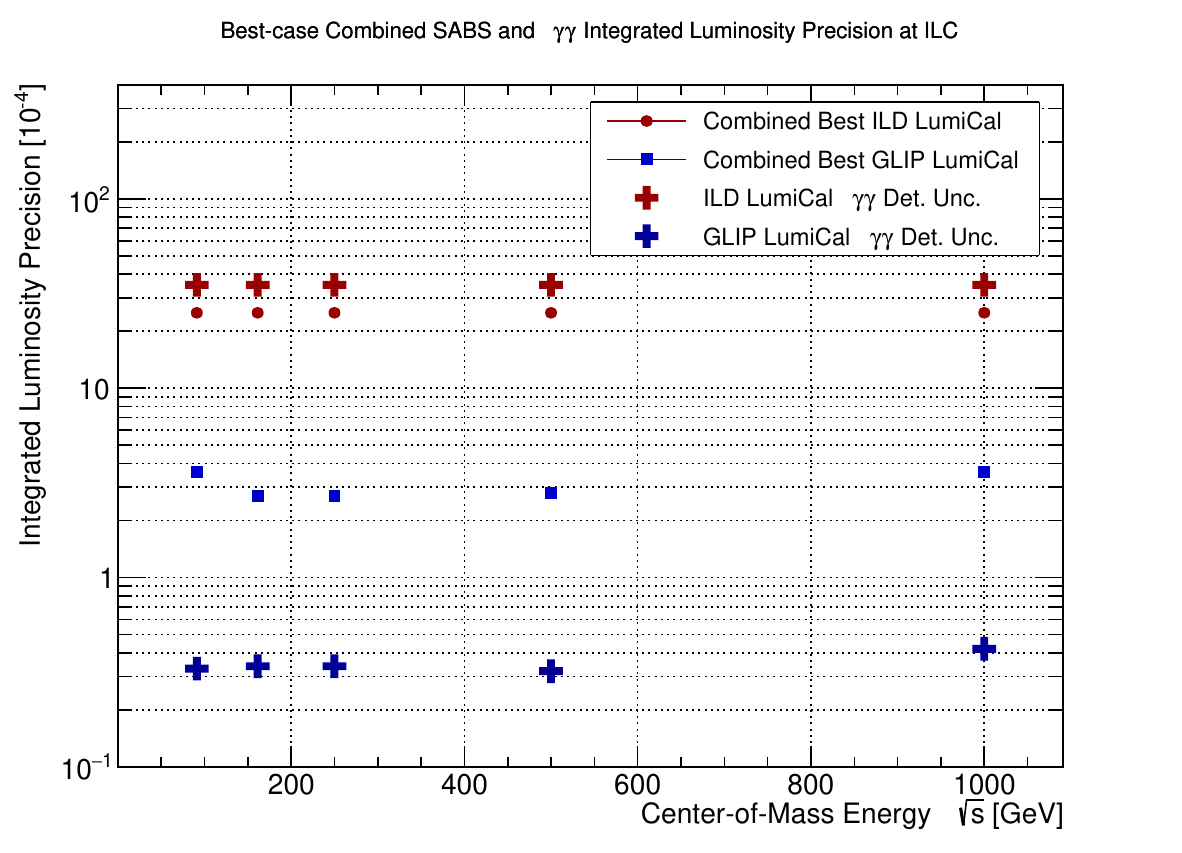}
\caption{Plot of the integrated luminosity precision from the combination of the best-case scenarios of SABS and diphotons for the two different LumiCal designs of the current ILD LumiCal and the GLIP LumiCal. We include cross marks for the $\gamma\gamma$ detector uncertainties of to show how they compare to the combined best case and illustrate how much each measurement could be improved by if the non-detector uncertainties were made negligible.}
\label{fig-CombinedLumi}       
\end{figure}

This section includes the tables of luminosity precision for each center-of-mass energy, tables~\ref{tab-LumiTable250}-~\ref{tab-LumiTable1000}. We also include two final plots. In figure~\ref{fig-BestWorse}, we provide the results given the worst-case and best-case scenarios for each LumiCal design; in figure~\ref{fig-CombinedLumi} we provide the results of the combination of the best-case scenarios of SABS and diphotons for each LumiCal design. The values used in this section are sourced from this chapter. We find that the main contribution to integrated luminosity precision is from non-detector effects. The electromagnetic beam deflection effect, if left uncorrected, is dominant and prevents achieving integrated luminosity precision below $\approx1$\%. For reaching precision on integrated luminosity of $10^{-3}$, and factors of $10^{-4}$, the current ILD LumiCal should be sufficient but only with improvements in the methodology of energy scale calibration and particle identification as outlined in section~\ref{sec-EneLumi} and section~\ref{sec-LumiPID}. This should be expected as the ILD LumiCal design is similar to OPAL's LumiCal, which achieved factors of $10^{-4}$ in integrated luminosity precision using \Gls{SABS}~\cite{OPALLumi}. To achieve precision on integrated luminosity of $10^{-4}$, and approach $10^{-5}$, the results indicate that it is a requirement to use an improved LumiCal design such as the GLIP LumiCal. With a design like the GLIP LumiCal we find that $10^{-4}$ precision on integrated luminosity is feasible at all center-of-mass energies at ILC from the Z-pole to 1~TeV. If the non-detector uncertainties are better controlled then a design like the GLIP LumiCal could see $4\times10^{-5}$ precision in integrated luminosity. This would expand the possible precision physics goals of future $\ee$ colliders. For this reason, we strongly recommend that an alternative LumiCal, like the GLIP LumiCal design, be developed and deployed at future $\ee$ colliders. The physics goals at the Z-pole demand, at minimum, $10^{-4}$ precision on integrated luminosity and the physics goals at other center-of-mass energies could be extended if the same level of precision on integrated luminosity was achieved at other center-of-mass energies.

\newpage

\begin{sidewaystable}[htbp]
\centering
\caption{Estimated \textbf{systematic uncertainties} on the integrated luminosity measurement using small-angle Bhabhas and diphotons at \textbf{ILC250} as measured in a luminosity calorimeter. Detector and non-detector sources of uncertainty are included but in separate sections. We provide the theory uncertainty, projected to their future expectations, in a separate section on their own. The final total has a best-case scenario and worst-case scenario quadrature sum of all contributions, not including theory. All numerical values are in parts per ten thousand ($10^{-4}$).\\ \footnotesize{*ILD LumiCal uses a different energy scale method than the GLIP LumiCal and has worse performance than the $\sqrt{s}$ method that the GLIP LumiCal uses so we quote the ILD LumiCal value for energy scale~\cite{Abramowicz_2010}.}}
\begin{tabular}{|l|c|c|c|c|l|}
\hline
\textbf{Non-Detector Uncertainties} & \textbf{LumiCal SABS} & \textbf{LumiCal $\gamma\gamma$} & \textbf{GLIP SABS} & \textbf{GLIP $\gamma\gamma$} & \textbf{Notes} \\
\hline
Upstream Material & 1 & 1 & 1 & 1 & Same for all~\cite{Sadeh2008} \\
Upstream Material & $\approx0$ & $\approx0$ & $\approx0$ & $\approx0$ & If corrected for, as done at LEP \\
Luminosity spectrum for 1\% & 0.1 & 0.1 & 0.1 & 0.1 & See section~\ref{sec-LumiSpec} \\
Luminosity spectrum for 0.1\% & 0.01 & 0.01 & 0.01 & 0.01 & See section~\ref{sec-LumiSpec} \\
Beam Deflection Effects & 270 & -- & 320 & -- & No M\o{}ller scattering run \\
Beam Deflection Effects & 0.13 & -- & 0.15 & -- & M\o{}ller scattering run, section~\ref{sec-Moller} \\
Polarization of beams & 0.043 & 20 & 0.043 & 20 & Using beam polarimeters \\
Polarization of beams & 0.007 & 1.7 & 0.007 & 1.7 & Using PFL method \\
Statistics (for 3 $\mathrm{ab}^{-1}$) & 0.05 & 2.5 & 0.05 & 1.5 & See section~\ref{sec-lumitheo} \\
\hline
\textbf{Worst-case Sum : } & \textbf{270} & \textbf{20.2} & \textbf{320} & \textbf{20.1} & Assuming uncorrelated \\
\textbf{Best-case Sum : } & \textbf{0.1} & \textbf{3.0} & \textbf{0.2} & \textbf{2.3} & Assuming uncorrelated \\
\hline
\textbf{Detector Uncertainties} & \textbf{LumiCal SABS} & \textbf{LumiCal $\gamma\gamma$} & \textbf{GLIP SABS} & \textbf{GLIP $\gamma\gamma$} & \textbf{Notes} \\
\hline
Angular acceptance, resolution & 1 & 0.1 & 3 & 0.1 & $\theta^{-3}$ vs. $\theta^{-1}$ dependence \\
Energy scale & 0.19 & 0.19 & 0.089 & 0.089 & Calibrate with $\sqrt{s}$ of tracker dimuons\\
Energy resolution bias & 1.6 & 1.6 & 0.1 & 0.1 & For control of 15\% \\
Position bias & 2.6 & 2.6 & 0.3 & 0.3 & For GLIP with 100 micron cells \\
PID and Contamination & 1.5 & 35 & 0.65 & 0.002 & See section~\ref{sec-LumiPID} \\
PID and Contamination & 0.01 & 35 & $\approx0$ & 0.002 & For SABS+4f method \\
\hline
\textbf{Sum : } & \textbf{3.2} & \textbf{35.1} & \textbf{3.0} & \textbf{0.34} & Using SABS+4f, uncorrelated sum \\
\hline
Theory Projection & 1.8 & 5 & 1.5 & 5 & See section~\ref{sec-lumitheo}. \\
\hline
Worst-case Total :  & 270 & 41 & 320 & 20 &  \\
Best-case Total :  & 3.2 & 35 & 3.0 & 2.3 &  \\
\hline
\textbf{Combined Best-case : } & & \textbf{25} & & \textbf{2.7} & Uncorrelated SABS and $\gamma\gamma$ sum\\
\hline
\end{tabular}
\label{tab-LumiTable250}
\end{sidewaystable}

\newpage

\begin{sidewaystable}[htbp]
\centering
\caption{Estimated \textbf{systematic uncertainties} on the integrated luminosity measurement using small-angle Bhabhas and diphotons at \textbf{ILCZ} as measured in a luminosity calorimeter. Detector and non-detector sources of uncertainty are included but in separate sections. We provide the theory uncertainty, projected to their future expectations, in a separate section on their own. The final total has a best-case scenario and worst-case scenario quadrature sum of all contributions, not including theory. All numerical values are in parts per ten thousand ($10^{-4}$).\\ \footnotesize{*ILD LumiCal uses a different energy scale method than the GLIP LumiCal and has worse performance than the $\sqrt{s}$ method that the GLIP LumiCal uses so we quote the ILD LumiCal value for energy scale~\cite{Abramowicz_2010}.}}
\begin{tabular}{|l|c|c|c|c|l|}
\hline
\textbf{Non-Detector Uncertainties} & \textbf{LumiCal SABS} & \textbf{LumiCal $\gamma\gamma$} & \textbf{GLIP SABS} & \textbf{GLIP $\gamma\gamma$} & \textbf{Notes} \\
\hline
Upstream Material & 1 & 1 & 1 & 1 & Same for all~\cite{Sadeh2008} \\
Upstream Material & $\approx0$ & $\approx0$ & $\approx0$ & $\approx0$ & If corrected for, as done at LEP \\
Luminosity spectrum for 1\% & 0.0001 & 0.0001 & 0.0001 & 0.0001 & See section~\ref{sec-LumiSpec} \\
Luminosity spectrum for 0.1\% & 0.00001 & 0.00001 & 0.00001 & 0.00001 & See section~\ref{sec-LumiSpec} \\
Beam Deflection Effects & 560 & -- & 660 & -- & No M\o{}ller scattering run \\
Beam Deflection Effects & -0.9 & -- & -1.1 & -- & M\o{}ller scattering run, section~\ref{sec-Moller} \\
Polarization of beams & 0.009 & 20 & 0.009 & 20 & Using beam polarimeters \\
Polarization of beams & 0.0007 & 1.7 & 0.0007 & 1.7 & Using PFL method \\
Statistics (for 0.1 $\mathrm{ab}^{-1}$) & 0.1 & 5 & 0.05 & 3.5 & See section~\ref{sec-lumitheo} \\
\hline
\textbf{Worst-case Sum : } & \textbf{560} & \textbf{20.7} & \textbf{660} & \textbf{20.4} & Assuming uncorrelated \\
\textbf{Best-case Sum : } & \textbf{0.91} & \textbf{5.3} & \textbf{1.1} & \textbf{3.9} & Assuming uncorrelated \\
\hline
\textbf{Detector Uncertainties} & \textbf{LumiCal SABS} & \textbf{LumiCal $\gamma\gamma$} & \textbf{GLIP SABS} & \textbf{GLIP $\gamma\gamma$} & \textbf{Notes} \\
\hline
Angular acceptance, resolution & 1 & 0.1 & 3 & 0.1 & $\theta^{-3}$ vs. $\theta^{-1}$ dependence \\
Energy scale & 0.0001 & 0.0001 & $\approx0$ & $\approx0$ & Calibrate with $\sqrt{s}$ of tracker dimuons\\
Energy resolution bias & 1.6 & 1.6 & 0.1 & 0.1 & For control of 15\% \\
Position bias & 2.6 & 2.6 & 0.3 & 0.3 & For GLIP with 100 micron cells \\
PID and Contamination & 1.5 & 40 & 2.0 & 0.008 & See section~\ref{sec-LumiPID} \\
PID and Contamination & 0.01 & 35 & $\approx0$ & 0.008 & For SABS+4f method \\
\hline
\textbf{Sum : } & \textbf{3.2} & \textbf{35.1} & \textbf{3.0} & \textbf{0.33} & Using SABS+4f, uncorrelated sum \\
\hline
Theory Projection & 1.8 & 5 & 1.5 & 5 & See section~\ref{sec-lumitheo}. \\
\hline
Worst-case Total :  & 560 & 41 & 660 & 20 &  \\
Best-case Total :  & 3.3 & 35 & 3.2 & 3.9 &  \\
\hline
\textbf{Combined Best-case : } & & \textbf{25} & & \textbf{3.6} & Uncorrelated SABS and $\gamma\gamma$ sum\\
\hline
\end{tabular}
\label{tab-LumiTableZ}
\end{sidewaystable}

\newpage

\begin{sidewaystable}[htbp]
\centering
\caption{Estimated \textbf{systematic uncertainties} on the integrated luminosity measurement using small-angle Bhabhas and diphotons at \textbf{ILC WW} as measured in a luminosity calorimeter. Detector and non-detector sources of uncertainty are included but in separate sections. We provide the theory uncertainty, projected to their future expectations, in a separate section on their own. The final total has a best-case scenario and worst-case scenario quadrature sum of all contributions, not including theory. All numerical values are in parts per ten thousand ($10^{-4}$).\\ \footnotesize{*ILD LumiCal uses a different energy scale method than the GLIP LumiCal and has worse performance than the $\sqrt{s}$ method that the GLIP LumiCal uses so we quote the ILD LumiCal value for energy scale~\cite{Abramowicz_2010}.}}
\begin{tabular}{|l|c|c|c|c|l|}
\hline
\textbf{Non-Detector Uncertainties} & \textbf{LumiCal SABS} & \textbf{LumiCal $\gamma\gamma$} & \textbf{GLIP SABS} & \textbf{GLIP $\gamma\gamma$} & \textbf{Notes} \\
\hline
Upstream Material & 1 & 1 & 1 & 1 & Same for all~\cite{Sadeh2008} \\
Upstream Material & $\approx0$ & $\approx0$ & $\approx0$ & $\approx0$ & If corrected for, as done at LEP \\
Luminosity spectrum for 1\% & 0.003 & 0.003 & 0.003 & 0.003 & See section~\ref{sec-LumiSpec} \\
Luminosity spectrum for 0.1\% & 0.0003 & 0.0003 & 0.0003 & 0.0003 & See section~\ref{sec-LumiSpec} \\
Beam Deflection Effects & 370 & -- & 440 & -- & No M\o{}ller scattering run \\
Beam Deflection Effects & -0.3 & -- & -0.35 & -- & M\o{}ller scattering run, section~\ref{sec-Moller} \\
Polarization of beams & 0.02 & 20 & 0.02 & 20 & Using beam polarimeters \\
Polarization of beams & 0.003 & 1.7 & 0.003 & 1.7 & Using PFL method \\
Statistics (for 0.5 $\mathrm{ab}^{-1}$) & 0.08 & 3.5 & 0.04 & 1.5 & See section~\ref{sec-lumitheo} \\
\hline
\textbf{Worst-case Sum : } & \textbf{370} & \textbf{20.3} & \textbf{440} & \textbf{20.1} & Assuming uncorrelated \\
\textbf{Best-case Sum : } & \textbf{0.31} & \textbf{3.9} & \textbf{0.35} & \textbf{2.3} & Assuming uncorrelated \\
\hline
\textbf{Detector Uncertainties} & \textbf{LumiCal SABS} & \textbf{LumiCal $\gamma\gamma$} & \textbf{GLIP SABS} & \textbf{GLIP $\gamma\gamma$} & \textbf{Notes} \\
\hline
Angular acceptance, resolution & 1 & 0.1 & 3 & 0.1 & $\theta^{-3}$ vs. $\theta^{-1}$ dependence \\
Energy scale & 0.13 & 0.13 & 0.068 & 0.068 & Calibrate with $\sqrt{s}$ of tracker dimuons\\
Energy resolution bias & 1.6 & 1.6 & 0.1 & 0.1 & For control of 15\% \\
Position bias & 2.6 & 2.6 & 0.3 & 0.3 & For GLIP with 100 micron cells \\
PID and Contamination & 1.5 & 35 & 1.5 & 0.005 & See section~\ref{sec-LumiPID} \\
PID and Contamination & 0.01 & 35 & $\approx0$ & 0.005 & For SABS+4f method \\
\hline
\textbf{Sum : } & \textbf{3.2} & \textbf{35.1} & \textbf{3.0} & \textbf{0.34} & Using SABS+4f, uncorrelated sum \\
\hline
Theory Projection & 1.8 & 5 & 1.5 & 5 & See section~\ref{sec-lumitheo}. \\
\hline
Worst-case Total :  & 370 & 40 & 440 & 20 &  \\
Best-case Total :  & 3.2 & 35 & 3.0 & 2.3 &  \\
\hline
\textbf{Combined Best-case : } & & \textbf{25} & & \textbf{2.7} & Uncorrelated SABS and $\gamma\gamma$ sum\\
\hline
\end{tabular}
\label{tab-LumiTableWW}
\end{sidewaystable}

\newpage

\begin{sidewaystable}[htbp]
\centering
\caption{Estimated \textbf{systematic uncertainties} on the integrated luminosity measurement using small-angle Bhabhas and diphotons at \textbf{ILC500} as measured in a luminosity calorimeter. Detector and non-detector sources of uncertainty are included but in separate sections. We provide the theory uncertainty, projected to their future expectations, in a separate section on their own. The final total has a best-case scenario and worst-case scenario quadrature sum of all contributions, not including theory. All numerical values are in parts per ten thousand ($10^{-4}$).\\ \footnotesize{*ILD LumiCal uses a different energy scale method than the GLIP LumiCal and has worse performance than the $\sqrt{s}$ method that the GLIP LumiCal uses so we quote the ILD LumiCal value for energy scale~\cite{Abramowicz_2010}.}}
\begin{tabular}{|l|c|c|c|c|l|}
\hline
\textbf{Non-Detector Uncertainties} & \textbf{LumiCal SABS} & \textbf{LumiCal $\gamma\gamma$} & \textbf{GLIP SABS} & \textbf{GLIP $\gamma\gamma$} & \textbf{Notes} \\
\hline
Upstream Material & 1 & 1 & 1 & 1 & Same for all~\cite{Sadeh2008} \\
Upstream Material & $\approx0$ & $\approx0$ & $\approx0$ & $\approx0$ & If corrected for, as done at LEP \\
Luminosity spectrum for 1\% & 0.5 & 0.5 & 0.5 & 0.5 & See section~\ref{sec-LumiSpec} \\
Luminosity spectrum for 0.1\% & 0.08 & 0.08 & 0.08 & 0.08 & See section~\ref{sec-LumiSpec} \\
Beam Deflection Effects & 160 & -- & 190 & -- & No M\o{}ller scattering run \\
Beam Deflection Effects & 0.22 & -- & 0.26 & -- & M\o{}ller scattering run, section~\ref{sec-Moller} \\
Polarization of beams & 0.16 & 20 & 0.16 & 20 & Using beam polarimeters \\
Polarization of beams & 0.03 & 1.7 & 0.03 & 1.7 & Using PFL method \\
Statistics (for 8.0 $\mathrm{ab}^{-1}$) & 0.06 & 3 & 0.03 & 1.8 & See section~\ref{sec-lumitheo} \\
\hline
\textbf{Worst-case Sum : } & \textbf{160} & \textbf{20.3} & \textbf{190} & \textbf{20.1} & Assuming uncorrelated \\
\textbf{Best-case Sum : } & \textbf{0.24} & \textbf{3.4} & \textbf{0.26} & \textbf{2.5} & Assuming uncorrelated \\
\hline
\textbf{Detector Uncertainties} & \textbf{LumiCal SABS} & \textbf{LumiCal $\gamma\gamma$} & \textbf{GLIP SABS} & \textbf{GLIP $\gamma\gamma$} & \textbf{Notes} \\
\hline
Angular acceptance, resolution & 1 & 0.1 & 3 & 0.1 & $\theta^{-3}$ vs. $\theta^{-1}$ dependence \\
Energy scale & 0.17 & 0.17 & 0.061 & 0.061 & Calibrate with $\sqrt{s}$ of tracker dimuons\\
Energy resolution bias & 1.6 & 1.6 & 0.1 & 0.1 & For control of 15\% \\
Position bias & 2.6 & 2.6 & 0.3 & 0.3 & For GLIP with 100 micron cells \\
PID and Contamination & 1.5 & 35 & 1.5 & 0.003 & See section~\ref{sec-LumiPID} \\
PID and Contamination & 0.01 & 35 & $\approx0$ & 0.003 & For SABS+4f method \\
\hline
\textbf{Sum : } & \textbf{3.2} & \textbf{35.1} & \textbf{3.0} & \textbf{0.32} & Using SABS+4f, uncorrelated sum \\
\hline
Theory Projection & 1.8 & 5 & 1.5 & 5 & See section~\ref{sec-lumitheo}. \\
\hline
Worst-case Total :  & 160 & 41 & 190 & 20 &  \\
Best-case Total :  & 3.2 & 35 & 3.0 & 2.5 &  \\
\hline
\textbf{Combined Best-case : } & & \textbf{25} & & \textbf{2.8} & Uncorrelated SABS and $\gamma\gamma$ sum\\
\hline
\end{tabular}
\label{tab-LumiTable500}
\end{sidewaystable}

\newpage

\begin{sidewaystable}[htbp]
\centering
\caption{Estimated \textbf{systematic uncertainties} on the integrated luminosity measurement using small-angle Bhabhas and diphotons at \textbf{ILC1000} as measured in a luminosity calorimeter. Detector and non-detector sources of uncertainty are included but in separate sections. We provide the theory uncertainty, projected to their future expectations, in a separate section on their own. The final total has a best-case scenario and worst-case scenario quadrature sum of all contributions, not including theory. All numerical values are in parts per ten thousand ($10^{-4}$).\\ \footnotesize{*ILD LumiCal uses a different energy scale method than the GLIP LumiCal and has worse performance than the $\sqrt{s}$ method that the GLIP LumiCal uses so we quote the ILD LumiCal value for energy scale~\cite{Abramowicz_2010}.}}
\begin{tabular}{|l|c|c|c|c|l|}
\hline
\textbf{Non-Detector Uncertainties} & \textbf{LumiCal SABS} & \textbf{LumiCal $\gamma\gamma$} & \textbf{GLIP SABS} & \textbf{GLIP $\gamma\gamma$} & \textbf{Notes} \\
\hline
Upstream Material & 1 & 1 & 1 & 1 & Same for all~\cite{Sadeh2008} \\
Upstream Material & $\approx0$ & $\approx0$ & $\approx0$ & $\approx0$ & If corrected for, as done at LEP \\
Luminosity spectrum for 1\% & 20 & 20 & 20 & 20 & See section~\ref{sec-LumiSpec} \\
Luminosity spectrum for 0.1\% & 2 & 2 & 2 & 2 & See section~\ref{sec-LumiSpec} \\
Beam Deflection Effects & 100 & -- & 118 & -- & No M\o{}ller scattering run \\
Beam Deflection Effects & 0.21 & -- & 0.25 & -- & M\o{}ller scattering run, section~\ref{sec-Moller} \\
Polarization of beams & 0.63 & 20 & 0.63 & 20 & Using beam polarimeters \\
Polarization of beams & 0.12 & 1.7 & 0.12 & 1.7 & Using PFL method \\
Statistics (for 8.0 $\mathrm{ab}^{-1}$) & 0.1 & 4 & 0.05 & 2.5 & See section~\ref{sec-lumitheo} \\
\hline
\textbf{Worst-case Sum : } & \textbf{102} & \textbf{28.5} & \textbf{120} & \textbf{28.4} & Assuming uncorrelated \\
\textbf{Best-case Sum : } & \textbf{2.0} & \textbf{4.8} & \textbf{2.0} & \textbf{3.6} & Assuming uncorrelated \\
\hline
\textbf{Detector Uncertainties} & \textbf{LumiCal SABS} & \textbf{LumiCal $\gamma\gamma$} & \textbf{GLIP SABS} & \textbf{GLIP $\gamma\gamma$} & \textbf{Notes} \\
\hline
Angular acceptance, resolution & 1 & 0.1 & 3 & 0.1 & $\theta^{-3}$ vs. $\theta^{-1}$ dependence \\
Energy scale & 0.71 & 0.71 & 0.26 & 0.26 & Calibrate with $\sqrt{s}$ of tracker dimuons\\
Energy resolution bias & 1.6 & 1.6 & 0.1 & 0.1 & For control of 15\% \\
Position bias & 2.6 & 2.6 & 0.3 & 0.3 & For GLIP with 100 micron cells \\
PID and Contamination & 1.5 & 35 & 1.5 & 0.005 & See section~\ref{sec-LumiPID} \\
PID and Contamination & 0.01 & 35 & $\approx0$ & 0.005 & For SABS+4f method \\
\hline
\textbf{Sum : } & \textbf{3.3} & \textbf{35.1} & \textbf{3.0} & \textbf{0.42} & Using SABS+4f, uncorrelated sum \\
\hline
Theory Projection & 1.8 & 5 & 1.5 & 5 & See section~\ref{sec-lumitheo}. \\
\hline
Worst-case Total :  & 102 & 45 & 120 & 28 &  \\
Best-case Total :  & 3.9 & 35 & 3.9 & 3.6 &  \\
\hline
\textbf{Combined Best-case : } & & \textbf{25} & & \textbf{3.7} & Uncorrelated SABS and $\gamma\gamma$ sum\\
\hline
\end{tabular}
\label{tab-LumiTable1000}
\end{sidewaystable}

\chapter{Theory and Motivation of Invisible + $\gamma$ ($X^0\gamma$) Events}
In this chapter we will explore events where the final state includes invisible particles, such as neutrinos, that are produced in association to photon(s). We use the label \Gls{neutral} for this type of event. These types of events are also known as single photon or mono-photon events when there is only a single photon in the event. They are also known as multi-photon or photonic events when there are numerous photons and nothing else in the event. These types of events are particularly useful for identifying and measuring invisible particles in the final state since there can be \Gls{ISR}. Since $\ee$ colliders have charged and low mass initial-state particles there can be considerable helicity conserving photon emission. The conservation of the helicity is important for ensuring that the photon emission does not significantly change the helicity of the underlying physics process. The measurement of the ISR can then be used to constrain the kinematics and particles of the final state that are otherwise unmeasured. Lepton colliders, especially $\ee$ colliders with precision measurements of their initial-state particles, are advantageous for this type of measurement because the initial state can be greatly constrained. Meaning that, if all ISR is measured, then the entirety of the kinematics of the initial state can be solved and, from it, even more of the final state can be constrained. 

Using $X^0\gamma$ events, various LEP experiments were able to measure neutrinos~\cite{ALEPH:monophoton,OPAL:monophoton,ALEPH:monophoton2,L3:monophoton,DELPHI:monophoton}. This method can also be used to directly measure the number of light neutrinos, $N_\nu$, at any center-of-mass energy at or above the Z-pole, as it requires neutrino-antineutrino production. It is less precise than the indirect measurement that is calculated using the Z boson invisible width and the total width of the Z boson that is done at the Z-pole. This method has also been tested for probing dark matter candidates, in an effective field theory setting, with the data from LEP~\cite{Fox_2011}. This analysis found that the LEP data excludes most dark matter, unless it is incredibly weakly interacting, with mass $<100$~GeV. This methodology and LEP data have also been used to probe invisible Supersymmetry (\Gls{SUSY}) particles. The SUSY neutral Majorana fermions, neutralinos ($\chi_i^0$), have been investigated as possible contributions to $X^0\gamma$~\cite{Ambrosanio_1996}. Neutralinos are also a candidate for dark matter and, in the cited analysis, they would interfere with any $X^0\gamma$ measurement of neutrinos. Extrapolated to a general case, any invisible particles that are present in a data sample and cannot be separated from neutrinos, would cause a deviation of $N_\nu$ from the standard model value of three. The presence of the standard model neutrino production also results in the neutralinos being incredibly difficult to measure using the $X^0\gamma$ method as the ISR photons for both are kinematically similar. The SUSY gravity superpartner, the gravitino ($\tilde{G}$), has also been investigated with LEP data as they could lead to events with photons and invisible, satisfying the $X^0\gamma$ criteria~\cite{Lopez_1997}. We will not discuss the gravitino further in this work. We now transition to motivating measurements of neutrinos and dark sector particles.

\section{Neutrinos and Dark Matter}\label{sec-Dark}
There are fundamental particles, measured and theoretical, that do not couple to the electromagnetic or strong force. These particles are typically referred to as the ``dark sector'' because they are either weakly interacting and neutral, and thus cannot interact electromagnetically, or they only interact gravitationally. These dark sector particles are often considered components of dark matter too, with differentiation between hot dark matter and cold dark matter. The components of cold dark matter are often referred to as weakly interacting massive particles (\Gls{WIMP}s) and, given the current cosmological measurements, are the dominant form of dark matter~\cite{Planck:2018vyg}. Opposingly, components of hot dark matter are typically stronger-interacting and/or lower mass than cold dark matter. In the Standard Model the closest particle to being like dark sector BSM particles is the neutrino and it is often considered a form of hot dark matter~\cite{Hannestad_2010}. In BSM models the dark sector particles can be bosons, as is the case for axions, Dirac fermions, such as massive or right-handed neutrinos, or, in the case of SUSY neutralinos, Majorana fermions. Depending on their mass and interaction strength, these new particles can be candidates for either hot dark matter or cold dark matter.

Astronomy and cosmology can measure the macroscopic effects of the dark sector by looking at the structure and bending of spacetime, from gravitational lensing and CMB measurements. The dark sector can also be measured from otherwise extreme environments like black holes and neutron stars, which are treated as sources of dark sector particles or interactions. From these astronomical and cosmological measurements, the existence of dark matter and dark energy has been established. We note that more recent modeling and measurements, done by two independent investigations of supernovae from the Pantheon+ dataset as well as results of the DES collaboration, indicate that dark energy is not as established as dark matter and is either incorrect or insufficient to explain reality~\cite{Seifert_2024,Lane_2024,des2025}. For this reason, we will focus our investigation of the dark sector, with respect to astrophysical measurements, on dark matter only.

In both cases there are terrestrial and collider measurements that can be done to constrain the dark sector and, more specifically, dark matter. There are numerous terrestrial experiments that precisely measure control volumes, usually large tanks of a homogeneous fluid, and look for anomalous decays or anomalous neutrino-like interactions. As of writing, the experiments which measure coherent scattering in control volumes are reaching sensitivity where astrophysical neutrinos, such as solar and atmospheric neutrinos, are the dominant signal source. This is considered the ``neutrino fog'' and was recently measured at XENON~\cite{XENON:2024hup}. This large background makes dark-matter measurement more difficult. We also note that the dark matter cross-section being probed at these experiments are orders-of-magnitude too small to be sensitive to the dark matter thermal annihilation rate. Existing measurements and modeling indicate $\langle \sigma v\rangle_{\rm th}\approx  10^{-25} - 10^{-23} \text{cm}^{3}/\text{s}$, while the most recent XENON measurement is $\langle \sigma v\rangle_{\rm th}\approx  10^{-33}\text{cm}^{3}/\text{s}$~\cite{Dutta_2023,XENON:2024hup}. Since XENON and PandaX are sensitive to coherent scattering and low energy neutrinos they will not be able to measure neutrino flavor or the number of neutrino species~\cite{XENON:2024ijk,PandaX:2024muv}. Other experiments, like IceCube and DUNE, measure higher energies and flavor, due to sensitivity to inelastic scattering, and can put further constraints on dark matter and neutrinos~\cite{icecubeNu,icecubeDM,DUNE:2020fgq}.

Experiments that measure neutrino oscillation, like DAYA bay and MINOS, or incoherent scattering, like IceCube and DUNE, are also sensitive to standard model and BSM neutrino couplings and mixings. As a generalization, the best measurements of these couplings and mixings for these experiments can reach $1\%$ precision~\cite{DUNE:2020fgq,IceCubeCollaboration:2021euf,MINOS:2020iqj,DayaBay:2022orm}. These experiments, while uniquely sensitive to oscillations, are less precise than collider experiments in measuring other aspects of neutrinos and do so at low center-of-mass energies. Two decades before the previous experiments, LEP collaborations achieved a precision of $\approx0.5\%$ on the number of light neutrino species and $\approx5\%$ on the neutrino vector coupling, $g_\text{V}$~\cite{LEPCombined}. In chapter~\ref{ch-Measure} we investigate the potential of future $\ee$ colliders on neutrinos and dark sector measurements. We now continue motivating and investigating $X^0\gamma$ events, neutrinos and the dark sector.

\subsection{Hot Dark Matter: Right-handed Neutrinos with the $\delta_R$ Model}

\begin{figure}[h]
\centering
\includegraphics[width=15cm,clip]{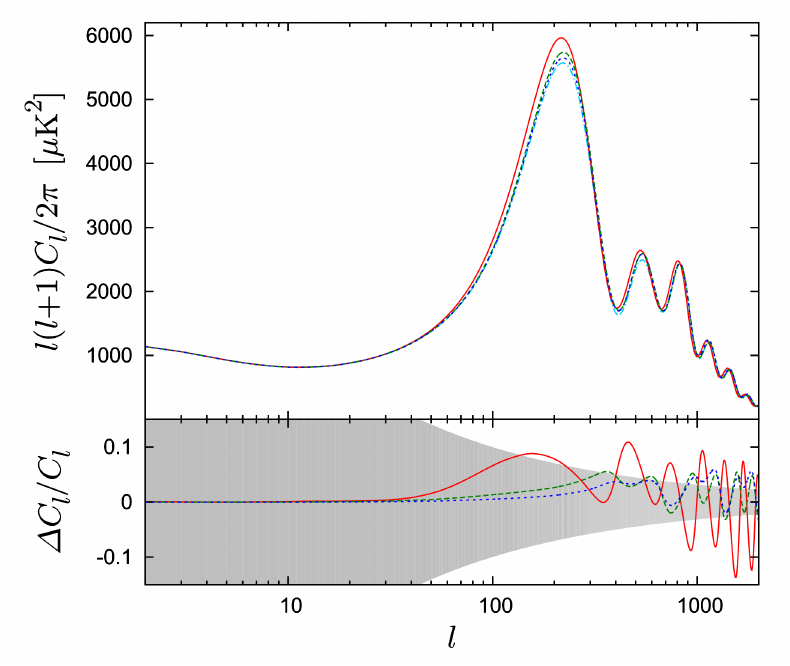}
\caption{(top)-CMB temperature anisotropy spectra for various mixtures of cold and hot dark matter.(bottom)-Differences
to standard $\mathrm{\Lambda}$CDM with expected variance in the shaded region. Standard $\mathrm{\Lambda}$CDM is in blue, $\mathrm{\Lambda}$CDM with neutrinos added is in red. Adding a modest source of axions, another kind of dark sector particle, is in green. A more extreme source of axions is in dark blue. Figure, with more details, from reference~\cite{Hannestad_2010}.}
\label{fig-CMBSpec}       
\end{figure}

Current cosmological measurements indicate that cold dark matter is dominant; this does not mean that hot dark matter does not exist. Measurements of hot dark matter allow constraints on neutrinos, under the assumption that they are standard model neutrinos. Using \Gls{CMB} angular multipole, $l$, power spectra data one can fit for contributions from neutrinos and dark matter. It is common for this power spectra to be converted to temperature values, hence the often quoted temperature of the CMB. The peaks of the CMB multipole power spectra is one of the classic discovery channels for dark matter and dark energy. Through the combination of CMB and optical telescope measurements, anisotropies in power spectra can be used to constrain the sum of neutrino masses to $<0.42$~$\text{eV}$, unless we introduce new particles that contribute to hot dark matter~\cite{Hannestad_2010}~\cite{Abell_n_2022}. The anisotropies in CMB measurements, as seen in figure~\ref{fig-CMBSpec}, prefer adding additional sources of hot dark matter, instead of having only $\mathrm{\Lambda}$CDM and neutrinos.

While axions, another type of dark sector particle, are a common suggestion, we will instead suggest the addition of right-handed neutrinos. Given existing measurements, we do not need to require the right-handed neutrinos to be sterile and instead there is still unmeasured parameter space for right-handed neutrinos to have nonzero but smaller coupling constants than standard model neutrinos and similar mass to standard model neutrinos~\cite{Dutra_2021}. Here we will assume they are comparable in mass to the Standard Model neutrinos and are, therefore, hot dark matter.

The Standard Model currently has non-zero couplings for the Z boson to right-handed fermions in all cases except for neutrinos. The zero coupling for neutrinos is motivated by the Z boson coupling form of
\begin{equation}\label{eqn-zcoup}
g_\text{Z} = T_3 - Q\sin^2{\theta_\text{W}}     
\end{equation}
where $T_3$ is the weak-isospin, which is zero for right-handed particles and $\frac{1}{2}$ for left-handed particles, and the charge, $Q$, is multiplied by a factor with the Weinberg angle, $\theta_\text{W}$. Since the neutrinos have no charge both terms in $g_\text{Z}$ for right-handed neutrinos are zero. This reality is troublesome for naturalness arguments. The lightest left-handed fermion state would be a neutrino with comparable to $\sim$eV mass while the lightest right-handed fermion would be an electron, with $\sim$100~keV mass. This is more problematic if we require the fermion be neutral, at which point the lightest neutral fermion with right-handed contributions is the neutron, with $\sim$1~GeV mass.

We propose a small correction to the Z boson so that the vertex of $\mathrm{Z}\rightarrow\nu_\text{R}\bar{\nu}_\text{L}$ is non-zero but small. We would keep the $\mathrm{W}^{\pm}$ coupling as zero. This scales with a factor that we denote as $\delta_\text{R}$ and refer to as the anomalous weak coupling. This associated model we refer to as the $\delta_\text{R}$ model. The resulting vertex factor Feynman rule is written
\begin{align}\label{eqn-FeynZR}
\begin{fmffile}{complex-a}
\begin{fmfgraph*}(100,100)
    \fmfleft{i1}
    \fmfright{o1,o2}
    \fmf{photon,label=$Z$}{i1,w1}
    \fmf{fermion,label=$\bar{\nu}_\text{L}$}{o1,w1}
    \fmf{fermion,label=$\nu_\text{R}$}{w1,o2}
    \fmfv{lab=$= \frac{ig\delta_\text{R}\gamma^{\mu}P_\text{R}}{\cos{\theta_\text{W}}}$,lab.dist=0.5w}{w1}
\end{fmfgraph*}
\end{fmffile}
\end{align}
with contributions such that it is similar to the typical vertex but for right-handed components and with the $\delta_\text{R}$ factor. Since there are already light right-handed states for the other fundamental fermions the naturalness argument would not require changing existing couplings for other Z boson interactions. This new term could also allow mixing of left-handed and right-handed neutrino states mediated by the Z boson.

\subsubsection{Existing Constraint on Anomalous Weak Coupling, $\delta_{R}$}\label{subsec-deltaR}

One clear way to constrain the anomalous weak coupling is to look at the decay widths of the Z boson. In calculating the total Z boson width, $\Gamma_{Z}$, we can break down contributions from the various decay channels. This leads to individual decay widths, $\Gamma_i$, for each decay channel such that

\begin{equation}\label{eqn-totwid}
    \Gamma_Z = \sum_{i}\Gamma_i = \Gamma_\text{had.} + \Gamma_\text{lep.} + \Gamma_\text{inv.}
\end{equation}

the individual decay widths are typically separated into their decay types. As seen in equation~\ref{eqn-totwid}, the decay channels are separated into hadronic ($\Gamma_\text{had.}$), charged lepton ($\Gamma_\text{lep.}$) and invisible ($\Gamma_\text{inv.}$) contributions. In the Standard Model the invisible decay width is from Z decays to neutrinos. The Standard Model only has left-handed neutrinos so there are only three invisible decay widths since Standard Model neutrinos are chargeless. The decay also does not have a significant number of higher-order corrections, since there cannot be emission of charged particles or photons, only Z and neutrinos. This also means that the main corrections are from the Z-$\nu$ loops that can occur. At even higher orders there can be additional real emission corrections, where a Z boson or even two additional neutrinos are produced, but the contribution from this is very small. At leading order the invisible width
\begin{equation}\label{eqn-nuwid}
    \Gamma_\text{inv.} = \frac{N_\nu}{6\sqrt{2}\pi}(g_\text{L}^2 + g_\text{R}^2)G_\text{F}m_\text{Z}^3 \approx 497.65 \pm 0.03~\text{MeV}
\end{equation}
depends on the number of light neutrinos, the Fermi constant, the Z boson mass and the left-handed and right-handed couplings of the neutrinos. Here we are assuming that the couplings are only with the Z boson. By including higher-order effects, the Standard Model invisible width is given as $
\Gamma_\text{SM,inv.} = 501.44\pm0.04~\text{MeV}$~\cite{Carena:2003aj}.

Precision measurements of the invisible width of the Z boson have been done at LEP and, recently, at the LHC~\cite{OPAL:1994kgw,L3:1998uub,CMS:2022ett}. For the LEP measurement, we include only the L3 and OPAL measurement as the ALEPH measurement was much less precise than the previous two. The resulting combined LEP measurement of invisible width is $\Gamma_\text{LEP,inv.} = 518\pm17~\text{MeV}$~\cite{OPAL:1994kgw,L3:1998uub}. From the LHC, the CMS experiment has measured the invisible width of the Z boson as $\Gamma_\text{LHC,inv.} = 523\pm16~\text{MeV}$~\cite{CMS:2022ett}. At LEP there was also an indirect measurement using the Z boson lineshape parameters~\cite{ALEPH:2005ab}. In this case we compute
\begin{equation}\label{eqn-nuwid2}
    \Gamma_\text{inv.} = \Gamma_\text{lep.}\sqrt{\frac{12\pi R_\text{lep.}}{\sigma_\text{had.}m_\text{Z}^2}} - R_\text{lep.} - 3 \approx 499.0 \pm 1.5~\text{MeV}
\end{equation}
the invisible width from the measured charged lepton width, the ratio of lepton width to hadron width, $R_\text{lep.}$, the hadronic cross-section, $\sigma_\text{had.}$. This is also done under the assumption of lepton universality, so each lepton flavor must have the same couplings. We combine the direct measurements into one experimental
result of $\Gamma_\text{Exp.,inv.} = 521\pm10~\text{MeV}$. The experimental result is 3.9\% higher than the Standard Model prediction and they are different by 2.0~$\sigma$.

We can compute what additional width,~$\Gamma_\text{R}$, the anomalous weak coupling would contribute since we know the coupling is $g_\text{R}=\delta_\text{R}$ and that the additional particles are neutrinos. We do not include higher-order corrections here. Starting with equation~\ref{eqn-nuwid} the total anomalous width
\begin{equation}\label{eqn-anowid}
   \Gamma_\text{R} = \frac{N_\nu\delta_\text{R}^2}{6\sqrt{2}\pi}G_{F}m_{Z}^3
\end{equation}
would then depend on the second power of the anomalous weak coupling. Here we have assumed, and will continue to assume going forward, that there are three flavors of right-handed neutrinos. The same flavors, and among them universality, exist as in the left-handed neutrinos. We can use equation~\ref{eqn-anowid} with the combined best direct measurement of the invisible width of the Z boson to constrain $\delta_\text{R}$ under the assumption that the anomalous width of the Z boson would need to be, at most, equal to the uncertainty of the invisible width. From these values, we find that $|\delta_{R}|=0.10\pm0.07$ is small but non-zero. Since equation~\ref{eqn-anowid} depends on $\delta_{R}^2$ we do not know the sign of $\delta_{R}$.

A similar analysis and extrapolation of this has been done prior but focusing on the $Z\rightarrow\gamma\nu\nu$ invisible state at lepton colliders~\cite{Carena:2003aj}. Such a final state would fall within the conditions of $X^0\gamma$ events. The photon emitted here is an \Gls{ISR} photon, being emitted by an initial state beam particle. By measuring the energy spectrum of the mono-photon signal one can constrain new physics and, likewise, $\delta_{R}$. By using the spectrum of photon energies, or with a transformation you can use the neutrino-antineutrino invariant mass as done in the reference, you are sensitive to both the W and Z boson exchange processes and their interference. Since each of these contributes a different dependence on the neutrino $g_\text{L}$ and $g_\text{R}$, fitting the spectrum can give a precise measurement, and BSM sensitivity, in the $Z\rightarrow\gamma\nu\nu$ components of $X^0\gamma$ events.

\begin{figure}[h]
\centering
\includegraphics[width=15cm,clip]{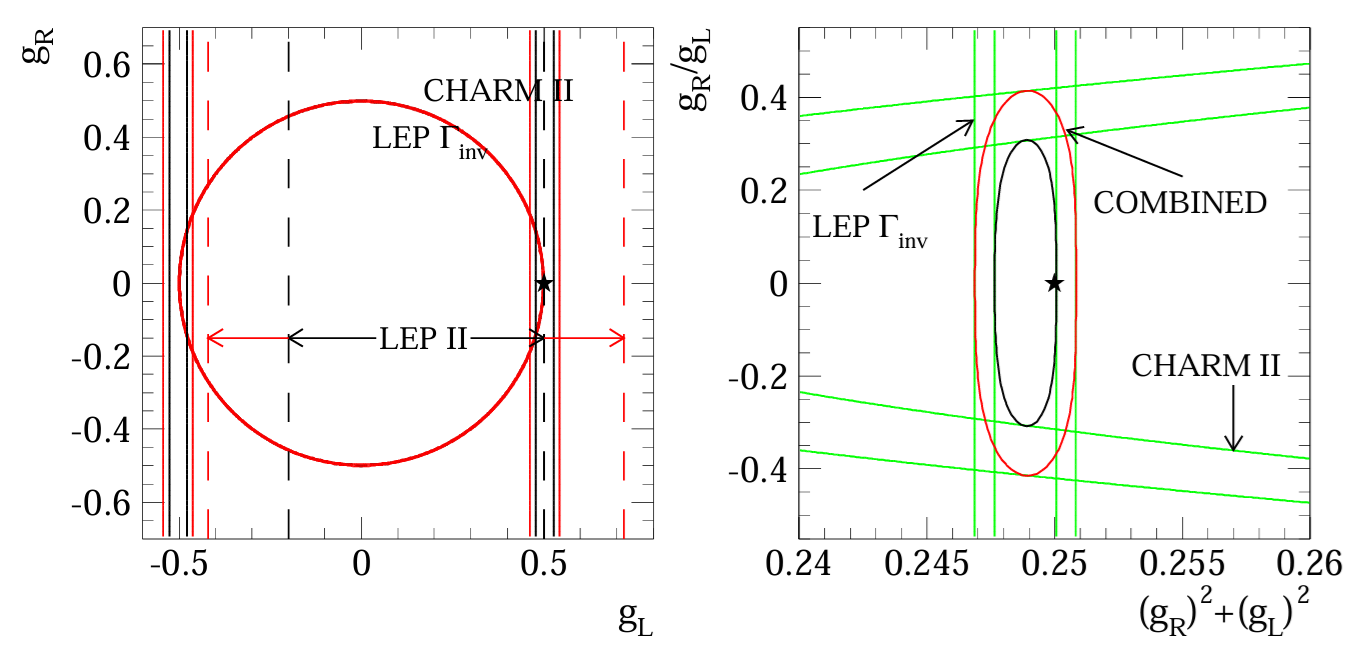}
\caption{(left)-Current constraint (one and two sigma confidence level contours) on $g_\text{L}$ and $g_\text{R}$ from the LEP (indirect) measurement of the invisible Z-width and the CHARM II experiment, on the $g_\text{L}$ × $g_\text{R}$-plane. The SM expectation is indicated by a star. (right)-The one and two sigma allowed regions
which are selected by combining the LEP and CHARM II results. Figure, with more details, from reference~\cite{Carena:2003aj}.}
\label{fig-grCon}       
\end{figure}

As seen in figure~\ref{fig-grCon}, the constraints on $\delta_\text{R}$ from the mono-photon signal and spectrum method indicates a similar magnitude of $\delta_\text{R}$ as the decay width method. Using a simple, though likely crude, method of combining the mono-photon method with the decay width method, we get $|\delta_\text{R}|=0.08\pm0.04$.

In summary, the current experimental results, from both LEP and LHC experiments, using decay width and mono-photon spectrum methods, prefer small but non-zero values of $\delta_R$. The statistical significance is not high, so it is important that more precise measurements of the invisible width of the Z and the Z boson with photon spectrum be done in future experiments.

\subsubsection{Effects on Neutrino Oscillations}

Here we investigate the ramifications of a non-zero $\delta_\text{R}$ in neutrino oscillation experiments. With the addition of the Feynman rule from equation~\ref{eqn-EffMix}, we may have neutrino weak-isospin mixing. This will be in addition to the already existing neutrino flavor mixing that is described by the PMNS matrix. We can construct an effect field theory (EFT) for this, in the same way as the Fermi interaction works. Such that
\begin{align}\label{eqn-EffMix}
\begin{fmffile}{complex-b}
\begin{fmfgraph*}(100,100)
    \fmfleft{i3,i4}
    \fmfright{o3,o4}
    \fmf{fermion,label=$\nu_{ai}$,lab.dist=-0.25w}{i3,w3}
    \fmf{fermion,label=$\bar{\nu}_{ai}$}{w3,i4}
    \fmf{fermion,label=$\bar{\nu}_{bj}$,lab.dist=-0.25w}{o3,w3}
    \fmf{fermion,label=$\nu_{bj}$}{w3,o4}
    \fmfv{decor.shape=circle,decor.filled=full,
decor.size=.05w,lab={}}{w3}
    \fmfv{lab=$\sim U_{ab}^2 \left[\delta_{ij}(1-\delta_{R}) + \delta_{R}\right]$,lab.dist=0.5w}{w3}
\end{fmfgraph*}
\end{fmffile}
\end{align}
the mixing of neutrino states can be expressed using this effective field theory vertex. The handedness of the initial neutrinos is written as $i$, while the final neutrinos handedness is written as $j$. A Kronecker delta, $\delta_{ij}$, is used so that weak-isospin violating and weak-isospin conserving can be included in one expression. The flavor of the neutrinos is written using the subscripts of $a$ and $b$. The flavor mixing given by the PMNS matrix, $U_{ab}$, is squared to reflect the reality that this vertex has two initial neutrinos mixing with two final neutrinos. We note that this EFT may be a starting point to building a more comprehensive field-theory description of neutrino mixing and oscillation. However, we leave further development of the EFT to future research.

Using equation~\ref{eqn-EffMix}, and assuming that the original 3x3 blocks are the same, we can expand the original PMNS matrix, we note as $U_{\nu}$, to a 6x6 representation, we note as $U_{\nu}'$. This new representation includes contributions from mixing flavor states that have weak-isospin violation. Using equation~\ref{eqn-EffMix} we know that the weak-isospin violating flavor mixing will pick up a factor of $\delta_{R}$ since that is the ratio of the weak-isospin violating coupling to the weak-isospin conserving coupling. The coefficient of the $\delta_{R}$ term is not currently known, so we will keep this case simple and assume it is unity.

We embed $U_{\nu}$ in the four quadrants of $U_{\nu}'$, with the mixing from weak-isospin violating components being the off-diagonal quadrants. We write, in block notation,
\begin{equation}\label{eqn-NewPMNS}
U_{\nu}'=
\begin{pmatrix}
U_{\nu} & \delta_{R} U_{\nu} \\
\delta_{R} U_{\nu} & U_{\nu}
\end{pmatrix}
\end{equation}
each block representing the original 3x3 PMNS matrix. It is assumed that $U_{\nu}'$ has a scalar multiplier to ensure that unitarity is not violated.

The computation for the mass eigen-states would then depend on the product of $U_{\nu}'$ with a vector containing both left-handed, $\nu_{L}$, and right-handed, $\nu_{R}$, flavor states. Such that
\begin{equation}\label{eqn-numass}
    \begin{pmatrix}
\nu_{mass} \\
\nu_{mass}' \\
\end{pmatrix} = 
\begin{pmatrix}
U_{\nu} & \delta_{R} U_{\nu} \\
\delta_{R} U_{\nu} & U_{\nu}
\end{pmatrix}
    \begin{pmatrix}
\nu_{L} \\
\nu_{R} \\
\end{pmatrix}
\end{equation}
the LHS of equation~\ref{eqn-numass} has the mass eigen-states for the currently known mass eigen-states, $\nu_{mass}$ and three additional mass eigen-states, $\nu_{mass}'$. With the 6x6 representation we can entertain the possibility of three additional mass eigen-states, however, here we will keep things simpler and assume that 
\begin{equation}\label{eqn-massCond}
    \nu_{mass}' = \nu_{mass}
\end{equation}
and thus there are only three unique mass eigen-states. This also means that the three currently measured neutrino mass eigen-states are superpositions of all six of the flavor and weak isospin states and that we do not expect more than three light neutrino species.

We will briefly entertain an example where we have an initial state that contains only left-handed electron neutrinos. We assume that the electron neutrino is the lightest mass state, in what is referred to as the normal or non-inverted mass hierarchy. With this initial state, we find $\nu_L \delta_{R} U_{\nu} = \delta_{R} U_{\nu}$ as our factor that oscillates from $\nu_e$ into $\nu_{R}$ since we only have left-handed electron neutrinos. However, we do not find that this results in an increasing loss of left-handed neutrinos to the right-handed sector. 

\begin{figure}[h]
\centering
\includegraphics[width=\textwidth, height=0.5\textheight,clip]{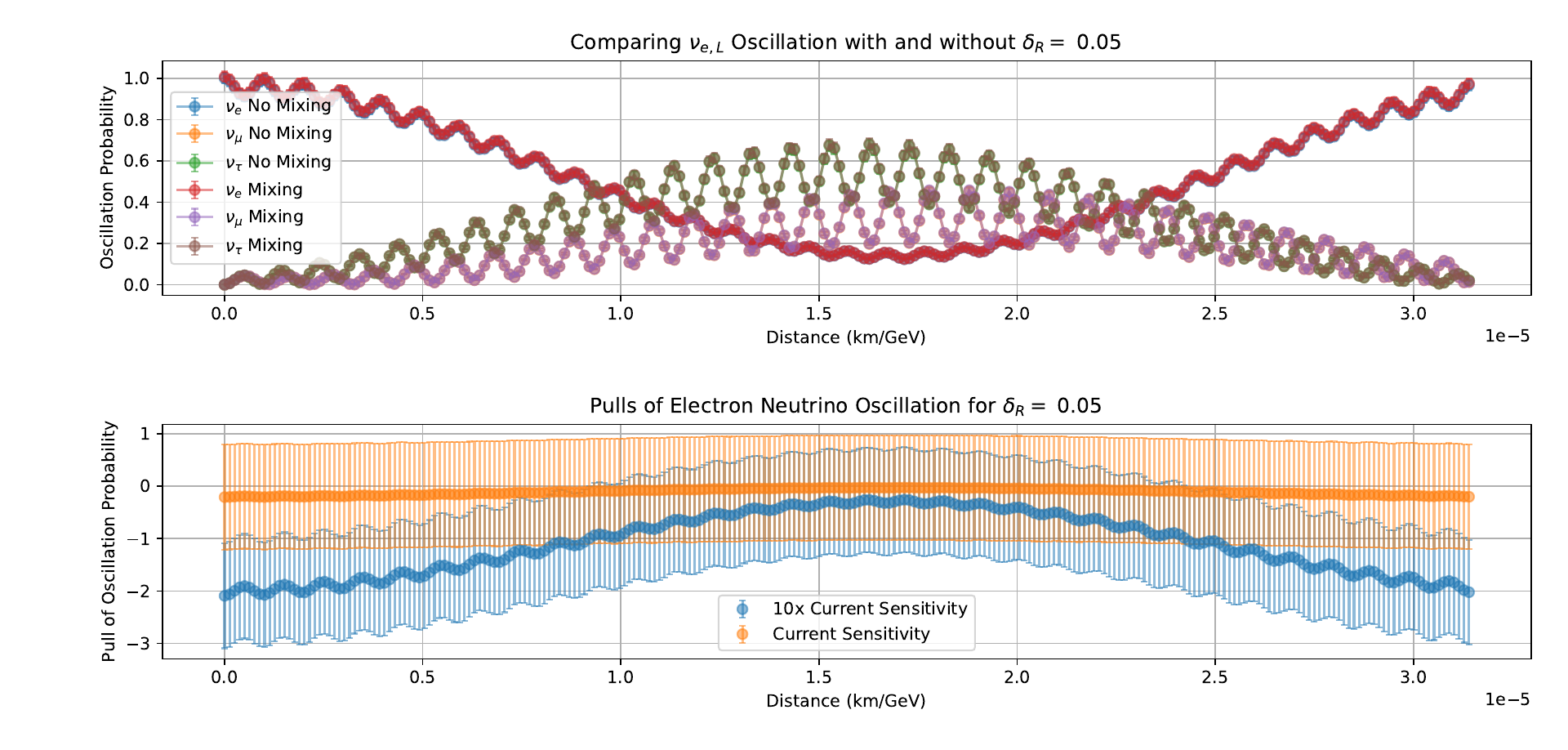}
\caption{(top)-Values from NuFit-6.0 were used to compute probabilities for neutrino oscillations from a starting source of left-handed electron neutrinos~\cite{Esteban_2024}.(bottom)-Pulls, using uncertainties extrapolated from NuFit-6.0 fits, of the electron neutrino oscillation for current and a possible future sensitivity.}
\label{fig-nuosc}       
\end{figure}

As seen in figure~\ref{fig-nuosc}, a small fraction, depending on $\delta_{R}$, of neutrinos become right-handed states and contribute as small corrections to the overall neutrino oscillations. For our test point of $\delta_R=~0.05$, we find that the change in electron neutrino oscillations, which is the best measured channel, is currently unmeasurable. Future experiments, with roughly 10x the sensitivity, may be able to see contributions from a non-zero $\delta_R$ that is not already constrained by collider measurements.

\subsection{Cold Dark Matter: Neutralinos from the MSSM}\label{sec-MSSM}

\subsubsection{MSSM Parameter Value Preamble}\label{sec-MSSMpre}
\begin{figure}[h]
\centering
\includegraphics[width=17cm,clip]{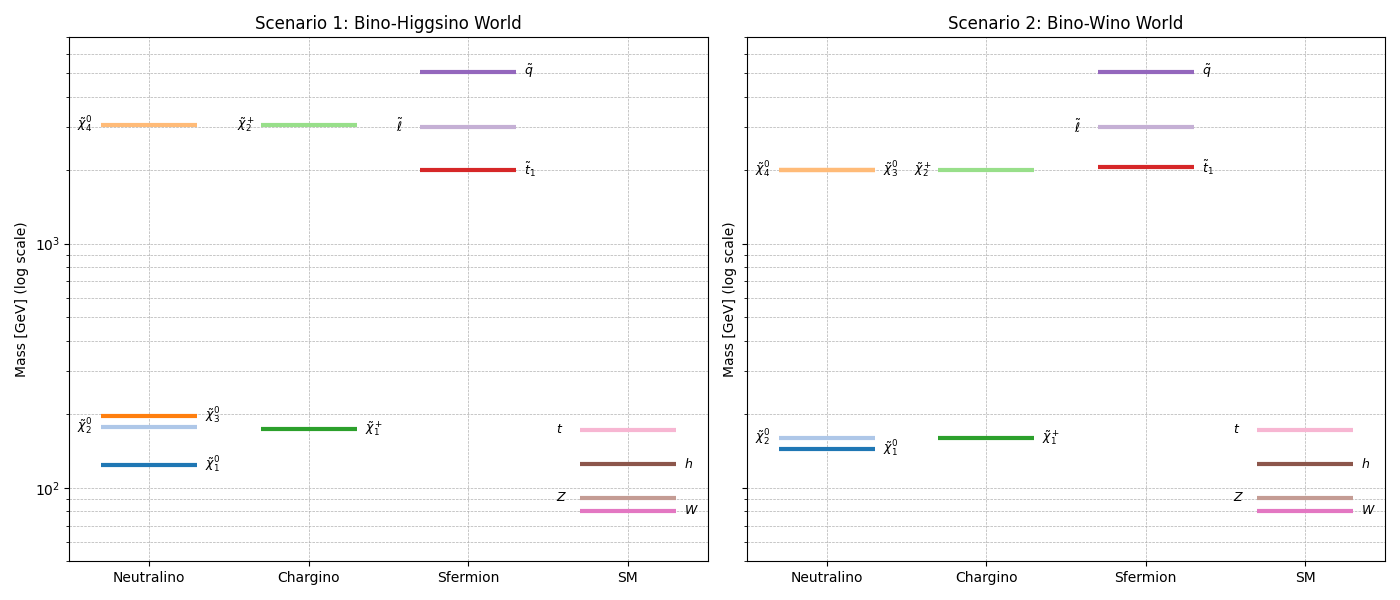}
\caption{Plot of the mass spectrum of the Bino-Wino and Bino-Higgsino MSSM models used in this work. We have grouped the squarks and sleptons together. The lightest stop is separated.}
\label{fig-MSSMSpec}       
\end{figure}

Before proceeding we note that we will entertain two MSSM models, as generated by SPheno, for the work done here~\cite{SPheno}. As a part of this we have required that the Higgs mass that SPheno calculates, from the input parameters, must be within the uncertainty of the current world best Higgs mass measurement. To achieve a valid Higgs mass we varied the trilinear couplings, parameters 11-13 in the SLHA EXTPAR block, and the lightest stop mass. We also set $\tan\beta=15$ for both models for the sake of consistency. Both MSSM models have been chosen to not be excluded by existing LEP and LHC results as calculated by hand and using SModelS v3.0.0~\cite{SModelSv3}. In particular, we have used micrOMEGAs v6.2.3 , which includes SModelS, to calculate both collider exclusion and direct detection exclusion~\cite{MICROMEGAS}. We note that the most relevant LEP exclusion comes from neutralino and chargino constraints, which require the neutralino mass to be $>\sim100$~GeV and the chargino mass to be $>\sim105$~GeV~\cite{Fox_2011,DELPHIChargino,ALEPHChargino,OPALChargino}. We have not put a requirement for the models in terms of direct detection exclusion. Exclusion values, reported in r-value, can be found in appendix~\ref{app-SModelS}. We have also required each MSSM model to be able to create the correct dark matter relic abundance, so that the lightest neutralino would satisfy WIMP dark matter constraints. The relic abundance, factoring in annihilation and co-annihilation, was computed using micrOMEGAs and compared to the most recent results from Planck~\cite{Planck:2018vyg}. Given the dark matter requirement and the collider constraints, and wanting fairly light neutralinos so they could be measured at ILC, we found that both models needed to have considerable fractions of Bino type neutralinos for the lightest neutralino. Beyond this, we found suitable models for Bino mixed with Wino and Bino mixed with Higgsino. We have dubbed these the Bino-Wino (BW) and Bino-Higgsino (BH) models. The relevant parameter values, typically reported in the SUSY Les Houches Accord (\Gls{SLHA}) format, can be found in appendix~\ref{app-SLHA-BW} and appendix~\ref{app-SLHA-BH} respectively. A plot of the mass spectrum of each model can be see in figure~\ref{fig-MSSMSpec}.

\subsubsection{MSSM Primer}\label{sec-MSSMprimer}
In addition to the high mass right-handed neutrinos discussed in the previous section, there are more exotic forms of cold dark matter, such as those from supersymmetry. Perhaps the most common addition to the standard model is introducing R-parity, the quantum number of supersymmetry, and the new particles that it comes with. This addition is motivated by the appealing naturalness of supersymmetry, the discovery of the Higgs boson, and the presence of differences between neutrino mass eigenstates and their flavor states. The latter two which have been measured recently and are predicted by supersymmetry models. The presence of the Higgs boson as a single mass eigenstate, that is distinct from its gauge eigenstate, is also something that is indicative of supersymmetry. Here we will focus on the Minimally Supersymmetry Standard Model (\Gls{MSSM}) as it is one of the most common supersymmetric models.

The R-parity of supersymmetric models can also ensure the stability of the lightest supersymmetric particle (LSP). Standard Model particles are assigned an R-parity of \( +1 \), while their supersymmetric partners, or sparticles, carry R-parity \( -1 \). This leads to three key consequences:
\begin{enumerate}
    \item \textbf{Sparticles are produced in pairs}, since R-parity conservation forbids single sparticle production in Standard Model processes and our experiments start with Standard Model particles.
    
    \item \textbf{The LSP is stable}, making it a natural dark matter candidate if it is also neutral.
    
    \item \textbf{Sparticles differ in spin by \( \frac{1}{2} \)} from their Standard Model counterparts. Fermions acquire boson partners known as sfermions, and bosons acquire fermion partners known as gauginos or higgsinos.
\end{enumerate}
In some supersymmetric models, the mass spectrum of sparticles is hierarchical, but not necessarily aligned with the Standard Model masses. Radiative corrections and soft-breaking terms often make the higgsino and the stop, the scalar partner of the top quark, among the lightest sparticles due to large Yukawa couplings and renormalization group effects. Due to these radiative corrections and renormalization terms, using existing software and calculators, like SPheno, is highly recommended for the process of generating plausible \Gls{SUSY} parameters.

\begin{figure}[h]
\centering
\includegraphics[width=15cm,clip]{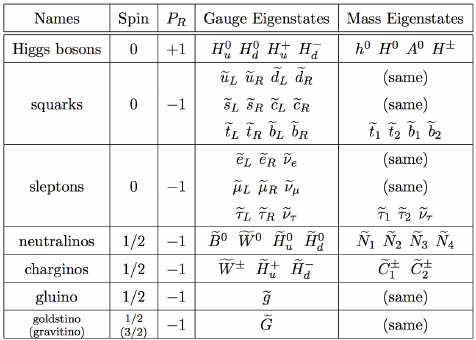}
\caption{Table of Supersymmetric particles broken down by their spin, R-parity, gauge state and mass state. Figure credit~\cite{Melzer_Pellmann_2014}.}
\label{fig-SUSYTab}       
\end{figure}

The fermionic superpartners of gauge and higgs bosons, called gauginos and higgsinos respectively, are typically Majorana fermions, meaning they can be their own antiparticles. These gauge eigenstates are not mass eigenstates, and must be diagonalized to obtain the physical mass eigenstates. This process is analogous to the mixing seen in the neutrino or quark flavor sectors. For example, the bino, wino, and higgsino gauge eigenstates mix to form four neutralino mass eigenstates and two chargino mass eigenstates that would represent what is observed in nature if supersymmetry is observable. We present figure~\ref{fig-SUSYTab}, for demonstrating the sparticles of the MSSM.
Since neutralinos would be neutral Majorana fermions, and likely be weakly interacting because they are neutral and massive, they are commonly considered \Gls{WIMP} dark matter candidates. Especially when the lightest neutralino, noted as $\chi_1^0$, is the LSP and therefore stable.

We will cover only relevant parameters since the MSSM, depending on formulation, can have over 120 parameters~\cite{BaerAndTata}. From the gaugino sector we have three gaugino masses that are written as $\mathrm{M_1}$ , $\mathrm{M_2}$ , $\mathrm{M_3}$. These are defined as their mass at the SUSY energy scale. Sometimes they are instead written as  $\mathrm{M_1'}$ , $\mathrm{M_2'}$ , $\mathrm{M_3'}$, which is their value at the Grand Unified Theory (\Gls{GUT}) energy scale. These masses can be positive or negative valued since they are Majorana masses. 

From the Higgs sector we have $\mu$ from the Higgs potential. This is similar to, but not exactly the same value as $\mu$ from the Vacuum Expectation Value (VEV) of the Standard Model Higgs potential, as the Higgs sector is slightly different in the MSSM. In particular the potential 
\begin{equation}\label{eqn-SMHiggsPot}
V(H) = \lambda(H^2-v^2)^2
\end{equation}
depends on the Higgs field, $H$, and the VEV. If we adjust the field in equation~\ref{eqn-SMHiggsPot} to this VEV, since potentials care not for absolute scales, then we obtain the adjusted field, $h$. The resulting adjusted potential
\begin{equation}\label{eqn-SMHiggsTri}
V(h) = \lambda h^4 + 4 \lambda v h^3 + 4 \lambda v^2 h^2
\end{equation}
has a mass term, the right most term. From equation~\ref{eqn-SMHiggsTri} we can derive the Higgs mass relation for the Standard Model as
\begin{equation}\label{eqn-SMHiggsMass}
    M_{H} = \sqrt{8\lambda v^2} = 4\mu
\end{equation}
and thus equation~\ref{eqn-SMHiggsMass} illustrates how the $\mu$ parameter, in the Standard Model, is simply a rescaling of the Higgs mass. 

In the MSSM the Higgs doublet of $\mathrm{\phi}$ is promoted to a left-chiral superfield of $\hat{H}_u$. However, this is insufficient for describing the Yukawa couplings as $\hat{H}_u$ only couples to positive weak-isospin particles. Therefore we observe that
\begin{equation}\label{eqn-MSSMDoublet}
    \phi \rightarrow \hat{H}_u \space, \space \hat{H}_d
\end{equation}
we must introduce a second doublet, the right-chiral superfield of $\hat{H}_d$. The superfields of equation~\ref{eqn-MSSMDoublet} are often referred to as the up-like Higgs superfield and the down-like Higgs superfield since they couple to up-like and down-like fermions respectively. The existence of these superfields also results in four new fermion fields, the four higgsinos of $\tilde{H}^0_u$, $\tilde{H}^0_d$, $\tilde{H}^+_u$, $\tilde{H}^+_d$. For the purpose of neutralino compostition only the neutral, $\tilde{H}^0_u$ and $\tilde{H}^0_d$, higgsinos contribute.

The resulting potential from this promotion is considerably more complicated than the Standard Model equivalent of equation~\ref{eqn-SMHiggsTri}. We will not write it out here as it is long and can be found in references~\cite{BaerAndTata}. The result is that $\mu$ in the MSSM need not be equivalent to $\mu$ in the Standard Model. It is also the case the there are now two VEVs, $v_u$ and $v_d$, one for each doublet. From these VEVs we can define
\begin{equation}\label{eqn-tanbeta}
    \mathrm{\tan\beta} = \frac{v_u}{v_d}
\end{equation}
as the ratio of the two VEVs. It is more common and simpler to use the MSSM value of $\mu$ and $\mathrm{\tan\beta}$ than to use $v_u$ and $v_d$. Even if an equivalent representation could be derived.

We will not write out the various terms of the MSSM Lagrangian but instead note that the neutralino sector of the Lagrangian will rely on the neutral higgsino terms, $\psi_u^0$ and $\psi_d^0$, and the neutral gaugino terms, $\lambda_0$ for bino-type and $\lambda_3$ for wino-type. The resulting Lagrangian can be simplified to
\begin{align}\label{eqn-LagNeu}
    \mathcal{L}_{\tilde{N}} &= -\frac{1}{2}\left[\bar{\psi}_{u}^0,\bar{\psi}_{d}^0,\bar{\lambda}_3,\bar{\lambda}_0\right]\mathcal{M}_{\tilde{N}}\begin{bmatrix}
           \psi_{u}^0 \\
           \psi_{d}^0 \\
           \lambda_3 \\
           \lambda_0
         \end{bmatrix}
  \end{align}
an expression that utilizes linear algebra with a newly defined neutralino mass matrix, $\mathcal{M}_{\tilde{N}}$. We can define the mass matrix from equation~\ref{eqn-LagNeu} as
\begin{equation}\label{eqn-NeuMass}
    \begin{bmatrix}
0 & \mu & -M_\text{Z} \sin\theta_\text{W} \cos\beta & M_\text{Z} \sin\theta_\text{W}\sin\beta\\
\mu & 0 & M_\text{Z} \cos\theta_\text{W}\cos\beta & -M_\text{Z} \cos\theta_\text{W}\sin\beta\\
-M_\text{Z} \sin\theta_\text{W} \cos\beta & M_\text{Z} \cos\theta_\text{W}\cos\beta & M_1 & 0\\
M_\text{Z} \sin\theta_\text{W}\sin\beta & -M_\text{Z} \cos\theta_\text{W}\sin\beta & 0 & M_2
\end{bmatrix}
\end{equation}
depending on four SUSY parameters, $M_1$, $M_2$, $\mu$ and $\mathrm{\tan\beta}$, as well as the electroweak Standard Model parameters of $M_\text{Z}$, the Z boson mass, and $\theta_\text{W}$, the Weinberg angle. 

Since the matrix of equation~\ref{eqn-NeuMass} is symmetric it is easy to solve for its eigenvectors and eigenvalues. The common practice is to use 
\begin{equation}\label{eqn-NeuMix}
    V_{\tilde N}^\dagger \mathcal{M}_{\tilde N}V_{\tilde N} = \mathcal{M}_\text{D}
\end{equation}
the unitary, and real orthogonal, matrix of $V_{\tilde N}$ that can be derived from the diagonalized matrix of eigenvalues, $\mathcal{M}_\text{D}$. The matrices of equation~\ref{eqn-NeuMix} are then used with $\mathcal{M}_\text{D}$ to define the four neutralino masses from the four eigenvalues and $V_{\tilde N}$ used to define the neutralino mass mixing. This is referred to as the neutralino mass mixing matrix. Since the eigenvalues can be Majorana masses they may have negative values. 

The mass hierarchy of the absolute value of the masses is used to define the neutralino ordering. Due to these factors, the neutralinos do not have consistent ordering or gaugino and higgsino content. There are also some contexts where the bino-type and wino-type content are remixed to be in terms of photino-type and zino-type. For our presentation of values in appendix~\ref{app-SLHA-BH} and appendix~\ref{app-SLHA-BW} we have chosen to use a consistent ordering of bino, wino, higgsino, higgsino for the matrices. Since the couplings of the neutralinos depend on their gaugino and higgsino content it is possible that neutralino cross-sections will vary by orders of magnitude for the same mass values in the same interactions. These are important considerations for generating possible MSSM neutralinos. As we will see in the next section, the presence of higgsino in the neutralinos tend to result in increased cross-sections. This is also reflective of the higher levels of exclusion that can be found in appendix~\ref{app-SModelS}, where direct detection experiments have excluded our Bino-Higgsino model more than our Bino-Wino model.

\subsubsection{Measuring Neutralinos with Monophotons}

In the MSSM the neutralinos have Feynman rules that describe vertices with the Z boson and Higgs boson such that
\begin{align}\label{eqn-FeynNeuZ}
\begin{fmffile}{complex-c}
\begin{fmfgraph*}(100,100)
    \fmfleft{i1}
    \fmfright{o1,o2}
    \fmf{photon,label=$Z$}{i1,w1}
    \fmf{fermion,label=$\tilde{N_i}$}{o1,w1}
    \fmf{photon}{o1,w1}
    \fmf{fermion,label=$\tilde{N_j}$}{w1,o2}
    \fmf{photon}{w1,o2}
    \fmfv{lab=$=g_{\text{Z}ij}$,lab.dist=0.4w}{w1}
\end{fmfgraph*}
\end{fmffile}
\end{align}
\begin{align}\label{eqn-FeynNeuH}
\begin{fmffile}{complex-d}
\begin{fmfgraph*}(100,100)
    \fmfleft{i1}
    \fmfright{o1,o2}
    \fmf{dashes,label=$H$}{i1,w1}
    \fmf{fermion,label=$\tilde{N_i}$}{o1,w1}
    \fmf{photon}{o1,w1}
    \fmf{fermion,label=$\tilde{N_j}$}{w1,o2}
    \fmf{photon}{w1,o2}
    \fmfv{lab=$=g_{\text{H}ij}$,lab.dist=0.4w}{w1}
\end{fmfgraph*}
\end{fmffile}
\end{align}
where the couplings are written as
\begin{align}\label{eqn-NeutCoup}
    \begin{gathered}
        g_{\text{Z}ij} \propto|N_{i,3}|^2-|N_{i,4}|^2 \\
        g_{\text{H}ij} \propto(N_{i,2}-\tan\theta_\text{W}N_{i,1})(N_{j,3}\sin\alpha_\text{H} + N_{j,4}\cos\alpha_\text{H})
    \end{gathered}
\end{align}
depending on the neutralino composition, $N_{i,n}$, and the Higgs mixing angle of $\alpha_\text{H}$. We use the subscript $i$ and $j$ to denote the neutralino order. This distinction is important as the neutralino mixing allows for $i\neq j$ and it is common for the largest cross-section for neutralino production to be when $i\neq j$. Typically, one of these channels will dominate over the other one depending on what the composition of the given neutralinos are. For the contribution of equation~\ref{eqn-FeynNeuH}, the initial state beam particles would need considerable mass to have a large cross-section with the Higgs propagator. While it may be possible to achieve this with heavy quark loops at hadron colliders, it is generally less accessible than equation~\ref{eqn-FeynNeuZ}, which has coupling that is independent the mass of the initial particles. This work also focuses on lepton colliders, so we will disregard contributions from the Higgs propagator.

At a lepton collider, the final state from equation~\ref{eqn-FeynNeuZ} or equation~\ref{eqn-FeynNeuH} could be invisible if we have the lightest neutralinos and they are the LSP. In the case of the heavier neutralinos then we would expect exotic decays, which often involve the stable neutralinos, but, due to their higher masses, lower cross-sections. Therefore, to measure these events, there needs to be other particles present. As with the direct invisible Z width discussed in SubSect.~\ref{subsec-deltaR}, we can measure the neutralinos using the recoil of \Gls{ISR}. It is also possible, as can be found in the branching ratios of the neutralinos in appendix~\ref{app-SLHA-BH} and appendix~\ref{app-SLHA-BW}, that the heavier neutralinos decay to photons and the LSP. So there are two possible sources of photons for constraining the invisible neutralino system.

The most dominant form of ISR will be the monophoton, as in one photon, radiation. The resulting diagram, for an electron positron collider,
\begin{align}\label{eqn-FeyneZNeu}
\begin{fmffile}{complex-f}
\begin{fmfgraph*}(100,100)
    \fmfleft{i1,i2}
    \fmfright{o1,o2,o3}
    \fmf{fermion,tension=2,label=$e^-$}{i1,w2}
    \fmf{photon,tension=0.0,label=$\gamma$}{w1,o3}
    \fmf{fermion,tension=4.0}{w2,w1}
    \fmf{fermion,tension=4.0,label=$e^+$}{w1,i2}
    \fmf{photon,tension=2.0,label=$Z$}{w2,w3}
    \fmf{fermion,tension=1,label=$\tilde{N_i}$}{o1,w3}
    \fmf{photon}{o1,w3}
    \fmf{fermion,tension=0.25,label=$\tilde{N_i}$}{w3,o2}
    \fmf{photon}{w3,o2}
\end{fmfgraph*}
\end{fmffile}
\end{align}
will have a Z boson propagator that utilizes the vertex from equation~\ref{eqn-FeynNeuZ}. The monophoton process will have backgrounds from Standard Model processes, namely neutrino-antineutrino production, but also from other SUSY processes. The supersymmetric partner to the electron, the selectron, can contribute
\begin{align}\label{eqn-FeynSelec}
\begin{fmffile}{complex-g}
\begin{fmfgraph*}(100,100)
    \fmfleft{i1,i2}
    \fmfright{o1,o2,o3}
    \fmf{fermion,tension=2,label=$e^-$}{i1,w3}
    \fmf{photon,tension=0.0,label=$\gamma$}{w1,o3}
    \fmf{fermion,tension=4.0}{w2,w1}
    \fmf{fermion,tension=4.0,label=$e^+$}{w1,i2}
    \fmf{dashes,label=$\tilde{e}$}{w2,w3}
    \fmf{fermion,tension=1,label=$\tilde{N_i}$}{o2,w2}
    \fmf{photon,tension=1}{o2,w2}
    \fmf{fermion,tension=1.25,label=$\tilde{N_i}$}{w3,o1}
    \fmf{photon,tension=1.25}{w3,o1}
\end{fmfgraph*}
\end{fmffile}
\end{align}
with a t-channel interaction. Since this is a selectron propagator the mass of the selectron is important. The expected cross-section for equation~\ref{eqn-FeynSelec}, as well as the interference effects it has with the main neutralino signal, is negligible for both of our models as we have chosen a large selectron mass, of order 1 TeV.

\begin{figure}[h]
\centering
\includegraphics[width=15cm,clip]{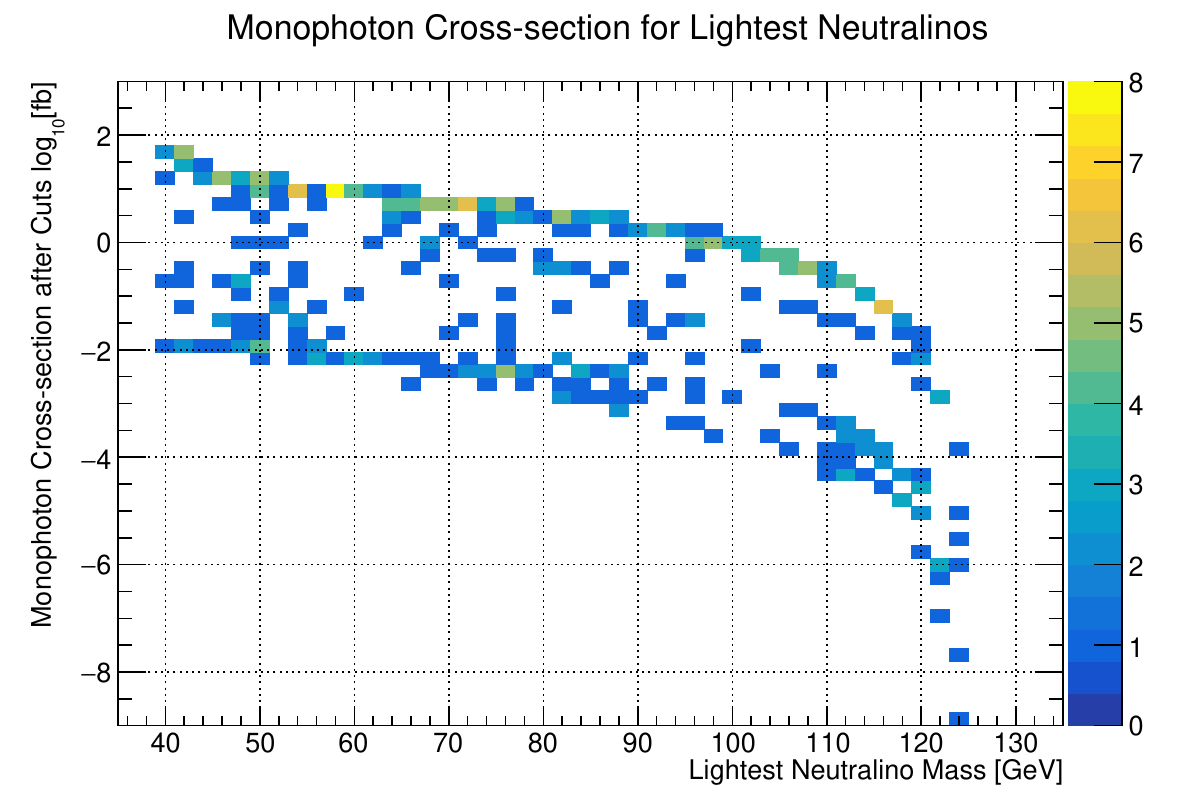}
\caption{Result of the monophoton signal associated with lightest neutralino production. The separation of the possible cross-sections is due to the different composition the neutralinos can have with respect to the wino, bino and higgsino.}
\label{fig-NeuXsec}       
\end{figure}

As a demonstration, we investigate how the monophoton cross-section, here only from ISR, changes as the neutralino mixing matrix composition changes. For this demonstration, we used a center-of-mass energy of 250~GeV and an unpolarized beam. Our methodology starts with randomly generating a neutralino mixing matrix from random values of $M_1$, $M_2$, $M_3$ and $\mu$ and then putting it in an \Gls{SLHA} file. We then use the SLHA file with the \Gls{MCEG} WHIZARD to estimate the expected cross-section for monophoton as measured in the acceptance of \Gls{ILD} ECAL and LumiCal. We use cuts done by previous monophoton analyzes, which utilize the $p_{t}$ of the photons measured~\cite{Kalinowski_2020}. The results, as seen in figure~\ref{fig-NeuXsec}, indicate that there are two dominant phases of neutralinos cross-section. This is expected for two reasons. First, the mixing of neutralinos will often prefer the neutralinos to be close to pure states of certain types of gaugino or higgsinos. Second, the higgsino-like neutralinos will have much larger cross-sections than the guagino-like neutralinos because the higgsino-like neutralinos have much larger couplings to the Z boson.

\chapter{Precision Measurements in $X^0\gamma$ Events}\label{ch-Measure}
\section{Simulating $X^0\gamma$ Events}\label{subsec-SimHeader}

In order to perform a plausible analysis of \Gls{neutral} events we need to use a particle physics event generator. Numerous event generators are available that can simulate the collisions of electrons and positrons. For this study we examine BabaYaga, WHIZARD, KKMC, and MadGraph5 ~\cite{Balossini_2008} ~\cite{Kilian_2011} ~\cite{Jadach_2023} ~\cite{Alwall_2014}. We provide table~\ref{tab-prog} as a reference for comparing these event generators in terms of $X^0\gamma$ processes and higher order corrections.

\begin{table}[h!]
    \centering
    \caption{Comparison of event generators relevant to simulating neutral events with photons. Each program was tested in this work to ensure function. Each program's files were scraped in order to determine the presence of effects and corrections possible in the program. This was done so program abilities that are self-reported by the authors, which are inherently biased, would not be relied upon.}
\label{tab-prog}    
    \begin{tabular}{|>{\raggedright\arraybackslash}m{2cm}|>{\centering\arraybackslash}m{2cm}|>{\centering\arraybackslash}m{2.25cm}|>{\centering\arraybackslash}m{1.5cm}|>{\centering\arraybackslash}m{1cm}|>{\centering\arraybackslash}m{1.5cm}|>{\centering\arraybackslash}m{1cm}|>{\centering\arraybackslash}m{1cm}|}
        \hline
        \textbf{Program} & \textbf{Version} & \textbf{Processes} & \textbf{Beam Pol. ?} & \textbf{Soft ISR?} & \textbf{QED?} & \textbf{EW?} & \textbf{QCD?} \\
        \hline
        Babayaga & fcc & $e^+e^-(\gamma)$ , $\gamma\gamma$ & No & Yes & NNLO$^\times$ & No & No \\
        \hline
        WHIZARD & 3.1.5 & All & Yes & No & NLO$^\dagger$ & No$^\dagger$ & No$^\dagger$ \\
        \hline
        KKMC & 5.00.02 & $f\bar{f}(\gamma)$ & No$^*$ & Yes & NNLO & NLO & No \\
        \hline
        MadGraph5 & 3.5.7 & All & No & No & NNLO & NLO & LO \\
        \hline
    \end{tabular}

        \small{$^\times$ Babayaga has NNLO corrections for Bhabha scattering but only NLO corrections for DiGamma.\\$^\dagger$ WHIZARD can use other programs, such as Recola or OpenLoops, to compute higher order corrections~\cite{Actis_2017} ~\cite{Denner_2018}~\cite{Buccioni_2019}.\\$^*$ KKMC originally had polarization dependent calculations but recent versions, devoted to FCC, have made the polarization functionality incorrect.}
\end{table}

\subsection{Simulating $X^0\gamma$ Part 1: Choosing an Event Generator}\label{subsec-Simulating}

As a first test of choosing an event generator we look at testing the relevance of soft radiative corrections. We use this term, soft radiative corrections, to refer to radiative photons that are emitted with energy that is less than 2\% of the total center-of-mass energy of the system. Previous studies investigated the performance of WHIZARD, as compared to KKMC with regard to the production of radiative Bhabhas, written as $e^+e^-(\gamma)$, and radiative neutrinos, written as $\nu\bar{\nu}(\gamma)$~\cite{Kalinowski_2020}. The study showed that, if the radiative effects are limited to hard radiative effects, noted by being greater than 5 GeV of $p_{T}$, then the simulators were equivalent. Considering this, we may forgo concern over soft radiative effects if we use a hard radiative cut and use event generators that do not have soft radiative effects without loss of relevancy.

Here we do a similar test, but with BabaYaga, known for its simulation of diphotons~\cite{CarloniCalame:2003yt}. Diphotons can be simulated in WHIZARD with the same settings as used in the radiative Bhabhas test. We can also limit the energy of radiative effects in BabaYaga to mimic the constraint used in WHIZARD. Since BabaYaga does not have a full simulation of beam polarization effects, we must also restrict WHIZARD to not use polarized beams. 

\begin{figure}[h!]
\centering
\includegraphics[width=15cm,clip]{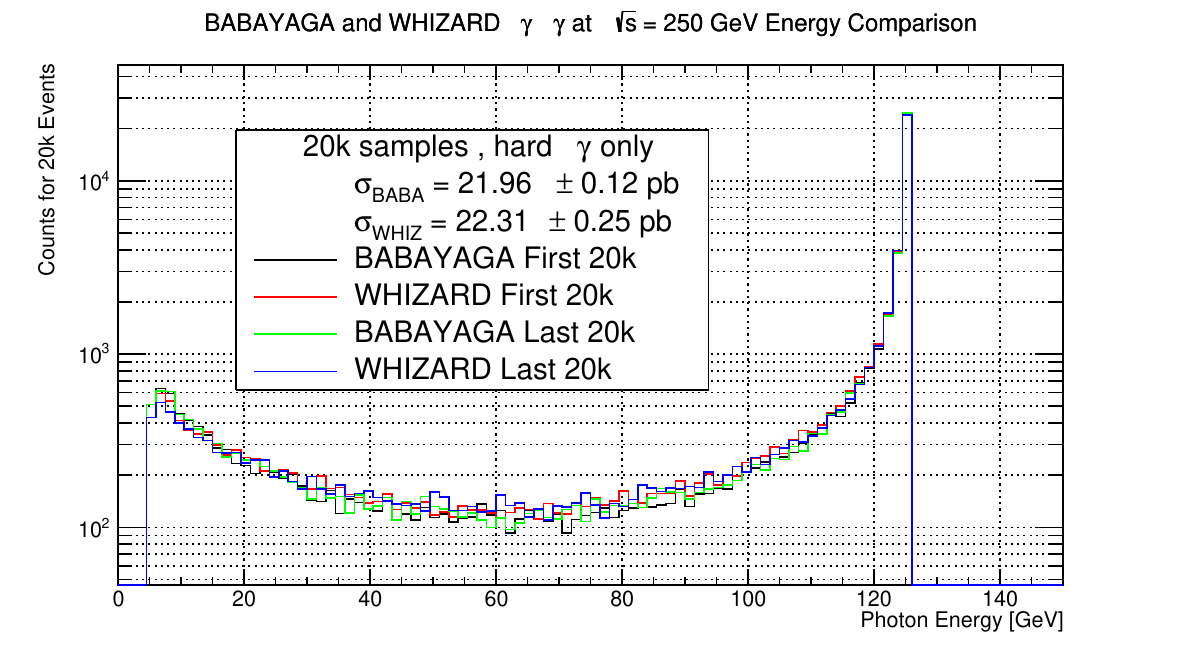}
\caption{Comparison of BabaYaga and WHIZARD production of diphotons with the constraints discussed in the text. The results indicate equivalent performance in terms of shape and integral. The performance of this was checked with different partitions of data and found to not change.}
\label{fig-specComp}       
\end{figure}

The test used 100k events at $\sqrt{s}$ of 250 GeV and with the inner acceptance set to 1 degree. The results of the photon energy spectrum, seen in figure~\ref{fig-specComp}, indicates that the energy spectra shapes are similar. We also find that the expected cross-sections are equivalent to each other and roughly 22 pb. Given these results, we are confident, assuming we are in the hard radiation regime, that WHIZARD is a suitable simulator for both radiative Bhabhas, radiative neutrinos, and diphotons. With performance equivalent comparable to BabaYaga and KKMC.
By using WHIZARD in place of BabaYaga, or the other event generators, we also gain access to polarization effects. Which, as will be shown, can have significant effects on experimental design and precision.

\subsection{Simulating $X^0\gamma$ Part2: Theory Precision}\label{subsec-theory}

Continuing with precision, we now investigate theory precision. Considering table~\ref{tab-prog}, we will look at using MadGraph5 as it has the most higher-order corrections. These corrections were made possible by MadGraph5's Standard Model loops model. For the DiGamma process this includes 90 higher-order diagrams. These corrections include loops that have both QED, electroweak and QCD contributions. 

\begin{align}\label{eqn-FeynDiGammaEWK}
\begin{fmffile}{complex-h}
\begin{fmfgraph*}(100,100)
    \fmfleft{i1,i2}
    \fmfright{o1,o2}
    \fmf{fermion,tension=1,label=$e^-$}{i1,w1}
    \fmf{photon,tension=1,label=$\gamma$}{w1,o1}
    \fmf{fermion,tension=2}{w1,w2}
    \fmf{photon,tension=0.5,label=$Z$}{w2,w3}
    \fmf{fermion,tension=1}{w2,w4}
    \fmf{fermion,tension=1.0,label=$e^+$}{w3,i2}
    \fmf{fermion,tension=1}{w4,w3}
    \fmf{photon,tension=1,label=$\gamma$}{w4,o2}
    \fmfv{lab=$\sim \frac{\alpha_W}{4\pi} \log^2\left(\frac{s}{M_Z^2}\right)$,label.angle=0,label.dist=0.5w}{w2}
\end{fmfgraph*}
\end{fmffile}
\end{align}

The introduction of electroweak loops introduces further changes due to diagrams like equation~\ref{eqn-FeynDiGammaEWK}. These can, and often do, have polarization dependence from the electroweak bosons~\cite{Ciafaloni_1999}~\cite{Kulesza:1999us}. These corrections usually have logarithmic forms and are referred to as Sudakov factors. The Sudakov factors can be renormalized by the mass of the loop particle. In this case the loop particle is a Z boson and therefore the Z boson mass is used. Following from references we expect that the correction

\begin{equation}\label{eqn-digammaEWK}
\delta_{\text{EW}}(P_-, P_+) = \left[ C_1 P_-P_+ + C_2 (1 - P_-P_+) \right] \frac{\alpha}{4\pi s_W^2} \log^2 \left( \frac{s}{M_Z^2} \right)
\end{equation}

from Z vertices depends on the logarithmic factor from equation~\ref{eqn-FeynDiGammaEWK}, as well as two factors, determined by model parameters and theory, of $C_1$ and $C_2$. From this we can estimate that the electroweak corrected polarization dependence will change by as much as 1.1\% at center-of-mass energy of 250 GeV. We note this as this means that any event generator of DiGammas that lacks electroweak loops cannot be more accurate than this. For example, the event generator BabaYaga is quoted as a 0.1\% accuracy simulation of DiGammas but the original work was for simulations less than 10 GeV center-of-mass energy~\cite{Balossini_2008}. The same level of accuracy cannot be expected when these electroweak and loop effects become larger.

We observe that the effects of these loop vertex corrections do not change the differential cross-section since they do not depend on the azimuthal or polar angles. As a general observation, these higher order contributions prefer left-handed fermions. As such, being able to change beam polarization will allow for better control and modeling of these precise effects.

In addition to the vertex loops like equation~\ref{eqn-FeynDiGammaEWK}, there are also box diagrams with electroweak corrections. For the Z box diagrams 

\begin{align}\label{eqn-FeynZBox}
    \begin{fmffile}{feynman1}
    \begin{fmfgraph*}(100,100)
        \fmfleft{i1,i2}
        \fmfright{o1,o2}
        \fmf{fermion}{i1,v1,v2}
        \fmf{photon}{v2,o1}
        \fmf{fermion}{v4,v3,i2}
        \fmf{photon}{v4,o2}
        \fmf{photon,lab=$Z$,lab.dist=-0.2w}{v1,v3}
        \fmf{fermion,lab=$e$,lab.dist=-0.2w}{v2,v4}
        \fmflabel{$e^-$}{i1}
        \fmflabel{$e^+$}{i2}
        \fmflabel{$\gamma$}{o1}
        \fmflabel{$\gamma$}{o2}
    \fmfv{lab=$\sim \frac{\alpha_W}{4\pi} \left[ \log^2\left(\frac{s+t}{M_Z^2}\right) - \log^2\left(\frac{t}{M_Z^2}\right)\right]$,label.angle=0,label.dist=0.5w}{v2}
    \end{fmfgraph*}
    \end{fmffile}
\end{align}

there is a dependence on the t-channel parameter $t$. The logarithmic factors in equation~\ref{eqn-FeynZBox} can be simplified to

\begin{equation}
    \left[ \log^2\left(\frac{s+t}{M_Z^2}\right) - \log^2\left(\frac{t}{M_Z^2}\right)\right] = \log\left(\frac{s}{M_Z^2}\right)\log\left(\frac{1+\cos(\theta)}{1-\cos(\theta)}\right) + const.
\end{equation}

which is dependent on the scattering polar angle. This changes the differential cross-section to prefer wider angles than the Born level cross-section. We can also extend this to the W box diagrams

\begin{align}\label{eqn-FeynWBox}
    \begin{fmffile}{feynman2}
    \begin{fmfgraph*}(100,100)
        \fmfleft{i1,i2}
        \fmfright{o1,o2}
        \fmf{fermion}{i1,v1}
        \fmf{photon}{v1,v2,o1}
        \fmf{fermion}{v3,i2}
        \fmf{photon}{v3,v4,o2}
        \fmf{fermion,lab=$\nu_e$,lab.dist=-0.2w}{v1,v3}
        \fmf{photon,lab=$W$,lab.dist=0.1w}{v2,v4}
        \fmflabel{$e^-$}{i1}
        \fmflabel{$e^+$}{i2}
        \fmflabel{$\gamma$}{o1}
        \fmflabel{$\gamma$}{o2}
    \fmfv{lab=$\sim \frac{\alpha_W}{4\pi} \left[ \log^2\left(\frac{s+t}{M_W^2}\right) - \log^2\left(\frac{t}{M_W^2}\right)\right]$,label.angle=0,label.dist=0.7w}{v3}
    \fmfv{lab=$+C_3\log^2\left(\frac{s+t}{M_W^2}\right)$,label.angle=0,label.dist=0.6w}{v2}
    \end{fmfgraph*}
    \end{fmffile}
\end{align}

which will only participate in the electron neutrino case and only for beams with left-handed weak isospin. Like $C_1$ and $C_2$, $C_3$ can be calculated from theory and other parameters. This additional term that depends on $C_3$ means that the W boson box diagrams will be dominant over the Z boson box diagrams when there are left-handed beams. Otherwise, the Z boson box diagrams will dominate. We also observe that, due to the similar dependence on $t$ as the Z box diagrams, we expect that the W box diagrams will prefer wide angles.

The contribution from these corrections was checked, in the unpolarized beams case, using MadGraph5. This is possible as there are models in MadGraph5 that allow for computing electroweak, loop and box corrections. From MadGraph5 we find that, in the unpolarized case and with these corrections, that the cross-section at center-of-mass energy of 250 GeV was 22.227 $\pm$ 0.002 pb. This gives the most accurate calculation of the cross-section of any of the methods used here. However, this calculation was not exhaustively checked, so it is likely less accurate than the accuracy reported by MadGraph5. Using the contributions from the cross-section as reported by MadGraph5, we also find that the box diagram contribution is roughly 0.1\% when using fully left-handed beams. Using this result we believe that it is reasonable to achieve, at least, a $10^{-4}$ theory precision on neutral events with photons cross-sections.

Four different event generators were tested, Babayaga, WHIZARD, KKMC and Madgraph5. A compilation of each event generator, its version used here, with features listed, can be found in table~\ref{tab-prog}. We find that there is no single event generator suitable for this study. The lack of support for polarized beams in the majority of event generators is troubling but we do not think it is representative of a lack of skill in the community. Current funding is mostly driven by the LHC, which does not give significant motivation for the development of polarized studies at lepton colliders because it is a hadron collider. If this were to change, we are hopeful that these issues would resolve too.

During this study, we found that MadGraph5 gave the most accurate unpolarized cross-section calculation but it was unable to reliably generate events. This, coupled with the fact that we want to use polarized beams in these studies, means that we must use WHIZARD. Even though WHIZARD lacks the higher-order corrections that are relevant to precision measurements of the neutral event processes.

For precision measurements, these are not corrections that can be ignored. Currently there is no particle physics event generator that has a dedicated, and vetted, environment for generating $X^0\gamma$ events with these kinds of corrections. It may be possible to use WHIZARD as it allows for the use of OpenLoops and its NLO corrections. Of which the OpenLoops library of $eevv\_ew$ contains such corrections for the DiGamma channel. It may also be feasible to use MadGraph5, but it is developed more so for hadron colliders and does not allow for polarized beams in its (N)NLO corrections. This is an area of research that requires development and work on theory and event generation that would warrant an entirely new thesis. Thus, it is outside the scope of this work.

\section{Measuring the $X^0\gamma$ Photon Spectra}\label{subsec-Measuring}

As discussed in subsect~\ref{subsec-deltaR}, monophoton measurements have been done at LEP in order to constrain the neutrino couplings. Here we will expand on this in a broader sense and look at the photon spectra of neutral events. By neutral events we mean events that have no reconstructed charged particles. This would mean all measured EM showers should be tagged as photons and there should be no hadronic showers or jets. Going forward we will refer to neutral events with photons as $X^0\gamma$ events, with $X^0$ referring to the neutral and/or unmeasured particle(s). Since there can be numerous sources of soft radiation we will restrict our photon spectra measurement to hard radiation with a $p_{T}$ cut of 1 GeV. We will use ILC250 in this section as the benchmark run. Unless otherwise noted, assume the run is for ILC250. We will also use detector effects per the ILD designs but with the exception of the LumiCal, where we will use the GLIP LumiCal design. To reiterate what was done in section~\ref{subsec-SimHeader}, we will use WHIZARD to generate events. We will also use CIRCE files, as generated from the output of GuineaPig++, to include beam effects in these events. The values use in GuineaPig++ can be found in appendix~\ref{app-GP}. Detector effects will be added with parametric models of the ILD and GLIP LumiCal detectors. We have also assumed here the particles generated are correctly identified 100\% of the time.

We propose a search of $X^0\gamma$ events as a way for test precision Standard Model predictions in addition to looking for BSM physics. To explain, suppose that, after all Standard Model contributions to the $X^0\gamma$ photon spectra are accounted for, there is a residual between the data and model. Such an experimental finding would then provide two motivations. First, it puts focus on theorists to improve the Standard Model contributions to include more higher-order corrections and, thus, higher precision in theory. Second, if the higher precision Standard Model contributions fail, then it is a contrapositive proof of BSM physics. 

\begin{table}
\centering
\caption{A collection of the events that are expected to contribute to the $X^0\gamma$ spectra at $\sqrt{s}$ of 250~GeV. We distinguish muon and tau contributions from electron contributions by using $l$ to represent muons and taus only. Similarly, the subscript for neutrinos is distinguished in this way. We also separate the electron neutrinos into their different contributions since individual couplings will be measured later. An inner acceptance of $1^\circ$ is used for cross-sections. A cross-section precision of at least 1\% was required for each cross-section computation and done with 800~$\text{fb}^{\text{-1}}$ of events.$^\dagger$ In addition to the acceptance cuts, we require that any charged particle must exit down the beam pipe or be less than 1~GeV and that there must be at least one photon of 1~GeV or greater in the acceptance.}
\label{tab-phospec}       
\begin{tabular}{|c|c|c|}
\hline
\multicolumn{3}{|c|}{\textbf{$X^0\gamma$ Processes at $\sqrt{s} =\text{250}$~GeV , $(P_-,P_+)=(-80\%,+30\%)$}}\\
\hline
\textbf{Process} & \textbf{Cross-section (pb)} & \textbf{After Cuts (pb)} \\\hline
$e^+e^-\gamma^\dagger$ & 107  & 96.9  \\
$l^+l^-\gamma$ & -*  & $<$ 0.001 \\
$\nu_e\bar{\nu}_e\gamma$ , $W$ only& 8.34  & 8.34   \\
$\nu_e\bar{\nu}_e\gamma$ , $Z$ only& 2.83  & 2.83   \\
$\nu_e\bar{\nu}_e\gamma$ , Combined& 13.2 & 13.2   \\
$\nu_l\bar{\nu}_l\gamma$ & 5.60  & 5.60   \\
$\gamma\gamma$ & 38.9 & 38.9  \\
$ZZ\gamma$ & 0.180 & 0.101   \\
$ZH$ & 0.0144 & 0.00341   \\  
$\nu_e\bar{\nu}_eH$ & 0.00249 & 0.00249   \\  
\textbf{Total : } & \textbf{164.9} & \textbf{154.7} \\\hline
\end{tabular}
\\[5pt]
\small{*Lepton pair production was simulated in WHIZARD with cuts, so only the post-cut cross-section is recorded.}\\
\small{$^\dagger$Radiative Bhabhas were only generated to 1\% of the integrated luminosity of the other processes due to very long simulation times.}
\end{table}

To obtain data and models for pursuing a SM and BSM analysis we need to simulate $X^0\gamma$ events with an event generator. We use WHIZARD as it has been used in a similar study~\cite{Kalinowski_2020}. In this context WHIZARD has also been verified to be equivalent in output to KKMC, an older simulation package that was considered, and in many cases still considered, the most precise simulator for electron-positron interactions~\cite{Jadach_2023}. 

The contributions to $X^0\gamma$ events can be found in Table~\ref{tab-phospec}. All particles were required to have more than 1~GeV of energy, or 2\% of center-of-mass energy, otherwise they were considered unmeasured. We started with generating 800~$\invfb$ of data at 250~GeV with beam polarization of $(P_-,P_+)=(-80\%,+30\%)$. We allowed for WHIZARD to have up to two radiative corrections for each process. The cross-sections that are quoted are those that have a photon within the $1^\circ$ inner acceptance and no charged particles in the acceptance. These initial cuts lead to the charge lepton-antilepton channel having a negligible cross-section. Even after this cut Bhabha scattering, particularly as \Gls{BERB}, is the dominant channel before the ``after cuts'' found in Table~\ref{tab-phospec}. We then apply a cut to reject any events that have a charged particle within the inner acceptance. This is different from the previous cut as it includes particles generated from decays, such as the Z boson and Higgs bosons in some of the processes, and effects from the beam effects that are included from using CIRCE in WHIZARD. This reduces the BERB, $ZZ$ and $ZH$ cross-sections slightly. For this analysis we also assume that muons and taus are 100\% discriminated with photons.

\begin{figure}[h!]
\centering
\includegraphics[width=12cm, trim=3.5cm 6cm 3.5cm 6cm]{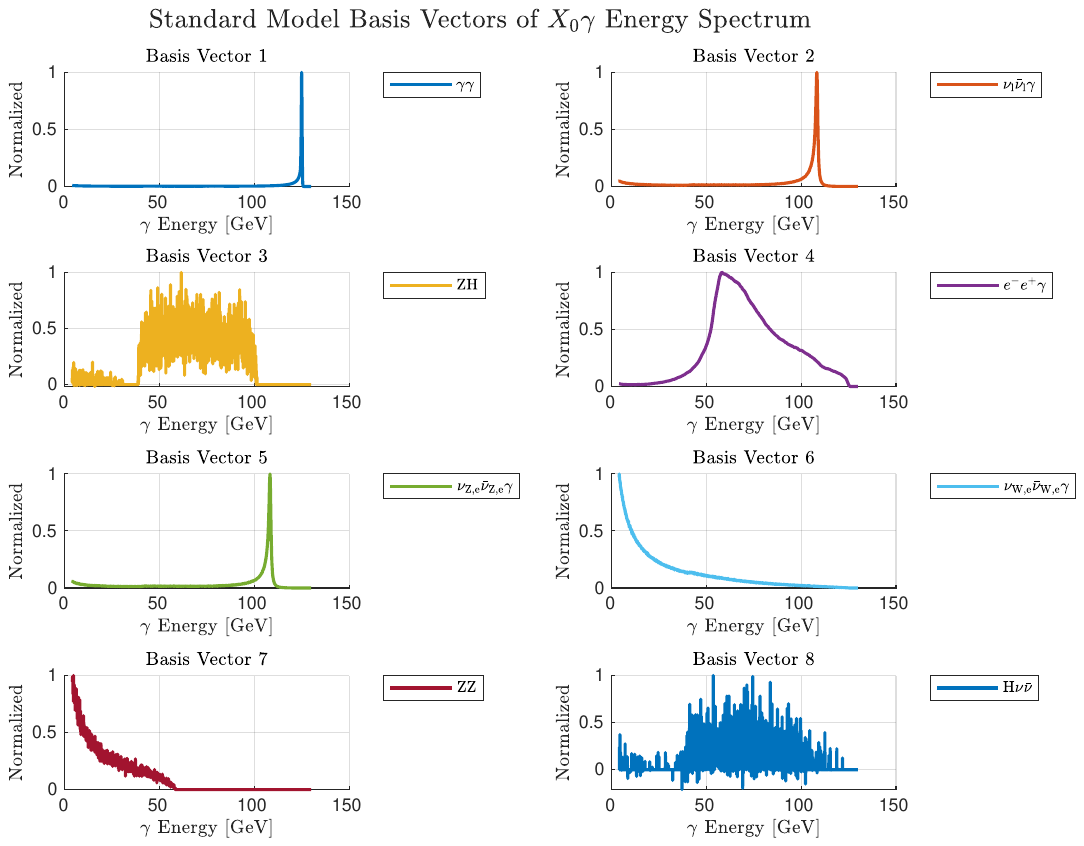}
\caption{Collection of the Standard Model processes that were used in the analysis of the $X_0\gamma$ events. This only shows the energy spectrum for the sake of brevity. These include up to two radiative corrections.}
\label{fig-basisvec}
\end{figure}

After generating the generator level data we apply a parametric model of ILD with a parametric model of GLIP LumiCal to create smeared detector level data. We then generate reconstruction level values for performing analsysis. We then conducted two rounds of analysis. the first round of analysis used only the photon energy spectrum, which includes all reconstructed photons in the event. The second round of analysis included additional reconstructed values. First we note $m_{\gamma\gamma}$, which is the invariant mass of the two highest $p_{\text{T}}$ photons in the event. This is particularly useful for the diphoton and $\text{ZH}$ events, where the invariant mass can be unique compared to other processes. This is because diphoton invariant mass peaks near the design center-of-mass energy and $\text{ZH}$ events include the possibility of Higgs decay to $\gamma\gamma$. The distribution of $p_\text{T,max}$, which is the highest $p_{\text{T}}$ photon of the event, was added. The motivation for this addition was the separation of events with high energy hard photons, comparable or greater than half the beam energy, from events where the hard photons are softer, being less than half the beam energy. This is particularly advantageous for $\text{ZZ}$ events, where the photons cannot have more than $\approx\sqrt{s}-2M_{\text{Z}}$ of energy. Two additional distributions, $\gamma_{\text{Z}}$ and $\gamma_{\text{H}}$, were included to help improve the Higgs and neutrino related measurements. These two distributions focus on a narrow window about the expected Z-pole radiative return photons and Higgs decay photons. The Z-pole radiative return photons have an energy of 
\begin{equation}\label{eqn-ZRadRet}
E_{\gamma\text{Z}} = \frac{s - M_{\text{Z}}^2}{2\sqrt{s}}
\end{equation}
or about 108 GeV at a center-of-mass energy of 250 GeV. We also included the missing energy along the longitudinal and transverse axes. For demonstration we include plots of the energy spectra as can be seen in figure~\ref{fig-basisvec}.

For fitting the photon spectra we use Singular Value Decomposition (SVD) Least squares to fit a given photon spectra, $P_\gamma(E)$, from the contributions expected from the Standard Model. Such that the spectra is
\begin{equation}\label{eqn-phospecfit}
    P_\gamma(E) = \sum_i^N a_\text{i} P_\text{i}(E)
\end{equation}
a sum of the constituent spectra, $P_{i}(E)$, with weights, $a_i$, determined by the least squares fit. Here $E$ is the energy after the detector smearing. In general it should be assumed for this work that the detector level values are used. The Standard Model expected spectra are generated from WHIZARD, but done individually so that sampling for said spectra can be controlled. The fit has been implemented in MATLAB as the matrix solvers in ROOT are unable to perform the decomposition for the spectra used here, probably because ROOT's matrix operations are outdated and insufficient. We also want to be able to add additional differentials to this so the final spectra we will have will be 
\begin{equation}\label{eqn-phospecfit2}
    P_\gamma(A_j) = \sum_j^M\sum_i^N a_\text{i} P_\text{i}(A_j)
\end{equation}
where $A_j$ are $j$ additional terms that we are fitting spectra over. We will do this in the second analysis, where we will add the additional reconstruction terms. For clarity here we have stiched all of our data histograms into one histogram to include all of our data with the various reconstructed values. This results in no fitting of covariances between the different reconstructed values and processes and instead everything is treated as independent.

The fitting of the generated data sample was initially performed using MatLab's Least Squares SVD fitting method with Trust Region bounds~\cite{MATLAB}. Each physics process was used as an input matrix, to be considered a part of the total matrix, for fitting purposes. As such, a basis of each physics process could be fitted for. The coefficients, which represented cross-sections after cuts, of each physics process was constrained to be between 50\% and 200\% of the expected value under the assumption of Standard Model convolved with cuts being correct. The initial fitting used the energy spectrum of $X_0\gamma$ photons as the matrix to be solved. Plots of these basis vectors for the matrix can be seen in figure~\ref{fig-basisvec}. Only using the energy spectrum resulted in poor fitting results for fitting the contributions from all but the diphoton and radiative Bhabha processes. The latter two processes were fitted to around 1\% and 0.1\% precision respectively.

\begin{figure}[h!]
\centering
\includegraphics[width=14cm, trim=0.5cm 3cm 0.0cm 3cm, clip]{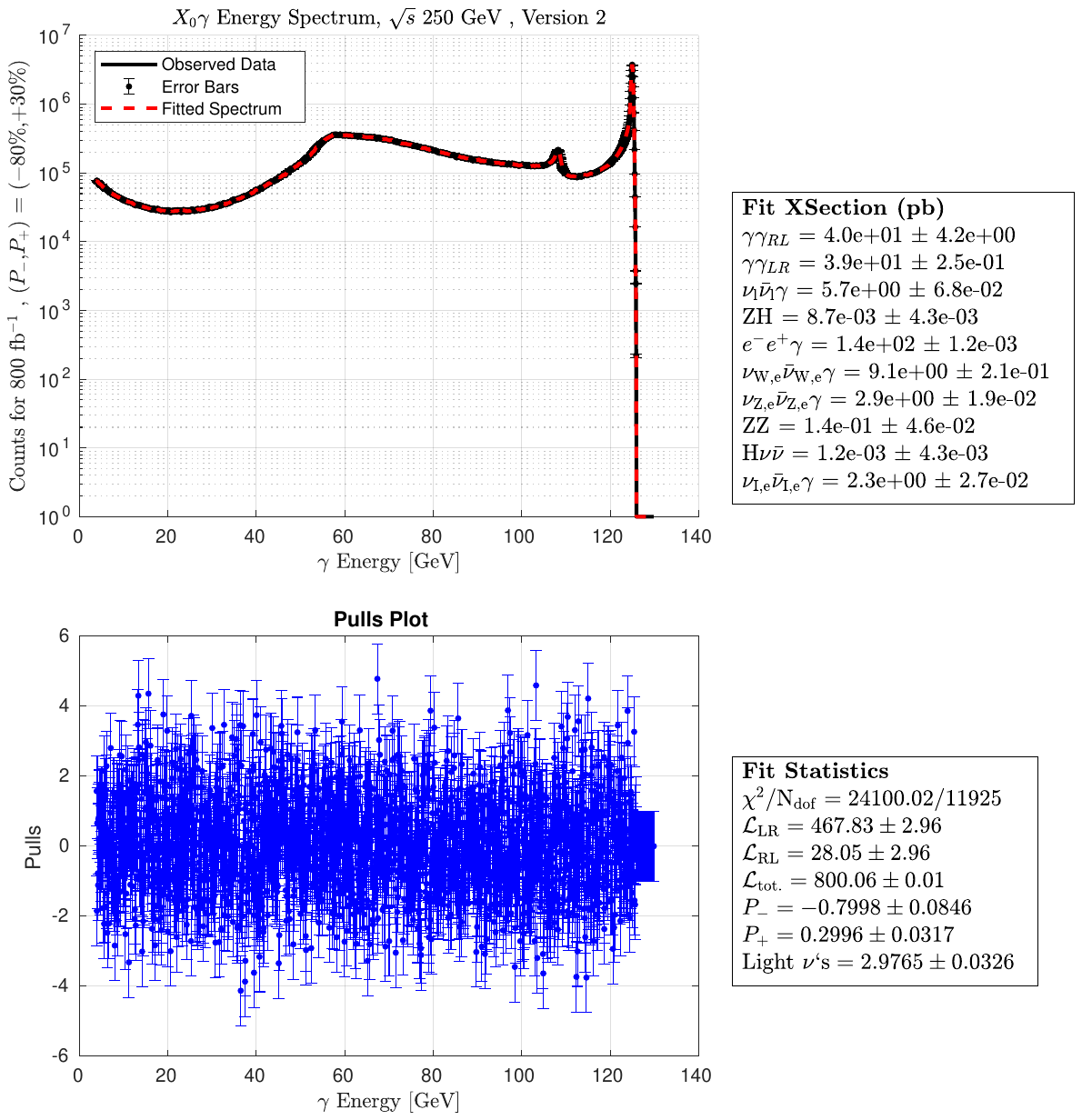}
\caption{Result of the SVD least squares fitting of the $X^0\gamma$ photon energy spectra. This fit includes fitting of quantities of $\gamma_{M}$, $\gamma_{max}$ and $\gamma_{Z}$ and $\gamma_{H}$ too. These quantities are described in Sect.~\ref{subsec-Measuring}. This version, includes full radiative corrections, not just fast simulation of ISR, up to two photons. It also includes beam effects from CIRCE2 and it removes fit variables for helicities in all but the DiGamma channel~\cite{Ohl:1996fi}. Where the ability to fit both helicity combinations is useful for determining luminosity and polarization. This does not include detector effects.}
\label{fig-gamspecfit1}       
\end{figure}

To improve fitting results, we added the previously discussed additional reconstruction values. The results of this second test, can be seen in figure~\ref{fig-gamspecfit1}.
As part of this second test, the helicity combinations of the diphoton process were fitted separately. This is made possible by using bounds on the fitted cross-section to be similar to the expected Standard Model values. If the fit tries to increase the LR or RL cross-section, to fit only with the one cross-section, since they are otherwise identical, it cannot proceed without violating bounds. These were included as sanity checks and not as a means to perform precision measurements. The integrated luminosities quoted are in units of $\invfb$. The p-value calculated from the $\chi^2$ per degrees of freedom indicates a poor fit. Upon additional work it was discovered that the poor fit of figure~\ref{fig-gamspecfit1} was from a miscalculation of the uncertainty in the bin errors of the histogram.

Since the BERB cross-section is polarization independent at leading order and the diphoton cross-section isn't, we can use the fitted cross-section values, assuming we already know the total integrated luminosity, to get measures of the different luminosity components. From these luminosity components, which are broken down by helicity combination, we can then estimate the beam polarization. The results of figure~\ref{fig-gamspecfit1} indicate that this approach is able to attain roughly 10\% precision on the luminosity components and beam polarization. These are not impressive measurements, but may be useful as sanity checks. In addition to these quantities, we are also able to infer the number of light neutrino species using methodology that will be further expanded on in section~\ref{sec-NeutrinoCoup}. To allow for this measurement, we have broken down the neutrino cross-sections into their constituent processes as depending on the W, Z, and WZ interference contributions by using WHIZARD to generate each of these processes individually, as well as in total. Finally, we note that we use a large number of bins as the SVD fit process loses quality when the number of bins decreases. We suspect that this is because the shape of the distributions are hard to differentiate when the number of bins is small.





\subsection{Measurement of Light $\nu$ and their Couplings}\label{sec-NeutrinoCoup}

For this section we will start with ILC250 with beam polarizations of $(P_-,P_+)=(-80\%,+30\%)$ as this polarization combination results in considerable W boson contributions. This contribution is key to measuring the W and WZ interference components for the studies that will follow. Starting with the fitting results of the previous section, we now have fits for the cross-sections of the various $X^0\gamma$ processes. For the electron neutrinos we also broke these down into the pure Z boson, pure W boson and W boson and Z boson interference terms. According to previous studies, having all three of these terms can enhance the sensitivity to measuring the left-handed and right-handed couplings of the electron neutrino~\cite{Carena:2003aj}. By using the cross-sections themselves we can derive
\begin{equation}\label{eqn-neu}
    \begin{gathered}
        \sigma_{\text{tot.}} = \sigma_{\mu} + \sigma_{\tau} + \sigma_\text{e} = 2\sigma_\text{ZZ} + \sigma_\text{e}\rightarrow \\
        \sigma_{\text{tot.}} - \sigma_{\mu} - \sigma_{\tau} = \sigma_\text{ZZ} + \sigma_\text{WW} + \sigma_\text{WZ} = \sigma_\text{e} \rightarrow \\
        \sigma_{\text{tot.}} \approx 3\sigma_\text{ZZ} + \sigma_\text{WW}+\sigma_\text{WZ}
    \end{gathered}
\end{equation}
that we can solve for the individual components of the electron neutrino cross-section, under the assumption that there is lepton universality. All that we need is the Z-like cross-section, from the previously performed fit which we label as $\sigma_\text{ZZ}$, the W-like cross-section, which we label as $\sigma_\text{WW}$, and the interference cross-section which we label as $\sigma_\text{WZ}$. 

\begin{figure}[h!]
\centering
\includegraphics[width=14cm]{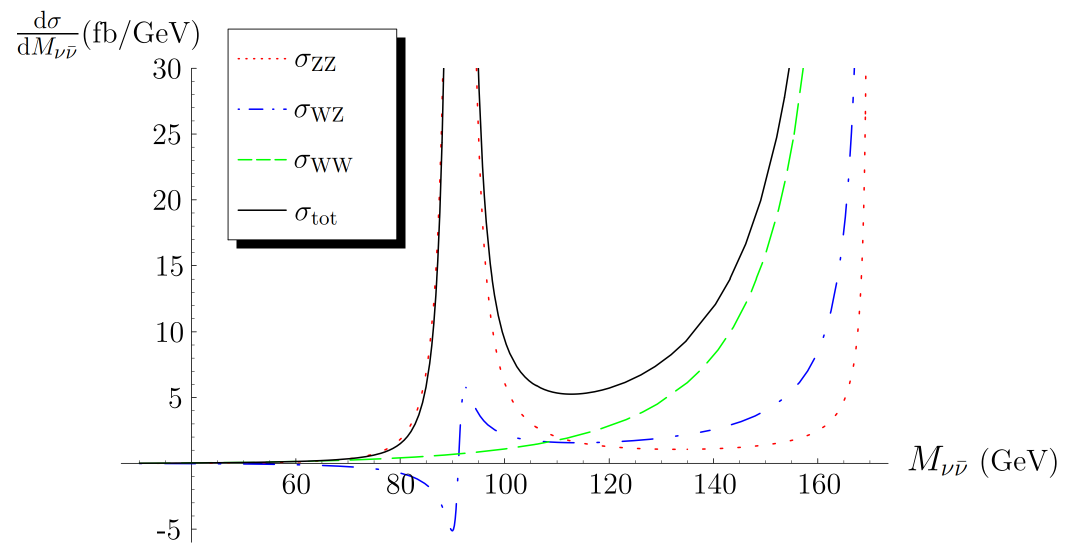}
\caption{The differential cross-section for $\ee \to \nu\bar{\nu}\gamma$ in terms of the dineutrino ($\nu\bar{\nu}$) invariant mass. Done for $\sqrt{s}=$ 170~GeV with an unpolarized collider and assuming that the number of neutrino species and currently assumed Standard Model neutrino couplings are correct. The sharp increase of the
differential cross section as the dineutrino mass approaches the center-of-mass energy is due to an infrared singularity as the radiated photon approaches zero energy. Figure credit~\cite{Carena:2003aj}.}
\label{fig-neumass}       
\end{figure}

\begin{figure}[h!]
\centering
\includegraphics[width=14cm]{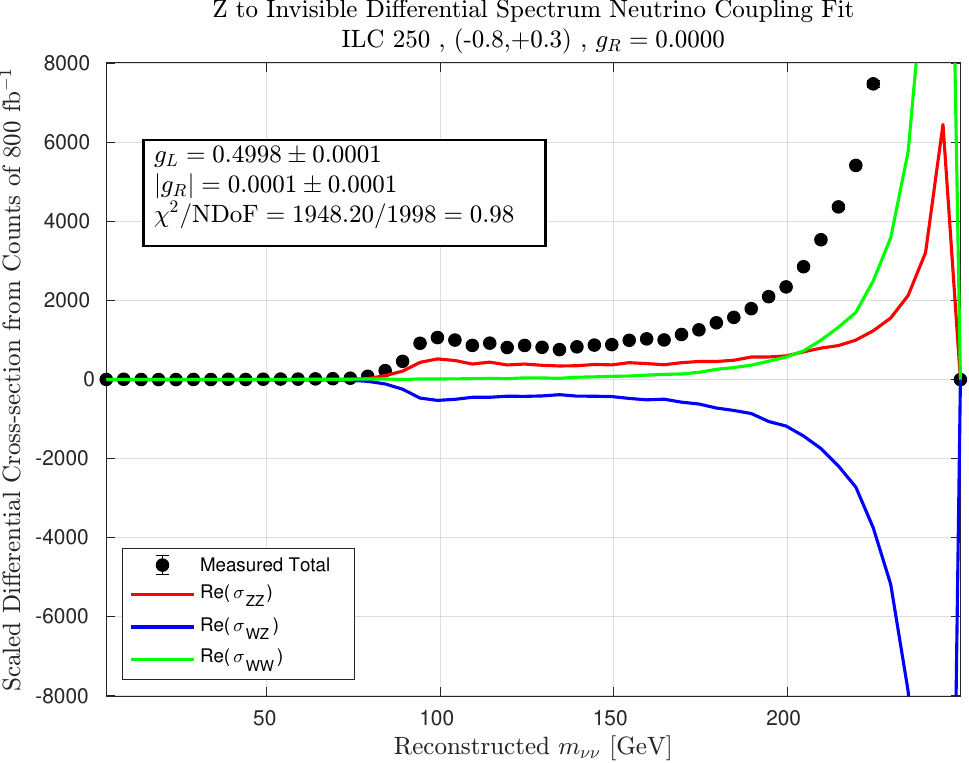}
\caption{The differential cross-section for $\ee \to \nu\bar{\nu}\gamma$ in terms of the dineutrino ($\nu\bar{\nu}$) invariant mass. Done for ILC at $\sqrt{s}=$ 250~GeV with polarized beams of (-80\%,+30\%) and assuming that the number of neutrino species and currently assumed Standard Model neutrino couplings are correct. We have also set the right-handed coupling for this to 0 and the left-handed coupling to +0.5. We have chosen to down-sample the number of bins to 50 for the purpose of presentation even though the original bins were 2000 bins in photon energy. Due to this, the binning here is not square but was interpolated to approximate being square and the number of degrees of freedom reported is larger than what can be seen. We include the measured differential cross-section with error-bars though the error bars are within the markers.}
\label{fig-neumassReal}       
\end{figure}

From the Standard Model we can rewrite equation~\ref{eqn-neu}
\begin{equation}\label{eqn-neuZ}
    \frac{d\sigma_{\nu,tot.}}{dz} \approx \sigma_\text{ZZ}(z)(g_\text{L}^2+g_\text{R}^2) + \sigma_\text{WZ}(z)g_\text{L} + \sigma_\text{WW}(z)
\end{equation}
as a differential cross-section in the energy fraction of the radiated photon, $z$. This includes the muon and tau contributions. This choice is, for the sake of this analysis, a necessity as we cannot differentiate the flavor content of the $\sigma_\text{ZZ}$ components. Equation~\ref{eqn-neuZ} gives us the dependence on the neutrino left-handed coupling, $g_\text{L}$, and the absolute value of right-handed coupling, $|g_{R}|$, since the term is squared. We will restrict our results here to require that $g_\text{L}$ and $g_\text{R}$ be real valued. If we use a form of equation~\ref{eqn-neuZ} that has been integrated over $z$, and therefore gives us total cross-sections of the individual components, we can then fit for $g_\text{L}$ and $|g_\text{R}|$. We implement a simple script of this using least-squares fitting and find
\begin{equation}\label{eqn-initialFit}
\begin{gathered}
    g_\text{L} = 0.500 \pm 0.009 \\
    |g_\text{R}| = 0 \pm 300 ,
\end{gathered}
\end{equation}
which shows that we can measure the left-handed coupling well but cannot measure the right-handed coupling at all. This result is expected as the information in equation~\ref{eqn-neuZ} that would allow for $g_\text{L}$ and $|g_\text{R}|$ to be differentiated is dependent on the value of $z$ or, more precisely, the dineutrino mass that can be calculated from the center-of-mass energy and $z$ under the assumption of single photon emission~\cite{Carena:2003aj}. This can be seen in figure~\ref{fig-neumass}, where the dineutrino mass for 170~GeV center-of-mass energy is shown with the differential cross-sections of the ZZ, WW, and WZ contributions.
Since the WZ term is an interference term once could argue that the representation is complex valued but here we are using real valued data so it is not complex valued. It should not be considered physical as the true cross-sections cannot be decomposed into these complex valued representations. By fitting as done in section~\ref{subsec-Measuring} we can solve for $\sigma_\text{WZ}$. Please note, that for this fit to work, the initial weighting of each contribution must be equal. If they are unequal, then one will introduce biases in the values of $g_\text{L}$ and $g_\text{R}$ that arise from the fact that the cross-sections are not equal. We note both of these things so that observers understand how to correctly do this method and why there are negative-valued cross-sections. To provide an additional example, but with data that has detector effects and beam effects, we provide figure~\ref{fig-neumassReal}.
Please note that the WZ component here is not the same as the theory WZ component. The reconstructed WZ component is derived from the difference of the ZZ, WW and total cross-section components. The theory WZ component is a calculated value from the interference terms in the matrix elements. They are similar but not equivalent. To reiterate, the values presented are also not theory expectations and are the results from fitting data per the methodology of section~\ref{subsec-Measuring}. We refer to the y-axis as the scaled differential cross-section from the counts as they are fitted values derived from the true experimental counts, and they are not true differential cross-sections. We retain the label for ease of comparison. The values in figure~\ref{fig-neumassReal} are also for ILC250, so the center-of-mass energy is higher than figure~\ref{fig-neumass}. In terms of sanity checks, we anticipate that the $\sigma_\text{ZZ}$ and $\sigma_\text{WZ}$ terms should both have inflection points near, but above, $m_\text{Z}$. We anticipate that the inflection point is above $m_\text{Z}$ for three reasons. There are radiative effects here that cause the maximum energy photon, which is the photon used for reconstructing $m_{\nu\nu}$, to be lower energy than what would be expected from equation~\ref{eqn-ZRadRet}. The luminosity spectrum influences the reconstructed dineutrino mass too as the center-of-mass energy is, on average, below the nominal center-of-mass energy. This then causes the reconstructed dineutrino mass to be higher than it would be if it was at the nominal center-of-mass energy. The $\sigma_\text{ZZ}$ and $\sigma_\text{WZ}$ contributions both contain an underlying exponential component that increases towards the nominal center-of-mass energy. This underlying exponential component causes the inflection point of the Z pole to shift towards higher dineutrino mass values.

We have also included the fit result for the couplings here. For this fit we have improved the simple script to handle the differential in dineutrino mass. This was done in MatLab using the least squares fit of equation~\ref{eqn-neuZ}~\cite{MATLAB}. The values for each component are taken from the cross-section fits and then injected with noise before doing the fit for the couplings. We find improved measurements
\begin{equation}\label{eqn-ILC250Fit}
\begin{gathered}
    g_\text{L} = 0.49983 \pm 0.00014 \\
    |g_\text{R}| = 0.00011 \pm 0.00007
\end{gathered}
\end{equation}
of the left-handed and right-handed neutrino couplings. By using the differential fitting we have improved both measurements and made the right-handed coupling measurement result useful. The measurement of the left-handed coupling was improved by a factor of $\approx60$. Both of these measurements outperform the current global fit world-best measurements of $g_\text{L} = 0.495 \pm 0.022$ and $g_\text{R} = 0.005 \pm 0.022$ by $\approx \times300$ for $g_\text{L}$ and $\approx \times 150$ for $|g_\text{R}|$~\cite{AtzoriCorona:2025xwr}. We note that the current global fit prefer a negative value of $g_\text{L}$ but our simulation was done with a WHIZARD model where $g_\text{L}$ was set to a positive value of $g_\text{L}$ so we flip the sign here to keep the comparison relevant. The current global best fits are also for a lower center-of-mass energy, $m_\text{Z}$, as compared to the value of 250~GeV used here. This discrepancy is important; if there are energetic runnings in the neutrino couplings from higher-order Standard Model corrections, or new physics corrections from BSM physics, then it will only appear in measurements of $g_\text{L}$ and $g_\text{R}$ at higher energies.

\begin{figure}[h!]
\centering
\includegraphics[width=16cm]{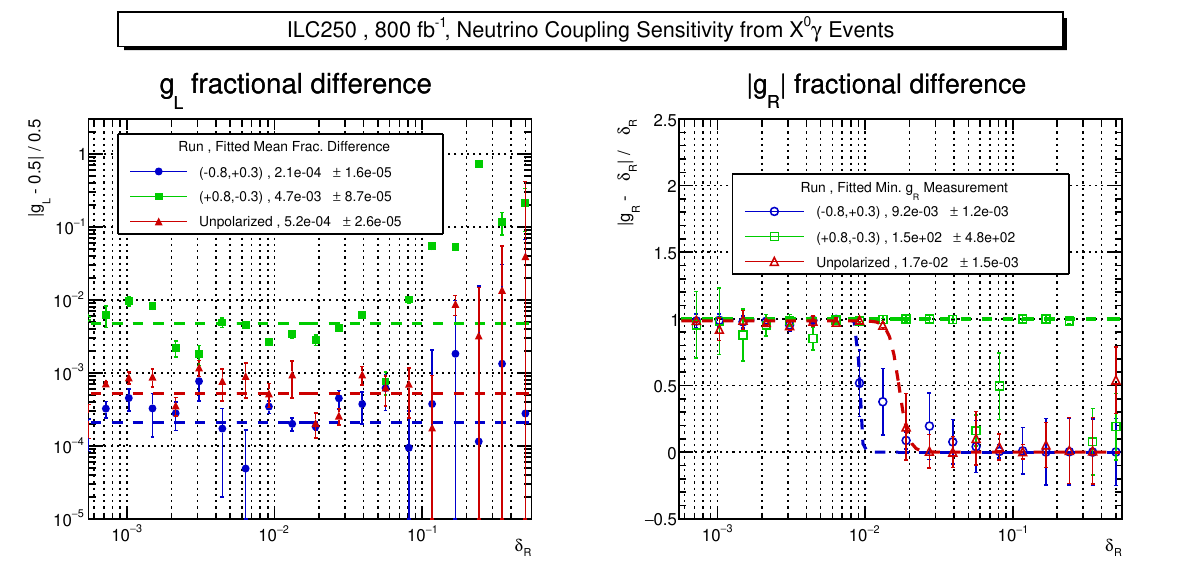}
\caption{Result of the fits of $g_\text{L}$ and $|g_\text{R}|$ for electron neutrinos from changing the true value of $|g_\text{R}|$, which we note as $\delta_R$. We use three different sets of values for beam polarization at ILC250 so that changes in sensitivity with changes in beam polarization can be seen. The left plot was fit for a constant value to evaluate the accuracy of the $g_\text{L}$ measurement. The right plot was fit to a sigmoid curve to find the point where sensitivity is lost for measuring $|g_\text{R}|$. All samples reflect 800~$\invfb$ of data.}
\label{fig-gRgL}       
\end{figure}

We can repeat this process but change the value of $g_\text{R}$ for the Z boson in WHIZARD using the Z$'$ model~\cite{Kilian_2011}. By using the Z$'$ model we have access to edit all of the couplings of the Z boson. Normally, these couplings are not easily tunable in WHIZARD and instead are hard-coded. We choose values of $g_\text{R}$ to be positive in the work done here. We then repeat the process done earlier in this section and in section~\ref{subsec-Measuring} such that we can fit the cross-sections and couplings. For reference, we expect the ZZ component of the neutrino cross-section to increase by $g_\text{L}^2+g_\text{R}^2$ since the Standard Model value of $g_\text{R}$ is zero and the increase follows equation~\ref{eqn-neuZ}. The relation for the increase
\begin{equation}\label{eqn-ZRatio}
    \frac{ZZ}{ZZ_{\text{R}=0}} = \frac{g_\text{L}^2+g_\text{R}^2}{g_\text{L}^2} = 1+\left(\frac{g_\text{R}}{g_\text{L}} \right)^2
\end{equation}
indicates that the change in the ZZ component is dependent on the ratio of $g_\text{R}$ and $g_\text{L}$. As an example, using $g_\text{R}$ of 0.05 and $g_\text{L}$ of 0.5 in equation~\ref{eqn-ZRatio} gives a ratio of 1.01, a 1\% increase. Since this change in the cross-section component is small, we must be sure that the WHIZARD simulation is done to sufficiently high integration statistics so that the values are accurate. For the sake of plotting and making things clearer, we will note the non-zero true value of the right-handed coupling as $\delta_R$. A plot of the fractional differences of the fitted $g_\text{L}$ and $g_\text{R}$ to their true values can be found in figure~\ref{fig-gRgL}.
We have provided three cases of 800~$\invfb$ data samples with varying beam polarizations of unpolarized, (-80\%,+30\%) and (+80\%,-30\%). For the plots of the fractional difference of $g_\text{L}$ we have used a fit to a constant value to evaluate the accuracy of the $g_\text{L}$ measurement. We find that the accuracy of the $g_\text{L}$ measurement is greatest for the (-80\%,+30\%) beam polarization data sample, being $\approx\times2.5$ times better than unpolarized and $\approx\times20$ better than the (+80\%,-30\%) data sample. For the $|g_\text{R}|$ data we have performed a sigmoid curve fit to determine the middle point of the s-curve, which we will treat as the point that the measurement loses sensitivity. From the sigmoid curve fits we find that the (-80\%,+30\%) beam polarization data sample is superior, being able to measure values of $|g_\text{R}|$ down to $9.2\times10^{-3}$. This is roughly twice as sensitive as the unpolarized data sample which can measure $|g_\text{R}|$ to $17\times10^{-3}$. The results of the $g_\text{L}$ fit has a caveat; the accuracy of the $g_\text{L}$ measurement changes significantly based on the value of $\delta_R$ when $\delta_R$ approaches the same value as $g_\text{L}$. When $\delta_R$ is large the fit of the neutrino couplings has trouble differentiating the couplings. We also suspect that this reflects that a high value of $\delta_R$ will lead to the ZZ component of equation~\ref{eqn-neuZ} becoming larger and more dominant which previous studies have found makes it harder to measure $|g_\text{R}|$ and the sign of $g_\text{L}$~\cite{Carena:2003aj}. So there are two processes at play that degrade performance in this extreme case of large $\delta_R$. We also observe that the sensitivity to $|g_\text{R}|$ is essentially nonexistent in the (+80\%,-30\%) data sample. We expect that this is from the comparatively low statistics of the W boson in this data sample. When the W boson statistics are small the ZZ term of equation~\ref{eqn-neuZ} dominates the WZ and WW terms. This then means that equation~\ref{eqn-neuZ} is effectively a fit of just $|g_\text{L}|$. The sensitivity to both $|g_\text{R}|$ and the sign of $g_\text{L}$ is lost. We did not check for the sign of $g_\text{L}$ here and kept it constant, so future studies should investigate whether this hypothesis is true or not.

\begin{figure}[h!]
\centering
\includegraphics[width=16cm]{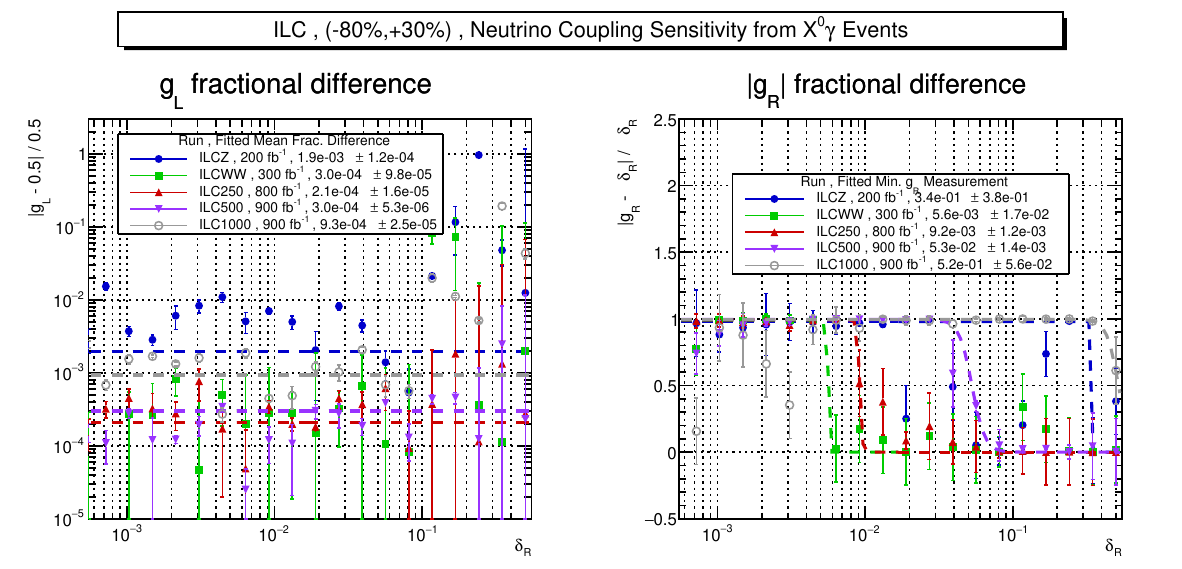}
\caption{Result of the fits of $g_\text{L}$ and $g_\text{R}$ for electron neutrinos from changing the true value of $|g_\text{R}|$, which we note as $\delta_R$. We use the five different values of center-of-mass energy at ILC as outline in previous studies in the work. The left plot was fit for a constant value to evaluate the accuracy of the $g_\text{L}$ measurement. The right plot was fit to a sigmoid curve to find the point where sensitivity is lost for measuring $|g_\text{R}|$.}
\label{fig-gRgLAll}       
\end{figure}

\begin{figure}[h!]
\centering
\includegraphics[width=14cm]{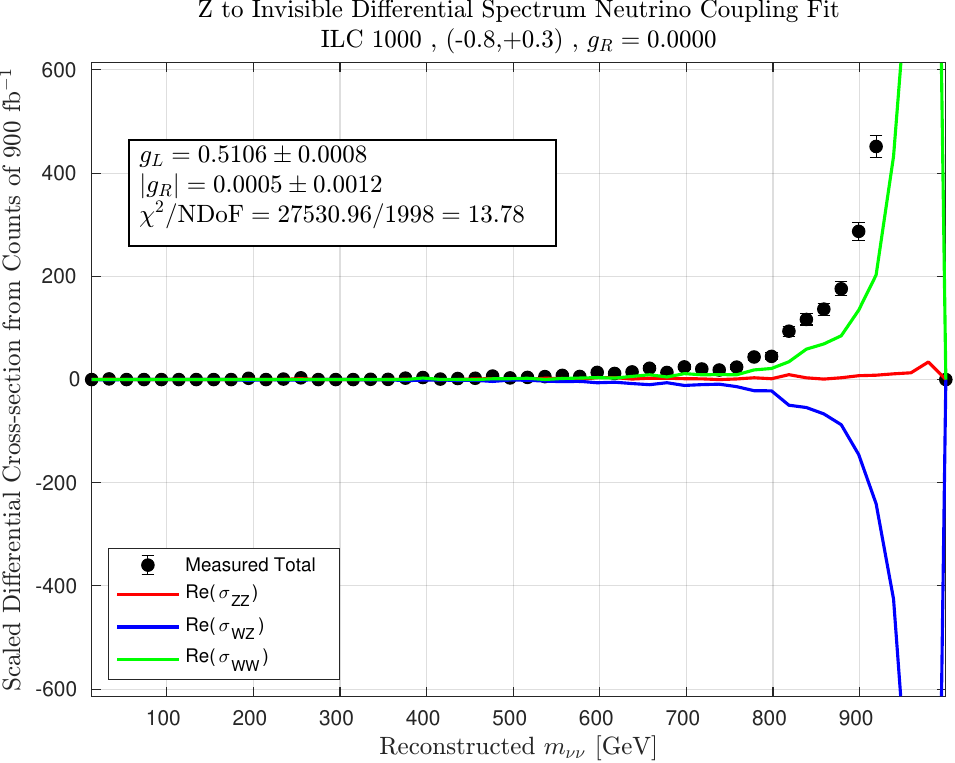}
\caption{The differential cross-section for $\ee \to \nu\bar{\nu}\gamma$ in terms of the dineutrino ($\nu\bar{\nu}$) invariant mass. Done for ILC at $\sqrt{s}=$ 1~TeV with polarized beams of (-80\%,+30\%) and assuming that the number of neutrino species and currently assumed Standard Model neutrino couplings are correct. We have also set the right-handed coupling for this to 0 and the left-handed coupling to +0.5. We have chosen to down-sample the number of bins to 50 for the purpose of presentation even though the original bins were 2000 bins in photon energy. Due to this, the binning here is not square but was interpolated to approximate being square and the number of degrees of freedom reported is larger than what can be seen. We include the measured differential cross-section with error-bars though the error bars are within the markers.}
\label{fig-ILC1000Neu}       
\end{figure}

\begin{figure}[h!]
\centering
\includegraphics[width=16cm]{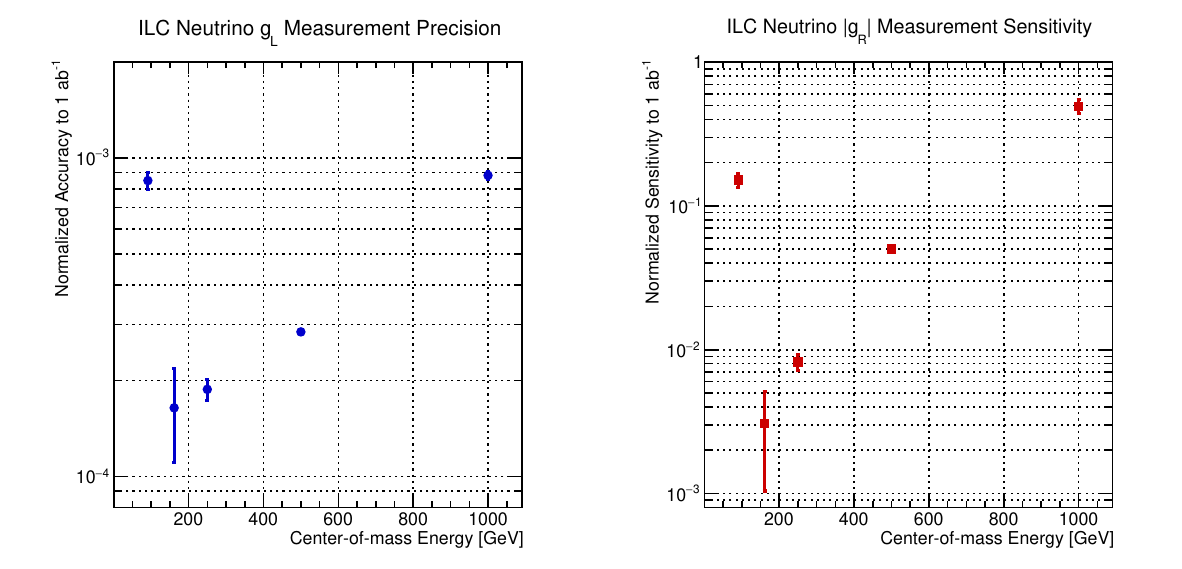}
\caption{Result of the fits of $g_\text{L}$ and $|g_\text{R}|$ for electron neutrinos from changing the true value of $|g_\text{R}|$ as computed by the fits of figure~\ref{fig-gRgLAll}. This is done for the various ILC center-of-mass energies used in this work. The y-axis has been normalized to reflect a sample size of 1~$\invab$, despite the original sample sizes being smaller. This is under the assumption that the measurements are statistics limited. The ILC250 and ILCWW perform best for both couplings.}
\label{fig-gRgLFair}       
\end{figure}

We extend the neutrino coupling measurement to other center-of-mass energies at ILC using the methods of section~\ref{subsec-Measuring} and the previous part of this section. The result, found in figure~\ref{fig-gRgLAll}, indicates that the WW threshold run is optimal to precisely measure both $g_\text{L}$ and $|g_\text{R}|$.
The results of figure~\ref{fig-gRgLAll} also indicate that the measurement of $|g_\text{R}|$ using $X^0\gamma$ events and this methodology degrades significantly at the Z pole and at 1~TeV center-of-mass energy. We suspect that this reflects the nature of computing the dineutrino invariant mass from the radiative photon. Using equation~\ref{eqn-ZRadRet} we can estimate that the radiative photon will be less than 1~GeV at the Z pole. Meaning that the radiative photon of the dineutrino system will be effectively unmeasurable. At the opposite extreme, at center-of-mass energy of 1~TeV the radiative photon energy will be 496~GeV, which is 99.2\% of the beam energy. This means that the identification of the dineutrino radiative photon will be in the same regime as beam energy processes, such as \Gls{SABS} and \gls{diphoton}s. As seen in figure~\ref{fig-ILC1000Neu}, the dineutrino mass spectrum for the WW contribution begins to dominate at higher energies too, meaning that the ZZ and WZ contributions will be dominated and significantly smeared out of the reconstructed dineutrino mass spectrum.
We also observe that the fit performs poorly in this regime as the ZZ component as the ZZ term is small and unable to be fit well. For these reasons, we expect that this result, that the intermediate energies of 250~GeV and the WW threshold are better for neutrino coupling measurements, is plausible. Finally, to provide a more fair comparison, we provide figure~\ref{fig-gRgLFair}, were we have compiled the $g_\text{L}$ and $|g_\text{R}|$ fitted result and then adjusted them to reflect an equal integrated luminosity of 1~$\invab$.
For the sake of providing a simpler comparison, we have normalized the results of each ILC run to 1~$\invab$ of integrated luminosity. We have also assumed that the original measurements of the neutrino couplings were statistics limited, and thus these projections reflect improvements from increased statistics. We find that the ILC250 and ILC W threshold runs are best suited for the measurement of electron neutrino $g_\text{L}$ and $|g_\text{R}|$ couplings.

\begin{figure}[h!]
\centering
\includegraphics[width=14cm]{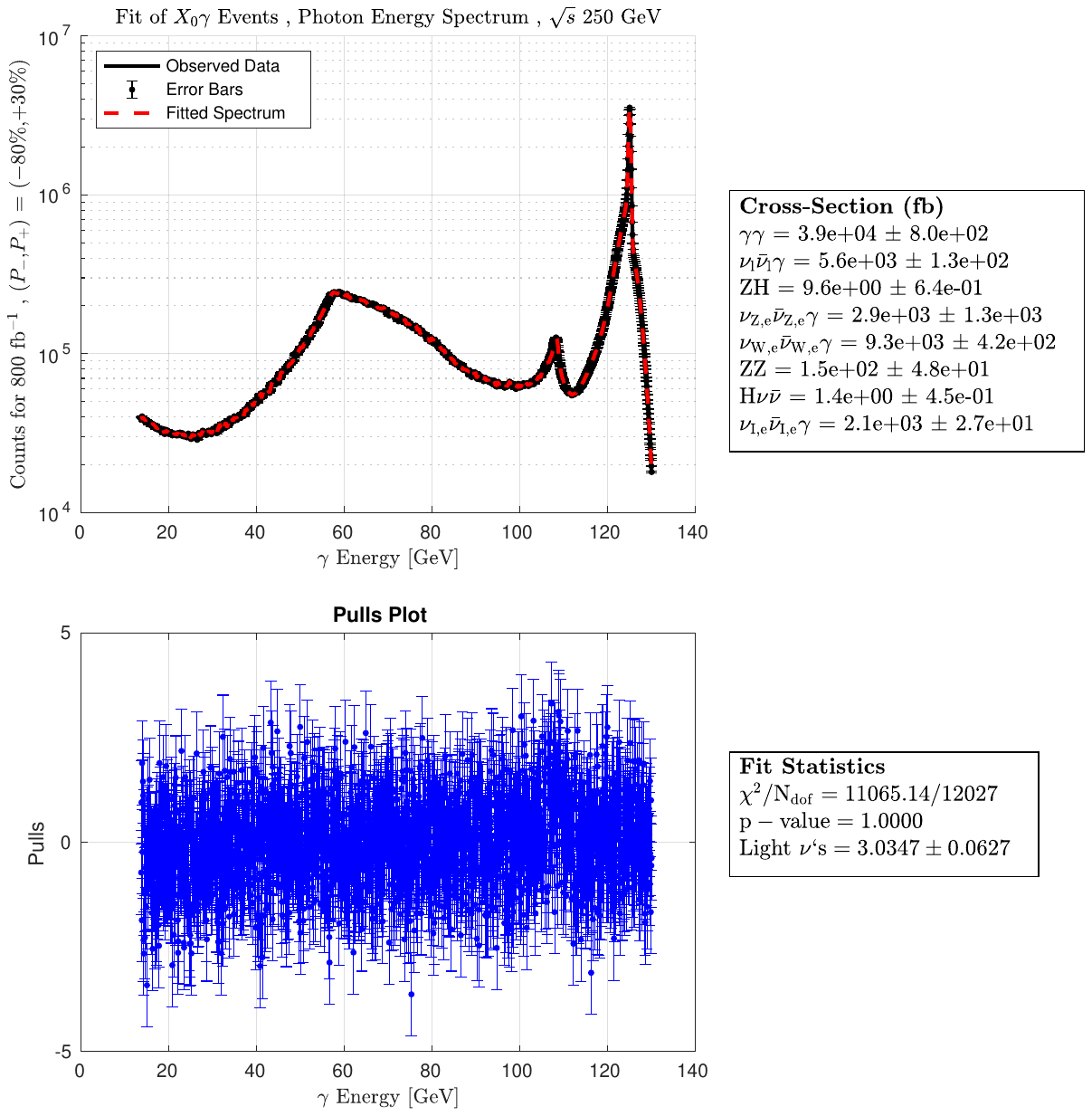}
\caption{Plot of the photon energy spectrum, which includes all photons in a $X^{0}\gamma$ event, after being fit using the SVD fit method of section~\ref{subsec-Measuring}. The fitted cross-sections, with uncertainties, are provided to the right. The quality of fit is included on the bottom right alongside the estimate of light neutrino species calculated according to equation~\ref{eqn-NumNeu}.}
\label{fig-FitNumNeu}       
\end{figure}

In addition to measuring the neutrino couplings it is advantageous to measure $N_\nu$, the number of light neutrino species. Not only is this value important for constraining the number of light neutrino species, it is also a test of lepton universality and of the pure left-handed neutrinos of the Standard Model. If there are light right-handed neutrinos the measurement of $N_\nu$ will not be exactly 3, and instead be a larger value~\cite{Carena:2003aj}. If there is no lepton universality then the value of $N_\nu$ as measured from different sources of neutrinos will be different. At future lepton colliders the measurement is inclusive, since the neutrinos are unmeasured, and assuming lepton universality, the number of light neutrinos
\begin{equation}\label{eqn-NumNeu}
    N_\nu = A_{\text{cor.}}\frac{\sigma_{\text{inv.},Z}}{2\sigma_{ll}}
\end{equation}
can be computed from the ratio of the invisible cross-section to any of the lepton cross-sections that come from the Z boson and the correction value of $A_{\text{cor.}}$. Since the dielectron cross-section has contamination with Bhabha scattering and the high rates of FSR, which will adjust the cross-section, it is not desirable. Dimuons are a plausible denominator process, as they are easily measured and tagged and have far less FSR. Tau-antitau is another candidate process that would have even less FSR but may suffer from correctly tagging and measuring. So we will restrict ourselves to dimuons for this study. Using WHIZARD we can simulate and compute what the value of dimuon and dineutrino cross-sections are when only the Z propagator is used by using the propagator restriction flag in WHIZARD. The result is that the Z only dimuon cross-section at ILC250 is 2.232 pb and the Z only, single flavor, dineutrino cross-section is 2.858 pb. This leads to a $A_{\text{cor.}}$ value of $\approx0.78$. In an experimental application this would need to be measured by looking at radiative return dimuons and estimating the degree of FSR while counting the number of radiative return dimuon events. From this we use the methodology outlined in section~\ref{subsec-Measuring} to fit for the cross-sections and then compute the number of light neutrinos from equation~\ref{eqn-NumNeu}. We propagate uncertainties based on the value of the cross-section uncertainties. The result, seen in figure~\ref{fig-FitNumNeu}, shows that this method recovers a value of the number of light neutrinos that is comparable to three.
This measurement of the number of light neutrinos is comparable to the LEP direct measurement but worse than the LEP indirect measurement, but this is to be expected~\cite{Janot_2020}. The amount of Z-like statistics is far smaller for a 250~GeV data sample compared to one at the Z pole.

To summarize, there is great potential for neutrino measurements at future $\ee$ colliders using $X^0\gamma$ events. For measurements at the Z pole there is great promise to improve upon the existing $N_\nu$ measurement. For higher center-of-mass energy runs, where the contributions of the W boson become considerable, there is greater sensitivity to the individual $g_\text{L}$ and $|g_\text{R}|$ couplings due to the ability to reconstruct the ZZ, WZ and WW components of the dineutrino mass spectrum. This sensitivity is found to be quite dependent on the beam polarization; the measurements with $(P_-,P_+)=(-80\%,+30\%)$ were able to measure $g_\text{L}$ and $g_\text{R}$ to, at least, 1\%, but this sensitivity was lost with $(P_-,P_+)=(+80\%,-30\%)$. This highlights the importance of beam polarization for future experiments with respect to neutrino measurements. If one can go to higher energies with polarized beams then even small deviations of $g_\text{L}$ and $|g_\text{R}|$ can be measured, as well as any running in these couplings from increasing center-of-mass energy. From the running and precision measurements one could then shed light on BSM physics.

\subsection{Measuring MSSM Neutralinos}\label{sec-DarkSect2}

For this section we will use the Bino-Wino and Bino-Higgsino MSSM models as discussed in section~\ref{sec-MSSM}. Due to the parameter values of these models the Bino-Wino model will have two neutralinos ($\chi_1^0,\chi_2^0$) that are accessible at ILC and the Bino-Higgsino model will have three neutralinos ($\chi_1^0,\chi_2^0,\chi_3^0$) that are accessible at ILC. The relevant parameter values and decays can be found in appendix~\ref{app-SLHA-BW} and appendix~\ref{app-SLHA-BH} respectively. More in depth documentation of values can be found on the GitHub relevant to this work~\cite{github}.

We can repeat the process of section~\ref{subsec-Measuring} but introduce neutralinos from the MSSM in order to act as a source of unknown new physics. These models are both an interesting model to choose and investigate. The Bino-Higgsino model is expected to have a larger cross-section, but has very few decays to photons so will look mostly like neutrino-antineutrino production. Making it hard to differentiate from the neutrino-antineutrino background. The Bino-Wino model is expected to have a small cross-section but the large branching ratio to photons gives a unique event signature. One that should be easier to identify amid the neutrino background.

For handling decays we have written a program to generate decays products using the branching ratios and 2-to-3 kinematics for fermions. This was done as WHIZARD is unable to compute the decays to photon(s) with neutralinos for an unknown reason. For the decays we are using, we then apply the cuts of section~\ref{subsec-Measuring}, which involves rejecting events that have charged particles reconstructed in the detector. Due to the neutralino masses used here the photons emitted from $\chi_2^0$ decays in the Bino-Wino model occupy an intermediate energy range at ILC250, comparable to half the beam energy.

To aid in the identification of the neutralino events we have used C5 to construct a BDT from the reconstructed values of the $X^0\gamma$ events, including neutralinos~\cite{C5}. We started by tagging the neutrino events as 'neutrino-like', the BERB events as 'BERB', the diphoton events as '$\gamma\gamma$', the ZH events as 'Higgs-like', the ZZ events as 'ZZ-like', the pure LSP events as '$\chi_1^0$-like' and events with heavier neutralinos as '$\chi_{2,3}^0$-like'. We then used the BDT to correctly tag events according to the above classifiers. We gave the BDT detector level reconstructed values of photon count, missing transverse momentum, missing longitudinal momentum, energy of most energetic photon, energy of second most energetic photon, polar angle of most energetic photon, polar angle of second most energetic photon, acollinearity of two most energetic photons, acoplanarity of two most energetic photons, invariant mass of two most energetic photons. 

The resulting confusion matrix from C5 after using 10 boosting trials for the BDT and passing cross-validation on a second dataset, can be seen in table~\ref{tab-confusion}. The confusion matrices for all models and energies can be found in appendix~\ref{app-Conf}.
\begin{table}[h]
\centering
\caption{Confusion matrix: columns indicate true class, rows indicate predicted class. This is for MSSM Bino-Wino world for ILC1000.}
\begin{tabular}{c|rrrrrrr}
\textbf{True \textbackslash\ Predicted} 
  & \textbf{$\gamma\gamma$} 
  & \textbf{$\nu\bar{\nu}$} 
  & \textbf{Higgs-like} 
  & \textbf{BERB} 
  & \textbf{ZZ-like} 
  & \textbf{$\chi^0_1$-like} 
  & \textbf{$\chi^0_{2,3}$-like} \\
\hline
\textbf{$\gamma\gamma$}       & 3,220,624 & 437,375 &         0 &       0 &      0 &     0 &    0 \\
\textbf{$\nu\bar{\nu}$}       &    30,313 & 6,220,924 &   1,836 &       0 &      0 &     0 &    0 \\
\textbf{Higgs-like}           &        19 &    3,698 &  27,451 &       0 &      0 &     0 &    0 \\
\textbf{BERB}                 &         2 &        0 &       0 &  12,248 &      0 &     0 &    0 \\
\textbf{ZZ-like}              &       585 &   64,075 &      85 &       0 &      0 &     0 &    0 \\
\textbf{$\chi^0_1$-like}      &         0 &        0 &       0 &       0 &      0 &   325 &    0 \\
\textbf{$\chi^0_{2,3}$-like}  &        11 &   11,517 &       0 &       0 &      0 &     0 &    6 \\
\end{tabular}
\end{table}\label{tab-confusion}

The majority of neutralino events get mis-classified as $\nu\bar{\nu}$. This makes sense as the neutrino background is numerous and both typically come from Z boson propagators and \Gls{ISR} to the Z.

\begin{figure}[h!]
\centering
\includegraphics[width=14cm]{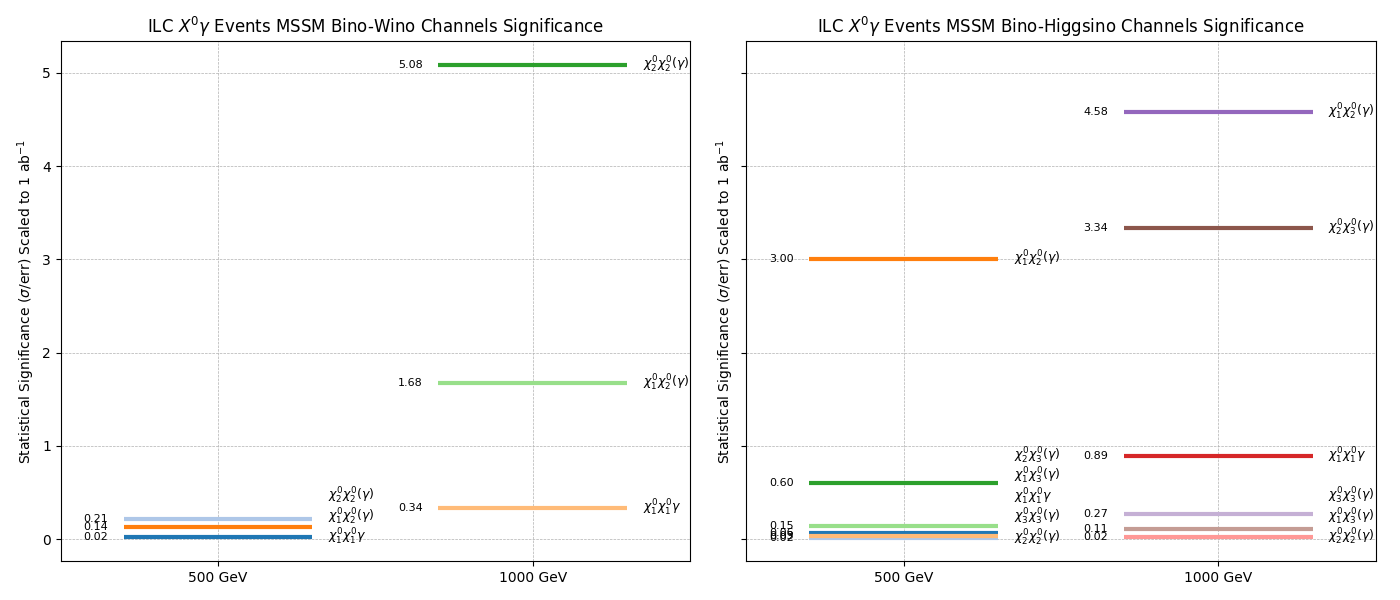}
\caption{Plot of the statistical significance, in terms of numbers of $\sigma$ or $\sigma/\delta\sigma$, for neutralinos at ILC using $X^0\gamma$ events. The significance values here have been scaled to 1~$\invab$.}
\label{fig-NeutDisc}       
\end{figure}

After using C5 to construct a BDT to identify neutralinos we then fed the BDT tagging result into the SVD fit as an additional basis vector. This greatly improved the performance of the SVD fit for identifying the smaller cross-section processes, most notably the Higgs-like and Neutralino-like processes. We then repeated the same process as before to estimate the Neutralino cross-sections for both of the MSSM models. From these we calculated the discovery significance, in number of $\sigma$, and plotted them in figure~\ref{fig-NeutDisc}.

\begin{figure}[h!]
\centering
\includegraphics[width=14cm]{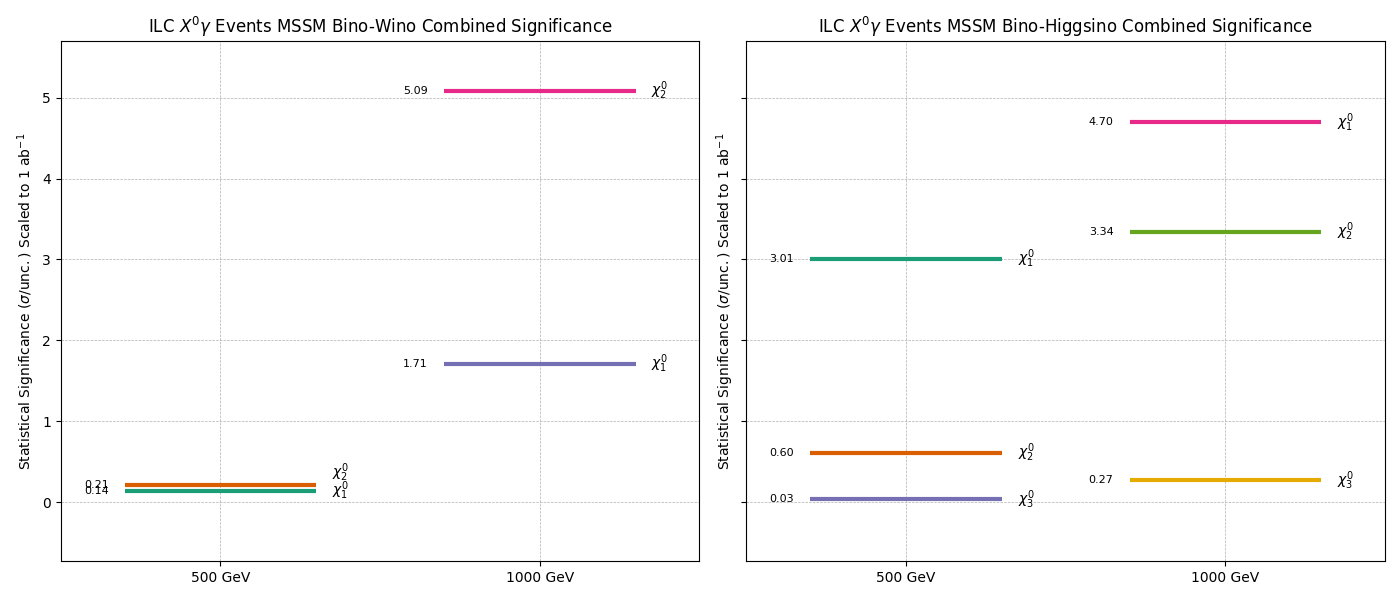}
\caption{Plot of the combined statistical significance, in terms of numbers of $\sigma$ or $\sigma/\delta\sigma$, for neutralinos at ILC using $X^0\gamma$ events. The significance values here have been scaled to 1~$\invab$.}
\label{fig-NeutDisc2}       
\end{figure}

Combining the channels that have a certain neutralino species into one neutralino discovery statistic gives the values seen in figure~\ref{fig-NeutDisc2}.
We find that, for the values tested here, there is a similar level of significance for the Bino-Wino and Bino-Higgsino models. The $\chi_2^0$ is the easiest to discover in the Bino-Wino model, likely due to its photon decay, and $\chi_1^0$ is the easiest to discover in the Bino-Higgsino model, likely due to having the smallest mass and largest cross-sections.

\subsection{Example of Model Agnostic Rejection of SM with MSSM Neutralinos}\label{sec-DarkSect}

\begin{figure}[h!]
\centering
\includegraphics[width=14cm]{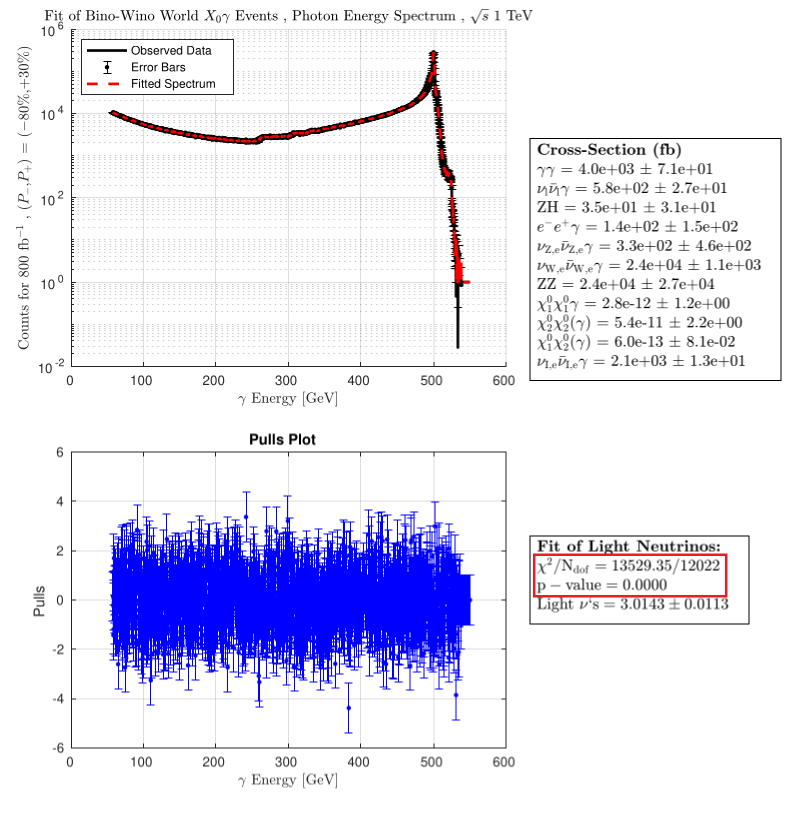}
\caption{Plot of the photon energy spectrum, which includes all photons in a $X^{0}\gamma$ event, including MSSM neutralinos from the Bino-Wino scenario as outlined in section~\ref{sec-DarkSect}. This was then fit using the SVD fit method of section~\ref{subsec-Measuring} but here we have effectively turned off the neutralino parameters by forcing them to be of the order of 1~ab. The fitted cross-sections, with uncertainties, are provided to the right. The quality of fit is included on the bottom right alongside the estimate of light neutrino species calculated according to equation~\ref{eqn-NumNeu}.}
\label{fig-RejectSM}       
\end{figure}

We now consider the case in which we do not know the MSSM model but the neutralino decays are still in the data. Here we use the Bino-Wino model for the demonstration. For the SVD fit method we can make a mock attempt of this by forcing the fitted neutralino cross-sections to be of the order of $10^{-6}$~pb. We then repeat the same process as before but now the fit will be unable to find an adequate value for the neutralino cross-sections. We expect that this will degrade the p-value and, potentially, other parts of the fit parameters. The results, seen in figure~\ref{fig-RejectSM}, find a p-value that has been degraded enough that it can be rejected at a 100\% confidence interval.

\begin{figure}[h!]
\centering
\includegraphics[width=14cm]{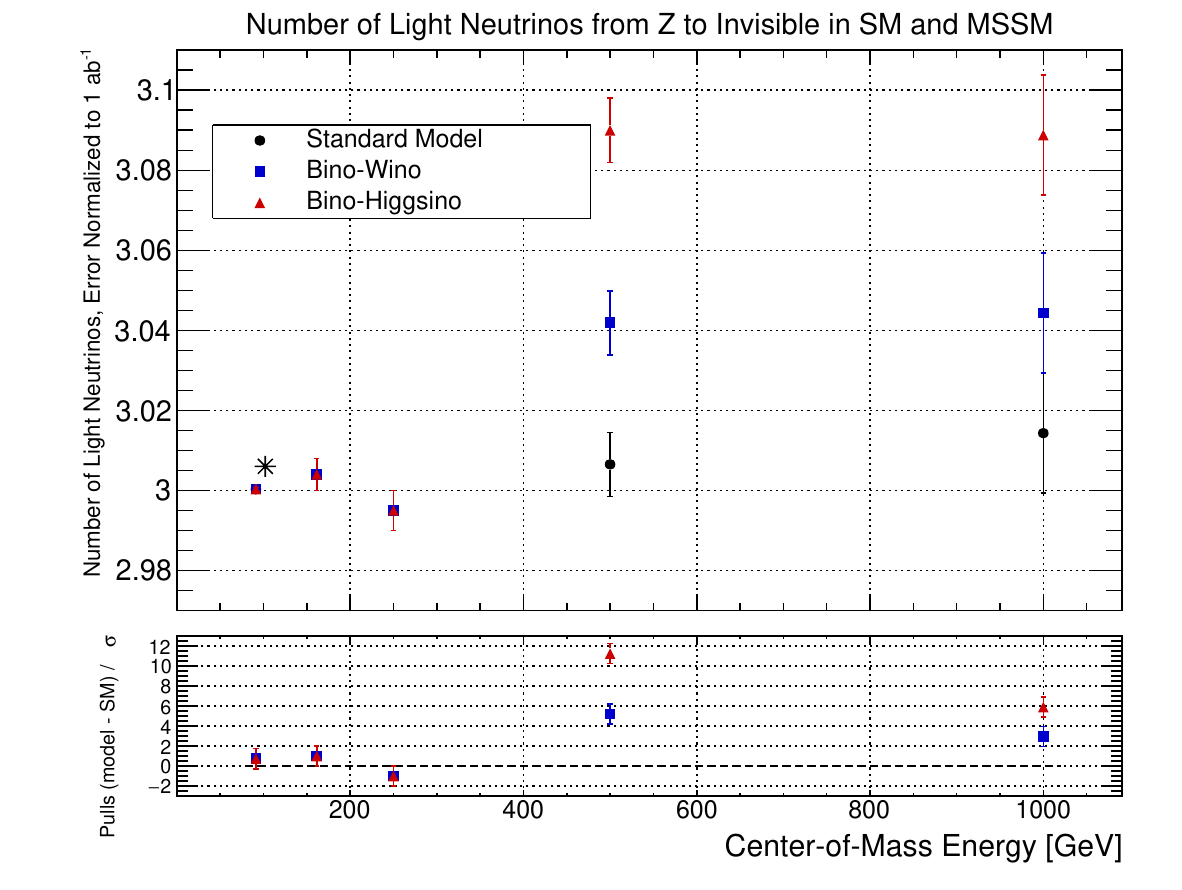}
\caption{Plot of the direct measurement of the number of light neutrino species using $X^0\gamma$ events and the methodology of section~\ref{sec-NeutrinoCoup}. We put a * on the values for the Z pole as they are taken from a reference that estimates the performance of using the indirect invisible Z boson width method~\cite{Carena:2003aj}. We find that the neutrino measurement shifts as neutralinos are produced.}
\label{fig-MSSMNeu}       
\end{figure}

In addition to the degradation in the p-value we also found that the measurement of the number of light neutrinos has lost accuracy. Considering this we repeated this process for both MSSM models to see how the measured number of light neutrinos would change. The result, seen in figure~\ref{fig-MSSMNeu}, shows how the measurement of the number of light neutrino species deviates away from the Standard Model once production of neutralinos is possible.
This gives a promising method of indirect measurement of dark matter for future experiments in the event that the dark matter is otherwise impossible to measure or separate from neutrinos. In such a case one would measure a deviation from the Standard Model. Due to how previous work has shown how $N_\nu$ and $g_\text{L}$ and $g_\text{R}$ are related, we this would also cause a shift in the measurement of the neutrino couplings too. So a precision measurement of the neutrino couplings at all center-of-mass energies is not only worthwhile for precision Standard Model measurements, but for constraining dark matter. Given these results we have two proofs that, in the presence of MSSM neutralinos, that the Standard Model is insufficient. Not only did the p-value of the SVD fit get worse, the neutrino measurements changed in accuracy. Both methods did not require assuming a certain BSM model and one was only chosen for the sake of demonstration.


\chapter{Outlook and Conclusion}
In this work we have significantly advanced both the precision measurement capabilities and the potential for BSM physics discovery at future $\ee$ Higgs factories made possible by novel instrumentation, with the GLIP LumiCal of chapter~\ref{ch-FCAL}, and innovative analysis techniques, such as the SVD fit method of chapter~\ref{ch-Measure}. The key contributions of this work can be summarized as follows. 

Firstly, we designed and demonstrated a highly granular luminosity calorimeter concept, the Granular Long Instrument for Precision (GLIP) LumiCal, optimized for diphoton-based luminosity measurements. By leveraging a high-granularity silicon-tungsten sampling structure and large transverse and longitudinal size, the GLIP LumiCal is capable of achieving precision measurement of diphoton events at $\approx35$~ppm precision on integrated luminosity for all energies of future $\ee$ colliders. This level of precision in integrated luminosity would allow for the next generation of lepton colliders to make incredibly precise measurements. By comparison, existing LumiCal designs, which are similar to the ILD LumiCal tested here and used in both linear and circular colliders, are limited to $\approx3\times10^{-4}$ on integrated luminosity precision for SABS and $\approx30\times10^{-4}$ on integrated luminosity precision for diphotons. 

Unfortunately, we find that the largest sources of uncertainty for integrated luminosity precision are not from the LumiCal designs and, instead, are from other sources. Particularly troublesome for using SABS is the beam deflection effect, which may limit integrated luminosity precision to the percent level. We propose a way of using M{\o}ller scattering to diagnose the beam deflection effect and reduce its uncertainty to the $10^{-4}$ level. Beyond this, the beam polarization and statistical uncertainty are the largest contribution of uncertainty for diphotons, contribution factors of $10^{-4}$ each. We propose to minimize these both by using particle physics processes, using the PFL method and simply taking more data than is currently planned.

\begin{figure}[h]
\centering
\includegraphics[width=14cm,clip]{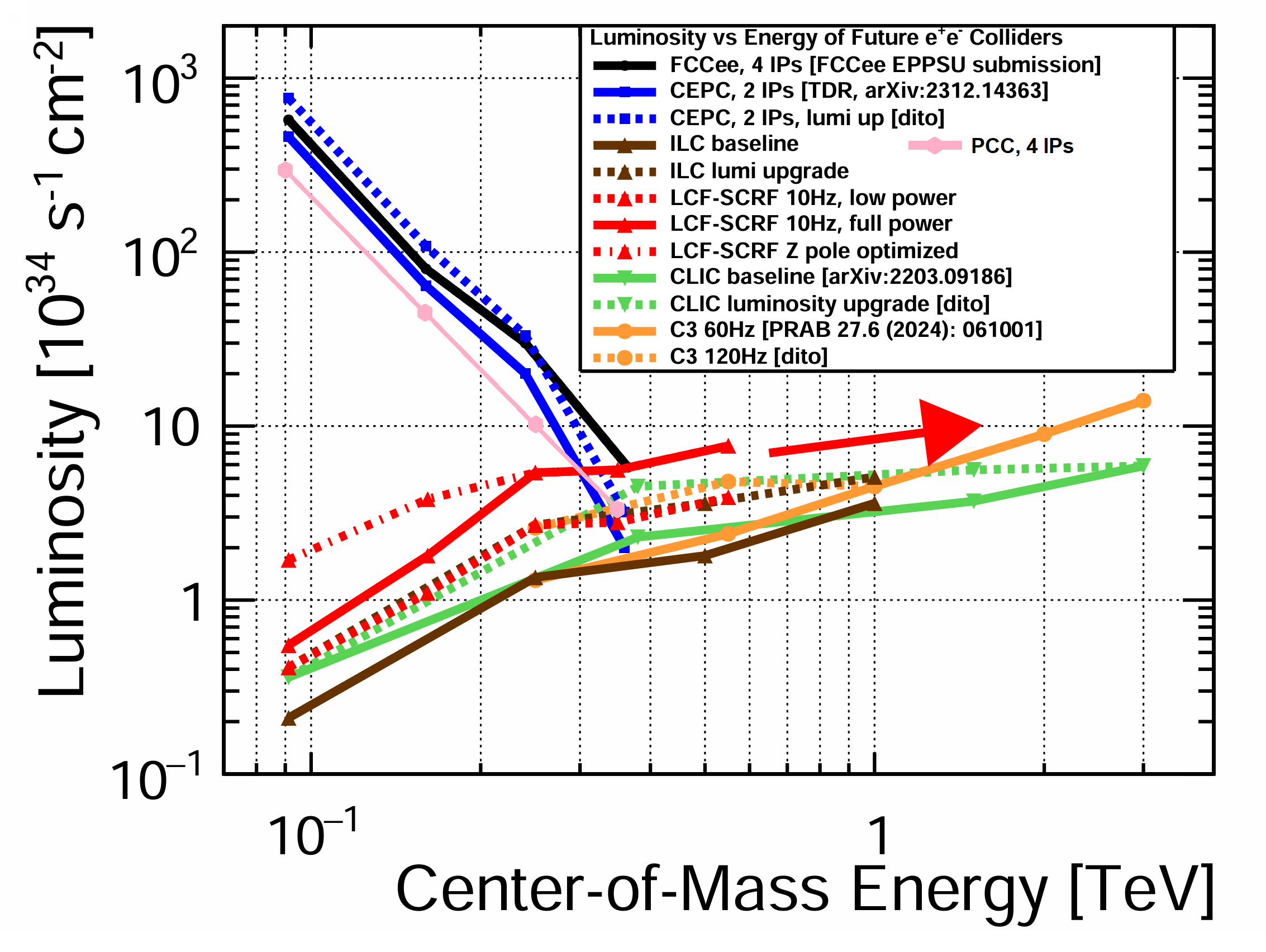}
\caption{LCVision instantaneous luminosity plot adjusted to include the PCC proposal~\cite{LCVision}}
\label{fig-newlumi}       
\end{figure}

\begin{figure}[h]
\centering
\includegraphics[width=14cm,clip]{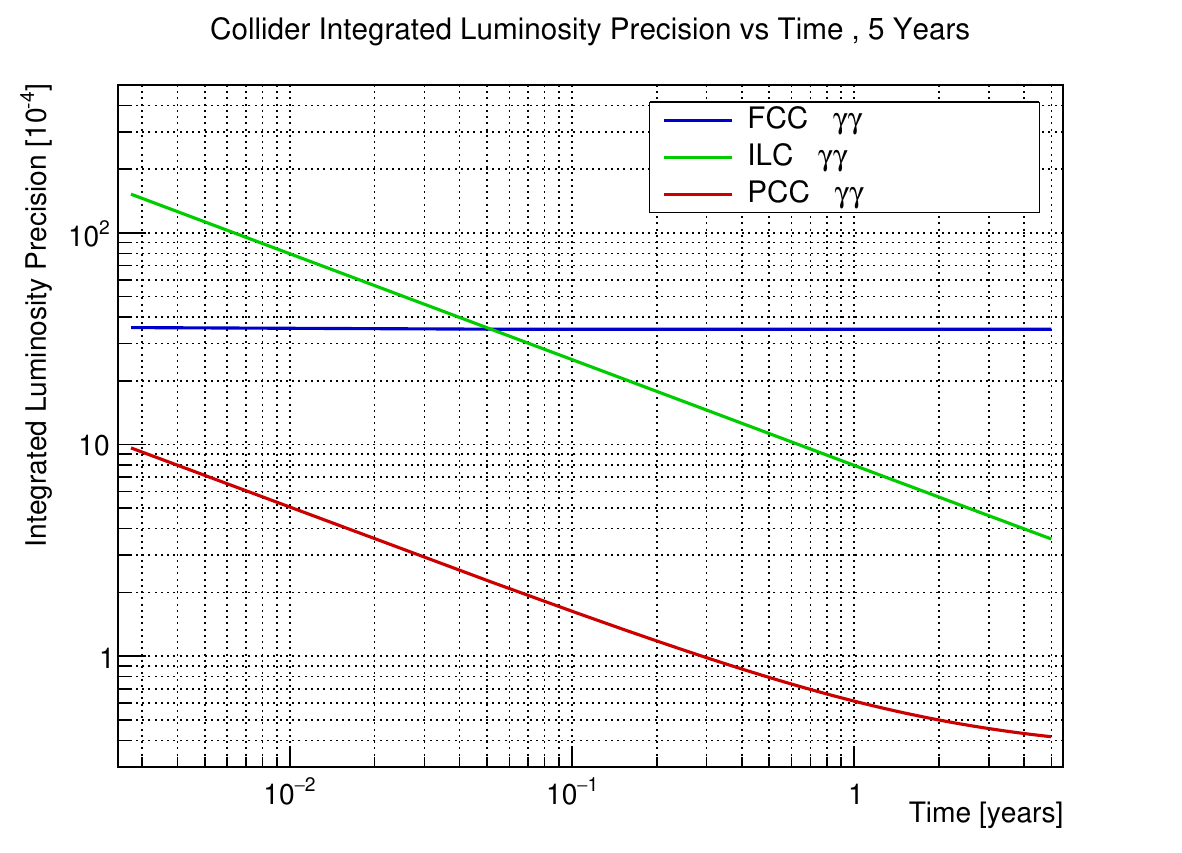}
\caption{Simulation of the level of precision on integrated luminosity using diphotons over time for various experiments at the Z pole.}
\label{fig-preccomp}       
\end{figure}

Due to constraints of the forward region geometry of circular colliders, \Gls{FCC} and \Gls{CEPC}, being at wider angles and smaller due to accelerator electromagnets, an upgrade to a highly granular design like the GLIP LumiCal is only possible at the currently proposed linear colliders. Due to this we propose that the circular colliders move their magnets back by roughly 1~m. We refer to this proposed alteration of FCC as Precision Circular Collider (PCC). Due to the relationship between the inner magnet distance and the collider luminosity we expect that this would decrease the instantaneous luminosity by roughly half. As seen in figure~\ref{fig-newlumi}, this is still larger than what is achieved at linear colliders when the center-of-mass energy is below 300~GeV.
To demonstrate the value of this, we then simulated how the precision on integrated luminosity using diphotons would accumlate over time for FCC, PCC and ILC at the Z pole under the assumption that FCC would have a LumiCal like the ILD LumiCal and PCC and ILC would both have a LumiCal like GLIP LumiCal. The result, seen in figure~\ref{fig-preccomp}, shows how FCC quickly becomes systematics limitted while PCC and ILC slowly improve towards their limits.
As a part of this we also note that existing feedback on the GLIP LumiCal design from circular collider authors has noted this and requested a shorter design or a design that uses novel technology to further reduce the GLIP LumiCal size. We did not consider novel technology and focused on existing calorimeter technologies. Beyond reducing the number of radiation lengths in the GLIP LumiCal design, which section~\ref{sec-EneRes} found should be no shorter than $34~X_0$, we did not explore GLIP LumiCal longitudinal segmentation either. Considering this, and the currently existing proposals, \textbf{$\mathbf{10^{-4}}$ precision on integrated luminosity via the GLIP LumiCal is only accessible at current linear collider proposals.}

The potential for a linear collider facility is also greater than a circular collider facility as a linear collider facility can undergo technology upgrades to achieve even higher luminosities and center-of-mass energies since it does not have as hard a physical limitation as circular colliders do from synchrotron radiation. As of writing, the ILC accelerator concept, which uses superconducting RF technology, is mature and could be confidently deployed and achieve gradients of $\sim30$~MV/m~\cite{LCVision}. It would be feasible, and has been proposed, that the site could later be upgraded to drive-beam technology, as CLIC proposes to use, to reach 100s of MV/m~\cite{LCVision}. This would push collision luminosity even higher, as well as center-of-mass energies, and turn the facility into a true precision at 1~TeV scale experiment. There also exists work on the use of plasma wake-field acceleration (PWFA) for future $\ee$ colliders, which would push gradients into the range of $>1$~GV/m~\cite{geddes2022reportsnowmass21accelerator}. A second accelerator technology upgrade, to use PWFA, could push the collision luminosity even higher and make center-of-mass energies up to 30~TeV possible~\cite{geddes2022reportsnowmass21accelerator}. Assuming one can continue developing accelerator technologies and making marginal improvements to the facility site, be it through energy or length upgrades, a linear collider facility could persist for numerous generations while continuously providing precision physics measurements and technological innovations.

Returning to the work done here, we have devised an SVD-based fit applied to differential distributions of $X^0\gamma$ data to enable fitting of the cross-sections of processes that contribute to $X^0\gamma$ events. This method allows for the effective deconstruction of the constituents of the neutrino cross-sections so that the left-handed and right-handed couplings can then be differentially fit in terms of the dineutrino mass. We have found that this leads to an increase in precision of the left-handed and right-handed neutrino couplings. It also allows for discovery of right-handed neutrino couplings down to $\approx1$\% coupling of their left-handed counter parts. This holds great promise for improving existing precision electroweak measurements. We also explored the BSM discovery potential afforded by this approach. We implemented a search for MSSM neutralinos, particularly the lightest neutralinos since they are dark matter candidates, using a combination of the SVD technique and machine learning via C5's BDT. By training a BDT classifier on kinematic features and incorporating its output into the SVD fit, we demonstrated enhanced sensitivity to elusive $X^0\gamma$ final states, with the neutralinos being our candidate example. We also showed that this approach can find potential signals of neutralinos, or any BSM physics, from fluctuations in the Standard Model measurements and Standard Model fit quality without \emph{a priori} BSM model-specific assumptions.

\newpage

\begin{appendices}
\chapter{GuineaPig++ Settings}\label{app-GP}
On the following page we provide a table of some of the simulation values used with GuineaPig++ for the various ILC runs simulated here. The table is not exhaustive, in order to keep things concise and focused.

\begin{sidewaystable}[htbp]
\centering
\caption{Table of GuineaPig++ simulation values. The units for these are in the units that GuineaPig++ uses, so they are not otherwise provided with the exception of the spread along the z-axis. For more detail see the acc.dat files on the GitHub~\cite{github}.}
\begin{tabular}{|l|c|c|c|c|c|}
\hline
\textbf{Value} & \textbf{ILC Z} & \textbf{ILC WW} & \textbf{ILC250} & \textbf{ILC500} & \textbf{ILC1000} \\
\hline
\textbf{Beam Energy [GeV]} & 91.2 & 161.4 & 250 & 500 & 1000 \\
\textbf{Beam Spread ($e^-$,~$e^+$) \%} & (0.2,~0.2) & (0.203,~0.203) & (0.190,~0.152) & (0.15,~0.15) & (0.15,~0.15) \\
\textbf{($\beta_x$,~$\beta_y)$} & (18.0,~0.39) & (15.8,~0.4) & (13.00,~0.41) & (44.9,~0.995) & (22.500,~0.244) \\
\textbf{Normalized Emittance ($\epsilon_x$,~$\epsilon_y)$} & (6.2,~0.0485) & (5.6,~0.0415) & (5.000,~0.035) & (5.000,~0.035) & (5.00,~0.03) \\
\textbf{Spread Along Z $\sigma_\text{z}$ [$\mu$m]} & 615 & 350 & 300 & 300 & 337.5 \\
\textbf{Waist as fraction of $\sigma_\text{z}$ ($w_x$,~$w_y$)} & (0,~1.3) & (0,~1.1) & (0,~1.1) & (0,~1.1) & (0,~1.1) \\
\textbf{Beam Angles} & all zero & all zero & all zero & all zero & all zero \\
\textbf{Grid (nx,~ny,~nz,~nt)} & (1024,~256,~64,~6) & (1024,~256,~64,~6) & (1024,~256,~64,~6) & (1024,~256,~64,~6) & (1024,~256,~64,~6) \\
\textbf{Number of Macro-particles} & 600k & 500k & 1M & 500k & 500k \\
\textbf{Cuts as fraction of $\sigma$ (x,y,z)} & (15,~30,~3.5) & (15,~30,~3.5) & (100,~30,~3.5) & (15,~30,~3.5) & (15,~30,~3.5) \\
\hline
\end{tabular}
\label{tab-GPTab}
\end{sidewaystable}

\chapter{CIRCE Settings}\label{app-CIRCE}
Below are the settings used in CIRCE to generate the CIRCE files from the output lumi.ee files from GuineaPig++.
\begin{figure}[h]
\centering
\begin{verbatim}
{ file = "ZPole_ee.circe"
{ design = "ILC"
roots = 91.2
scale = 45.6
bins = 100
{ pid/1 = electron
pid/2 = positron
pol = 0
events = "lumi.ee.outfile"
columns = 2
lumi = 8.008e33
min = 0
max = 1.05
fix = *
iterations = 10
smooth = 5 [0.00,1.05) [0.00,1.05)
smooth = 5 [1.05] [0,1.05)
smooth = 5 [0,1.05) [1.05]
} } }
\end{verbatim}
\caption{CIRCE input code as would be used with \textit{circe2\_tool.opt}. This example is for the ILC Z run.}
\label{fig-CIRCECode}
\end{figure}

\chapter{MSSM Bino-Wino World SLHA Values}\label{app-SLHA-BW}
Here we provide tables of values from the Supersymmetry Les Houches Accord (SLHA) file used in the neutralino and MSSM Bino-Wino world simulation done using WHIZARD in section~\ref{sec-DarkSect}~\cite{Skands_2004}. The mass spectrum, mixing matrices, and decays were calculated using SPheno~\cite{SPheno}. A script was written to automatically generate the bulk of the formatting and content of these tables and can be found on the thesis GitHub~\cite{github}. Some editing was done to improve presentation. The full SLHA file can also be found there. We have edited the script output to only include neutralino relevant values, such as the mixing matrix and decays of the two lightest neutralinos, as they are relevant to this work.

\begin{table}[h!]
\centering
\begin{tabular}{|c|c|l|}
\hline
Index & Value & Comment \\
\hline
3 & 1.50000000E+01 & $\tan\beta$ \\
\hline
\end{tabular}
\caption{SLHA Block: MINPAR}
\end{table}

\begin{table}[h!]
\centering
\begin{tabular}{|c|c|l|}
\hline
Index & Value & Comment \\
\hline
1 & 1.45E+02 & $m_1$ \\
2 & 1.46E+02 & $m_2$ \\
3 & 3.00E+03 & $m_3$ \\
11 & 2.20E+03 & $A_\text{t}$ \\
12 & 2.20E+03 & $A_\text{b}$ \\
13 & 2.20E+03 & $A_\tau$ \\
23 & 2.00E+03 & $\mu$ \\
26 & 3.00E+03 & $m_\text{A}$ \\
31 & 3.00E+03 & $m_{\text{L},11}$ \\
32 & 3.00E+03 & $m_{\text{L},22}$ \\
33 & 3.00E+03 & $m_{\text{L},33}$ \\
34 & 3.00E+03 & $m_{\text{E},11}$ \\
35 & 3.00E+03 & $m_{\text{E},22}$ \\
36 & 3.00E+03 & $m_{\text{E},33}$ \\
41 & 5.00E+03 & $m_{\text{Q},11}$ \\
42 & 5.00E+03 & $m_{\text{Q},22}$ \\
43 & 5.00E+03 & $m_{\text{Q},33}$ \\
44 & 5.00E+03 & $m_{\text{U},11}$ \\
45 & 5.00E+03 & $m_{\text{U},22}$ \\
46 & 1.90E+03 & $m_{\text{U},33}$ \\
47 & 5.00E+03 & $m_{\text{D},11}$ \\
48 & 5.00E+03 & $m_{\text{D},22}$ \\
49 & 5.00E+03 & $m_{\text{D},33}$ \\
\hline
\end{tabular}
\caption{SLHA Block: Input parameters , EXTPAR}
\end{table}

\begin{table}[h!]
\centering
\begin{tabular}{|c|c|l|}
\hline
Index & Value & Comment \\
\hline
6 & 1.73200000E+02 & t \\
23 & 9.11876000E+01 & Z \\
24 & 8.03908250E+01 & $\text{W}^{\pm}$ \\
25 & 1.25157644E+02 & h \\
35 & 1.01877470E+03 & H \\
36 & 1.02016000E+03 & A \\
37 & 1.02216316E+03  & $H^\pm$ \\
1000001 & 5.06602425E+03 & $~d_L$ \\
2000001 & 5.05764419E+03 & $~d_R$ \\
1000002 & 5.06558168E+03 & $~u_L$ \\
2000002 & 5.05746128E+03 & $~u_R$ \\
1000003 & 5.06604017E+03 & $~s_L$ \\
2000003 & 5.05762856E+03 & $~s_R$ \\
1000004 & 5.06558145E+03 & $~c_L$ \\
2000004 & 5.05746075E+03 & $~c_R$ \\
1000005 & 5.04253471E+03 & $~b_1$ \\
2000005 & 5.05904367E+03 & $~b_2$ \\
1000006 & 2.05243381E+03 & $~t_1$ \\
2000006 & 5.04750787E+03 & $~t_2$ \\
1000011 & 3.00501168E+03 & $~e_L$ \\
2000011 & 3.00492449E+03 & $~e_R$ \\
1000012 & 3.00362299E+03  & $~\nu_{eL}$ \\
1000013 & 3.00445613E+03 & $~\mu_L$ \\
2000013 & 3.00547424E+03 & $~\mu_R$ \\
1000014 & 3.00362088E+03 & $~\nu_{\mu L}$ \\
1000015 & 2.99598030E+03 & $~\tau_1$ \\
2000015 & 3.01175439E+03 & $~\tau_2$ \\
1000016 & 3.00289332E+03 & $~\nu_{\tau L}$ \\
1000021 & 3.31001771E+03 & $~g$ \\
1000022 & 1.44118895E+02 & $~\chi_1^0$ \\
1000023 & 1.60299916E+02 & $~\chi_2^0$ \\
1000025 &-2.00875008E+03 & $~\chi_3^0$ \\
1000035 & 2.00978081E+03 & $~\chi_4^0$ \\
1000024 & 1.60447615E+02 & $~\chi_1^\pm$ \\
1000037 & 2.01036219E+03  & $~\chi_2^\pm$ \\
\hline
\end{tabular}
\caption{SLHA Block: Mass spectrum , MASS}
\end{table}

\begin{table}[h!]
\centering
\begin{tabular}{|c|c|l|}
\hline
Index & Value & Comment \\
\hline
1 1 & 9.99478569E-01 & $N_{11}$, Bino \\
1 2 & -2.16858795E-02 & $N_{12}$, Wino \\
1 3 & 2.36947185E-02 & $N_{13}$, Higgsino \\
1 4 & -3.29758286E-03 & $N_{14}$, Higgsino \\
2 1 & 2.26066309E-02 & $N_{21}$, Bino \\
2 2 & 9.98986785E-01 & $N_{22}$, Wino  \\
2 3 & -3.85070059E-02 & $N_{23}$, Higgsino \\
2 4 & 5.61728041E-03 & $N_{24}$, Higgsino \\
3 1 & 1.39050491E-02 & $N_{31}$, Bino \\
3 2 & -2.35720552E-02 & $N_{32}$, Wino  \\
3 3 & -7.06524359E-01 & $N_{33}$, Higgsino \\
3 4 & -7.07159344E-01 & $N_{34}$, Higgsino \\
4 1 & 1.83896970E-02 & $N_{41}$, Bino \\
4 2 & -3.16146161E-02 & $N_{42}$, Wino  \\
4 3 & -7.06242947E-01 & $N_{43}$, Higgsino \\
4 4 & 7.07024211E-01 & $N_{44}$, Higgsino \\
\hline
\end{tabular}
\caption{SLHA Block: Neutralino mixing matrix , NMIX}
\end{table}

\begin{table}[h!]
\centering
\begin{tabular}{|c|c|l|}
\hline
Index & Value & Comment \\
\hline
1 1 & -9.98477054E-01 & $U_{11}$  \\
1 2 & 5.51685878E-02 & $U_{12}$ \\
2 1 & 5.51685878E-02  & $U_{21}$ \\
2 2 & 9.98477054E-01 & $U_{22}$  \\
\hline
\end{tabular}
\caption{SLHA Block: Chargino U mixing matrix , UMIX}
\end{table}

\begin{table}[h!]
\centering
\begin{tabular}{|c|c|l|}
\hline
Index & Value & Comment \\
\hline
1 1 & -9.99967579E-01 & $V_{11}$  \\
1 2 & 8.05238010E-03 & $V_{12}$ \\
2 1 & 8.05238010E-03 & $V_{21}$ \\
2 2 & 9.99967579E-01 & $V_{22}$  \\
\hline
\end{tabular}
\caption{SLHA Block: Chargino V mixing matrix , VMIX}
\end{table}

\begin{table}[h!]
\centering
\begin{tabular}{|c|c|l|}
\hline
Parent ID or Daughter ID & Width [GeV] or Branching Ratio & Comment \\
\hline
1000022 & 0.00000000E+00 & $~\chi_1^0$ decays \\
\hline
1000023 & 2.75728266E-12 & $~\chi_2^0$ decays \\
22 & 3.14057464E-01 & $BR(~\chi_2^0 \to ~\chi_1^0 ~\gamma)$ \\
2 & 1.14107295E-01 & $BR(~\chi_2^0 \to ~\chi_1^0 \bar{u}      u)$ \\
1 & 1.50366873E-01 & $BR(~\chi_2^0 \to ~\chi_1^0 \bar{d}      d)$ \\
4 & 1.05133956E-01 & $BR(~\chi_2^0 \to ~\chi_1^0 \bar{c}      c)$ \\
3 & 1.50310282E-01 & $BR(~\chi_2^0 \to ~\chi_1^0 \bar{s}      s)$ \\
5 & 6.92478900E-02 & $BR(~\chi_2^0 \to ~\chi_1^0 \bar{b}      b)$ \\
11 & 1.37255593E-02 & $BR(~\chi_2^0 \to ~\chi_1^0 e^+      e^-)$ \\
13 & 1.37265807E-02 & $BR(~\chi_2^0 \to ~\chi_1^0 \mu^+     \mu^-)$ \\
15 & 1.37770120E-02 & $BR(~\chi_2^0 \to ~\chi_1^0 \tau+    \tau-)$ \\
12 & 5.55470876E-02 & $BR(~\chi_2^0 \to ~\chi_1^0 \bar{\nu}\nu)$ \\
\hline
\end{tabular}
\caption{SLHA Block: Two lightest neutralino decays as calculated by SPheno}
\end{table}

\chapter{MSSM Bino-Higgsino World SLHA Values}\label{app-SLHA-BH}
Here we provide tables of values from the Supersymmetry Les Houches Accord (SLHA) file used in the neutralino and MSSM Bino-Higgsino world simulation done using WHIZARD in section~\ref{sec-DarkSect}~\cite{Skands_2004}. The mass spectrum, mixing matrices, and decays were calculated using SPheno~\cite{SPheno}. A script was written to automatically generate the bulk of the formatting and content of these tables and can be found on the thesis GitHub~\cite{github}. Some editing was done to improve presentation. The full SLHA file can also be found there. We have edited the script output to only include neutralino relevant values, such as the mixing matrix and decays of the three lightest neutralinos, as they are relevant to this work. For the third lightest neutralino decays we have chosen to make the notation more compact by combining similar decay processes as there are numerous decay processes for the third lightest neutralino. Note that, for the $\chi_3^0$ decays to charginos we have used $\bar{f}f$ but the pairs are not neutral matter with anti-matter pairs, as one might find in a photon or Z boson mediated process. Instead they are charged left-handed with right-handed matter with anti-matter pairs, as is the case for W boson mediated processes. As an example, $e^+\nu_e$ is one of the possible decay channels we list as $\bar{f}f$.

\begin{table}[h!]
\centering
\begin{tabular}{|c|c|l|}
\hline
Index & Value & Comment \\
\hline
3 & 1.50000000E+01 & $\tan\beta$ \\
\hline
\end{tabular}
\caption{SLHA Block: MINPAR}
\end{table}

\begin{table}[h!]
\centering
\begin{tabular}{|c|c|l|}
\hline
Index & Value & Comment \\
\hline
1 & 1.45E+02 & $m_1$ \\
2 & 3.00E+03 & $m_2$ \\
3 & 3.00E+03 & $m_3$ \\
11 & 2.60E+03 & $A_\text{t}$ \\
12 & 2.60E+03 & $A_\text{b}$ \\
13 & 2.60E+03 & $A_\tau$ \\
23 & 1.70E+02 & $\mu$ \\
26 & 3.00E+03 & $m_\text{A}$ \\
31 & 3.00E+03 & $m_{\text{L},11}$ \\
32 & 3.00E+03 & $m_{\text{L},22}$ \\
33 & 3.00E+03 & $m_{\text{L},33}$ \\
34 & 3.00E+03 & $m_{\text{E},11}$ \\
35 & 3.00E+03 & $m_{\text{E},22}$ \\
36 & 3.00E+03 & $m_{\text{E},33}$ \\
41 & 5.00E+03 & $m_{\text{Q},11}$ \\
42 & 5.00E+03 & $m_{\text{Q},22}$ \\
43 & 5.00E+03 & $m_{\text{Q},33}$ \\
44 & 5.00E+03 & $m_{\text{U},11}$ \\
45 & 5.00E+03 & $m_{\text{U},22}$ \\
46 & 1.90E+03 & $m_{\text{U},33}$ \\
47 & 5.00E+03 & $m_{\text{D},11}$ \\
48 & 5.00E+03 & $m_{\text{D},22}$ \\
49 & 5.00E+03 & $m_{\text{D},33}$ \\
\hline
\end{tabular}
\caption{SLHA Block: Input parameters , EXTPAR}
\end{table}

\begin{table}[h!]
\centering
\begin{tabular}{|c|c|l|}
\hline
Index & Value & Comment \\
\hline
6 & 1.73200000E+02 & t \\
23 & 9.11876000E+01 & Z \\
24 & 8.04087148E+01 & $\text{W}^{\pm}$ \\
25 & 1.25336140E+02 & h \\
35 & 1.01901921E+03 & H \\
36 & 1.02016000E+03 & A \\
37 & 1.02237054E+03  & $H^\pm$ \\
1000001 & 5.06908927E+03 & $~d_L$ \\
2000001 & 5.05763913E+03 & $~d_R$ \\
1000002 & 5.06865355E+03 & $~u_L$ \\
2000002 & 5.05745154E+03 & $~u_R$ \\
1000003 & 5.06908891E+03 & $~s_L$ \\
2000003 & 5.05763852E+03 & $~s_R$ \\
1000004 & 5.06865314E+03 & $~c_L$ \\
2000004 & 5.05745116E+03 & $~c_R$ \\
1000005 & 5.04891828E+03 & $~b_1$ \\
2000005 & 5.05618202E+03 & $~b_2$ \\
1000006 & 2.00857233E+03 & $~t_1$ \\
2000006 & 5.05128643E+03 & $~t_2$ \\
1000011 & 3.02277714E+03 & $~e_L$ \\
2000011 & 3.00494526E+03 & $~e_R$ \\
1000012 & 3.02141573E+03  & $~\nu_{eL}$ \\
1000013 & 3.02277610E+03 & $~\mu_L$ \\
2000013 & 3.00493928E+03 & $~\mu_R$ \\
1000014 & 3.02141326E+03 & $~\nu_{\mu L}$ \\
1000015 & 3.003149970E+03 & $~\tau_1$ \\
2000015 & 3.02189420E+03 & $~\tau_2$ \\
1000016 & 3.02052509E+03 & $~\nu_{\tau L}$ \\
1000021 & 3.31001150E+03 & $~g$ \\
1000022 & 1.23626359E+02 & $~\chi_1^0$ \\
1000023 &-1.77711270E+02 & $~\chi_2^0$ \\
1000025 & 1.96418156E+02 & $~\chi_3^0$ \\
1000035 & 3.06070114E+03 & $~\chi_4^0$ \\
1000024 & 1.74044490E+02 & $~\chi_1^\pm$ \\
1000037 & 3.06086675E+03  & $~\chi_2^\pm$ \\
\hline
\end{tabular}
\caption{SLHA Block: Mass spectrum , MASS}
\end{table}

\begin{table}[h!]
\centering
\begin{tabular}{|c|c|l|}
\hline
Index & Value & Comment \\
\hline
1 1 & -8.21047709E-01 & $N_{11}$, Bino \\
1 2 & 9.82081747E-03 & $N_{12}$, Wino \\
1 3 & -4.58288726E-01 & $N_{13}$, Higgsino \\
1 4 & 3.40228826E-01 & $N_{14}$, Higgsino \\
2 1 & -9.32278383E-02 & $N_{21}$, Bino \\
2 2 & 1.59555767E-02 & $N_{22}$, Wino \\
2 3 & 6.95876151E-01 & $N_{23}$, Higgsino \\
2 4 & 7.11906154E-01 & $N_{24}$, Higgsino \\
3 1 & -5.63195373E-01 & $N_{31}$, Bino \\
3 2 & -1.77535750E-02 & $N_{32}$, Wino \\
3 3 & 5.52917254E-01 & $N_{33}$, Higgsino \\
3 4 & -6.13822688E-01 & $N_{34}$, Higgsino \\
4 1 & 4.48016970E-04 & $N_{41}$, Bino \\
4 2 & -9.99666835E-01 & $N_{42}$, Wino \\
4 3 & -3.21499370E-03 & $N_{43}$, Higgsino \\
4 4 & 2.56062767E-02 & $N_{44}$, Higgsino \\
\hline
\end{tabular}
\caption{SLHA Block: Neutralino mixing matrix , NMIX}
\end{table}

\begin{table}[h!]
\centering
\begin{tabular}{|c|c|l|}
\hline
Index & Value & Comment \\
\hline
1 1 & -4.54827927E-03 & $U_{11}$ \\
1 2 & 9.99989657E-01 & $U_{12}$ \\
2 1 & 9.99989657E-01  & $U_{21}$ \\
2 2 & 4.54827927E-03 & $U_{22}$ \\
\hline
\end{tabular}
\caption{SLHA Block: Chargino U mixing matrix , UMIX}
\end{table}

\begin{table}[h!]
\centering
\begin{tabular}{|c|c|l|}
\hline
Index & Value & Comment \\
\hline
1 1 & -3.62299835E-02 & $V_{11}$ \\
1 2 & 9.99343479E-01 & $V_{12}$ \\
2 1 & 9.99343479E-01 & $V_{21}$ \\
2 2 & 3.62299835E-02 & $V_{22}$ \\
\hline
\end{tabular}
\caption{SLHA Block: Chargino V mixing matrix , VMIX}
\end{table}

\begin{table}[h!]
\centering
\begin{tabular}{|c|c|l|}
\hline
Parent ID or Daughter ID & Width [GeV] or Branching Ratio & Comment \\
\hline
1000022 & 0.00000000E+00 & $~\chi_1^0$ decays \\
\hline
1000023 & 1.14537313E-04 & $~\chi_2^0$ decays \\
22 & 1.20527779E-03 & $BR(~\chi_2^0 \to ~\chi_1^0 ~\gamma)$ \\
2 & 1.17196602E-01 & $BR(~\chi_2^0 \to ~\chi_1^0 \bar{u}      u)$ \\
1 & 1.52226833E-01 & $BR(~\chi_2^0 \to ~\chi_1^0 \bar{d}      d)$ \\
4 & 1.16826967E-01 & $BR(~\chi_2^0 \to ~\chi_1^0 \bar{c}      c)$ \\
3 & 1.52223427E-01 & $BR(~\chi_2^0 \to ~\chi_1^0 \bar{s}      s)$ \\
5 & 1.45324008E-01 & $BR(~\chi_2^0 \to ~\chi_1^0 \bar{b}      b)$ \\
11 & 3.50299405E-02 & $BR(~\chi_2^0 \to ~\chi_1^0 e^+      e^-)$ \\
13 & 3.50287489E-02 & $BR(~\chi_2^0 \to ~\chi_1^0 \mu^+     \mu^-)$ \\
15 & 3.46947116E-02 & $BR(~\chi_2^0 \to ~\chi_1^0 \tau+    \tau-)$ \\
12 & 2.10233221E-01 & $BR(~\chi_2^0 \to ~\chi_1^0 \bar{\nu}\nu)$ \\
\hline
1000025 & 2.52154324E-05 & $~\chi_3^0$ decays \\
22 & 4.76374943E-03 & $BR(~\chi_3^0 \to ~\chi_1^0 ~\gamma)$ \\
1,2,3,4,5 & 4.30781034E-01 & $BR(~\chi_3^0 \to ~\chi_1^0 \bar{q}      q)$ \\
11,13,15 & 6.48588081E-02 & $BR(~\chi_3^0 \to ~\chi_1^0 l^{+}      l^{-})$ \\
12 & 1.32171547E-01 & $BR(~\chi_3^0 \to ~\chi_1^0 \bar{\nu}      \nu)$ \\
1,2,3,4,5& 3.45650791E-02 & $BR(~\chi_3^0 \to ~\chi_2^0 \bar{q}      q)$ \\
11,13,15 & 5.64087599E-03 & $BR(~\chi_3^0 \to ~\chi_2^0 l^{+}      l^{-})$ \\
12 & 1.14763142E-02 & $BR(~\chi_3^0 \to ~\chi_2^0 \bar{\nu}      \nu)$ \\
1,2,3,4,5,11,12,13,14,15,16 & 3.15669902E-01 & $BR(~\chi_3^0 \to ~\chi_1^\pm \bar{f}      f)$ \\
\hline
\end{tabular}
\caption{SLHA Block: Three lightest neutralino decays as calculated by SPheno}
\end{table}

\chapter{Decision Tree Confusion Matrices}\label{app-Conf}
In this appendix we provide table representations for the confusion matrices, as generated by C5.0, used for the four fermion and luminosity process separation, as noted in section~\ref{sec-LumiPID}, as well as the neutralino searches in section~\ref{sec-DarkSect}~\cite{C5}. The correct method to read these tables is as follows, the column is the true classification and the row is the predicted classification. The diagonal elements represent correctly classified events. The final data sample of a given event type would be from the row of that event type. This includes contamination, wherein incorrect event types are included in the data sample. The contamination events come from events along the row that are not the diagonal element.

\section{Raw Luminosity Confusion Matrices}\label{app-conf-rawff}
We first present the raw confusion matrices for separating four-fermion and the luminosity channels of \gls{diphoton}s and \Gls{SABS} wherein 100k events for each channel were used before cuts. Further discussion can be found in section~\ref{sec-LumiPID}.
\begin{table}[h!]
\centering
\caption{Confusion matrix: columns indicate true class, rows indicate predicted class. This is for luminosity channel and four-fermion separation at an $\ee$ collider at \textbf{Z-pole} using the GLIP LumiCal acceptance of $1^\circ\to6^\circ$.}
\begin{tabular}{c|rrr}
\textbf{True \textbackslash\ Predicted} 
  & {$\gamma\gamma$} 
  & {SABS} 
  & {Four-fermion} \\
\hline
{$\gamma\gamma$}   & 35,472 & 11,333 &  2,593 \\
{SABS}             &  5,339 & 40,281 &  3,615 \\
{Four-fermion}     &    907 &  1,128 & 45,368 \\
\end{tabular}
\label{tab:confusion_ILCZ_FF}
\end{table}

\begin{table}[h!]
\centering
\caption{Confusion matrix: columns indicate true class, rows indicate predicted class. This is for luminosity channel and four-fermion separation at an $\ee$ collider at \textbf{WW threshold} using the GLIP LumiCal acceptance of $1^\circ\to6^\circ$.}
\begin{tabular}{c|rrr}
\textbf{True \textbackslash\ Predicted} 
  & {$\gamma\gamma$} 
  & {SABS} 
  & {Four-fermion} \\
\hline
{$\gamma\gamma$}   & 35,562 & 10,716 &  3,108 \\
{SABS}             &  5,596 & 39,280 &  4,128 \\
{Four-fermion}     &  1,205 &  1,073 & 46,289 \\
\end{tabular}
\label{tab:confusion_ILCWW_FF}
\end{table}

\begin{table}[h!]
\centering
\caption{Confusion matrix: columns indicate true class, rows indicate predicted class. This is for luminosity channel and four-fermion separation at an $\ee$ collider at center-of-mass energy of \textbf{250~GeV} using the GLIP LumiCal acceptance of $1^\circ\to6^\circ$.}
\begin{tabular}{c|rrr}
\textbf{True \textbackslash\ Predicted} 
  & {$\gamma\gamma$} 
  & {SABS} 
  & {Four-fermion} \\
\hline
{$\gamma\gamma$}   & 34,372 & 12,056 &  3,053 \\
{SABS}             &  3,920 & 41,452 &  3,963 \\
{Four-fermion}     &  1,352 &  1,266 & 45,749 \\
\end{tabular}
\label{tab:confusion_ILC250_FF}
\end{table}

\begin{table}[h!]
\centering
\caption{Confusion matrix: columns indicate true class, rows indicate predicted class. This is for luminosity channel and four-fermion separation at an $\ee$ collider at center-of-mass energy of \textbf{500~GeV} using the GLIP LumiCal acceptance of $1^\circ\to6^\circ$.}
\begin{tabular}{c|rrr}
\textbf{True \textbackslash\ Predicted} 
  & {$\gamma\gamma$} 
  & {SABS} 
  & {Four-fermion} \\
\hline
{$\gamma\gamma$}   & 35,723 & 10,779 &  3,535 \\
{SABS}             &  5,063 & 39,184 &  5,023 \\
{Four-fermion}     &  2,062 &  2,291 & 43,914 \\
\end{tabular}
\label{tab:confusion_ILC500_FF}
\end{table}

\begin{table}[h!]
\centering
\caption{Confusion matrix: columns indicate true class, rows indicate predicted class. This is for luminosity channel and four-fermion separation at an $\ee$ collider at center-of-mass energy of \textbf{1000~GeV} using the GLIP LumiCal acceptance of $1^\circ\to6^\circ$.}
\begin{tabular}{c|rrr}
\textbf{True \textbackslash\ Predicted} 
  & {$\gamma\gamma$} 
  & {SABS} 
  & {Four-fermion} \\
\hline
{$\gamma\gamma$}   & 34,900 & 10,519 &  4,076 \\
{SABS}             &  4,678 & 39,305 &  5,108 \\
{Four-fermion}     &  2,363 &  2,218 & 44,858 \\
\end{tabular}
\label{tab:confusion_ILC1000_FF}
\end{table}

\section{Weighted Luminosity Confusion Matrices for GLIP LumiCal}\label{app-conf-glip}
Below are the tables for separation of the \gls{diphoton} and \Gls{SABS} luminosity channels from four-fermion production after normalizing for cross-section and weighting for particle identification using the GLIP LumiCal design. The values for particle identification can be found in section~\ref{sec-PID} while discussion and values for cross-sections and generation of the confusion matrices can be found in section~\ref{sec-LumiPID}.

\begin{table}[h!]
\centering
\caption{Confusion matrix at \textbf{Z-pole}, normalized by cross-section and folded with photon and electron particle identification probabilities from section~\ref{sec-PID} using the GLIP LumiCal design.}
\begin{tabular}{c|rrr}
\textbf{Predicted \textbackslash\ True} & $\gamma\gamma$ & SABS & Four-fermion \\
\hline
$\gamma\gamma$   & 80.42 & $1.1\times10^{-4}$ & $2.92 \times 10^{-8}$ \\
SABS             & $1.94 \times 10^{-6}$ & 40112.16 & 4.07 \\
Four-fermion     & $3.29 \times 10^{-7}$ & 1123.27 & 51.07 \\
\end{tabular}
\label{tab:confusion_91}
\end{table}

\begin{table}[h!]
\centering
\caption{Confusion matrix at \textbf{WW threshold}, normalized by cross-section and folded with photon and electron particle identification probabilities from section~\ref{sec-PID} using the GLIP LumiCal design.}
\begin{tabular}{c|rrr}
\textbf{Predicted \textbackslash\ True} & $\gamma\gamma$ & SABS & Four-fermion \\
\hline
$\gamma\gamma$   & 66.20 & $1.1\times10^{-4}$ & $4.89 \times 10^{-8}$ \\
SABS             & $1.67 \times 10^{-6}$ & 39113.78 & 6.49 \\
Four-fermion     & $3.59 \times 10^{-7}$ & 1068.46 & 72.74 \\
\end{tabular}
\label{tab:confusion_161}
\end{table}

\begin{table}[h!]
\centering
\caption{Confusion matrix at center-of-mass energy of \textbf{250~GeV}, normalized by cross-section and folded with photon and electron particle identification probabilities from section~\ref{sec-PID} using the GLIP LumiCal design.}
\begin{tabular}{c|rrr}
\textbf{Predicted \textbackslash\ True} 
  & $\gamma\gamma$ 
  & SABS
  & Four-fermion \\
\hline
$\gamma\gamma$   & 61.82 & $1.2\times10^{-4}$ & $3.68 \times 10^{-8}$ \\
SABS             & $1.13 \times 10^{-6}$ & 41294.41 & 4.77 \\
Four-fermion     & $3.89 \times 10^{-7}$ & 1261.19 & 55.10 \\
\end{tabular}
\label{tab:confusion_ILC250_PID_xsecNorm}
\end{table}

\begin{table}[h!]
\centering
\caption{Confusion matrix at center-of-mass energy of \textbf{500~GeV}, normalized by cross-section and folded with photon and electron particle identification probabilities from section~\ref{sec-PID} using the GLIP LumiCal design.}
\begin{tabular}{c|rrr}
\textbf{Predicted \textbackslash\ True} & $\gamma\gamma$ & SABS & Four-fermion \\
\hline
$\gamma\gamma$   & 93.21 & $1.1\times10^{-4}$ & $4.91 \times 10^{-8}$ \\
SABS             & $2.11 \times 10^{-6}$ & 38996.11 & 6.97 \\
Four-fermion     & $8.61 \times 10^{-7}$ & 2280.01 & 60.94 \\
\end{tabular}
\label{tab:confusion_500}
\end{table}

\begin{table}[h!]
\centering
\caption{Confusion matrix at center-of-mass energy of \textbf{1000~GeV}, normalized by cross-section and folded with photon and electron particle identification probabilities from section~\ref{sec-PID} using the GLIP LumiCal design.}
\begin{tabular}{c|rrr}
\textbf{Predicted \textbackslash\ True} & $\gamma\gamma$ & SABS & Four-fermion \\
\hline
$\gamma\gamma$   & 60.09 & $1.1\times10^{-4}$ & $3.17 \times 10^{-8}$ \\
SABS             & $1.29 \times 10^{-6}$ & 39175.38 & 3.97 \\
Four-fermion     & $6.51 \times 10^{-7}$ & 2210.69 & 34.86 \\
\end{tabular}
\label{tab:confusion_1000}
\end{table}

\section{Weighted Luminosity Confusion Matrices for ILD LumiCal}\label{app-conf-lumical}
Below are the tables for separation of the \gls{diphoton} and \Gls{SABS} luminosity channels from four-fermion production after normalizing for cross-section and weighting for particle identification using the ILD LumiCal design. The values for particle identification can be found in section~\ref{sec-PID} while discussion and values for cross-sections and generation of the confusion matrices can be found in section~\ref{sec-LumiPID}. We use the same raw values as in section~\ref{app-conf-glip} so that comparisons are only reflective of the particle identification effects.

\begin{table}[h!]
\centering
\caption{Confusion matrix at \textbf{Z-pole}, normalized by cross-section and folded with photon and electron particle identification probabilities from section~\ref{sec-PID} using the ILD LumiCal design.}
\begin{tabular}{c|rrr}
\textbf{Predicted \textbackslash\ True} & $\gamma\gamma$ & SABS & Four-fermion \\
\hline
$\gamma\gamma$   & 79.64 & 0.28 & $7.3 \times 10^{-5}$ \\
SABS             & 0.037 & 35849.84 & 3.64 \\
Four-fermion     & 0.006 & 1003.91 & 45.64 \\
\end{tabular}
\label{tab:confusion_91_alt}
\end{table}

\begin{table}[h!]
\centering
\caption{Confusion matrix at \textbf{WW Threshold}, normalized by cross-section and folded with photon and electron particle identification probabilities from section~\ref{sec-PID} using the ILD LumiCal design.}
\begin{tabular}{c|rrr}
\textbf{Predicted \textbackslash\ True} & $\gamma\gamma$ & SABS & Four-fermion \\
\hline
$\gamma\gamma$   & 65.55 & 0.27 & $1.2 \times 10^{-4}$ \\
SABS             & 0.032 & 34957.54 & 5.80 \\
Four-fermion     & 0.007 & 954.92 & 65.01 \\
\end{tabular}
\label{tab:confusion_161_alt}
\end{table}

\begin{table}[h!]
\centering
\caption{Confusion matrix at \textbf{250~GeV}, normalized by cross-section and folded with photon and electron particle identification probabilities from section~\ref{sec-PID} using the ILD LumiCal design.}
\begin{tabular}{c|rrr}
\textbf{Predicted \textbackslash\ True} & $\gamma\gamma$ & SABS & Four-fermion \\
\hline
$\gamma\gamma$   & 61.22 & 0.30 & $9.2 \times 10^{-5}$ \\
SABS             & 0.021 & 36906.46 & 4.27 \\
Four-fermion     & 0.007 & 1127.17 & 49.24 \\
\end{tabular}
\label{tab:confusion_250_alt}
\end{table}

\begin{table}[h!]
\centering
\caption{Confusion matrix at \textbf{500~GeV}, normalized by cross-section and folded with photon and electron particle identification probabilities from section~\ref{sec-PID} using the ILD LumiCal design.}
\begin{tabular}{c|rrr}
\textbf{Predicted \textbackslash\ True} & $\gamma\gamma$ & SABS & Four-fermion \\
\hline
$\gamma\gamma$   & 92.29 & 0.27 & $1.2 \times 10^{-4}$ \\
SABS             & 0.040 & 34852.37 & 6.23 \\
Four-fermion     & 0.016 & 2037.74 & 54.47 \\
\end{tabular}
\label{tab:confusion_500_alt}
\end{table}

\begin{table}[h!]
\centering
\caption{Confusion matrix at \textbf{1000~GeV}, normalized by cross-section and folded with photon and electron particle identification probabilities from section~\ref{sec-PID} using the ILD LumiCal design.}
\begin{tabular}{c|rrr}
\textbf{Predicted \textbackslash\ True} & $\gamma\gamma$ & SABS & Four-fermion \\
\hline
$\gamma\gamma$   & 59.51 & 0.26 & $7.9 \times 10^{-5}$ \\
SABS             & 0.024 & 35012.60 & 3.55 \\
Four-fermion     & 0.012 & 1975.78 & 31.16 \\
\end{tabular}
\label{tab:confusion_1000_alt}
\end{table}

\section{MSSM Neutralino Confusion Matrices}\label{app-conf-mssm}
We have provided, for completion, a copy of table~\ref{tab-confusion}, even though it is already present in section~\ref{sec-DarkSect}. We first present the confusion matrices for Bino-Higgsino world and then the confusion matrices for Bino-Wino world.

\begin{table}[h!]
\centering
\caption{Confusion matrix: columns indicate true class, rows indicate predicted class. This is for MSSM Bino-Higgsino world for ILC500.}
\begin{tabular}{c|rrrrrrr}
\textbf{True \textbackslash\ Predicted} 
  & {$\gamma\gamma$} 
  & {$\nu\bar{\nu}$} 
  & {Higgs-like} 
  & {BERB} 
  & {ZZ-like} 
  & {$\chi^0_1$-like} 
  & {$\chi^0_{2,3}$-like} \\
\hline
{$\gamma\gamma$}       & 9,979,997 & 786,285 &       0 &      0 &     0 &     0 &      43 \\
{$\nu\bar{\nu}$}       &    41,231 &36,164,044 &  5,017 &      0 &     0 &     0 &   3,754 \\
{Higgs-like}           &        28 &    16,239 & 67,190 &      0 &     0 &     0 &      0 \\
{BERB}                 &        20 &        28 &     0  & 364,422 &   0 &     0 &      0 \\
{ZZ-like}              &        88 &   105,259 &     6  &      0 &     0 &     0 &     13 \\
{$\chi^0_1$-like}      &         0 &       596 &     0  &      0 &     0 &     0 &      0 \\
{$\chi^0_{2,3}$-like}  &         0 &    40,998 &     1  &      0 &     0 &     0 &  22,773 \\
\end{tabular}
\label{tab:confusion_ILC500_BH}
\end{table}

\begin{table}[h!]
\centering
\caption{Confusion matrix: columns indicate true class, rows indicate predicted class. This is for MSSM Bino-Higgsino world for ILC1000.}
\begin{tabular}{c|rrrrrrr}
\textbf{True \textbackslash\ Predicted} 
  & {$\gamma\gamma$} 
  & {$\nu\bar{\nu}$} 
  & {Higgs-like} 
  & {BERB} 
  & {ZZ-like} 
  & {$\chi^0_1$-like} 
  & {$\chi^0_{2,3}$-like} \\
\hline
{$\gamma\gamma$}       & 3,218,750 & 439,152 &     0 &     0 &     0 &     0 &    26 \\
{$\nu\bar{\nu}$}       &    28,267 & 6,235,369 & 1,751 &     0 &     0 &     0 &   639 \\
{Higgs-like}           &        23 &     3,643 & 27,435 &    0 &     0 &     0 &     0 \\
{BERB}                 &         2 &         1 &     0 & 12,227 &     0 &     0 &     0 \\
{ZZ-like}              &       550 &    64,121 &    87 &     0 &     0 &     0 &     9 \\
{$\chi^0_1$-like}      &         4 &     1,242 &     1 &     0 &     0 &     0 &     0 \\
{$\chi^0_{2,3}$-like}  &        51 &    14,102 &     3 &     0 &     0 &     0 &  3,984 \\
\end{tabular}
\label{tab:confusion_ILC1000_BH}
\end{table}

\begin{table}[h!]
\centering
\caption{Confusion matrix: columns indicate true class, rows indicate predicted class. This is for MSSM Bino-Wino world for ILC500.}
\begin{tabular}{c|rrrrrrr}
\textbf{True \textbackslash\ Predicted} 
  & {$\gamma\gamma$} 
  & {$\nu\bar{\nu}$} 
  & {Higgs-like} 
  & {BERB} 
  & {ZZ-like} 
  & {$\chi^0_1$-like} 
  & {$\chi^0_{2,3}$-like} \\
\hline
{$\gamma\gamma$}       & 9,979,450 & 786,884 &     0 &      0 &     0 &     0 &     0 \\
{$\nu\bar{\nu}$}       &    39,795 &36,168,151 & 6,351 &      0 &     0 &     0 &     0 \\
{Higgs-like}           &        38 &    14,404 & 68,957 &     0 &     0 &     0 &     0 \\
{BERB}                 &        23 &        23 &     0  & 364,444 &   0 &     0 &     0 \\
{ZZ-like}              &         1 &     2,548 &     0  &      0 &     0 &     0 &     0 \\
{$\chi^0_1$-like}      &         0 &         3 &     0  &      0 &     0 &     0 &     0 \\
{$\chi^0_{2,3}$-like}  &         0 &       310 &     0  &      0 &     0 &     0 &     0 \\
\end{tabular}
\label{tab:confusion_ILC500_BW}
\end{table}

\begin{table}[h]
\centering
\caption{Confusion matrix: columns indicate true class, rows indicate predicted class. This is for MSSM Bino-Wino world for ILC1000.}
\begin{tabular}{c|rrrrrrr}
\textbf{True \textbackslash\ Predicted} 
  & {$\gamma\gamma$} 
  & {$\nu\bar{\nu}$} 
  & {Higgs-like} 
  & {BERB} 
  & {ZZ-like} 
  & {$\chi^0_1$-like} 
  & {$\chi^0_{2,3}$-like} \\
\hline
{$\gamma\gamma$}       & 3,220,624 & 437,375 &         0 &       0 &      0 &     0 &    0 \\
{$\nu\bar{\nu}$}       &    30,313 & 6,220,924 &   1,836 &       0 &      0 &     0 &    0 \\
{Higgs-like}           &        19 &    3,698 &  27,451 &       0 &      0 &     0 &    0 \\
{BERB}                 &         2 &        0 &       0 &  12,248 &      0 &     0 &    0 \\
{ZZ-like}              &       585 &   64,075 &      85 &       0 &      0 &     0 &    0 \\
{$\chi^0_1$-like}      &         0 &        0 &       0 &       0 &      0 &   325 &    0 \\
{$\chi^0_{2,3}$-like}  &        11 &   11,517 &       0 &       0 &      0 &     0 &    6 \\
\end{tabular}
\end{table}\label{tab-ConBW1000}

\chapter{micrOMEGAs and SModelS Results}\label{app-SModelS}
Here we provide a table with exclusion values, represented with r-values, from the output of micrOMEGAs and SModelS. Both of these  looks for the exclusion of BSM physics models against existing LHC and other particle physics analyses~\cite{SModelSv3}. We put a cut on the r-value to only include results with an r-value $\geq10^{-4}$ so that only results that are relevant to the MSSM model used here are presented. We put an exception on this for the direct detection results so that they are always presented. The process name and analyses are provided for the sake of reference. Since micrOMEGAs reports direct detection in terms of confidence percentage we have used a script, which can be found in the MSSM directory on GitHub, to convert confidence percentage to r-value~\cite{github}. We also used micrOMEGAs to compute precision standard model and astroparticle signatures but we have chosen to omit these for the sake of brevity and lack of relation to the surrounding work. Both models were modified to ensure that the relic abundance of dark matter is within 10\% of results from experimental measurements, particularly those from the Planck mission~\cite{Planck:2018vyg}. So there is no quoted exclusion for dark matter relic abundance as it is considered an already tuned and accepted value. These steps were done for both MSSM models of the Bino-Wino world scenario and the Bino-Higgsino world scenario.
\begin{table}[h!]
\centering
\begin{tabular}{|c|c|l|}
\hline
\multicolumn{3}{|c|}{\textbf{Bino-Wino World Constraints from Current Analyses}} \\
\hline
TXName or Measurement & Analysis or Experiment & r-value \\
\hline
TChiWZoff & CMS-SUS-18-004~\cite{CMS-SUS-18-004} & 0.9778\\
TChiWZoff & CMS-SUS-17-004~\cite{CMS-SUS-17-004} & 0.6533\\
TChiWZoff & ATLAS-SUSY-2018-16~\cite{ATLAS:2019lng} & 0.5358 \\
TChiWZoff & CMS-SUS-16-048~\cite{CMS-SUS-16-048} & 0.1713 \\
T1 & ATLAS-SUSY-2016-07~\cite{ATLAS-SUSY-2016-07} & 0.0386 \\
TChiWZoff & CMS-SUS-16-039-agg~\cite{CMS:2018Agg} & 0.0075 \\
T2tt & CMS-SUS-19-006-agg~\cite{CMS-SUS-19-006-agg} & 0.0030 \\
TChiWZoff & ATLAS-SUSY-2013-12~\cite{ATLAS-SUSY-2013-12} & 0.0030 \\
Direct Detection & DarkSide-50~\cite{Agnes_2018}~\cite{Belanger:2020gnr} & 0.0006 \\
TChiWZoff & CMS-SUS-16-039~\cite{CMS:2018Agg} & 0.0001 \\
Direct Detection & IceCube~\cite{Abbasi_2009}~\cite{B_langer_2015} & <0.0001 \\
\hline
\end{tabular}
\caption{SModelS v3.0.0 Results for MSSM Bino-Wino world model tested in this work, only keeping results that are equal to or above an r-value of $10^{-4}$. We also include direct detection and annihilation measurements, as calculated by micrOMEGAS v6.2.3 here to provide a comprehensive exclusion results.}
\end{table}

\begin{table}[h!]
\centering
\begin{tabular}{|c|c|l|}
\hline
\multicolumn{3}{|c|}{\textbf{Bino-Higgsino World Constraints from Current Analyses}} \\
\hline
TXName or Measurement & Analysis or Experiment & r-value \\
\hline
Direct Detection & DarkSide-50~\cite{Agnes_2018}~\cite{Belanger:2020gnr} & >2.0 \\
Direct Detection & IceCube~\cite{Abbasi_2009}~\cite{B_langer_2015} & >2.0 \\
TChiWZoff & ATLAS-SUSY-2019-09~\cite{ATLAS:2021moa} & 0.8420\\
TChiWZoff & CMS-SUS-17-004~\cite{CMS-SUS-17-004} & 0.3756\\
TChiWZoff & ATLAS-SUSY-2018-05-ewk~\cite{ATLAS:2022zwa} & 0.1924\\
TChiZoff & ATLAS-SUSY-2018-16~\cite{ATLAS:2019lng} & 0.0423 \\
TChiWZoff & CMS-SUS-16-048~\cite{CMS-SUS-16-048} & 0.0164 \\
T1 & ATLAS-SUSY-2016-07~\cite{ATLAS-SUSY-2016-07} & 0.0150 \\
T1 & ATLAS-SUSY-2013-02~\cite{ATLAS:2014jxt} & 0.0036 \\
TChiWZoff & CMS-SUS-16-039-agg~\cite{CMS:2018Agg} & 0.0024 \\
TChiWWoff & ATLAS-SUSY-2013-11~\cite{ATLAS:2014zve} & 0.0019 \\
TChiWZoff & ATLAS-SUSY-2018-16~\cite{ATLAS:2019lng} & 0.0004 \\
TChiWZoff & ATLAS-SUSY-2018-05~\cite{ATLAS:2022zwa} & 0.0001\\
\hline
\end{tabular}
\caption{SModelS v3.0.0 Results for MSSM Bino-Higgsino world model tested in this work, only keeping results that are equal to or above an r-value of $10^{-4}$. We also include direct detection and annihilation measurements, as calculated by micrOMEGAS v6.2.3 here to provide a comprehensive exclusion results.}
\end{table}

\printglossary

\end{appendices}

\begin{center}
\line(1,0){250}
\end{center}

\vspace{1.0 cm}


\global\long\def\bibname{References}%

\bibliographystyle{JHEP}
\bibliography{Masters_v4}

\providecommand{\href}[2]{#2}\begingroup\raggedright\begin{thebibliography}{100}

\bibitem{Myers:1991ym}
S.~Myers, \emph{{The {LEP} collider, from design to approval and commissioning}}, \href{http://dx.doi.org/10.5170/CERN-1991-008}{\emph{CERN-YELLOW} {\bf 1991-008} (10, 1991) }.

\bibitem{CERN:1991gja}
$\text{The LHC Study Group}$, \emph{{Design Study of the Large Hadron Collider ({LHC}): A Multiparticle Collider in the {LEP} Tunnel}}, \href{http://dx.doi.org/10.5170/CERN-1991-003}{\emph{CERN-YELLOW} {\bf 91-03} (5, 1991) }.

\bibitem{LEPHiggs}
{\scshape LEP Working Group for Higgs boson searches, ALEPH, DELPHI, L3, OPAL} collaboration, R.~Barate et~al., \emph{{Search for the standard model {Higgs} boson at {LEP}}}, \href{http://dx.doi.org/10.1016/S0370-2693(03)00614-2}{\emph{Phys. Lett. B} {\bf 565} (2003) 61--75}, [\href{https://arxiv.org/abs/hep-ex/0306033}{{\tt hep-ex/0306033}}].

\bibitem{ZurbanoFernandez:2020cco}
I.~Zurbano~Fernandez et~al., \emph{{High-Luminosity Large Hadron Collider ({HL-LHC}): Technical design report}}, \href{http://dx.doi.org/10.23731/CYRM-2020-0010}{\emph{CERN} {\bf 10/2020} (12, 2020) }.

\bibitem{ATLAS:2022hro}
{\scshape ATLAS} collaboration, G.~Aad et~al., \emph{{Luminosity determination in $pp$ collisions at $\sqrt{s}=13$ TeV using the {ATLAS} detector at the {LHC}}}, \href{http://dx.doi.org/10.1140/epjc/s10052-023-11747-w}{\emph{Eur. Phys. J. C} {\bf 83} (2023) 982}, [\href{https://arxiv.org/abs/2212.09379}{{\tt 2212.09379}}].

\bibitem{CMS:2012qbp}
{\scshape CMS} collaboration, S.~Chatrchyan et~al., \emph{{Observation of a New Boson at a Mass of 125 GeV with the {CMS} Experiment at the {LHC}}}, \href{http://dx.doi.org/10.1016/j.physletb.2012.08.021}{\emph{Phys. Lett. B} {\bf 716} (2012) 30--61}, [\href{https://arxiv.org/abs/1207.7235}{{\tt 1207.7235}}].

\bibitem{ATLAS:2012yve}
{\scshape ATLAS} collaboration, G.~Aad et~al., \emph{{Observation of a new particle in the search for the Standard Model {Higgs} boson with the {ATLAS} detector at the {LHC}}}, \href{http://dx.doi.org/10.1016/j.physletb.2012.08.020}{\emph{Phys. Lett. B} {\bf 716} (2012) 1--29}, [\href{https://arxiv.org/abs/1207.7214}{{\tt 1207.7214}}].

\bibitem{KLEIN_1926}
O.~Klein, \emph{The atomicity of electricity as a quantum theory law}, \href{http://dx.doi.org/10.1038/118516a0}{\emph{Nature} {\bf 118} (Oct, 1926) 516–516}.

\bibitem{Pardo_2018}
K.~Pardo, M.~Fishbach, D.~E. Holz and D.~N. Spergel, \emph{Limits on the number of spacetime dimensions from gw170817}, \href{http://dx.doi.org/10.1088/1475-7516/2018/07/048}{\emph{Journal of Cosmology and Astroparticle Physics} {\bf 2018} (July, 2018) 048–048}.

\bibitem{schwarzschild}
K.~Schwarzschild, \emph{On the gravitational field of a mass point according to einstein's theory}, {\emph{Math.Phys.} (1916) 189--196}, [\href{https://arxiv.org/abs/physics/9905030}{{\tt physics/9905030}}].

\bibitem{Hawking:1974rv}
S.~W. Hawking, \emph{{Black hole explosions}}, \href{http://dx.doi.org/10.1038/248030a0}{\emph{Nature} {\bf 248} (1974) 30--31}.

\bibitem{EHT}
{\scshape The Event Horizon Telescope} collaboration, K.~e.~a. Akiyama, \emph{First m87 event horizon telescope results. viii. magnetic field structure near the event horizon}, \href{http://dx.doi.org/10.3847/2041-8213/abe4de}{\emph{The Astrophysical Journal Letters} {\bf 910} (Mar., 2021) L13}.

\bibitem{Giddings_2018}
S.~B. Giddings and D.~Psaltis, \emph{Event horizon telescope observations as probes for quantum structure of astrophysical black holes}, \href{http://dx.doi.org/10.1103/physrevd.97.084035}{\emph{Physical Review D} {\bf 97} (Apr., 2018) }.

\bibitem{Planck:2018vyg}
{\scshape Planck} collaboration, N.~Aghanim et~al., \emph{{Planck 2018 results. VI. Cosmological parameters}}, \href{http://dx.doi.org/10.1051/0004-6361/201833910}{\emph{Astron. Astrophys.} {\bf 641} (2020) A6}, [\href{https://arxiv.org/abs/1807.06209}{{\tt 1807.06209}}].

\bibitem{Jungman:1995df}
G.~Jungman, M.~Kamionkowski and K.~Griest, \emph{{Supersymmetric dark matter}}, \href{http://dx.doi.org/10.1016/0370-1573(95)00058-5}{\emph{Phys. Rept.} {\bf 267} (1996) 195--373}, [\href{https://arxiv.org/abs/hep-ph/9506380}{{\tt hep-ph/9506380}}].

\bibitem{Bertone:2004pz}
G.~Bertone, D.~Hooper and J.~Silk, \emph{{Particle dark matter: Evidence, candidates and constraints}}, \href{http://dx.doi.org/10.1016/j.physrep.2004.08.031}{\emph{Phys. Rept.} {\bf 405} (2005) 279--390}, [\href{https://arxiv.org/abs/hep-ph/0404175}{{\tt hep-ph/0404175}}].

\bibitem{Clifton:2011jh}
T.~Clifton, P.~G. Ferreira, A.~Padilla and C.~Skordis, \emph{{Modified Gravity and Cosmology}}, \href{http://dx.doi.org/10.1016/j.physrep.2012.01.001}{\emph{Phys. Rept.} {\bf 513} (2012) 1--189}, [\href{https://arxiv.org/abs/1106.2476}{{\tt 1106.2476}}].

\bibitem{PhysRevD.103.122002}
{\scshape LIGO Scientific and Virgo} collaboration, R.~e.~a. Abbott, \emph{Tests of general relativity with binary black holes from the second ligo-virgo gravitational-wave transient catalog}, \href{http://dx.doi.org/10.1103/PhysRevD.103.122002}{\emph{Phys. Rev. D} {\bf 103} (Jun, 2021) 122002}.

\bibitem{CMSDarkMatter}
{\scshape CMS} collaboration, S.~e.~a. Chatrchyan, \emph{Search for invisible decays of {Higgs} bosons in the vector boson fusion and associated zh production modes}, \href{http://dx.doi.org/10.1140/epjc/s10052-014-2980-6}{\emph{The European Physical Journal C} {\bf 74} (Aug., 2014) }.

\bibitem{XENON:2018voc}
{\scshape XENON} collaboration, E.~Aprile et~al., \emph{{Dark Matter Search Results from a One Ton-Year Exposure of XENON1T}}, \href{http://dx.doi.org/10.1103/PhysRevLett.121.111302}{\emph{Phys. Rev. Lett.} {\bf 121} (2018) 111302}, [\href{https://arxiv.org/abs/1805.12562}{{\tt 1805.12562}}].

\bibitem{RefPDG}
{\scshape Particle Data Group} collaboration, R.~L. Workman et~al., \emph{{Review of Particle Physics}}, \href{http://dx.doi.org/10.1093/ptep/ptac097}{\emph{PTEP} {\bf 2022} (2022) 083C01}.

\bibitem{L3:1990jlf}
{\scshape L3} collaboration, B.~Adeva et~al., \emph{{A Test of QCD based on four jet events from Z0 decays}}, \href{http://dx.doi.org/10.1016/0370-2693(90)90043-6}{\emph{Phys. Lett. B} {\bf 248} (1990) 227--234}.

\bibitem{Dainton:2006wd}
J.~B. Dainton, M.~Klein, P.~Newman, E.~Perez and F.~Willeke, \emph{{Deep inelastic electron-nucleon scattering at the {LHC}}}, \href{http://dx.doi.org/10.1088/1748-0221/1/10/P10001}{\emph{JINST} {\bf 1} (2006) P10001}, [\href{https://arxiv.org/abs/hep-ex/0603016}{{\tt hep-ex/0603016}}].

\bibitem{Surrow:2007zz}
B.~Surrow, \emph{{Low-x physics at a future electron-ion collider (EIC) facility}}, \href{http://dx.doi.org/10.1088/1742-6596/110/2/022049}{\emph{J. Phys. Conf. Ser.} {\bf 110} (2008) 022049}.

\bibitem{CMS:2019ekd}
{\scshape CMS} collaboration, A.~M. Sirunyan et~al., \emph{{Measurements of the {Higgs} boson width and anomalous $HVV$ couplings from on-shell and off-shell production in the four-lepton final state}}, \href{http://dx.doi.org/10.1103/PhysRevD.99.112003}{\emph{Phys. Rev. D} {\bf 99} (2019) 112003}, [\href{https://arxiv.org/abs/1901.00174}{{\tt 1901.00174}}].

\bibitem{Barish:1977qk}
S.~J. Barish et~al., \emph{{Study of Neutrino Interactions in Hydrogen and Deuterium. 1. Description of the Experiment and Study of the Reaction Neutrino d --\ensuremath{>} mu- p p(s)}}, \href{http://dx.doi.org/10.1103/PhysRevD.16.3103}{\emph{Phys. Rev. D} {\bf 16} (1977) 3103}.

\bibitem{Sirunyan_2019}
{\scshape CMS} collaboration, A.~Sirunyan et~al., \emph{Search for mssm {Higgs} bosons decaying to $\mu^+\mu^-$ in proton-proton collisions at $\sqrt{s}=13$ tev}, \href{http://dx.doi.org/10.1016/j.physletb.2019.134992}{\emph{Physics Letters B} {\bf 798} (Nov., 2019) 134992}.

\bibitem{atlas_muon2019}
{\scshape ATLAS} collaboration, M.~Aaboud et~al., \emph{Search for scalar resonances decaying into $\mu^+\mu^-$ in events with and without b-tagged jets produced in proton-proton collisions at $\sqrt{s}=13$ tev with the {ATLAS} detector}, \href{http://dx.doi.org/10.1007/jhep07(2019)117}{\emph{Journal of High Energy Physics} {\bf 2019} (July, 2019) }.

\bibitem{Aad_2022}
{\scshape ATLAS} collaboration, G.~Aad et~al., \emph{Direct constraint on the {Higgs}–charm coupling from a search for {Higgs} boson decays into charm quarks with the {ATLAS} detector}, \href{http://dx.doi.org/10.1140/epjc/s10052-022-10588-3}{\emph{The European Physical Journal C} {\bf 82} (Aug., 2022) }.

\bibitem{Borexino:2017fbd}
{\scshape Borexino} collaboration, M.~Agostini et~al., \emph{{Limiting neutrino magnetic moments with Borexino Phase-II solar neutrino data}}, \href{http://dx.doi.org/10.1103/PhysRevD.96.091103}{\emph{Phys. Rev. D} {\bf 96} (2017) 091103}, [\href{https://arxiv.org/abs/1707.09355}{{\tt 1707.09355}}].

\bibitem{L3:1998uub}
{\scshape L3} collaboration, M.~Acciarri et~al., \emph{{Determination of the number of light neutrino species from single photon production at {LEP}}}, \href{http://dx.doi.org/10.1016/S0370-2693(98)00519-X}{\emph{Phys. Lett. B} {\bf 431} (1998) 199--208}.

\bibitem{ALEPH:2005ab}
{\scshape ALEPH, DELPHI, L3, OPAL, SLD, LEP Electroweak Working Group, SLD Electroweak Group, SLD Heavy Flavour Group} collaboration, S.~Schael et~al., \emph{{Precision electroweak measurements on the $Z$ resonance}}, \href{http://dx.doi.org/10.1016/j.physrep.2005.12.006}{\emph{Phys. Rept.} {\bf 427} (2006) 257--454}, [\href{https://arxiv.org/abs/hep-ex/0509008}{{\tt hep-ex/0509008}}].

\bibitem{Rossi:2014nea}
G.~Rossi, C.~Y\`eche, N.~Palanque-Delabrouille and J.~Lesgourgues, \emph{{Constraints on dark radiation from cosmological probes}}, \href{http://dx.doi.org/10.1103/PhysRevD.92.063505}{\emph{Phys. Rev. D} {\bf 92} (2015) 063505}, [\href{https://arxiv.org/abs/1412.6763}{{\tt 1412.6763}}].

\bibitem{Graham_2015}
P.~W. Graham, D.~E. Kaplan and S.~Rajendran, \emph{Cosmological relaxation of the electroweak scale}, \href{http://dx.doi.org/10.1103/physrevlett.115.221801}{\emph{Physical Review Letters} {\bf 115} (Nov., 2015) }.

\bibitem{Espinosa_2015}
J.~Espinosa, C.~Grojean, G.~Panico, A.~Pomarol, O.~Pujolàs and G.~Servant, \emph{Cosmological {Higgs}-axion interplay for a naturally small electroweak scale}, \href{http://dx.doi.org/10.1103/physrevlett.115.251803}{\emph{Physical Review Letters} {\bf 115} (Dec., 2015) }.

\bibitem{Carena:2003aj}
M.~Carena, A.~de~Gouvea, A.~Freitas and M.~Schmitt, \emph{{Invisible Z boson decays at e+ e- colliders}}, \href{http://dx.doi.org/10.1103/PhysRevD.68.113007}{\emph{Phys. Rev. D} {\bf 68} (2003) 113007}, [\href{https://arxiv.org/abs/hep-ph/0308053}{{\tt hep-ph/0308053}}].

\bibitem{Garcia_Cely_2013}
C.~Garcia~Cely, A.~Ibarra, E.~Molinaro and S.~Petcov, \emph{{Higgs} decays in the low scale type i see-saw model}, \href{http://dx.doi.org/10.1016/j.physletb.2012.11.026}{\emph{Physics Letters B} {\bf 718} (Jan., 2013) 957–964}.

\bibitem{OPALGamGam}
{\scshape OPAL} collaboration, G.~Abbiendi et~al., \emph{Multi-photon production in ee collisions at $\sqrt{s} = $ 181-209 gev}, \href{http://dx.doi.org/10.1140/epjc/s2002-01074-5}{\emph{The European Physical Journal C} {\bf 26} (Jan., 2003) 331–344}.

\bibitem{ALEPH:2013dgf}
{\scshape ALEPH, DELPHI, L3, OPAL, LEP Electroweak} collaboration, S.~Schael et~al., \emph{{Electroweak Measurements in Electron-Positron Collisions at W-Boson-Pair Energies at {LEP}}}, \href{http://dx.doi.org/10.1016/j.physrep.2013.07.004}{\emph{Phys. Rept.} {\bf 532} (2013) 119--244}, [\href{https://arxiv.org/abs/1302.3415}{{\tt 1302.3415}}].

\bibitem{Giudice:1998ck}
G.~F. Giudice, R.~Rattazzi and J.~D. Wells, \emph{{Quantum gravity and extra dimensions at high-energy colliders}}, \href{http://dx.doi.org/10.1016/S0550-3213(99)00044-9}{\emph{Nucl. Phys. B} {\bf 544} (1999) 3--38}, [\href{https://arxiv.org/abs/hep-ph/9811291}{{\tt hep-ph/9811291}}].

\bibitem{Eboli:1998vg}
O.~J.~P. Eboli, M.~C. Gonzalez-Garcia, S.~M. Lietti and S.~F. Novaes, \emph{{Bounds on {Higgs} and gauge boson interactions from {LEP}-2 data}}, \href{http://dx.doi.org/10.1016/S0370-2693(98)00752-7}{\emph{Phys. Lett. B} {\bf 434} (1998) 340--346}, [\href{https://arxiv.org/abs/hep-ph/9802408}{{\tt hep-ph/9802408}}].

\bibitem{QuantumILC}
A.~Freitas, K.~Hagiwara, S.~Heinemeyer, P.~Langacker, K.~Moenig, M.~Tanabashi et~al., \emph{Exploring quantum physics at the {ILC}},  \href{https://arxiv.org/abs/1307.3962}{{\tt 1307.3962}}.

\bibitem{L3:2004ulv}
{\scshape L3} collaboration, P.~Achard et~al., \emph{{Measurement of triple gauge boson couplings of the $W$ boson at {LEP}}}, \href{http://dx.doi.org/10.1016/j.physletb.2004.02.045}{\emph{Phys. Lett. B} {\bf 586} (2004) 151--166}, [\href{https://arxiv.org/abs/hep-ex/0402036}{{\tt hep-ex/0402036}}].

\bibitem{asner2018ilchiggswhitepaper}
{\scshape the ILC International Development Team and the ILC community} collaboration, D.~Asner et~al., \emph{{ILC} higgs white paper},  \href{https://arxiv.org/abs/1310.0763}{{\tt 1310.0763}}.

\bibitem{Binoth:2005ua}
T.~Binoth, M.~Ciccolini, N.~Kauer and M.~Kramer, \emph{{Gluon-induced WW background to {Higgs} boson searches at the {LHC}}}, \href{http://dx.doi.org/10.1088/1126-6708/2005/03/065}{\emph{JHEP} {\bf 03} (2005) 065}, [\href{https://arxiv.org/abs/hep-ph/0503094}{{\tt hep-ph/0503094}}].

\bibitem{PDG2024}
{\scshape Particle Data Group} collaboration, S.~Navas et~al., \emph{{Review of particle physics}}, \href{http://dx.doi.org/10.1103/PhysRevD.110.030001}{\emph{Phys. Rev. D} {\bf 110} (2024) 030001}.

\bibitem{ATLAS2018}
{\scshape ATLAS} collaboration, M.~Aaboud et~al., \emph{{Measurement of the effective leptonic weak mixing angle using electron and muon pairs from $Z$-boson decay in the {ATLAS} experiment at $\sqrt s = 8$ TeV}}, .

\bibitem{Shalaev2018}
V.~Shalaev, I.~Gorbunov and S.~Shmatov, \emph{{Measurement of the Forward-Backward Asymmetry in the Drell-Yan Dilepton Production in Proton-Proton Collisions at the {CMS} Experiment at the {LHC}}}, \href{http://dx.doi.org/10.1051/epjconf/201817704010}{\emph{EPJ Web Conf.} {\bf 177} (2018) 04010}.

\bibitem{OPAL:2000ufp}
{\scshape OPAL} collaboration, G.~Abbiendi et~al., \emph{{Precise determination of the Z resonance parameters at {LEP}: 'Zedometry'}}, \href{http://dx.doi.org/10.1007/s100520100627}{\emph{Eur. Phys. J. C} {\bf 19} (2001) 587--651}, [\href{https://arxiv.org/abs/hep-ex/0012018}{{\tt hep-ex/0012018}}].

\bibitem{Janot_2020}
P.~Janot and S.~Jadach, \emph{Improved {Bhabha} cross section at {LEP} and the number of light neutrino species}, \href{http://dx.doi.org/10.1016/j.physletb.2020.135319}{\emph{Physics Letters B} {\bf 803} (Apr., 2020) 135319}.

\bibitem{ATLAS:2024vqf}
{\scshape ATLAS} collaboration, G.~Aad et~al., \emph{{Differential cross-sections for events with missing transverse momentum and jets measured with the {ATLAS} detector in 13 TeV proton-proton collisions}}, \href{http://dx.doi.org/10.1007/JHEP08(2024)223}{\emph{JHEP} {\bf 08} (2024) 223}, [\href{https://arxiv.org/abs/2403.02793}{{\tt 2403.02793}}].

\bibitem{Esteban_2024}
I.~Esteban, M.~C. Gonzalez-Garcia, M.~Maltoni, I.~Martinez-Soler, J.~P. Pinheiro and T.~Schwetz, \emph{Nufit-6.0: updated global analysis of three-flavor neutrino oscillations}, \href{http://dx.doi.org/10.1007/jhep12(2024)216}{\emph{Journal of High Energy Physics} {\bf 2024} (Dec., 2024) }.

\bibitem{deblas2024globalsmeftfitsfuture}
J.~de~Blas, Y.~Du, C.~Grojean, J.~Gu, V.~Miralles, M.~E. Peskin et~al., \emph{Global smeft fits at future colliders},  \href{https://arxiv.org/abs/2206.08326}{{\tt 2206.08326}}.

\bibitem{Litvinenko:2022qbd}
V.~N. Litvinenko, N.~Bachhawat, M.~Chamizo-Llatas, Y.~Jing, F.~M\'eot, I.~Petrushina et~al., \emph{{The ReLiC: Recycling Linear $e^+e^-$ Collider}},  \href{https://arxiv.org/abs/2203.06476}{{\tt 2203.06476}}.

\bibitem{Assmann:1994wt}
R.~Assmann et~al., \emph{{Lepton beam polarization at {LEP}}}, \href{http://dx.doi.org/10.1063/1.48860}{\emph{AIP Conf. Proc.} {\bf 343} (1995) 219--229}.

\bibitem{Abramowicz_2017}
{\scshape The Compact Linear Collider (CLIC) Study} collaboration, H.~Abramowicz et~al., \emph{{Higgs} physics at the clic electron–positron linear collider}, \href{http://dx.doi.org/10.1140/epjc/s10052-017-4968-5}{\emph{The European Physical Journal C} {\bf 77} (July, 2017) }.

\bibitem{Behnke2013}
T.~Behnke, J.~E. Brau, B.~Foster, J.~Fuster, M.~Harrison, J.~M. Paterson et~al., \emph{The international linear collider technical design report - volume 1: Executive summary},  \href{https://arxiv.org/abs/1306.6327}{{\tt 1306.6327}}.

\bibitem{Brau2015}
T.~Barklow, J.~Brau, K.~Fujii, J.~Gao, J.~List, N.~Walker et~al., \emph{{{ILC} Operating Scenarios}},  \href{https://arxiv.org/abs/1506.07830}{{\tt 1506.07830}}.

\bibitem{Bechtle2024}
P.~Bechtle, S.~Heinemeyer, J.~List, G.~Moortgat-Pick and G.~Weiglein, \emph{{Physics case for an e+e\ensuremath{-} collider at 500 GeV and above}}, \href{http://dx.doi.org/10.1051/epjconf/202431501001}{\emph{EPJ Web Conf.} {\bf 315} (2024) 01001}, [\href{https://arxiv.org/abs/2410.16191}{{\tt 2410.16191}}].

\bibitem{Fujii2019}
{\scshape LCC Physics Working Group} collaboration, K.~Fujii et~al., \emph{{Tests of the Standard Model at the International Linear Collider}},  \href{https://arxiv.org/abs/1908.11299}{{\tt 1908.11299}}.

\bibitem{LCVision}
{\scshape Linear Collider Vision} collaboration, D.~Atti\'e et~al., \emph{{A Linear Collider Vision for the Future of Particle Physics}},  \href{https://arxiv.org/abs/2503.19983}{{\tt 2503.19983}}.

\bibitem{CLIC2018}
{\scshape The Compact Linear Collider (CLIC) study} collaboration, {Charles, T.K. and others}, \emph{Cern yellow reports: Monographs, vol 2 (2018): The compact linear e+e- collider (clic) : 2018 summary report},  2018.
\newblock 10.23731/CYRM-2018-002.

\bibitem{Aicheler2012}
{\scshape The Compact Linear Collider (CLIC) study} collaboration, M.~Aicheler et~al., \emph{{A Multi-TeV Linear Collider Based on CLIC Technology}: {CLIC Conceptual Design Report}}, .

\bibitem{CLICdetector}
{\scshape CLICdp} collaboration, D.~Arominski et~al., \emph{{A detector for CLIC: main parameters and performance}},  \href{https://arxiv.org/abs/1812.07337}{{\tt 1812.07337}}.

\bibitem{Abada2019}
{\scshape FCC} collaboration, A.~Abada et~al., \emph{{FCC-ee: The Lepton Collider}: {Future Circular Collider Conceptual Design Report Volume 2}}, \href{http://dx.doi.org/10.1140/epjst/e2019-900045-4}{\emph{Eur. Phys. J. ST} {\bf 228} (2019) 261--623}.

\bibitem{CEPC2018}
{\scshape CEPC Study Group} collaboration, M.~Dong et~al., \emph{{CEPC Conceptual Design Report: Volume 2 - Physics \& Detector}},  \href{https://arxiv.org/abs/1811.10545}{{\tt 1811.10545}}.

\bibitem{ILD2020}
{\scshape ILD Concept Group} collaboration, H.~Abramowicz et~al., \emph{{International Large Detector: Interim Design Report}},  \href{https://arxiv.org/abs/2003.01116}{{\tt 2003.01116}}.

\bibitem{ILC2013}
{\scshape ILC} collaboration, H.~Baer et~al., \emph{{The International Linear Collider Technical Design Report - Volume 2: Physics}},  \href{https://arxiv.org/abs/1306.6352}{{\tt 1306.6352}}.

\bibitem{Wilson2023}
G.~W. Wilson, \emph{{Further investigation of dilepton-based center-of-mass energy measurements at e$^{+}$e$^{-}$ colliders}},  in \emph{{International Workshop on Future Linear Colliders}}, 8, 2023.
\newblock \href{https://arxiv.org/abs/2308.10414}{{\tt 2308.10414}}.

\bibitem{Madison2022}
B.~Madison and G.~W. Wilson, \emph{{Center-of-mass energy determination using $\mathrm{e}^{+} \mathrm{e}^{-} \to \mu^{+} \mu^{-} (\gamma)$ events at future $\mathrm{e}^{+} \mathrm{e}^{-}$ colliders}},  in \emph{{Snowmass 2021}}, 9, 2022.
\newblock \href{https://arxiv.org/abs/2209.03281}{{\tt 2209.03281}}.

\bibitem{Madison2023}
B.~Madison, \emph{{Using the GP2X framework for center-of-mass energy precision studies at $e^+e^-$ {Higgs} factories}},  in \emph{{International Workshop on Future Linear Colliders}}, 8, 2023.
\newblock \href{https://arxiv.org/abs/2308.09676}{{\tt 2308.09676}}.

\bibitem{Wilson2016}
G.~W. Wilson, \emph{{Updated Study of a Precision Measurement of the W Mass from a Threshold Scan Using Polarized $\rm{e}^-$ and $\rm{e}^+$ at {ILC}}},  in \emph{{International Workshop on Future Linear Colliders}}, 3, 2016.
\newblock \href{https://arxiv.org/abs/1603.06016}{{\tt 1603.06016}}.

\bibitem{Barklow2017}
T.~Barklow, K.~Fujii, S.~Jung, M.~E. Peskin and J.~Tian, \emph{{Model-Independent Determination of the Triple {Higgs} Coupling at e+e- Colliders}}, \href{http://dx.doi.org/10.1103/PhysRevD.97.053004}{\emph{Phys. Rev. D} {\bf 97} (2018) 053004}, [\href{https://arxiv.org/abs/1708.09079}{{\tt 1708.09079}}].

\bibitem{torndal2023}
J.~M. Torndal and J.~List, \emph{{Higgs} self-coupling measurement at the international linear collider},  \href{https://arxiv.org/abs/2307.16515}{{\tt 2307.16515}}.

\bibitem{Moortgat_Pick_2008}
G.~Moortgat-Pick et~al., \emph{Polarized positrons and electrons at the linear collider}, \href{http://dx.doi.org/10.1016/j.physrep.2007.12.003}{\emph{Physics Reports} {\bf 460} (May, 2008) 131–243}.

\bibitem{Kuraev:1985hb}
E.~A. Kuraev and V.~S. Fadin, \emph{{On Radiative Corrections to e+ e- Single Photon Annihilation at High-Energy}}, {\emph{Sov. J. Nucl. Phys.} {\bf 41} (1985) 466--472}.

\bibitem{BHLUMI}
S.~Jadach, W.~Płaczek, E.~Richter-Was, B.~Ward and Z.~Was, \emph{Upgrade of the monte carlo program bhlumi for bhabha scattering at low angles to version 4.04}, \href{http://dx.doi.org/https://doi.org/10.1016/S0010-4655(96)00156-7}{\emph{Computer Physics Communications} {\bf 102} (1997) 229--251}.

\bibitem{BHWIDE}
S.~Jadach, W.~Płaczek and B.~Ward, \emph{Bhwide 1.00: O($\alpha$) yfs exponentiated monte carlo for bhabha scattering at wide angles for {LEP}1/slc and {LEP}2}, \href{http://dx.doi.org/10.1016/s0370-2693(96)01382-2}{\emph{Physics Letters B} {\bf 390} (Jan., 1997) 298–308}.

\bibitem{OPALLumi}
{\scshape OPAL} collaboration, G.~Abbiendi~et al., \emph{Precision luminosity for z $^0$ lineshape measurements with a silicon-tungsten calorimeter}, \href{http://dx.doi.org/10.1007/s100520000353}{\emph{The European Physical Journal C} {\bf 14} (June, 2000) 373–425}.

\bibitem{Brock:1996ty}
{\scshape L3} collaboration, I.~C. Brock et~al., \emph{{Luminosity Measurement in the L3 Detector at {LEP}}}, \href{http://dx.doi.org/10.1016/S0168-9002(96)00734-6}{\emph{Nucl. Instrum. Meth. A} {\bf 381} (1996) 236--266}.

\bibitem{grahFCAL}
{\scshape FCAL} collaboration, C.~Grah, M.~Idzik, R.~Ingbir, W.~Lange, A.~Levy, K.~Mönig et~al., \emph{Report for the {ILC} detector r \& d panel instrumentation of the very forward region},  2007.

\bibitem{YFS1961}
D.~R. Yennie, S.~C. Frautschi and H.~Suura, \emph{{The infrared divergence phenomena and high-energy processes}}, \href{http://dx.doi.org/10.1016/0003-4916(61)90151-8}{\emph{Annals Phys.} {\bf 13} (1961) 379--452}.

\bibitem{Hollik:1988ii}
W.~F.~L. Hollik, \emph{{Radiative Corrections in the Standard Model and their Role for Precision Tests of the Electroweak Theory}}, \href{http://dx.doi.org/10.1002/prop.2190380302}{\emph{Fortsch. Phys.} {\bf 38} (1990) 165--260}.

\bibitem{OPALALpha}
{\scshape OPAL} collaboration, \emph{Measurement of the running of the qed couplingin small-angle {Bhabha} scattering at {LEP}}, \href{http://dx.doi.org/10.1140/epjc/s2005-02389-3}{\emph{The European Physical Journal C} {\bf 45} (Jan., 2006) 1–21}.

\bibitem{babayaga}
G.~Balossini, C.~Bignamini, C.~Carloni~Calame, G.~Montagna, O.~Nicrosini and F.~Piccinini, \emph{Photon pair production at flavour factories with per mille accuracy}, \href{http://dx.doi.org/10.1016/j.physletb.2008.04.007}{\emph{Physics Letters B} {\bf 663} (May, 2008) 209–213}.

\bibitem{Jadach:2018jjo}
S.~Jadach, W.~P\l{}aczek, M.~Skrzypek, B.~F.~L. Ward and S.~A. Yost, \emph{{The path to 0.01\% theoretical luminosity precision for the FCC-ee}}, \href{http://dx.doi.org/10.1016/j.physletb.2019.01.012}{\emph{Phys. Lett. B} {\bf 790} (2019) 314--321}, [\href{https://arxiv.org/abs/1812.01004}{{\tt 1812.01004}}].

\bibitem{SLD:1994kvj}
{\scshape SLD} collaboration, K.~Abe et~al., \emph{{Polarized Bhabha scattering a precision measurement of the electron neutral current couplings}}, \href{http://dx.doi.org/10.1103/PhysRevLett.74.2880}{\emph{Phys. Rev. Lett.} {\bf 74} (1995) 2880--2884}, [\href{https://arxiv.org/abs/hep-ex/9410009}{{\tt hep-ex/9410009}}].

\bibitem{Montagna:1998kp}
G.~Montagna, O.~Nicrosini, F.~Piccinini and G.~Passarino, \emph{{TOPAZ0 4.0: A New version of a computer program for evaluation of deconvoluted and realistic observables at {LEP}-1 and {LEP}-2}}, \href{http://dx.doi.org/10.1016/S0010-4655(98)00080-0}{\emph{Comput. Phys. Commun.} {\bf 117} (1999) 278--289}, [\href{https://arxiv.org/abs/hep-ph/9804211}{{\tt hep-ph/9804211}}].

\bibitem{Beenakker:1990mb}
W.~Beenakker, F.~A. Berends and S.~C. van~der Marck, \emph{{Large angle Bhabha scattering}}, \href{http://dx.doi.org/10.1016/0550-3213(91)90328-U}{\emph{Nucl. Phys. B} {\bf 349} (1991) 323--368}.

\bibitem{Peskin:1995ev}
M.~E. Peskin and D.~V. Schroeder, \emph{{An Introduction to quantum field theory}}.
\newblock Addison-Wesley, Reading, USA, 1995, \href{http://dx.doi.org/10.1201/9780429503559}{10.1201/9780429503559}.

\bibitem{Sudakov1954}
V.~V. Sudakov, \emph{{Vertex parts at very high-energies in quantum electrodynamics}}, {\emph{Sov. Phys. JETP} {\bf 3} (1956) 65--71}.

\bibitem{Adachi_2025}
{\scshape Belle II} collaboration, I.~Adachi, L.~Aggarwal, H.~Ahmed, J.~K. Ahn, H.~Aihara, N.~Akopov et~al., \emph{Measurement of the integrated luminosity of data samples collected during 2019-2022 by the belle ii experiment}, \href{http://dx.doi.org/10.1088/1674-1137/ad806c}{\emph{Chinese Physics C} {\bf 49} (Jan., 2025) 013001}.

\bibitem{Berends:1987ab}
F.~A. Berends, W.~L. van Neerven and G.~J.~H. Burgers, \emph{{Higher Order Radiative Corrections at {LEP} Energies}}, \href{http://dx.doi.org/10.1016/0550-3213(88)90313-6}{\emph{Nucl. Phys. B} {\bf 297} (1988) 429}.

\bibitem{Jadach_2023}
S.~Jadach, B.~Ward, Z.~Was, S.~Yost and A.~Siodmok, \emph{Multi-photon monte carlo event generator kkmcee for lepton and quark pair production in lepton colliders}, \href{http://dx.doi.org/10.1016/j.cpc.2022.108556}{\emph{Computer Physics Communications} {\bf 283} (Feb., 2023) 108556}.

\bibitem{Kilian_2011}
W.~Kilian, T.~Ohl and J.~Reuter, \emph{Whizard—simulating multi-particle processes at {LHC} and {ILC}}, \href{http://dx.doi.org/10.1140/epjc/s10052-011-1742-y}{\emph{The European Physical Journal C} {\bf 71} (Sept., 2011) }.

\bibitem{Abramowicz_2010}
H.~Abramowicz, A.~Abusleme, K.~Afanaciev, J.~Aguilar, P.~Ambalathankandy, P.~Bambade et~al., \emph{Forward instrumentation for {ILC} detectors}, \href{http://dx.doi.org/10.1088/1748-0221/5/12/p12002}{\emph{Journal of Instrumentation} {\bf 5} (Dec., 2010) P12002–P12002}.

\bibitem{bozovicjelisavcic2014}
I.~Bozovic-Jelisavcic, S.~Lukic, M.~Pandurovic and I.~Smiljanic, \emph{Precision luminosity measurement at {ILC}},  2014.

\bibitem{smiljanić2024metrology}
I.~Smiljanić, I.~Božović, G.~Kačarević, M.~Radulović and J.~Stevanović, \emph{Metrology requirements for the integrated luminosity measurement using small angle {Bhabha} scattering at {ILC}},  2024.

\bibitem{SBoogert_2009}
S.~Boogert, A.~F. Hartin, M.~Hildreth, D.~Käfer, J.~List, T.~Maruyama et~al., \emph{Polarimeters and energy spectrometers for the {ILC} beam delivery system}, \href{http://dx.doi.org/10.1088/1748-0221/4/10/P10015}{\emph{Journal of Instrumentation} {\bf 4} (oct, 2009) P10015}.

\bibitem{Madison:2024jak}
B.~Madison and G.~Wilson, \emph{{Novel position reconstruction methods for highly granular electromagnetic calorimeters}}, \href{http://dx.doi.org/10.1051/epjconf/202431503007}{\emph{EPJ Web Conf.} {\bf 315} (2024) 03007}.

\bibitem{ATLAS:2022rkn}
{\scshape ATLAS} collaboration, \emph{{Graph Neural Network Jet Flavour Tagging with the {ATLAS} Detector}}, .

\bibitem{Duarte:2018ite}
J.~Duarte et~al., \emph{{Fast inference of deep neural networks in FPGAs for particle physics}}, \href{http://dx.doi.org/10.1088/1748-0221/13/07/P07027}{\emph{JINST} {\bf 13} (2018) P07027}, [\href{https://arxiv.org/abs/1804.06913}{{\tt 1804.06913}}].

\bibitem{CRimbault_2007}
C.~Rimbault, P.~Bambade, K.~Mönig and D.~Schulte, \emph{Impact of beam-beam effects on precision luminosity measurements at the {ILC}}, \href{http://dx.doi.org/10.1088/1748-0221/2/09/P09001}{\emph{Journal of Instrumentation} {\bf 2} (sep, 2007) P09001}.

\bibitem{Voutsinas_2020}
G.~Voutsinas, E.~Perez, M.~Dam and P.~Janot, \emph{Beam-beam effects on the luminosity measurement at {LEP} and the number of light neutrino species}, \href{http://dx.doi.org/10.1016/j.physletb.2019.135068}{\emph{Physics Letters B} {\bf 800} (Jan., 2020) 135068}.

\bibitem{Voutsinas_2019}
G.~Voutsinas, E.~Perez, M.~Dam and P.~Janot, \emph{Beam-beam effects on the luminosity measurement at fcc-ee}, \href{http://dx.doi.org/10.1007/jhep10(2019)225}{\emph{Journal of High Energy Physics} {\bf 2019} (Oct., 2019) }.

\bibitem{Guinea-PIG}
D.~Schulte, \emph{{Study of Electromagnetic and Hadronic Background in the Interaction Region of the TESLA Collider}}.
\newblock PhD thesis, Hamburg U., 1997.

\bibitem{James:1994vla}
F.~James, \emph{{MINUIT Function Minimization and Error Analysis: Reference Manual Version 94.1}}, .

\bibitem{Zhang:2024wob}
W.~Zhang, T.~Grismayer and L.~O. Silva, \emph{{Anomalous pinch in electron-electron beam collision}},  \href{https://arxiv.org/abs/2412.09398}{{\tt 2412.09398}}.

\bibitem{Karl:424633}
R.~Karl, \emph{{F}rom the {M}achine-{D}etector {I}nterface to {E}lectroweak {P}recision {M}easurements at the {ILC} — {B}eam-{G}as {B}ackground, {B}eam {P}olarization and {T}riple {G}auge {C}ouplings}.
\newblock Dissertation, Universität Hamburg, Hamburg, 2019.
\newblock 10.3204/PUBDB-2019-03013.

\bibitem{GWILCCon21}
G.~Wilson, ``Center-of-mass energy determination using dimuon events at {ILC}.'' \url{https://agenda.linearcollider.org/event/9211/contributions/49276/}.

\bibitem{BMLCWS24}
B.~Madison, ``Photon and electron reconstruction in an ultra-high granularity luminosity calorimeter.'' \url{https://agenda.linearcollider.org/event/10134/contributions/54678/}.

\bibitem{FCAL:2017uhw}
{\scshape FCAL} collaboration, H.~Abramowicz et~al., \emph{{Measurement of shower development and its Moli\textbackslash{}`ere radius with a four-plane LumiCal test set-up}}, \href{http://dx.doi.org/10.1140/epjc/s10052-018-5611-9}{\emph{Eur. Phys. J. C} {\bf 78} (2018) 135}, [\href{https://arxiv.org/abs/1705.03885}{{\tt 1705.03885}}].

\bibitem{Aplin:2013oca}
S.~Aplin, M.~Boronat, D.~Dannheim, J.~Duarte, F.~Gaede, A.~Ruiz-Jimeno et~al., \emph{{Forward tracking at the next $e^{+}e^{-}$ collider part II: experimental challenges and detector design}}, \href{http://dx.doi.org/10.1088/1748-0221/8/06/T06001}{\emph{JINST} {\bf 8} (2013) T06001}, [\href{https://arxiv.org/abs/1303.3187}{{\tt 1303.3187}}].

\bibitem{Magnan:2017exp}
{\scshape CMS} collaboration, A.~M. Magnan, \emph{{HGCAL: a High-Granularity Calorimeter for the endcaps of CMS at {HL-LHC}}}, \href{http://dx.doi.org/10.1088/1748-0221/12/01/C01042}{\emph{JINST} {\bf 12} (2017) C01042}.

\bibitem{Fermi-LAT:2009ihh}
{\scshape Fermi-LAT} collaboration, W.~B. Atwood et~al., \emph{{The Large Area Telescope on the Fermi Gamma-ray Space Telescope Mission}}, \href{http://dx.doi.org/10.1088/0004-637X/697/2/1071}{\emph{Astrophys. J.} {\bf 697} (2009) 1071--1102}, [\href{https://arxiv.org/abs/0902.1089}{{\tt 0902.1089}}].

\bibitem{Verkerke:2003ir}
W.~Verkerke and D.~P. Kirkby, \emph{{The RooFit toolkit for data modeling}}, {\emph{eConf} {\bf C0303241} (2003) MOLT007}, [\href{https://arxiv.org/abs/physics/0306116}{{\tt physics/0306116}}].

\bibitem{RefWig}
R.~Wigmans, \emph{{Calorimetry: Energy measurement in particle physics}}, vol.~107.
\newblock 2000, \href{http://dx.doi.org/10.1093/oso/9780198786351.001.0001}{10.1093/oso/9780198786351.001.0001}.

\bibitem{Serpukhov-Brussels-AnnecyLAPP:1981yhn}
{\scshape Serpukhov-Brussels-Annecy(LAPP)} collaboration, F.~Binon et~al., \emph{{HODOSCOPE GAMMA SPECTROMETER GAMS-200}}, \href{http://dx.doi.org/10.1016/0029-554X(81)90261-5}{\emph{Nucl. Instrum. Meth.} {\bf 188} (1981) 507}.

\bibitem{C5}
M.~Kuhn and R.~Quinlan, \emph{C50: C5.0 Decision Trees and Rule-Based Models}, 2025.

\bibitem{FCCeeCoMPrecision}
A.~Blondel, P.~Janot, J.~Wenninger, R.~Aßmann, S.~Aumon, P.~Azzurri et~al., \emph{Polarization and centre-of-mass energy calibration at fcc-ee},  2019.

\bibitem{Ohl:1996fi}
T.~Ohl, \emph{{CIRCE version 1.0: Beam spectra for simulating linear collider physics}}, \href{http://dx.doi.org/10.1016/S0010-4655(96)00167-1}{\emph{Comput. Phys. Commun.} {\bf 101} (1997) 269--288}, [\href{https://arxiv.org/abs/hep-ph/9607454}{{\tt hep-ph/9607454}}].

\bibitem{ParzenKDE}
E.~Parzen, \emph{On estimation of a probability density function and mode}, {\emph{The Annals of Mathematical Statistics} {\bf 33} (1962) 1065--1076}.

\bibitem{RomanoKDE}
J.~P. Romano, \emph{On weak convergence and optimality of kernel density estimates of the mode}, {\emph{The Annals of Statistics} {\bf 16} (1988) 629--647}.

\bibitem{VenterMode}
J.~H. Venter, \emph{On estimation of the mode}, {\emph{The Annals of Mathematical Statistics} {\bf 38} (1967) 1446--1455}.

\bibitem{EfronBootstrap}
B.~Efron, \emph{Bootstrap methods: Another look at the jackknife}, {\emph{The Annals of Statistics} {\bf 7} (1979) 1--26}.

\bibitem{ILCSnowmass}
{\scshape ILC International Development Team} collaboration, A.~Aryshev et~al., \emph{{The International Linear Collider: Report to Snowmass 2021}},  \href{https://arxiv.org/abs/2203.07622}{{\tt 2203.07622}}.

\bibitem{MadisonMasters}
B.~Madison, \emph{Center of Mass Energy Measurements using Dimuons at {ILC}}.
\newblock Doctorate comprehensive exam, University of Kansas, Lawrence, Kansas, 2022.

\bibitem{Smiljanic:2024twn}
I.~Smiljani\'c, I.~Bo\v{z}ovi\'c, G.~Ka\v{c}arevi\'c, M.~Radulovi\'c and J.~Stevanovi\'c, \emph{{Metrology Requirements for the Integrated Luminosity Measurement Using Small-Angle {Bhabha} Scattering at {ILC}}}, \href{http://dx.doi.org/10.1093/ptep/ptaf015}{\emph{PTEP} {\bf 2025} (2025) 023H03}, [\href{https://arxiv.org/abs/2407.03024}{{\tt 2407.03024}}].

\bibitem{CMS:2014pgm}
{\scshape CMS} collaboration, S.~Chatrchyan et~al., \emph{{Description and performance of track and primary-vertex reconstruction with the {CMS} tracker}}, \href{http://dx.doi.org/10.1088/1748-0221/9/10/P10009}{\emph{JINST} {\bf 9} (2014) P10009}, [\href{https://arxiv.org/abs/1405.6569}{{\tt 1405.6569}}].

\bibitem{ATLAS:2010nca}
{\scshape ATLAS} collaboration, \emph{{Alignment Performance of the {ATLAS} Inner Detector Tracking System in 7 TeV proton-proton collisions at the {LHC}}}, .

\bibitem{Sadeh2008}
I.~Sadeh, \emph{Influence of the shape of the beampipe on the luminosity measurement at the {ILC}},  in \emph{Proceedings of the Workshop of the Collaboration on Forward Calorimetry at {ILC}} (I.~Bozovic~Jelisavcic, ed.), (Vinča Institute of Nuclear Sciences Belgrade, Serbia), p.~58–63, 2008.

\bibitem{Jadach:2021ayv}
S.~Jadach, W.~P\l{}aczek, M.~Skrzypek and B.~F.~L. Ward, \emph{{Study of theoretical luminosity precision for electron colliders at higher energies}}, \href{http://dx.doi.org/10.1140/epjc/s10052-021-09860-9}{\emph{Eur. Phys. J. C} {\bf 81} (2021) 1047}.

\bibitem{Carloni_Calame_2019}
C.~M. Carloni~Calame, M.~Chiesa, G.~Montagna, O.~Nicrosini and F.~Piccinini, \emph{Electroweak corrections to $e^+e^-\to\gamma\gamma$ as a luminosity process at fcc-ee}, \href{http://dx.doi.org/10.1016/j.physletb.2019.134976}{\emph{Physics Letters B} {\bf 798} (Nov., 2019) 134976}.

\bibitem{Jadach:1990zf}
S.~Jadach, E.~Richter-Was, B.~F.~L. Ward and Z.~Was, \emph{{Analytical O(alpha) distributions for Bhabha scattering at low angles}}, \href{http://dx.doi.org/10.1016/0370-2693(91)91754-J}{\emph{Phys. Lett. B} {\bf 253} (1991) 469--477}.

\bibitem{Dam:2021sdj}
M.~Dam, \emph{{Challenges for FCC-ee luminosity monitor design}}, \href{http://dx.doi.org/10.1140/epjp/s13360-021-02265-3}{\emph{Eur. Phys. J. Plus} {\bf 137} (2022) 81}, [\href{https://arxiv.org/abs/2107.12837}{{\tt 2107.12837}}].

\bibitem{Sailer:2009zz}
A.~P. Sailer, \emph{{Studies of the measurement of differential luminosity using Bhabha events at the International Linear Collider}},  other thesis, 4, 2009.
\newblock 10.3204/DESY-THESIS-2009-011.

\bibitem{Poss:2013oea}
S.~Poss and A.~Sailer, \emph{{Luminosity Spectrum Reconstruction at Linear Colliders}}, \href{http://dx.doi.org/10.1140/epjc/s10052-014-2833-3}{\emph{Eur. Phys. J. C} {\bf 74} (2014) 2833}, [\href{https://arxiv.org/abs/1309.0372}{{\tt 1309.0372}}].

\bibitem{BaBar:2023upu}
{\scshape BaBar} collaboration, J.~P. Lees et~al., \emph{{Precision $e^-$ Beam Polarimetry at an $e^+e^-$ B Factory using Tau-Pair Events}}, \href{http://dx.doi.org/10.1103/PhysRevD.108.092001}{\emph{Phys. Rev. D} {\bf 108} (2023) 092001}, [\href{https://arxiv.org/abs/2308.00774}{{\tt 2308.00774}}].

\bibitem{karl2017polarimetryilc}
R.~Karl and J.~List, \emph{Polarimetry at the ilc},  2017.

\bibitem{GWWTalk}
G.~W. Wilson, \emph{Beam polarization measurement using single bosons with missing energy},  2012.

\bibitem{GrahamWW}
G.~W. Wilson, \emph{Updated study of a precision measurement of the w mass from a threshold scan using polarized $\rm{e}^-$ and $\rm{e}^+$ at {ILC}},  2016.

\bibitem{SLD:2000leq}
{\scshape SLD} collaboration, K.~Abe et~al., \emph{{A High precision measurement of the left-right Z boson cross-section asymmetry}}, \href{http://dx.doi.org/10.1103/PhysRevLett.84.5945}{\emph{Phys. Rev. Lett.} {\bf 84} (2000) 5945--5949}, [\href{https://arxiv.org/abs/hep-ex/0004026}{{\tt hep-ex/0004026}}].

\bibitem{fermisV}
J.~Hilgart, R.~Kleiss and F.~Le~Diberder, \emph{{An Electroweak Monte Carlo for four fermion production}}, \href{http://dx.doi.org/10.1016/0010-4655(93)90175-C}{\emph{Comput. Phys. Commun.} {\bf 75} (1993) 191--218}.

\bibitem{koralw}
S.~Jadach, W.~Placzek, M.~Skrzypek, B.~F.~L. Ward and Z.~Was, \emph{{Monte Carlo program KoralW 1.42 for all four-fermion final states in e+ e- collisions}}, \href{http://dx.doi.org/10.1016/S0010-4655(99)00219-2}{\emph{Comput. Phys. Commun.} {\bf 119} (1999) 272--311}, [\href{https://arxiv.org/abs/hep-ph/9906277}{{\tt hep-ph/9906277}}].

\bibitem{ALEPH:monophoton}
{\scshape ALEPH} collaboration, A.~Heister et~al., \emph{{Single photon and multiphoton production in $e^{+} e^{-}$ collisions at $\sqrt{s}$ up to 209-GeV}}, \href{http://dx.doi.org/10.1140/epjc/s2002-01129-7}{\emph{Eur. Phys. J. C} {\bf 28} (2003) 1--13}.

\bibitem{OPAL:monophoton}
{\scshape OPAL} collaboration, R.~Akers et~al., \emph{{Measurement of single photon production in e+ e- collisions near the Z0 resonance}}, \href{http://dx.doi.org/10.1007/BF01571303}{\emph{Z. Phys. C} {\bf 65} (1995) 47--66}.

\bibitem{ALEPH:monophoton2}
{\scshape ALEPH} collaboration, R.~Barate et~al., \emph{{Single photon and multiphoton production in $e^{+} e^{-}$ collisions at a center-of-mass energy of 183-GeV}}, \href{http://dx.doi.org/10.1016/S0370-2693(98)00468-7}{\emph{Phys. Lett. B} {\bf 429} (1998) 201--214}.

\bibitem{L3:monophoton}
{\scshape L3} collaboration, P.~Achard et~al., \emph{{Single photon and multiphoton events with missing energy in $e^{+} e^{-}$ collisions at LEP}}, \href{http://dx.doi.org/10.1016/j.physletb.2004.01.010}{\emph{Phys. Lett. B} {\bf 587} (2004) 16--32}, [\href{https://arxiv.org/abs/hep-ex/0402002}{{\tt hep-ex/0402002}}].

\bibitem{DELPHI:monophoton}
{\scshape DELPHI} collaboration, W.~Adam et~al., \emph{{Search for anomalous production of single photons at $\sqrt{s}$ = 130-GeV and 136-GeV}}, \href{http://dx.doi.org/10.1016/0370-2693(96)00671-5}{\emph{Phys. Lett. B} {\bf 380} (1996) 471--479}.

\bibitem{Fox_2011}
P.~J. Fox, R.~Harnik, J.~Kopp and Y.~Tsai, \emph{Lep shines light on dark matter}, \href{http://dx.doi.org/10.1103/physrevd.84.014028}{\emph{Physical Review D} {\bf 84} (July, 2011) }.

\bibitem{Ambrosanio_1996}
S.~Ambrosanio, \emph{Single-photon signal from neutralinos at lep2}, \href{http://dx.doi.org/10.1016/0550-3213(96)00421-x}{\emph{Nuclear Physics B} {\bf 478} (Oct., 1996) 46–58}.

\bibitem{Lopez_1997}
J.~L. Lopez, D.~V. Nanopoulos and A.~Zichichi, \emph{Single-photon signals at cern lep in supersymmetric models with a light gravitino}, \href{http://dx.doi.org/10.1103/physrevd.55.5813}{\emph{Physical Review D} {\bf 55} (May, 1997) 5813–5825}.

\bibitem{Hannestad_2010}
S.~Hannestad, A.~Mirizzi, G.~G. Raffelt and Y.~Y. Wong, \emph{Neutrino and axion hot dark matter bounds after wmap-7}, \href{http://dx.doi.org/10.1088/1475-7516/2010/08/001}{\emph{Journal of Cosmology and Astroparticle Physics} {\bf 2010} (Aug., 2010) 001–001}.

\bibitem{Seifert_2024}
A.~Seifert, Z.~G. Lane, M.~Galoppo, R.~Ridden-Harper and D.~L. Wiltshire, \emph{Supernovae evidence for foundational change to cosmological models}, \href{http://dx.doi.org/10.1093/mnrasl/slae112}{\emph{Monthly Notices of the Royal Astronomical Society: Letters} {\bf 537} (Nov., 2024) L55–L60}.

\bibitem{Lane_2024}
Z.~G. Lane, A.~Seifert, R.~Ridden-Harper and D.~L. Wiltshire, \emph{Cosmological foundations revisited with pantheon+}, \href{http://dx.doi.org/10.1093/mnras/stae2437}{\emph{Monthly Notices of the Royal Astronomical Society} (Nov., 2024) }.

\bibitem{des2025}
{\scshape DES} collaboration, T.~M.~C. Abbott, M.~Acevedo, M.~Adamow, M.~Aguena, A.~Alarcon, S.~Allam et~al., \emph{Dark energy survey: implications for cosmological expansion models from the final des baryon acoustic oscillation and supernova data},  \href{https://arxiv.org/abs/2503.06712}{{\tt 2503.06712}}.

\bibitem{XENON:2024hup}
{\scshape XENON} collaboration, E.~Aprile et~al., \emph{{First Search for Light Dark Matter in the Neutrino Fog with XENONnT}}, \href{http://dx.doi.org/10.1103/PhysRevLett.134.111802}{\emph{Phys. Rev. Lett.} {\bf 134} (2025) 111802}, [\href{https://arxiv.org/abs/2409.17868}{{\tt 2409.17868}}].

\bibitem{Dutta_2023}
K.~Dutta, A.~Ghosh, A.~Kar and B.~Mukhopadhyaya, \emph{Mev to multi-tev thermal wimps: most conservative limits}, \href{http://dx.doi.org/10.1088/1475-7516/2023/08/071}{\emph{Journal of Cosmology and Astroparticle Physics} {\bf 2023} (Aug., 2023) 071}.

\bibitem{XENON:2024ijk}
{\scshape XENON} collaboration, E.~Aprile et~al., \emph{{First Indication of Solar B8 Neutrinos via Coherent Elastic Neutrino-Nucleus Scattering with XENONnT}}, \href{http://dx.doi.org/10.1103/PhysRevLett.133.191002}{\emph{Phys. Rev. Lett.} {\bf 133} (2024) 191002}, [\href{https://arxiv.org/abs/2408.02877}{{\tt 2408.02877}}].

\bibitem{PandaX:2024muv}
{\scshape PandaX} collaboration, Z.~Bo et~al., \emph{{First Indication of Solar B8 Neutrinos through Coherent Elastic Neutrino-Nucleus Scattering in PandaX-4T}}, \href{http://dx.doi.org/10.1103/PhysRevLett.133.191001}{\emph{Phys. Rev. Lett.} {\bf 133} (2024) 191001}, [\href{https://arxiv.org/abs/2407.10892}{{\tt 2407.10892}}].

\bibitem{icecubeNu}
R.~Abbasi, Y.~Abdou, M.~Ackermann, J.~Adams, M.~Ahlers, K.~Andeen et~al., \emph{Measurement of atmospheric neutrino oscillation parameters using convolutional neural networks with 9.3 years of data in icecube deepcore},  \href{https://arxiv.org/abs/2405.02163}{{\tt 2405.02163}}.

\bibitem{icecubeDM}
R.~Abbasi, Y.~Abdou, M.~Ackermann, J.~Adams, M.~Ahlers, K.~Andeen et~al., \emph{Search for dark matter from the center of the earth with ten years of icecube data},  \href{https://arxiv.org/abs/2412.12972}{{\tt 2412.12972}}.

\bibitem{DUNE:2020fgq}
{\scshape DUNE} collaboration, B.~Abi et~al., \emph{{Prospects for beyond the Standard Model physics searches at the Deep Underground Neutrino Experiment}}, \href{http://dx.doi.org/10.1140/epjc/s10052-021-09007-w}{\emph{Eur. Phys. J. C} {\bf 81} (2021) 322}, [\href{https://arxiv.org/abs/2008.12769}{{\tt 2008.12769}}].

\bibitem{IceCubeCollaboration:2021euf}
{\scshape IceCube} collaboration, R.~Abbasi et~al., \emph{{All-flavor constraints on nonstandard neutrino interactions and generalized matter potential with three years of IceCube DeepCore data}}, \href{http://dx.doi.org/10.1103/PhysRevD.104.072006}{\emph{Phys. Rev. D} {\bf 104} (2021) 072006}, [\href{https://arxiv.org/abs/2106.07755}{{\tt 2106.07755}}].

\bibitem{MINOS:2020iqj}
{\scshape MINOS+, Daya Bay} collaboration, P.~Adamson et~al., \emph{{Improved Constraints on Sterile Neutrino Mixing from Disappearance Searches in the MINOS, MINOS+, Daya Bay, and Bugey-3 Experiments}}, \href{http://dx.doi.org/10.1103/PhysRevLett.125.071801}{\emph{Phys. Rev. Lett.} {\bf 125} (2020) 071801}, [\href{https://arxiv.org/abs/2002.00301}{{\tt 2002.00301}}].

\bibitem{DayaBay:2022orm}
{\scshape Daya Bay} collaboration, F.~P. An et~al., \emph{{Precision Measurement of Reactor Antineutrino Oscillation at Kilometer-Scale Baselines by Daya Bay}}, \href{http://dx.doi.org/10.1103/PhysRevLett.130.161802}{\emph{Phys. Rev. Lett.} {\bf 130} (2023) 161802}, [\href{https://arxiv.org/abs/2211.14988}{{\tt 2211.14988}}].

\bibitem{LEPCombined}
A.~Collaboration, D.~Collaboration, L.~Collaboration, O.~Collaboration and the LEP Electroweak Working~Group, \emph{Combination procedure for the precise determination of z boson parameters from results of the lep experiments},  \href{https://arxiv.org/abs/hep-ex/0101027}{{\tt hep-ex/0101027}}.

\bibitem{Abell_n_2022}
G.~F. Abellán, Z.~Chacko, A.~Dev, P.~Du, V.~Poulin and Y.~Tsai, \emph{Improved cosmological constraints on the neutrino mass and lifetime}, \href{http://dx.doi.org/10.1007/jhep08(2022)076}{\emph{Journal of High Energy Physics} {\bf 2022} (Aug., 2022) }.

\bibitem{Dutra_2021}
M.~Dutra, V.~Oliveira, C.~A. de~S.~Pires and F.~S. Queiroz, \emph{A model for mixed warm and hot right-handed neutrino dark matter}, \href{http://dx.doi.org/10.1007/jhep10(2021)005}{\emph{Journal of High Energy Physics} {\bf 2021} (Oct., 2021) }.

\bibitem{OPAL:1994kgw}
{\scshape OPAL} collaboration, R.~Akers et~al., \emph{{Measurement of single photon production in e+ e- collisions near the Z0 resonance}}, \href{http://dx.doi.org/10.1007/BF01571303}{\emph{Z. Phys. C} {\bf 65} (1995) 47--66}.

\bibitem{CMS:2022ett}
{\scshape CMS} collaboration, A.~Tumasyan et~al., \emph{{Precision measurement of the Z boson invisible width in pp collisions at s=13 TeV}}, \href{http://dx.doi.org/10.1016/j.physletb.2022.137563}{\emph{Phys. Lett. B} {\bf 842} (2023) 137563}, [\href{https://arxiv.org/abs/2206.07110}{{\tt 2206.07110}}].

\bibitem{SPheno}
W.~Porod and F.~Staub, \emph{Spheno 3.1: extensions including flavour, cp-phases and models beyond the mssm}, \href{http://dx.doi.org/10.1016/j.cpc.2012.05.021}{\emph{Computer Physics Communications} {\bf 183} (Nov., 2012) 2458–2469}.

\bibitem{SModelSv3}
M.~M. Altakach, S.~Kraml, A.~Lessa, S.~Narasimha, T.~Pascal, C.~Ramos et~al., \emph{{SModelS} v3: Going beyond z2 topologies},  2024.

\bibitem{MICROMEGAS}
G.~Alguero, G.~Bélanger, F.~Boudjema, S.~Chakraborti, A.~Goudelis, S.~Kraml et~al., \emph{micromegas 6.0: N-component dark matter}, \href{http://dx.doi.org/10.1016/j.cpc.2024.109133}{\emph{Computer Physics Communications} {\bf 299} (June, 2024) 109133}.

\bibitem{DELPHIChargino}
{\scshape DELPHI} collaboration, J.~Abdallah et~al., \emph{{Searches for supersymmetric particles in e+ e- collisions up to 208-GeV and interpretation of the results within the MSSM}}, \href{http://dx.doi.org/10.1140/epjc/s2003-01355-5}{\emph{Eur. Phys. J. C} {\bf 31} (2003) 421--479}, [\href{https://arxiv.org/abs/hep-ex/0311019}{{\tt hep-ex/0311019}}].

\bibitem{ALEPHChargino}
{\scshape ALEPH} collaboration, A.~Heister et~al., \emph{{Search for charginos nearly mass degenerate with the lightest neutralino in e+ e- collisions at center-of-mass energies up to 209-GeV}}, \href{http://dx.doi.org/10.1016/S0370-2693(02)01584-8}{\emph{Phys. Lett. B} {\bf 533} (2002) 223--236}, [\href{https://arxiv.org/abs/hep-ex/0203020}{{\tt hep-ex/0203020}}].

\bibitem{OPALChargino}
{\scshape OPAL} collaboration, G.~Abbiendi et~al., \emph{{Search for chargino and neutralino production at s**(1/2) = 192-GeV to 209 GeV at LEP}}, \href{http://dx.doi.org/10.1140/epjc/s2004-01758-8}{\emph{Eur. Phys. J. C} {\bf 35} (2004) 1--20}, [\href{https://arxiv.org/abs/hep-ex/0401026}{{\tt hep-ex/0401026}}].

\bibitem{Melzer_Pellmann_2014}
I.~Melzer-Pellmann and P.~Pralavorio, \emph{Lessons for susy from the {LHC} after the first run}, \href{http://dx.doi.org/10.1140/epjc/s10052-014-2801-y}{\emph{The European Physical Journal C} {\bf 74} (May, 2014) }.

\bibitem{BaerAndTata}
H.~Baer and X.~Tata, \emph{{WEAKSCALE SUPERSYMMETRY From Superfields to Scattering Events}}.
\newblock Cambridge University Press, Shaftesbury Road, Cambridge CB2 8EA, United Kingdom, 2022.

\bibitem{Kalinowski_2020}
J.~Kalinowski, W.~Kotlarski, P.~Sopicki and A.~F. Żarnecki, \emph{Simulating hard photon production with whizard}, \href{http://dx.doi.org/10.1140/epjc/s10052-020-8149-6}{\emph{The European Physical Journal C} {\bf 80} (July, 2020) }.

\bibitem{Balossini_2008}
G.~Balossini, C.~Bignamini, C.~Carloni~Calame, G.~Montagna, O.~Nicrosini and F.~Piccinini, \emph{Photon pair production at flavour factories with per mille accuracy}, \href{http://dx.doi.org/10.1016/j.physletb.2008.04.007}{\emph{Physics Letters B} {\bf 663} (May, 2008) 209–213}.

\bibitem{Alwall_2014}
J.~Alwall, R.~Frederix, S.~Frixione, V.~Hirschi, F.~Maltoni, O.~Mattelaer et~al., \emph{The automated computation of tree-level and next-to-leading order differential cross sections, and their matching to parton shower simulations}, \href{http://dx.doi.org/10.1007/jhep07(2014)079}{\emph{Journal of High Energy Physics} {\bf 2014} (July, 2014) }.

\bibitem{Actis_2017}
S.~Actis, A.~Denner, L.~Hofer, J.-N. Lang, A.~Scharf and S.~Uccirati, \emph{R e c o l a—recursive computation of one-loop amplitudes}, \href{http://dx.doi.org/10.1016/j.cpc.2017.01.004}{\emph{Computer Physics Communications} {\bf 214} (May, 2017) 140–173}.

\bibitem{Denner_2018}
A.~Denner, J.-N. Lang and S.~Uccirati, \emph{Recola2: Recursive computation of one-loop amplitudes 2}, \href{http://dx.doi.org/10.1016/j.cpc.2017.11.013}{\emph{Computer Physics Communications} {\bf 224} (Mar., 2018) 346–361}.

\bibitem{Buccioni_2019}
F.~Buccioni, J.-N. Lang, J.~M. Lindert, P.~Maierhöfer, S.~Pozzorini, H.~Zhang et~al., \emph{Openloops 2}, \href{http://dx.doi.org/10.1140/epjc/s10052-019-7306-2}{\emph{The European Physical Journal C} {\bf 79} (Oct., 2019) }.

\bibitem{CarloniCalame:2003yt}
C.~M. Carloni~Calame, G.~Montagna, O.~Nicrosini and F.~Piccinini, \emph{{The BABAYAGA event generator}}, \href{http://dx.doi.org/10.1016/j.nuclphysbps.2004.02.008}{\emph{Nucl. Phys. B Proc. Suppl.} {\bf 131} (2004) 48--55}, [\href{https://arxiv.org/abs/hep-ph/0312014}{{\tt hep-ph/0312014}}].

\bibitem{Ciafaloni_1999}
P.~Ciafaloni and D.~Comelli, \emph{Sudakov effects in electroweak corrections}, \href{http://dx.doi.org/10.1016/s0370-2693(98)01541-x}{\emph{Physics Letters B} {\bf 446} (Jan., 1999) 278–284}.

\bibitem{Kulesza:1999us}
A.~Kulesza and W.~J. Stirling, \emph{{Sudakov logarithm resummation for vector boson production at hadron colliders}}, \href{http://dx.doi.org/10.1088/0954-3899/26/5/319}{\emph{J. Phys. G} {\bf 26} (2000) 637--642}, [\href{https://arxiv.org/abs/hep-ph/9912300}{{\tt hep-ph/9912300}}].

\bibitem{MATLAB}
T.~M. Inc., \emph{Matlab version: 9.13.0 (r2024b)},  2024.

\bibitem{AtzoriCorona:2025xwr}
M.~Atzori~Corona, M.~Cadeddu, N.~Cargioli, F.~Dordei, C.~Giunti and C.~A. Ternes, \emph{{The Standard Model tested with neutrinos}},  \href{https://arxiv.org/abs/2504.05272}{{\tt 2504.05272}}.

\bibitem{github}
B.~Madison, \emph{\href{https://github.com/BrendonMadison/ThesisRepo}{ThesisRepo}},  Apr., 2025.
\newblock 10.5281/zenodo.1234.

\bibitem{geddes2022reportsnowmass21accelerator}
C.~Geddes, M.~Hogan, P.~Musumeci and R.~Assmann, \emph{Report of snowmass 21 accelerator frontier topical group 6 on advanced accelerators},  \href{https://arxiv.org/abs/2208.13279}{{\tt 2208.13279}}.

\bibitem{Skands_2004}
P.~Skands, B.~Allanach, H.~Baer, C.~Balazs, G.~Belanger, F.~Boudjema et~al., \emph{Susy les houches accord: Interfacing susy spectrum calculators, decay packages, and event generators}, \href{http://dx.doi.org/10.1088/1126-6708/2004/07/036}{\emph{Journal of High Energy Physics} {\bf 2004} (July, 2004) 036–036}.

\bibitem{CMS-SUS-18-004}
{\scshape CMS} collaboration, A.~Tumasyan, W.~Adam, J.~W. Andrejkovic, T.~Bergauer, S.~Chatterjee, M.~Dragicevic et~al., \emph{Search for supersymmetry in final states with two or three soft leptons and missing transverse momentum in proton-proton collisions at $ \sqrt{s} $ = 13 tev}, \href{http://dx.doi.org/10.1007/jhep04(2022)091}{\emph{Journal of High Energy Physics} {\bf 2022} (Apr., 2022) }.

\bibitem{CMS-SUS-17-004}
{\scshape CMS} collaboration, A.~M. Sirunyan, A.~Tumasyan, W.~Adam, F.~Ambrogi, E.~Asilar, T.~Bergauer et~al., \emph{Combined search for electroweak production of charginos and neutralinos in proton-proton collisions at $ \sqrt{s}=13 $tev}, \href{http://dx.doi.org/10.1007/jhep03(2018)160}{\emph{Journal of High Energy Physics} {\bf 2018} (Mar., 2018) }.

\bibitem{ATLAS:2019lng}
{\scshape ATLAS} collaboration, G.~Aad et~al., \emph{{Searches for electroweak production of supersymmetric particles with compressed mass spectra in $\sqrt{s}=$ 13 TeV $pp$ collisions with the {ATLAS} detector}}, \href{http://dx.doi.org/10.1103/PhysRevD.101.052005}{\emph{Phys. Rev. D} {\bf 101} (2020) 052005}, [\href{https://arxiv.org/abs/1911.12606}{{\tt 1911.12606}}].

\bibitem{CMS-SUS-16-048}
{\scshape CMS} collaboration, A.~Sirunyan, A.~Tumasyan, W.~Adam, F.~Ambrogi, E.~Asilar, T.~Bergauer et~al., \emph{Search for new physics in events with two soft oppositely charged leptons and missing transverse momentum in proton–proton collisions at $\sqrt{s}=13$ tev}, \href{http://dx.doi.org/10.1016/j.physletb.2018.05.062}{\emph{Physics Letters B} {\bf 782} (July, 2018) 440–467}.

\bibitem{ATLAS-SUSY-2016-07}
{\scshape ATLAS} collaboration, M.~Aaboud et~al., \emph{{Search for squarks and gluinos in final states with jets and missing transverse momentum using 36 fb$^{-1}$ of $\sqrt{s}=13$ TeV pp collision data with the ATLAS detector}}, \href{http://dx.doi.org/10.1103/PhysRevD.97.112001}{\emph{Phys. Rev. D} {\bf 97} (2018) 112001}, [\href{https://arxiv.org/abs/1712.02332}{{\tt 1712.02332}}].

\bibitem{CMS:2018Agg}
{\scshape CMS} collaboration, A.~M. Sirunyan, A.~Tumasyan, W.~Adam, F.~Ambrogi, E.~Asilar, T.~Bergauer et~al., \emph{Search for electroweak production of charginos and neutralinos in multilepton final states in proton-proton collisions at $ \sqrt{s}=13 $ tev}, \href{http://dx.doi.org/10.1007/jhep03(2018)166}{\emph{Journal of High Energy Physics} {\bf 2018} (Mar., 2018) }.

\bibitem{CMS-SUS-19-006-agg}
{\scshape CMS} collaboration, A.~M. Sirunyan, A.~Tumasyan, W.~Adam, F.~Ambrogi, T.~Bergauer, J.~Brandstetter et~al., \emph{{Search for supersymmetry in proton-proton collisions at 13 TeV in final states with jets and missing transverse momentum}}, \href{http://dx.doi.org/10.1007/JHEP10(2019)244}{\emph{JHEP} {\bf 10} (2019) 244}, [\href{https://arxiv.org/abs/1908.04722}{{\tt 1908.04722}}].

\bibitem{ATLAS-SUSY-2013-12}
{\scshape ATLAS} collaboration, G.~Aad et~al., \emph{{Search for direct production of charginos and neutralinos in events with three leptons and missing transverse momentum in $\sqrt{s} =$ 8TeV $pp$ collisions with the ATLAS detector}}, \href{http://dx.doi.org/10.1007/JHEP04(2014)169}{\emph{JHEP} {\bf 04} (2014) 169}, [\href{https://arxiv.org/abs/1402.7029}{{\tt 1402.7029}}].

\bibitem{Agnes_2018}
P.~Agnes, I.~Albuquerque, T.~Alexander, A.~Alton, G.~Araujo, D.~Asner et~al., \emph{Low-mass dark matter search with the darkside-50 experiment}, \href{http://dx.doi.org/10.1103/physrevlett.121.081307}{\emph{Physical Review Letters} {\bf 121} (Aug., 2018) }.

\bibitem{Belanger:2020gnr}
G.~Belanger, A.~Mjallal and A.~Pukhov, \emph{{Recasting direct detection limits within micrOMEGAs and implication for non-standard Dark Matter scenarios}}, \href{http://dx.doi.org/10.1140/epjc/s10052-021-09012-z}{\emph{Eur. Phys. J. C} {\bf 81} (2021) 239}, [\href{https://arxiv.org/abs/2003.08621}{{\tt 2003.08621}}].

\bibitem{Abbasi_2009}
{\scshape IceCube} collaboration, R.~Abbasi, Y.~Abdou, M.~Ackermann, J.~Adams, M.~Ahlers, K.~Andeen et~al., \emph{Limits on a muon flux from neutralino annihilations in the sun with the icecube 22-string detector}, \href{http://dx.doi.org/10.1103/physrevlett.102.201302}{\emph{Physical Review Letters} {\bf 102} (May, 2009) }.

\bibitem{B_langer_2015}
G.~Bélanger, J.~D. Silva, T.~Perrillat-Bottonet and A.~Pukhov, \emph{Limits on dark matter proton scattering from neutrino telescopes using micromegas}, \href{http://dx.doi.org/10.1088/1475-7516/2015/12/036}{\emph{Journal of Cosmology and Astroparticle Physics} {\bf 2015} (Dec., 2015) 036–036}.

\bibitem{ATLAS:2021moa}
{\scshape ATLAS} collaboration, G.~Aad et~al., \emph{{Search for chargino\textendash{}neutralino pair production in final states with three leptons and missing transverse momentum in $\sqrt{s} = 13$~TeV pp collisions with the ATLAS detector}}, \href{http://dx.doi.org/10.1140/epjc/s10052-021-09749-7}{\emph{Eur. Phys. J. C} {\bf 81} (2021) 1118}, [\href{https://arxiv.org/abs/2106.01676}{{\tt 2106.01676}}].

\bibitem{ATLAS:2022zwa}
{\scshape ATLAS} collaboration, G.~Aad et~al., \emph{{Searches for new phenomena in events with two leptons, jets, and missing transverse momentum in 139~fb$^{-1}$ of $\sqrt{s}=13$~TeV $pp$ collisions with the ATLAS detector}}, \href{http://dx.doi.org/10.1140/epjc/s10052-023-11434-w}{\emph{Eur. Phys. J. C} {\bf 83} (2023) 515}, [\href{https://arxiv.org/abs/2204.13072}{{\tt 2204.13072}}].

\bibitem{ATLAS:2014jxt}
{\scshape ATLAS} collaboration, G.~Aad et~al., \emph{{Search for squarks and gluinos with the ATLAS detector in final states with jets and missing transverse momentum using $\sqrt{s}=8$ TeV proton--proton collision data}}, \href{http://dx.doi.org/10.1007/JHEP09(2014)176}{\emph{JHEP} {\bf 09} (2014) 176}, [\href{https://arxiv.org/abs/1405.7875}{{\tt 1405.7875}}].

\bibitem{ATLAS:2014zve}
{\scshape ATLAS} collaboration, G.~Aad et~al., \emph{{Search for direct production of charginos, neutralinos and sleptons in final states with two leptons and missing transverse momentum in $pp$ collisions at $\sqrt{s} =$ 8 TeV with the ATLAS detector}}, \href{http://dx.doi.org/10.1007/JHEP05(2014)071}{\emph{JHEP} {\bf 05} (2014) 071}, [\href{https://arxiv.org/abs/1403.5294}{{\tt 1403.5294}}].

\bibitem{GEANT4}
S.~Agostinelli et~al., \emph{Geant4—a simulation toolkit}, \href{http://dx.doi.org/https://doi.org/10.1016/S0168-9002(03)01368-8}{\emph{Nuclear Instruments and Methods in Physics Research Section A: Accelerators, Spectrometers, Detectors and Associated Equipment} {\bf 506} (2003) 250--303}.

\end{thebibliography}\endgroup

\end{document}